\RequirePackage{fix-cm}
\documentclass[twocolumn,epjc3]{svjour3}  
\usepackage[T1]{fontenc}
\smartqed  
\RequirePackage{graphicx}
\RequirePackage{subfiles,currfile,cite,xspace,tikz,ragged2e}
\RequirePackage{cancel}
\RequirePackage{placeins}
\RequirePackage{booktabs}
\RequirePackage{slashbox}
\RequirePackage{amsmath,amssymb}
\RequirePackage{rotating}
\RequirePackage{tabulary}
\RequirePackage[colorlinks, urlcolor=blue, linkcolor=blue, citecolor=blue, plainpages=false]{hyperref}
\PassOptionsToPackage{unicode}{hyperref}
\PassOptionsToPackage{naturalnames}{hyperref}
\RequirePackage{float}
\floatstyle{plaintop}
\restylefloat{table}
\RequirePackage{xargs}
\RequirePackage{xstring}
\RequirePackage{lineno}
\RequirePackage{siunitx}
\RequirePackage{multirow,colortbl}
\RequirePackage{float}
\newcommand{\mc}{\textrm{MuC}}
\newcommand{\lrD}{~\!\overset{\leftrightarrow}{\hspace{-0.1cm}D}\!}
\newcommand{\lrDa}{~\!\overset{\leftrightarrow}{\hspace{-0.1cm}D}\!^{\!~a}}
\newcommand{\hc}{\mathrm{h.c.}}
\journalname{Eur. Phys. J. C}
\begin{document}










\onecolumn
\noindent
{\bf\LARGE Towards a Muon Collider \\}
\newline

\noindent
\begin{flushleft}
{\bf 
Carlotta~Accettura$^{1}$, Dean~Adams$^{2}$, Rohit~Agarwal$^{3}$, Claudia~Ahdida$^{1}$, Chiara~Aim{\`e}$^{4,5}$, Nicola~Amapane$^{6,7}$, David~Amorim$^{1}$, Paolo~Andreetto$^{8}$, Fabio~Anulli$^{9}$, Robert~Appleby$^{10}$, Artur~Apresyan$^{11}$
% ...
}
\end{flushleft}
\vspace*{2mm}  

\begin{flushleft}
{\em\small
$^{1}$ Organisation Européenne pour la Recherche Nucléaire (CERN), CH-1211 Genève 23, Geneva, Switzerland \\ 
$^{2}$ STFC Rutherford Appleton Laboratory (RAL), Harwell Oxford, United Kingdom \\ 
$^{3}$ Computer Science Department, University of California (UC), Berkeley, CA 94720-1776, United States \\ 
$^{4}$ Dipartimento di Fisica, Universit{\`a} di Pavia, Corso Strada Nuova 65, I-27100 Pavia, Italy \\ 
$^{5}$ INFN Sezione di Pavia, via Bassi 6, I-27100 Pavia, Italy \\ 
$^{6}$ INFN Sezione di Torino,  via Giuria 1, I-10125 Torino, Italy \\ 
$^{7}$ Dipartimento di Fisica, Universit{\`a} di Torino, Via Giuria 1, I-10125 Torino, Italy \\ 
$^{8}$ INFN Sezione di Padova, Via Marzolo 8, I-35131 Padova, Italy \\ 
$^{9}$ INFN Sezione di Roma, Piazzale Aldo Moro, 2, I-00185 Roma, Italy \\ 
$^{10}$ School of Physics and Astronomy, University of Manchester, Oxford Road, Manchester M13 9PL, United Kingdom \\ 
$^{11}$ Fermi National Accelerator Laboratory, Batavia, IL 60510,  United States \\ 
% ...
}
\end{flushleft}


\vspace{20pt}

\begin{abstract}
A muon collider would enable the big jump ahead in energy reach that is needed for a fruitful exploration of fundamental interactions. The challenges of producing muon collisions at high luminosity and 10~TeV centre of mass energy are being investigated by the recently-formed International Muon Collider Collaboration. This Review summarises the status and the recent advances on muon colliders design, physics and detector studies. The aim is to provide a global perspective of the field and to outline directions for future work.
\end{abstract}

\setcounter{tocdepth}{3}
\tableofcontents

\twocolumn

\clearpage

\section{Introduction}\label{IntroSect}

Colliders are microscopes that explore the structure and the interactions of particles at the shortest possible length scale. Their goal is not to chase discoveries that are inevitable or perceived as such based on current knowledge. On the contrary, their mission is to explore the unknown in order to acquire radically novel knowledge.

The current experimental and theoretical situation of particle physics is particularly favourable to collider exploration. No inevitable discovery diverts our attention from pure exploration, and we can focus on the basic questions that best illustrate our ignorance. Why is electroweak symmetry broken and what sets the scale? Is it really broken by the Standard Model Higgs or by a more rich Higgs sector? Is the Higgs an elementary or a composite particle? Is the top quark, in light of its large Yukawa coupling, a portal towards the explanation of the observed pattern of flavor? Is the Higgs or the electroweak sector connected with dark matter? Is it connected with the origin of the asymmetry between baryons and anti-baryons in the Universe?

The next collider should deepen our understanding of the questions above, and offer broad and varied opportunities for exploration to enable radically unexpected discoveries. A comprehensive exploration must exploit the complementarity between energy and precision. Precise measurements allow us to study the dynamics of the particles we already know, looking for the indirect manifestation of yet unknown new physics. With a very high energy collider we can access the new physics particles directly. These two exploration strategies are normally associated with two distinct machines, either colliding electrons/positrons ($ee$) or protons ($pp$). 

With muons instead, both strategies can be effectively pursued at a single collider that combines the advantages of $ee$ and of $pp$ machines. Moreover, the simultaneous availability of energy and precision offers unique perspectives of indirect sensitivity to very heavy new physics, as well as unique perspectives for the characterisation of new heavy particles discovered at the muon collider itself. 

This is the picture that emerges from the investigations of the muon colliders physics potential performed so far, to be reviewed in this document in Sections~\ref{ch2_phys_opp} and~\ref{sec:phys_studies}. These studies identify a Muon Collider (MuC), with 10~TeV energy or more in the centre of mass and sufficient luminosity, as an ideal tool for a substantial ambitious jump ahead in the exploration of fundamental particles and interactions. Assessing its technological feasibility is thus a priority for the future of particle physics.

\paragraph{Muon collider concept}
\

\begin{figure}
\centerline{\includegraphics[width=8cm]{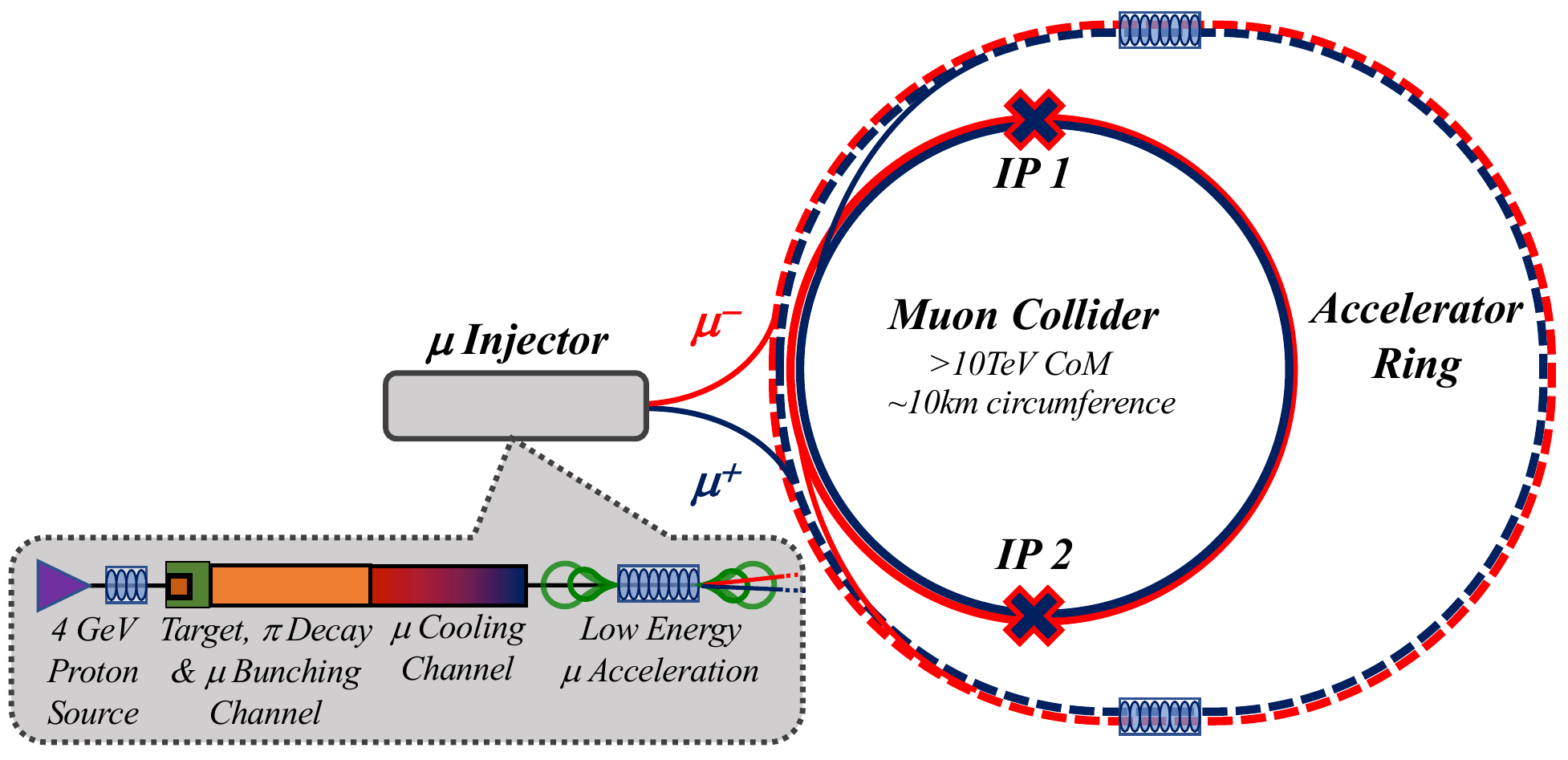}}
\caption{A conceptual scheme of the muon collider.}
\label{f:design}
\end{figure}

Initial ideas for muon colliders were proposed long ago~\cite{Tikhonin:initial,Budker:initial,Neuffer:1979gq,Cline:1980sa,Skrinsky:1981ht,Neuffer:1983jr}. Subsequent studies culminated in the Muon Accelerator Program (MAP) in the US (see \cite{Ankenbrandt:1999cta,Palmer:2014nza,Boscolo:2018ytm,Neuffer:2017hnu} and \cite{Delahaye:2019omf,Long:2020wfp} for an overview). The MAP concept for the muon collider facility is displayed in Figure~\ref{f:design}. The proton complex produces a short, high-intensity proton pulse that hits the target and produces pions. The decay channel guides the pions and collects the muons from their decay into a bunching and phase rotator system to form a muon beam. Several cooling stages then reduce the longitudinal and transverse emittance of the beam using a sequence of absorbers and radiofrequency (RF) cavities in a high magnetic field. A system of a linear accelerators (linac) and two recirculating linacs accelerate the beams to 60~GeV. They are followed by one or more rings to accelerate them to higher energy, for instance one to 300~GeV and one to 1.5~TeV, in the case of a 3~TeV centre of mass energy MuC. In the 10~TeV collider an additional ring from 1.5 to 5~TeV follows. These rings can be either fast-pulsed synchrotrons or Fixed-Field Alternating gradients (FFA) accelerators. Finally, the beams are injected at full energy into the collider ring. Here, they will circulate to produce luminosity until they are decayed. Alternatively they can be extracted once the muon beam current is strongly reduced by decay. There are wide margins for the optimisation of the exact energy stages of the acceleration system, taking also into account the possible exploitation of the intermediate-energy muon beams for muon colliders of lower centre of mass energy. 

The concept developed by MAP provides the baseline for present and planned work on muon colliders, reviewed in Section~\ref{sec_fac}. Three main reasons sparked this renewed interest in muon colliders. First, the focus on high collision energy and luminosity where the muon collider is particularly promising and offers the perspective of revolutionising particle physics. Second, the advances in technology and muon colliders design. Third, the difficulty of envisaging a radical jump ahead in the high-energy exploration with $ee$ or $pp$ colliders.

In fact, the required increase of energy and luminosity in future high-energy frontier colliders poses severe challenges~\cite{Shiltsev:2019rfl,Roser:2022sht}. Without breakthroughs in concept and in technologies, the cost and use of land as well as the power consumption are prone to increase to unsustainable levels.

The muon collider promises to overcome these limitations and allow to push the energy frontier strongly. Circular electron-positron colliders are limited in energy by the emission of synchrotron radiation that increases strongly with energy. Linear colliders overcome this limitation but require the beam to be accelerated to full energy in a single pass through the main linac and allow to use the beams only in a single collision~\cite{Stapnes:2019dcu}. The high mass of the muon mitigates synchrotron radiation emission, allowing them to be accelerated in many passes through a ring and to collide repeatedly in another ring. This results in cost effectiveness and compactness combined with a luminosity per beam power that roughly increases linearly with energy. Protons can be also accelerated in rings and made to collide with very high energy. However, protons are composite particles and therefore only a small fraction of their collision energy is available to probe short-distance physics through the collisions of their fundamental constituents. The effective energy reach of a muon collider thus corresponds to the one of a proton collider of much higher centre of mass energy. This concept is illustrated more quantitatively in Section~\ref{whym}.

Currently, the limit of the energy reach for muon colliders has not been identified. Ongoing studies focus on a 10~TeV design with an integrated luminosity goal of $10 ~ \rm ab^{-1}$. This goal is expected to provide a good balance between an excellent physics case and affordable cost, power consumption and risk. Once a robust design has been established at 10~TeV---including an estimate of the cost, power consumption and technical risk---other, higher energies will be explored.

The 2020 Update of the European Strategy for Particle Physics (ESPPU) recommended, for the first time in Europe, an R{\&}D programme on muon colliders design and technology. This led to the formation of the International Muon Collider Collaboration (IMCC)~\cite{IMCC} with the goal of initiating this programme and informing the next ESPPU process on the muon collider feasibility perspectives. This will enable the next ESPPU and other strategy processes to judge the scientific justification of a full Conceptual Design Report (CDR) and demonstration programme.

The European Roadmap for Accelerator R\&D~\cite{Adolphsen:2022ibf}, published in 2021, includes the muon collider. The report is based on consultations of the community at large, combined with the expertise of a dedicated Muon Beams Panel. It also benefited from significant input from the MAP design studies and US experts. The report assessed the challenges of the muon collider and did not identify any insurmountable obstacle. However, the muon collider technologies and concepts are less mature than those of electron-positron colliders. Circular and linear electron-positron colliders already have been constructed and operated but the muon collider would be the first of its kind. The limited muon lifetime gives rise to several specific challenges including the need of rapid production and acceleration of the beam. These challenges and the solutions under investigation are detailed in Section~\ref{sec_fac}.

The Roadmap describes the R\&D programme required to develop the maturity of the key technologies and the concepts in the coming few years. This will allow the assessment of realistic luminosity targets, detector backgrounds, power consumption and cost scale, as well as whether one can consider implementing a MuC at CERN or elsewhere. Mitigation strategies for the key technical risks and a demonstration programme for the CDR phase will also be addressed. The use of existing infrastructure, such as existing proton facilities and the LHC tunnel, will also be considered. This will allow the next strategy process to make an informed choice on the future strategy. Based on the conclusions of the next strategy processes in the different regions, a CDR phase could then develop the technologies and the baseline design to demonstrate that the community can execute a successful MuC project.

Important progress in the past gives confidence that this goal can be achieved and that the programme will be successful. In particular, the developments of superconducting magnet technology has progressed enormously and high-temperature superconductors have become a practical technology used in industry. Similarly, RF technology has progressed in general and experiments demonstrated the solution of the specific muon collider challenge---operating RF cavities in very high magnetic fields---that previously has been considered a showstopper. Component designs have been developed that can cool the initially diffuse beam and accelerate it to multi-TeV energy on a time scale compatible with the muon lifetime. However, a fully integrated design has yet to be developed and further development and demonstration of technology are required. 

The technological feasibility of the facility is one vital component of the muon collider programme, but the planning of the facility exploitation is equally important. This includes the assessment of the muon collider potential to address physics questions, as well as the design of novel detectors and reconstruction techniques to perform experiments with colliding muons.

\paragraph{The path to a new generation of experiments}
\

The main challenge to operating a detector at a muon collider is the fact that muons are unstable particles. As such, it is impossible to study the muon interactions without being exposed to decays of the muons forming the colliding beams. From the moment the collider is turned on and the muon bunches start to circulate in the accelerator complex, the products of the in-flight decays of the muon beams and the results of their interactions with beam line material, or the detectors themselves, will reach the experiments contributing to polluting the otherwise clean collision environment. The ensemble of all these particles is usually known as ``Beam Induced Backgrounds'', or BIB. The composition, flux, and energy spectra of the BIB entering a detector is closely intertwined with the design of the experimental apparatus, such as the beam optics that integrate the detectors in the accelerator complex or the presence of shielding elements, and the collision energy. However, two general features broadly characterise the BIB: it is composed of low-energy particles with a broad arrival time in the detector.

The design of an optimised muon collider detector is still in its infancy, but the work has initiated and it is reviewed in Section~\ref{sec:detectorandreconstruction}. It is already clear that the physics goals will require a fully hermetic detector able to resolve the trajectories of the outgoing particles and their energies. While the final design might look similar to those experiments taking data at the LHC, the technologies at the heart of the detector will have to be new. The large flux of BIB particles sets requirements on the need to withstand radiation over long periods of time, and the need to disentangle the products of the beam collisions from the particles entering the sensitive regions from uncommon directions calls for high-granularity measurements in space, time and energy. The development of these new detectors will profit from the consolidation of the successful solutions that were pioneered for example in the High Luminosity LHC upgrades, as well as brand new ideas. New solutions are being developed for use in the muon collider environment spanning from tracking detectors, calorimeters systems to dedicated muon systems. The whole effort is part of the push for the next generation of high-energy physics detectors, and new concepts targeted to the muon collider environment might end up revolutionising other future proposed collider facilities as well.

Together with a vibrant detector development program, new techniques and ideas needs to be developed in the interpretation of the energy depositions recorded by the instrumentation. The contributions from the BIB add an incoherent source of backgrounds that affect different detector systems in different ways and that are unprecedented at other collider facilities. The extreme multiplicity of energy depositions in the tracking detectors create a complex combinatorial problem that challenges the traditional algorithms for reconstructing the trajectories of the charged particles, as these were designed for collisions where sprays of particles propagate outwards from the centre of the detector. At the same time, the potentially groundbreaking reach into the high-energy frontier will lead to strongly collimated jets of particles that need to be resolved by the calorimeter systems, while being able to subtract with precision the background contributions. The challenging environment of the muon collider offers fertile ground for the development of new techniques, from traditional algorithms to applications of artificial intelligence and machine learning, to brand new computing technologies such as quantum computers.

\paragraph{Muon collider plans}
\

The ongoing reassessment of the muon collider design and the plans for R\&{D} allow us to envisage a possible path towards the realisation of the muon collider and a tentative technically-limited timeline, displayed in Figure~\ref{fig_muon:RDtimeline}.

The goal~\cite{Delahaye:2019omf,Long:2020wfp} is a muon collider with a centre of
mass energy of 10~TeV or more (a \mbox{$10^+$\,TeV}~MuC). Passing this energy threshold enables, among other things, a vast jump ahead in the search for new heavy particles relative to the LHC. The target integrated luminosity is obtained by considering the cross-section of a typical $2\to2$ scattering processes mediated by the electroweak interactions, $\sigma\sim 1~{\rm{fb}}\cdot(10~{\rm{TeV}})^2/E_{\rm{cm}}^2$. In order to measure such cross-sections with good (percent-level) precision and to exploit them as powerful probes of short distance physics, around ten thousand events are needed. The corresponding integrated luminosity is
\begin{equation}\label{lums}
\displaystyle
\mathfrak{L}_{\rm{int}}=10\,{\rm{ab}}^{-1}\left(\frac{E_{\rm{cm}}}{10\,{\rm{TeV}}}\right)^2\,.
\end{equation}
The luminosity requirement grows quadratically with the energy in order to compensate for the cross-section decrease. We will see in Section~\ref{sec_fac} that achieving this scaling is indeed possible at muon colliders.

Assuming a muon collider operation time of $10^7$ seconds per year, and one interaction point, eq.~(\ref{lums}) corresponds to an  instantaneous luminosity 
\begin{equation}\label{lumsin}
\displaystyle
\mathfrak{L}=\frac{5~{\rm{years}}}{\rm{time}}
\left(\frac{E_{\rm{cm}}}{10\,{\rm{TeV}}}\right)^22\cdot 10^{35}{\rm{cm}}^{-2}{\rm{s}}^{-1}\,.
\end{equation}
The current design target parameters (see Table~\ref{MC:t:param}) enable to collect the required integrated luminosity in a 5-year run, ensuring an appealingly compact temporal extension to the muon collider project even in its data taking phase. Furthermore this ambitious target leaves space to increase the integrated luminosity by running longer or by foreseeing a second interaction point. One could similarly compensate for a possible instantaneous luminosity reduction in the final design.

\paragraph{Muon collider stages}
\

In order to design the path towards a \mbox{$10^+$\,TeV} MuC, one could exploit the possibility of building it in stages. In fact, the design of many elements of the facility is simply independent of the collider energy. Once built and exploited for a lower $E_{\rm{cm}}$ MuC, they can thus be reused for a higher energy collider. This applies to the muon source and cooling, and to the accelerator complex as well because energy staging is anyway required for fast acceleration. Only the final collision ring of the lower $E_{\rm{cm}}$ collider could not be reused. However because of its limited size it is a minor addition to the total cost.

A staged approach has several advantages. First, it spreads the total cost over a longer time period and reduces the initial investment. This could enable a faster financing for the first energy stage and accelerate the whole project. Furthermore the reduced energy of the first stage allows, if needed, to make compromises on technologies that might not yet be fully developed, avoiding potential delays. In particular completing construction in 2045 as foreseen in Figure~\ref{fig_muon:RDtimeline} could be optimistic for a \mbox{$10^+$\,TeV} MuC, but realistic for a first lower-energy collider at few TeV. A centre of mass energy $E_{\rm{cm}}=3$~TeV is being tentatively considered for the first stage. This matches, with a much more compact design, the maximal $e^+e^-$ energy that could be achieved by the last stage of the CLIC linear collider~\cite{Stapnes:2019dcu}.

The 3~TeV stage of the muon collider offers amazing opportunities for progress with respect to the LHC and its high-luminosity successor (HL-LHC)~\cite{ZurbanoFernandez:2020cco}. These opportunities include a determination of the Higgs trilinear coupling, extended sensitivity to Higgs and top quark compositeness and to extended Higgs sectors. A selected summary of available studies is reported in Section~\ref{sec:phys_studies}. On the other hand, the physics potential of the \mbox{$10^+$\,TeV} collider is much superior to the one of the 3~TeV collider. The higher energy stage will radically advance the knowledge acquired with the first stage operation. Additionally, the energy upgrade would enable to investigate new physics discoveries or tensions with the SM that might emerge at the first stage.


The 3~TeV stage, following eq.~(\ref{lumsin}), would collect $0.9\simeq 1$~ab$^{-1}$ luminosity (with one detector) in five years of full luminosity, after an initial ramp-up phase of two to three years. In the most optimistic scenario the construction of the first stage will proceed rapidly. The first stage will terminate after seven years to leave space to the second stage with radically improved physics performances. If the second stage is instead delayed, the one at 3~TeV could run longer. The optimistic and pessimistic scenarios thus foresee 1 and  2~ab$^{-1}$ at 3~TeV, respectively.

\paragraph{Other muon colliders}
\

The tentative staging scenario detailed above should serve as the baseline for future investigations of alternative plans. In particular, one could consider the possibility of a first stage of much lower energy than 3~TeV, to be possibly built on an even shorter time scale. However, it is worth remarking in this context that the quadratic luminosity scaling with energy in eqs.~(\ref{lums}) and (\ref{lumsin}) is not only the aspirational target, but it is also the natural scaling of the luminosity at muon colliders. By following the scaling for low $E_{\rm{cm}}$ one immediately realises that the luminosity of a muon collider at order 100~GeV energy can not be competitive with the one of an $e^+e^-$ circular or linear collider. For instance this implies that while there is evidently a compelling physics case for a leptonic ``Higgs factory'' at around 250~GeV energy, a muon collider would be probably unable to collect the high luminosity needed for a successful program of Higgs coupling measurements, while this is possible for $e^+e^-$ colliders. In general, the luminosity scaling suggests that a physics-motivated first stage of the muon collider should either exploit some peculiarity of the muons that make $\mu^+\mu^-$ collisions more useful than $e^+e^-$ collisions, or target the TeV energy that is hard to reach with $e^+e^-$.

The possibility of operating a first muon collider at the Higgs pole $E_{\rm{cm}}=m_H=125$~GeV has been discussed extensively in the literature. The idea here is to exploit the large Yukawa coupling of the muon, much larger than the one of the electron, in order to produce the Higgs boson in the $s$-channel and study its lineshape. The physics potential of such Higgs-pole muon collider will be described in Section~\ref{sec:phys_studies}. The major results would be a rather precise and direct determination of the Higgs width and an astonishingly accurate measurement of the Higgs mass. However, the Higgs is a rather narrow particle, with a width over mass ratio $\Gamma_H/m_H$ as small as $3\cdot10^{-5}$. The muon beams would thus need a comparably small energy spread $\Delta E/E\hspace{-2pt}=\hspace{-2pt}3\cdot10^{-5}$ for the programme to succeed. Engineering such tiny energy spread might perhaps be possible. However it poses a challenge for the facility design that is peculiar to the Higgs-pole collider and of no relevance for higher energies, where a much higher spread is perfectly adequate for physics. For this reason, the Higgs-pole muon collider is not currently considered in the staging plan and further study is needed. 

Further work is also needed to assess the possible relevance of a muon collider at the $t{\overline{t}}$ production threshold $E_{\rm{cm}}\simeq 343$~GeV, aimed at measuring the top quark mass with precision. The top threshold could be reached also with $e^+e^-$ colliders. However the $e^+e^-$ Higgs factory at 250~GeV, to be possibly built before the muon collider, might not be easily and quickly upgradable to 343~GeV. Moreover, the naturally small (permille-level) beam energy spread and the reduction of initial state radiation effects give an advantage to muons over electrons for the threshold scan. Such ``first muon collider'' was proposed long ago~\cite{Berger:1997xu,Barger:1997yk}. Its modern relevance stems from the need of an improved top mass determination for establishing the possible instability of the SM Higgs potential~\cite{Franceschini:2022veh}. We will not discuss this option further and we refer the reader to the literature.

\paragraph{This Review}
\

A muon collider could be a sustainable innovative device for a big ambitious jump ahead in fundamental physics exploration. It is a long-term project, but with a tight schedule and a narrow temporal window of opportunity. The initiated work must continue in the next decade, fostered by a positive recommendation of the 2023 US Particle Physics Prioritization Panel (P5) and the next Update of the European Strategy for Particle Physics foreseen in 2026/2027. Progress must be made by then on the perspectives for a muon collider to be built and operated, for the outcome of its collisions to be recorded, interpreted and exploited to advance physics knowledge. This offers stimulating challenges for accelerator, experimental and theoretical physics. 

Muon colliders require innovative research in each of these three directions. The novelty of the theme and the lack of established solutions enable a high rate of progress, but it also requires that the three directions advance simultaneously because progress in one motivates and supports work in the others. Furthermore, exploiting synergies between accelerator, experimental and theoretical physics is of utmost importance at this initial stage of the muon collider project design.

In this spirit, the present Review summarises the state of the art and the recent progress in all these three areas, and outlines directions for future work. The aim is to provide a global perspective on muon colliders.

This Review is organised as follows. Section~\ref{ch2_phys_opp} summarises the key exploration opportunities offered by very high energy muon colliders and illustrates the potential for progress on selected physics questions. We also outline the challenges for the theoretical predictions needed to exploit this potential. These challenges constitute in fact a tremendous opportunity to advance knowledge of SM physics in a regime where the electroweak bosons are relatively light particles, entailing the emergence of the novel phenomenon of electroweak radiation. Section~\ref{sec_fac} describes the challenges and the opportunities of muon colliders for accelerator physics. It reviews the basic principles for the design of the muon production, cooling and fast acceleration systems. The required R\&{D}, and a tentative staging plan and timeline, are also outlined. Section~\ref{sec:detectorandreconstruction} describes the experimental conditions that are expected at the muon collider and the ongoing work on the design of the detector and of the event reconstruction software. We devote Section~\ref{sec:phys_studies} to selected muon collider sensitivity projection studies. The \mbox{$10^+$\,TeV} MuC is the main focus, but some opportunities of the 3~TeV stage are also described, as well as those of the Higgs-pole collider. A summary of the perspectives and opportunities for future work on muon colliders is reported in Section~\ref{ConclSect}.

\clearpage

\section{Physics opportunities}\label{ch2_phys_opp}

\subsection{Why muons?}\label{whym}

Muons, like protons, can be made to collide with a centre of mass energy of $10$~TeV or more in a relatively compact ring, without fundamental limitations from synchrotron radiation. However, being point-like particles, unlike protons, their nominal centre of mass collision energy $E_{\rm{cm}}$ is entirely available to produce high-energy reactions that probe length scales as short as $1/E_{\rm{cm}}$. The relevant energy for proton colliders is instead the centre of mass energy of the collisions between the partons that constitute the protons. The partonic collision energy is distributed statistically, and approaches a significant fraction of the proton collider nominal energy with very low probability. A muon collider with a given nominal energy and luminosity is thus evidently way more effective than a proton collider with comparable energy and luminosity.

\begin{figure*}[t]
\begin{minipage}{0.48\textwidth}
\begin{center}
\includegraphics[width=\textwidth]{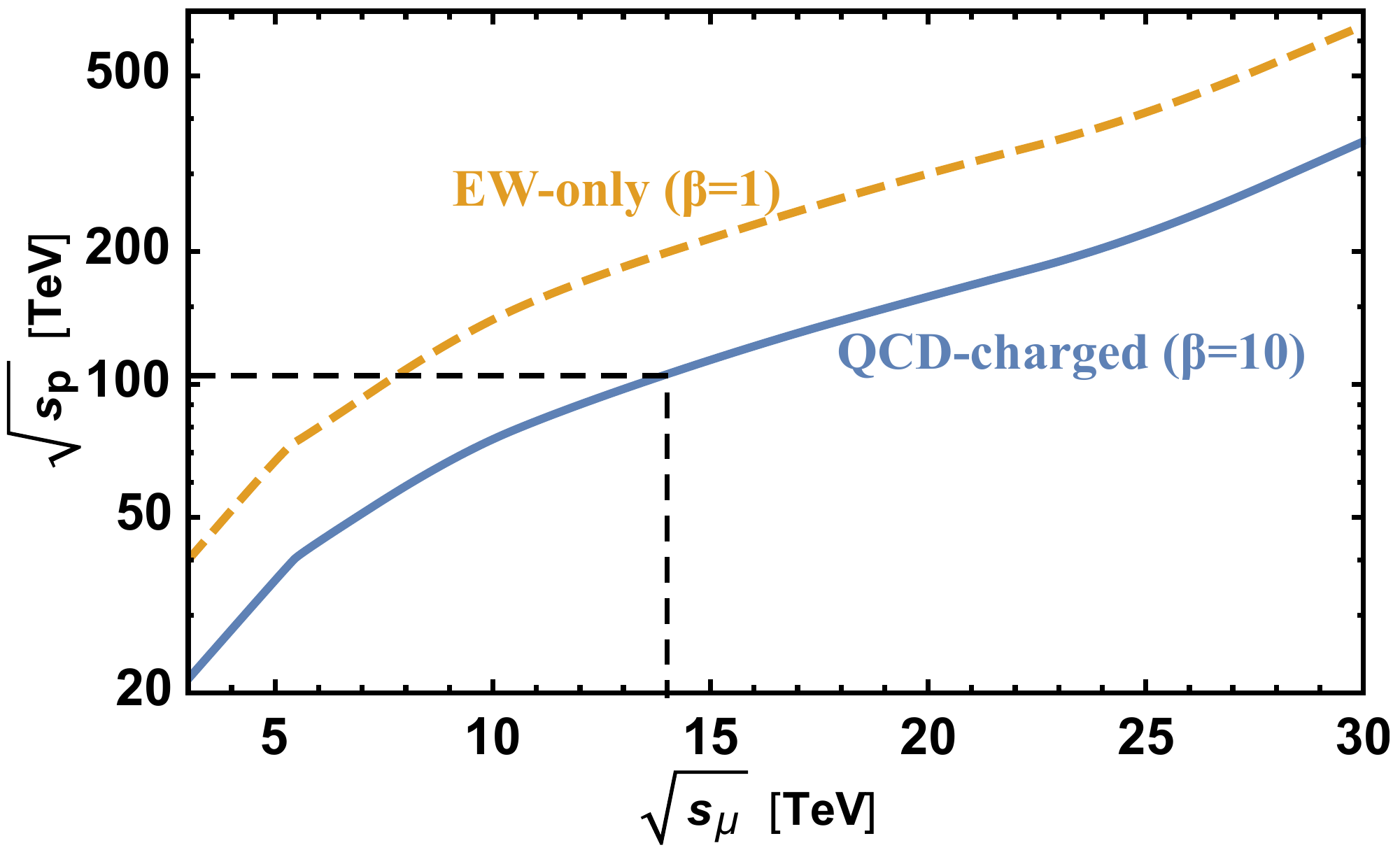}
\end{center}
\end{minipage}
\hfill
\begin{minipage}{0.47\textwidth}
\begin{center}
\includegraphics[width=\textwidth]{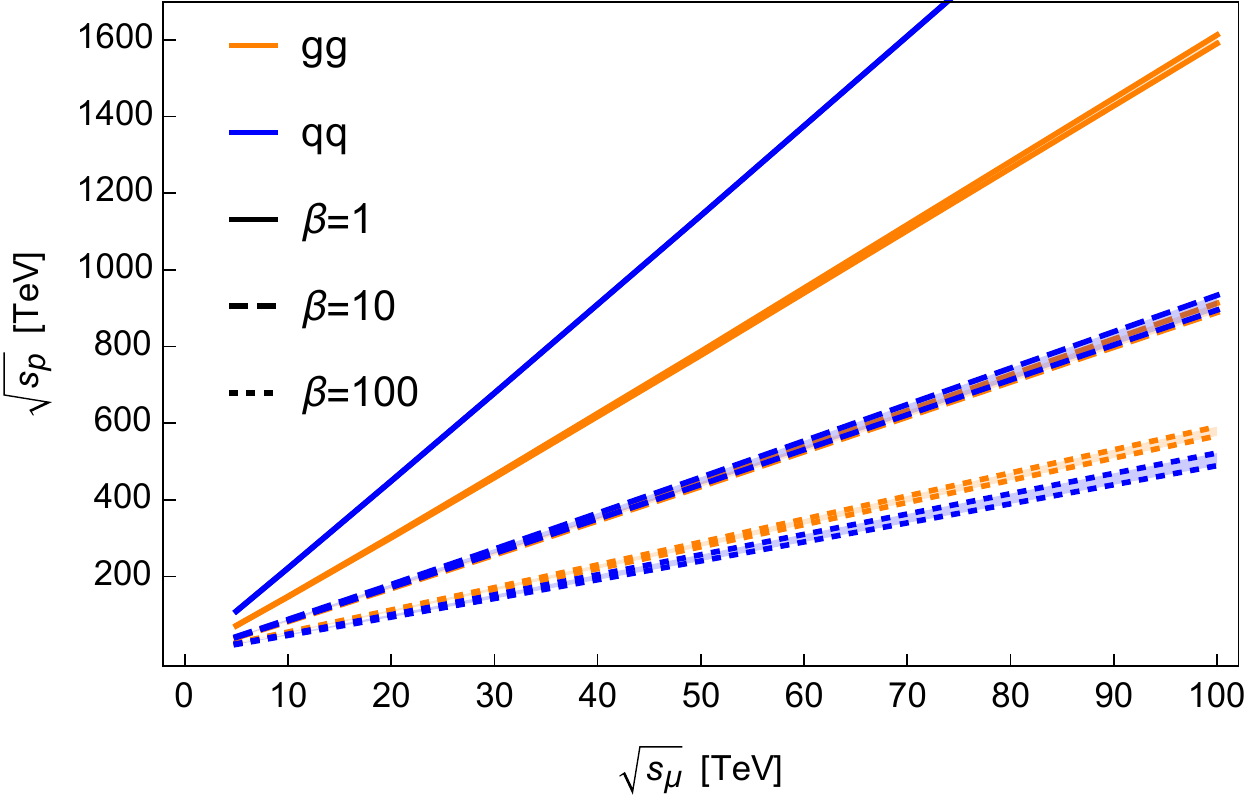}
\end{center}
\end{minipage}
\caption{Equivalent proton collider energy. The left plot~\cite{Delahaye:2019omf}, assumes that $qq$ and~$gg$~partonic initial states both contribute to the production. In the right panel~\cite{AlAli:2021let}, production from $qq$ and from $gg$ are considered separately.
\label{pvsm}}
\end{figure*}

This concept is made quantitative in Figure~\ref{pvsm}. The figure displays the center of mass energy \parbox[c][0pt]{19pt}{${\sqrt{s\,}}_{\hspace{-2pt}p}$} that a proton collider must possess to be ``equivalent'' to a muon collider of a given energy \parbox[c][0pt]{54pt}{$E_{\rm{cm}}=\sqrt{s\,}_{\hspace{-2pt}\mu}$}. Equivalence is defined~\cite{Delahaye:2019omf,Costantini:2020stv,AlAli:2021let} in terms of the pair production cross-section for heavy particles, with mass close to the muon collider kinematical threshold of \parbox[c][0pt]{32pt}{$\sqrt{s\,}_{\hspace{-2pt}\mu}/2$}. The equivalent \parbox[c][0pt]{19pt}{${\sqrt{s\,}}_{\hspace{-2pt}p}$} is the proton collider centre of mass energy for which the cross-sections at the two colliders are equal. 

The estimate of the equivalent \parbox[c][0pt]{19pt}{${\sqrt{s\,}}_{\hspace{-2pt}p}$} depends on the relative strength $\beta$ of the heavy particle interaction with the partons and with the muons. If the heavy particle only possesses electroweak quantum numbers, $\beta=1$ is a reasonable estimate because the particles are produced by the same interaction at the two colliders. If instead it also carries QCD colour, the proton collider can exploit the QCD interaction to produce the particle, and a ratio of $\beta=10$ should be considered owing to the large QCD coupling and colour factors. The orange line on the left panel of Figure~\ref{pvsm}, obtained with $\beta=1$, is thus representative of purely electroweak particles. The blue line, with $\beta=10$, is instead a valid estimate for particles that also possess QCD interactions, as it can be verified in concrete examples.

The general lesson we learn from the left panel of Figure~\ref{pvsm} (orange line) is that at a proton collider with around $100$~TeV energy the cross-section for processes with an energy threshold of around $10$~TeV is quite smaller than the one of a muon collider (MuC) operating at \parbox[c][0pt]{54pt}{$E_{\rm{cm}}={\sqrt{s\,}}_{\hspace{-2pt}\mu}$} $\sim10$~TeV. The gap can be compensated only if the process dynamics is different and more favourable at the proton collider, like in the case of QCD production. The general lesson has been illustrated for new heavy particles production, where the threshold is provided by the particle mass. But it also holds for the production of light SM particles with energies as high as $E_{\rm{cm}}$, which are very sensitive indirect probes of new physics. This makes exploration by high energy measurements more effective at muon than at proton colliders, as we will see in Section~\ref{HEM}. Moreover the large luminosity for high energy muon collisions produces the copious emission of effective vector bosons. In turn, they are responsible at once for the tremendous direct sensitivity of muon colliders to ``Higgs portal'' type new physics and for their excellent perspectives to measure single and double Higgs couplings precisely as we will see in Section~\ref{dirr} and~\ref{VBF}, respectively.

On the other hand, no quantitative conclusion can be drawn from Figure~\ref{pvsm} on the comparison between the muon and proton colliders discovery reach for the heavy particles. That assessment will be performed in the following section based on available proton colliders projections.

\subsection{Direct reach}\label{dirr}

\begin{figure*}
\begin{minipage}{0.52\textwidth}
\begin{center}
\includegraphics[width=\textwidth]{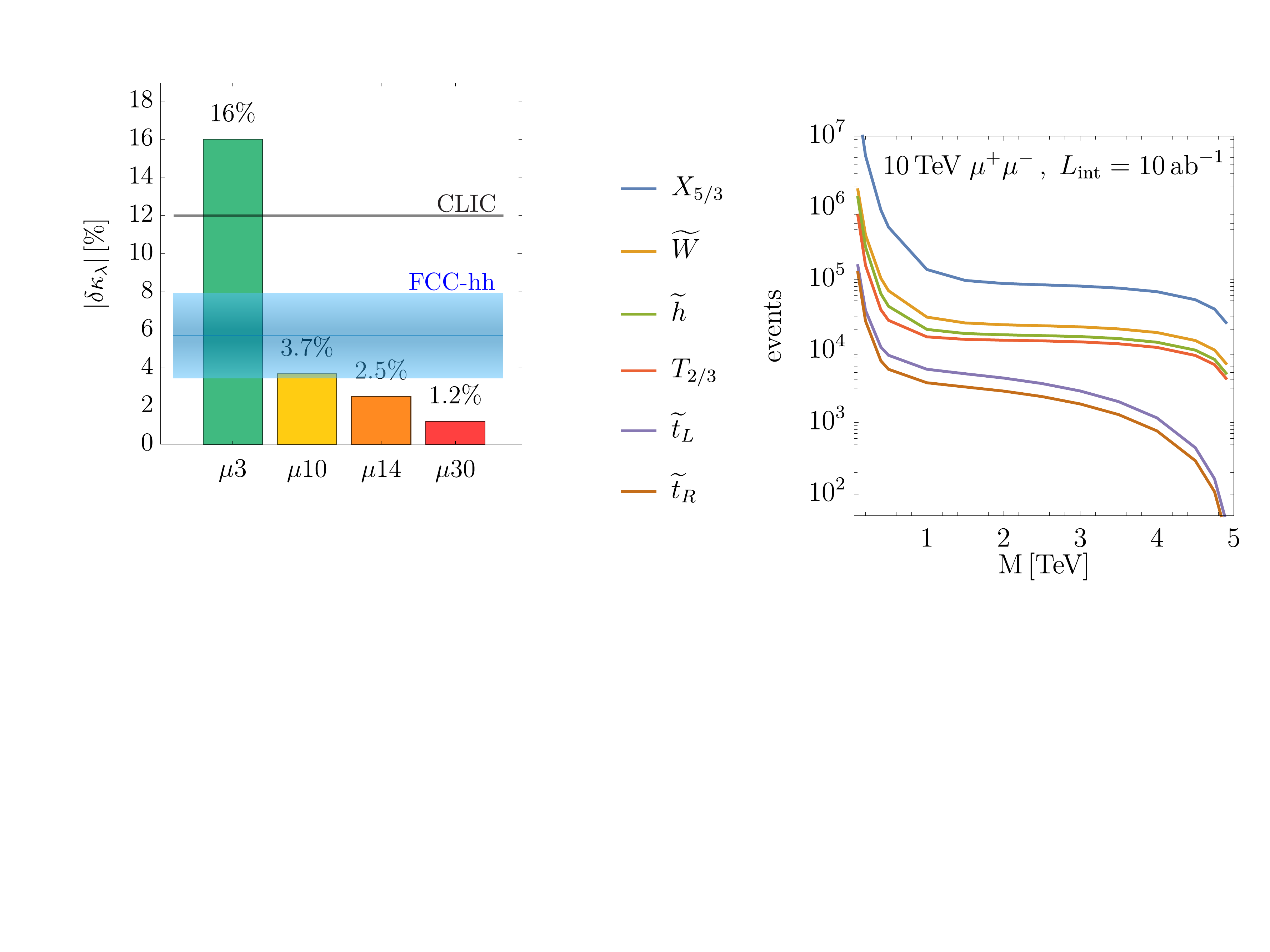}
\end{center}
\end{minipage}
\hfill
\begin{minipage}{0.38\textwidth}
\begin{center}
\includegraphics[width=\textwidth]{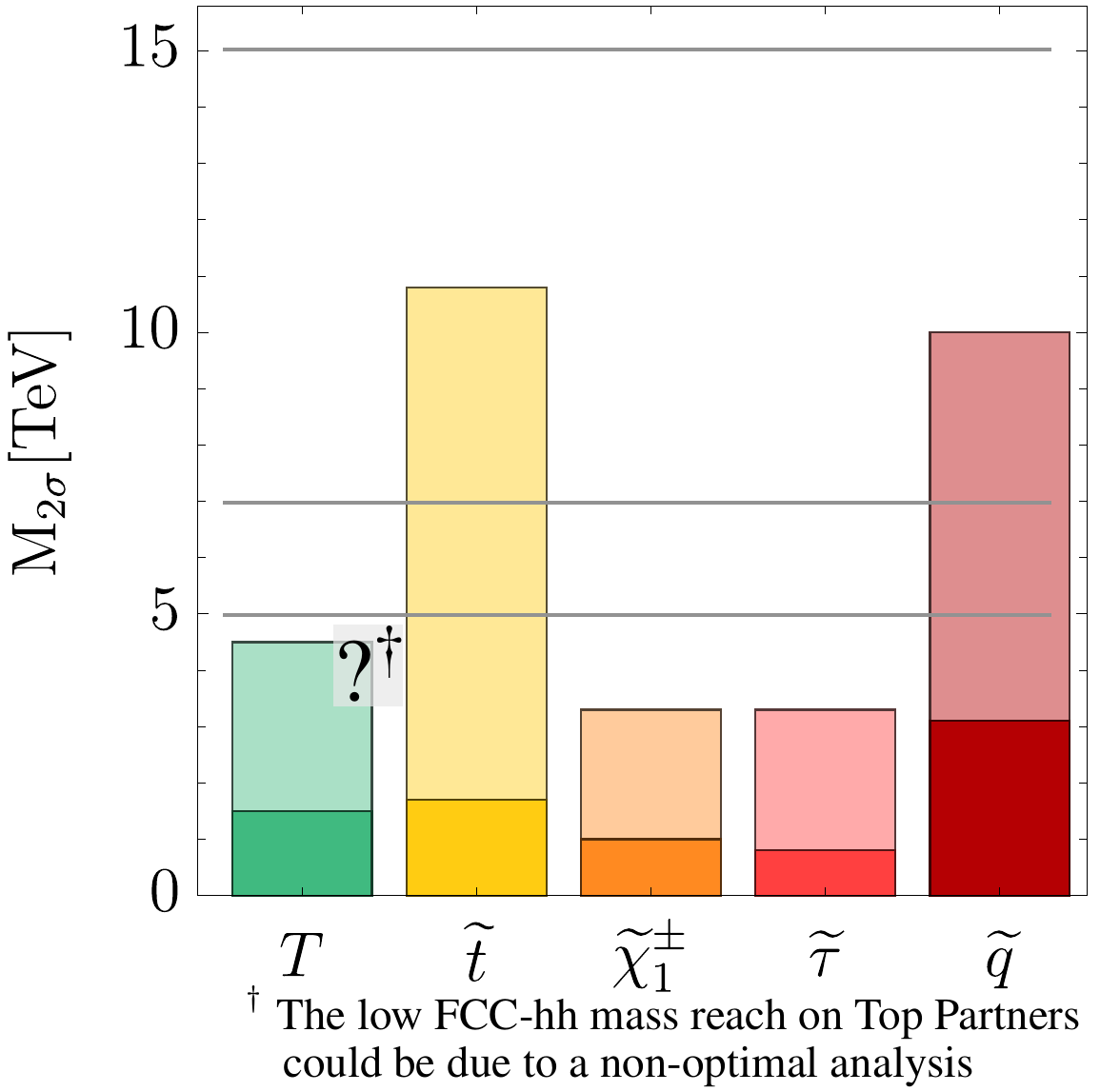}
\end{center}
\end{minipage}
\caption{Left panel: the number of expected events (from Ref.~\cite{Buttazzo:2020uzc}) at a $10$~TeV MuC, with  $10$~ab$^{-1}$ luminosity, for several BSM particles. Right panel: $95\%$~CL mass reach, from Ref.~\cite{EuropeanStrategyforParticlePhysicsPreparatoryGroup:2019qin}, at the HL-LHC (solid bars) and at the FCC-hh (shaded bars). The tentative discovery reach of a 10, 14 and 30~TeV MuC are reported as horizontal lines.
\label{dr}}
\end{figure*}

The left panel of Figure~\ref{dr} displays the number of expected events, at a $10$~TeV MuC with $10$~ab$^{-1}$ integrated luminosity, for the pair production due to electroweak interactions of Beyond the Standard Model (BSM) particles with variable mass M. The particles are named with a standard BSM terminology, however the results do not depend on the detailed BSM model (such as Supersymmetry or Composite Higgs) in which these particles emerge, but only on their Lorentz and gauge quantum numbers. The dominant production mechanism at high mass is the direct $\mu^+\mu^-$ annihilation, whose cross-section flattens out below the kinematical threshold at ${\rm{M}}=5$~TeV. The cross-section increase at low mass is due to the production from effective vector boson annihilation.

The figure shows that with the target luminosity of $10$~ab$^{-1}$ a $10$~TeV MuC can produce the BSM particles abundantly. If they decay to energetic and detectable SM final states, the new particles can be definitely discovered up to the kinematical threshold. Taking into account that the entire target integrated luminosity will be collected in $5$ years, a few month run could be sufficient for a discovery. Afterwards, the large production rate will allow us to observe the new particles decaying in multiple final states and to measure kinematical distributions. We will thus be in the position of characterising the properties of the newly discovered states precisely. Similar considerations hold for muon colliders with higher $E_{\rm{cm}}$, up to the fact that the kinematical mass threshold obviously grows to $E_{\rm{cm}}/2$. Notice however that the production cross-section decreases as $1/E_{\rm{cm}}^2$.\footnote{The scaling is violated by the vector boson annihilation channel, which however is relevant only at low mass.} 
Therefore, we obtain as many events as in the left panel of Figure~\ref{dr} only if the integrated luminosity grows quadratically with the energy as in eq.~(\ref{lums}). A luminosity that is lower than this by a factor of around $10$ would not affect the discovery reach, but it might reduce the potential for characterising the discoveries.

The direct reach of muon colliders vastly and generically exceeds the sensitivity of the High-Luminosity LHC (HL-LHC). This is illustrated by the solid bars on the right panel of Figure~\ref{dr}, where we report the projected HL-LHC mass reach~\cite{EuropeanStrategyforParticlePhysicsPreparatoryGroup:2019qin} on several BSM states. The $95\%$~CL exclusion is reported, instead of the discovery, as a quantification of the physics reach. Specifically, we consider Composite Higgs fermionic top-partners $T$ (e.g., the \parbox[c][0pt]{21pt}{$X_{5/3}$} and the \parbox[c][0pt]{19pt}{$T_{2/3}$}) and supersymmetric particles such as stops~\raisebox{2pt}{\parbox[c][0pt]{6pt}{${\widetilde{t}}$}}, charginos \raisebox{2pt}{\parbox[c][0pt]{14pt}{${\widetilde{\chi}}_1^\pm$}}, stau leptons~\raisebox{2pt}{\parbox[c][0pt]{6pt}{${\widetilde{\tau}}$}} and squarks~\raisebox{1pt}{\parbox[c][0pt]{6pt}{${\widetilde{q}}$}}. For each particle we report the highest possible mass reach, as obtained in the configuration for the BSM particle couplings and decay chains that maximises the hadron colliders sensitivity. The reach of a $100$~TeV proton-proton collider (FCC-hh) is shown as shaded bars on the same plot. The muon collider reach, displayed as horizontal lines for $E_{\rm{cm}}=10$, $14$ and~$30$~TeV, exceeds the one of the FCC-hh for several BSM candidates and in particular, as expected, for purely electroweak charged states. It should be noted that detailed muon collider sensitivity projections for the BSM candidates in Figure~\ref{dr} have not been performed yet. In general, a relatively limited literature exists on direct new physics searches at the MuC~\cite{Buttazzo:2018qqp,Ruhdorfer:2019utl,Liu:2021jyc,Asadi:2021gah,Huang:2021nkl,Liu:2021akf,Qian:2021ihf,Bandyopadhyay:2021pld,Capdevilla:2020qel,Capdevilla:2021rwo,Dermisek:2021ajd,Capdevilla:2021kcf,Homiller:2022iax,Bottaro:2022one,Bottaro:2021srh,Mekala:2023diu,Li:2023tbx,Kwok:2023dck,Jueid:2023zxx}. More studies would be desirable also to offer targets to the design of the detector.

\begin{figure*}
\begin{minipage}{0.5\textwidth}
\begin{center}
\includegraphics[width=1\textwidth]{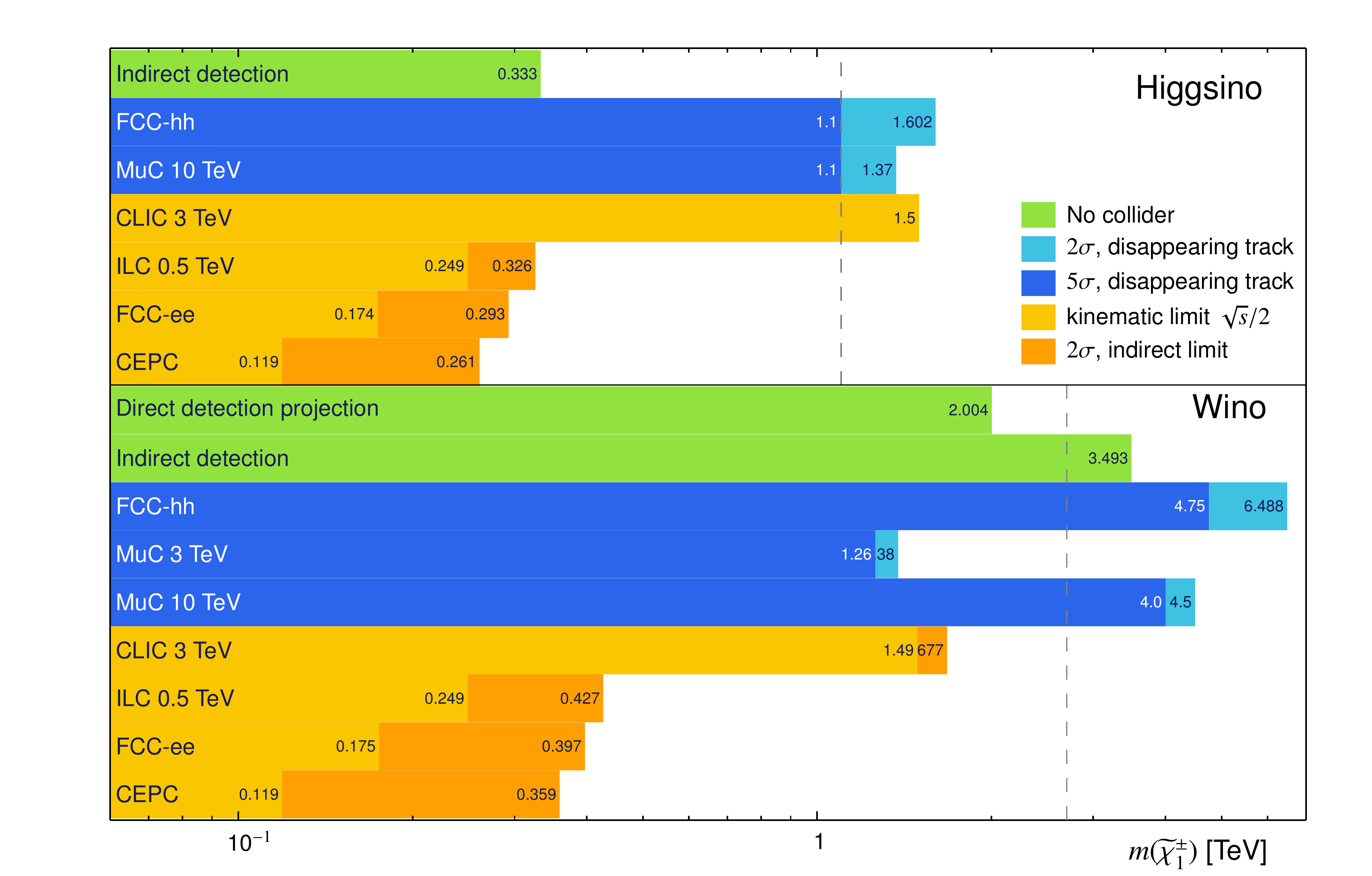}
\vspace{-4pt}
\end{center}
\end{minipage}
\hfill
\begin{minipage}{0.51\textwidth}
\begin{center}
\includegraphics[width=\textwidth]{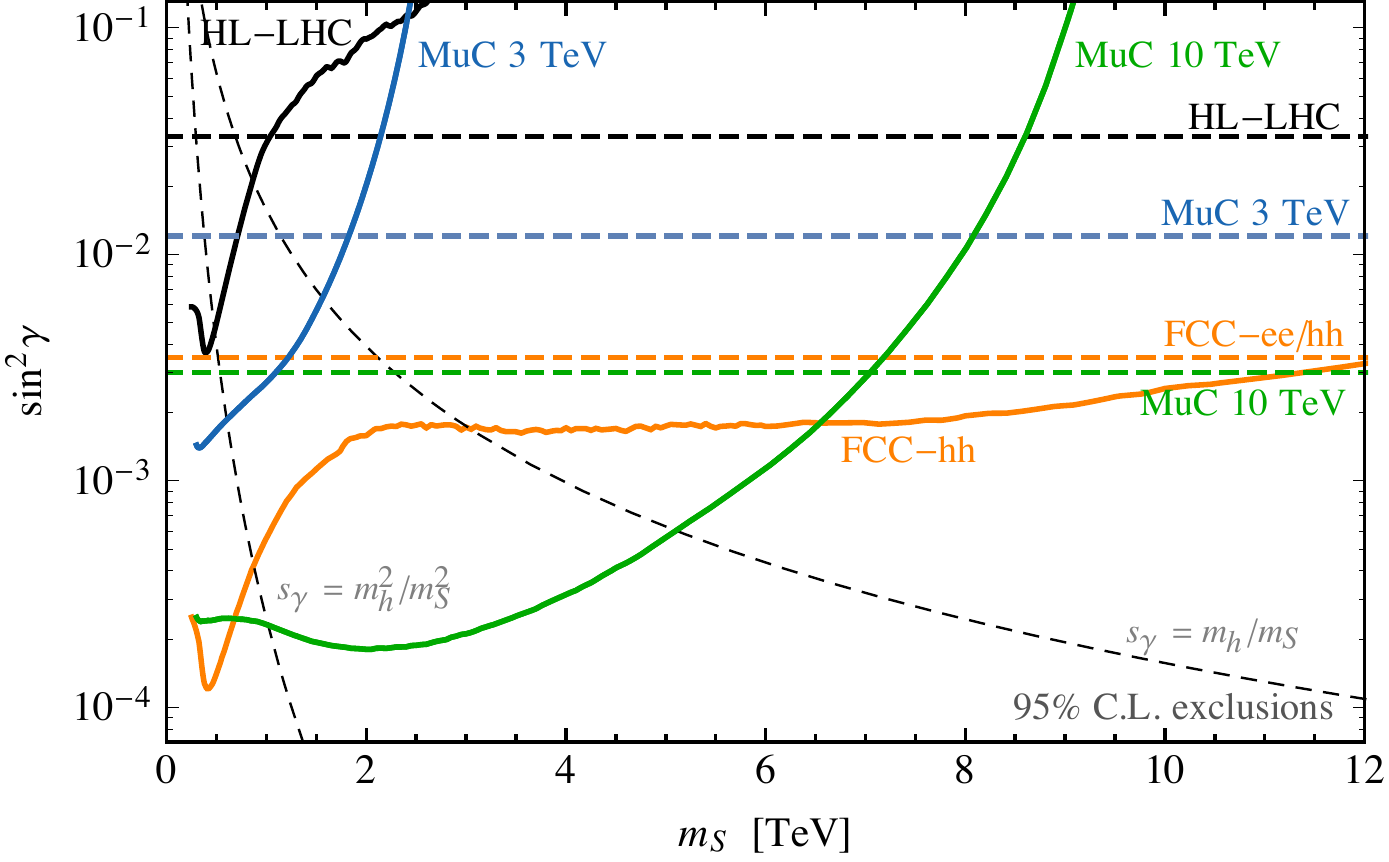}
\vspace{-18pt}
\end{center}
\end{minipage}
\caption{Left panel: exclusion and discovery mass reach on Higgsino and Wino dark matter candidates at muon colliders from disappearing tracks, and at other facilities. The plot is adapted from Ref.~\cite{Capdevilla:2021fmj}. Right: exclusion contour~\cite{AlAli:2021let} for a scalar singlet of mass $m_\phi$ mixed with the Higgs boson with strength $\sin\gamma$. More details in Section~\ref{sec:Higgs}.
\label{dmfig}}
\end{figure*}

Several interesting BSM particles do not decay to easily detectable final states, and an assessment of their observability requires dedicated studies. A clear case is the one of minimal WIMP Dark Matter (DM) candidates. The charged state in the DM electroweak multiplet is copiously produced, but it decays to the invisible DM plus a soft undetectable pion, owing to the small mass-splitting. WIMP DM can be studied at muon colliders in several channels (such as mono-photon) without directly observing the charged state~\cite{Han:2020uak,Bottaro:2021snn}. Alternatively, one can instead exploit the disappearing tracks produced by the charged particle~\cite{Capdevilla:2021fmj}. The result is displayed on the left panel of Figure~\ref{dmfig} for the simplest candidates, known as Higgsino and Wino. A $10$~TeV muon collider reaches the ``thermal'' mass, marked with a dashed line, for which the observed relic abundance is obtained by thermal freeze out. Other minimal WIMP candidates become kinematically accessible at higher muon collider energies~\cite{Han:2020uak,Bottaro:2021snn}. Muon colliders could actually even probe some of these candidates when they are above the kinematical threshold, by studying their indirect effects on high-energy SM processes~\cite{DiLuzio:2018jwd,Franceschini:2022sxc}. A more extensive overview of the muon collider potential to probe WIMP DM is provided in Section~\ref{secdm}.

New physics particles are not necessarily coupled to the SM by gauge interaction. One setup that is relevant in several BSM scenarios (including models of baryogenesis, dark matter, and neutral naturalness; see Section~\ref{sec:Higgs}) is the ``Higgs portal'' one, where the BSM particles interact most strongly with the Higgs field. By the Goldstone Boson Equivalence Theorem, Higgs field couplings are interactions with the longitudinal polarisations of the SM massive vector bosons $W$ and $Z$, which enable Vector Boson Fusion (VBF) production of the new particles. A muon collider is extraordinarily sensitive to VBF production, owing to the large luminosity for effective vector bosons. This is illustrated on the right panel of Figure~\ref{dmfig}, in the context of a benchmark model~\cite{Buttazzo:2018qqp,AlAli:2021let} (see also \cite{Ruhdorfer:2019utl,Liu:2021jyc}) where the only new particle is a real scalar singlet with Higgs portal coupling. The coupling strength is traded for the strength of the mixing with the Higgs particle, $\sin\gamma$, that the interaction induces. The scalar singlet is the simplest extension of the Higgs sector. Extensions with richer structure, such as involving a second Higgs doublet, are a priori easier to detect as one can exploit the electroweak production of the new charged Higgs bosons, as well as their VBF production. See Refs.~\cite{Han:2021udl,Chakrabarty:2014pja,Kalinowski:2020rmb,Rodejohann:2010jh,Li:2023ksw} for dedicated studies, and Section~\ref{sec:Higgs} for a review.

In several cases the muon collider direct reach compares favourably to the one of the most ambitious future proton collider project. This is not a universal statement, in particular at a muon collider it is obviously difficult to access heavy particles that carry only QCD interactions. One might also expect a muon collider of $10$~TeV to be generically less effective than a $100$~TeV proton collider for the detection of particles that can be produced singly. For instance, for additional $Z'$ massive vector bosons, that can be probed at the FCC-hh well above the $10$~TeV mass scale. We will see in Section~\ref{HEM} that the situation is slightly more complex and that, in the case of $Z'$s, a $10$~TeV MuC sensitivity actually exceeds the one of the FCC-hh in most of the parameter space (see the right panel of Figure~\ref{fig:HEM}).

\subsection{A vector bosons collider}\label{VBF}

When two electroweak charged particles like muons collide at an energy much above the electroweak scale $m_{{\rm{\textsc{w}}}}\sim100~$GeV, they have a high probability to emit electroWeak (EW) radiation. There are multiple types of EW radiation effects that can be observed at a muon collider and play a major role in muon collider physics. Actually we will argue in Section~\ref{EWRadiation} that the experimental observation and the theoretical description of these phenomena emerges as a self-standing reason of interest in muon colliders. 

\begin{figure*}
\centering{
\begin{minipage}{0.35\textwidth}
\vspace{-10pt}
\includegraphics[width=\textwidth]{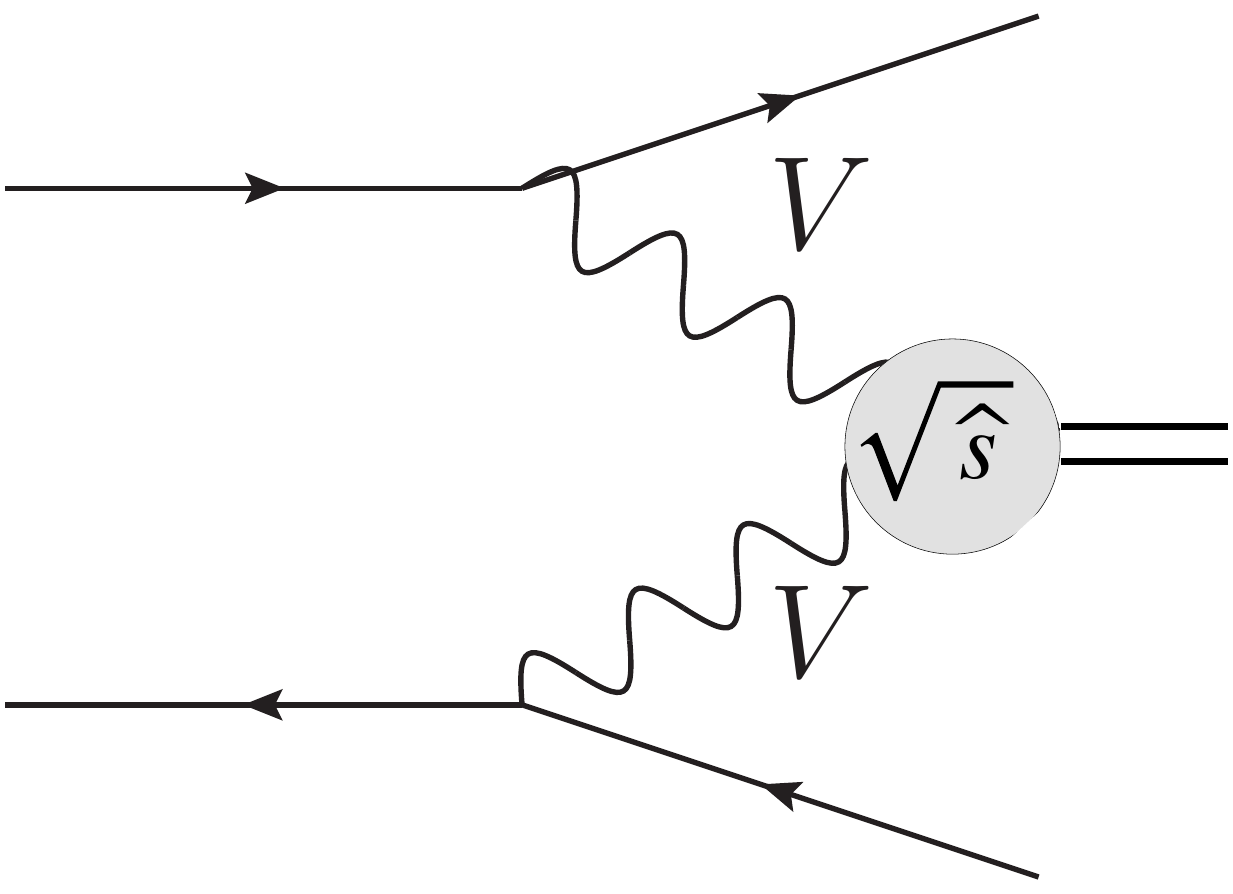}
\end{minipage}
\hspace{30pt}
\begin{minipage}{0.5\textwidth}
\vspace{0pt}
\includegraphics[width=0.95\textwidth]{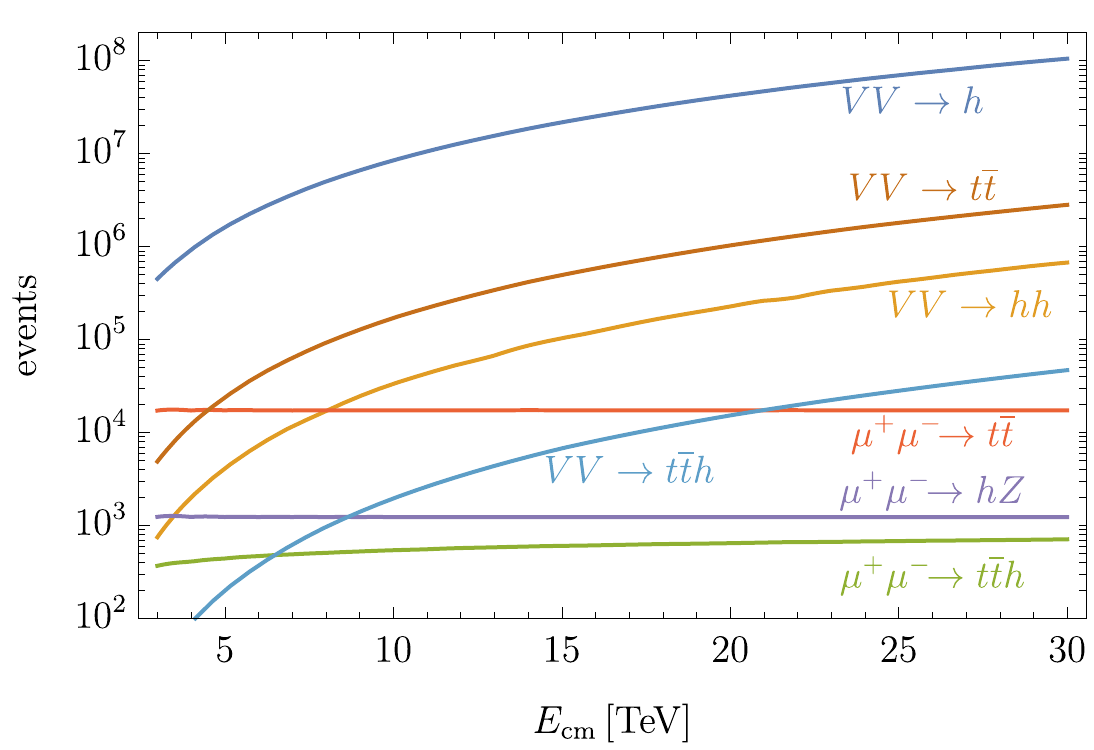}
\end{minipage}}
\caption{Left panel: schematic representation of vector boson fusion or scattering processes. The collinear $V$ bosons emitted from the muons participate to a process with hardness \mbox{$\sqrt{\hat{s}}\ll E_{\rm{cm}}$}. Right panel: number of expected events for selected SM processes at a muon collider with variable $E_{\rm{cm}}$ and luminosity scaling as in eq.~(\ref{lums}). 
\label{fig:EV}}
\end{figure*}

Here we focus on the practical implications~\cite{Delahaye:2019omf,Han:2020uid,Costantini:2020stv,Han:2020pif,Buttazzo:2020uzc,AlAli:2021let,Forslund:2022xjq} of the collinear emission of nearly on-shell massive vector bosons, which is the analog in the EW context of the Weizs\"{a}cker--Williams emission of photons. The vector bosons $V$ participate, as depicted in Figure~\ref{fig:EV}, to a scattering process with a hard scale \parbox[c][0pt]{15pt}{$\sqrt{\hat{s}}$} that is much lower than the muon collision energy \parbox[c][0pt]{20pt}{$E_{\rm{cm}}$}. The typical cross-section for $VV$ annihilation processes is thus enhanced by \parbox[c][0pt]{30pt}{$E_{\rm{cm}}^2/{\hat{s}}$}, relative to the typical cross-section for $\mu^+\mu^-$ annihilation, whose hard scale is instead $E_{\rm{cm}}$. The emission of the $V$ bosons from the muons is suppressed by the EW coupling, but the suppression is mitigated or compensated by logarithms of the separation between the EW scale and $E_{\rm{cm}}$ (see~\cite{Han:2020uid,Costantini:2020stv,AlAli:2021let} for a pedagogical overview). The net result is a very large cross-section for VBF processes that occur at $\sqrt{\hat{s}}\sim m_{\rm{\textsc{w}}}$, with a tail in {$\sqrt{\hat{s}}$} up to the TeV scale.

We already emphasised (see Figure~\ref{dr}) the importance of VBF for the direct production of new physics particles. The relevance of VBF for probing new physics indirectly simply stems for the huge rate of VBF SM processes, summarised on the right panel of Figure~\ref{fig:EV}. In particular we see that a $10$~TeV muon collider produces ten million Higgs bosons, which is around $10$ times more than future $e^+e^-$ Higgs factories. Since the Higgs bosons are produced in a relatively clean environment, without large physics backgrounds from QCD, a $10$~TeV muon collider (over-)qualifies as a Higgs factory~\cite{Forslund:2022xjq,AlAli:2021let,Han:2020pif,Bartosik:2019dzq,Bartosik:2020xwr}. Unlike $e^+e^-$ Higgs factories, a muon collider also produces Higgs pairs copiously, enabling accurate and direct measurements of the Higgs trilinear coupling~\cite{Costantini:2020stv,Han:2020pif,Buttazzo:2020uzc} and possibly also of the quadrilinear coupling~\cite{Chiesa:2020awd}. 

\begin{figure*}
\begin{minipage}{0.55\textwidth}
\renewcommand{\arraystretch}{.89}
\setlength{\arrayrulewidth}{.2mm}
\setlength{\tabcolsep}{0.6 em}
\begin{center}
\begin{tabular}{c|c|c|c}
\hline
& \multicolumn{1}{c|}{\makebox[35pt]{\small{HL-LHC}}} & \makebox[35pt]{\small{HL-LHC}} & \makebox[35pt]{\small{HL-LHC}} \\[0pt]
& \multicolumn{1}{c|}{\ } & \multicolumn{1}{l|}{\makebox[35pt]{+\small{$10\,\textrm{{TeV}}$}}} & \multicolumn{1}{l}{\makebox[35pt]{+\small{$10\,\textrm{{TeV}}$}}}  \\[-2pt]
\ & \ & \multicolumn{1}{c|}{\ } & \multicolumn{1}{l}{\hspace{0.5pt}+
{$e e$}} \\ 
\hline
$\kappa_W$ & 1.7 & 0.1 & 0.1 \\ \hline
$\kappa_Z$ & 1.5 & 0.4 & 0.1 \\ \hline
$\kappa_g$ & 2.3 & 0.7 & 0.6 \\ \hline
$\kappa_{\gamma}$ & 1.9 & 0.8 & 0.8 \\ \hline
$\kappa_{Z\gamma}$ & 10 & 7.2 & 7.1 \\ \hline 
$\kappa_c$ & - & 2.3 & 1.1 \\ \hline
$\kappa_b$ & 3.6 & 0.4 & 0.4\\ \hline
$\kappa_{\mu}$ & 4.6 & 3.4 & 3.2 \\ \hline
$\kappa_{\tau}$ & 1.9 & 0.6 & 0.4 \\ \hline\hline
$\kappa_t^*$ & 3.3 & 3.1 & 3.1 \\ \hline
\multicolumn{4}{l}{ {\scriptsize $^*$ No input used for the MuC}}\\
\end{tabular}
\end{center}
\end{minipage}
\hfill
\begin{minipage}{0.45\textwidth}
\vspace{0pt}
\includegraphics[width=.93\textwidth]{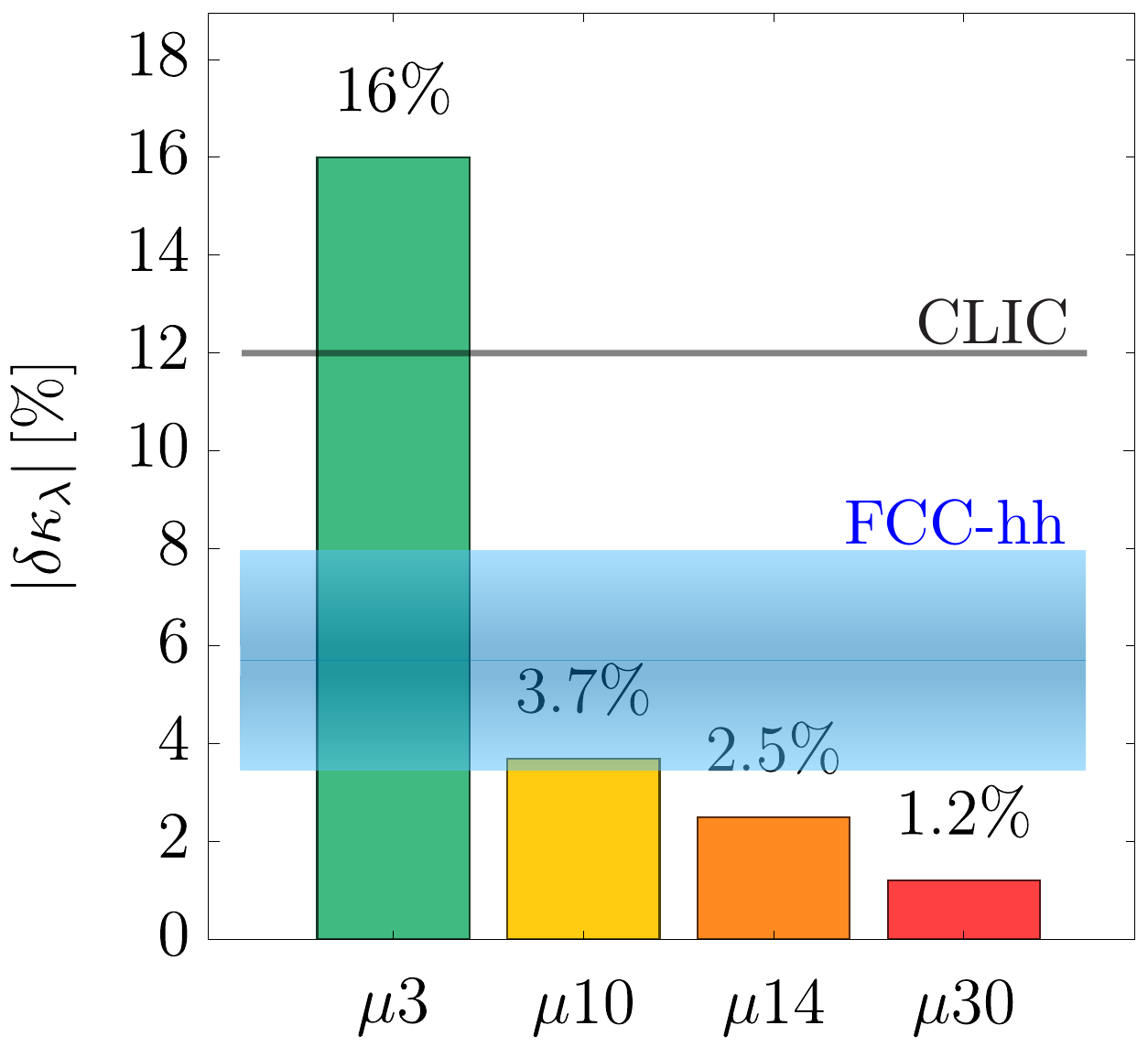}
\end{minipage}
\caption{Left panel: $1\sigma$ sensitivities (in \%) from a 10-parameter fit in the $\kappa$-framework at a $10$~TeV MuC with $10$~ab$^{-1}$, compared with HL-LHC. The effect of measurements from a $250$~GeV $e^+e^-$ Higgs factory is also reported. Right panel: sensitivity to $\delta\kappa_\lambda$ for different $E_{\rm{cm}}$. The luminosity is as in eq.~(\ref{lums}) for all energies, apart from $E_{\rm{cm}}\hspace{-2pt}=\hspace{-2pt}3$~TeV, where doubled luminosity (of 2~ab$^{-1}$) is assumed. More details in Section~\ref{sec:Higgs}.
\label{tab:higgscouplingfit}}
\end{figure*}

The opportunities for Higgs physics at a muon collider are summarised extensively in Section~\ref{sec:Higgs}. In Figure~\ref{tab:higgscouplingfit} we report for illustration the results of a 10-parameter fit to the Higgs couplings in the $\kappa$-framework at a $10$~TeV MuC, and the sensitivity projections on the anomalous Higgs trilinear coupling $\delta\kappa_\lambda$. The table shows that a $10$~TeV MuC will improve significantly and broadly our knowledge of the properties of the Higgs boson. The combination with the measurements performed at an $e^+e^-$ Higgs factory, reported on the third column, does not affect the sensitivity to several couplings appreciably, showing the good precision that a muon collider alone can attain. However, it also shows complementarity with an $e^+e^-$ Higgs factory program. 

On the right panel of the figure we see that the performances of muon colliders in the measurement of $\delta\kappa_\lambda$ are similar or much superior to the one of the other future colliders where this measurement could be performed. In particular, CLIC measures $\delta\kappa_\lambda$ at the $10\%$ level~\cite{deBlas:2018mhx}, and the FCC-hh sensitivity ranges from $3.5$ to $8\%$ depending on detector assumptions~\cite{Mangano:2020sao}. A determination of $\delta\kappa_\lambda$ that is way more accurate than the HL-LHC projections is possible already at a low energy stage of a muon collider with $E_{\rm{cm}}=3$~TeV as discussed in Section~\ref{sec:Higgs}.

The potential of a muon collider as a vector boson collider has not been explored fully. In particular a systematic investigation of vector boson scattering processes, such as $WW\hspace{-3pt}\to\hspace{-3pt} WW$, has not been performed. The key role played by the Higgs boson to eliminate the energy growth of the corresponding Feynman amplitudes could be directly verified at a muon collider by means of differential measurements that extend well above one TeV for the invariant mass  of the scattered vector bosons. Along similar lines, differential measurements of the $WW\hspace{-3pt}\to\hspace{-3pt} HH$ process has been studied in~\cite{Buttazzo:2020uzc,Han:2020pif} (see also~\cite{Costantini:2020stv}) as an effective probe of the composite nature of the Higgs boson, with a reach that is comparable or superior to the one of Higgs coupling measurements. A similar investigation was performed in~\cite{AlAli:2021let,Costantini:2020stv} (see also~\cite{Costantini:2020stv}) for $WW\hspace{-3pt}\to\hspace{-3pt} t{\overline{t}}$, aimed at probing Higgs-top interactions.

\subsection{High-energy measurements}\label{HEM}

Direct $\mu^+\mu^-$ annihilation, such as $HZ$ and $t{\overline{t}}$ production, displays a number of expected events of the order of several thousands, reported in Figure~\ref{fig:EV}. These are much less than the events where a Higgs or a $t{\overline{t}}$ pair are produced from VBF, but they are sharply different and easily distinguishable. The invariant mass of the particles produced by direct annihilation is indeed sharply peaked at the collider energy $E_{\rm{cm}}$, while the invariant mass rarely exceeds one tenth of $E_{\rm{cm}}$ in the VBF production mode. 

The good statistics and the limited or absent background thus enables few-percent level measurements of SM cross sections for hard scattering processes of energy $E_{\rm{cm}}=10$~TeV at the 10~TeV MuC. An incomplete list of the many possible measurements is provided in Ref.~\cite{Chen:2022msz}, including the resummed effects of EW radiation on the cross section predictions. It is worth emphasising that also charged final states such as $WH$ or $\ell\nu$ are copiously produced at a muon collider. The electric charge mismatch with the neutral $\mu^+\mu^-$ initial state is compensated by the emission of soft and collinear $W$ bosons, which occurs with high probability because of the large energy.

High energy scattering processes are as unique theoretically as they are experimentally~\cite{Delahaye:2019omf,Buttazzo:2020uzc,Chen:2022msz}. They give direct access to the interactions among SM particles with $10$~TeV energy, which in turn provide indirect sensitivity to new particles at the $100$~TeV scale of mass. In fact, the effects on high-energy cross sections of new physics at energy $\Lambda\gg E_{\rm{cm}}$ generically scale as $(E_{\rm{cm}}/\Lambda)^2$ relative to the SM. Percent-level measurements thus give access to $\Lambda\sim100$~TeV. This is an unprecedented reach for new physics theories endowed with a reasonable flavor structure. Notice in passing that high-energy measurements are also useful to investigate flavor non-universal phenomena, as we will see in Section~\ref{muonspec}.

This mechanism is not novel. Major progress in particle physics always came from raising the available collision energy, producing either direct or indirect discoveries. Among the most relevant discoveries that did not proceed through the resonant production of new particles, there is the one of the inner structure of nucleons. This discovery could be achieved~\cite{nobel} only when the transferred energy in electron scattering could reach a significant fraction of the proton compositeness scale $\Lambda_{\rm{\textsc{qcd}}}=1/r_{p}=300$~MeV. Proton-compositeness effects became sizeable enough to be detected at that energy, precisely because of the quadratic enhancement mechanism we described above.

\begin{figure*}
\begin{minipage}{0.47\textwidth}
\includegraphics[width=0.96\textwidth]{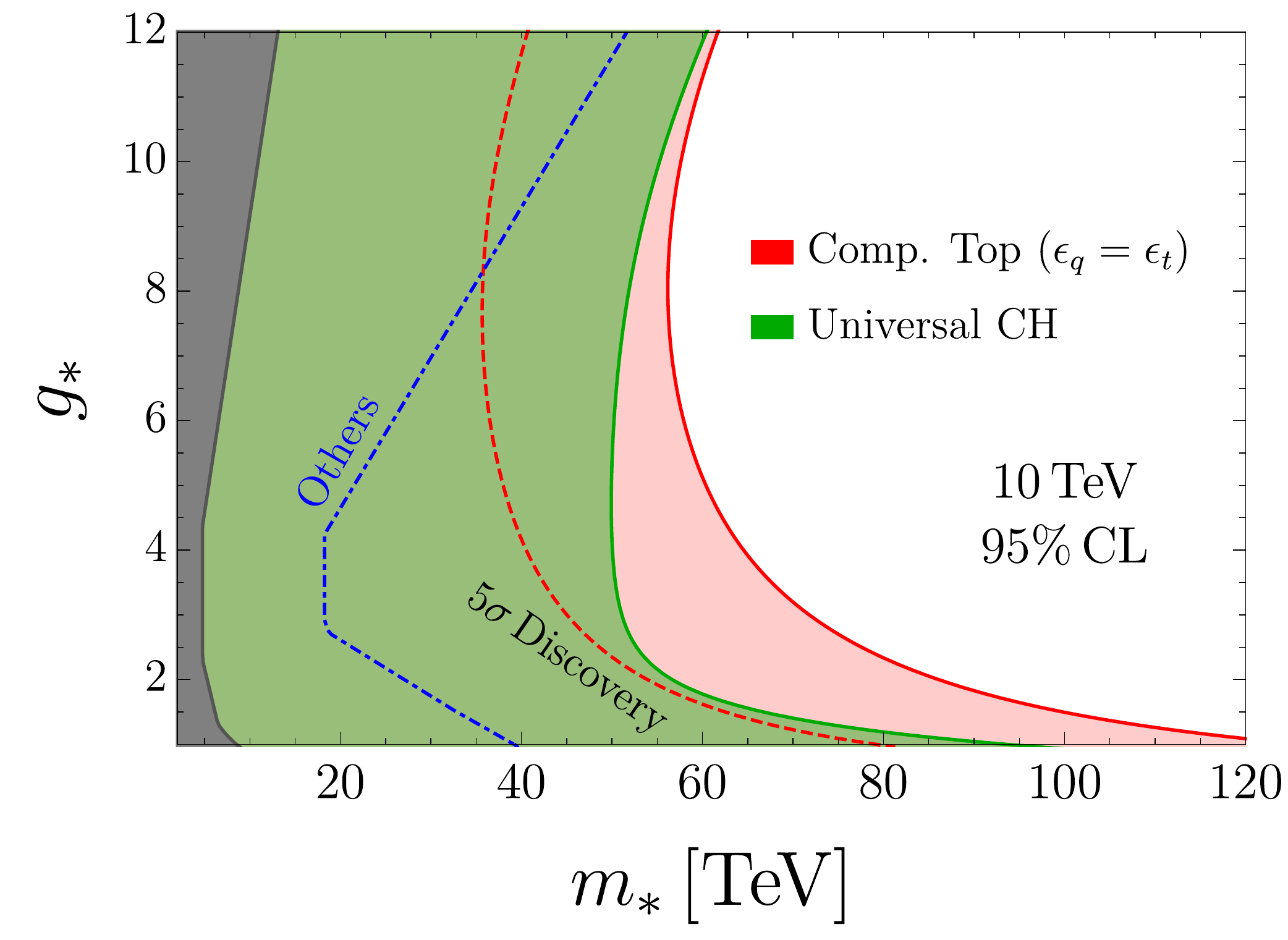}
\end{minipage}
\hfill
\begin{minipage}{0.47\textwidth}
\vspace{0pt}
\includegraphics[width=0.96\textwidth]{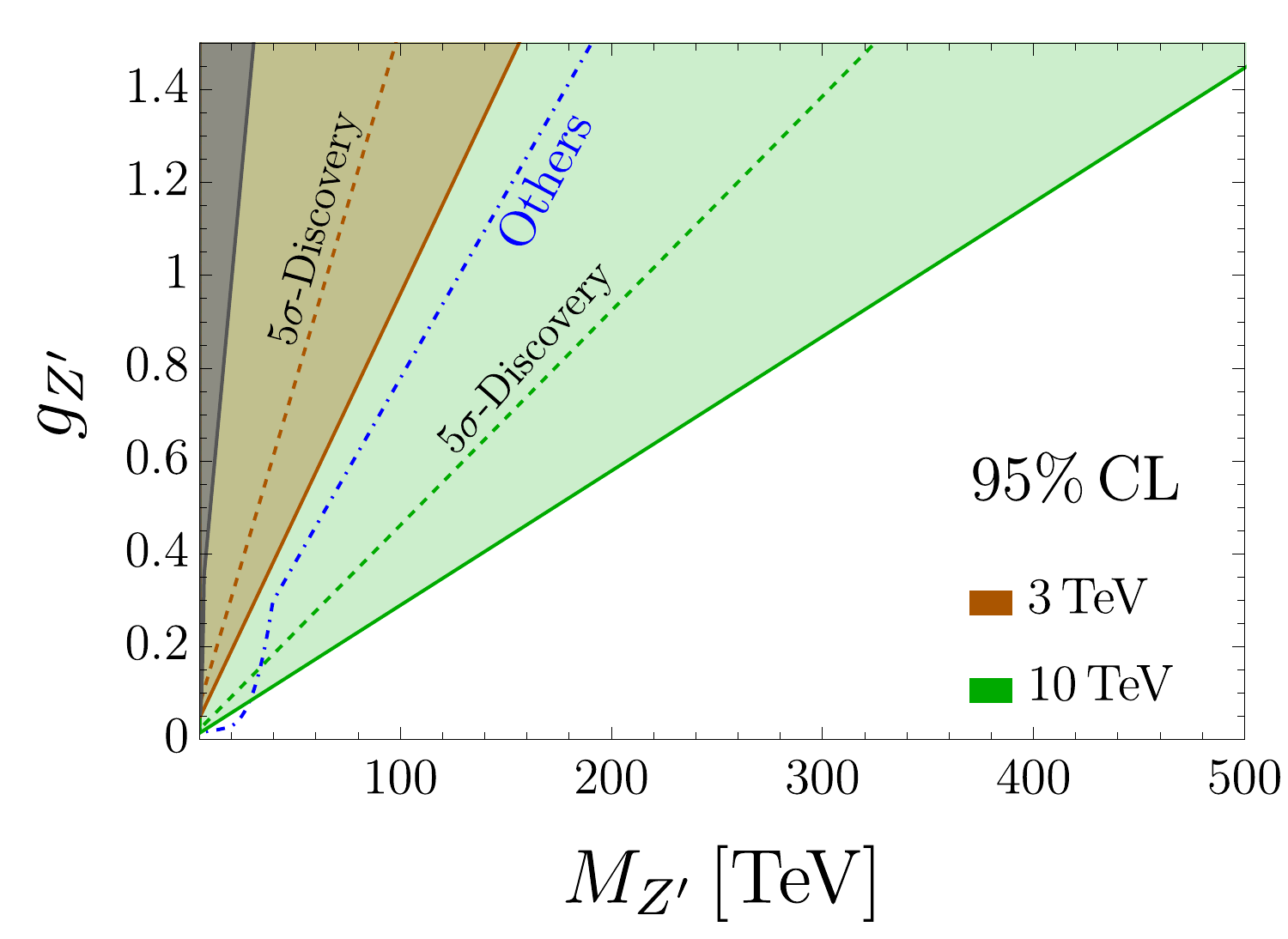}
\end{minipage}
\caption{Left panel: $95\%$ reach on the Composite Higgs scenario from high-energy measurements in di-boson and di-fermion final states~\cite{Chen:2022msz}. The green contour display the sensitivity from ``Universal'' effects related with the composite nature of the Higgs boson and not of the top quark. The red contour includes the effects of top compositeness. Right panel: sensitivity to a minimal $Z'$~\cite{Chen:2022msz}. Discovery contours at $5\sigma$ are also reported in both panels. 
\label{fig:HEM}}
\end{figure*}

Figure~\ref{fig:HEM} illustrates the tremendous reach on new physics of a $10$~TeV MuC with $10$~ab$^{-1}$ integrated luminosity. The left panel (green contour) is the sensitivity to a scenario that explains the microscopic origin of the Higgs particle and of the scale of EW symmetry breaking by the fact that the Higgs is a composite particle. In the same scenario the top quark is likely to be composite as well, which in turn explains its large mass and suggest a ``partial compositeness'' origin of the SM flavour structure. Top quark compositeness produces additional signatures that extend the muon collider sensitivity up to the red contour. The sensitivity is reported in the plane formed by the typical coupling $g_*$ and of the typical mass $m_*$ of the composite sector that delivers the Higgs. The scale $m_*$ physically corresponds to the inverse of the geometric size of the Higgs particle. The coupling $g_*$ is limited from around $1$ to $4\pi$, as in the figure. In the worst case scenario of intermediate $g_*$, a $10$~TeV MuC can thus probe the Higgs radius up to the inverse of $50$~TeV, or discover that the Higgs is as tiny as $(35$~TeV$)^{-1}$. The sensitivity improves in proportion to the centre of mass energy of the muon collider. 

The figure also reports, as blue dash-dotted lines denoted as ``Others'', the envelop of the $95\%$~CL sensitivity projections of all the future collider projects that have been considered for the~2020 update of the European Strategy for Particle Physics, summarised in Ref.~\cite{EuropeanStrategyforParticlePhysicsPreparatoryGroup:2019qin}. These lines include in particular the sensitivity of very accurate measurements at the EW scale performed at possible future $e^+e^-$ Higgs, electroweak and Top factories. These measurements are not competitive because new physics at $\Lambda\sim100$~TeV produces unobservable one part per million effects on $100$~GeV energy processes. High-energy measurements at a $100$~TeV proton collider are also included in the dash-dotted lines. They are not competitive either, because the effective parton luminosity at high energy is much lower than the one of a $10$~TeV MuC, as explained in Section~\ref{whym}. For example the cross-section for the production of an $e^+e^-$ pair with more than $9$~TeV invariant mass at the FCC-hh is only $40$~ab, while it is $900$~ab at a $10$~TeV muon collider. Even with a somewhat higher integrated luminosity, the FCC-hh just does not have enough statistics to compete with a $10$~TeV MuC.

The right panel of Figure~\ref{fig:HEM} considers a simpler new physics scenario, where the only BSM state is a heavy $Z'$ spin-one particle. The ``Others'' line also includes the sensitivity of the FCC-hh from direct $Z'$ production. The line exceeds the $10$~TeV MuC sensitivity contour (in green) only in a tiny region with $M_{Z'}$ around $20$~TeV and small $Z'$ coupling. This result substantiates our claim in Section~\ref{dirr} that a reach comparison based on the $2\to 1$ single production of the new states is simplistic. Single $2\to 1$ production couplings can produce indirect effect in $2\to 2$ scattering by the virtual exchange of the new particle, and the muon collider is extraordinarily sensitive to these effects. Which collider wins is model-dependent. In the simple benchmark $Z'$ scenario, and in the motivated framework of Higgs compositeness that future colliders are urged to explore, the muon collider is just a superior device.

We have seen that high energy measurements at a muon collider enable the indirect discovery of new physics at a scale in the ballpark of $100$~TeV. However the muon collider also offers amazing opportunities for direct discoveries at a mass of several TeV, and unique opportunities to characterise the properties of the discovered particles, as emphasised in Section~\ref{dirr}. High energy measurements will enable us take one step further in the discovery characterisation, by probing the interactions of the new particles well above their mass. For instance in the Composite Higgs scenario one could first discover Top Partner particles of few TeV mass, and next study their dynamics and their indirect effects on SM processes. This might be sufficient to pin down the detailed theoretical description of the newly discovered sector, which would thus be both discovered and theoretically characterised at the same collider. Higgs coupling determinations and other precise measurements that exploit the enormous luminosity for vector boson collisions, described in Section~\ref{VBF}, will also play a major role in this endeavour. 

We can dream of such glorious outcome of the project, where an entire new sector is discovered and characterised in details at the same machine, only because energy and precision are simultaneously available at a muon collider. 


\subsection{Electroweak radiation}\label{EWRadiation}

The novel experimental setup offered by lepton collisions at $10$~TeV energy or more outlines possibilities for theoretical exploration that are at once novel and speculative, yet robustly anchored to reality and to phenomenological applications. 

The muon collider will probe for the first time a new regime of EW interactions, where the scale $m_{{\rm{\textsc{w}}}}\hspace{-2pt}\sim\hspace{-2pt}100~$GeV of EW symmetry breaking plays the role of a small IR scale, relative to the much larger collision energy. This large scale separation triggers a number of novel phenomena that we collectively denote as ``EW radiation'' effects. Since they are prominent at muon collider energies, the comprehension of these phenomena is of utmost importance not only for developing a correct physical picture but also to achieve the needed accuracy of the theoretical predictions.

The EW radiation effects that the muon collider will observe, which will play a crucial role in the assessment of its sensitivity to new physics, can be broadly divided in two classes. 

The first class includes the emission of low-virtuality vector bosons from the initial muons. It  effectively makes the muon collider a high-luminosity vector boson collider, on top of a very high-energy lepton-lepton machine. The compelling associated physics studies described in Section~\ref{VBF} pose challenges for  fixed-order theoretical predictions and Monte Carlo event generation even at tree-level, owing to the sharp features of the Monte Carlo integrand induced by the large scale separation and the need to correctly handle QED and weak radiation at the same time, respecting EW gauge invariance. Strategies to address these challenges are available in  {\texttt{WHIZARD}}~\cite{Kilian:2007gr}, they have been recently implemented in {\texttt{MadGraph5\_aMC@NLO}}~\cite{Costantini:2020stv,Ruiz:2021tdt} and applied to several phenomenological studies in the muon collider context. Dominance of such initial-state collinear radiation will eventually require a systematic theoretical modelling in terms of EW Parton Distribution Function where multiple collinear radiation effects are resummed. First studies show that EW resummation effects can be significant at a 10~TeV MuC~\cite{Han:2020uid}.

The second class of effects are the virtual and real emissions of soft and soft-collinear EW radiation. They affect most strongly the measurements performed at the highest energy, described in Section~\ref{HEM}, and impact the corresponding cross-section predictions at order one~\cite{Chen:2022msz}. They also give rise to novel processes such as the copious production of charged hard final states out of the neutral $\mu^+\mu^-$ initial state, and to new opportunities to detect new short distance physics by studying, for one given hard final state, different patterns of radiation emission~\cite{Chen:2022msz}. The deep connection with the sensitivity to new physics makes the study of EW radiation an inherently multidisciplinary enterprise that overcomes the traditional separation between ``SM background'' and ``BSM signal'' studies. 

At very high energies EW radiation displays similarities with QCD and QED radiation, but also remarkable differences that pose profound theoretical challenges. 

First, being EW symmetry broken at low energy, different particles in the same EW multiplet---i.e., with different “EW color” like the $W$ and the $Z$---are distinguishable.
In particular the beam particles (e.g., charged left-handed leptons) carry definite colour thus violating the KLN theorem assumptions. Therefore, no cancellation takes place between virtual and real radiation contributions, regardless of the final state observable inclusiveness~\cite{Ciafaloni:2000df,Ciafaloni:2000rp}. Furthermore, the EW colour of the final state particles can be measured, and it must be measured for a sufficiently accurate exploration of the SM and BSM dynamics. For instance, distinguishing the top from the bottom quark or the $W$ from the $Z$ boson (or photon) is necessary to probe accurately and comprehensively new short-distance physical laws that can affect the dynamics of the different particles differently. When dealing with QCD and QED radiation only, it is sufficient instead to consider ``safe'' observables where QCD/QED radiation effects can be systematically accounted for and organised in well-behaved (small) corrections. The relevant observables for EW physics at high energy are on the contrary dramatically affected by EW radiation effects. 

Second, in analogy with QCD and unlike QED, for EW radiation the IR scale is physical. However, at variance with QCD, the theory is weakly-coupled at the IR scale, and the EW ``partons'' do not ``hadronise''. EW~showering therefore always ends at virtualities of order 100~GeV, where  heavy EW states $(t,W,Z,H)$ coexist with light SM ones, and then decay.  Having a complete and consistent description of the evolution from high virtualities where EW symmetry is restored, to the weak scale where it is broken, to GeV scales, including also leading QED/QCD effects in all regimes is a new challenge~\cite{Han:2021kes}. 

Such a strong phenomenological motivation, and the peculiarities of the problem, stimulate work and offer a new perspective on resummation and showering techniques, or more in general trigger theoretical progress on IR physics. Fixed-order calculations will also play an important role. Indeed while the resummation of the leading logarithmic effects of radiation is mandatory at muon collider energies~\cite{Chen:2022msz,Bauer:2018xag}, subleading logarithms could perhaps be included at fixed order. Furthermore one should eventually develop a description where resummation is merged with fixed-order calculations in an exclusive way, providing the most accurate predictions in the corresponding regions of the phase space, as currently done for QCD computations. 

A significant literature on EW radiation exists, starting from the earliest works on double-logarithm resummations based on Asymptotic Dynamics~\cite{Ciafaloni:2000df,Ciafaloni:2000rp} or on the IR evolution equation~\cite{Fadin:1999bq,Melles:2000gw}. The factorisation of virtual massive vector boson emissions, leading to the notion of effective vector boson is also known since long~\cite{Kane:1984bb,Dawson:1984gx,Chanowitz:1985hj,Kunszt:1987tk}. More recent progress includes resummation at the next to leading log in the Soft-Collinear Effective Theory framework~\cite{Chiu:2007yn,Chiu:2007dg,Chiu:2009ft,Manohar:2014vxa,Manohar:2018kfx}, the operatorial definition of the distribution functions for EW partons~\cite{Bauer:2017isx,Fornal:2018znf,Bauer:2018xag} and the calculation of the corresponding evolution, as well as the calculation of the EW splitting functions~\cite{Chen:2016wkt} for EW showering and the proof of collinear EW emission factorisation~\cite{Borel:2012by,Wulzer:2013mza,Cuomo:2019siu}. Additionally, fixed-order virtual EW logarithms are known for generic process at the $1$-loop order~\cite{Denner:2000jv,Denner:2001gw} and are implemented in {\texttt{Sherpa}}~\cite{Bothmann:2020sxm} and  {\texttt{MadGraph5\_aMC@NLO}}~\cite{Pagani:2021vyk}. Exact EW corrections at NLO are available in an automatic form for arbitrary processes in the SM, for example in the {\texttt{MadGraph5\_aMC@NLO}}~\cite{Frederix:2018nkq} and in the {\texttt{Sherpa+Recola}}~\cite{Biedermann:2017yoi} packages or using {\texttt{WHIZARD+Recola}}~\cite{Bredt:2022dmm}. Implementations of EW showering are also available through a limited set of splittings in {\texttt{Pythia 8}}~\cite{Christiansen:2014kba,Christiansen:2015jpa} and in a complete way in {\texttt{Vincia}}~\cite{Brooks:2021kji}.

\begin{figure}[t]
\begin{center}
\includegraphics[width=0.5\textwidth]
{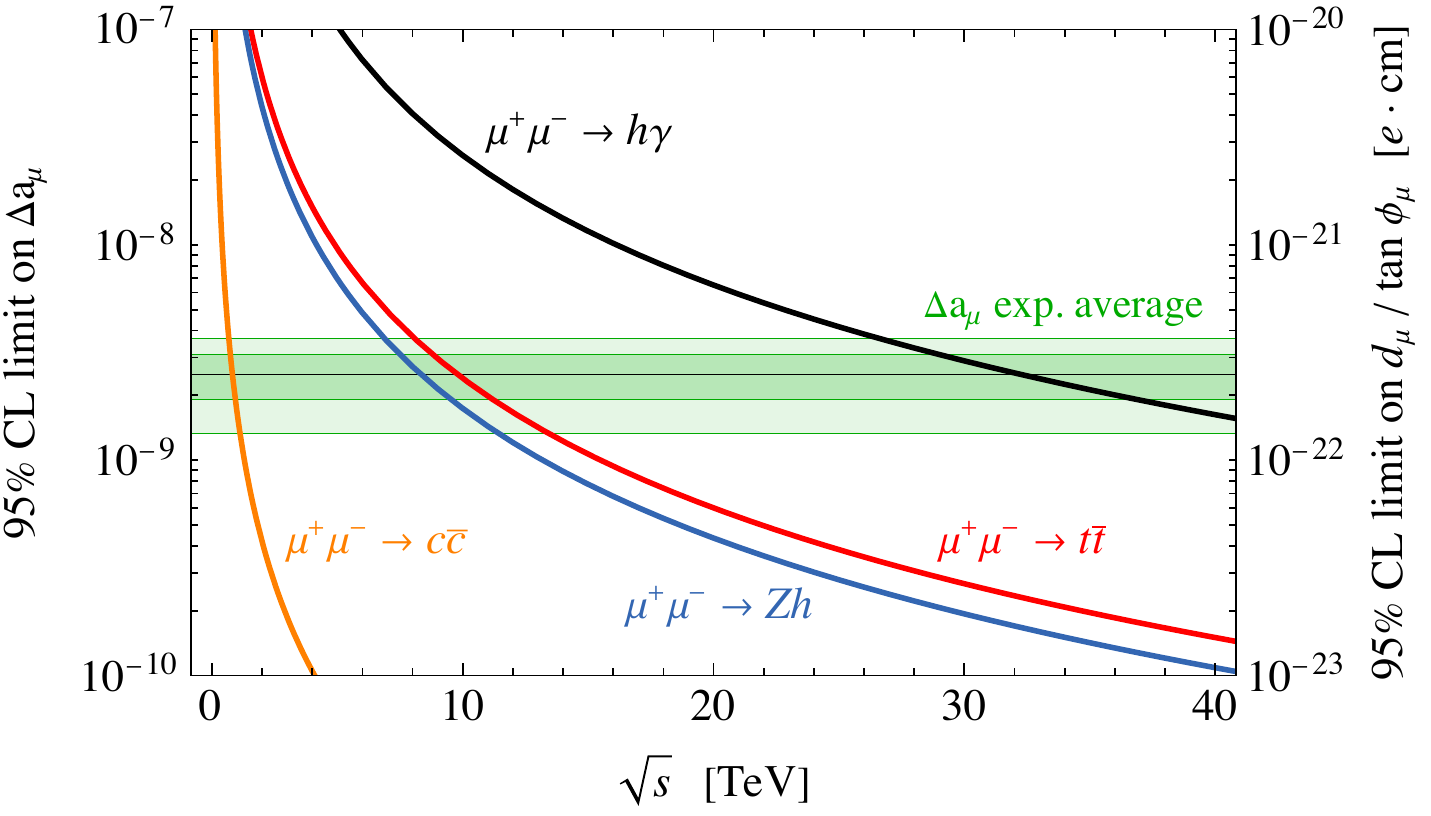}
\end{center}
\caption{Summary, from Section~\ref{muonspec} of the muon collider perspectives to probe the muon $g$-2 anomaly.
\label{fan}}
\end{figure}

While we are still far from an accurate systematic understanding of EW radiation, the present-day knowledge is sufficient to enable rapid progress in the next few years. The outcome will be an indispensable toolkit for muon collider predictions. Moreover, while we do expect that EW radiation phenomena can in principle be described by the Standard Model, they still qualify as  ``new phenomena'' until when we will be able to control the accuracy of the predictions and verify them experimentally. Such investigation is a self-standing reason of scientific interest in the muon collider project.

\subsection{Muon-specific opportunities\label{muspec}}

In the quest for generic exploration, engineering collisions between muons and anti-muons is in itself a unique opportunity. The concept can be made concrete by considering scenarios where the sensitivity to new physics stems from colliding muons, rather than electrons or other particles. An overview of such ``muon-specific'' opportunities is provided in Section~\ref{muonspec} based on the available literature~\cite{AlAli:2021let,Chakrabarty:2014pja,Capdevilla:2020qel,Buttazzo:2020ibd,Yin:2020afe,Capdevilla:2021rwo,Dermisek:2021ajd,Dermisek:2021mhi,Capdevilla:2021kcf,Huang:2021biu,Asadi:2021gah,Huang:2021nkl,Homiller:2022iax,Casarsa:2021rud,Han:2021lnp,Azatov:2022itm,Liu:2021akf,Cesarotti:2022ttv,Altmannshofer:2022xri,Bause:2021prv,Qian:2021ihf,Allanach:2022blr,Bandyopadhyay:2021pld,Dermisek:2022aec,Paradisi:2022vqp,Ruzi:2023atl,Qian:2022wxa}. A brief discussion is reported below.

It is worth emphasising in this context that lepton flavour universality is not a fundamental property of Nature. Therefore new physics could exist, coupled to muons, that we could not yet discover using electrons. In fact, it is not only conceivable, but even expected that new physics could couple more strongly to muons than to electrons. Even in the SM lepton flavour universality is violated maximally by the Yukawa interaction with the Higgs field, which is larger for muons than for electrons. New physics associated to the Higgs or to flavour could follow the same pattern, offering a competitive advantage to muon over electron collisions at similar energies. The comparison with proton colliders is less straightforward. By the same type of considerations one expects larger couplings with quarks, especially with the ones of the second and third generation. This expectation should be folded in with the much lower luminosity for heavier quarks at proton colliders than for muons at a muon collider. The perspectives of muon versus proton colliders are model-dependent and of course strongly dependent on the energy of the muon and of the proton collider.

Recently experimental anomalies in $g$-2 and in $B$-meson physics measurements triggered numerous studies of muon-philic new physics. These results provide interesting quantitative illustrations of the generic added value for exploration of a collider that employs second-generation particles. They show the muon collider potential to probe new physics that is presently untested because it couples mostly to muons. These models, and others with the same property, will still exist---though in a slightly different region of their parameter space---even if the anomalies will be explained by SM physics as the most recent LHCb results suggest for the $B$-meson anomalies~\cite{LHCb:2022qnv,LHCb:2022zom}.

Illustrative results are reported in Figure~\ref{fan}, displaying the minimal muon collider energy that is needed to probe different types of new physics potentially responsible for the $g$-2 anomaly. The solid lines correspond to limits on contact interaction operators due to unspecified new physics, that contribute at the same time to the muon $g$-2 and to high-energy scattering processes. Semi-leptonic muon-charm (muon-top) interactions that can account for the $g$-$2$ discrepancy can be probed by di-jets at a $3$~TeV ($10$~TeV) MuC, whereas a 30~TeV collider could even probe a tree-level contribution to the muon electromagnetic dipole operator directly through $\mu\mu\to h\gamma$. These sensitivity estimates are agnostic on the specific new physics model responsible for the anomaly. Explicit models typically predict light particles that can be directly discovered at the muon collider, and correlated deviations in additional observables. We will see in Section~\ref{muonspec} that a complete coverage of several models that accommodate the current discrepancy is possible already at a $3$~TeV MuC, and a collider of tens of TeV could provide a full-fledged no-lose theorem.


\clearpage


\section{Facility}\label{sec_fac}
\begin{figure*}
        \includegraphics[width=0.75\textwidth, trim={0cm 0 0 0}]{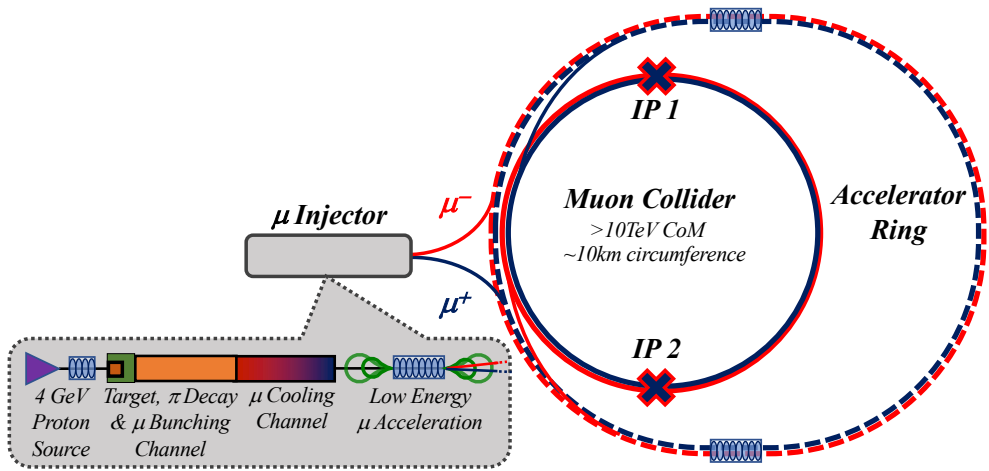}
        \centering\textbf{}
\caption{A conceptual scheme for a muon collider.}
\label{fig_muon:sketch}
\end{figure*}

The Muon Accelerator Programme (MAP) has developed a concept for the muon collider, shown in Figure~\ref{fig_muon:sketch}. This concept serves as the starting point for the baseline concept and a seed for the tentative parameters in the design studies initiated by the International Muon Collider Collaboration (IMCC)~\cite{IMCC}. The tentative target parameters are reported in Table~\ref{MC:t:param}.

This section describes the physical principles that motivate the present baseline design and outlines promising avenues that may yield improved performances or efficiency. Technical issues are also highlighted. Section~\ref{sec:fac_design_overview} provides an overview of the overall design concept. An approximate expression for luminosity is derived in some detail, as this motivates many of the design choices. Consideration of attainable energy, facility scale and power requirements are described. Possible upgrade schemes and timescales are outlined. The detailed description of design concepts and technical issues surrounding each subsystem are reported in Sections~\ref{sec:fac_proton_driver} -- \ref{sec:fac_collider}, describing the proton source, target, front end, muon cooling system, acceleration and collider in turn. The concepts and technologies developed for the muon collider will require technical demonstration, to be achieved by a number of demonstrator facilities described in Section~\ref{sec:fac_demonstrators}. Section~\ref{sec:fac_synergies} is devoted to the synergies between the R\&{D} programme for the muon collider and the one of other muon-beam facilities. We summarise our conclusions in Section~\ref{sec:conc_out}.

\subsection{Design overview}
\label{sec:fac_design_overview}
Most muon collider designs foresee that muons are created as a product of a high power proton beam incident on a target. Most muons are produced by decay of pions created in the target. In order to capture a large number of negatively and positively charged secondary particles over a broad range of momenta, a solenoid focusing system is used rather than the more conventional horn-type focusing. Following the target the beam is cleaned and most pions decay leaving a beam composed mostly of muons. The muons are captured longitudinally in a sequence of RF cavities arranged to manipulate the single short bunch with large energy spread into a series of bunches each with a much smaller energy spread. 

Following creation of a bunch train, the beam is split into the different charge species and each charge species is cooled separately by a 6D ionisation cooling system. Ionisation cooling increases the beam brightness and hence luminosity. An initial cooling line reduces the muon phase space volume sufficiently that each bunch train can be remerged into a single bunch. Further cooling then reduces the longitudinal and transverse beam size. Final cooling systems for each charge species results in a beam suitable for acceleration and collision.

Ionisation cooling is chosen as it operates on a time scale that is competitive with the muon lifetime. Despite the short time scale, a significant number of muons are lost due to muon decay as well as transmission losses. Nonetheless the increased beam brightness provided by the cooling system yields a significant increase in luminosity.

The beam is then accelerated. Rapid acceleration is required in order to maintain an acceleration time that is much shorter than the muon lifetime in the laboratory frame. Satisfactory yields may be achieved by leveraging the muon lifetime increase during acceleration due to Lorentz time dilation to maximise the acceleration efficiency $\eta_\tau$. The number of muons $N$ changes with time $t$ according to
\begin{equation}
\frac{dN}{dt} = -\frac{m_\mu N c^2}{E \tau_\mu}\,,
\end{equation}
where $m_\mu$, $E$ and $\tau_\mu$ are the muon mass, energy and lifetime respectively. Assuming the muons are travelling, near to the speed of light $c$, through a mean field gradient $\bar{V}$, their energy changes as $dE/dt=e\bar{V}c$, where $e$ is the muon charge. We thus find
\begin{equation}
\frac{dN}{dE} = - \frac{N}{\delta_\tau E} \,,
\end{equation}
where $\delta_\tau = e\bar{V} \tau_\mu/m_\mu c$ is the mean change in energy in one muon lifetime, normalised to the muon rest energy. Integrating yields the acceleration efficiency
\begin{equation}
\eta_\tau =  \frac{N_\pm}{N_{0\pm}} =  \prod_i \left(\frac{E_{i+1}}{E_{i}}\right)^{-1/\delta_{\tau,i}}\,,
\end{equation}
where the product is taken over all accelerator subsystems. Mean gradients $\bar{V} \sim O(1-10)$ MV/m are possible, with higher gradients available in the early parts of the accelerator chain, yielding $\delta_\tau \sim O(10)\gg1$. Muons can thus be accelerated faster than they decay, entailing a limited loss of muons during the acceleration process. 

At low energy rapid acceleration is achieved using a linear accelerator, in order to maximise the average accelerating gradient in this relatively short section. At higher energies recirculation may be used to improve the system efficiency, for example in a dogbone recirculator. Finally, a sequence of pulsed synchrotrons bring the beam up to final energy. Synchrotrons that employ a combination of fixed high-field, superconducting dipoles and lower field pulsed dipoles are under study. The muons are eventually transferred into a low circumference collider ring where collisions occur.

\subsubsection*{Luminosity}
The muon collider benefits from significant luminosity even at high energies. Many of the design parameters for a muon collider are driven by the need to achieve a good luminosity. An approximate expression for luminosity may be derived to inform design choices and highlight the critical parameters for optimisation. In particular proton sources are a relatively well-known technology, with examples such as SNS and JPARC in a similar class to the proton driver required for the muon collider. Muon beam facilities comparable to the muon collider have instead never been constructed. In order to quantify the required performance for a muon collider facility, it is convenient to express the luminosity in terms of the proton source parameters and muon facility performance indicators, for example the final muon energy, muon collection efficiency and muon beam quality.

\begin{table*}
\begin{center}
  \begin{tabular}{|*6{c|}}
    \hline
    Parameter & Symbol & Unit &\multicolumn{3}{c|}{Target value} \\
    \hline
    Centre-of-mass energy & $E_\text{cm}$ & \si{\tera\electronvolt} & 3 & 10 & 14 \\
    Luminosity & $\mathfrak{L}$ & \SI{e34}{\per\square\centi\meter\per\second} & 2 & 20 & 40 \\
    Collider circumference& $C_\text{coll}$ & \si{\kilo\meter} & 4.5 & 10 & 14 \\
    \hline
    Muons/bunch & $N_{\pm}$ & \num{e12} & 2.2 & 1.8 & 1.8 \\
    Repetition rate & $f_\text{r}$ & \si{\hertz} & 5 & 5 & 5 \\
    Total beam power  & $P_{_-}+P_{_+}$ & \si{\mega\watt} & 5.3  & 14 &20 \\
    Longitudinal emittance& $\varepsilon_\text{l}$ & \si{\mega\electronvolt\meter} & 7.5 & 7.5 & 7.5 \\
    Transverse emittance& $\varepsilon_\perp$ & \si{\micro\meter} & 25 & 25 & 25 \\
    \hline
    IP bunch length& $\sigma_z $ & \si{\milli\meter} & 5 & 1.5 & 1.1 \\
    IP beta-function& $\beta^*_{\perp} $ & \si{\milli\meter} & 5 & 1.5 & 1.1 \\
    IP beam size& $\sigma_{\perp} $ & \si{\micro\meter} & 3 & 0.9 & 0.6 \\
    \hline
  \end{tabular}
\end{center}
\caption{\label{MC:t:param}Tentative target parameters for MuCs of different
  energies based on the MAP design with modifications. \\[-3pt] \ }
\end{table*}

In each beam crossing in a collider the integrated luminosity increases by \cite{Wolski:2014lba}
\begin{equation}
\Delta \mathfrak{L} = \frac{N_{+,j} N_{-,j}}{4\pi\sigma^2_\perp}\,,
\end{equation}
where $N_{\pm,j}$ are the number of muons in each positively and negatively charged bunch on the $j^{th}$ crossing and $\sigma_\perp$ is the geometric mean of the horizontal ($x$) and vertical ($y$) RMS beam sizes, assumed to be the same for both charge species.

The number of particles in each beam on the $j^{th}$ crossing decreases due to muon decay as 
\begin{equation}
N_{\pm,j} = N_{\pm} \exp(-2 \pi R j/(c\gamma \tau_\mu))\,,
\end{equation}
where $R$ is the collider radius and $\gamma$ the Lorentz factor of the muons. If the facility has a repetition rate of $f_r$ acceleration cycles per second and $n_b$ bunches circulate in the collider, the luminosity will be
\begin{equation}
\mathfrak{L} = f_r n_b \frac{N_{+} N_{-}}{4\pi\sigma^2_\perp} \sum^{j_\text{max}}_{j=0} \exp\left(-\frac{4\pi R}{\gamma c \tau_\mu} j\right)\,.
\end{equation}
For the designs discussed here the muon passes around the collider ring many times ($j_\text{max}\to\infty$) so we can sum the geometric series. Furthermore, $2 \pi R/(c\gamma \tau_\mu)\ll1$, therefore to a good approximation
\begin{equation}
\mathfrak{L} \approx f_r n_b \frac{N_{+} N_{-}}{(4\pi)^2\sigma^2_\perp} \frac{\gamma c \tau_\mu}{R}\,.
\end{equation}
The average collider radius $R$, in terms of the average bending field $\bar{B}$, is $R = p/(e\bar{B}) \approx \gamma m_\mu c / (e\bar{B})$ and
\begin{equation}
\mathfrak{L} \approx f_r n_b \frac{N_{+} N_{-}}{(4\pi)^2\sigma^2_\perp} \frac{\tau_\mu e \bar{B}}{m_\mu}.
\end{equation}

The transverse beam size $\sigma_\perp$ may be expressed in terms of the beam quality (emittance) and the focusing provided by the magnets. $\varepsilon_\perp$ and $\varepsilon_l$ are the normalised emittances in transverse and longitudinal coordinates; a small $\varepsilon$ indicates a beam occupying a small region in position and momentum phase space. To a good approximation $\varepsilon$ is conserved during acceleration. The degree to which the beam is focused is denoted by the lattice Twiss parameter $\beta^*_\perp$. For a short bunch
\begin{equation}\label{sp0}
\sigma_\perp = \sqrt{\frac{m_\mu c \beta^*_\perp \varepsilon_\perp}{p}}\,.
\end{equation}
Stronger lenses create a tighter focus and make the beam size smaller at the interaction point, reducing $\beta^*_\perp$. The minimum beam size is practically limited by the ``hourglass effect''; when the focal length of the lensing system is much shorter than the length of the beam itself, the average beam size at the crossing is dominated by particles that are not at the focus \cite{Chao:2013rba}. For example, when the RMS bunch length is not zero, but $\sigma_z = \beta^*_\perp$, eq.~(\ref{sp0}) is replaced by 

\begin{equation}
\sigma_\perp = \sqrt{\frac{m_\mu c \sigma_z \varepsilon_\perp}{p f_{hg}}}\,,
\end{equation}
with a hourglass factor $f_{hg} \approx 0.76$. The RMS longitudinal emittance is $\varepsilon_l = \gamma m_\mu c^2 \sigma_\delta \sigma_z$ where $\sigma_\delta$ is the fractional RMS energy spread, so the luminosity may be expressed as
\begin{equation}
\mathfrak{L} \approx \frac{e \tau_\mu}{(4 \pi m_\mu c)^2} \frac{f_{hg} \sigma_\delta \bar{B}}{\varepsilon_\perp \varepsilon_L} {E_\mu}^2 N_+ N_- n_b f_r \,,
\end{equation}
where $E_\mu=\gamma m_\mu c^2$ is the energy of the collding muons.

Naively, the number of muons reaching the accelerator may be obtained from the number and energy of protons, i.e. from the proton beam power. This assumes proton energy is fully converted to pions and the capture and beam cooling systems have no losses. In reality pion production is more complicated; practical constraints such as pion reabsorption, other particle production processes and geometrical constraints in the target have a significant effect. Decay and transmission losses occur in the ionisation cooling system that significantly degrades the efficiency.

The final number of muons per bunch in the collider, $N_\pm$, can be related to the proton beam power on target $P_p$ and the conversion efficiency per proton per unit energy $\eta_\pm$ by 
\begin{equation}
N_{\pm} =\frac{\eta_\tau \eta_\pm P_p}{n_b f_r}.
\end{equation}

Overall the luminosity may be expressed as
\begin{equation}
\label{lumform}
\mathfrak{L} \approx 
\underbrace{
    \frac{e \tau_\mu}{(4 \pi m_\mu c)^2}
}_{K_L} 
\frac{f_{hg} \sigma_\delta \bar{B}}{\varepsilon_\perp \varepsilon_L n_b f_r}
\underbrace{
    \eta_+ \eta_- (\eta_\tau P_p E_\mu)^2 
    \vphantom{\frac{e \tau_\mu }{(m_\mu)^2}} 
}_{P_+ P_-}\,,
\end{equation}
where $K_L = 4.38 \times 10^{36} \mathrm{ \: MeV \: MW^{-2} \: T^{-1} \: s^{-2}} $ and $P_\pm$ is the muon beam power per species.

This luminosity dependence yields a number of consequences. The luminosity improves approximately with the square of energy at fixed average bending field. We thus find the desired scaling in eq.~(\ref{lums}) that entails, as discussed in the previous section, a constant rate for very massive particles pair-production, as well as a growing VBF rate for precision measurements. The quadratic scaling of the luminosity with energy is peculiar of muon colliders and it is not present, for example, in a linear collider. This is because the beam can be recirculated many times through the interaction point and beamstrahlung has a negligible affect on the focusing that may be achieved at the interaction point of the muon collider. This yields an improvement in power efficiency with energy.

The luminosity is highest for collider rings having strong dipole fields (large $\bar{B}$), so that the circumference is smaller and muons can pass through the interaction region many times before decaying. For this reason a separate collider ring with the highest available dipole fields is proposed after the final acceleration stage, as in Figure~\ref{fig_muon:sketch}.

The luminosity is highest for a small number of very high intensity bunches. The MAP design demanded a single muon bunch of each charge, which yields the highest luminosity per detector. Such a design would enable detectors to be installed at two interaction points. 

The luminosity decreases linearly with the facility repetition rate, assuming a fixed proton beam power. For the baseline design, a low repetition rate has been chosen relative to equivalent pulsed proton sources.

The luminosity decreases with the product of the transverse and longitudinal emittance. It is important to achieve a low beam emittance in order to deliver satisfactory luminosity, while maintaining the highest possible efficiency $\eta_\pm$ of converting protons to muons.

\begin{figure}
\centering
        \includegraphics[width=0.45\textwidth, trim={0 0 0 0}]{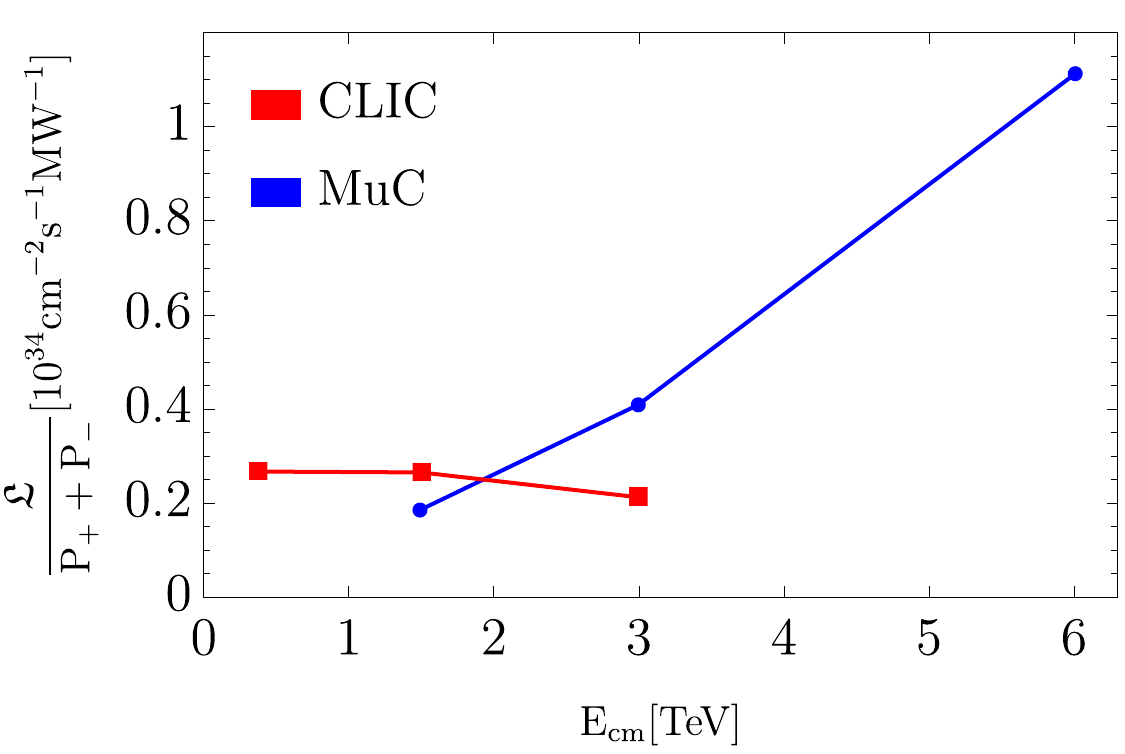}
        \centering

\caption{MuC luminosity normalised to the muon beam power and compared to CLIC, for different beam energies.}
\label{fig_muon:power_luminosity}
\end{figure}

Based on these considerations, an approximate guide to the luminosity normalised to beam power is shown in Figure~\ref{fig_muon:power_luminosity} and compared with the one of CLIC.

\subsubsection*{Facility size}

The geometric dimensions of the MuC depend on future technology and design choices. Some indication of the dimensions can be estimated. The facility scale is expected to be driven by the pulsed synchrotrons in the acceleration system.

The rapid pulsing required in the synchrotrons precludes the use of high-field ramped superconducting magnets such as those used in the LHC. Static high-field superconducting dipoles are proposed combined with rapidly pulsed low-field dipoles. As the beam accelerates the pulsed dipoles are ramped, enabling variation of the mean dipole field. The static dipoles provide a relatively compact and efficient bend. 

Preliminary estimates indicate that acceleration up to 3~TeV centre-of-mass energy, assuming 10~T static dipoles and pulsed dipoles with a field swing $\pm$1.8~T, would require a ring of circumference around 10~km. Around 60~\% of the ring is estimated to be required for pulsed and static bending dipoles. Acceleration up to 10~TeV centre-of-mass energy would require 16~T static dipoles, which are only expected to become available later in the century,  and an approximately 70~\% dipole packing fraction. A 10~TeV facility could be implemented as an upgrade to the 3~TeV facility, as discussed below. Estimates indicate that a ring circumference of up to 35~km may be required. Tuning to accommodate the beam into existing infrastructure such as the 26.7~km circumference LHC tunnel is possible, for example by changing the energy swing in each ring so that more or less space is required for pulsed dipoles. Options that have a fixed dipole field that varies radially in the same superconducting magnet are under study, which may enable an increased average dipole field to be considered.

\subsubsection*{Wall-plug power requirements}
The power usage of future accelerator facilities, often referred to as the `wall-plug power', is of great concern and in future may be a stronger practical constraint than the financial cost. The goal is to remain at a wall-plug power consumption for the 10~TeV MuC well below the level estimated for CLIC at 3~TeV (550~MW) or FCC-hh (560~MW). This seems readily achievable; the facility length is considerably shorter than other proposed colliders. Reuse of magnets and RF should make the facility more efficient than linear colliders while the lower energy requirement should result in lower power requirements than hh colliders. The design has to advance more to assess the power consumption scale in a robust fashion.

Muon beam production imposes a fixed power consumption requirement. In particular, the muon cooling system requires several GeV of acceleration in normal conducting cavities at high gradient. Additional power consumption arises from the proton source and cooling plant for the target and other cryogenic systems.

A number of key components drive the power consumption as one extends to high energy:
\begin{itemize}
\item The power loss in the fast-ramping magnets of the pulsed synchrotron and their power converter.
\item The cryogenics system that cools the superconducting
magnets in the collider ring. This depends on the efficiency of shielding the magnets from the muon decay-induced heating.
\item The cryogenics power to cool the superconducting magnets and RF cavities in the pulsed synchrotrons.
\item The power to provide the RF for accelerating cavities in the pulsed synchrotrons.
\end{itemize}
The first contribution requires particular study as it depends on unprecedented large-scale fast ramping systems. The second and third contributions require optimisation of the volume reserved for shielding as compared to magnetised volume. The contributions can be estimated reliably following a suitable design and optimisation of the relevant equipment.

\begin{figure*}[t]
\begin{minipage}{0.47\textwidth}
\begin{center}
\includegraphics[width=\textwidth,trim={100 0  170 0}, clip]{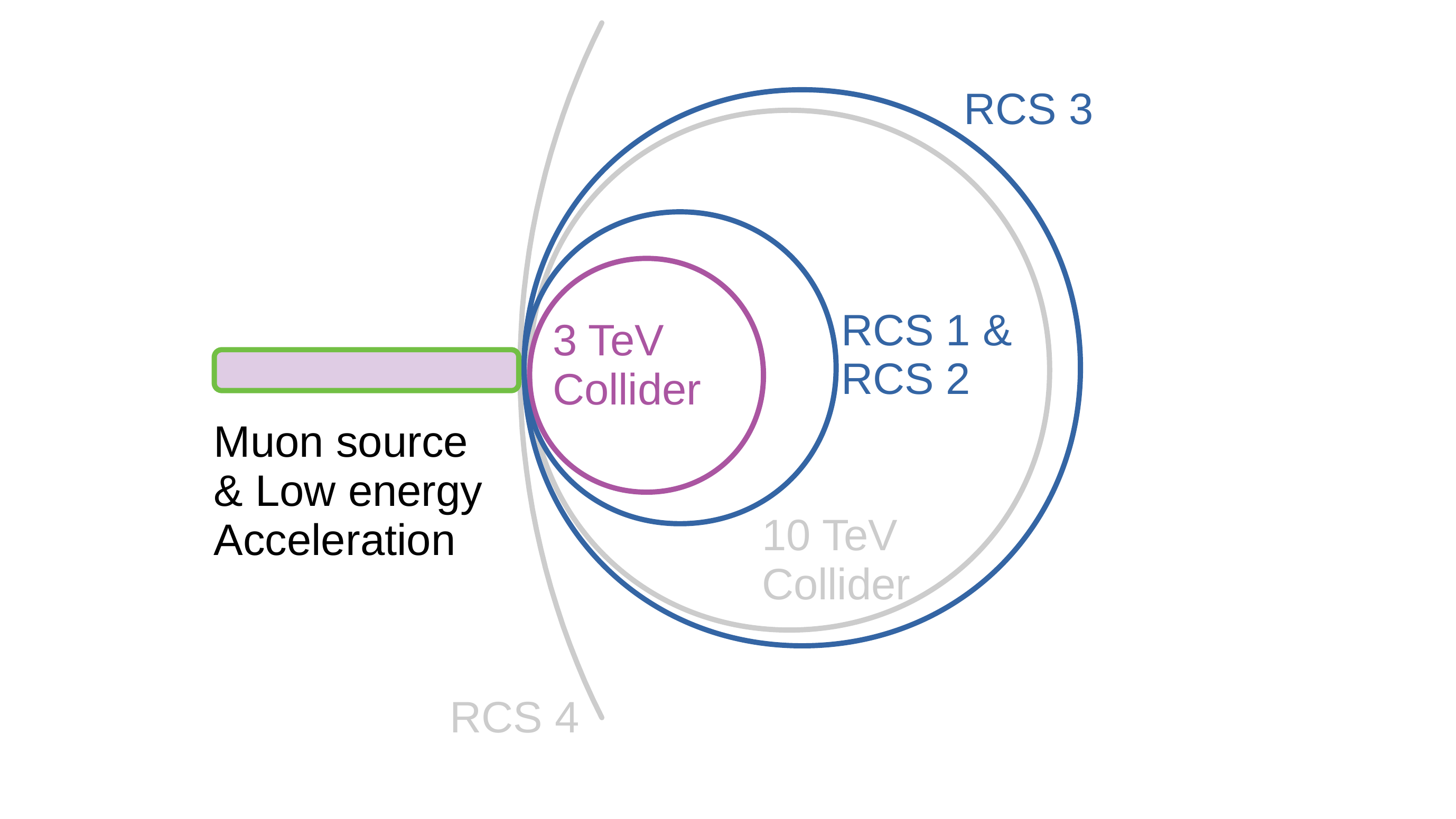}
\end{center}
\end{minipage}
\hfill
\begin{minipage}{0.47\textwidth}
\begin{center}
\includegraphics[width=\textwidth,trim={110 0  160 0}, clip]{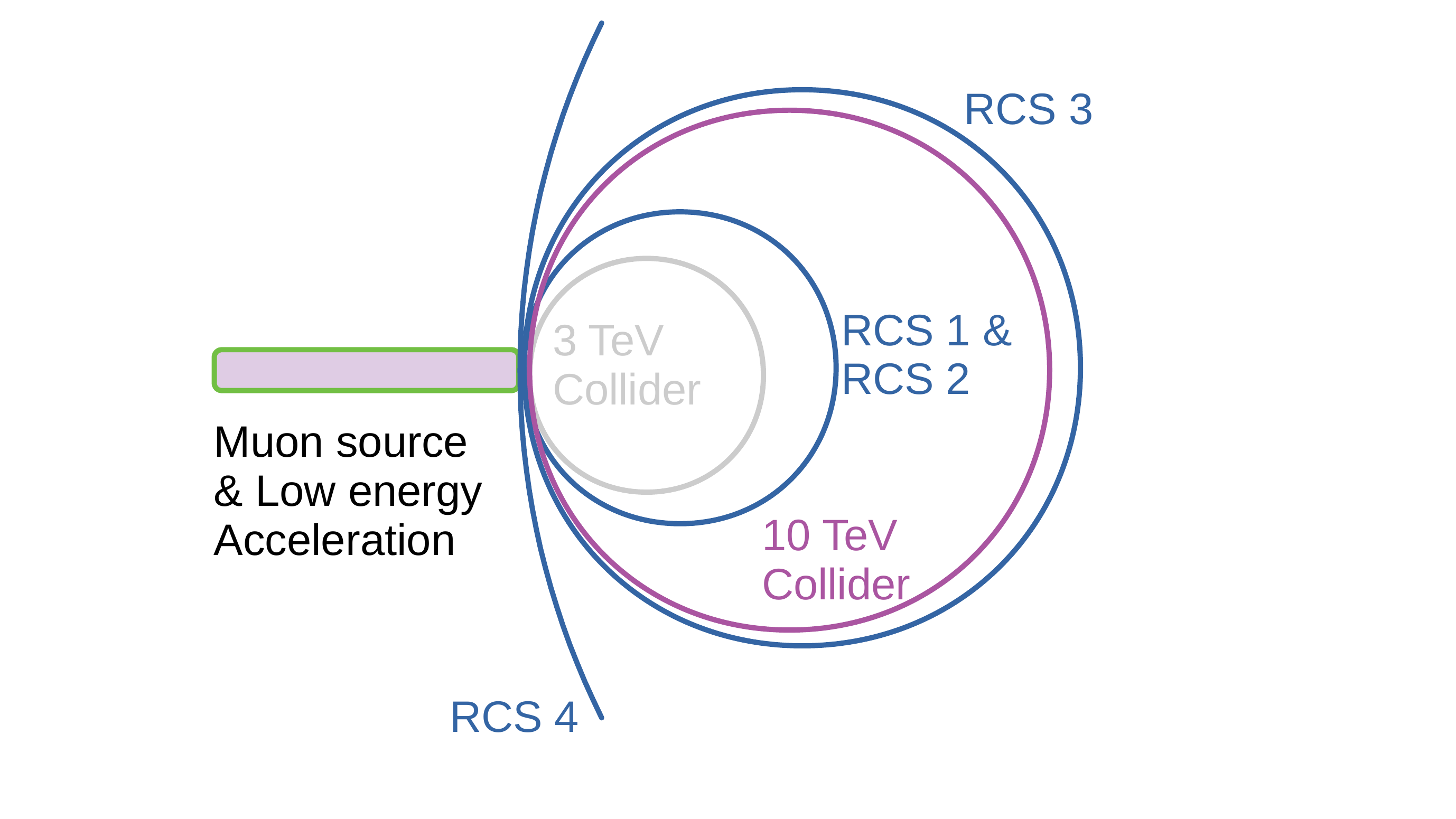}
\end{center}
\end{minipage}
\caption{A schematic of a possible staged approach to a muon collider. The first stage, shown on the left, would produce collisions at 3 TeV center-of-mass energy while the second would produce collisions at 10 TeV centre-of-mass energy. Sections of the facility that are not required are shown in grey.
\label{fig_muon:upgrade}}
\end{figure*}

In summary power consumption follows a relation
\begin{equation}
P = P_{src} + P_{linac} + P_{rf} + P_{rcs}+P_{coll}\,,
\end{equation}
where $P_{src}$ is the power needed for the muon source, which is constant with energy, $P_{linac}$ is the power requirement for the linac, which is fixed by the transition energy between the linacs and RCS (Rapid Cycling Synchrotron). $P_{rf}$ is the power requirement for RCS and collider RF cavities, which is approximately proportional to the beam energy, $P_{rcs}$ is the power requirement for the RCS magnets which is likely to rise slowly with energy due to the lower losses associated with slower ramping at high energy and $P_{coll}$ is the power requirement for the collider ring, which is approximately proportional to the collider length i.e. proportional to the beam energy. Overall, the power requirement is expected to rise slightly less than linearly with energy and hence the wall-plug power is expected to be approximately proportional to the beam power.

\subsubsection*{Upgrade scheme}

The muon collider can be implemented as a staged concept providing a road
toward higher energies. One such staging scenario is shown in Figure~\ref{fig_muon:upgrade}. The accelerator chain can be expanded by an additional accelerator ring for each energy stage. A new collider ring is required for each energy stage. It may be possible to reuse the magnets and other equipment of the previous collider ring, for example in the new accelerator ring.

Currently, the focus of studies is on 10~TeV. This energy is well beyond the 3~TeV of the third and highest energy stage of CLIC, the highest energy $\mathrm{e^+e^-}$ collider proposal to reach a mature design. A potential intermediate energy of 3~TeV is envisaged at this moment. Its physics case is similar to the final stage of CLIC and it is expected that this stage would roughly cost half as much as the 10~TeV stage~\cite{Roser:2022sht}.

A 3~TeV stage is less demanding in several technological areas. It may not be necessary to implement a mechanical neutrino flux mitigation system in the collider ring arcs; moving the beam inside of the magnet apertures may be sufficient. The final focus magnets require an aperture and a gradient comparable to the values for HL-LHC. In general, it will be possible to implement larger margins in the design at 3~TeV. The operational experience will then allow to accept smaller margins at 10~TeV.

The strength of the collider ring dipoles is crucial for determining the ring size and luminosity. The cost optimum is given by the magnet and tunnel cost; it is possible that cheaper, well established magnet technology---such as NbTi at this moment--- would result in a lower cost even if the tunnel has to be longer. For fixed beam current, the luminosity is proportional to the magnet field and is one aspect of the optimisation.

The Snowmass 2021 international Collider Implementation Task Force has undertaken direct comparison of all future collider proposals and concepts in terms of luminosity, environmental impact, power consumption, R\&D duration, time to construct, and cost, and its report~\cite{Roser:2022sht} indicates that a multi-TeV MuC is potentially the most promising facility for a 10+ TeV energy  range, with, among other factors, the lowest cost range.
Once the cost scale of the muon collider concept is more precisely known and once the mitigation of the technical challenges are well defined, the energy staging may be reviewed taking into account the physics case and additional considerations from the site and reuse of existing equipment and infrastructure, such as the LHC tunnel, making possible a high-impact and sustainable physics programme with each upgrade manageable in terms of cost and technical feasibility.

\begin{figure*}
        \includegraphics[width=0.7\textwidth, trim={0 0 0 0}]
        {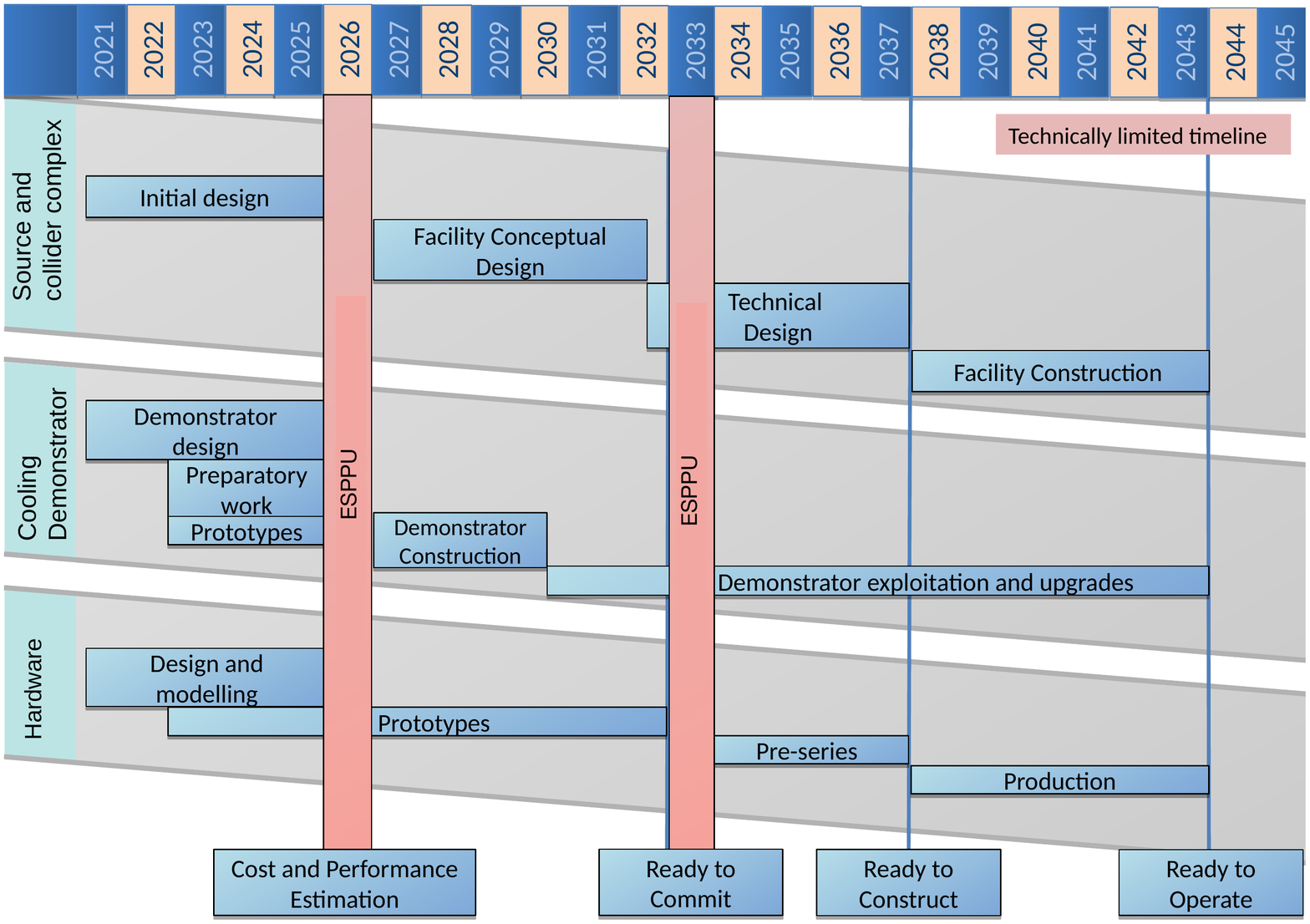}
        \centering\textbf{}

\caption{A technically limited timeline for the muon collider R\&D programme that would see a 3 TeV muon collider constructed in the 2040s.}
\label{fig_muon:RDtimeline}
\end{figure*}

\subsubsection*{Timeline}
A muon collider with a centre-of-mass energy around 3~TeV could be delivered on a time scale compatible with the end of operation of the HL-LHC. A technically limited timeline is shown in Figure~\ref{fig_muon:RDtimeline} and discussed in greater detail in \cite{Schulte:2022cbw}. The muon collider R\&D programme will consist of the initial phase followed by the conceptual and the technical design phases. The initial phase will establish the potential of the muon collider and the required R\&D programme for the subsequent phases. 

The performance and cost of the facility would be established in detail. A programme of test stands and prototyping of equipment would be performed over a five-year period, including a cooling cell prototype and the possibility of beam tests in a cooling demonstrator. This programme is expected to be consistent with the development of high field solenoid and dipole magnets that could be exploited for both the final stages of cooling and the collider ring development. A technical design phase would follow in the early 2030s with a continuing programme focusing on prototyping and pre-series development before production for construction begins in the mid-2030s, to enable delivery of a 3~TeV MuC by 2045. The programme is flexible, in order to match the prioritisation and timescales defined by the next ESPPU, the Particle Physics Projects Prioritization Panel (P5) in the US and equivalent processes.

\subsubsection*{Principal technical challenges}

The timeline described above is technically limited by the time required to address a number of key technical challenges.

\begin{itemize}
    \item The collider can potentially produce a high neutrino flux that might lead to neutron showering far from the collider. A scheme is under study to ensure that the effect is negligible.
    \item Beam impurities such as products of muon decay may strike the detector causing beam-induced background. The detector and machine need to be simultaneously optimised in order to ensure that the physics reach is not limited by this effect.
    \item The collider ring and the acceleration system that follows the muon cooling can limit the energy reach. These systems have not been studied for \SI{10}{TeV} or higher energy. The collider ring design impacts the neutrino flux and the design of the machine-detector interface.
    \item The production of a high-quality muon beam is required to achieve the desired luminosity. Optimisation and improved integration are required to achieve the performance goal, while maintaining low power consumption and cost. The source performance also impacts the high-energy design.
\end{itemize}

The technology options and mitigation measures that can address these challenges are described in some detail below. Further dedicated studies of the key technologies would provide a robust quantitative assessment of their maturity and technical risk, for example based on Technical Readiness Levels (TRL) as described in~\cite{Roser:2022sht}.

\subsection{Proton driver}
\label{sec:fac_proton_driver}
The MuC proton driver has similarities with existing and proposed high intensity proton facilities such as those used for neutron and neutrino production. 
The main parameters of the MuC proton source are listed in Table~\ref{MC:t:proton_param}. The technology choices for the MuC, in particular for acceleration and bunch compression, have not yet been determined. 

The main part of the proton source follows existing pulsed proton driver designs, for example similar to JPARC, Fermilab or ISIS. H$^-$ ions are created in an ion source, accelerated through a radiofrequency quadrupole followed by a series of drift-tube linacs. Acceleration proceeds through a linear accelerator using conventional RF cavities before the ions are injected into a ring using charge-exchange injection and phase space painting. In some designs, the protons are further accelerated using a Rapid Cycling Synchrotron or Fixed Field Altenating Gradient accelerator while in others protons are accelerated to the top energy in the linac and the final ring is used only for accumulation of the protons. Uniquely for the muon collider, it is desirable to compress the protons into a very short bunch with RMS length 1-3~ns, which may require an additional ring. However, bunches with lengths of the order of 30~ns would reduce the produced muon yield only by up to 50\%~\cite{Sayed:2013cdq,Sayed:2014wya}. The bunch is then extracted and transferred onto the pion production target where the short proton bunch in turn creates a short pion bunch, which is important for capture of the resultant beam.

\begin{table}
\caption{Typical proton source parameters. The parameters are indicative.\\[-4pt] \ }
\label{MC:t:proton_param}
\begin{center}
  \begin{tabular}{|c|c|c|}
    \hline
    Parameter & Unit & Target value \\
    \hline
    Beam power & \si{MW} & 2-4 \\
    Energy & \si{GeV} & 5-30 \\
    Repetition Rate & \si{Hz} & 5-10 \\
    RMS bunch length & \si{ns} & 1-3 \\
    \hline
  \end{tabular}
\end{center}
\end{table}

\subsubsection*{Technical issues and required R\&D}
The technology choice for the MuC proton driver relies on successfully managing the heat deposition in the injection foil in the accumulator ring, limits on beam intensity due to space charge and the successful compression of the proton bunch. The MuC uses a beam that has a higher intensity at lower repetition rate than comparable machines, resulting in a space-charge dominated facility. At higher energy space charge limitations may be relaxed, but care must be taken to avoid uncontrolled stripping of the  H$^-$ beam and excessive heat deposition that can damage the foil.

In order to achieve the high current and short bunches in the presence of space charge, the MAP scheme used a high harmonic lattice with a series of extraction lines to bring multiple proton bunches onto the target, each extraction line having a different time-of-flight, enabling the required proton beam current to be brought onto the target within a short time.

The possibility to stack bunches in longitudinal space, for example using an FFA ring that has a naturally large momentum acceptance \cite{Ferger:1963yza}, may enable repetition rates lower than the baseline 5 Hz. Time-compression of a beam having larger momentum spread would be more challenging.

Bunch compression has been performed, for example at ISIS and the SPS, yielding short bunches, but not as short as those required for the muon collider. Simulations indicate that such a compression is possible \cite{Schonauer:2000be, Aiba:2008zzc, Lebedev:2009zzd}.

In order to deliver a self-consistent design, simulations must be performed for each parameter set. Many potential host sites have existing facilities which could be reused, given appropriate consideration of the muon collider requirements.

\subsection{Pion production target and active handling region}
The MuC design calls for a target immersed in a magnetic field of 15 -- 20~T  where muons, pions, kaons and other secondary particles are produced \cite{Sayed:2014wya}. Unlike more conventional horn focusing arrangements the solenoid field captures secondary particles of both signs over a wide range of momenta. The pions and kaons decay to produce muons. In order to create a beam having the smallest emittance, a very short proton bunch is required having a small spot size. Typical proton bunch sizes considered are a few mm across with lengths 1 -- 3~ns RMS. The resultant secondary particles have a large transverse and longitudinal momentum spread, but initially the position spread is the same as the proton beam. The time spread of the beam is approximately given by the proton bunch length. Some non-relativistic secondary particles are produced which will pass through the target region more slowly than the protons, resulting in a slight bunch lengthening.

Following the target the magnetic field is tapered to values sustainable over long distances. If the field strength is tapered adiabatically along the beam line, the relativistic magnetic moment
\begin{equation}
I_0 = \frac{p_\perp^2}{m_\mu qB}\,,
\end{equation}
is an invariant. As the field $B$ is reduced, the transverse momentum also reduces so that $I_0$ does not change. Magnetic fields conserve total momentum, so reduction in transverse momentum $p_\perp$ must lead to an appropriate increase in longitudinal momentum. For particles that initially have small transverse momentum, the longitudinal momentum growth will be small, while for those with large transverse momentum, the growth will be large. Overall, the longitudinal momentum spread increases while the transverse momentum spread decreases. This is equivalent to an exchange from transverse emittance to longitudinal emittance. The longitudinal emittance is already large, so the relative increase is small, while the decrease in transverse emittance is beneficial and results in a large flux of muons compared to horn geometries.

The beam leaving the target area has a large longitudinal and transverse emittance. A number of undesirable particles are captured in the beam that could be lost in an uncontrolled manner further downstream. In order to handle these unwanted particles, a beam cleaning system is foreseen. A solenoid-style double chicane is used to remove particles having too high momentum. The chicane has a toroidal solenoid field which induces a vertical dispersion in the beam. The vertical dispersion is exactly cancelled by a reverse bend yielding, for particles lower than the maximum momentum, a virtually unperturbed beam. High momentum particles strike the walls of the chicane where they are collected on a dedicated collimator system.

A number of low momentum beam impurities remain in the beamline. Protons arising from spallation of the target would create an unacceptable radiation hazard if they were lost in subsequent systems. A Beryllium plug is placed immediately after the chicane to remove these low momentum protons. The mean energy loss is given by \cite{ParticleDataGroup:2022pth}
\begin{equation}
\frac{dE}{dx} = \frac{Z}{A}\frac{K q^2}{\beta^2 \rho} \left[0.5\,\ln\left(2 m_e \beta^2 \gamma^2 \frac{T_{max}}{I^2}\right) - \beta^2\right]\,.
\end{equation}
Due to their larger mass, protons have much lower $\beta$ and less kinetic energy compared to muons following the chicane and so stop in a shorter distance. By choosing an appropriate thickness, protons are stopped while most muons are able to pass through. A low Z material such as beryllium is considered as it causes less scattering in the muon beam for a given proton stopping power.


\subsubsection*{Alternative schemes}
In addition to a long graphite target of the type used in pion production targets for neutrino beams, moving targets have been considered. The required dimensions of the target make a rotating wheel such as the one in use at PSI challenging to implement. Liquid eutectic targets or fluidised powder jets have been considered, travelling coaxially with the proton beam.

Focusing using a horn may also be considered, as used in other pion production targets. A horn target is a well-known technology but it is only capable of capturing a single sign of pion. Two targets would likely be required, leading to an increased demand on the proton driver.

\subsubsection*{Technical issues and required R\&D}
The target region is technically challenging. The high beam power and short proton pulse length will create a significant instantaneous shock in the target, which may lead to mechanical damage. The target will become heated by the proton beam and active cooling is expected to be required. Long term radiation damage will degrade the mechanical qualities of the target necessitating regular replacement. Calculations indicate that a graphite target, similar to the targets used at existing neutrino sources, should survive for an adequate period in the adverse conditions.

The high field in the target region, in the presence of a radiation source, is challenging to achieve. Thick shielding will be required to prevent damage to the insulation and superconductor in the coil and excessive heating of the cryogenic materials. A large aperture is required to accommodate the shielding, and this in turn poses challenges for the magnet. A magnet bore diameter in the range of 2 m would be required in case of coils built with low-temperature superconductor, and operated with liquid helium.  Alternative conductor configurations, based on high-temperature superconductor with large operating margin may be operated at higher temperature with improved efficiency. This may provide means to significantly decrease the size and cost of the system. The target magnet would be in a similar class to those proposed for fusion facilities.  

Handling of the spent proton beam requires significant care. The remnant beam power is expected to be beyond the capabilities of a collimation scheme to manage. A system for removing these primary protons to a beam dump must be considered. Such a system has to extract the protons without adversely affecting the pion and muon transport in the line.


Even following the removal of primary protons, significant radiation will impinge on the chicane aperture. In this region only modest fields are required but nonetheless an appropriate shielding and collimation scheme must be designed.

The Low EMittance Muon Accelerator (LEMMA) is an alternative scheme to produce a muon beam with a very small emittance \cite{Boscolo:2018tlu, Boscolo:2020xfv, Amapane:2021npr}. An injector complex produces a high-current positron beam \cite{Amapane:2019oog}. The positrons impact a target with an energy of 45 GeV, sufficient to produce muon pairs by annihilating with the electrons of the target. This scheme can produce small emittance muon beams. However, it is difficult to achieve a high muon beam current and hence competitive luminosity. Novel ideas are required to overcome this limitation.

\subsection{Muon front end}
Following the target, the muon front end system captures the beam into a train of bunches suitable for ionisation cooling. This is achieved in three stages. First the beam is allowed to drift longitudinally. Faster particles reach the subsequent RF system first. Once a suitable time-energy correlation has developed, RF cavities are placed sequentially with adiabatically increasing voltage as the beam passes along the line. Because the beam is not captured longitudinally, the RF frequency of subsequent cavities is lower so that the evolving buckets are correctly phased. This adiabatic capture results in microbunches forming in the bunch train. Once suitable microbunches have been formed, the cavities are dephased so that higher energy, early bunches, are at a decelerating phase. Lower energy, later bunches are at an accelerating phase. At the end of the capture system the resulting microbunches have the same energy, with spacings appropriate for 325 MHz RF cavities. 

Following the capture system a charge selection system is required to split the beam into a positive and negative muon line. A second solenoid chicane system is envisaged. Unlike in the beam cleaning system, the bunch structure in the beam would be maintained by this system, requiring appropriate RF cavity placement.

\subsubsection*{Alternative schemes}
Direct capture into 44 MHz RF buckets has been considered. These lower frequency RF systems may have a larger muon energy and time acceptance than those at few 100 MHz, despite the lower voltages which may be achieved before breakdown occurs. Use of a single RF frequency would be simpler to construct and operate than the system described above, requiring only a single RF frequency. However, such a system would yield bunches having a larger longitudinal emittance, which would need to be cooled. The efficacy of the cooling system is directly related to the voltage available, which would be lower for 44 MHz than 200-300 MHz proposed in the design described above.

\subsubsection*{Technical issues and required R\&D}
The front end system design itself is mature, although re-optimisation is likely to be required as the other facility parameters are developed. The charge selection system has received only preliminary conceptual work and a full design is required. Use of many RF frequencies would require a dedicated power source for each frequency, which may be costly to implement.

\subsection{Muon cooling}
\label{sec:fac_cooling}
The beam arising from the muon front end occupies a large volume in position-momentum phase space. According to Liouville's theorem, in a non-dissipative system, phase space volume is conserved. In order to achieve a satisfactory luminosity it is necessary to reduce the phase space volume of the muon beam, a process known as beam cooling. Typical cooling techniques require time scales that are not competitive with the muon life time. Ionisation cooling, a relatively novel technique, is proposed to reduce the phase space volume of the muon beam.

\subsubsection*{Transverse cooling}
In muon ionisation cooling, muons are passed through an energy-absorbing material. The transverse and longitudinal momentum of the muon beam is reduced, reducing the phase space volume occupied by the beam in the direction transverse to the beam motion. The muon beam is subsequently re-accelerated in RF cavities. The cooling effect is reduced by multiple Coulomb scattering of the muons off nuclei in the absorber, which tends to increase the transverse momentum of the beam. By using a material having low atomic number and focusing the beam tightly, the effect of scattering can be minimised. 

The Muon Ionisation Cooling Experiment (MICE) demonstrated the principle of transverse ionisation cooling~\cite{MICE:2019jkl}. In MICE, the particles were passed through a solenoid focusing system. The particles were incident on various configurations of absorbers and focusing arrangements. Configurations with a lithium hydride cylinder and a liquid hydrogen-filled thin-walled vessel were compared with configurations having an empty vessel and no absorber at all. The particle position and momenta were measured upstream and downstream of the focus and the distance from the beam cores were calculated in normalised phase space coordinates. By examining the behaviour of the ensemble of particles, an increase in phase space density in the beam core could be identified when an absorber was installed, indicating ionisation cooling had taken place. No such increase was measured when no absorber was installed, as expected. The results were consistent with simulation.

Phase space volume is conveniently quantified by the beam RMS emittance, given by
\begin{equation}
\varepsilon = \sqrt[n]{|\mathbf{V}|}/m_\mu\,,
\end{equation}
where $\mathbf{V}$ is the matrix of covariances of the phase space vector, having elements $v_{ij} = \mathrm{Cov}(u_i, u_j)$. $\vec{u}$ is the $n$-dimensional phase space vector under consideration. Typically either the four-dimensional transverse vector, $(x, p_x, y, p_y)$ or the two-dimensional longitudinal vector, $(c(t-t_0), \delta)$, is considered, where $x$ and $y$ are the horizontal and vertical positions relative to the beam axis, $p_x$ and $p_y$ are the corresponding momenta, $t-t_0$ is the time relative to some reference trajectory's time $t_0$ and $\delta = E-E_0$ is the energy relative to the reference trajectory's energy $E_0$. This quantity is proportional to the content of an elliptical contour in phase space that is one standard deviation from the reference trajectory. In the limit that the beam is paraxial, it is a conserved quantity in the absence of dissipative forces.

Skrinsky et al \cite{Skrinsky:1981ht} first introduced the concept of ionisation cooling. Neuffer  \cite{Neuffer:1983jr} derived equations that characterise the emittance change on passing through an absorber in terms of the beam energy, the normalised beam size and the properties of the energy absorbing material. The change in transverse emittance on passing through an axially symmetric absorber with radiation length $L_R$ is
\begin{equation}
\frac{d\varepsilon_n}{dz} \approx \frac{1}{\beta^2 E} \left<\frac{dE}{dz}\right>(1-\frac{g_L}{2})\varepsilon_n + \frac{(13.6 \,{\rm{MeV}})^2}{2\,m_\mu L_R} \frac{\beta_\perp}{\beta^3 E}\,.
\label{eq:emittance_change}
\end{equation}
Longitudinal cooling, discussed below, may be achieved by arranging for a correlation between energy and energy loss and is parameterised by $g_L$, the longitudinal partition function. In the absence of longitudinal cooling $g_L$ may be assumed to be 0. Assuming that beam energy is continuously replaced by RF cavities the emittance change is 0 at the equilibrium emittance
\begin{equation}
\varepsilon_{n,eqm} \approx \frac{1}{2m_\mu} \frac{(13.6 \,{\rm{MeV}})^2}{L_R} \frac{\beta_\perp}{\beta \left<\frac{dE}{dz}\right>(1-\frac{g_L}{2})}\,,  \label{eq:eqm_emittance}
\end{equation}
where emittance growth due to scattering and emittance reduction due to ionisation are equal.

Two principal cooling stages are proposed for the MuC. In the first stage, known as rectilinear cooling, muons are cooled both transversely and longitudinally. In the second stage, known as final cooling, muons are cooled transversely and heated longitudinally.

Both cooling systems use solenoids with approximately cylindrically symmetric beams in order to provide the tight focus required for satisfactory cooling performance. The focusing is characterised by the beam transverse $\beta_\perp$ function. This is the variance of the beam size, normalised to the beam emittance
\begin{equation}
\beta_\perp = \frac{p_z \mathrm{Var}(x)}{m_\mu c\, \varepsilon_n} = \frac{p_z \mathrm{Var}(y)}{m_\mu c \,\varepsilon_n}\,.
\end{equation}
In a solenoid, $\beta_\perp$ evolves according to
\begin{equation}
2 \beta_\perp \beta_\perp'' - \beta_\perp'^2 + 4 \beta_\perp^2 \kappa^2 - 4(1+\mathcal{L}^2) = 0\,.
\label{eq:beta_sol}
\end{equation}
Here $\mathcal{L}$ is the normalised canonical angular momentum. $\kappa^2$ is the solenoid focusing strength, where
\begin{equation}
\kappa = \frac{qc B_z(r=0, z)}{2p_z}\,.
\end{equation}
Particles in the solenoid may be considered as oscillators with angular phase advance of the oscillator over a distance $z_0$
\begin{equation}
\phi = \int_0^{z_0} \frac{1}{\beta_\perp}dz\,.
\end{equation}
Oscillations are known as betatron oscillations.

Wang and Kim \cite{Wang:2001vt} showed that for periodic systems, solutions to eq.~(\ref{eq:beta_sol}) can be related to the Fourier coefficients $\vartheta_n$ of $\kappa$
\begin{equation}
\beta_\perp(z=0) \approx \frac{z_0}{\pi} \frac{\sin(\sqrt{\vartheta_0}\pi)}{\sqrt{\vartheta_0}\sin\mu}
\left[1+\sum^{\infty}_{n=1}\frac{Re[\vartheta_n]}{n^2-\vartheta_0}\right]\,.
\end{equation}
Explicitly $\vartheta_n$ are defined by
\begin{equation}
\left(\frac{L}{\pi}\right)^2 \kappa^2(z) = \sum^{\infty}_{-\infty} \vartheta_n e^{i 2n\pi s/L}\,.
\end{equation}
There are values of $\kappa$ where the solenoid focusing is unstable known as stop-bands. In these regions, particle motion follows hyperbolae in phase space and the phase advance is complex. These regions can be expressed in terms of the Fourier coefficients
\begin{equation}
\left|\sqrt{\vartheta_0}-n+\frac{5}{16}\left|\frac{\vartheta_n}{\vartheta_0}\right|^2\right|<\frac{1}{2}\left|\frac{\vartheta_n}{\vartheta_0}\right|\,.
\end{equation}
Ionisation cooling systems are often characterised in terms of the phase advance and the nearby stop bands. The pass band in which the cooling cell operates can be selected for properties of acceptance and achievable $\beta_\perp$. The pass band is determined in practice by scaling the average $B^2_z$ and cell length in relation to the beam central $p_z$. The properties can be tuned by adjusting the harmonic content of $B^2_z(z)$. The rectilinear cooling system has been developed with particular attention to the resonance structure in order to minimise $\beta_\perp$.

\subsubsection*{Longitudinal cooling}
The process described above only reduces transverse emittance. In order to reach suitable luminosity, both transverse and longitudinal beam emittance must be reduced. This may be achieved by exchanging emittance from longitudinal to transverse phase space.

Emittance exchange is achieved in two steps. First a position-energy correlation is introduced into the muon beam using a dipole. Higher energy particles have a larger bending radius, so that a correlation is introduced between transverse position and energy. An appropriately arranged wedge-shaped absorber is used so that the higher energy particles pass through a thicker part of the wedge, losing more energy. In this way the energy spread is reduced and the position spread is increased; the emittance is moved from longitudinal space to transverse space.

The horizontal dispersion can be related to the beam by defining 
\begin{equation}
D_x = \frac{1}{p}\mathrm{Var}(x, E)\,,
\end{equation}
with a similar relationship for vertical dispersion. The change in longitudinal emittance is 
\begin{equation}
\frac{d\varepsilon_L}{dz} = -\frac{g_L}{\beta^2 E} \frac{dE}{dz} \varepsilon_L + \frac{\beta_\phi}{2} \frac{d Var(E)}{dz}\,,
\end{equation}
with $\beta_\phi$ the longitudinal Twiss parameter,
\begin{equation}
g_L=\frac{D_x}{\rho(x=0)} \frac{d\rho}{dx}\,,
\end{equation}
and $\rho$ the effective line density of the absorber at different transverse positions. $\rho$ may be adjusted by using variable density materials but more commonly wedge-shaped absorbers are assumed.


\subsubsection*{Rectilinear cooling}
The rectilinear cooling channel employs solenoids with weak dipoles that yield the tight focusing and dispersion required to provide a satisfactory cooling performance. In this lattice the focusing effect of the solenoids is dominant compared to weak dipole focusing and the transverse optics approach discussed above is appropriate.

In order to create a cooling channel having a suitable performance, it is necessary to maintain a sufficient dynamic aperture (DA), so that the beam is not lost, while decreasing the emittance as quickly as possible using low $\beta_\perp$ in order to prevent significant losses through muon decay. Typically, the requirement for large DA and small $\beta_\perp$ are in tension. The process of tapering the cooling channel is employed where the DA and absorber $\beta_\perp$ is successively reduced as the emittance decreases in order to provide a satisfactory cooling performance.

Two working points have been chosen, known as A--type and B--type lattices. A--type lattices operate in the first stability region, with phase advance below the $\pi$ stop-band. These regions typically are characterised by larger dynamic aperture and larger $\beta_\perp$ at the absorber, which is suitable for the initial part of the cooling channel.

B--type lattices operate in the second stability region. In order to increase the phase advance, stronger solenoids are required. The harmonic content of the lattice is chosen to optimise the positioning of $\pi$ and $2\pi$ stop bands; initially the stop bands are placed far apart, in order to maximise the momentum acceptance and transverse DA. As the transverse and longitudinal beam emittance is reduced the harmonic content of the lattice is adjusted to move the stop bands closer together in order to optimise the focusing strength. It will be further noted that $\beta_\perp$ scales with $BL$. By reducing the cell length and employing stronger magnets a lower equilibrium emittance may be achieved.

In \cite{Stratakis:2014nna} a dipole field is introduced by small tilts to the focusing solenoids. \cite{Witte:2014gba} proposed using solenoids with additional dipole coils to achieve the same effect. The dipole field introduces a weak dispersion which, when combined with wedge-shaped absorbers, leads to longitudinal emittance reduction. 

\subsubsection*{Bunch merge}
Between the A--type and B--type lattices, a bunch merge system is employed. The beam emittance is sufficiently small that the train of 21 bunches created in the front end can be compressed into one single bunch which can then be further cooled. This improves the luminosity, which is inversely proportional to the number of bunches as previously shown.


Bunch merge is achieved by first rotating the bunch in time-energy space using a sequence of RF cavities having different voltage and frequency. As in the phase rotation scheme described above, the early bunch is slowed while the late bunch is accelerated. The bunches then naturally merge, yielding seven bunches.

Each of the bunches is kicked into one of seven separate beamlines using a kicker that has a rotating transverse field. Each beamline has a different length so that the bunches, when they reach the end, are coincident. The beamlines are terminated by a funnel so that each individual bunch is arranged adjacent to the other six bunches in phase space, yielding a single combined bunch having large longitudinal and transverse emittance.

\subsubsection*{Final cooling}
Final cooling is achieved by means of a series of high field solenoids, in which very tight focusing may be achieved. Liquid hydrogen absorbers in the solenoids provide cooling. Particles experience many betatron oscillations in each solenoid. Deviations in momentum and transverse parameters can change the number of betatron oscillations, which causes emittance growth. To minimise this emittance growth the beam within the uniform solenoid field is matched so that $\beta_\perp'$, and hence $\beta_\perp''$, is close to 0. By reference to eq.~(\ref{eq:beta_sol})
\begin{equation}
\beta_\perp = \frac{\sqrt{1+\mathcal{L}^2}}{\kappa}\,,
\end{equation}
so the equilibrium transverse emittance that may be reached is inversely proportional to $B_z$ and proportional to $p_z$.

The transverse emittance that is reached in the final cooling system determines the overall luminosity of the complex. The smallest $\beta_\perp$, and hence highest fields, must be used. In order to provide the tightest focus possible the beam momentum is decreased to approximately 70 MeV/c. As the beam momentum decreases, longitudinal emittance growth due to the natural curvature of the Bethe-Bloch relationship becomes stronger. This effect would be further enhanced by a large energy spread in the incident beam. Maintaining a large momentum acceptance is very challenging in lattices with such a large phase advance per cell. For these reasons the energy spread must be minimised.

In order to minimise the growth of the longitudinal emittance  and maintain satisfactory transmission, the beam is lengthened and energy spread is reduced using a phase rotation system between each solenoid. A relatively weaker, higher aperture solenoid field maintains transverse containment. The beam is allowed to drift longitudinally so that faster particles move ahead of the beam and slower particles lag behind. The beam passes through accelerating electric fields where slower, later particles undergo an accelerating field and faster, earlier particles undergo a decelerating field. The phase rotation enables the energy spread to stay roughly constant and a time spread increase.

Some longitudinal emittance growth inevitably occurs in the final cooling system.  Later cooling cells require low frequency RF or induction-based acceleration in order to contain the full beam.

\subsubsection*{Alternative concepts}
Several alternatives to the scheme outlined above have been proposed. The rectilinear cooling scheme is itself an enhanced version of earlier ring cooler concepts. Ring coolers were rejected owing to issues surrounding injection and extraction of large emittance beams. Rings cannot take advantage of the $\beta_\perp$ tapering. Small circumference rings require tight bending fields that significantly perturb the transverse optics and are challenging to extract and inject from. Higher circumferences are possible, but only a small number of turns may be achieved before the beam approaches equilibrium emittance, making the additional challenge of a ring geometry less appealing. Higher bending radius `Guggenheim' \cite{Stratakis:2013xfa} and `helical' cooling schemes \cite{Yonehara:2018mlw} have been considered, with concomitantly higher dispersion. The scheme presented here achieved the best performance.

A dual-sign precooler, known as an `HFoFo' lattice, has also been considered. In the HFoFo lattice, positive and negative muons have dispersion that is partially aligned even in the same magnetic lattice. A wedge shaped absorber may be placed in such a lattice that cools both positively- and negatively- charged muons at the same time. Such a scheme is attractive to consider as it can cool the beam before charge separation. Charge separation is likely to be easier for lower beam emittances. Collective effects such as beam loading will be more severe in such a lattice due to the higher number of particles passing through the system and must be carefully considered.

In Parametric Resonance Ionisation Cooling (PIC), very low $\beta_\perp$ is achieved by driving the beam near to resonances rather than using high-field solenoids. The DA would normally be poor in such a lattice, but higher order corrections may enable an enhanced DA.

Frictional cooling is of interest both for a muon collider and also for existing muon facilities. At low energies, below $\beta\gamma \approx 0.1$, the energy loss becomes greater for higher energy particles. This leads directly to longitudinal cooling in addition to transverse cooling. Additional consideration must be made to account for losses that may occur due to $\mu^+$ and $\mu^-$ capture on electrons and nuclei respectively. A muon cooling scheme operating at these energies has been demonstrated \cite{Antognini:2020uyp}.


\subsubsection*{Technical issues and R\&D}
\paragraph{Rectilinear cooling}
In order to maintain the tight focusing required to yield good cooling performance in the rectilinear cooling system, a compact lattice is required with large real-estate RF gradient. This results in a short lattice with large forces between adjacent solenoids and RF operating near to the break down limit while immersed in a strong solenoid field. Integration of the various components of a cooling cell, considering the necessary mechanical support and the many vacuum and cryogenic interfaces, will pose significant engineering challenges. 

The operation of RF cavities in a solenoid field poses specific challenges. The solenoid field guides electrons that are emitted at one location of the cavity surface to another location on the opposing wall and leads to localised heating that can result in breakdown and cavity damage. Operation of copper cavities in 3 T  field showed a maximum useable gradient of only 10 MV/m.

Three approaches to overcome this obstacle are known:
\begin{itemize}
\item Use of lower-Z materials such as beryllium to limit the energy loss density.
\item The use of high-pressure hydrogen gas inside the cavity. In this case the mean free path of the electrons is limited and does not allow them to gain   enough energy to ionise the gas or to produce a breakdown.
\item The use of very short RF pulses to limit the duration of the heat load in the cavity.
\end{itemize}
The first two techniques have been experimentally verified in MUCOOL with a field of about 3 T (limited by the solenoid). They yielded a gradient of 50 MV/m in a beryllium cavity under vacuum and 65 MV/m in a molybdenum cavity with hydrogen \cite{Bowring:2018smm, Freemire:2017htp},  demonstrating no degradation in achievable field in the presence of an applied magnetic field.

Systematic studies in a new test stand are required to further develop the technologies. It will also be important to test the cavities in the actual field configuration of the cooling cell, which differs from a homogeneous longitudinal field. Finally, given the number of components required, the system optimisation will require designs that require minimal amount of material and suitable for medium scale production. Compact solenoid coil windings are hence of specific interest for this part of the complex. 

\paragraph{Final cooling}
In the final muon cooling system, solenoids with the highest practical field are needed. A design based on 30 T solenoids --- a value that has already been exceeded in a user facility using high-temperature superconductor --- demonstrated that an emittance about a factor two above the
target can be achieved \cite{Sayed:2015zta}; it should be noted that the study aimed at this larger target. Several options to improve the emittance will be studied. Solenoids with field of 30 T are close to become commercially available, and fields above 40T are planned at several high magnetic field user facilities. This development would benefit the muon collider, and activities are directed towards conceptual design of cooling solenoids in this range of field and higher, exploring the performance limits in the operating conditions of an accelerator. Operating the cooling at lower beam energy may also make cooling to lower emittance possible; preliminary studies indicate that 30 T may be sufficient to reach the emittance target.

The use of liquid hydrogen absorbers is technically challenging in the presence of high beam currents. The instantaneous beam power is large enough to cause significant heating. The associated pressure increase may damage the absorber windows. The use of solid absorbers or hybrid absorbers may mitigate this issue.

\subsection{Acceleration}
The baseline design for acceleration is to use a series of linacs and recirculating linacs at sub-100 GeV energies followed by a series of pulsed synchrotrons.

\subsubsection*{Linacs and recirculating linacs}
Initial acceleration is performed using a sequence of linear accelerators. The beam at the end of the final cooling system has low energy and large time spread owing to the phase rotation system discussed previously. This makes rapid acceleration challenging. A design for this acceleration region does not exist and is being pursued. 

As the beam becomes relativistic a conventional linac may be used. A linac can give very high real-estate gradients enabling rapid acceleration, which is crucial in the initial stages where the muon beam is not sufficiently time--dilated.

Above a few GeV, such a linac would become very costly. Efficient use of the RF system may be made by recirculating the beam through the linac using recirculating arcs, yielding a ``dogbone'' shaped accelerator (Figure~\ref{fig_muon:sketch})~\cite{Bogacz:2017iia}. The arcs have fixed magnetic fields. A conventional focusing solution requires a different arc for each beam momentum. FFA-type focusing may be employed so that beams having different focusing strengths are directed along the same arc. Muons of different momenta pass through quadrupole magnets, offset from the quadrupole centre so that the magnets deliver a bending field that is stronger for higher momenta particles. 

Care must be taken to ensure that the muon beam remains synchronised with RF cavities, which places constraints on the length of the arcs. Negatively and positively charged beams must be injected with a half-wave phase delay relative to each other so that they are correctly phased for acceleration, and the arcs must return the bunch with a half-wave phase delay so that, when travelling in the opposite direction the beam is still accelerated. The transverse focusing must also have appropriate symmetries so that both positively and negatively charge muons are contained.

A similar system has been demonstrated in practice at CBETA \cite{Bartnik:2020pos} using an electron beam. CBETA used a racetrack-style layout, and only electrons were accelerated. In CBETA the beam was brought into a linear accelerator. A spreader magnet diverted beams of different momenta into short delay lines on each turn. The beams were subsequently recombined and recirculated through an FFA magnet system. A further spreader, delay line and recombiner was placed prior to entry back into the RF linac. The beam recirculated in this way for four turns during which the beam was accelerated and four further turns during which the beam was decelerated.

\subsubsection*{Rapid cycling synchrotrons}

At higher energy a RCS becomes possible and, owing to the larger number of recirculations through each cavity, is cost effective compared to recirculating linacs. Four RCS are envisaged, accelerating to around 300, 750, 1500 and 5000~GeV respectively. Even at these energies, the muon lifetime constrains the ramp times to be a few ms or less.

In order to maintain a large average dipole field, the higher energy rings are designed with a combination of pulsed dipoles and superconducting fixed field dipoles. In this arrangement, care must be taken to ensure that the beam excursion in the fixed dipoles maintains the beam within the aperture and the path length deviation is not large enough to cause the beam to lose phase with the RF cavities.

\subsubsection*{Alternative concepts}
Acceleration using Fixed Field Alternating Gradient accelerators (FFAs) is an interesting alternative to RCS. FFAs have a dipole field that increases with position, either vertically or radially, rather than with time and so do not require fast ramping, reducing the overall power consumption. As the field increases spatially, FFA magnets are naturally focusing (bending) or defocusing (reverse bending) in the horizontal plane with the opposite focusing properties in the vertical plane. In order to provide alternating gradient, a mix of bending and reverse bending magnets are required. FFAs can be designed that move the beam horizontally or vertically; a vertical orbit excursion FFA would have a path length that does not vary with energy and so is isochronous in the relativistic limit. In addition, non-linear FFA lattices that employ gradient, edge and weak focusing to achieve simultaneous control of betatron tunes and time of flight have also been studied recently. So far, however, these have not been applied to muon acceleration for a muon collider. Acceleration of electrons was demonstrated in the EMMA FFA as a scaled representation of muon acceleration \cite{Machida:2012zz}.

FFAs often use reverse bending to enable vertical focusing, which can decrease the efficiency in their use of dipoles. Horizontal orbit excursion FFAs must have a limited orbit excursion so that the time-of-flight does not vary significantly during the acceleration cycle in order that the beam remains phased with the RF.

The beam moves across the RF cavity during acceleration, which limits the size, frequency and efficiency of the cavities. Fixed field or pulsed dispersion suppressors have been proposed to reduce the impact, but no design yet exists.

\subsubsection*{Technical issues and required R\&D}
The RCS must be able to ramp extremely quickly while maintaining synchronisation with the RF system.  Acceleration cycles as short as a few 100~$\mu$s and ramp rates O(kT/s) are envisaged, around 100 times faster than existing RCS. The design of the resistive pulsed magnets is optimized to obtain a minimum stored energy, thus reducing the power required for ramping. This still reaches values of the order of several tens of GW. Fast pulsed power supplies may be considered that can ramp on this time scale, but in order to maintain power and cost efficiency resonant circuits are required. The total large power of a single accelerator stage is expected to be divided into several independent sectors. Since resonant circuits discharges can have large uncertainties, additional active power converters may be required to guarantee the controllability of the field ramps among sectors and follow the approximately constant energy change produced by the acceleration system. 
Efficient energy recovery is a must, implying that magnet losses should be kept to small fraction of the stored energy. This can be achieved using thin laminations of low-hysteresis and high-resistivity magnetic alloys, whereby material characterization in the representative range of frequencies and field swings will still be needed. 

Heating of the magnets arising from muon decay products must be successfully managed. This may require a modest amount of shielding in the bore of the superconducting dipoles, in turn increasing the magnet bore requirement.

\subsection{Collider}
\label{sec:fac_collider}
The collider ring itself must be as low circumference as possible in order to maintain the highest possible luminosity. High field bending magnets are desirable. In order to deliver a collider in a timely manner, 10~T dipoles are assumed for a 3~TeV collider. 11~T dipoles are planned for the HL LHC upgrade. 16~T dipoles, under study for FCC-hh, are assumed for a 10 TeV collider. At present the strongest accelerator-style dipoles have a field of 14.5~T.

The collider ring requires a small beta-function at the collision point, resulting in significant chromaticity that needs to be compensated. A short bunch is required to reduce the ``hour-glass-' effect, previously discussed,
 whereby the $\beta^*$ of the bunch is degraded to be the average $\beta^*$ along the bunch at collision. 

The bunch length may be exchanged with energy spread using RF cavities, respecting longitudinal emittance conservation. Additional energy spread makes focusing at the interaction point more challenging; in general higher energy particles have a longer focal length than lower energy particles in a quadrupole focusing system, a feature known as chromatic aberration. This limitation may be mitigated by using sextupole magnets that have stronger focusing on one side of the magnet than the other. By arranging dipoles near the sextupoles, such that higher energy particles are aligned with the stronger focusing region, correction of these chromatic aberrations is possible. 

The minimum bunch length is limited by momentum compaction. Particles having a larger momentum may have a different path length $l$ than particles having a lower momentum. Hence the higher energy particles have a longer time-of-flight around the ring, and this practically limits the minimum bunch length. In order to deliver sufficiently short bunches, the momentum compaction factor, $dl/dp$, must be almost 0. This can be achieved by careful consideration of dispersion and focusing around the collider. 

Overall, a solution for 3 TeV has been developed and successfully addressed these challenges. A design of 10 TeV is one of the key ongoing efforts.

\subsubsection*{Alternative concepts}
Use of combined function skew quadrupole magnets for bending has been proposed to reduce the momentum compaction factor in the ring. Such an arrangement would introduce dispersion in the vertical plane by appropriate matching of the dipole and skew-quadrupole fields, enabling significant reduction in the momentum compaction factor. Further reduction may be achieved by including higher order multipoles.


\subsubsection*{Technical issues and required R\&D}
\paragraph{Neutrino flux}

The decay of muons in the collider ring produces neutrinos that will exit from the ground hundreds of kilometers away from the collider. Since they are very energetic, these neutrinos have a non-negligible probability to interact in material near to the Earth's surface producing secondary particle showers. A study is underway to ensure that this effect does not entail any noticeable addition to natural radioactivity and that the environmental impact of the muon collider is negligible, similar, for instance, to the impact from the LHC. 

The flux density arising from the collider ring arcs will be reduced to a negligible level by deforming the muons trajectory, achieving a wide enough angular spread of the neutrinos. Wobbling of the muon beam within the beam pipe would be enough for 1.5~TeV muon beam energy. At 5~TeV muon beam energy, the beam line components in the arcs may have to be placed on movers to deform the ring periodically in small steps such that the muon beam direction would change over time. Studies must be performed to address the mechanical aspects of the solution and its impact on the beam operation. Similarly the flux densities arising from the straight sections at the interaction points are addressed in the study by optimising the location and orientation of the collider.

In order to predict the environmental impact and to design suitable methods for demonstrating compliance, detailed studies of the expected neutrino and secondary-particle fluxes are being performed with the FLUKA Monte Carlo particle transport code~\cite{Ahdida:2022gjl, Battistoni:2015epi}. The preliminary results confirm that the effective doses generated by the neutrinos are dominated by neutrons and electromagnetic showers produced in the material located close to the ground level. By contrast, the radiological impact due to activation of materials in the area of concern, such as soil, water, or air, has been found to be negligible.

Accidental beam loss will have a significant impact on the surrounding equipment but also may create a particle shower. The impact of accidental beam loss can be mitigated by placing the tunnel sufficiently deep. In this case a lost muon beam would not be able to penetrate the Earth sufficiently to escape from the surface.

\paragraph{Beam induced background}
Muon beam decay produces a significant flux of secondary and tertiary particles in the detector. The current solution to mitigate such a background flux, initially proposed by MAP, consists of two tungsten cone-shaped shields (nozzles) around the beampipe, with the origin in proximity of the interaction point. A study of the impact of the background on the detector performance has been performed and is discussed in Section~\ref{sec:environment}.

Further optimisation is foreseen of the Interaction Region (IR) together with the shielding elements to reduce their dimensions, therefore increasing the
detector acceptance in the forward region. A proper combined optimisation of the system of detector, shielding and IR will enable reduction of peak background increasing the detector performance in the forward region.
The lessons learned in design of the 3~TeV system will serve as a starting point for the 10~TeV case.


\paragraph{Magnets and shielding}
The high-energy electrons and positrons arising from muon decay and striking the collider ring magnets can cause radiation damage and unwanted heat load. The decay electrons and positrons mostly strike the inner wall of the chamber due to their lower magnetic rigidity. Synchrotron radiation is emitted by the electrons and positrons and the resultant photons strike both sides of the aperture. The heating can be mitigated with sufficient tungsten shielding; a successful design has been developed at 3 TeV. First studies
at 10 TeV indicate that the effect is comparable to 3
TeV, since the power per unit length of the particle loss
remains similar.


The shielding requires a substantial aperture in the superconducting magnets. The limit for the dipole field is thus given by the maximum stress that the conductor can withstand (mechanics), and by the stored magnetic energy (quench protection), rather than by the maximum field that it can support. Novel concepts such as stress-managed coils will allow mechanical challenges to be addressed. Demonstration of such concepts is hence crucial to the feasibility of the collider magnets. Alternative schemes are considered, based on high-temperature superconductors, exploiting the same compact winding features planned to be developed for the high- and ultra-high-field cooling solenoids. Such development would benefit from similar activities in the field of superconducting motors and generators, whose pole windings have similar geometry, albeit at reduced dimension.

\subsection{Technical demonstrators}
\label{sec:fac_demonstrators}

Demonstrations are required both for the muon source and the high energy complex.
\begin{itemize}
\item The compact nature of the muon cooling system, high gradients and solenoids of relatively high field  poses some unique challenges that require demonstration. 
\item The high-power target presents a number of challenges that should be evaluated using irradiation facilities or single impact beam tests.
\item The issues in the high energy complex arise from the muon lifetime. Fast acceleration systems and appropriate handling of decay products result in unique challenges for the equipment.
\end{itemize}
The following new facilities are foreseen.

\paragraph{Ionisation cooling demonstrator}

\begin{figure*}
\centering
\includegraphics[width=0.8\textwidth, trim={0cm 0cm 0 0cm}]{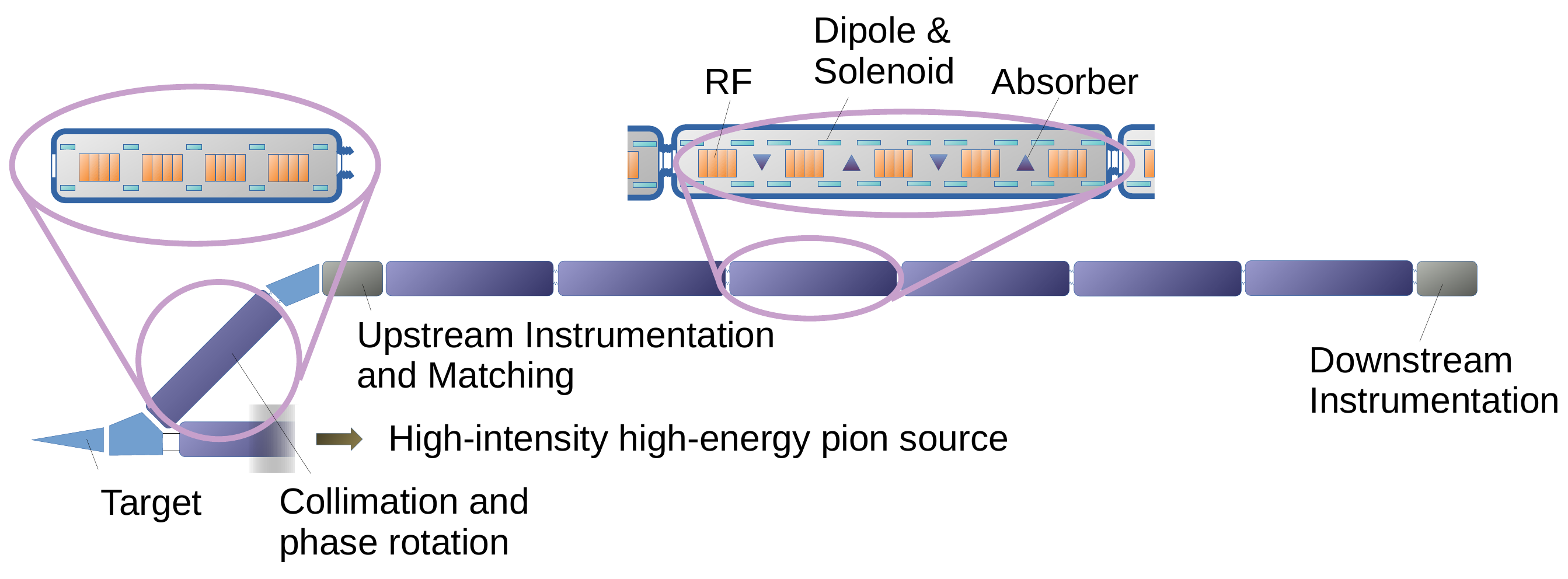}
\caption{Schematic of the muon cooling demonstrator. \label{fig_muon:demo}}
\end{figure*}

MICE has delivered the seminal demonstration of transverse ionisation cooling. In order to prove the concept for a muon collider further tests must be performed to demonstrate the 6D cooling principle at low emittance and including re-acceleration through several cooling cells. 

A schematic of a Muon Cooling Demonstrator is shown in Figure~\ref{fig_muon:demo}. A proton beam strikes a target creating pions. Pions with momenta 100-300 MeV/c are brought into a beam preparation line where RF cavities are used to develop a pulsed beam and muons having large transverse emittance are scraped from the beam using collimators. The resultant high emittance beam is brought back onto the beamline and transported through a high-fidelity instrumentation system before passing through a number of cooling cells. Finally the beam is delivered onto a downstream instrumentation system.

Many of the challenges in delivering such a facility are associated with integration issues of the magnets, absorbers and RF cavities. For example, operation of normal conducting cavities near to superconducting magnets may compromise the cryogenic performance of the magnet. Installation of absorbers, particularly using liquid hydrogen, may be challenging in such compact assemblies. In order to understand and mitigate the associated risks, an offline prototype cooling system will be required. Such a system will require an assembly and testing area, with access to RF power and support services. This could be integrated with the demonstrator facility, which will need an area for staging and offline testing of equipment prior to installation on the beamline. 

The possibility to perform intensity studies with a muon beam are limited. In the first instance such effects will be studied using simulation tools. If such studies reveal potential technical issues, beam studies in the presence of a high intensity source will be necessary, for example using a proton beam.

\paragraph{Ionisation cooling RF development}
The cooling systems require normal-conducting RF cavities that can operate with high gradient in strong magnetic fields without breakdown. No satisfactory theory exists to model the breakdown. Considerable effort was made by MAP to develop high-gradient RF cavities. Two test cavities have been developed. The first cavity was filled with gas at very high pressure. The second cavity used beryllium walls. Both tests presented promising results. Operation of normal-conducting RF cavities at liquid nitrogen temperature has been demonstrated to reduce multipacting. In order to test the concepts above and others further, a dedicated test facility is required. An RF source having high peak power at the appropriate frequency and a large aperture solenoid that can house the RF cavity will be needed. No such facility exists at present.

\paragraph{Cooling magnet tests}
In order to improve the rectilinear cooling channel performance, high field magnets are required with opposing-polarity coils very close together. The possibility to implement high-field magnets (including those based on HTS) needs to be investigated, with appropriate design studies leading to the construction of high-field solenoid magnets having fields in the range 20 T to 25 T. Very high field magnets are required for the final cooling system. In this system, the ultimate transverse emittance is reached using focusing in the highest-field magnets. As a first step, a 30 T magnet, corresponding to the MAP baseline, would be designed and constructed. Feasibility studies towards a 50 T magnet would also be desirable, which may include material electro-mechanical characterisation at very high field as well as technology demonstration at reduced scale. These very demanding magnets are envisaged to be developed separately to the cooling demonstrator. Eventually they could be tested in beam if it was felt to be a valuable addition to the programme. In order to support this magnet R\&D, appropriate facilities will be required. Testing of conductors requires a suitable test installation, comprising high field magnets, variable temperature cryogenics and high-current power supplies. Magnet development and test will also require these facilities in addition to access to appropriate coil and magnet manufacturing capabilities.

\paragraph{Acceleration RCS magnets}
Acceleration within the short muon lifetime is rather demanding. The baseline calls for magnets that can cycle through several T on a time scale of a few ms. A resonant circuit is the best suited solution to power the magnets for good energy storage efficiency. The design of the magnet and powering system will be highly integrated, and work on scaled prototypes is anticipated. Superconducting RCS magnets may offer higher field reach than normal-conducting magnets, but are challenging to realise owing to heating arising from energy dissipation in the conductor during cycling (AC loss). This heating can lead to demands on the cryogenic systems that outweigh the benefits over normal-conducting magnets. Recent prototypes have been developed using HTS that can operate at higher temperatures, and in configurations leading to lower AC losses, yielding improved performance \cite{Piekarz:2019yis,Piekarz:2021mna}. In order to continue this research, magnet tests with rapid pulsed power supplies and cryogenic infrastructure will be required.

\paragraph{Effects of radiation in material}
The high beam power incident on the target and its surroundings is very demanding. Practical experience from existing facilities coupled with numerical studies indicate that there will be challenges in terms of target temperature and lifetime. Instantaneous shock load on the target will also be significant. Tests are foreseen to study behaviour of target material under beam in this instance. Tests are desirable both for instantaneous shock load and target lifetime studies.
Additionally, the effect of radiation in the target
region on the superconducting materials (LTS and HTS) and insulators is an important parameter. Additional studies may be required taking into account the magnet arrangement, conductor design and estimates of radiation levels. In order to realise such tests, facilities having both instantaneous power and integrated protons on target equivalent to the proton beam parameters assumed for this study are desirable. The database of radiation effects on superconductors (HTS) and insulators also requires an extension to cover the projected conditions in the target area.
\paragraph{Superconducting RF cavities}
Development of efficient superconducting RF with large accelerating gradient is essential for the high energy complex. Initially, the work will focus on cavity design; however eventually a high gradient prototype at an appropriate frequency will be required. In order to realise such a device, appropriate superconducting cavity production and test facilities will be required including surface preparation techniques and a capability for high power tests.

\subsection {Start-to-end facility simulations}
\label{sec:s2e_simulation}

As part of the design process for the muon collider, detailed studies of the various subsystems would be followed by an assessment of the overall performance of the whole facility. Individual components will be modelled to understand the impact of misalignments and imperfections. Such simulations would help to build a comprehensive understanding of the effects induced by deviations from the ideal running conditions. These tolerance specifications would inform and guide the technical R\&D programme.

The assessment may be refined by modelling the whole system in a start-to-end simulations, from the proton driver to the collider ring. 
This could be performed using a set of codes and a systematic procedure to transfer the results from one code to another including data format conversions and coordinate system transformations.

\subsection {Synergies with other concepts or existing facilities}
\label{sec:fac_synergies}

\begin{figure}
\centering\includegraphics[width=0.45\textwidth]{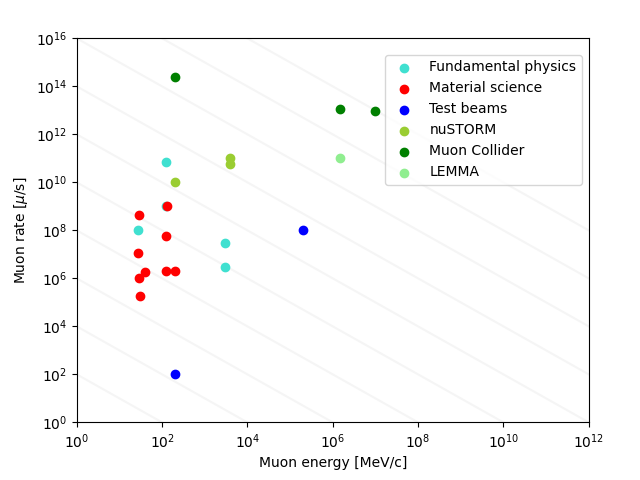}
\caption{Muon energy and rate of different muon facilities. \label{fig_muon:power}}
\end{figure}

\begin{figure*}
\begin{tikzpicture}[scale=0.05]
\draw[blue!80, rounded corners=3pt, fill=blue!10] (10,-2) -- (10,-5) -- (0,-7) -- (-1,3) -- (10,5) -- (10, 2);
\node[] at (5,-12) {Target};

\shade[bottom color=blue!20,top color=orange!20] (10,-2) -- (21,7) -- (21,11) -- (10,2);
\shade[bottom color=orange!20,top color=orange!5] (70,11) -- (70,7) -- (21, 7) -- (21, 11);
\draw[blue!80] (10,2) -- (20,11) -- (70,11) -- (70,7) -- (21, 7) -- (15, 2);
\node[] at (115,9) {MuC cooling demonstrator};
\draw[thick,->, color=red] (30,15) -- (35,15);
\node[text=red] at (40,15) {$\mu$};

\shade[top color=orange!20, bottom color=orange!5] (50, 0) -- (260,0) -- (260,-4) -- (50, -4);
\draw[blue!80, fill=orange!20] (50,-2) arc (110:270:30);
\draw[blue!80, fill=white] (60,-4) arc (90:270:26);
\draw[blue!80] (60, 0) -- (260,0);
\draw[blue!80] (60, -4) -- (260,-4);
\draw[blue, fill=orange!20] (260,-60) arc (-90:90:30);
\draw[blue, fill=white] (260,-56) arc (-90:90:26);
\shade[top color=orange!5, bottom color=orange!20] (60, -56) -- (260,-56) -- (260,-60) -- (60, -60);
\draw[blue!80] (60, -56) -- (260,-56);
\draw[blue!80] (60, -60) -- (260,-60);
\draw[thick,->, color=red] (160,-10) -- (165,-10);
\node[text=red] at (170,-10) {$\mu$};
\draw[thick,<-, color=red] (160,-65) -- (165,-65);
\node[text=red] at (170,-65) {$\mu$};
\node[] at (170,-80) {nuSTORM Storage ring};

\shade[bottom color=blue!5,top color=blue!20] (15, 2) -- (25,3.5) -- (30,4) -- (35,3.5) -- (50,2) -- (60,0) -- (50,-2) -- (35,-0.5) -- (30,0) -- (25,-0.5) -- (10, -2);
\draw[blue!80, rounded corners=3pt] (15, 2) -- (25,3.5) -- (30,4) -- (35,3.5) -- (50,2) -- (60,0);
\draw[blue!80, rounded corners=3pt] (10, -2) -- (25,-0.5) -- (30,0) -- (35,-0.5) -- (50,-2);
\draw[thick,->, color=blue] (20,-5) -- (25,-5);
\node[text=blue] at (30,-5) {$\pi$};
\node[] at (60,-10) {OCR};

\draw[gray, dashed] (270,-2) -- (310, -2);
\draw[blue!80, fill=gray!50, rounded corners=3pt] (310,-6) rectangle (320,4);
\draw[thick,->, color=gray] (285,2) -- (290,2);
\node[text=gray] at (290,10) {$\nu_{\mu}$, $\nu_{e}$, };
\node[] at (315,-10) {Detector};

\end{tikzpicture}
\vspace{-20pt}
\caption{Schematic showing nuSTORM including the muon cooling demonstrator for the muon collider. \label{fig_muon:nustorm}}
\end{figure*}
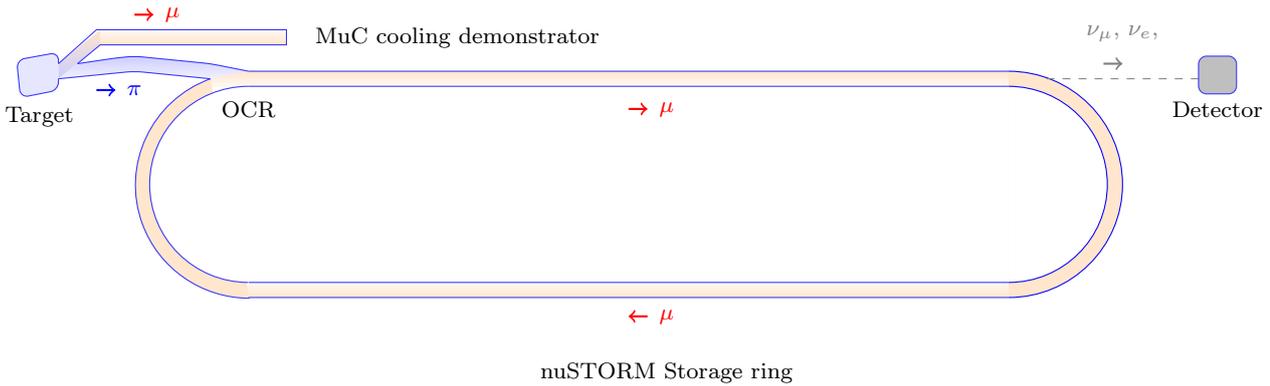

The ambitious programme of R\&D necessary to deliver the muon collider has the potential to enhance the science that can be done at other muon-beam facilities. The progress in other accelerator facilities will also benefit the design and construction of the muon collider in the future.

nuSTORM \cite{Ahdida:2020whw} and ENUBET \cite{Delogu:2022yqp} offer world-leading precision in the measurement of neutrino cross sections and exquisite sensitivity to sterile neutrinos and physics beyond the Standard Model. nuSTORM in particular will require capture and storage of a high-power pion and muon beam and management of the resultant radiation near to superconducting magnets. In nuSTORM, multi-GeV pions are brought from a target and injected into a racetrack-shaped storage ring. The storage ring is tuned to capture muons arising from pions that decay in the first straight. Remnant pions are extracted to a beam dump, while the muons are circulated many times. The apparatus would deliver a `flash' of neutrinos from the initial pion decay followed by a well-characterised neutrino beam arising from the decay of the circulating muon beam. The momentum of the circulating muon beam, and hence the resultant neutrinos, would be tunable, enabling characterisation of neutrino interactions with the detector over a broad range of momenta.

The muon rate and energy for nuSTORM as compared to the muon collider and other muon beamlines is shown in Figure~\ref{fig_muon:power}. The target and capture system for nuSTORM and ENUBET may also provide a testing ground for the technologies required at the muon collider and as a possible source of beams for the essential 6D cooling-demonstration experiment, for example as in the schematic shown in Figure~\ref{fig_muon:nustorm}.

The ongoing LBNF and T2HK projects and their future upgrades will develop graphite targets to sustain the bombardment of MW-level proton beams, which may also lead to a solution for the muon production target for the muon collider. The next generation searches for charged lepton flavour violation exploit high-power proton beams impinging on a solid target placed within a high-field solenoid, such as COMET at J-PARC and Mu2e at FNAL. The technological issues of target and muon capture for these experiments are similar to those present in the muon collider design.

 The potential to deliver high quality muon beams could enhance the capabilities of muon sources such as those at PSI, J-PARC and ISIS. The use of frictional cooling to deliver ultra-cold positive and negative   muon beams is under study at PSI and may be applicable to the muon collider.

High-power proton accelerators are in use throughout the world, accelerating protons using linacs and accumulation in fixed energy rings or prior to further acceleration using rapid cycling synchrotrons. Proton drivers having power ranging from hundreds of kW to multiple MW are used as spallation neutron sources at SNS, J-PARC, ESS, PSI, ISIS and CSNS and neutrino sources at FNAL and J-PARC proton accelerator complexes as well accelerator-driven systems such as CiADS and MYRRHA. Many of the accelerator technologies required for the muon collider proton beam and for rapid acceleration are in use or under development at these facilities. For example FFAs have been proposed as a route to attain high proton beam power for secondary particle sources such as neutron spallation sources, owing to the potential for high repetition rate and lower wall plug power compared to other accelerator schemes. 

The underlying technologies required for the muon collider are also of interest in many scientific fields. The delivery of high field solenoid magnets is of great interest to fields as wide ranging as particle physics, accelerator science and imaging technology. Operation of RF cavities with high gradient is of interest to the accelerator community.

\subsection{Outlook}\label{sec:conc_out}
The muon collider presents enormous potential for fundamental physics at the energy frontier.
Previous studies  have demonstrated feasibility of many critical components  of the facility. Several proof-of-principle experiments and component tests like MICE,
CBETA, EMMA and the MUCOOL programme, have been carried out to practically demonstrate the underlying
technologies. 
Bright muon beams are also the basis of the nuSTORM facility. This experiment could share a large part of the complex with a cooling demonstrator.

The muon collider is a novel concept and is not as mature as the other high-energy lepton
collider options. However, it promises a
unique opportunity to deliver physics reach at the energy frontier on a cost, power consumption and time
scale that might improve significantly on other energy-frontier colliders. At this stage, building
upon significant prior work, no insurmountable technological issues were identified.
Therefore a development path can address the major challenges and deliver
a 3 TeV muon collider by 2045. 

A global assessment has identified the R\&D effort that is essential to
address these challenges before the next regional strategy processes to a level that allows estimation of the performance,
cost and power consumption with adequate certainty.   Execution of this R\&D is required in order to
maintain the timescale described in this document.  Ongoing developments in underlying technologies
will be exploited as they arise in order to ensure the best possible performance.  This R\&D effort will
allow future strategy processes to make fully informed recommendations.  Based on the subsequent decisions, a significant ramp-up of resources could
be made to accomplish construction by 2045 and exploit the enormous potential of the muon collider.

\clearpage


\section{Particle detectors and event reconstruction}
\label{sec:detectorandreconstruction}

The unstable nature of muons makes the beam-induced background (BIB) a much more challenging issue at a muon collider than it is at facilities that use stable-particle beams. For instance, with $2.2 \cdot 10^{12}$ muons per bunch a \qty{1.5}{\TeV} muon beam leads to about $2 \cdot 10^5$ muon decays per meter in a 3~TeV MuC with parameters as in Table~\ref{MC:t:param}. The interactions of the decay products with the accelerator lattice produce even larger amounts of particles that eventually reach the detector, making the reconstruction of clean $\mu^+ \mu^-$ collision events nearly impossible without a dedicated BIB mitigation. 

Muon collider detectors and event reconstruction techniques therefore need to be designed specifically to cope with the presence of the continuous flux of secondary and tertiary particles from the BIB. This section reviews the state-of-the-art design studies, and it is organised as follows. In Section~\ref{sec:environment} we describe the muon collision environment based on simulations of the BIB fluxes and composition reaching the detector. A tentative detector model is employed. The software setup used for the detector response simulation is described in Section~\ref{sec:software}. Section~\ref{sec:technology} presents promising technologies that could be employed in the tracking detector, the calorimeter systems, and dedicated muon spectrometers. General considerations regarding trigger systems and data acquisition are also discussed. Section~\ref{sec:performance} describes the status of development of the reconstruction algorithms and their expected performance for the basic objects needed to carry out a comprehensive physics programme. The reconstruction of other objects, as $\tau$ leptons or missing momentum, is still in progress, but it is expected to pose challenges similar to the ones that have been already solved. A discussion of the special challenges and opportunities for progress that are posed by the forward region of the detector is reported in Section~\ref{sec:fwdandlumi}. The summary and conclusions are presented in Section~\ref{sec:outlook}.

\subsection{Collision environment}
\label{sec:environment}

The BIB creates a large particle flux that interacts with the detector elements. On top of a detector model, its detailed simulation would require the design of the machine interaction region and of the Machine–Detector Interface (MDI). In fact, the BIB emerges from a chain of interactions with the material that composes these elements, entailing a strong dependence on their configuration of the BIB composition, flux, and energy spectra~\cite{Mokhov:2011zzd,Mokhov:2014hza,Bartosik:2019dzq,Lucchesi:2020dku}. The interaction region and MDI design also offers opportunities for the mitigation of the level of BIB that reaches the detector. 

Since no final design of these elements nor of the detector is available, and even the conceptual design of the collider facility is ongoing, current studies are based on tentative configurations, described below.

\begin{figure}[t]
    \begin{center}
        \centering
        \includegraphics[trim={0.8cm 0.8cm 0.8cm 0.8cm},clip,width=0.5\textwidth]{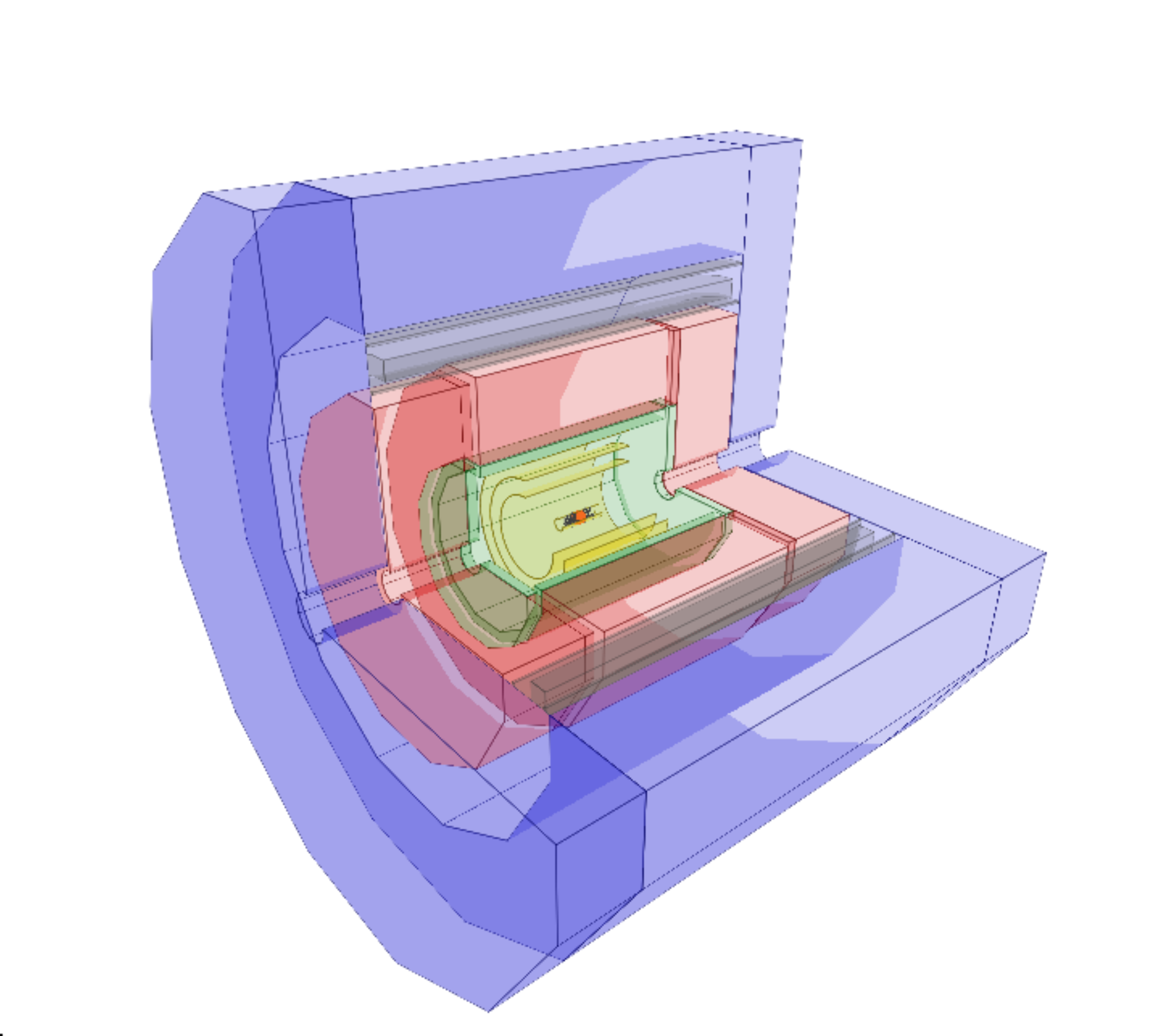}
    \end{center}
    \caption{Illustration of the full detector, from the \textsc{Geant4} model. Different colours represent different sub-detector systems: the innermost region, highlighted in the yellow shade, represents the tracking detectors. The green and red elements represent the calorimeter system, while the blue outermost shell represents the magnet return yoke instrumented with muon chambers. The space between the calorimeters and the return yoke is occupied by a 3.57~T solenoid magnet.}
    \label{fig:detector}
\end{figure}

\subsubsection*{Detector model}

The design of a dedicated experiment is still in its preliminary phase, but some general conclusions can be already drawn. Given the breadth of the expected physics programme, a hermetic detector with angular coverage as close as possible to $4\pi$ is required. The detector will feature a cylindrical layout and will include: an inner tracking detector immersed in a magnetic field; a set of calorimeter systems designed to fully contain the products of the muon collisions; and an external muon spectrometer.

\begin{figure*}
    \centering
    \includegraphics[width=0.7\textwidth]{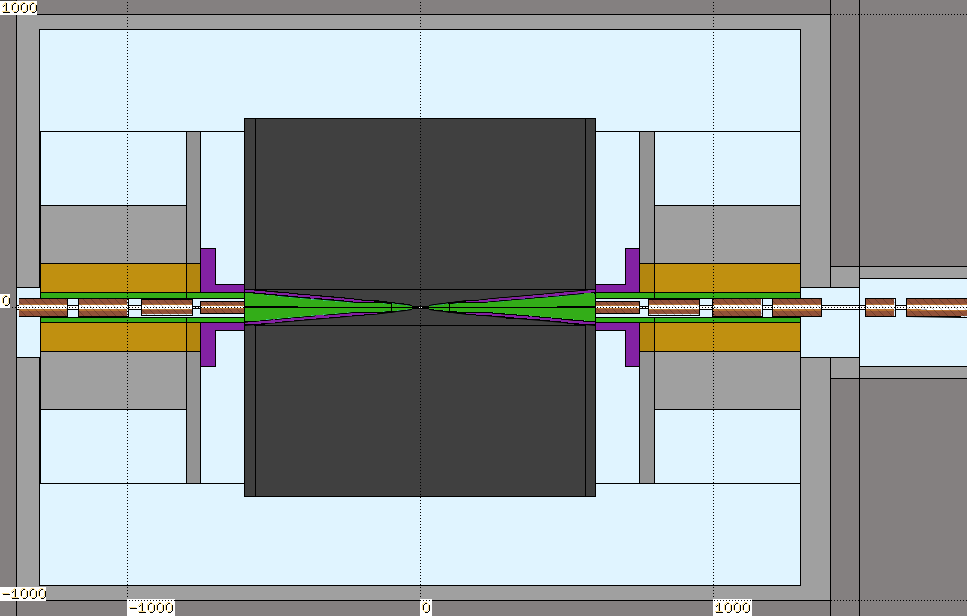}
    \includegraphics[width=.7\textwidth, trim={-10 0 0 -10}]{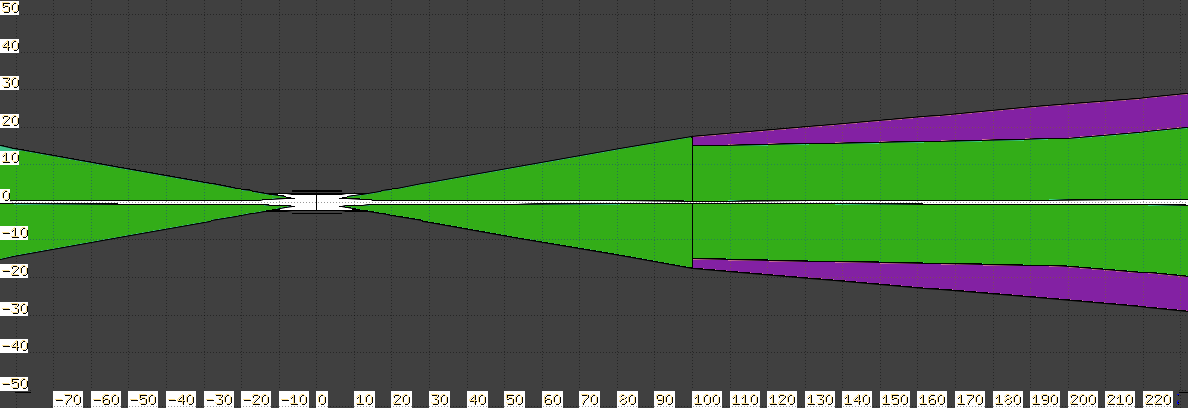}
    \caption{Cross-sectional view of the MDI as designed by the MAP collaboration for a $\sqrt{s} = $1.5~TeV MuC and visualised with FLUKA. Distinct colours represent different materials of the MDI: tungsten (green), borated polyethylene (dark magenta), iron (dark yellow), and concrete (gray). The black box in the center encloses the detector volume, which is excluded from the standalone BIB simulation process. Dimensions are reported in centimeters.}
    \label{fig:bib-mdi}
\end{figure*}

The tentative detector model we consider, referred to as Muon Collider Detector (MCD), is based on the CLICdet concept~\cite{CLICdp:2017vju,CLICdp:2018vnx,ILDConceptGroup:2020sfq,ILC:2007vrf}. The innermost system consists of a full-silicon tracking detector divided in three sub-detectors: the Vertex Detector, the Inner and the Outer Tracker. The tracking detector is surrounded by a calorimeter system that consists of an electromagnetic calorimeter (ECAL) and a hadronic calorimeter (HCAL), and is immersed in a magnetic field of 3.57~T provided by a solenoid with an inner bore of 3.5~m. Finally, the outermost part of the detector features a magnet iron yoke designed to contain the return flux of the magnetic field and is instrumented with muon chambers. The full detector is shown in Figure~\ref{fig:detector}. 
The most relevant modifications to the CLICdp detector are in the tracker topology. They are introduced for the installation of two double–cone shielding absorbers made of tungsten with a borated polyethylene (BCH2) coating and having an opening angle of $10^{\circ}$, referred to as ``nozzles''. The nozzles are located inside the detector in the forward regions\footnote{A right-handed coordinate system with its origin at the nominal interaction point in the centre of the detector is used. Cylindrical coordinates $(r,\phi)$ are used in the transverse plane, $\phi$ being the azimuthal angle around the z-axis and $\theta$ the polar angle with respect to the z-axis.} along the beam axis in the region between 6 and 600 cm away from the Interaction Point (IP), as displayed in Figure~\ref{fig:bib-mdi}.

The installation of the nozzles was proposed by the MAP collaboration~\cite{Mokhov:2011zzd} in order to mitigate the BIB effects. These nozzles, assisted by the magnetic field induced by a solenoidal magnet encasing the innermost detector region, could trap most of the electrons arising from muon decays close to the IP, as well as most of incoherent $e^{+}e^{-}$ pairs generated at the IP. With this sophisticated shielding in the MDI region, a total BIB reduction of more than three orders of magnitude was obtained~\cite{Mokhov:2011zzd}. The exact shape and positioning of the nozzles, including the \qty{10}{\degree} opening angle and \qty{12}{\centi\metre} distance between the tips, was optimised specifically for the MAP design of a MuC with $\sqrt{s} =$~\qty{1.5}{\TeV} energy in the centre of mass. Their re-optimisation will be an important component of future work on the design of the MDI for the 3 and 10~TeV colliders.

\subsubsection*{Characterisation of BIB}

\begin{table*}[!ht]
    \centering
        \caption{Multiplicities of different types of particles after the shielding structure, therefore arriving on the detector surface. A single bunch crossing with $2\cdot 10^{12}$ muons is considered. In all cases, the MAP 1.5~TeV collider design and optimised MDI is assumed.}
    \label{tab:BIB_com_energy}
    \vspace{2mm}
    \begin{tabular}{|l|c|c|c|c|c|} \hline
         Monte Carlo simulator                          & MARS15             & MARS15            & FLUKA             & FLUKA             & FLUKA \\ \hline
         Beam energy [GeV]                              & 62.5               & 750               & 750               & 1500              & 5000  \\
         $\mu$ decay length [m]                         & $3.9\cdot 10^5$    & $46.7\cdot 10^5$  & $46.7\cdot 10^5$  & $93.5\cdot 10^5$  & $311.7\cdot 10^5$\\
         $\mu$ decay/m/bunch                            & $51.3\cdot 10^5$   & $4.3\cdot 10^5$   & $4.3\cdot 10^5$   & $2.1\cdot 10^5$   & $0.64\cdot 10^5$ \\
         Photons ($E_{\gamma}>0.1$~MeV)                 & $170\cdot 10^6$    & $86\cdot 10^6$    & $51\cdot 10^6$    & $70\cdot 10^6$    & $107\cdot 10^6$  \\
         Neutrons ($E_{n}> 1$~MeV)                      & $65\cdot 10^6$     & $76\cdot 10^6$    & $110\cdot 10^6$   & $91\cdot 10^6$    & $101\cdot 10^6$ \\ 
         Electrons \& positrons ($E_{e^{\pm}}>0.1$~MeV) & $1.3\cdot 10^6$    & $0.75\cdot 10^6$  & $0.86\cdot 10^6$  & $1.1\cdot 10^6$   & $0.92\cdot 10^6$\\
         Charged hadrons ($E_{h^{\pm}}>0.1$~MeV)        & $0.011\cdot 10^6$  & $0.032\cdot 10^6$ & $0.017\cdot 10^6$ & $0.020\cdot 10^6$ & $0.044\cdot 10^6$ \\
         Muons ($E_{\mu^{\pm}}>0.1$~MeV)                & $0.0012\cdot 10^6$ & $0.0015\cdot 10^6$& $0.0031\cdot 10^6$& $0.0033\cdot 10^6$& $0.0048\cdot 10^6$\\
\hline
    \end{tabular}
\end{table*}

Detailed BIB simulations were first performed in the context of the MAP studies, employing the MARS15 \cite{Mokhov:2017klc} Monte Carlo software. These are based on the accelerator lattice and interaction regions designed by MAP for a \qty{1.5}{\TeV} MuC. The previously-mentioned optimised MDI design was based on these simulations. A Higgs-pole muon collider with \qty{62.5}{\GeV} energy beams was also considered by MAP. 

The BIB simulation for the \qty{1.5}{\TeV} collider have been repeated in~\cite{Collamati:2021sbv}, using the Monte Carlo multi-particle transport code FLUKA~\cite{Ahdida:2022gjl,Battistoni:2015epi}. The complex FLUKA geometry was assembled by means of the LineBuilder program~\cite{Mereghetti:2012zz} using the optics file provided by the MAP collaboration. The accelerator elements have been defined in the FLUKA Elements Database following the information contained in this file and in MAP publications~\cite{Alexahin:2011zz, DiBenedetto:2018cpy}. The particles induced by the muon decays are collected at the outer surface of the MDI and before entering the detector volume, which is represented by a black box on Figure~\ref{fig:bib-mdi}. This will allow to later simulate their interaction with the detector together with particles from the $\mu^+\mu^-$ collision. The ``BIB sample'' that we describe here thus refers to the collection of particles originating from the muon decays before any interaction with the detector material.

The results obtained by FLUKA are in good agreement with the ones from MAP as shown in Table~\ref{tab:BIB_com_energy} and discussed in Ref.~\cite{Collamati:2021sbv} in more detail. The FLUKA simulation is then repeated with the same setup for the higher energy beams of the 3 and 10~TeV MuC~\cite{Calzolari:2022lgu}. This corresponds to assuming that the interaction region and the MDI are the same at all energies, which is not fully realistic but sufficient for a first assessment of the BIB levels dependence on the collider energy. Furthermore, the findings of~\cite{Collamati:2021sbv} confirm the major role that is played by the nozzles in determining the particles fluxes that arrive on the detector surface. Their optimisation for the 3 and 10~TeV MuC could thus reduce the estimated BIB levels strongly. Studies for the 10~TeV collider option showed that lattice design choices such as combined function magnets in the final focus region or a larger $L^{*}$ provide instead only a limited potential for reducing the BIB~\cite{Calzolari:2022lgu}.

Table~\ref{tab:BIB_com_energy} (see also~\cite{Collamati:2021sbv}) displays a moderate dependence of the BIB multiplicities on the collider energy. In what follows we will thus employ \qty{1.5}{\TeV} BIB simulation results, being confident that no dramatic changes are expected at higher energies. The results below are obtained for a single beam travelling counterclockwise starting \qty{200}{\m} away from the IP. The other beam will have the mirrored effect owing to the symmetric nature of the BIB due to $\mu^+$ and $\mu^-$ decays.

The most important BIB property is that it is composed of a large number of particles with low energy, thanks to the MDI mitigation action, and it is characterised by a broad arrival time in the detector. More specifically, around $4 \cdot 10^{8}$ low-momentum particles exit the MDI in a single bunch crossing depositing energy to the detector in a diffused manner. 
There is a substantial spread in the arrival time of the BIB particles with respect to the bunch crossing, ranging from a few nanoseconds for electrons and photons to microseconds for neutrons, due to their smaller velocity.

Each of these aspects has different implications for the BIB signatures in different parts of the detector, which depend on the position, spatial granularity and timing capabilities of the corresponding sensitive elements. Thus, a careful choice of detector technologies and reconstruction techniques allows to mitigate the negative BIB effects, as demonstrated in later sections.

\begin{figure*}[t]
\centering
\includegraphics[width=0.5\textwidth]{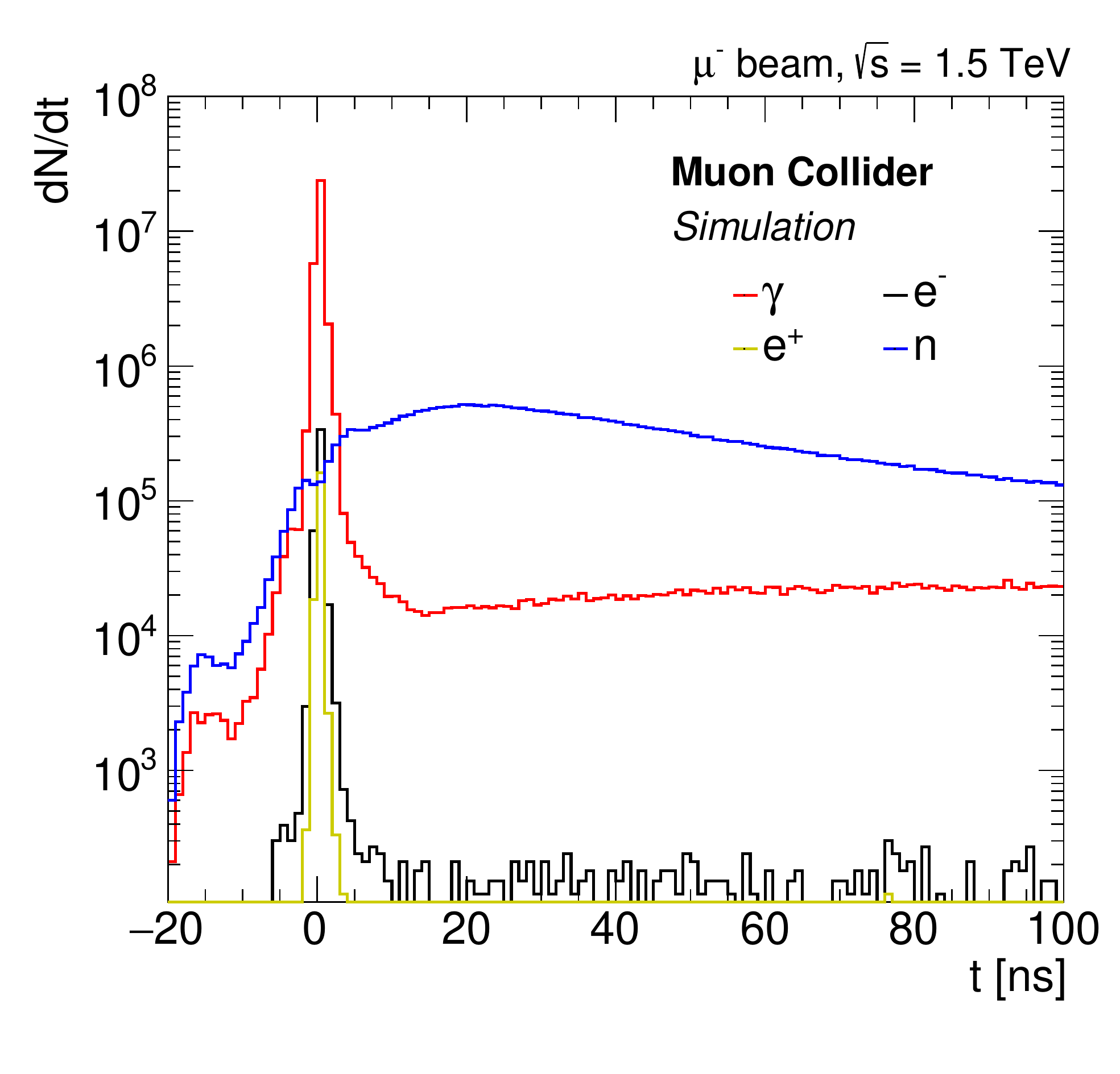}\includegraphics[width=0.5\textwidth]{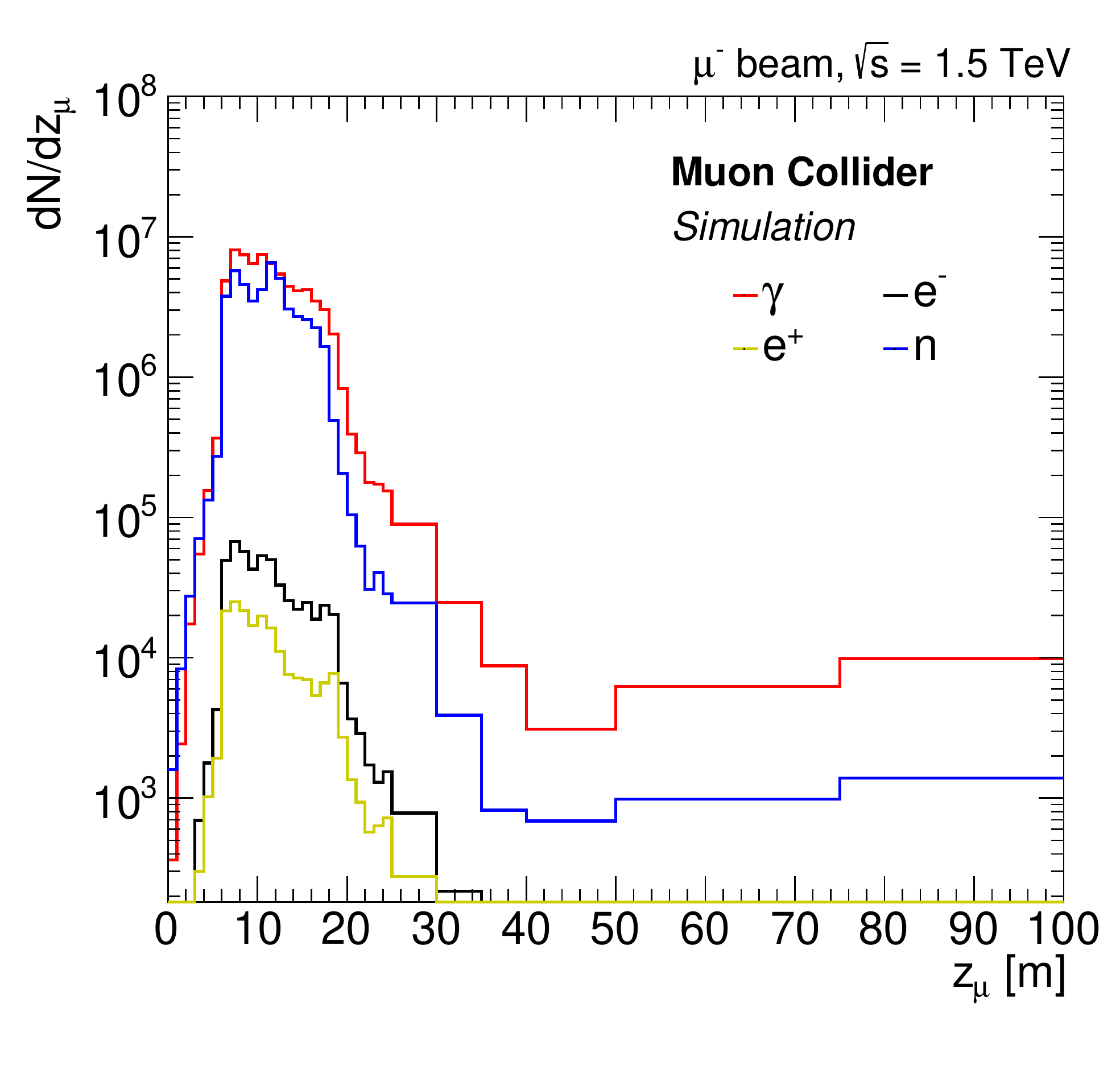}
\caption{\label{fig:bibtime} Time distribution of BIB particles exiting the machine (left) and longitudinal distribution of primary $\mu^-$ decay generating BIB particles exiting the machine (right). The results are based on the  FLUKA simulation, considering the primary $\mu^-$ decays within \qty{100}{\m} from the IP. }
\end{figure*}

The time at which the BIB particles exit the machine in the interaction region is spread over several tens of ns, but the major  concentration is around the beam crossing time ($t = 0$), as shown by the left panel of Figure~\ref{fig:bibtime}. This distribution suggests that the use of time-sensitive detectors would allow to suppress a large fraction of the background. The right panel of Figure~\ref{fig:bibtime} reports the longitudinal distribution of primary $\mu^-$ decays generating the most relevant BIB components: neutrons, photons and electrons/positrons. Simulations show that to correctly account for the secondary $\mu^\pm$, it is necessary to consider primary decays up $\sim\qty{100}{\m}$ from the IP.

\begin{figure*}[t]
\centering
\includegraphics[width=0.5\textwidth]{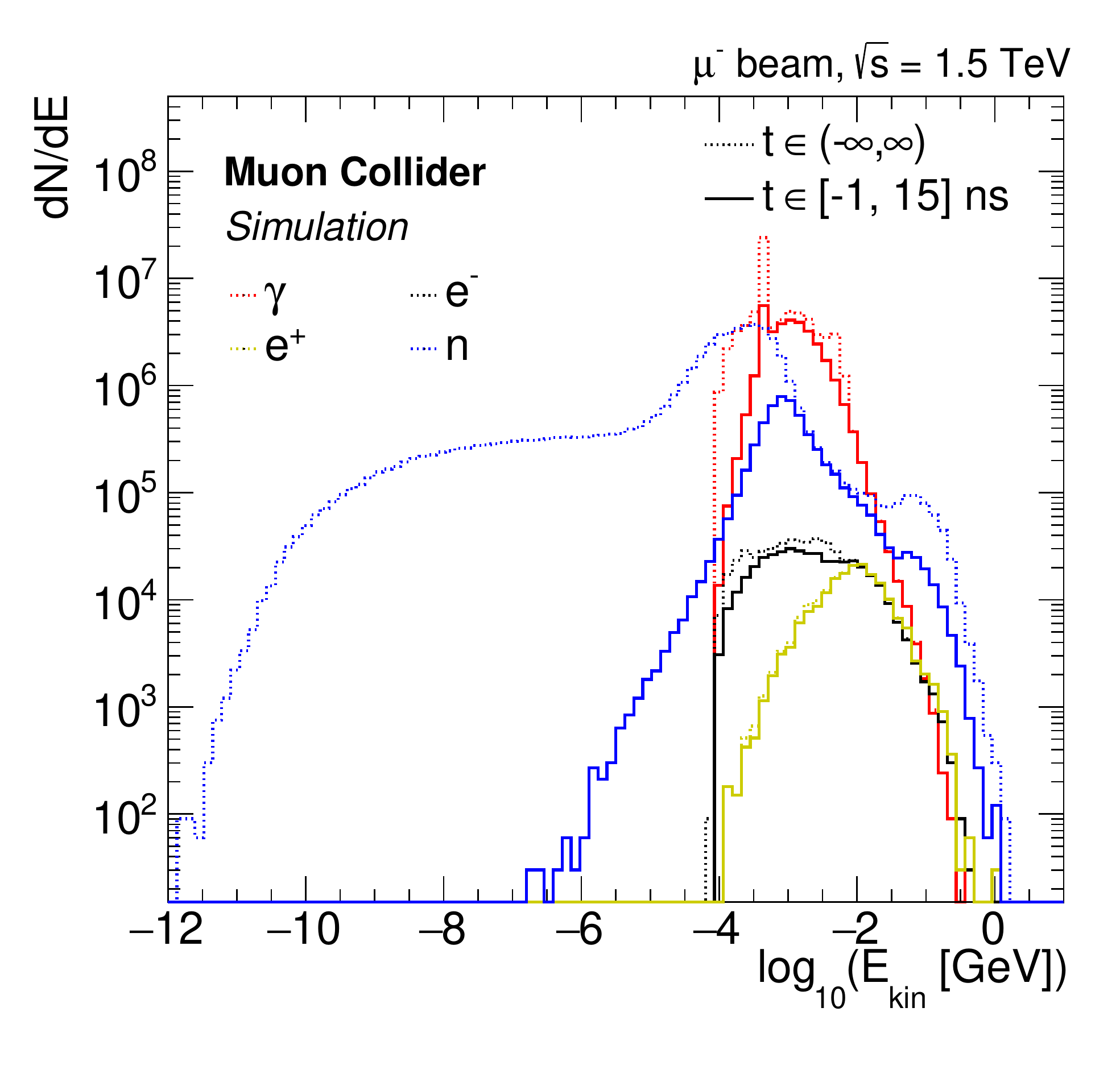}\includegraphics[width=0.5\textwidth]{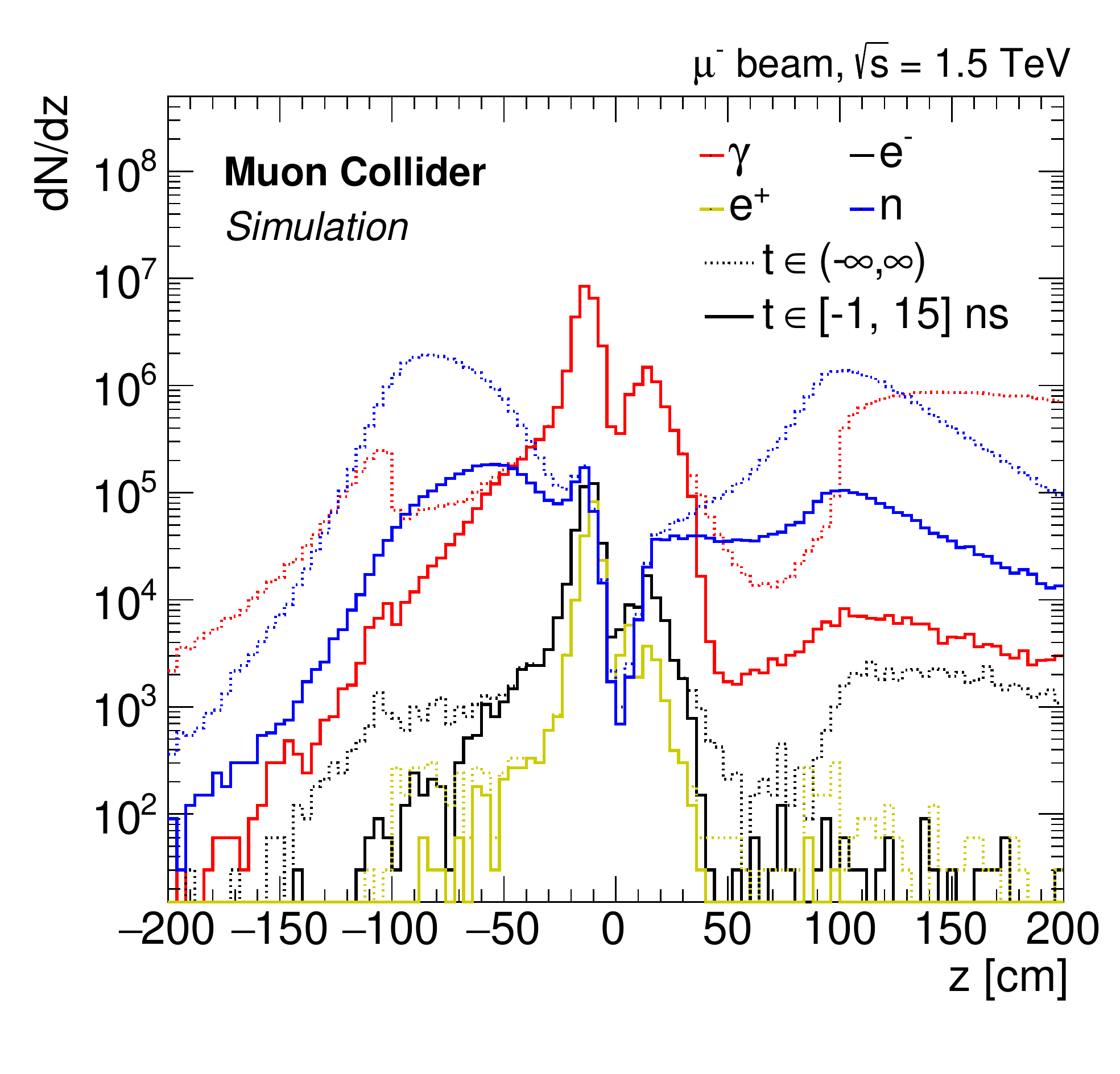}
\caption{\label{fig:bibenergy} Lethargy plot (left) and longitudinal exit coordinate distribution (right) of BIB particles, by particle type. No time cut is applied to distributions represented in dotted lines while in solid lines only particles exiting the machine between -1 and \qty{15}{\ns} are considered. The results are based on the FLUKA simulation, considering primary $\mu^-$ within \qty{100}{\m} from the IP.}
\end{figure*}

The kinetic energy distribution of most relevant BIB particle types is reported in Figure~\ref{fig:bibenergy}. Energy cutoffs have been applied in the simulation at 100 keV for $\gamma$, $e^\pm$, $\mu^\pm$, charged hadrons and at 10$^{-14}$ GeV for neutrons. The shielding nozzles strongly suppress the high energy BIB component, making the fraction of particles entering the detector volume with kinetic energy above few GeVs negligible. Only charged hadrons and secondary muons can reach higher energies, but their rate is of the order of 10$^{4}$ and 10$^{3}$, with respect to 10$^{7}$ photons, neutrons and 10$^{5}$ electrons, positrons. The longitudinal exit coordinate distribution displayed in the figure shows that most BIB particles enter the detector with a large longitudinal displacement from the collision region. This suggest that detectors with excellent pointing capabilities would allow to strongly suppress these background contributions.

\subsubsection*{Radiation levels}
The BIB distributions and rates are crucial to quantify the radiation levels and in turn the requirements on the detector components. The FLUKA BIB sample and the detector model previously described are employed.
The simulation~\cite{Collamati:2021sbv} used in fact a simplified detector geometry. The calorimeters, magnetic coils, and the return yoke were approximated with cylindrical elements with densities and material composition based on the averages from the full geometry. The magnetic field was assumed to be uniform. The silicon layers composing the inner tracker were instead included with exact dimensions.

\begin{figure*}[h]
    \begin{center}
        \centering
\includegraphics[width=0.8\textwidth]{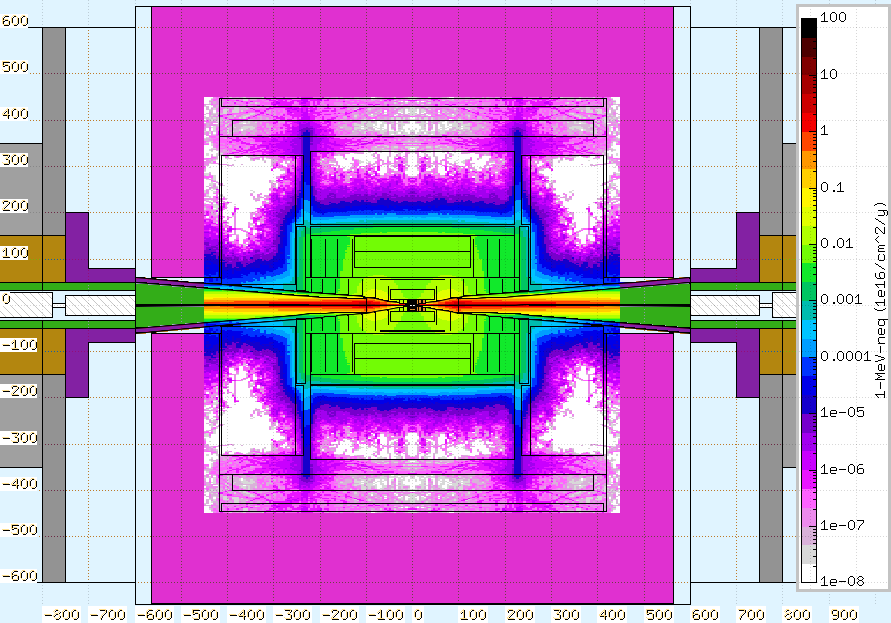}
    \end{center}
    \caption{Map of the 1-MeV-neq fluence in the detector region, shown as a function of the position along the beam axis and the radius. The map is normalised to one year of operation (200 days/year) and a collision rate of 100~kHz.}
    \label{fig:fluence}
\end{figure*}

\begin{figure*}[h]
    \begin{center}
        \centering
        \includegraphics[width=.8\textwidth]{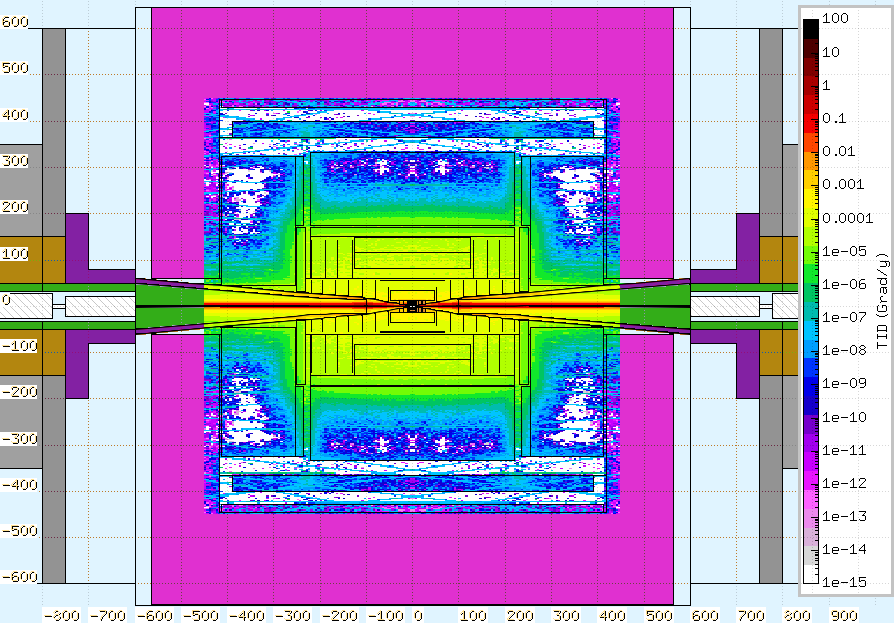}
    \end{center}
    \caption{Map of the TID in the detector region, shown as a function of the position along the beam axis and the radius. The map is normalised to one year of operation (200 days/year) and a collision rate of 100~kHz.}
    \label{fig:tid}
\end{figure*}

Figures~\ref{fig:fluence} and~\ref{fig:tid} display respectively the expected 1~MeV neutron equivalent fluence (1-MeV-neq) and the total ionising dose (TID) in the detector region, shown as a function of the beam axis z and the radial distance $r$ from the beam axis.
The normalisation for the dose maps is computed considering that the muon collisions are expected to happen at the maximum rate of 100~kHz, corresponding to the minimum time between crossings of 10~$\mu$s. With a single bunch collider operation scheme, this in turn corresponds to a minimal collider ring length of 2.5~km.
%
Assuming 200 days of operation during a year, the 1-MeV-neq fluence is expected to be $\sim 10^{14-15}$~cm$^{-2}$y$^{-1}$ in the region of the tracking detector and of $\sim 10^{14}$~cm$^{-2}$y$^{-1}$ in the electromagnetic calorimeter, with a steeply decreasing radial dependence beyond it. The total ionising dose is $\sim10^{-3}$~Grad/y on the tracking system and $\sim10^{-4}$~Grad/y on the electromagnetic calorimeter.

\subsection{Detector simulation software}
\label{sec:software}

The full simulation of a $\mu^+ \mu^-$ collision event involves several stages going from the generation of input particles, the simulation of their interactions with the detector material and of the detector response.

The first stage corresponds to the generation of all particles entering the detector. This stage is handled by standalone software, such as FLUKA or MARS15 for the BIB particles as previously described and Monte Carlo event generators for the $\mu^+ \mu^-$ scattering process. 

The input particles are then propagated through the detector material and their interactions with the passive and sensitive material of the detector are simulated with the 
\textsc{Geant4}~\cite{GEANT4:2002zbu} software.
The iLCSoft framework~\cite{ilcsoft}, previously used by CLIC~\cite{Linssen:2012hp} and now forked for developments of muon collider studies~\cite{mucolsoft}, is used for this and all further processing stages.

The detector response and event reconstruction are handled inside the modular Marlin framework~\cite{marlin}.
The detector geometry is defined using the DD4hep detector description toolkit~\cite{dd4hep}, which provides a consistent interface with both the \textsc{Geant4} and Marlin environments. The response of each sensitive detector element to the corresponding energy deposits returned by \textsc{Geant4} is simulated by dedicated digitisation modules implemented as individual Marlin processors.

The tracking detectors use Gaussian smearing functions to account for the spatial and time resolutions of the hits registered on the sensor surface. Acceptance time intervals, individually configured for each detector, are used for replicating the finite readout time windows in the electronics of a real detector and to reject hits from from out-of-time BIB particles.

The result of this simplified approach is a one-to-one correspondence between the \textsc{Geant4} hits and digitised hits, which ignores the effect of charge distribution across larger area due to the Lorentz drift and shallow crossing angles with respect to the sensor surface.
These effects are taken into account in the more realistic tracker digitisation software that is currently under development and will allow stronger BIB suppression based on cluster-shape analysis.

The ECAL and HCAL detectors are digitised using realistic segmentation of sensitive layers into cells by summing all energy deposits in a single cell over the configured integration time of \qty{\pm 250}{\pico\second}.
The time of the earliest energy deposit is consequently assigned to the whole digitised hit. The same digitisation approach is used also for the Muon Detector.

More details about the software structure and computational optimisation methods used for simulating the very large number of BIB particles are given in Ref.~\cite{Bartosik:2021bjh}. For the interested user, pointers to the documentation of the software stack, tutorials and other tools are available at the MuC software project page~\cite{mucolsoft}.

\subsection{Detector technologies}
\label{sec:technology}

The simulation workflow described in the previous sections enables a first assessment of the challenges for the various detector systems and of the required technologies, which are described in the present section.

\subsubsection*{Tracking systems}

The ability to reconstruct trajectories of charged particles in the tracking system and to measure their parameters with high precision is essential at the muon collider experiments. 
In the expected operating conditions, high performance tracking is necessary to achieve good efficiency and resolution for reconstructing charged leptons, jets, energy sums, displaced vertices originating from the heavy flavour hadron decays, as well as potential new phenomena. 

\begin{figure}[!ht]
\center
\includegraphics[width=0.5\textwidth, trim= {35 55 0 0 } , clip]{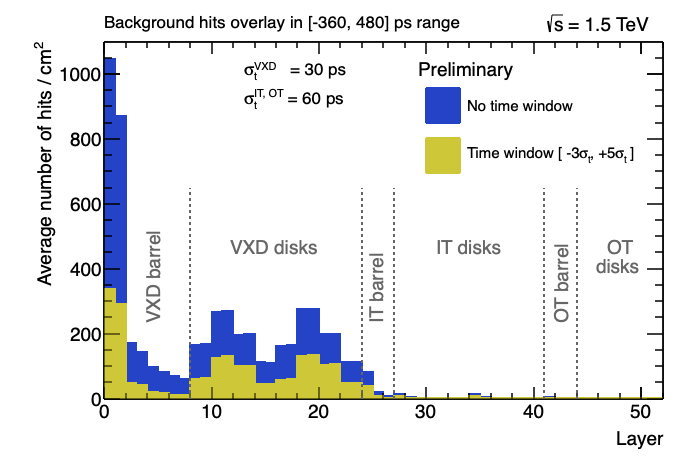}
\caption{Average hit density per bunch crossing in the tracker as a function of the detector layer.}
\label{fig:HitMultTracker}
\end{figure}

The BIB represents a significant hurdle for tracking, both in terms of the data volumes generated by the tracker as well as by introducing a combinatorial challenge in the track reconstruction. In each bunch crossing, BIB particles on average generate 500\,000 hits in the most inner layer of the tracker, located just few centimetres away from the interaction point. This corresponds to hit density of up to 1000 hits/cm$^2$. However, the density does decrease rapidly as a function of the radial distance from the beam-line, as shown in Figure~\ref{fig:HitMultTracker}. 

It is clear from these numbers that high granularity of silicon pixels is necessary in order to achieve hit occupancy level of a few \%. In addition, various handles to reduce the BIB should be explored for both on- and off-detector filtering.
Possible filtering schemes include:
\begin{itemize}
\item \textbf{Timing}: Removing hits incompatible with the main bunch crossing time could reduce the data load by about a factor of 3. Timing information will eventually be used in the event reconstruction, but an initial on-detector filtering could be implemented as well.
\item \textbf{Clustering}: Pixel clustering to reduce the number of single pixels to be read out. This requires more on-detector processing and results in more bits per cluster and a higher power budget, but can reduce the number of hits read out. Selection requirements can also be applied to the cluster shape. The effectiveness needs to be assessed for each BIB cluster type.
\item \textbf{Energy Deposition}: Each of the backgrounds has a characteristic energy deposition signature. For example neutrons have low, localised energy deposits. On-detector filters could efficiently exploit this quantity.
\item  \textbf{Correlation Between Layers}: This is a powerful handle for background rejection. However, an implementation may be complex and costly, doubling the number of channels. For on-detector filtering, it also requires transfer of data between layers in a very busy environment.
\item \textbf{Local Track Angle}: Track angle measurement can be made in a single detector if the thickness/pitch ratio distributes the signal over several pixels. This avoids the complexity of inter-detector connections and could provide a monolithic solution~\cite{Lipton:2022njd,Lipton:2019drv}.
\item \textbf{Pulse Shape}: Signals from BIB can come with a variety of angles and may not give the deposit profile and pulse shape of a typical minimum ionising particle (MIP).  Appropriate pulse processing, such as multiple sampling, RC-CR filters, zero crossing, or delay line clipping can be used to reduce the data load.
\end{itemize}

The basic trade-offs are between the complexity, power, and mass needed to implement a on-detector filter, and the benefit of reduced data rate. Particular caution should be taken when it comes to on-detector filtering: overly aggressive front-end filtering schemes can introduce irrecoverable inefficiencies and biases in track reconstruction and can limit acceptance for some beyond the Standard Model signatures, such as those of long-lived particles. 

A study was conducted in simulation to determine the granularity and timing requirements for the tracker sensors in order to reduce the hit occupancy to under the 1\% target level. In this study, the pixel size and per hit timing resolution were independently varied in each layer of the detector. The detector hits were integrated in the time period of 1~ns following the bunch crossing. It was found that for the vertex detector granularity of $25 \times 25$~$\mu$m$^{2}$ and time resolution of 30~ps are needed to achieve the desired occupancy goal. The inner tracker was relying on asymmetric macropixels with $50 \times 100$~$\mu$m$^{2}$ size and 60~ps timing resolution were sufficient to satisfy the requirement, while the outer tracker assumed either macropixels or microstrips with the size of 50~$\mu$m$ \times 10$~mm and a 60~ps time resolution. It is thus evident that R\&D efforts towards 4D tracking are necessary to achieve the required spacial granularity and timing resolution. 

Silicon-based sensors have come to dominate the technology for collider detector tracking systems. This is likely to continue into the muon collider era. In the past decade there have been a number of technological developments that promise to achieve many of the capabilities discussed in the previous section. They address different aspects of the needs for space and time resolution, pattern recognition, electronics integration, radiation hardness, and low cost.
A description of promising technologies for achieving such goals is provided below. 

\subsubsection*{Monolithic Devices (CMOS MAPS)}
\begin{figure*}[h]
\centering
\includegraphics[width=0.95\textwidth]{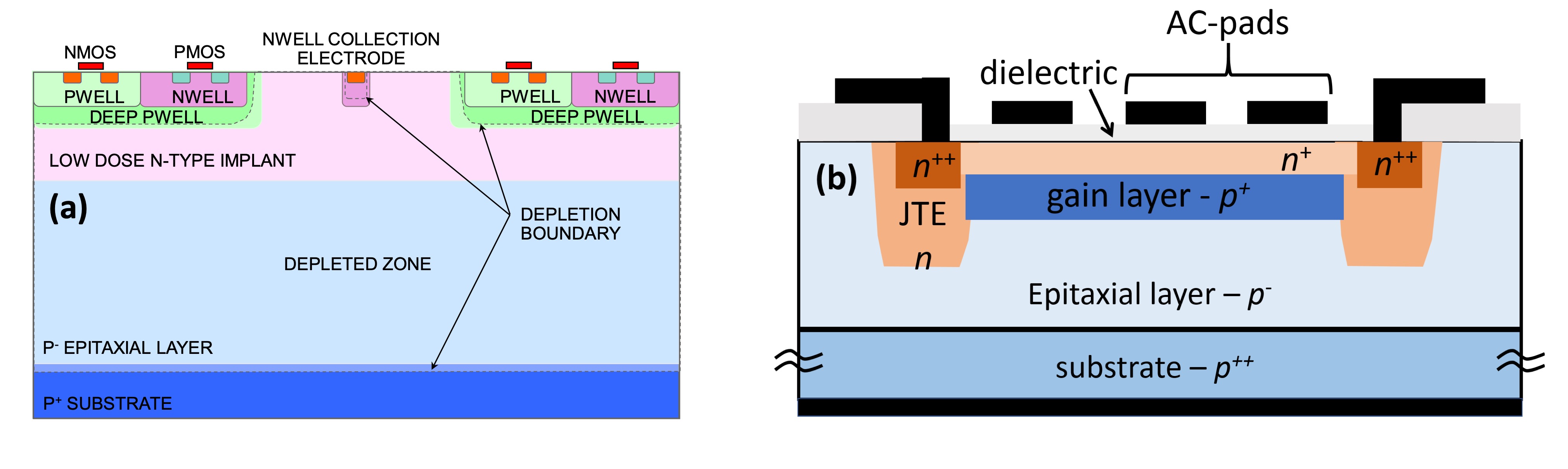}
\caption{a) Cross section of the MALTA~\cite{Cardella:2019ksc} CMOS sensor showing the implant structure.  The low dose n-implant provides an electrode that improves the uniformity of the p-epitaxial drift field.  b) Cross section of an AC coupled LGAD~\cite{Giacomini:2019kqz}.  The Junction Termination Edge (JTE in the figure), which would normally separate each pixel, is only needed at the edge of the device, providing near 100\% fill factor.}
\label{CMOS_LGAD}
\end{figure*}

CMOS Monolithic Active Pixel Sensors (MAPS) are based on standard CMOS process flows with thick (20--50~$\mu$m) epitaxial layers. Charge is collected from electron-hole pairs generated in the epitaxy. There is a small area n-type collection electrode with the CMOS circuitry embedded in a deep $p$-well to avoid parasitic charge collection by the CMOS transistors. The geometry of the electrodes and associated circuitry means that the epitaxy is difficult to deplete evenly. The first such devices used diffusion rather than drift to collect charge.  Recent prototypes, shown in Figure~\ref{CMOS_LGAD}a, have added a deep, lightly doped, n-layer below the p-well to provide a more uniform drift field in the epitaxy~\cite{Cardella:2019ksc}. 
The proximity of the CMOS transistors to the epitaxy means these devices have significant sensitivity to analog-digital cross-talk. It is likely that the CMOS analog section would have to be 3D stacked with digital TDC, ADC and I/O tiers to achieve adequate isolation and overall functionality.  Radiation hardness needs to be studied and improved.  In particular the effects of doping evolution in the epitaxy will affect fields and operation as the device ages. This is a particular problem with the large ratio of collection node to collection area which forces a difficult geometry for the collection field. The size of CMOS sensors is limited by the typically 2x3 cm CMOS reticule. To achieve a large area device the reticules must be ``stitched'' using an additional metal layer or tiled while minimising dead area at the edges.

\subsubsection*{Devices with Intrinsic Gain}
In sensors based on the Low Gain Avalanche Diode (LGAD) design, the initial charge created by an impinging particle is amplified in a ``gain layer'' by a factor of about 10--30. The resulting current signal is large and fast, enabling a 20--30~ps time resolution. An interesting feature of LGAD-based sensors is that the associated front-end requires less power since the sensor provides the first amplification stage. 
The LGAD  design is relatively new and is undergoing rapid development.
The design of the current generation of LGADs being used for endcap timing layers in the CMS and ATLAS HL-LHC upgrades has now been superseded. This first generation of devices suffers from limited fill-factor due to edge field limiting structures and moderate radiation hardness. Recent works have combined internal gain with the novel resistive read-out design, reaching a design (the so-called RSD, Resistive Silicon Detectors) with a 100\% fill factor and excellent spatial precision even with pixels with a large pitch (a precision of less than 5\% of the pitch)~\cite{Tornago:2020otn}. Parallel to this development, new studies suggest possible improvements in the radiation resistance of the LGAD design. The two most promising are: (i) the insertion of an additional layer of carbon to reduce acceptor removal in the gain implant and (ii) the design of the gain implant using both acceptor and donor dopings so that acceptor removal is compensated by donor removal, extending the radiation hardness of the gain implant. 
The resistive charge-sharing design also allows for sparser read-out geometries, limiting the density of analog channels. 

For the muon collider, being a high-rate environment, the novel DC-coupled design of the resistive read-out technique might be very beneficial, as it allows for a faster recovery time. 
Detailed studies of AC-RSD and DC-RSD will be needed, optimising the electrode density and geometry in conjunction with charge deposition characteristics of the photon, electron/positron, and neutron backgrounds. Dopant removal is the primary limitation to radiation hardness, and continued study is needed to understand the practical restrictions imposed by this effect. The total current going into detector bias will be larger in devices with gain and may represent a significant fraction of the total power.

\subsubsection*{Hybrid Small Pixel devices}
\begin{figure*}[h]
\centering
\includegraphics[width=0.95\textwidth]{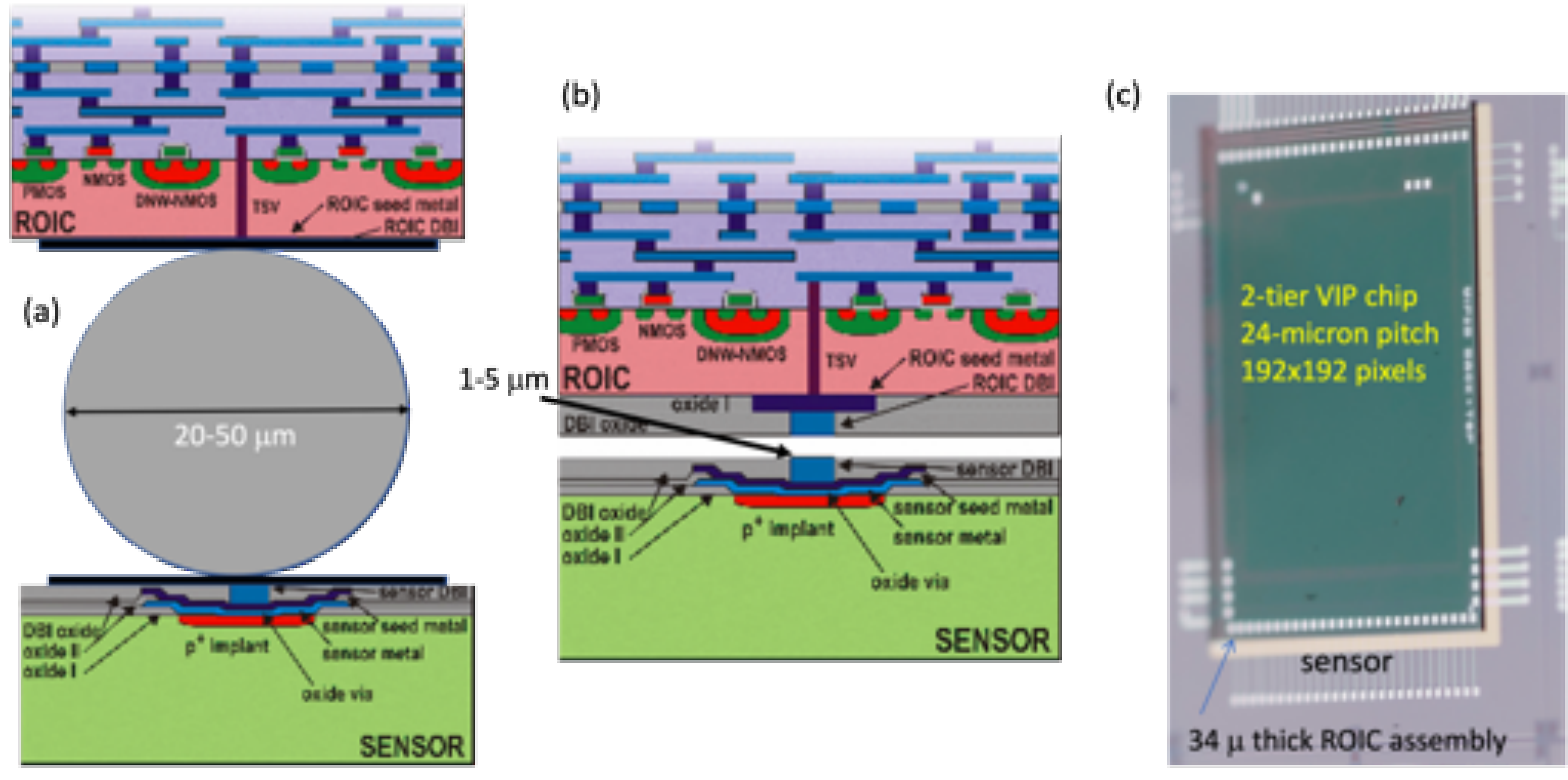}
\caption{a) A typical bump bonded sensor/readout chip geometry. The spacing is determined by the size of the bump and under-bump metalization pad.  b) A hybrid bonded sensor/readout stack with the pitch limited by the micron-level hybrid metalization imbedded in the top oxide layer.  c) An example of a three-tier hybrid bonded stack with separate analog and digital readout layers\cite{Lipton:2015vca}. The readout pitch is 24 microns and the readout stack thickness is 35$\mu$.}
\label{fig:Bump_Hybrid}
\end{figure*}

These are standard ``hybrid'' pixel diode sensors without gain. However fast timing and good position resolution can be achieved with fine pixel pitch and low input capacitance. Bump bonding using solder, indium, or copper, is limited to $\approx$15-50 micron interconnect pitch. Three-dimensional hybrid bonding~\footnote{The electronics industry has adopted ``hybrid bonding'' as the generic term for oxide bonded stacks of wafers or chips with embedded metal to achieve top to bottom electrical connection. This technology is commonly used in cell phone image sensors.}, illustrated in Figure~\ref{fig:Bump_Hybrid}, can achieve both $<5~\mu$m pitch and low enough interconnect capacitance to meet noise and power limitations. The sensors have the advantage of being intrinsically radiation hard with signal/noise large enough to provide 20--30~ps time resolution. It will likely be necessary a design where ADCs and TDC service multiple small pixels to reduce the density of ADCs and TDCs, thus reducing both the in-pixel density and power. 

\subsubsection*{Intelligent Sensors}
The different characteristic signals generated by electromagnetic, neutron, and charged hadron backgrounds and signal MIPs prompts consideration of more ``intelligent'' sensors that can separate the BIB from the signal. An example is the current 2-layer track trigger design for CMS at the LHC where low $p_{\textrm{T}}$ tracks are filtered out by comparing hits on separated sensor layers. Such multi-layer designs are limited by the complex interconnection and data transmission paths needed to communicate between sensor layers. However, for a device where the thickness/pixel pitch ratio is large enough, the pixel pulse shapes and cluster patterns will be very different for MIPs and BIB hits. This information can be used for a prompt local filter to reject BIB. Radiation-induced traps will cause the pulse shapes and induced current patterns to change during the lifetime of the detector, possibly necessitating changes in algorithms.

Appropriate information density can be achieved in small pixel devices or double-sided LGADs~\cite{Lipton:2022njd}. In the double sided LGAD fast timing signals are read on a top, larger pitch layer coupled to the gain layer. Charge deposition patterns and timing are reconstructed on a bottom, pixelated layer. Other concepts can be explored where the very different pulse shapes and patterns can be used to separate signal from BIB, perhaps incorporating on-chip machine-learning techniques.

\subsubsection*{Power Considerations}
This section describes a tentative estimate of the power constraints on the tracker based on extrapolations of the existing technologies. The study focuses on the vertex detector and assumes a design with $25\times25$ $\mu$m$^2$ pixels with four barrel layers and four endcap disks on each side, as previously described. Conventional scaled CMOS electronics~\cite{OConnor:1999isd} and possible extrapolations of optical-based data transmission are also assumed. New technologies might change the picture completely. 

For conventional CMOS-based amplifiers and a conventional silicon detector operating with no internal gain the front-end power will be determined by the capacitive load on the front-end and the desired signal/noise and rise time. For example, a simple SPICE model of a preamplifier loaded with 100~fF capacitance provides 4~ps time jitter for 1 $\mu$A bias current and  45~ps for a bias current of 100~nA. A time jitter of $< 30$~ps can be achieved in a conventional sensor with 50~$\mu$m thickness and front-end current of less than 250~pA if the detector capacitance is carefully controlled using 3D interconnections. If the sensor is based on LGAD-like internal gain the signal presented to the preamplifier can be $10-20$ times larger. For the same signal over noise, this could reduce the front-end transductance and associated drain current by roughly the square root of the gain. A conventional CMOS amplifier is expected to draw about 450~W of power into the vertex detector for the analog bias.

In addition to front-end power there is the power necessary to bias the detector. This can become significant for heavily irradiated detectors. If a HL-LHC-like operating scenario is considered, the final depletion voltage can be as high as  500~V (depending on the technology chosen). Under these conditions the vertex sensor bias power is about 100~W.

It is also useful to estimate the detector data load. Using the simulated layer occupancies in the vertex detector, a total rate of hits of $15 \times 10^{13}$ bits per second (b/s) for the vertex detector is obtained. More detail about the estimate is provided at the end of the section.
The power needed per bit for the Low Power Gigabit Transceiver (lpGBT)~\cite{Guettouche:2022wzn} is about 41~pJoule/bit. Power efficient optical transmission is the subject of intense study by the semiconductor industry and 10~pJoule/bit is assumed as a conservative estimate for the future power consumption. This gives us a data transmission power of about 1.5~kW. Attojoule/bit levels, albeit before radiation damage considerations, appear feasible in the near future~\cite{7805240}, further reducing the penalty for reading the full event. Each link must be capable of transmitting at a rate of about $20$~Gb/s to limit the number of physical optical connections to one per module. Higher speeds (if available) will lead to a reduction of the overall number of optical links in the system. 

\FloatBarrier
\subsubsection*{Calorimeter systems}

The measurement of physics processes at the energy frontier requires excellent energy and spatial resolutions to resolve the structure of collimated high-energy jets. Jet reconstruction, and the improvement of the reconstructed jet energy resolution, is the driving theme of ongoing R\&D activities in high energy physics, and a muon collider will not diverge from this general theme. Future lepton colliders aim at separating W and Z bosons in the dijet channel, which requires a 3-4\% jet energy resolution for jets above 100~GeV.

\begin{figure}[!ht]
\center
\includegraphics[width=0.425\textwidth, trim = {65 55 0 10} ,clip]{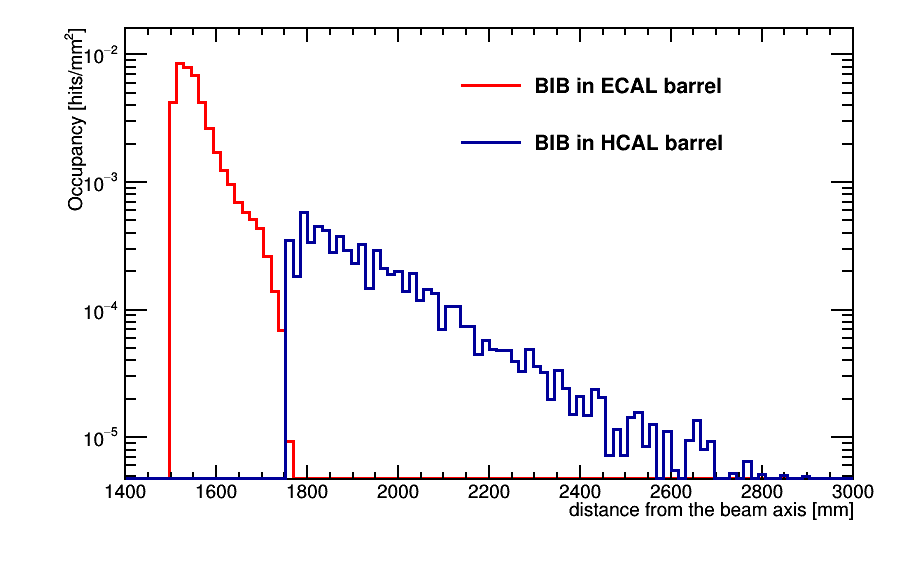}
\caption{BIB hit occupancy in the calorimenter barrel region in a single bunch-crossing.}
\label{fig:calo_occupancy_barrel}
\end{figure}

\begin{figure}[!ht]
\center
\includegraphics[width=0.4226\textwidth,trim = {65 45 0 0} ,clip]{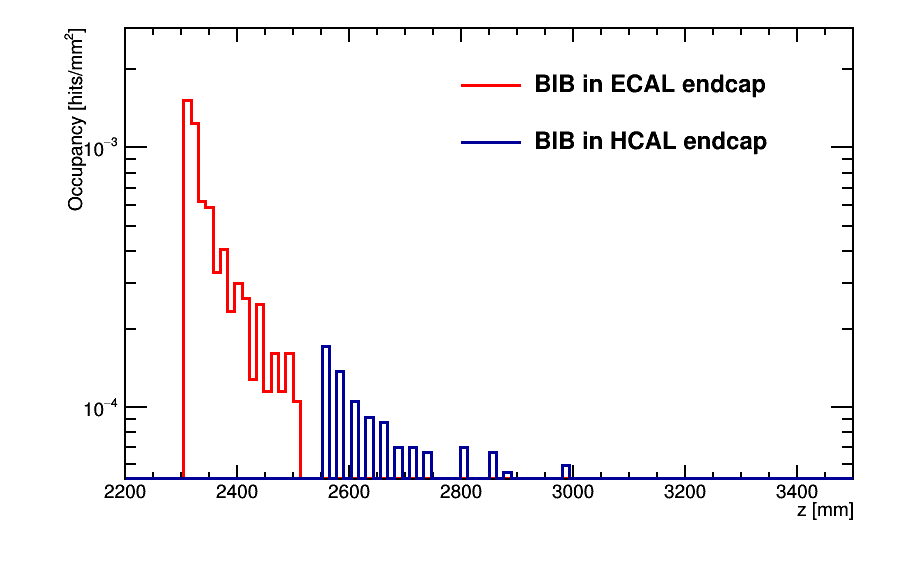}
\caption{BIB hit occupancy in the calorimeter endcap region in a single bunch-crossing.}
\label{fig:calo_occupancy_endcap}
\end{figure}

\begin{figure}[!t]
\center
\includegraphics[width=0.48\textwidth,trim = {14 9 0 0} ,clip]{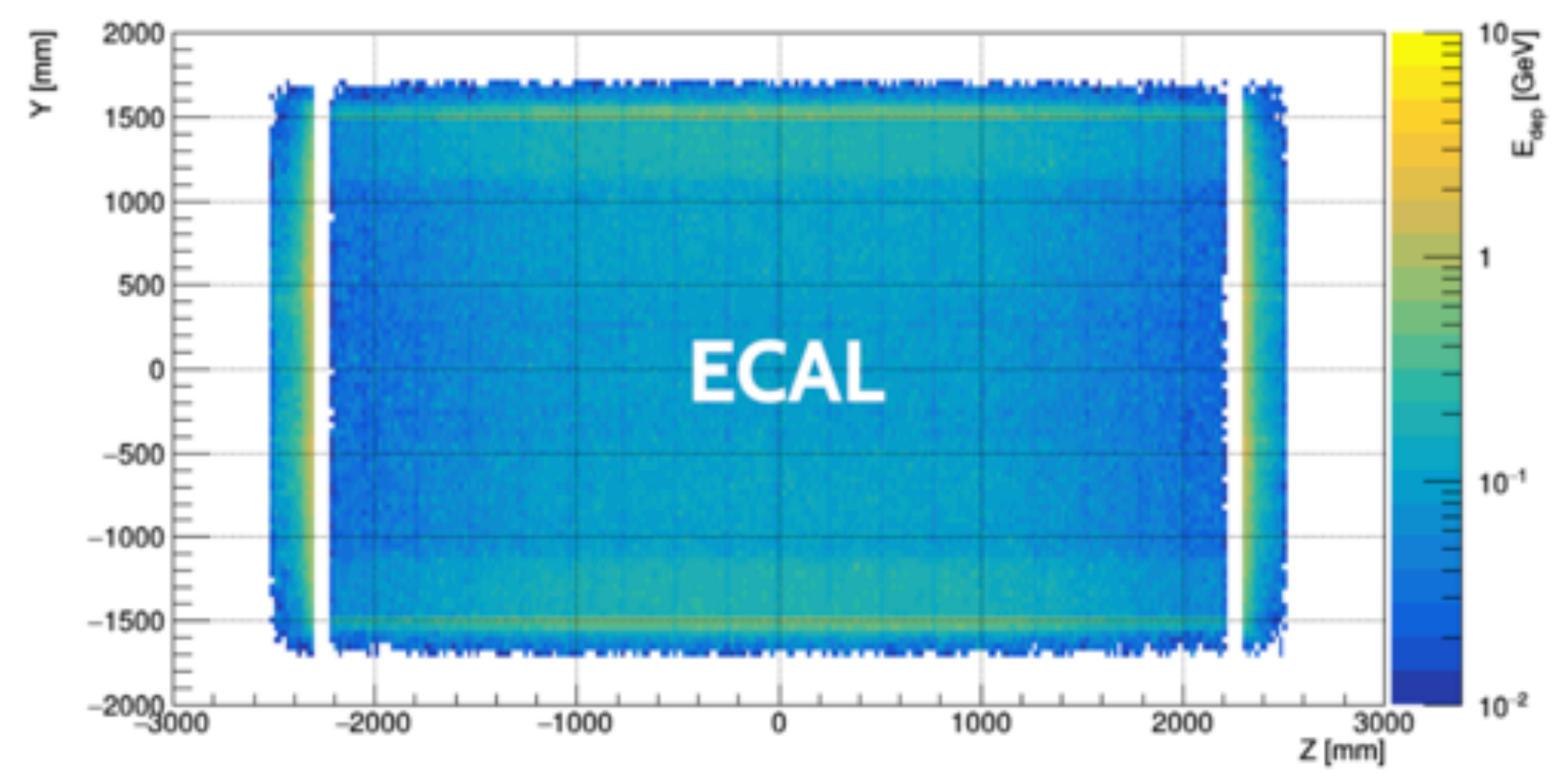}   \caption{Energy deposited by the BIB in a single bunch-crossing in the ECAL.}
\label{fig:ecal_deposit}
\end{figure}

\begin{figure}[!t]
\center
\includegraphics[width=0.48\textwidth,trim = {12 9 0 0} ,clip]{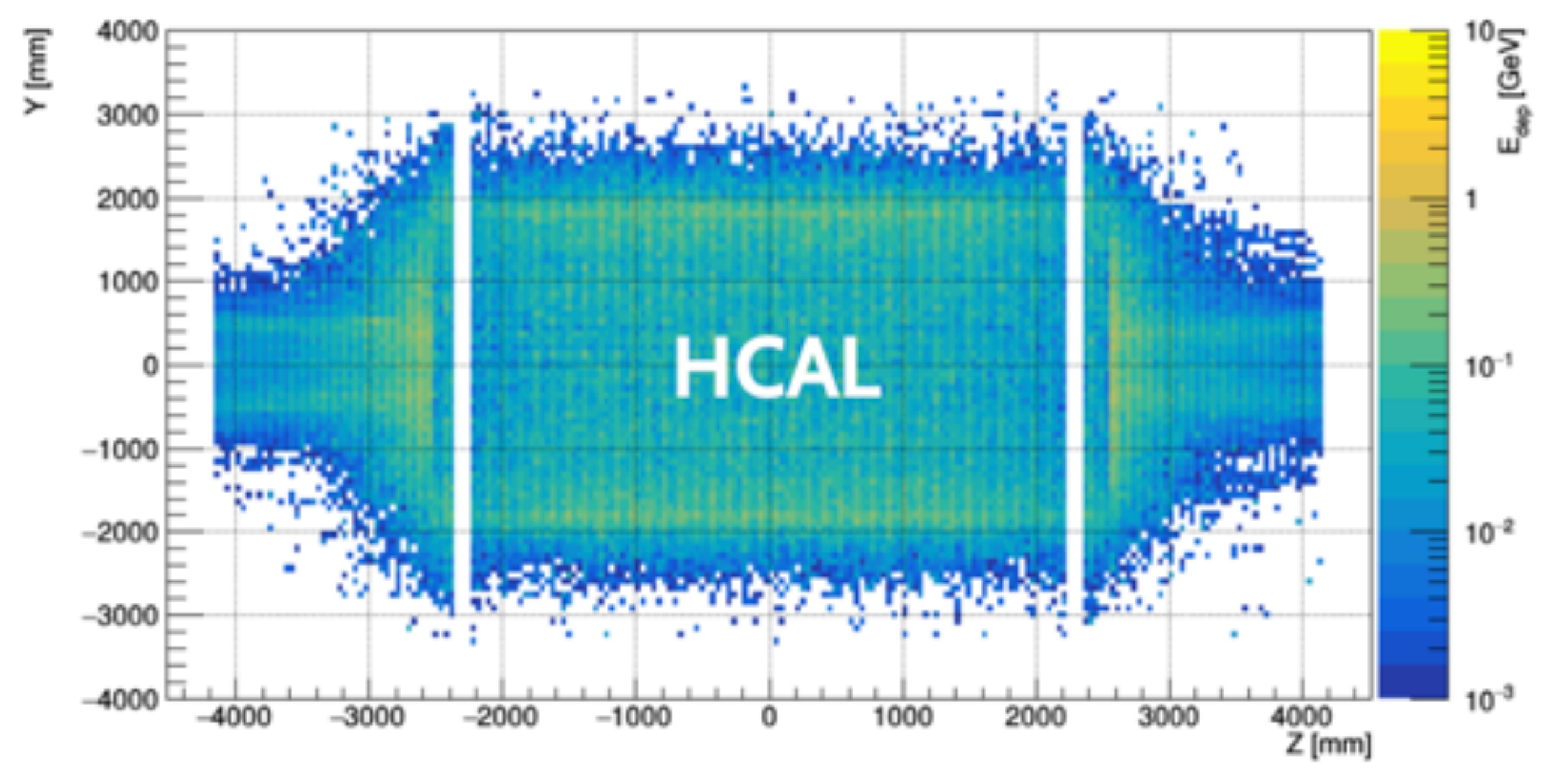}
\caption{Energy deposited by the BIB in a single bunch-crossing in the HCAL.}
\label{fig:hcal_deposit}
\end{figure}

In a multi-TeV muon collider, the calorimeter system has to operate in the intense flux of low energy particles arising from the BIB. The BIB in the calorimeter region is mainly formed by photons (96\%) and neutrons (4\%). In the current detector layout, a flux of about 300 particles per cm$^2$ is present at the ECAL barrel surface, with an average photon energy of about 1.7~MeV. Given the high flux, several particles may overlap in a single cell, resulting in a hit where their energy is summed up. The occupancy, defined as the number of hits per mm$^2$ in a calorimeter layer, is shown as a function of the calorimeter depth in Figure~\ref{fig:calo_occupancy_barrel} and of the z coordinate in Figure~\ref{fig:calo_occupancy_endcap}. The simulation shows how the ECAL system absorbs most of the BIB radiation, resulting in a significantly lower occupancy for the HCAL system.
The ECAL is expected to receive approximately 100~krad/y of total ionising dose and a $10^{13-14}$~cm$^{-2}$~1-MeV-neq fluence. The spatial distributions of the energy deposited in ECAL and HCAL in a single bunch-crossing are shown respectively in Figure~\ref{fig:ecal_deposit} and Figure~\ref{fig:hcal_deposit}.

Similarly to what can be done in the tracker, the time of arrival of particles in a calorimeter cell can be exploited to discriminate the BIB contributions from the primary interactions. At the same time, the BIB particles are expected to deposit most of their energy in the innermost layers of the calorimeter,  with particles from the primary interaction propagating deeper in the detector.

The technology and the design of the calorimeters should be chosen to reduce the effect of the BIB, while keeping good physics performance. Several requirements can be inferred:
\begin{itemize}
    \item \textbf{High granularity} to reduce the overlap of BIB particles in the same calorimeter cell. The overlap can produce hits with an energy similar to the signal, making harder to distinguish it from the BIB;
    \item \textbf{Good timing} to reduce the out-of-time component of the BIB. An acquisition time window of about $\Delta t = 300$~ps could be applied to remove most of the BIB, while preserving most of the signal. This means that a time resolution in the order of $\sigma_t = 100$~ps (from $\Delta t \approx 3 \sigma_t$) should be achieved;
    \item \textbf{Longitudinal segmentation} to discriminate between the different energy profiles of signal processes and the BIB. A fine segmentation of the calorimeter can help in distinguishing the signal showers from the fake showers produces by the BIB;
    \item \textbf{Good energy resolution} of ${10\%}/{\sqrt{E}}$ in the ECAL system is expected to be enough to obtain good physics performance, as has been already demonstrated for conceptual particle flow calorimeters.
\end{itemize}

The requirements imposed by the need to house the calorimeter systems within a large magnetic coil tend to disfavour designs fully based on homogeneous calorimetry. Sampling calorimeters based on alternating dense passive materials, such as copper, steel, or tungsten, and active readout materials, such as plastic scintillators, silicon, or gaseous detectors are likely to be employed, at least in the HCAL.
Two major approaches are being pursued to exploit sampling calorimeters and improve upon the current generation of collider experiments: multi-readout (dual or triple)~\cite{Lee:2017shn,INFNRD-FA:2020fiu,Ferrari:2018noa} and particle flow~\cite{Thomson:2009rp} calorimetry.
The first approach focuses on reducing the fluctuations in the hadronic shower reconstruction, which are the main responsible for the deterioration in the determination of the jet energy. This goal is achieved by measuring independently the electromagnetic and the non-electromagnetic components of a hadronic shower, thus allowing to correct event-by-event for the different response of the calorimeter to various particle species. 
The second approach focuses on the reconstruction of the four-momenta of every particle recorded by the detector. This method exploits tracking information and requires a detector with extreme granularity, combined with powerful reconstruction algorithms aimed at resolving each particle's trajectory through the whole detector.

\subsubsection*{Dual-readout calorimetry}
The energy resolution of a calorimeter system is affected by fluctuations in the energy deposited in its active elements. When measuring hadronic showers, the fluctuations in the electromagnetic component of a shower represent the dominant contribution to the total resolution.
To minimise these fluctuations, Dual-Readout calorimetry aims at measuring both the scintillation light component (sensitive to both hadrons and e.m. particles) and the Cherenkov light component (sensitive only to relativistic e.m. particles). Notable examples of implementations of this concept can be found in the work of the DREAM/RD52 collaboration~\cite{Wigmans:2007es} and the proposed IDEA~\cite{FCC:2018evy,CEPCStudyGroup:2018rmc} detector for FCCee/CepC.
In these calorimeters, signals are generated in scintillating fibers, which measure the deposited energy, and in clear plastic PMMA or quartz fibers, which are sensitive to the Cherenkov light. A large number of such fibers are embedded in a fully projective lead or copper absorber structure. This detector is not longitudinally segmented: e.m. and hadronic showers can be distinguished using short and long fibers in the calorimeter, and longitudinal information could be extracted by reading out the fibers on both ends. Past results~\cite{Lee:2017xss,Lee:2017shn} have demonstrated to reach an energy resolution for charged pions of 34\%/$\sqrt{\textrm{GeV}}$ and the final goal of 30\%/$\sqrt{\textrm{GeV}}$ seems well within reach. The main challenges appear in the handling of the high number of fibers and SiPMs which are of the order of few 10$^8$ and constitute an important fraction of the technology cost. 
The implementation of a third fiber material, or alternatively the time readout of the scintillation fibers, can be used to measure the MeV-scale neutrons produced in a hadronic shower and suppress the fluctuations arising from binding energy loss~\cite{Lee:2017oye}. This latter method is referred to as triple readout.
The absence of longitudinal segmentation is likely the limiting factor for deploying such design in a muon collider detector. However, hybrid designs composed of an ECAL made of crystals (such as LYSO or PbWO$_{4}$) and the hadronic section based on the DREAM fiber prototype have been also proposed~\cite{Lucchini:2020bac,Gaudio:2011zzb}. These designs need to demonstrate the feasibility of dual readout concepts in the crystal matrix.

\subsubsection*{Particle flow calorimetry}
In recent years the concept of high granularity particle flow calorimetry~\cite{Brient:2001fow} has been developed in the context of the proposed International Linear Collider (ILC). The CALICE collaboration is the major developer of calorimeter concepts and technologies for highly granular detectors for particle flow. The goal of particle flow calorimeters is to build an image of the showers induced by the various jet-fragments to allow the correct matching of these showers with the charged particles measured in the tracker. This in turn enables to correctly identify and measure the energy of the showers induced by neutral hadrons.

\subsubsection*{Silicon-based sampling}
With a radiation length ($X_0$) of 0.35~cm and an interaction length ($\lambda_I$) of 9.95~cm, tungsten is an ideal absorber for an electromagnetic sampling calorimeter. Silicon sensors can be used as active elements to achieve a high channel granularity and longitudinal segmentation. Moreover, state-of-the-art silicon sensors can sustain the high radiation dose of the expected BIB. Analogous technologies are adopted by LHC experiments upgrades~\cite{CMS:2017jpq}, and considered by the CLIC collaboration. The CLIC ECAL barrel, on which the current muon collider detector design is based, is composed of 64M sensors sampling 40 layers. Future developments should implement a precise timing measurement in these sensors ($<$100 ps) in order to make them usable at a muon collider. Although the high granularity is a clear advantage, the associated number of electronic read-out channels is a non-trivial technological problem. Moreover the cost for such a system exceeds the cost of other solutions. 

\subsubsection*{Scintillator-based sampling}
Plastic scintillators can be used for a high-granularity detector. Small tiles or strips of scintillating material can be produced with a typical size of 1-5~cm$^{2}$ for e.m. applications and about 10~cm$^{2}$ for hadronic applications. The typical thickness of the active layer is 0.3-0.5~cm, which makes it possible to design detectors with a longitudinal segmentation of several tens of layers. Each calorimeter cell can be read out via a silicon-based photo-detector mounted directly on the scintillating element~\cite{Sefkow:2018rhp}. Passive absorbers, such as steel, can be intertwined with the sensitive layers.
The high granularity that can be achieved allows to implement compensation in hadronic showers in the reconstruction software using energy density techniques.

\subsubsection*{Micro-pattern gaseous detector-based sampling}
Calorimeters with gaseous detectors as active element can reach a higher granularity with respect to more traditional scintillator-based calorimeters. Furthermore, a 1x1~cm$^2$ pad is economically affordable with respect to 3x3~cm$^2$ scintillator tiles. The CALICE collaboration has been studying the performance of digital~\cite{Adams:2016por} and semi-digital~\cite{Baulieu:2015pfa} hadronic calorimeters. Besides having a high granularity, gaseous detectors have the advantage to be radiation hard, leading to simple calibration procedures. Resistive-Plate Chambers (RPCs) have been chosen historically because of their low cost to instrument large area, because of their intrinsic digital nature and because of the rather low rate expected at future electron-positron colliders, which were the main target of these designs. Recently, Micro-Pattern Gaseous Detectors (MPGDs) have been proposed since they will likely outperform RPCs for this task because of their higher rate capability, their operation with environmental-friendly gases, their good energy resolution (of about 20\%), high detector stability and low pad multiplicity. In the last decade resistive Micromegas were developed and tested for calorimetry at ILC~\cite{Chefdeville:2021ava} and for tracking in high-rate environments~\cite{Alviggi:2020hoy} at LHC, while new resistive detectors like the $\mu$RWELL~\cite{Bencivenni:2014exa} and RP-WELL~\cite{Rubin:2013jna} were developed, the latter being actively pursued for SDHCAL calorimeter of CepC detector~\cite{ShakedRenous:2020xeu}. 
It is expected that a good time resolution will play an important role in helping to match reconstructed charged particle trajectories and calorimeter energy deposits. Single-layer RPCs can reach time-resolutions of few hundred ps, but rely on gases with high global warming potential that will be phased out in the near future. Replacement gases are currently under investigation. Micro-Pattern Gaseous Detectors typically have a time resolution of the order of few ns, with R\&D ongoing to bring this to sub-ns levels.

\subsubsection*{Other promising calorimeter technologies}

Calorimeters are generally divided in two categories, homogeneous and sampling. The best compromise between the two technologies is sought in order to optimise the experimental requirements and minimise the drawbacks associated with the limitations of standard solutions. The most recent technological developments allow this rigid distinction to be abandoned in favour of novel architectures: the Crilin calorimeter~\cite{Ceravolo:2022rag} is a semi-homogeneous calorimeter based on Lead Fluoride (PbF$_2$) crystals read out by surface mounted UV extended Silicon Photomultipliers (SiPMs).

\subsubsection*{Crilin: a CRystal calorImeter with Longitudinal INformation}

The Crilin calorimeter can be segmented longitudinally as a function of the energy of the particles and of the background level, thanks to its modular design which enables a high degree of reconfigurability. The Crilin R\&D proposal embeds a modular architecture based on stackable submodules composed of matrices of  crystals, in which each crystal is individually read out by two series of two UV-extended surface mount SiPMs.  Crystal dimensions are $10\times10\times40$~mm$^3$ and the surface area of each SiPM is 4$\times$4~mm$^2$, so as to closely match the crystal surface.

\begin{figure}[h!]
    \centering
    \includegraphics[width=0.45\textwidth]{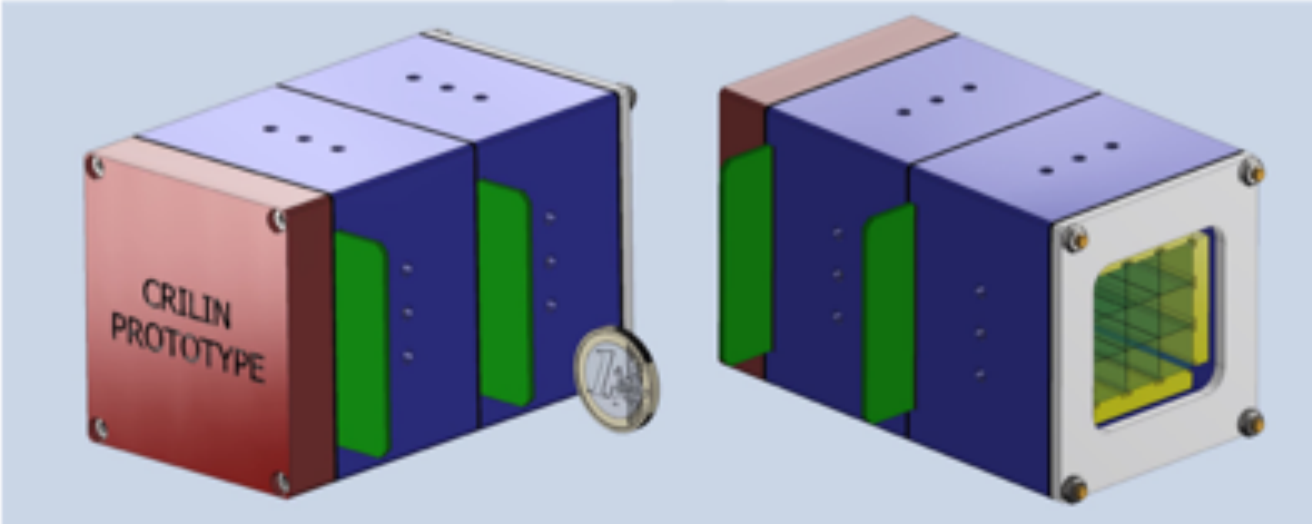}
    \caption{CAD model of  Crilin Prototype.}
    \label{fig:x1}
\end{figure}

In the current design, the prototype consists of two submodules, each composed of a 3-by-3 crystals matrix. The submodules are arranged in a series and assembled together by screws, resulting in a compact and small calorimeter, shown in Figure~\ref{fig:x1}.

Each crystal matrix is housed in a light-tight case which also embeds the front-end electronic boards and the cooling system. The on-detector electronics and the SiPMs must be cooled during operation, so as to improve and stabilise the performance of SiPMs against irradiation. The Crilin design is capable of removing the heat load due to the increased photosensor currents after exposure to the expected $2-5 \cdot 10^{13}$~1-MeV-neq~cm$^{-2}$/year fluence, along with the power dissipated by the amplification circuitry. The total heat load was estimated as 350~mW per channel. The Crilin cooling system, which is based on conduction and forced convection of nitrogen, will provide the optimum operating temperature for the electronics and SiPMs at around $0^{\circ}$ C. Gas fluxing will also prevent any condensation on SiPM or crystal surfaces.

\FloatBarrier
\subsubsection*{Muon systems}
A compact detector can be obtained using an iron yoke to concentrate the magnetic flux return from the solenoid, instrumented with several layers of muon detectors.

\begin{figure}[!t]
    \center
    \includegraphics[width=0.48\textwidth, trim = {10 6 0 32} , clip]{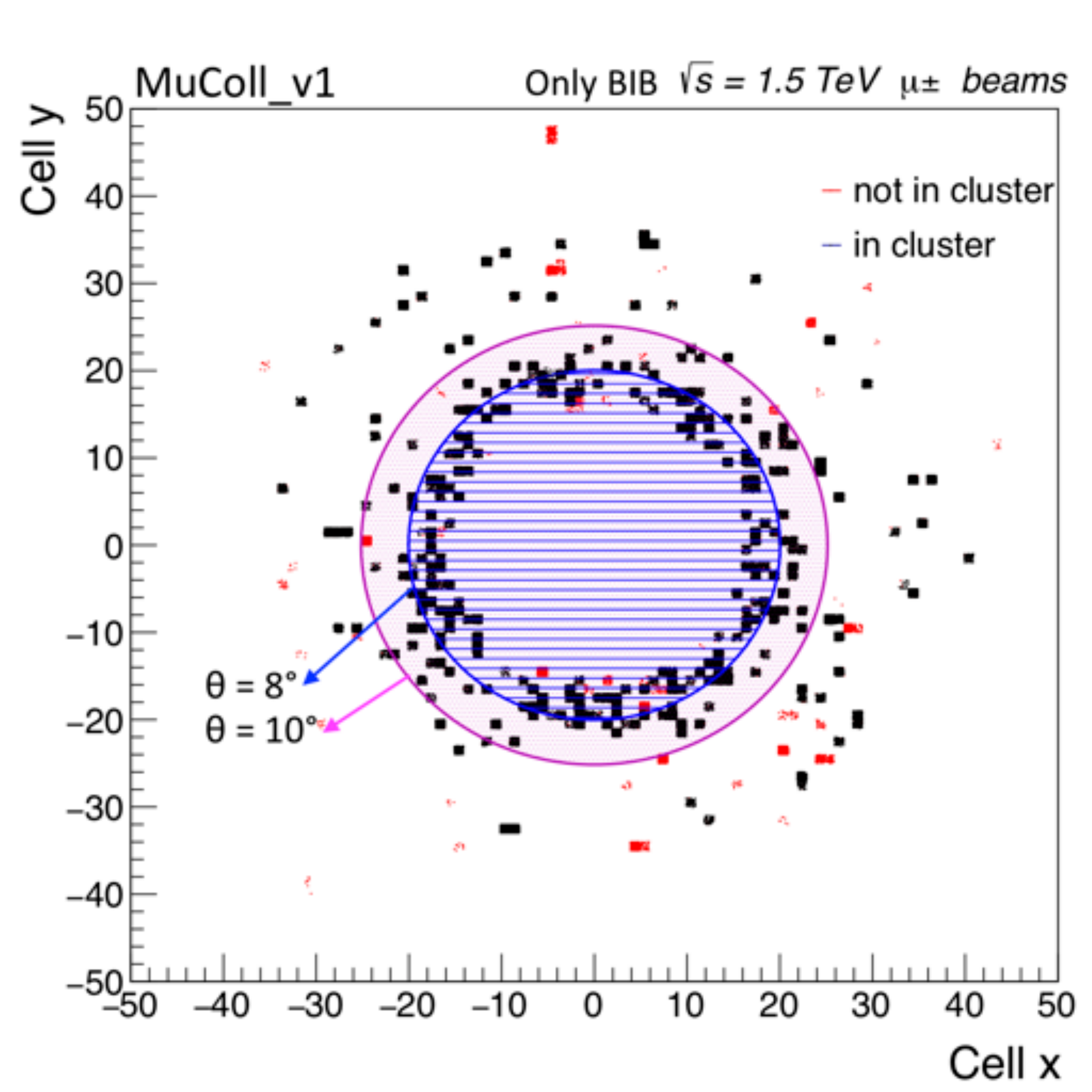}
    \caption{BIB muon hit spatial distribution in the first layer of the muon endcap. The detector hits not associated to a cluster are shown by the red markers. The blue circle corresponds to region $\theta<8^{\circ}$, while the purple to $\theta<10^{\circ}$.}
    \label{fig:BIB2fig}
\end{figure}

\begin{figure}[!t]
    \center
    \includegraphics[trim={0cm 0cm 0cm 2cm},clip,width=0.5\textwidth]{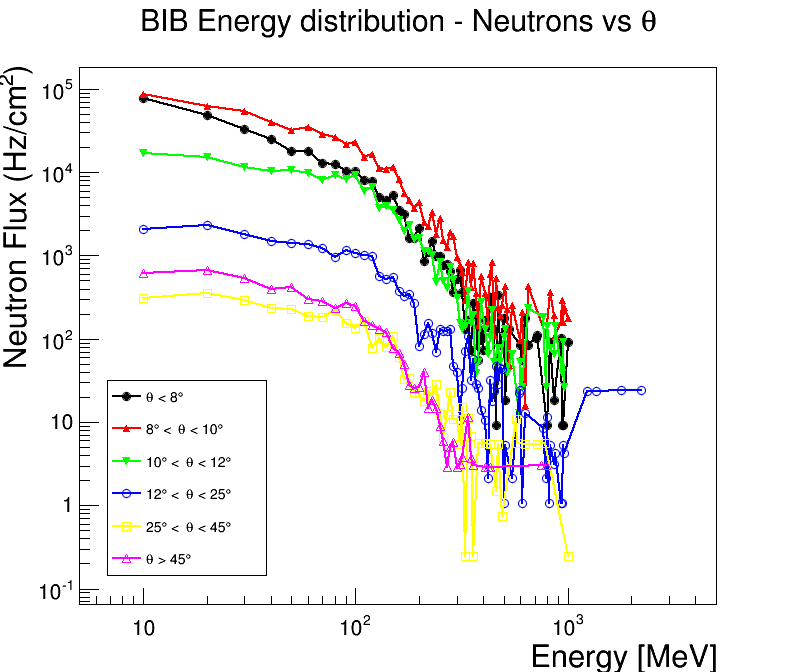}
    \caption{Energy distribution of neutrons from BIB. Colours represent different geometrical regions of the muon system.}
    \label{fig:BIB_15TeV_flux_neutron}
\end{figure}

The BIB hits in the muon system are concentrated around the beam axis in the endcaps, as shown in Figure~\ref{fig:BIB2fig}.
The BIB in the muon system is mainly composed of high energy neutrons and photons. The neutron energy, shown in Figure~\ref{fig:BIB_15TeV_flux_neutron}, ranges from \SI{10}{MeV} up to \SI{2.5}{GeV}, with the majority of the flux in the region $E < \SI{100}{MeV}$. 

\begin{figure}[!h]
    \center
    \includegraphics[trim={0cm 0cm 0cm 2cm},clip,width=0.5\textwidth]{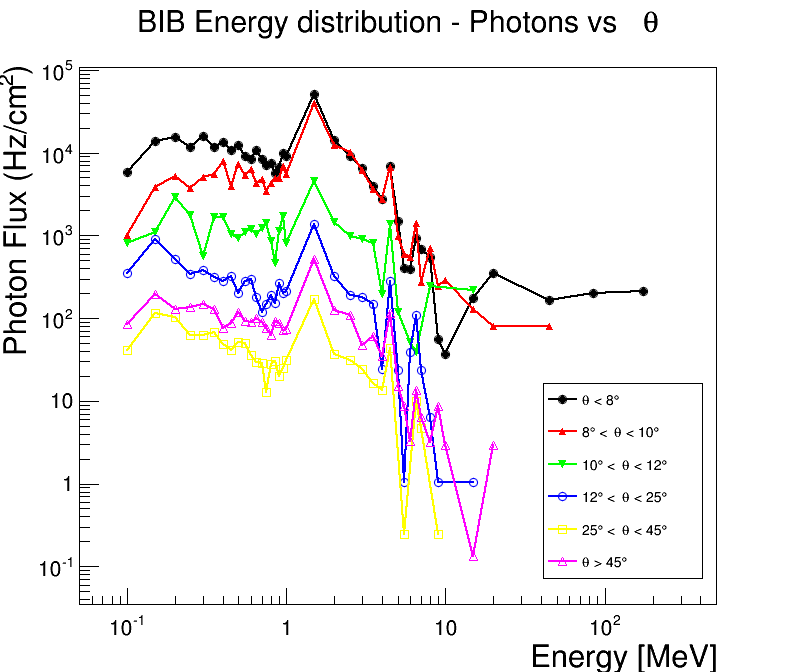}
    \caption{Energy distribution of photons from BIB. Colours represent different geometrical regions of the muon system.}
    \label{fig:BIB_15TeV_flux_photon}
\end{figure}

The photon energy, shown in Figure~\ref{fig:BIB_15TeV_flux_photon}, instead ranges between \SI{100}{keV} and \SI{200}{MeV}, with the majority of the flux in the region $E < $\SI{10}{MeV}.

The colours represent different geometrical regions of the detector, based on the polar coordinate $\theta$: the fluxes are higher in the inner part of the endcap, at lower $\theta$ and closer to the beam line, and then lower in the outer regions. 

While at high energies it is expected for the muon momentum resolution to be dominated by the measurements performed by the inner tracking detectors, with the muon detectors providing the additional muon identification, sensitivity studies pointed out that some technologies, such as RPCs, are already at the limit of their current rate capability in the most forward regions. These results, together with preliminary requirements on the spatial ($\approx$\SI{100}{\micro \meter}) and time resolution (below \SI{1}{ns}), suggest the need for gaseous detectors R\&D. Classical well-known MPGDs, such as GEMs or Micromegas, are characterised by an excellent spatial resolution, but do not match the demanding request on the timing resolution. R\&D on new generation MPGDs are still at the initial phase, but are obtaining promising results for a future implementation. A possible muon system design can be a heterogeneous detector, composed of layers of different technologies to optimise the timing and tracking performance. Moreover, an excellent spatial resolution would give the possibility to use the standalone muon objects to seed the global muon track reconstruction.

\subsubsection*{Detectors with high spatial resolution}
\label{detec_space}
Classical gaseous detectors, Multi-Wire Proportional Chambers (MWPCs), reach spatial resolutions of the order of few mm (dominated by the mechanical limitation in the wire spacing). Resolutions of \SI{50}{\micro m} to \SI{100}{\micro m} are obtained measuring precisely the drift time (Drift Tubes, DT) or by patterning the cathode combined with precise charge measurement (Cathode Strip Chambers, CSCs), however all wire-based detectors have intrinsic rate limitations due to the slow evacuation of ions. This limitation has been overcome in Micro-pattern Gaseous Detectors, where the electrodes are created using photo-lithographic techniques, which allows the reduction of the electrode spacing of at least one order of magnitude, resulting in fast ion evacuation combined with high spatial resolution.

A Gas Electron Multiplier (GEM) consists of a thin polymer foil (often \SI{50}{\micro m} polyimide), cladded on both sides with a thin layer of copper (\SI{5}{\micro m}), chemically perforated with a high density of holes, typically of 100/mm$^2$~\cite{Sauli:1997qp}. Applying a potential difference between the top and bottom electrodes, a high electric field is formed in the holes where electron multiplication can take place. Several GEM-foils can be stacked on top of each other leading to detectors with high gain ($>10^5$) and low discharge probability ($<10^{-10}$). GEM detectors are used in different collider experiments~\cite{Ketzer:2001dt}, mainly for tracking and triggering purposes. Spatial resolution down to \SI{50}{\micro \meter} is possible, with a time resolution that depends on the gas mixture used: 7--10~ns in Ar:CO2~\cite{CMSMuon:2019pzw} down to 3.5~ns when CF$_4$ is used in the gas mixture~\cite{Alfonsi:2004gh}. Rate capabilities up to 100 kHz/cm$^2$ have been assessed.

Micromegas are parallel-plate chambers where the amplification takes  place in a thin gap, separated from the conversion region by a fine metallic mesh~\cite{Giomataris:1995fq}. Micromegas are used in collider experiments~\cite{Thers:2001qs} and a spark-protected evolution with resistive strips will be used mainly for tracking in the upgraded forward muon system of the ATLAS experiment~\cite{Kawamoto:2013udg}. A spatial resolution of \SI{80}{\micro \meter}, a time resolution of 7-\SI{10}{ns}, and a rate capability up to \SI{100}{kHz/cm^2} have been achieved.

The Micro-resistive Well ($\mu$-RWELL) is a single amplification stage resistive MPGD~\cite{Bencivenni:2014exa}. Such technology is reliable, since the presence of the resistive layer assures a very low discharge rate quenching the spark amplitude. It can achieve a rate capability up to 10~MHz/cm$^2$ with a detection efficiency of the order of 97-98\%~\cite{Bencivenni:2019wxr}. Typical spatial resolution is $<$\SI{60}{\micro \meter}, time resolution measured with CF$_4$ gas mixture was measured to be below \SI{6}{ns}, time resolution in Ar:CO$_2$ mixtures is expected to be similar to the triple-GEM detectors (7-10\,ns).

\subsubsection*{Detectors with sub-ns timing resolution}
\label{detec_timing}

MRPC~\cite{CerronZeballos:1995iy} have acquired solidity and importance in the High Energy Physics domain where both high efficiency and  good time resolution are demanding. A time resolution of about \SI{60}{ps} and $95\%$ efficiency has been obtained~\cite{ALICE:2008ngc,Llope:2005yw} for a detector composed of 10 gas gaps $250~\mu$m size arranged in a double stack design using floating soda-lime glass (bulk resistivity $ 5 \times 10^{12}\, \Omega\, {\rm{cm}}$).  A recirculating gas mixture (C$_2$H$_2$F$_4$:SF$_6$ 97:3) allows operation in avalanche mode with electric field of approximately $100$~kV/cm. The technology was operated~\cite{Friese:2006dj} in high particle fluxes (though for small-size detector units) in a series of tests for the endcap upgrade of the STAR ToF and for the mini-CBM  experiment at the GSI/SIS8 Synchrotron.  Data with thinner standard float glass show rate capability of some~kHz/cm$^2$ at time resolutions of 70--80~ps.

The rate capability is limited by the current flowing through the resistive plates and hence a step forward to increase this value, at a constant front end electronics threshold, would be to use low resistivity glass or to decrease their thickness. Lower resistivity glass ($ 10^{10}\,\Omega\,$cm) have been used in tests beam~\cite{Wang:2019jjz} and showed rate capability of 35~kHz/cm$^2$ with efficiency above $90\%$ and time resolutions below \SI{80}{ps}. A still better behaviour can be seen by lowering the glass resistivity by one order of magnitude, currently R\&D is concentrated on establishing \SI{20}{ps} time resolution at rates of 100~kHz/cm$^2$.

Although the MRPC have shown excellent timing resolution over large area and R\&D for larger rate capability and even better timing is underway, this detector technology can not be considered for future collider experiments with the current gas mixture which has a high Global Warming Potential (GWP). The use of freons will be gradually phased out by 2030. While for standard (High Pressure Laminate, HPL) RPCs encouraging results have been obtained to replace the freons with alternative gases as Hydrofluoroolefine (HFO-1234ze), MRPC performances have yet to be proven with those new gas mixtures. Moreover MRPC performance relies on a non-negligible fraction of SF$_6$ which has even higher GWP with respect to the freons. R\&D should start urgently in order to propose these detectors for future use.

The time resolution of a classical MPGD is dominated by the fluctuations on the position on the first ionisation cluster in the drift gap.
The contribution to the time resolution of the drift velocity ($v_d$) is then given by $\sigma_t = (\lambda v_d)^{-1}$ where $\lambda$ is the average number of primary clusters generated by an ionising particle inside the gas per unit length~\cite{DeOliveira:2015bda}. Therefore a better time resolution is expected with a faster mixture. However, even with a fast gas mixture, as for example Ar:CO$_2$:CF$_4$, classical MPGDs usually cannot reach time resolution better than few ns.

A possible approach aimed at improving the time resolution is the one followed by the PicoSec Collaboration~\cite{Bortfeldt:2019ecw}: the proposed detector is a standard Micromegas (MM) with a drift gap reduced to \SI{200}{\micro \meter} in order to minimise the possibility of primary ionisation. Particles pass instead through a Cherenkov radiator placed on top of the MM, where they produce Cherenkov photons, which are then converted by a photocathode and enter in the drift region, removing the fluctuations on the position of the first ionisation cluster.
Preliminary results of the first prototypes of few~cm$^2$ proved the reliability of the principle with a time resolution of \SI{25}{ps} measured with a Ne/C$_2$H$_6$/CF$_4$ gas mixture. Further studies of this new technology will be focused on: proper stability of the detectors, choice of the materials and geometry, radiation hardness and gas mixture (avoid use of CF$_4$ and flammable gases).

An alternative approach is represented by the Fast Timing Micropattern (FTM) gas detector~\cite{DeOliveira:2015bda}. Here the drift gap is segmented in $N$ thinner fully resistive drift and amplification stages, which are in competition between each other when the detector is fully efficient (sum of drift gaps $\geq$ \SI{1.5}{mm}). The fastest stage is the one that determines the timing of the signal, thus reducing the time resolution of the detector of a factor $N$ with respect to a standard MPGD. The first prototype, made of just two stages, obtained a time resolution of \SI{2}{ns} with pion beam~\cite{Abbaneo:2017tat}. The current R\&D is focused on improving the quality of the resistive layers used for the amplification stage and establishing the detection principle with multiple layers on a small scale prototype~\cite{Colaleo:2019tmy}, and assessing the technology on larger scale prototypes. The main limitations come from the quality of the detector elements with resistive layers, which is a technology currently developed, and the single-stage reachable gain; interesting results have been recently obtained with a Ne/iC$_4$H$_{10}$\,95/5 gas mixture~\cite{Pellecchia:2021sip}.

\subsubsection*{Technology comparison}
In order to understand the response of the muon detectors to the BIB particles, the detector sensitivities have been studied with a standalone \textsc{Geant4} simulation.

The sensitivity is defined as the probability for a BIB particle to generate a visible signal in the detector. It is computed as the ratio $s = N/M$, where $N$ is the number of events in which at least one charged particle reaches a sensitive gas gap, while $M$ is the number of incident particles. 

\begin{figure}[!ht]
    \center
    \includegraphics[trim={1cm 1cm 0cm 2cm},clip,width=0.5\textwidth]{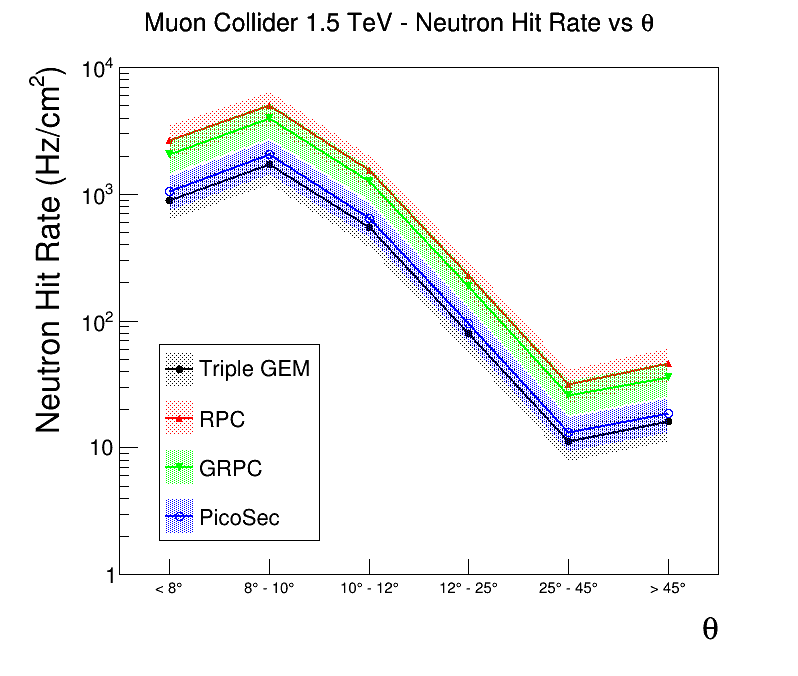}
    \caption{Estimated hit rate from neutrons at a \SI{3}{TeV} muon collider. Different colours represent different gaseous detector technologies considered: triple GEM, standard HPL RPC, glass RPCs (GRPC) and PicoSec. The shaded bands represent the statistical uncertainty from the simulated events.}
    \label{fig:BIB_15TeV_hitrates_neutron}
\end{figure}

\begin{figure}[!ht]
    \center
    \includegraphics[trim={1cm 1cm 0cm 2cm},clip,width=0.5\textwidth]{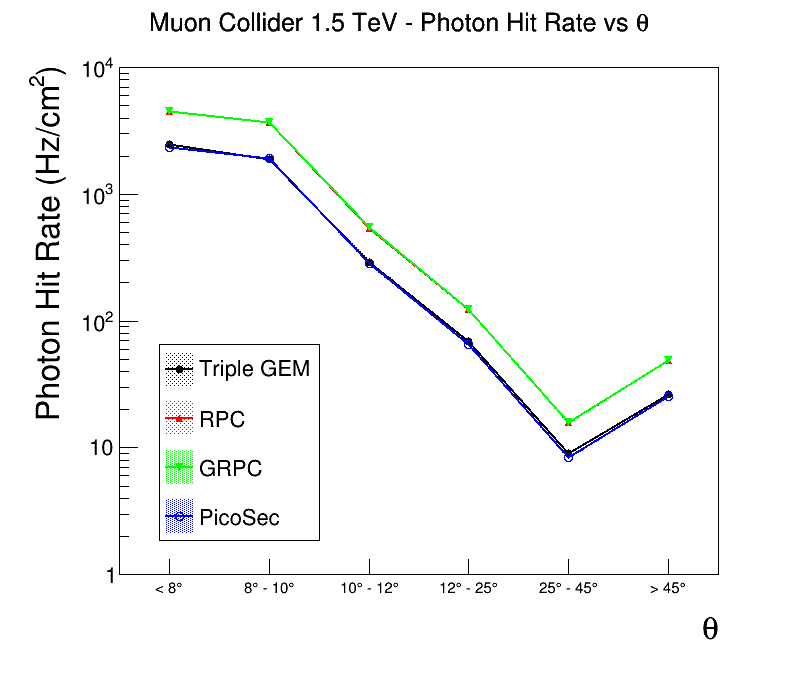}
    \caption{Estimated hit rate from photons at a \SI{3}{TeV} muon collider. Different colours represent different gaseous detector technologies considered: triple GEM, standard HPL RPC, glass RPCs (GRPC) and PicoSec. The shaded bands represent the statistical uncertainty from the simulated events.}
    \label{fig:BIB_15TeV_hitrates_photon}
\end{figure}

The hit rate (R) is then obtained from the flux ($\Phi$) as $R = s \times \Phi$ for each energy value and particle type. The estimated rate is shown in Figure~\ref{fig:BIB_15TeV_hitrates_neutron} for neutron and in Figure~\ref{fig:BIB_15TeV_hitrates_photon} photons as a function of the angular coordinate $\theta$ for the different detector technologies considered.

The difference between the technologies is mainly due to a different material composition of the detectors: in general Micro-Pattern Gaseous Detectors, i.e. Triple-GEM and PicoSec, result in having a lower hit rates when compared to RPC.

\subsubsection*{Trigger and data acquisition}

Experiments at a high energy muon collider are expected to operate at instantaneous luminosity levels of $10^{34}-10^{35}$ cm$^{-2}$ s$^{-1}$.

The Trigger and Data Acquisition (TDAQ) systems of the future muon collider experiments will be required to perform partial or full reconstruction of every collision event in order to identify and store events of interest to the physics programme. Given that realising the muon collider will take time, it is way too early to define the TDAQ strategy. Future advancements in the data transmission and processing technologies can substantially alter the vision of what a TDAQ system at the muon collider would look like. However, an initial estimate of the data rates and processing needs helps to outline possible options and strategies, in particular when put in the context of today’s technologies. 

Trigger and DAQ strategy taken by different collider experiments varies a lot and depends on the luminosity and complexity of their collision events. Experiments such as ATLAS and CMS, utilise hardware triggers~\cite{ATLAS:2016wtr,CMS:2016ngn} that rely on a subset of the detector subsystems for initial filtering of the events. This is followed by a High Level Trigger (HLT) farm where further processing and filtering takes place using more complete event information. The LHCb experiment, operating at lower luminosity and with smaller event size, recently opted for a so-called ``triggerless'' or ``streaming'' approach~\cite{LHCbCollaboration:2014vzo,Colombo:2018upq}, which eliminates need for a hardware trigger and where all collision data is streamed at 40~MHz directly to a HLT farm for event reconstruction. Similarly, electron-positron collider experiments~\cite{Behnke:2013lya,CLICdp:2018vnx} typically adopt a triggerless readout scheme due to the relative cleanliness of events when compared to hadron colliders. A streaming approach offers a number of advantages: the availability of the full event data typically translates into a better trigger decision, it is easier to support and upgrade software triggers, simplified design of the detector front-end, etc. However, the presence of large BIB at the muon collider may be prohibitive for a full triggerless TDAQ scheme. In the following, an initial estimate of the data rates is provided to show that from the data rates/volumes consideration a streaming DAQ implementation is feasible. Early estimates of the event processing time are also provided and compared to those anticipated at the HL-LHC experiments.

The amount of data acquired by the muon collider experiments is expected to be dominated by the tracker and calorimeter systems. For the silicon tracker, the event size and data rates are estimated by acquiring an average number of hits per event from simulation and multiplying it by the 100 kHz event rate. The average hit multiplicity as a function of the tracker layer can be found in Figure~\ref{fig:HitMultTracker}. In this estimation, it is assumed that each hit consists of 32 bits to encode charge, position, and time information and that zero-suppression is applied in the detector front-end. The hits are integrated in the time period of 1~ns following the bunch crossing, which allows to preserve good efficiency for hits from particles originating in the hard scattering but rejects a significant fraction of the BIB. No filtering based on the hit direction information is applied to avoid possible biases in the online selection. An additional safety factor of 2 is embedded in the calculation. With these assumptions, the tracker event size is estimated to be 40 Megabytes (MB) and the data rate from the tracker to be 30 Terabits per second (Tb/s). It should be noted that the numbers are dominated by the BIB hits. A similar approach is applied for calculating event data rates originating in the calorimeter. Here, the ECAL dominates with approximately 90 million channels and average occupancy of about 10$^{-3}$ hits per mm$^{2}$. A minimum energy threshold of 0.2~MeV is required in order for the hit to be read out and hits are assumed to be 20 bits wide. The HCAL contribution is small, less than 10\% of the ECAL.  After applying a safety factor of 2, calorimeter event size is estimated to be 40 MB and similar to that for the tracker. The full data rate corresponding to the sum of the tracker and calorimeter rates is therefore about 60 Tb/s, which is a factor of few larger than HLT input of LHCb experiment in Run-3~\cite{LHCbCollaboration:2014vzo} and comparable to the HLT input of CMS experiment in HL-LHC~\cite{Collaboration:2759072}. Therefore, from the data volumes point of view, a streaming operation at 100~kHz appears to be feasible. It should be emphasised that the rates are directly proportional to the bunch crossing frequency and can be much larger with a smaller collider ring or a multi-bunch operation scheme, in which case the strategy may have to be re-evaluated.

Another important parameter to consider is the HLT output rate to offline storage. Here again different approaches can be taken. One approach is to eliminate most of the events using filtering done in the HLT, but preserve full raw event information for the ones that pass the filter.  HL-LHC experiments assume HLT output bandwidth of approximately 60 GB/s. This would translate into 750 Hz of full event content sent to the storage. For comparison, single Higgs production at $\sqrt{s}=10$~TeV is expected to have a much lower rate of less than 0.1~Hz, and WW production via vector boson fusion will be at 1~Hz level. Storing full event content allows for later reprocessing of the data in order to improve the performance or reconstruct novel signatures. However, this approach would require about 4~PB of storage per day of running. It will produce a total dataset similar in size to that of ATLAS and CMS in HL-LHC, but dominated by BIB hits that are not interesting from the physics point of view. An alternative approach based on reducing event content by filtering out hits and clusters clearly unassociated with the hard scattering may be more prudent. Here one may choose to preserve the output rate of 750~kHz and reduce the total dataset size. Alternatively, one can aim for the same total dataset size but increase the rate of events, for example if on average 99\% of BIB hits in each event can be filtered out, every produced event at 100~kHz can be sent to the storage. 

In addition to data rates, it is also important to take into account time needed to reconstruct each event at the HLT.  Long processing times lead to unmanageable large farm requirements, which in turn is difficult to procure, maintain, and operate. HL-LHC experiments project processing times of about one second per event with tails extending as far as a minute. At the muon collider, the offline event reconstruction is currently dominated by charged particle tracking and takes up to few minutes per event, which is significantly slower than a good HLT target of few seconds. To estimate the amount of time needed at the HLT, a preliminary reconstruction of charged tracks was attempted in the outer layers where the BIB density is less severe. An initial estimation was performed by taking hits from the outer six barrel layers and applying an algorithm based on a three-dimensional Hough Transform in the ($\phi$, $\theta$, $p_{\rm T}$) parameter space. This approach assumed that all tracks would come from the origin. The preliminary analytical calculations suggest that, even with a three-dimensional array finely subdivided into 240 million cells, with the currently estimated BIB rates, tens of purely combinatorial track candidates with $p_{\rm T} > 2.5$~GeV would be found, precluding any possibility to use that information as an effective event filter. However, if an independent filter based on timing (or pointing, etc ...) is applied to lower the BIB hit rate by a factor four, requiring a single candidate track with $p_{\rm T} > 2.5$~GeV would give us the needed event rejection factor of about 50. In this case, 44 million cells in the Hough transform array would have to be filled for each beam crossing. Doing this in few seconds requires a powerful CPU, but is not out of reach. Future CPUs and accelerators (GPU, FPGA, etc), as well as algorithm improvements can further improve timing. The bottom line is that this is a crucial area of development and careful attention needs to be paid to this area in order to keep the processing time under control.

\begin{figure}[!t]
\center
\includegraphics[width=0.42\textwidth]{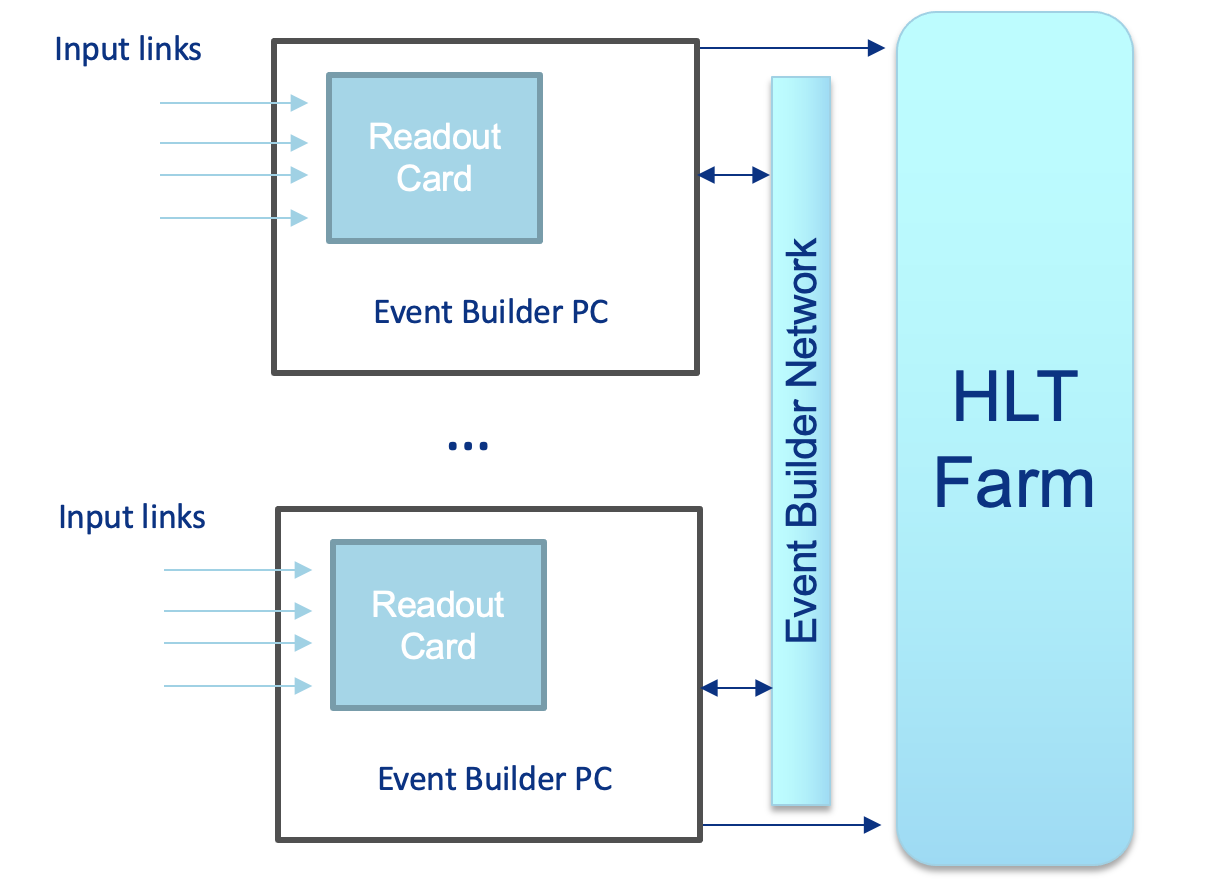}
\includegraphics[width=0.42\textwidth]{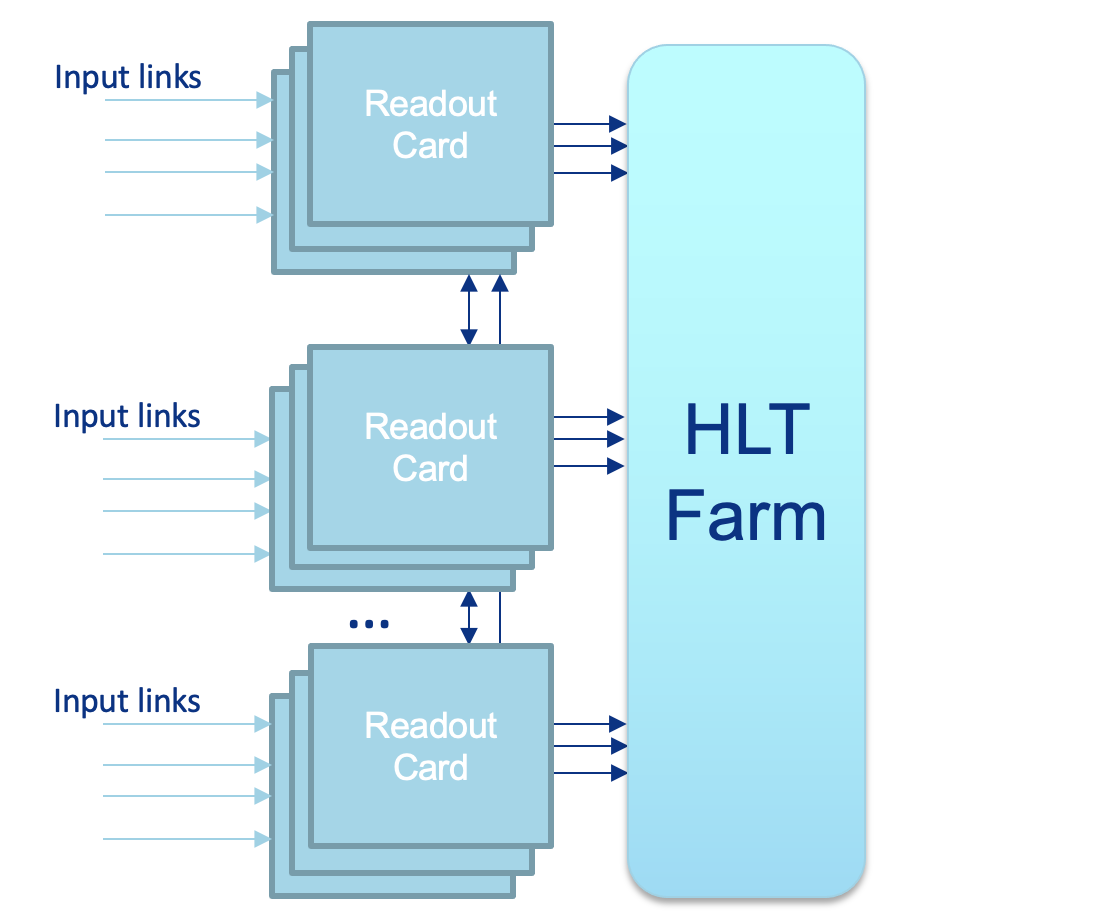}
\caption{Possible TDAQ system architectures using (top) an LHCb-like approach with a software Event Builder or (bottom) hardware boards to structure event data and pass them to a high-level trigger farm.}
\label{fig:TDAQArch_LHCb_HW}
\end{figure}

Finally, it may be useful to sketch how a potential TDAQ system could look and estimate its size. The number of input links is an important parameter for this purpose. The lpGBT developed for HL-LHC are expected to provide bandwidth of up to 10~Gb/s per link. It is reasonable to assume that links with at least twice higher speeds (20~Gb/s) will be available on the timescale of 20 years. Under the assumption that data from multiple detector modules can be combined into a single 20~Gb/s link, close to 10,000 of such links will be needed to bring the data from the detector to the electronics area. The back-end system will consist of few hundred readout boards used to receive and format the data. A full-mesh hardware or a software Event Builder network with aggregate bandwidth of about 60~Tb/s will be required, for which both custom and industry provided solutions can be considered. Note that this number can be much smaller if a more aggressive filtering scheme than the one described above is utilised in the front-end. For example, preliminary studies indicate that filtering tracker hits based on their pointing information can yield up to a factor of eight reduction in the data rates. The required output bandwidth to storage will depend on the chosen filtering strategy at the HLT, but should not exceed 100~Gb/s even in the most aggressive scenarios. A schematic illustration of two alternative architectures is in Figure~\ref{fig:TDAQArch_LHCb_HW}. 

To summarise, despite the large presence of BIB, preliminary estimates based on simulation indicate that a streaming DAQ architecture can provide an attractive solution for the future muon collider experiments. With improvements to the tracking speed, such a solution can likely be realised with technologies available today. Future advancements (e.g. higher speed optical links, fast processors, etc) are likely to result in a smaller and/or more performing DAQ system. Work should be invested in improving  HLT reconstruction algorithms and exploiting GPU/FPGA/ASIC acceleration schemes with the aim to bring per-event processing time down to a few second level. 

\FloatBarrier
\subsection{Reconstruction performance studies}
\label{sec:performance}

This section describes the current status of the event reconstruction performance of a multi-purpose MCD with the layout described in Section~\ref{sec:environment}. The study focuses on the reconstruction algorithms and performance of high-p$_{\rm T}$ objects that will be needed for a successful physics programme. The results are built upon and significantly extend the ones reported in Ref.~\cite{Bartosik:2020xwr}. The current work aims at showing that satisfactory performance can be achieved in a muon collider environment. Further optimisation of the detector design and reconstruction algorithms are expected to significantly improve the presented performance and will be subject of future work.

The results are obtained using detailed simulations of the MCD, using the software framework described in Section~\ref{sec:software}. All simulations, unless otherwise stated, include overlaying on top of the desired $\mu^+\mu^-$ collision the model of BIB described in Section~\ref{sec:environment}.
The BIB sample used for detector-performance studies is the one of the $\sqrt{s} =$~\qty{1.5}{\TeV} muon collider produced with the MARS15 software, because the FLUKA-based workflow was not fully validated when most of the results were obtained.

\subsubsection*{Charged particle reconstruction}

\begin{figure}[h]
    \centering
    \includegraphics[width=0.5\textwidth]{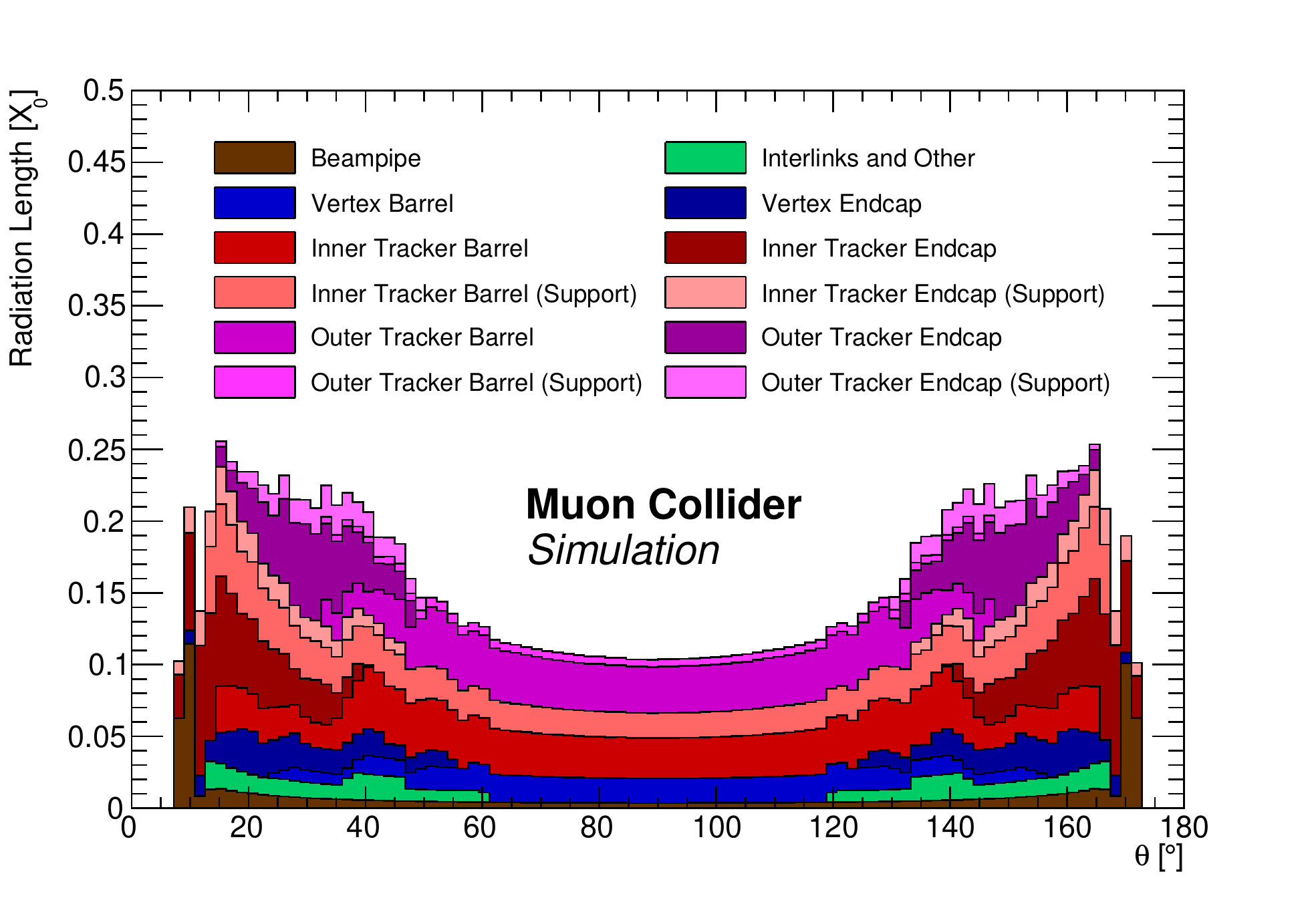}
    \caption{Radiation length of the tracking detectors, as seen along a line defined by the nominal interaction point and the polar angle $\theta$.}
    \label{fig:tracking:x0-tracker}
\end{figure}

\begin{figure}[h]
    \centering
    \includegraphics[width=0.5\textwidth]{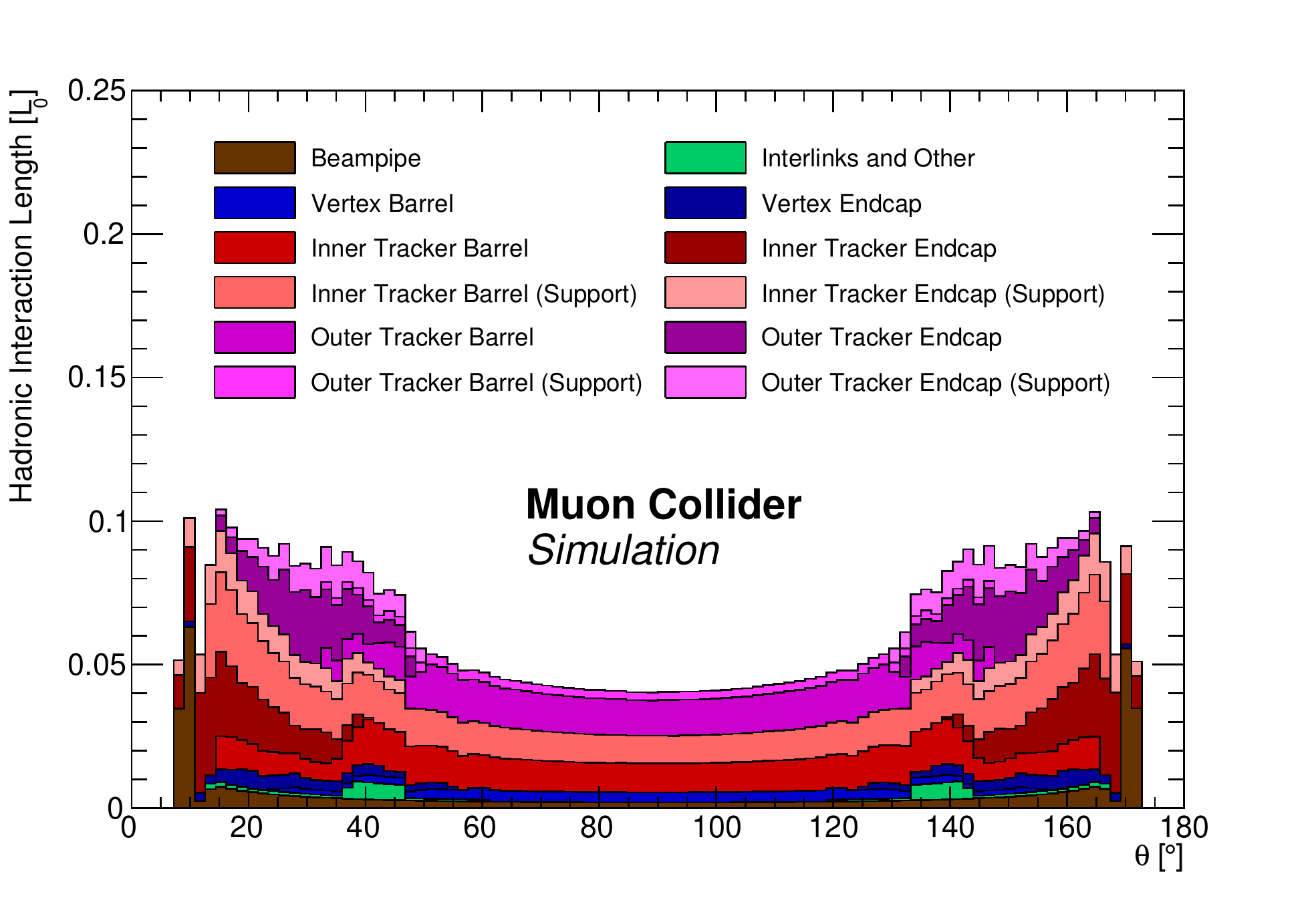}
    \caption{Hadronic interaction length of the tracking detectors, as seen along a line defined by the nominal interaction point and the polar angle $\theta$.}
    \label{fig:tracking:l0-tracker}
\end{figure}

In the magnetic field of the MCD, a charged particle will follow a helix trajectory aligned with the $z$ axis. The radius of curvature in the transverse plane is proportional to the transverse momentum of the particle and inversely proportional to the magnetic field.  Deviations from a perfect helix can occur due to multiple-scattering, ionising  energy losses and bremsstrahlung. The first two are a direct function of the detector material. The amount of radiation lengths and hadronic interaction lengths that a particle traverses through the tracking detector when starting from the nominal collision point are shown respectively in Figures~\ref{fig:tracking:x0-tracker} and~\ref{fig:tracking:l0-tracker}.

The charged particle trajectory is reconstructed from the spacepoints corresponding to the hit positions in the silicon tracking detectors. The reconstructed object is called a track. A track consists of a set of hits (one per layer) and five fitted parameters describing the helix. A track reconstruction algorithm can roughly be broken up into two steps: pattern recognition to identify the hits belonging to a single track and fitting the hit coordinates by a track model to deduce the relevant track parameters.

Track reconstruction in the MCD is complicated by the presence of a huge number of hits in the silicon sensor originating from the beam-induced background (BIB hits). The density of BIB hits is ten times larger than the expected contribution from pile-up events at a High Luminosity LHC detector. Table~\ref{tab:tracking:density} compares the hit density between the MCD, the ATLAS Inner TracKer (ITk)~\cite{ATLAS:2017svb, ATLAS:2017azf} and the ALICE ITS3~\cite{ALICE-PUBLIC-2018-013} upgrades for HL-LHC operation. The increase in possible hit combinations creates a challenge for the hit pattern recognition step. It is crucial to reduce the amount of hits given as input to a track reconstruction algorithm through alternate means, such as precision timing information. The BIB hits are out-of-time with hard collision hits after the time-of-flight correction has been applied. By applying a $-3\sigma/+5\sigma$ time window, the hit density can be reduced by a factor of two as seen in Figure~\ref{fig:HitMultTracker}.

\begin{table}[htb]
    \centering
    \caption{Comparison of the hit density in the tracking detector between a MuC with full BIB overlay, the ATLAS ITk and ALICE ITS3 upgrades for HL-LHC. The hit densities for the first and second layers of the vertex detectors are shown. The MCD hit densities are reported after timing cuts.}
    \label{tab:tracking:density}
    \vspace{0.3cm}
    \begin{tabular}{l|c|c|c}
        \textbf{Detector} & \multicolumn{3}{c}{\textbf{Hit Density [\unit{\mm^{-2}}]}} \\
        \textbf{Reference} & \textbf{MCD} & \textbf{ATLAS ITk}  & \textbf{ALICE ITS3} \\
        \hline
        \hline
         Pixel Layer 0 & 3.68 & 0.643  & 0.85 \\
         Pixel Layer 1 & 0.51 & 0.022  & 0.51 \\
    \end{tabular}
\end{table}

The spatial distribution of BIB-hits is also unique. They are different from hits created by pile-up collisions. Pile-up hits come from real charged particle tracks originating from multiple vertices in the collision region. On the other hand, BIB-hits come from a diffuse shower of soft particles originating from the nozzles. The compatibility of a track with a trajectory originating from the luminous region provides an important handle for differentiating ``real'' tracks of charged particles produced in the primary collision and ``fake'' tracks generated from random combinations of BIB-hits.

The remainder of this section describes three approaches that were studied for track reconstruction at the MCD. The first two use the Conformal Tracking (CT) algorithm developed for the clean environment of the electron-positron colliders~\cite{Brondolin:2019awm}. However in the presence of BIB, the CT algorithm takes weeks to reconstruct a single event and is impractical for large-scale production of simulated data. To ease the computational effort, the input hits are first reduced by either defining a Region of Interest or by exploiting the double-layered Vertex Detector to select only hit pairs pointing to the collision region. A third approach uses the Combinatorial Kalman Filter (CKF)~\cite{Billoir:1989mh, Billoir:1990we, Mankel:1997dy} algorithm developed for the busy environment of hadron colliders. It can perform track reconstruction in a reasonable time without requiring any additional filtering of input hits.

It should be noted that the CT and CKF algorithms have very different software implementations that are responsible for much of the difference in their performance. The CKF algorithm is implemented using the A Common Tracking Software (ACTS)~\cite{Ai:2021ghi} library that is heavily optimised for efficient computing. The same is not true for the CT algorithm implemented directly in iLCSoft with less emphasis on computational efficiency. It is possible that part of the computational improvements come from the code optimisation alone. For example, the ACTS Kalman Filter implementation is a factor 200 faster than the default iLCSoft implementation given the same inputs.  This demonstrates the advantage of an experiment-independent track reconstruction library developed by a dedicated team with strong computing expertise.

The expected performance is assessed using a sample of single muons originating from the interaction point. Two set of samples are used; one set is generated with the muon having a fixed momentum $(p)$ of $1$, $10$ and $100$~GeV and a uniform distribution in $\theta$. The second set is generated at discrete values of $\theta=13^\circ,30^\circ,89^\circ$ and uniform transverse momentum distribution in the $1-100$~GeV range. The chosen $\theta$ values correspond to particles expected to leave hits entirely in the endcap system, some in the barrel and some in the endcap, and entirely in the barrel region, respectively.

\subsubsection*{Conformal tracking in regions of interest}
\label{sec:trk-roi}
The Conformal Tracking algorithm is based on Ref.~\cite{Brondolin:2019awm} and its implementation in iLCSoft. The algorithm is designed and optimised to find charged particles in very clean environments, as the ones in $e^+e^-$ colliders. The conformal mapping technique~\cite{Hansroul:1988wa} is combined with cellular automaton approach~\cite{Glazov:1993ur} to increase the acceptance of non-prompt particles.  

Due to the very large hit multiplicity from BIB-hits, running such an algorithm for the full event is prohibitive in terms of CPU and memory resources. Instead, a region-of-interest approach is used, where hits to be considered are pre-selected based on existing objects reconstructed in either the calorimeters or the muon system. To assess tracking performance only hits within a cone of $\Delta R < 0.5$ around the signal muon were selected as input to the CT algorithm, where $\Delta R = \sqrt{\Delta \phi^2 + \Delta \eta^2}$.

\begin{figure}[t]
\centering
\includegraphics[width=0.49\textwidth, trim = {0 4 0 18 } , clip]{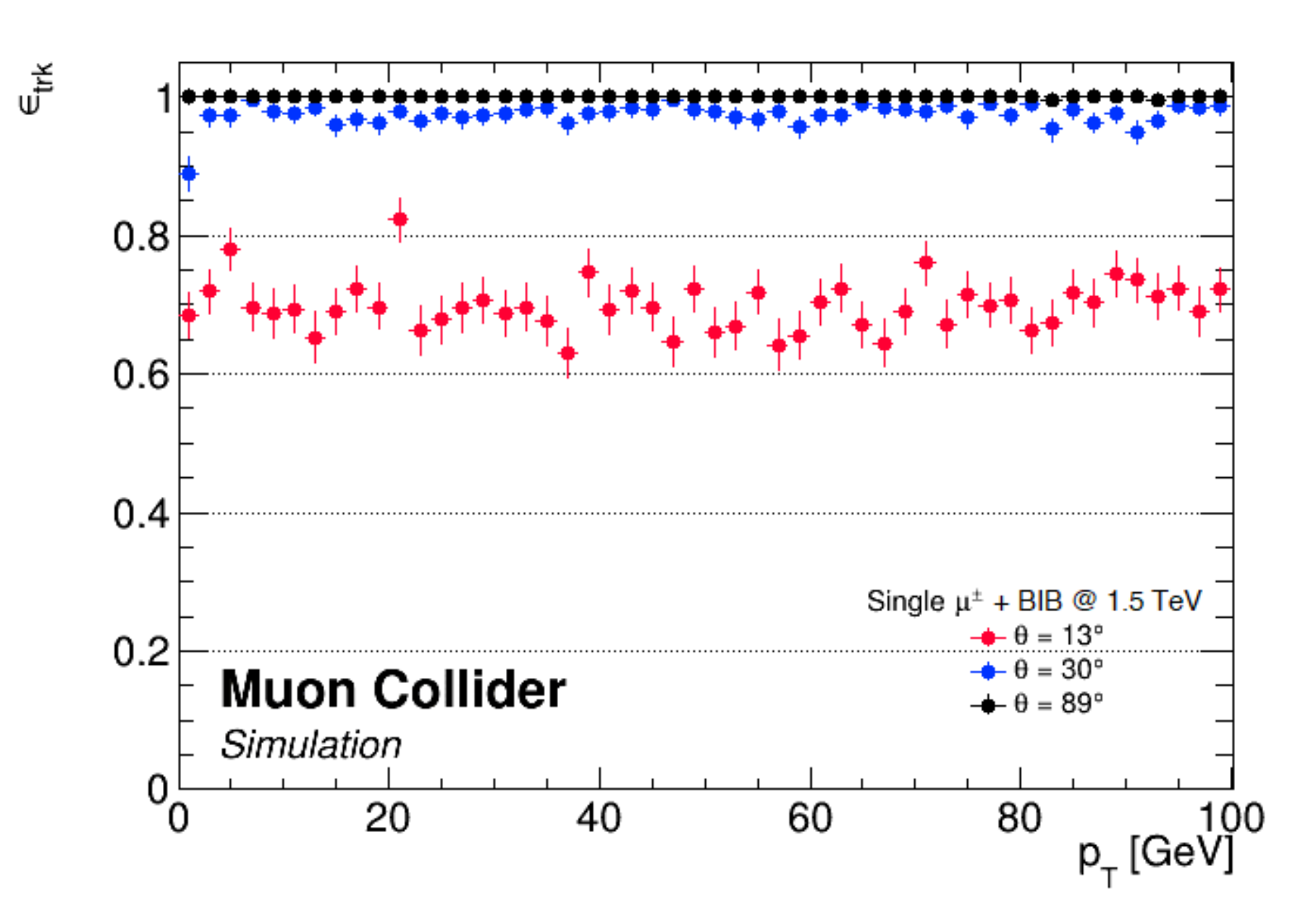}
\caption{Track reconstruction efficiency as a function of $p_{\rm T}$ for single-muon events overlaid with BIB.}
\label{fig:tracking:perf-roi_pteff}
\end{figure}

Figure~\ref{fig:tracking:perf-roi_pteff} shows the reconstruction efficiency as a function of $p_{\rm T}$; a muon is considered reconstructed if at least half of the hits associated to the track have been originated by the muon. Optimal reconstruction efficiency is achieved throughout the $p_{\rm T}$ spectrum, with a somewhat smaller efficiency for very forward particles due to their proximity to the nozzle and much larger expected occupancy in the region. The latter is expected to be recovered by a more dedicated tuning of the algorithm or by using one of the algorithms described in the next sections. 

\begin{figure}[t]
\centering
\includegraphics[width=0.49\textwidth, trim = {0 8 0 18 } , clip]{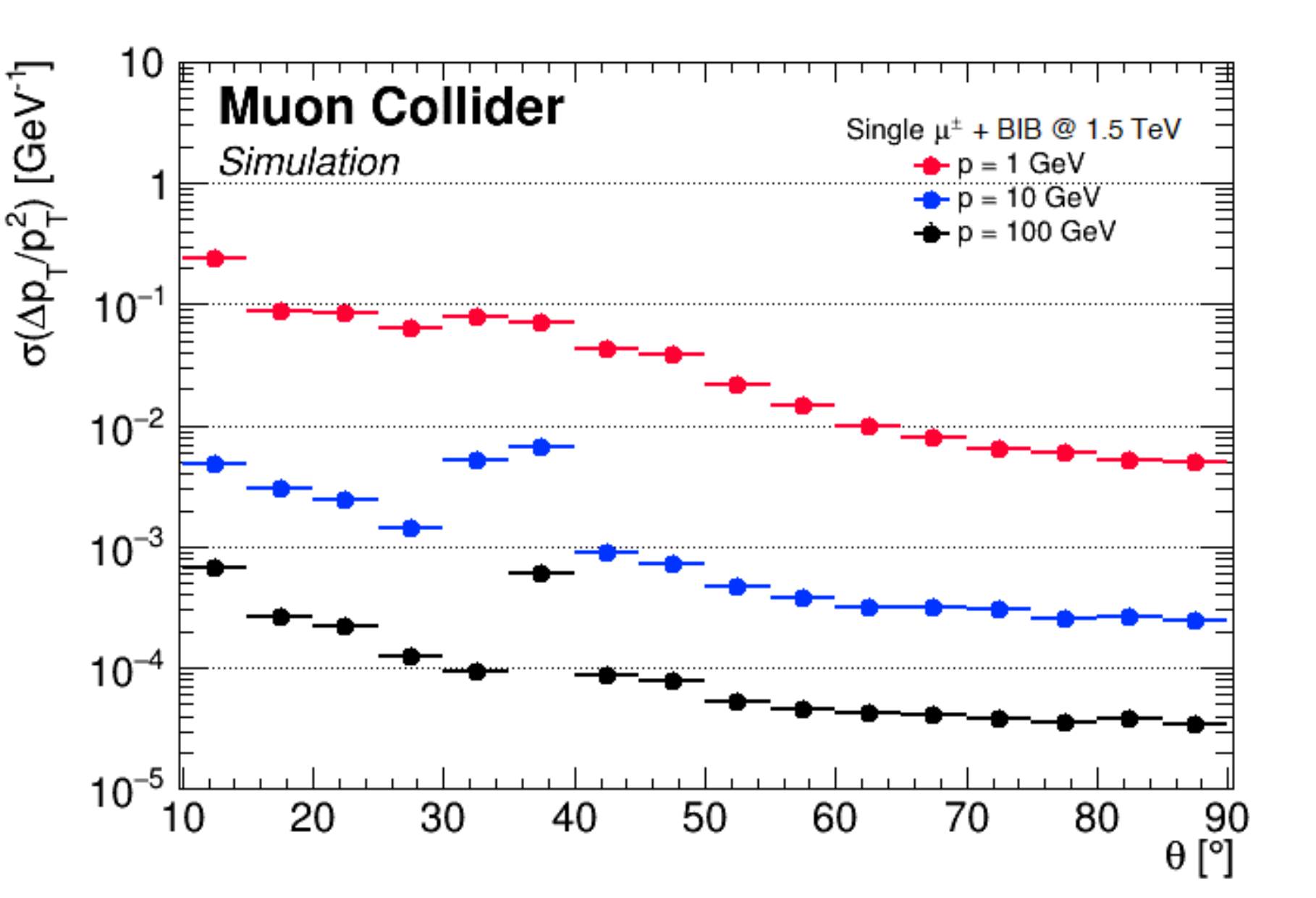}
\caption{Momentum resolution $\Delta p_{\rm T}/p^2_{\rm T}$ as a function of $\theta$ for single-muon events overlaid with BIB.}
\label{fig:tracking:perf-roi_resopt}
\end{figure}

Figure~\ref{fig:tracking:perf-roi_resopt} shows transverse momentum resolution as a function of polar angle $\theta$. The resolution is computed by comparing reconstructed and generated $p_{\rm T}$ and shown divided by $p^2_{\rm T}$. A localised degradation of the resolution can be seen around $\theta\approx 35^\circ$, corresponding to the barrel-endcap transition; in addition, the feature is enhanced by the non-physical lack of spread of the muon originating point. It is expected that a more realistic simulation of the luminous region as well as future optimisations of the tracker layout will mitigate such localised degradation to negligible levels.

\begin{figure}[t]
\centering
\includegraphics[width=0.49\textwidth, trim = {0 10 0 18 } , clip]{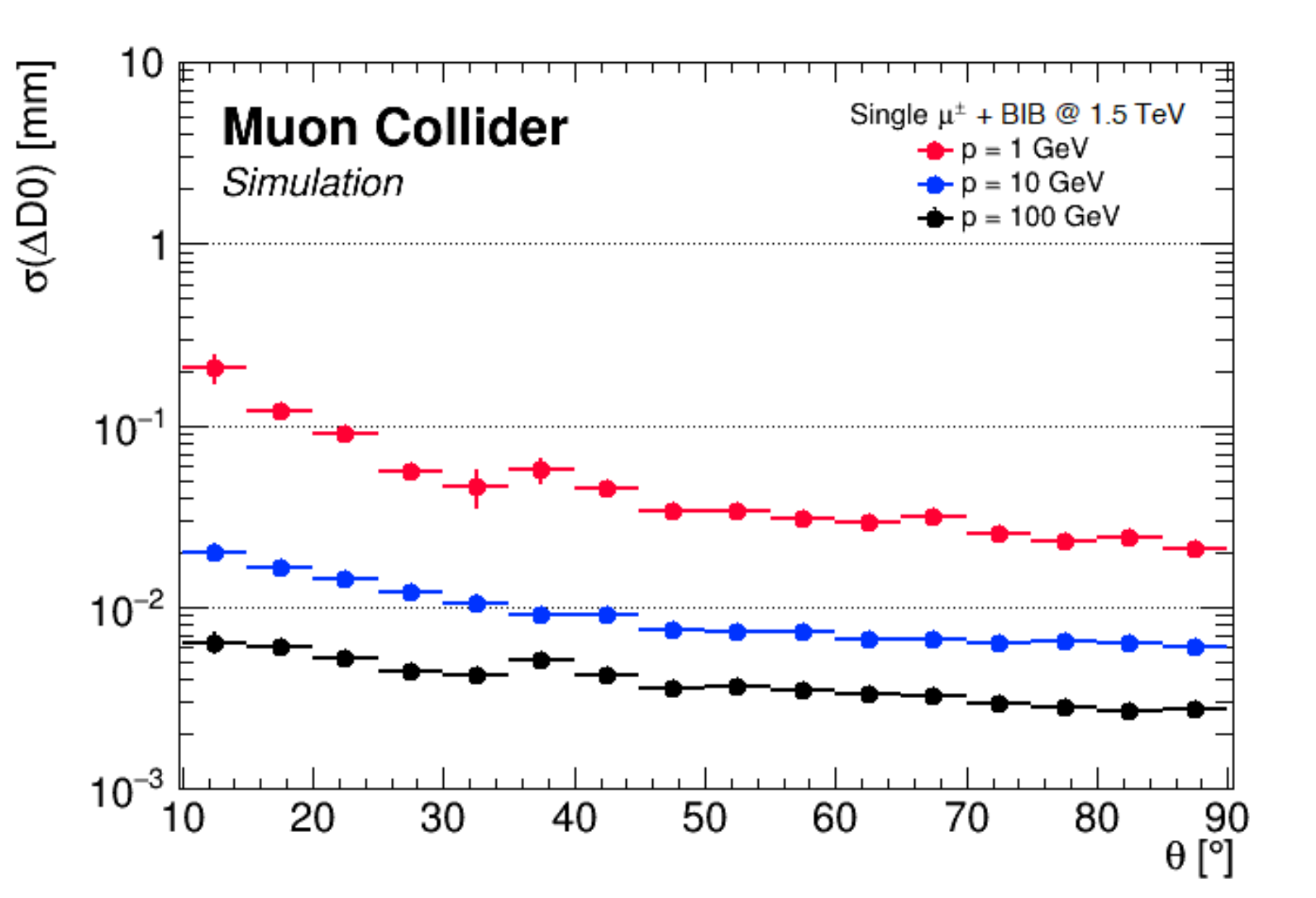}
\caption{Transverse impact parameter resolution as a function of $\theta$ for single muon events overlaid with BIB.}
\label{fig:tracking:perf-roi_d0}
\end{figure}

\begin{figure}[h]
\centering
\includegraphics[width=0.49\textwidth, trim = {0 10 0 18 } , clip]{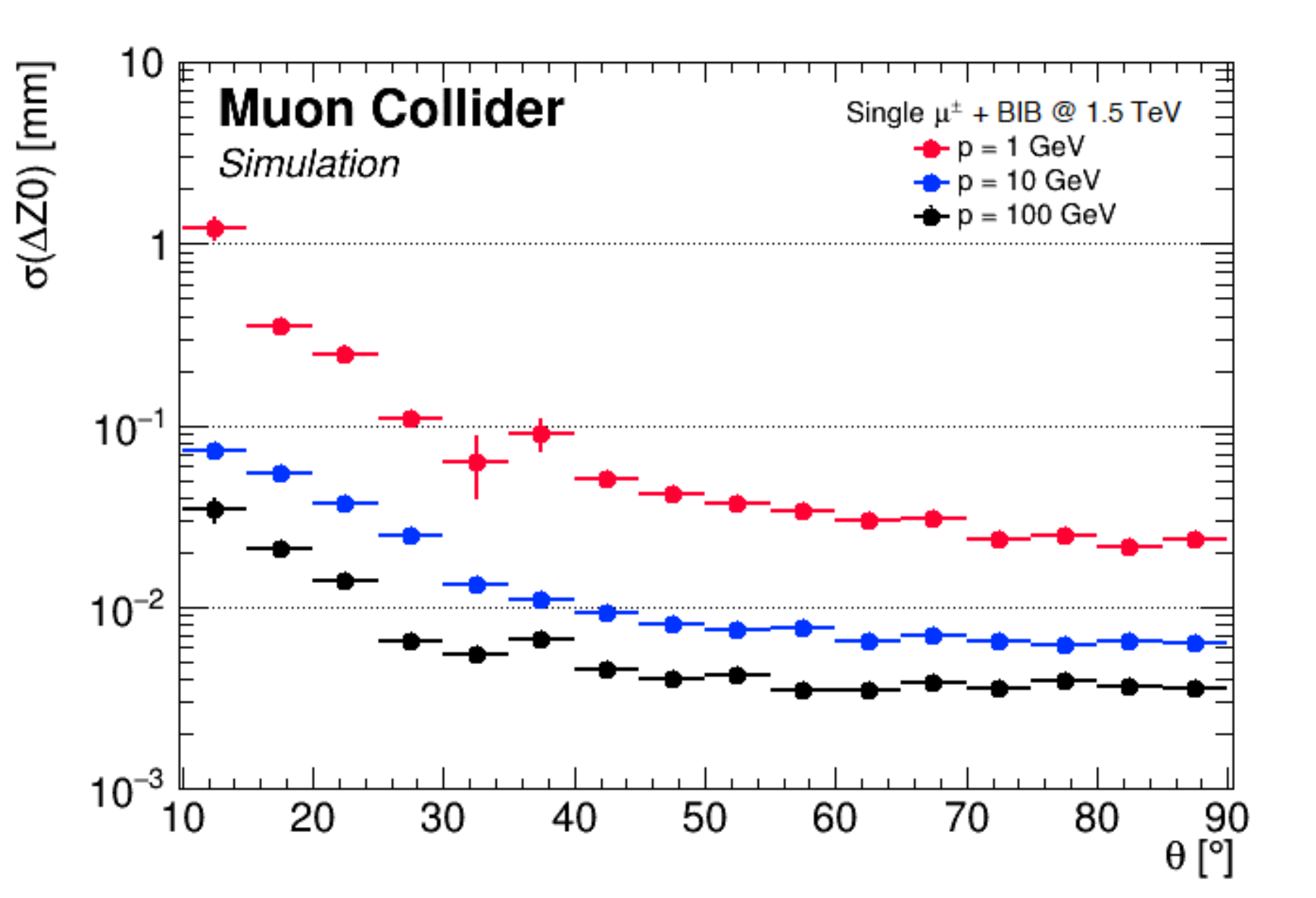}
\caption{Longitudinal impact parameter resolution as a function of $\theta$ for single muon events overlaid with BIB.}
\label{fig:tracking:perf-roi_z0}
\end{figure}

Figure~\ref{fig:tracking:perf-roi_d0} shows the resolution on the transverse impact parameter $D_0$, while Figure~\ref{fig:tracking:perf-roi_z0} the resolution on the longitudinal impact parameter $Z_0$ as a function of the polar angle $\theta$. Similarly to the case of the resolution on $p_{\rm T}$, the resolution on $D_0$ and $Z_0$ slightly degrades in the barrel-endcap transition region, around $\theta \approx 35^\circ$. 

\subsubsection*{Conformal tracking with double layers}
\label{sec:trk-dl}
\begin{figure}[t]
    \centering
    \includegraphics[width=0.5\textwidth]{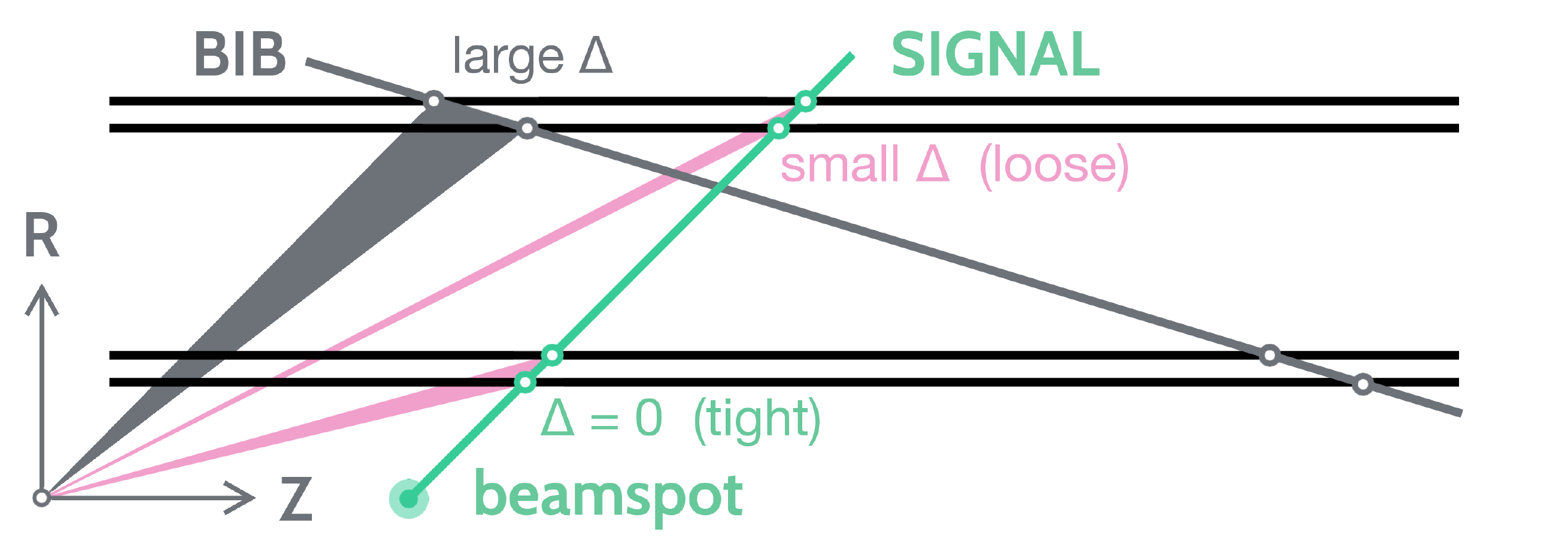}
    \caption{Illustration of the doublet-layer filtering used for the rejection of BIB-induced hits in the Vertex Detector. The horizontal black lines represent double layers of pixel sensors that are crossed by signal (green) and BIB (grey) particle tracks. Hit doublets created by BIB particles are characterised by larger angular difference than those created by signal particles, due to their shallow crossing angle and more displaced origin.}
    \label{fig:tracking:dl-howto}
\end{figure}

The Vertex Detector is constructed using double-layers (DL). A double-layer consists of two silicon detector layers separated by a small distance (\qty{2}{mm} for the MCD). This concept will also be used by the CMS Phase-II tracking detector\cite{CMS:2017lum} to reduce the hit combinatorics for a fast track reconstruction in their trigger system. It works by selecting only those hits that can form a pair with a hit from the neighbouring layer that is aligned with the IP. If there is no second hit in the double-layer to form a consistent doublet the hit is discarded. This approach is particularly effective for rejecting BIB hits, because BIB electrons are very likely to either stop in the first layer due to their very low momentum, or to cross the double-layer at shallow angles, creating doublets that are not aligned with the IP.
The DL filtering implemented in the simulation software is based on the angular distance between the two hits of a doublet when measured from the interaction point, as shown in Figure~\ref{fig:tracking:dl-howto}. For simplicity the two variables used for filtering are the polar ($\Delta\theta$) and azimuthal ($\Delta\phi$) angle differences.

In practice there are several limitations to the precision of alignment that can be imposed by the DL filtering while maintaining high efficiency for signal tracks.
The first is driven by the finite spatial resolution of the pixel sensors, which limits the minimum resolvable displacement between the two hits of a doublet. The sensor positions needs to be known beforehand and any uncertainty will result in an inefficiency. The latter point is also important for particles with non-negligible lifetime, such as $b$-meson decay products, that do not originate from the IP.

\begin{figure}[t]
\centering
\includegraphics[width=0.47\textwidth]{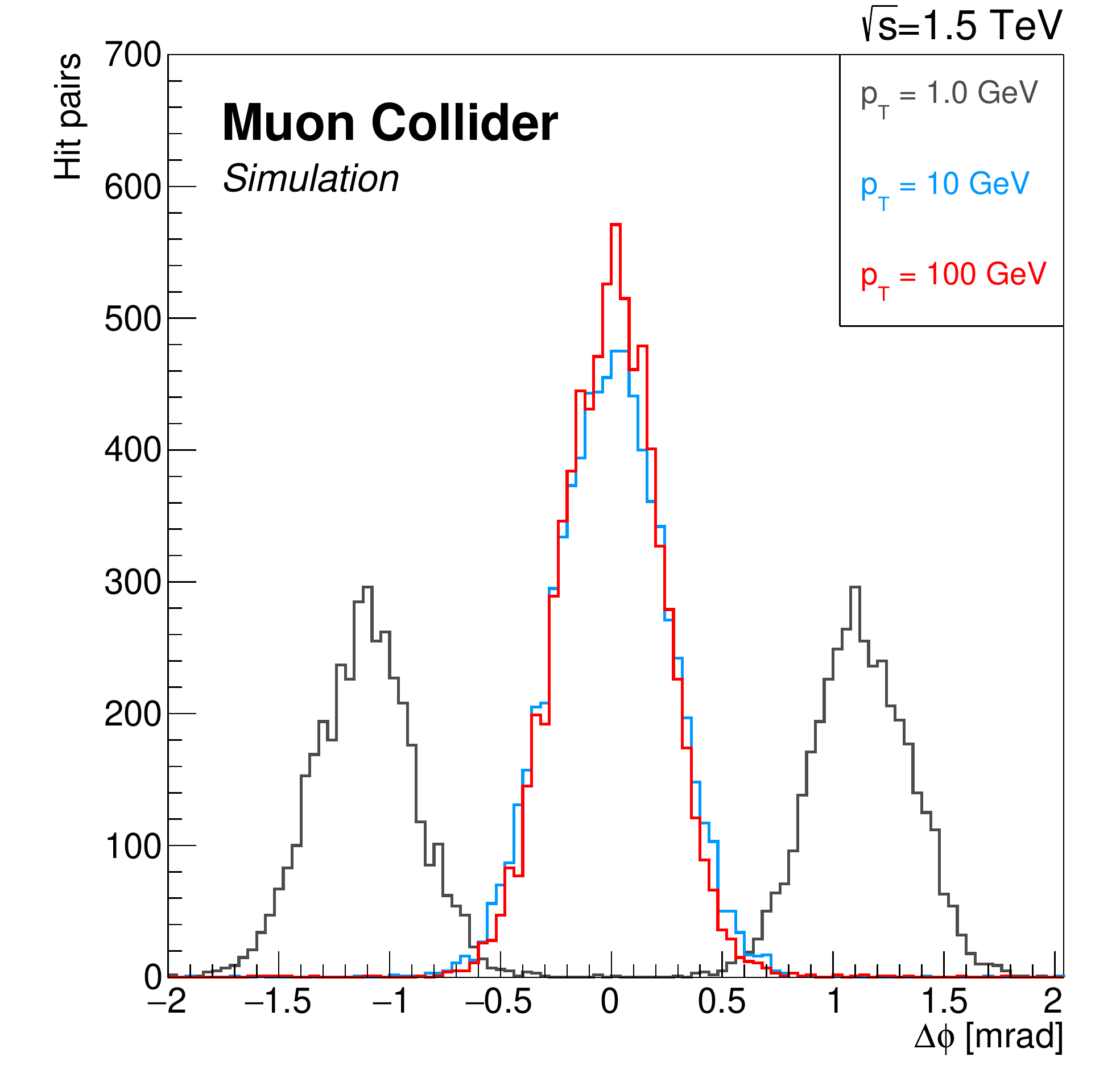}
\caption{Distribution of $\Delta\phi$ in hit doublets in the innermost double-layer of the Vertex Detector in the barrel region for muon tracks of different transverse momenta. Two separate peaks become visible for low-$p_{\rm T}$ tracks corresponding to $\mu^+$ and $\mu^-$.}
\label{fig:tracking:perf-dl-dphi}
\end{figure}
Figure~\ref{fig:tracking:perf-dl-dphi} shows the distributions of $\Delta\phi$ in the first layer of the vertex detector for single-muon events with a realistic beamspot spatial distribution. 
The bi-modal nature of the $\Delta\phi$ distribution for low energy muons is the result of the circular path that charged particle take in the transverse plane. This biases the DL filtering toward charge particles with higher $p_{\rm T}$ unless very loose selection criteria are used.

\begin{figure}[t]
\centering
\includegraphics[width=0.47\textwidth]{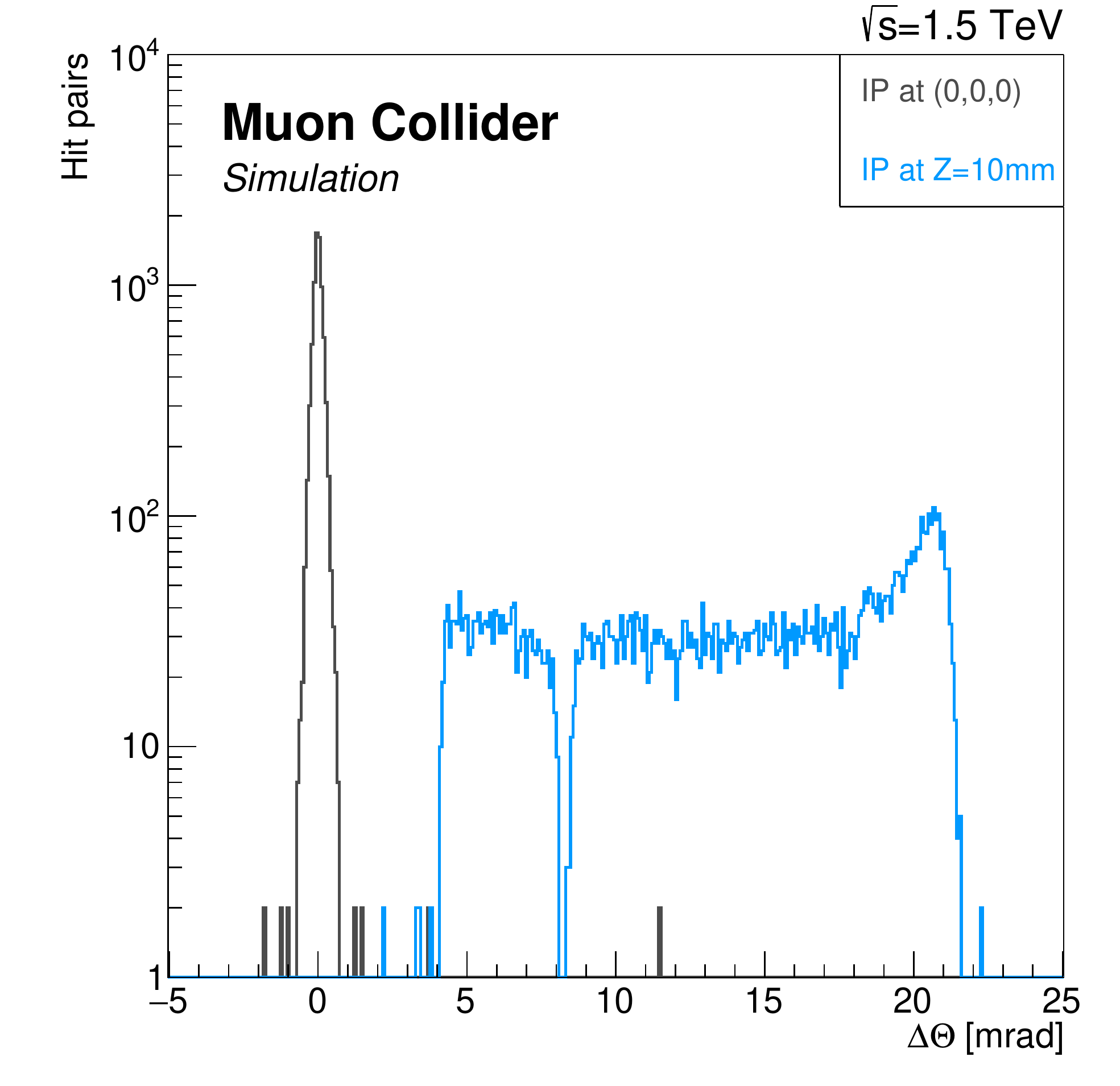}
\caption{Distribution of $\Delta\theta$ in hit doublets in the innermost double-layer of the Vertex Detector in the barrel region for muon tracks of $p_{\rm T} = \qty{1}{\GeV}$ varying the longitudinal displacement of the interaction point by \qty{10}{\milli\metre}.}
\label{fig:tracking:perf-dl-dtheta}
\end{figure}

\begin{figure}[h]
    \centering
    \includegraphics[width=0.45\textwidth]{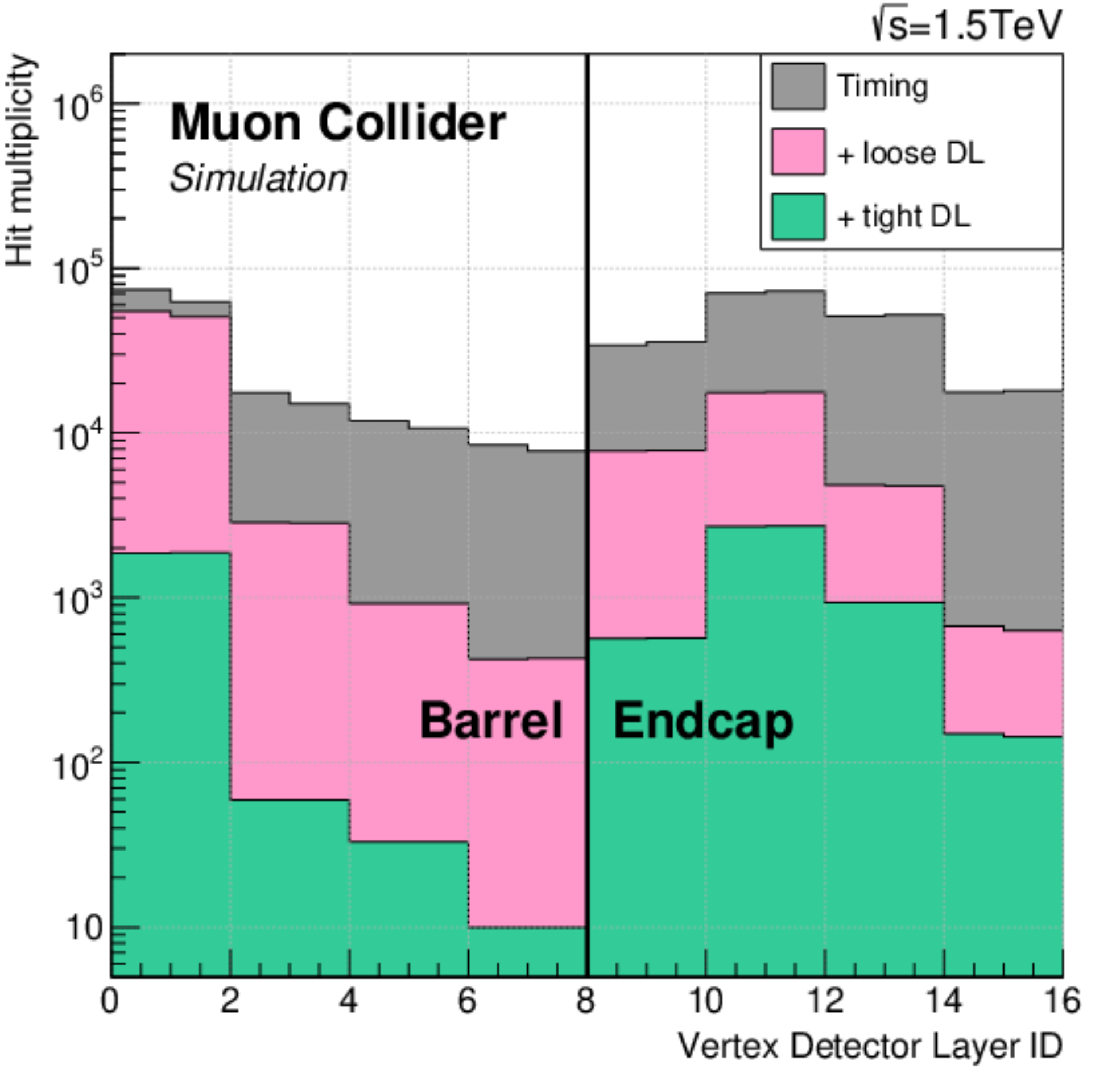}
    \caption{Expected reduction of hit multiplicity in the Vertex Detector achieved by applying the loose or tight double-layer filtering.}
    \label{fig:tracking:perf-dl-hits-reduction}
\end{figure}

The distribution of $\Delta\theta$ is shown instead in Figure~\ref{fig:tracking:perf-dl-dtheta}, comparing scenarios with different longitudinal displacements of the interaction point. 
Table~\ref{tab:tracking:dl-rejection} lists the two operating points ({\em loose} and {\em tight}) used to filter hits for the CT algorithm in two stages.
\begin{table*}[h]
    \centering

    \begin{tabular}{c|l||c|c|c|c||c|c|c|c}
    \multicolumn{2}{r||}{}       & \multicolumn{4}{c||}{\textbf{Barrel}} & \multicolumn{4}{c}{\textbf{Endcap}}\\
    \multicolumn{1}{c}{} & Layer IDs & 0,1 & 2,3 & 4,5 & 6,7 & 0,1 & 2,3 & 4,5 & 6,7\\
    \hline\hline
    \multirow{3}{.75in}{\centering \textbf{Loose DL} selections} 
    &Max. $\Delta\phi$ (mrad)    & 2.8 & 2.0 & 1.7 & 1.5 & 2.1 & 1.7 & 1.6 & 1.5 \\
    &Max. $\Delta\theta$ (mrad)  & 35 & 18 & 10 & 6.5 & 3.5 & 1.5 & 0.7 & 0.5  \\
    &Hit surival fraction        &\multicolumn{4}{c||}{55\%} &\multicolumn{4}{c}{18\%}\\
    \hline
    \multirow{3}{.75in}{\centering \textbf{Tight DL} selections} 
    &Max. $\Delta\phi$ (mrad)    & 3.0 & 2.0 & 1.6 & 1.5 & 2.2 & 1.8 & 1.7 & 1.6\\
    &Max. $\Delta\theta$ (mrad)  & 0.5 & 0.4 & 0.3 & 0.25 & 0.2 & 0.18 & 0.12 & 0.1 \\
    &Hit survival fraction       &\multicolumn{4}{c||}{2\%} & \multicolumn{4}{c}{2\%}\\
    \end{tabular}
    
    \caption{Angular selection on double-layers for hit suppression. A loose (left) and tight (right) selection is shown. The hit survival rate represents the fraction of hits (mostly from BIB) surviving the selections.}
    \label{tab:tracking:dl-rejection}
\end{table*}
Figure~\ref{fig:tracking:perf-dl-hits-reduction} shows the hit multiplicity (mostly BIB-hits) as a function of VXD layer. 

The {\em loose} working point targets high efficiency reconstruction of $p>\qty{1}{\GeV}$ muons with a realistic beam-spot size. It reduces the number of hits in the innermost double-layer by a factor of two. This reduces the CT reconstruction time from $\approx\qty{1}{week/event}$ to $\approx\qty{2}{days/event}$. While significant, this is still not practical for sample production. Instead the {\em loose} working point is used with a special CT configuration as a first stage of a two-stage reconstruction process. The first stage reconstructs high-$p_{\rm T}$ tracks to precisely determine the IP position. 

The {\em tight} working point is optimised for the scenario when the interaction point is precisely known. It has a hit survival rate of $\approx{}2\%$ in the inner-most layer. This reduces the hit multiplicity enough for the CT algorithm to complete in $\approx\qty{2}{min/event}$. It is used as the second stage of the track reconstruction algorithm, with the IP determined from the first-stage.

The two-stage doublet-layer filtering provides a computationally viable method to track reconstruction in the presence of the BIB. However its efficiency is very limited for reconstructing non-prompt particles, including those arising from $b$-meson decays.

\subsubsection*{Combinatorial Kalman filter}
\label{sec:trk-ckf}
The CKF algorithm seeded using hit triplets was implemented using the ACTS library v13.0.0. The implementation is a Marlin Processor that serves as a drop-in replacement for the existing tracking Processors. In addition to providing an alternate algorithm designed for large hit multiplicity, ACTS also provides a heavily optimised code for fast computation. The reconstruction strategy used in this section reconstructs tracks at the rate of \qty{4}{\min/event}. It provides the first practical and comprehensive tracking solution for the MCD.

The seeds for the CKF algorithm are formed from hit triplets in the four layers of the Vertex Detector. Only hits in the outer half of doublet layer are considered. Several heuristics are use to determine if each triplet is compatible with a track and can be used as a seed. The seeding algorithm is configured using the ACTS default values, with the exception of:
\begin{itemize}
    \item the radial distance between hits is between \qty{5}{\mm} and \qty{80}{\mm},
    \item the minimum estimated $p_{\rm T}$~is \qty{500}{\MeV},
    \item the maximum forward angle of \qty{80}{\degree},
    \item the extrapolated collision region is within \qty{1}{mm} of detector centre,
    \item the average radiation length per seed is 0.1,
    \item the allowed amount of scattering is \qty{50}{\sigma},
    \item the middle hits in each seed are unique.
\end{itemize} 

This configuration has not been fully optimised. For example, the size of the collision region is smaller than the expected beam-spot size. An initial implementation of track seeding from the Outer Tracker, followed by a track extrapolation towards the centre of the detector, has recently become available but has not yet studied in details. The lower BIB-hit multiplicity in this detector region could be exploited to loosen some requirements and improve the reconstruction efficiency for tracks originating far from the IP.

Around 150\,000 seeds are found per event. The efficiency of the seeding algorithm is found to be fully efficient for muons with $p_{\rm T}>2$~GeV, dropping to about 90\% in the region within \qty{10}{\degree} from the beam axis. Loosening the collision region definition increases the amount of fake seeds and reduces the seed finding efficiency due to the seed overlap removal. The latter can be addressed, at a cost in run time, by allowing multiple seeds to share the same middle hit.

\begin{figure}[t]
    \centering
    \includegraphics[width=0.486\textwidth, trim = {0 0 0 15 } , clip]{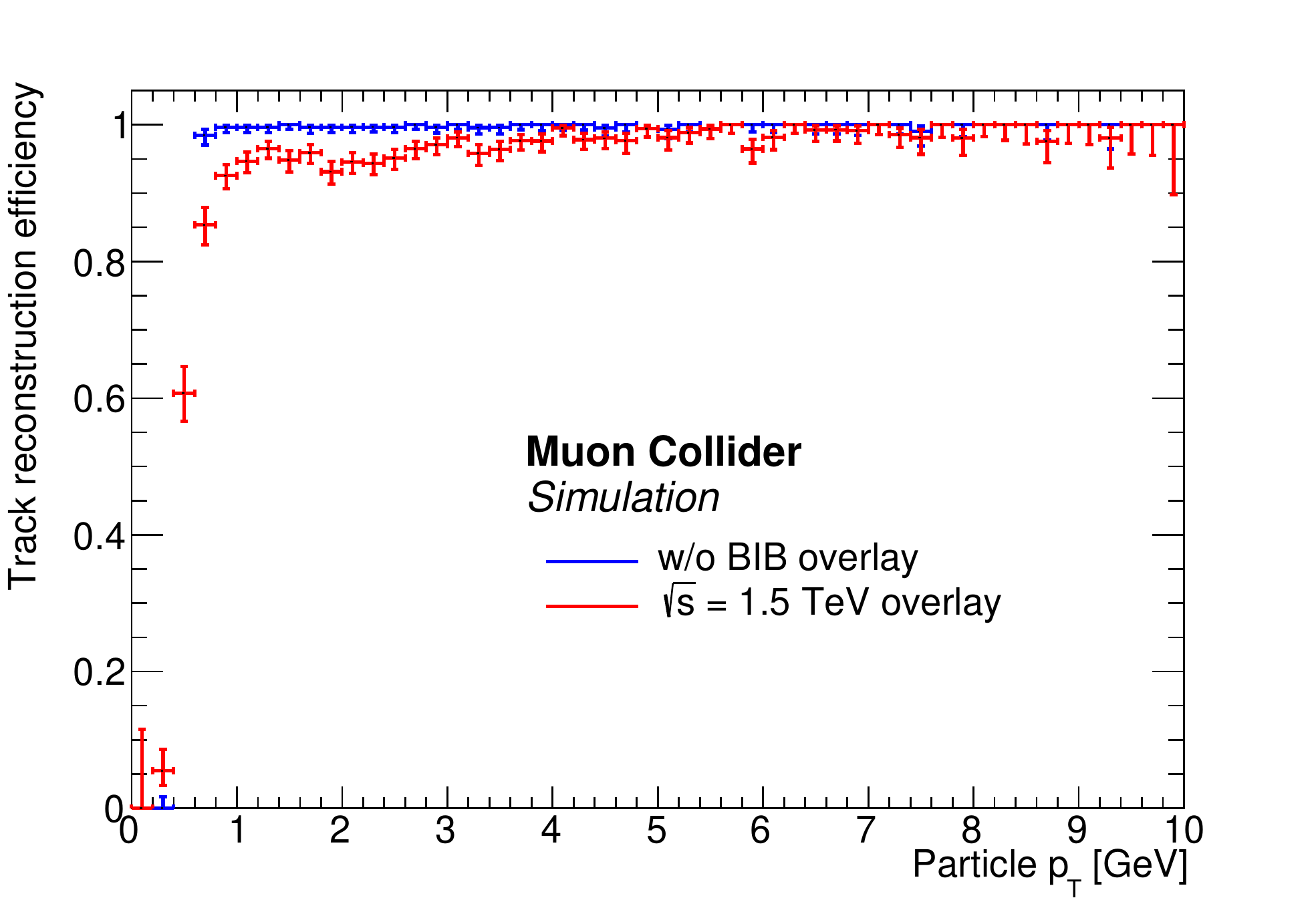}
    \caption{Track reconstruction efficiency for events containing a single muon with (red) and without (blue) BIB overlay as a function of the true muon $p_{\rm T}$.}
    \label{fig:trk-ckf-pteff}
\end{figure}

\begin{figure}[t]
    \centering
    \includegraphics[width=0.475\textwidth, trim = {0 0 0 25 } , clip]{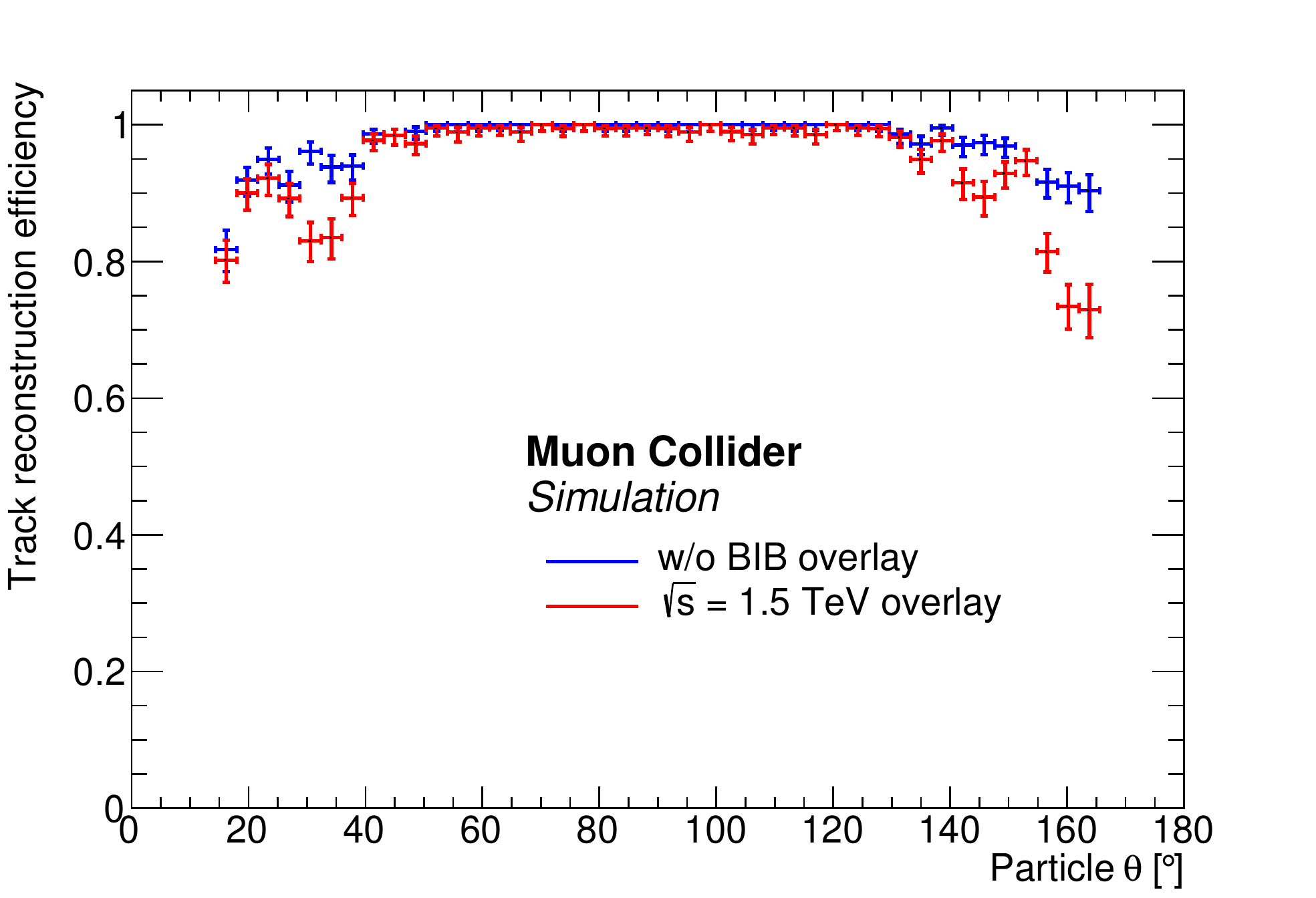}
    \caption{Track reconstruction efficiency for events containing a single muon with (red) and without (blue) BIB overlay as a function of the true muon $\theta$.}
    \label{fig:trk-ckf-thetaeff}
\end{figure}

The CKF is run inside-out, meaning that the track extrapolation starts from the inner-most seed hit and continues outward in the radial direction. The initial track parameters are estimated from the seed. The CKF algorithm has only two tunable parameters: the number of hits added for each layer, which is set to one, and the width of the hit search window at each layer, which is set to \qty{10}{\chi^2}. As with the seeding algorithm, these values were not yet optimised. The consideration of only a single hit at each layer means that the CKF algorithm will not branch to consider multiple track candidates for a single seed. A good track reconstruction efficiency is observed irrespective of this tight requirement, as demonstrated in Figures~\ref{fig:trk-ckf-pteff} and~\ref{fig:trk-ckf-thetaeff}, which show the track reconstruction efficiency as a function of particle $p_{\rm T}$ and $\theta$ for single muon events with $p=\qty{10}{\GeV}$. This underlines the difference between BIB and pile-up; the BIB-hits are a ``random'' background that is not compatible with the trajectory of a track.

\begin{figure}[t]
\centering
\includegraphics[width=0.48\textwidth, trim = {0 0 0 10 } , clip]{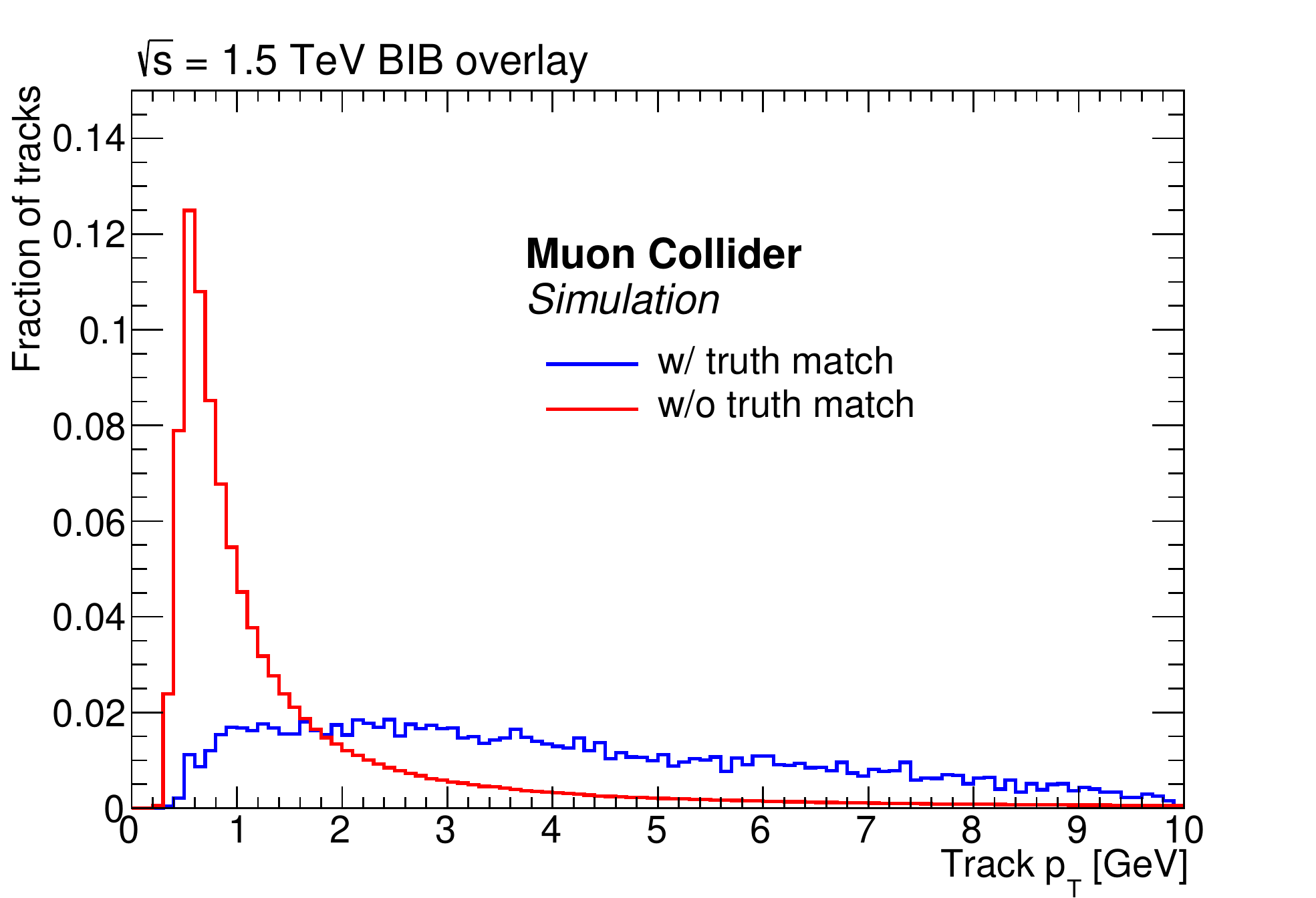}
\caption{Track $p_{\rm T}$ distributions for tracks with (blue) and without (red) a match to the true simulated tracks, for single muon events with BIB overlay.}
\label{fig:trk-ckf-pt}
\end{figure}

\begin{figure}[t]
\centering
\includegraphics[width=0.475\textwidth]{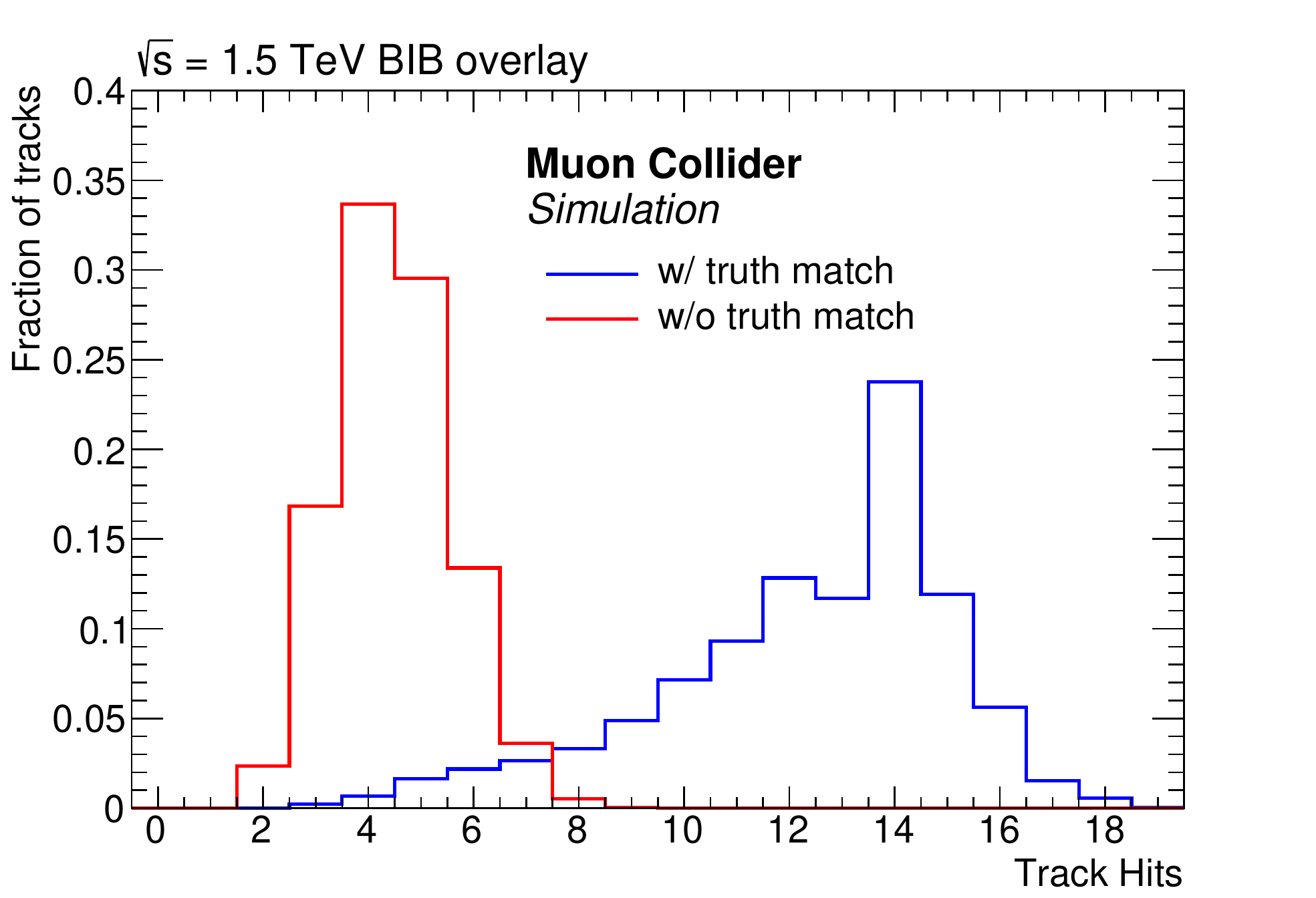}
\caption{Hit multiplicity $N_{\text{hit}}$ distribution for tracks with (blue) and without (red) a match to the true simulated tracks, for single muon events with BIB overlay.}
\label{fig:trk-ckf-nhits}
\end{figure}

Muons with $p_{\rm T}>\qty{2}{\GeV}$ are reconstructed with 90\% efficiency or greater even in the presence of BIB. There is a considerable amount of fake tracks ($\approx100,000$ per event). Figure~\ref{fig:trk-ckf-pt} compares the $p_{\rm T}$ distribution for real and fake tracks. Fake tracks are mostly at low $p_{\rm T}$.
Similarly, when considering the number of hits associated to a track, shown in Figure~\ref{fig:trk-ckf-nhits}, the fake tracks arising from the BIB are associated with a significantly smaller number of hits. The latter further underlines the randomness of the BIB-hits and provides a handle for reducing the fake rate.

\subsubsection*{Jets}

Jet reconstruction is one of the most difficult reconstruction tasks at a muon collider, since almost all sub-systems are involved, and the impact of the BIB is significant in all of them, with different features in different sub-detectors.
The jet reconstruction algorithm employed is described in this section, and its performance is discussed. The algorithm has been designed to reconstruct jets in the presence of the BIB, but it is far from being fully optimised, and further studies are needed in the future.
However, even at this very early stage, the jet reconstruction can achieve a decent performance.

The algorithm follows the principles of particle flow reconstruction as implemented in the PandoraPFA package, and comprises of the following steps:
\begin{enumerate}
    \item tracks are reconstructed using the CKF algorithm described in Section~\ref{sec:trk-ckf} and are required to have at least three hits in the Vertex Detector and at least two hits in the Inner Tracker;
    \item calorimeter hits are selected by requiring a hit time window and an energy threshold;
    \item tracks and calorimeter hits are used as inputs in the PandoraPFA~\cite{Marshall:2013bda} algorithm to obtain reconstructed particles;
    \item the reconstructed particles are clustered into jets with the $k_t$ algorithm.
\end{enumerate}

The reconstructed jets are then required so satisfy quality selections to reject fake jets arising from the spurious combination of BIB energy deposits. Finally, the energy of the jets passing the BIB-removal selection are calibrated.

The jet performance has been evaluated on simulated samples of $b\bar{b}$, $c\bar{c}$ and $q\bar{q}$ dijets, where $q$ stands for a light quark ($u$,$d$ or $s$). These samples have been generated with an almost uniform dijet $p_{\rm T}$ distribution from 20 to \qty{200}{\GeV}. Samples of $\mu^+ \mu^- \to H ( \to b\bar{b}) + X $ and $\mu^+ \mu^- \to Z ( \to b\bar{b}) + X $ at $\sqrt{s}=3$ TeV are also used to study the dijet invariant mass resolution. 

\subsubsection*{Calorimeter hit selection}

\begin{figure}[b]
\centering
\includegraphics[width=0.48\textwidth]{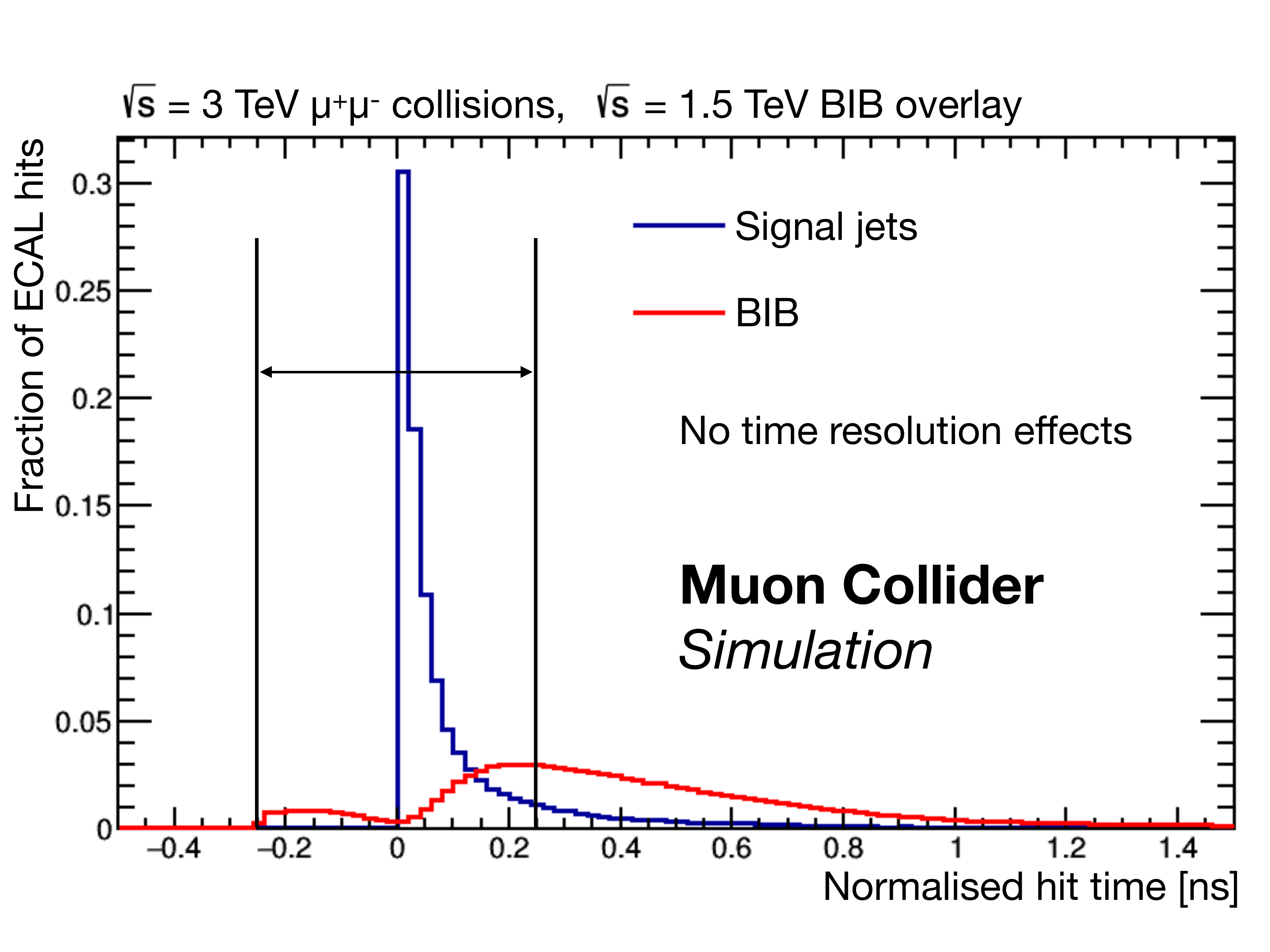}
\caption{Normalised hit time in ECAL barrel, for $b$-jets and BIB. Both distributions are normalised to the same area. The time is not smeared for the detector time resolution. The time window of \qty{\pm 250}{\pico\second} applied in the jet reconstruction is shown.}
\label{fig:calo_time}
\end{figure}

Calorimeter hits are filtered depending on the normalised hit time, defined as $t_N = t - t_0 - cD$, where $t$ is the absolute hit time, $t_0$ is the collision time, $c$ is the speed of light, and $D$ is the hit distance from the origin of the reference system.
A time window of $\pm 250$~ps is applied to remove most of the BIB hits but preserving the signal, as can be seen in Figure~\ref{fig:calo_time}. 

A time window of width $\Delta$ is generally applicable if the Full Width at Half Maximum (FWHM) of the time resolution of the calorimeter cell is below $\Delta/3$. Therefore, in this particular case, a FWHM of at least 167~ps is assumed, which should be achievable by state-of-the-art calorimeter technologies.
    
Several calorimeter hit energy thresholds have been tested, in order to reduce hits produced by BIB. The computing time of the jet algorithm exponentially grows with the number of calorimeter hits: therefore, with the current resources it is not possible to reduce the thresholds far below 2~MeV. A threshold of 2~MeV is hence applied to both ECAL and HCAL. 
This requirement reduces the average number of ECAL Barrel hits from 1.5 million to less than $10^4$.

\subsubsection*{Particle flow, jet clustering and fake jet removal} 
    
Calorimeter hits and tracks are given as input to the PandoraPFA algorithm, that produces as output reconstructed particles known as particle flow objects. The PandoraPFA algorithm is described in detail in Ref.\cite{Thomson:2009rp}.
The particle flow objects are then clustered into jets by the $k_T$ algorithm. A cone radius of $R=0.5$ is used.

\begin{figure}[b]
\centering
\includegraphics[width=0.47\textwidth]{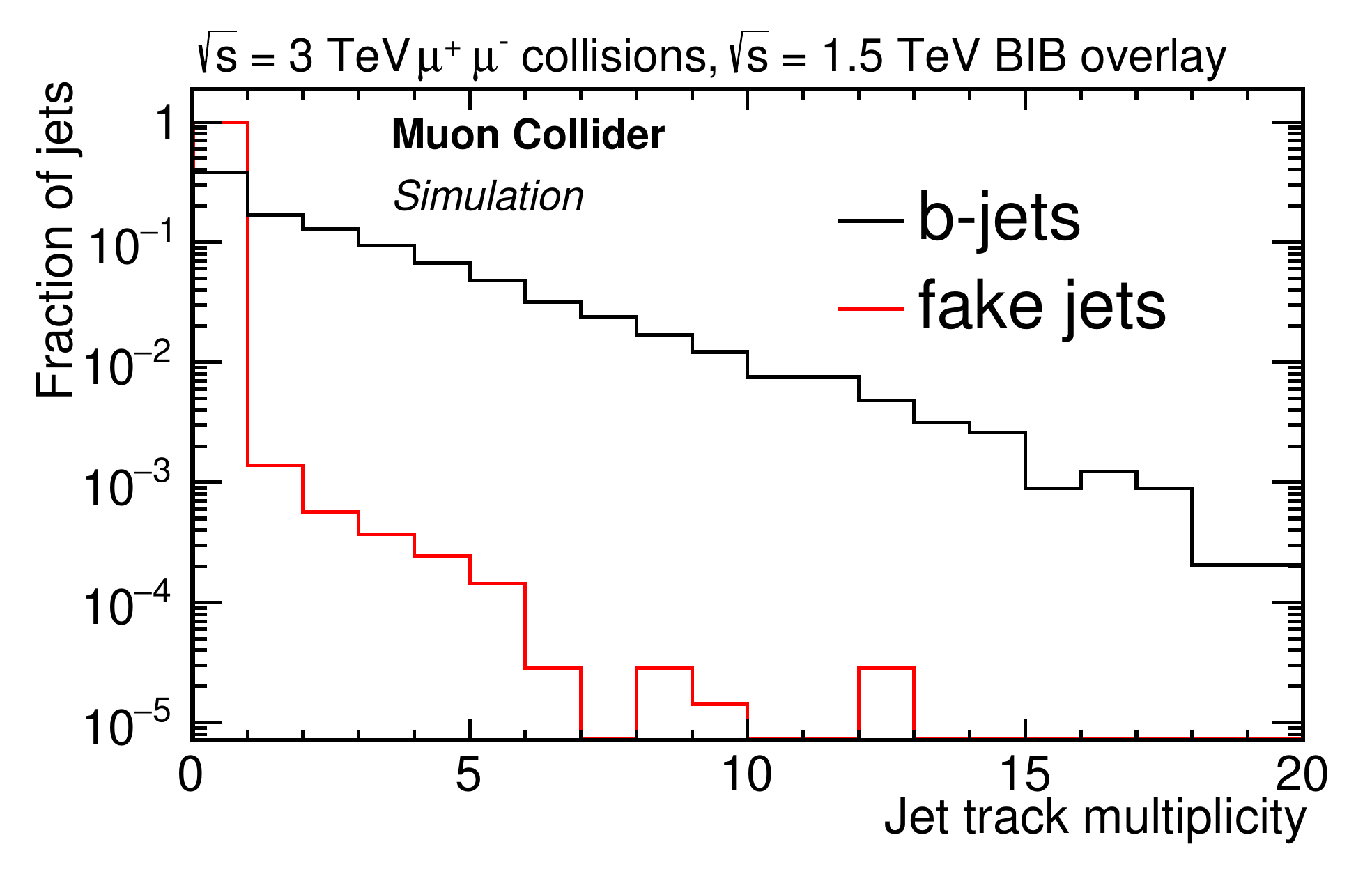}
\caption{Number of tracks associated to real $b$-jets and to fake jets from BIB.}
\label{fig:jet_ntracks}
\end{figure}

An average of 13 fake jets from BIB energy deposits per event is reconstructed, which need to be removed by applying additional quality criteria. The number of tracks in the jet has been found to be the most discriminating feature between real and fake jets, as shown in Figure~\ref{fig:jet_ntracks}.

\begin{figure}[b]
\centering
\includegraphics[width=0.5\textwidth]{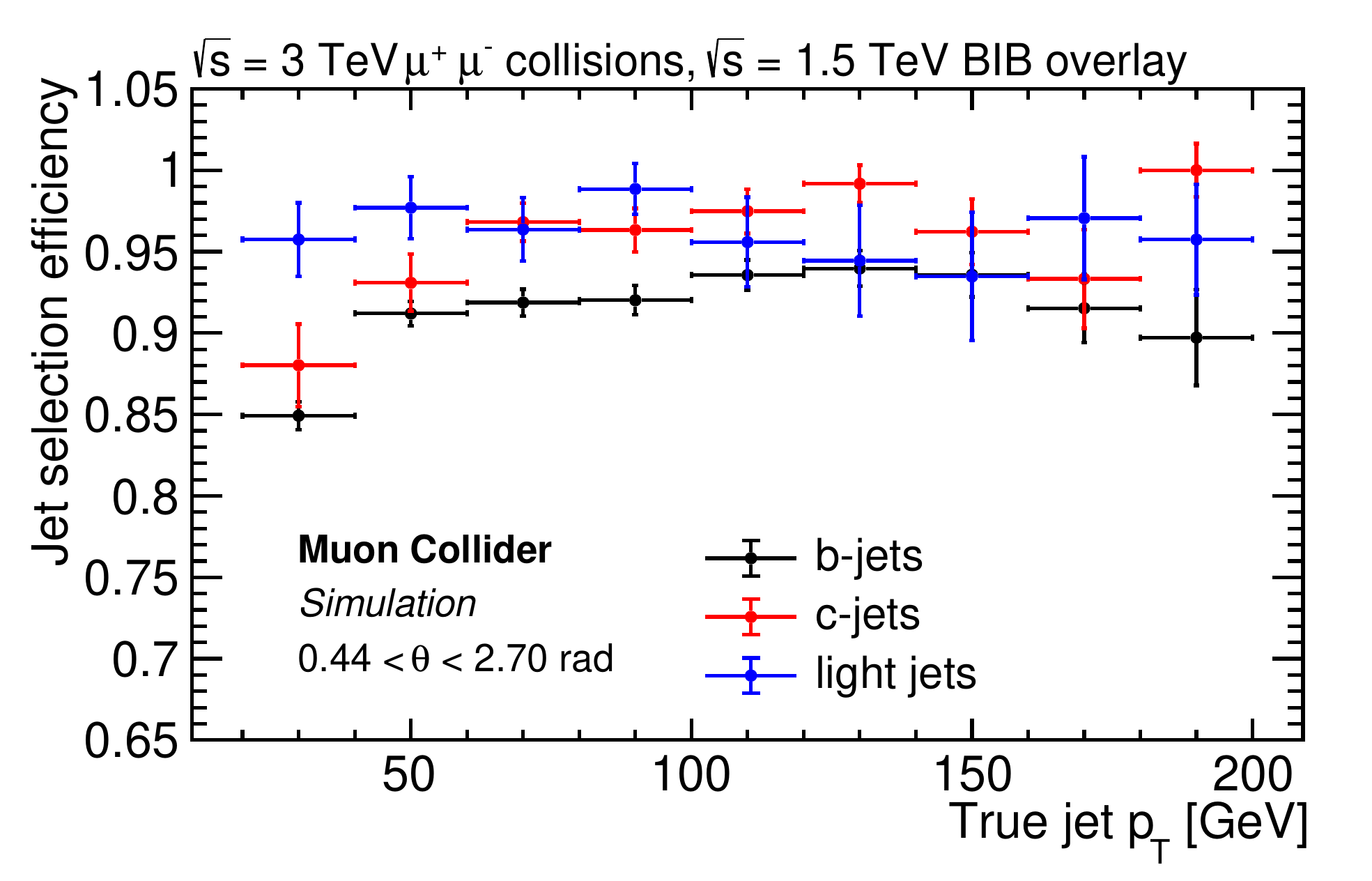}
\caption{Jet selection efficiency as a function jet $p_{\rm T}$ for $b$-jets, $c$-jets and light jets in the central region $|\eta|<1.5$. The differences between the jet flavours are mainly due to different jet $\eta$ distributions in the three samples.}
\label{fig:jet_pT_eff}
\end{figure}

Most of the fake jets from BIB have no tracks associated to them. Therefore, requiring at least one track allows to reduce the rate of fake jets by more than two orders of magnitude, with a moderate cost, at the level of 5--10\%, in terms of real jet selection efficiency. Figure~\ref{fig:jet_pT_eff} shows the jet selection efficiency as a function of the true jet $p_{\rm T}$ in the region $|\eta|<1.5$, for different jet flavours. The overall efficiency varies between 82$\%$ at low $p_{\rm T}$ to 95$\%$ at higher $p_{\rm T}$. 

\begin{figure}[b]
\centering
\includegraphics[width=0.48\textwidth]{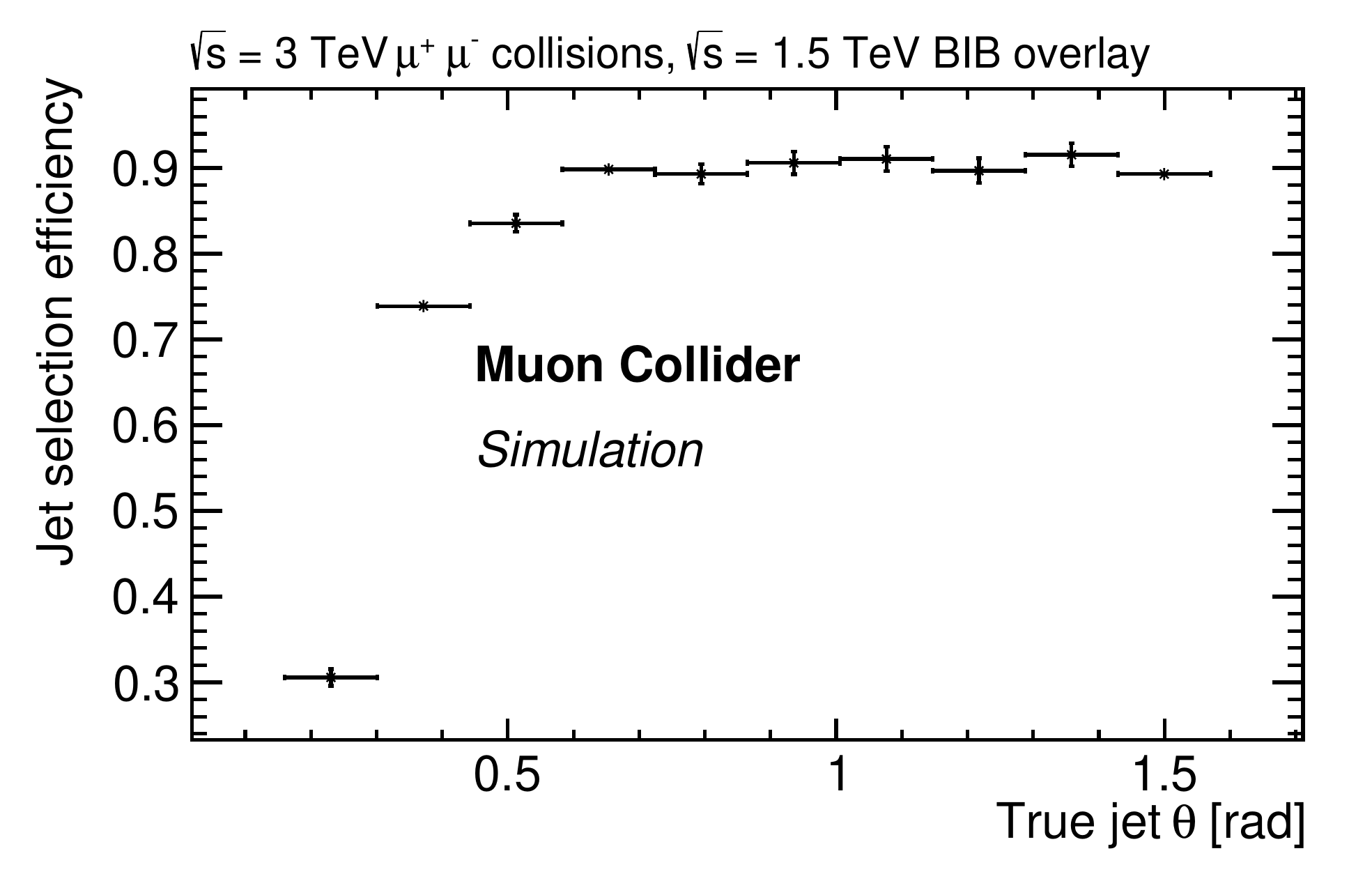}
\caption{Efficiency of jet selection as a function of truth-level jet $\theta$.}
\label{fig:bjet_eta_eff}
\end{figure}

The dependency of the selection efficiency on the jet polar angle $\theta$ is shown in Figure~\ref{fig:bjet_eta_eff} for a sample of $b$-jets. The efficiency is around 90$\%$ in the central region. A significant drop is observed for $|\theta|<0.5$, where the efficiency is below 30$\%$. This effect is mainly due to the track requirement, since many jets without reconstructed tracks are found in the forward region. 

\begin{figure}[t]
\centering
\includegraphics[width=0.5\textwidth]{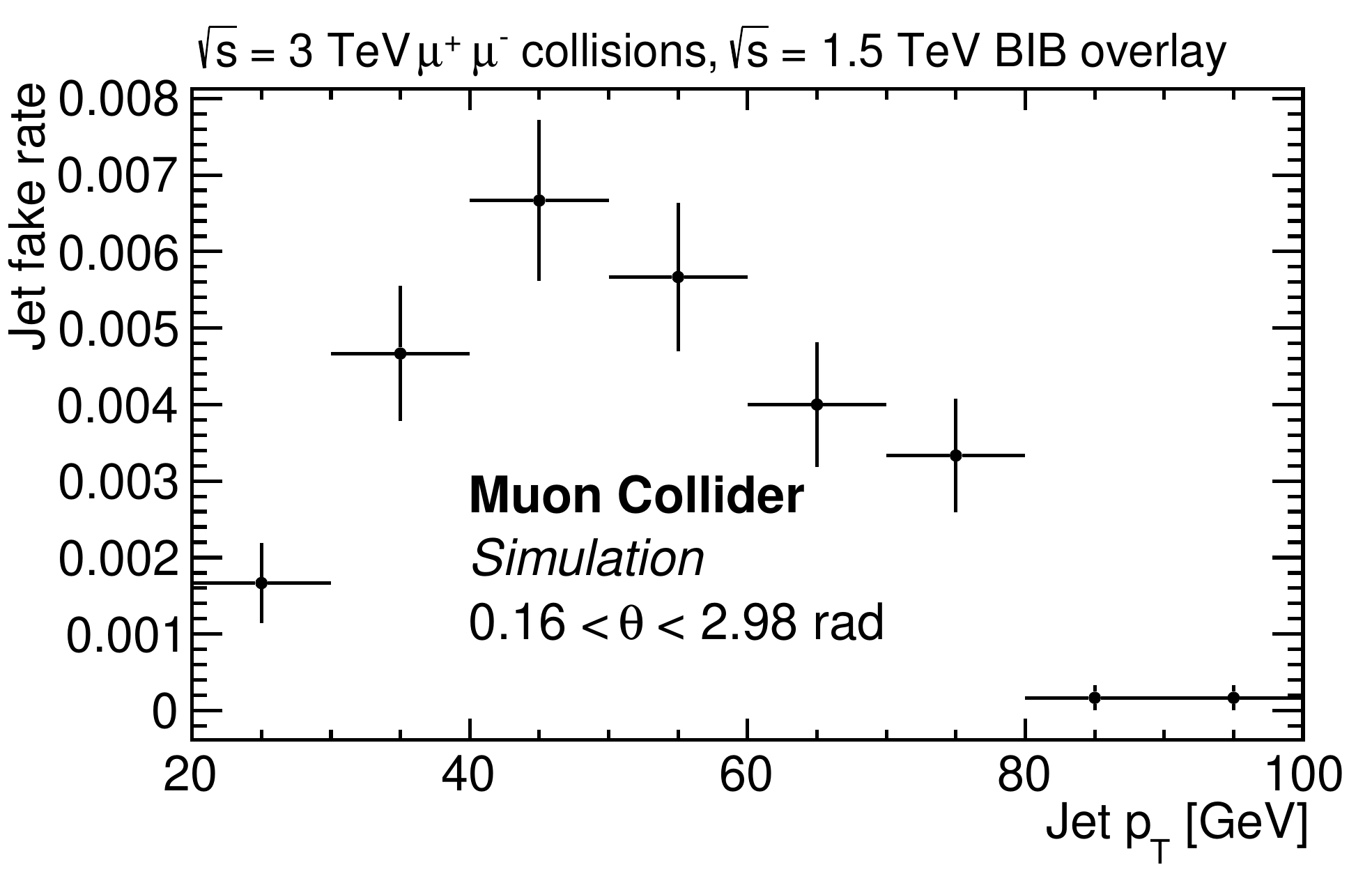}
\caption{Jet fake rate as a function of the jet $p_{\rm T}$, obtained after requiring at least one track associated to the jet.}
\label{fig:jet_fakes}
\end{figure}

Figure~\ref{fig:jet_fakes} shows the fake jet rate as a function of jet $p_{\rm T}$ for a sample of $b$-jets. The fake rate is defined as the average number of reconstructed fake jets that are not matched with a true particle originating from the IP per event, which is well below 1\%.

\subsubsection*{Jet momentum correction}

The jet 4-momentum is defined by the sum of 4-momenta of particles that belong to the jet. The jet axis is identified by the jet momentum direction. A fiducial region for jet reconstruction is defined by selecting jets with $|\eta|<2.5$.

In order to recover the energy lost by reconstruction inefficiencies, non sensitive material, as well as to take into account BIB contamination, a correction to the jet 4-momentum is applied. This correction has been determined by comparing the reconstructed jet $p_{\rm T}$ with the corresponding truth-level jet $p_{\rm T}$. Truth-level jets are defined as jet clustered by applying the $k_t$ algorithm to visible Monte Carlo particles. Reconstructed and truth-level jets are matched if their $\Delta R = \sqrt{(\Delta \eta)^2 + (\Delta \phi)^2} < 0.5$, where $\Delta \eta$ and $\Delta \phi$ are respectively the pseudo-rapidity and the azimuthal angle differences between the reconstructed-level jet axis and truth-level jet axis. If more than one reconstructed jet is matched to the same truth-level jet, then the one with lowest $\Delta R$ is chosen.

\begin{figure}[t]
    \centering
    \includegraphics[width=0.48\textwidth]{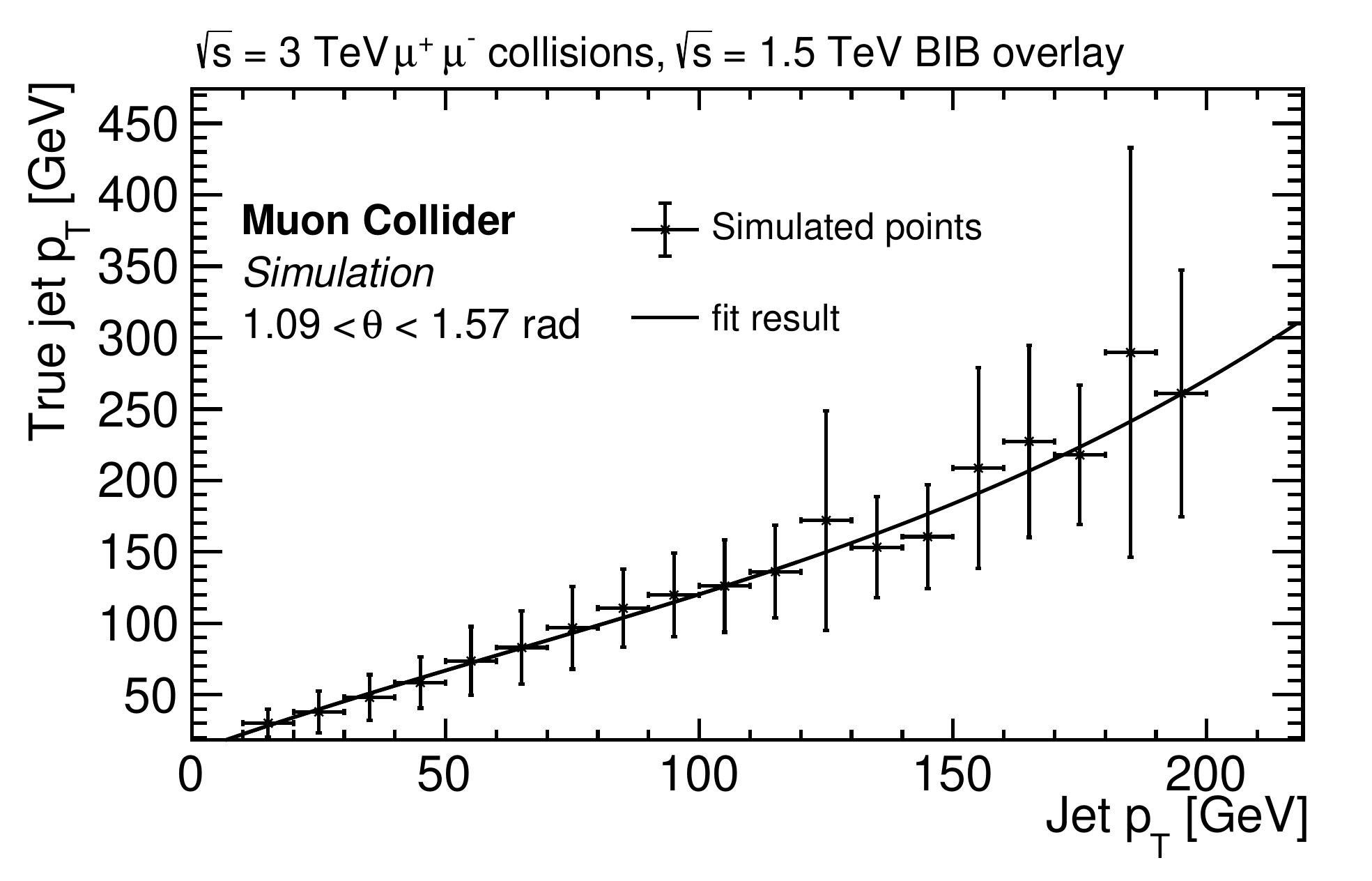}
    \caption{Transfer function used for jet momentum corrections. The average truth-level jet $p_{\rm T}$ as a function of the reconstructed jet $p_{\rm T}$ is shown, for $1.09<\theta<1.57$. The error bars represent the standard deviation of the truth-level jet $p_{\rm T}$ in each interval.}
    \label{fig:jec1}
\end{figure}

\begin{figure}[t]
    \centering
    \includegraphics[width=0.48\textwidth]{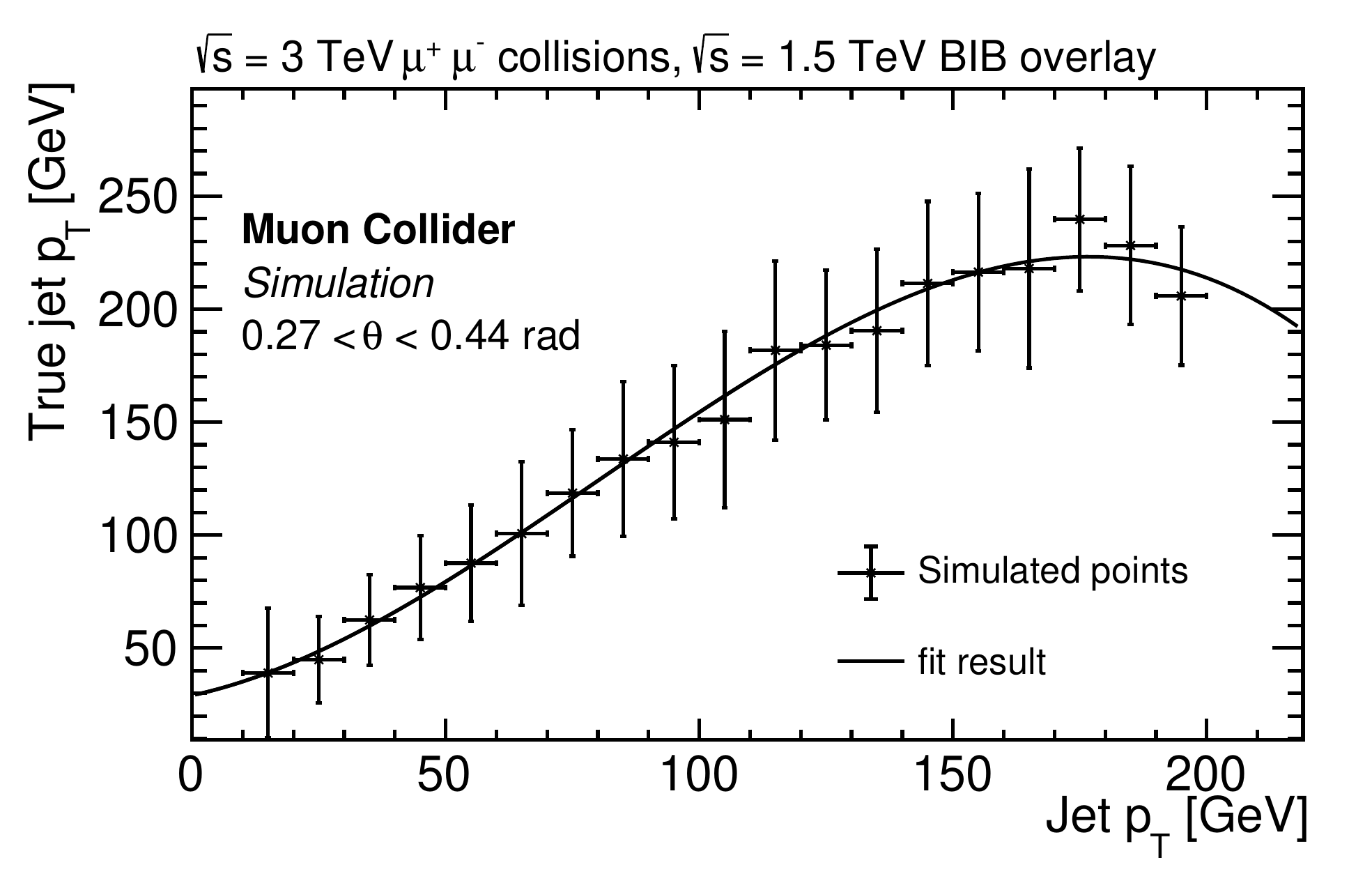}
    \caption{Transfer function used for jet momentum corrections. The average truth-level jet $p_{\rm T}$ as a function of the reconstructed jet $p_{\rm T}$ is shown, for $0.27<\theta<0.44$. The error bars represent the standard deviation of the truth-level jet $p_{\rm T}$ in each interval.}
    \label{fig:jec2}
\end{figure}

The correction is evaluated in five equal-width intervals of reconstructed $|\eta|$ between 0 and 2.5. Each pseudo-rapidity interval is further divided in 19 equal-width intervals of reconstructed $p_{\rm T}$ between 10 and 200~GeV. For each interval the average and standard deviation of the truth-level jet $p_{\rm T}$ distribution is calculated. Transfer functions are then obtained in each $\eta$ interval by fitting the average truth-level jet $p_{\rm T}$ as a function of reconstructed jet $p_{\rm T}$. Examples of transfer functions are shown in Figures~\ref{fig:jec1} and~\ref{fig:jec2}. These functions are then used to obtain the scale factor that is applied to each component of the reconstructed 4-momentum.

\subsubsection*{Jet reconstruction performance}

\begin{figure}[t]
\centering
\includegraphics[width=0.5\textwidth]{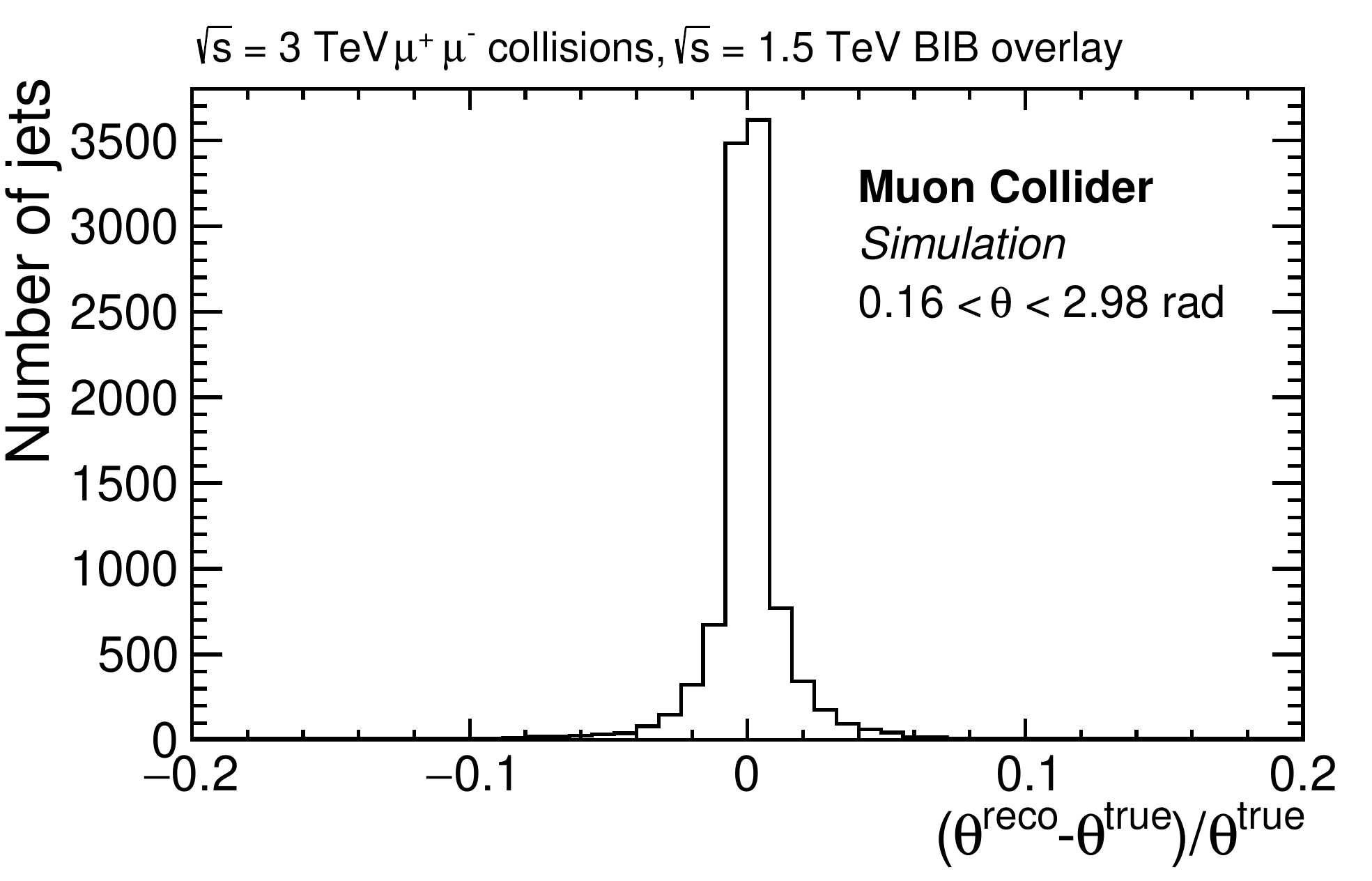}
\caption{Relative difference between reconstructed and true jet pseudo-rapidity.}
\label{fig:theta_reso}
\end{figure}

The jet reconstruction performance is evaluated with a simulated sample of dijets events.
The relative difference between reconstructed and true jet $\theta$ is shown in Figure~\ref{fig:theta_reso}: the standard deviation of this distribution, directly related to the jet-axis angular resolution, is 14$\%$.

\begin{figure}[t]
\centering
\includegraphics[width=0.48\textwidth]{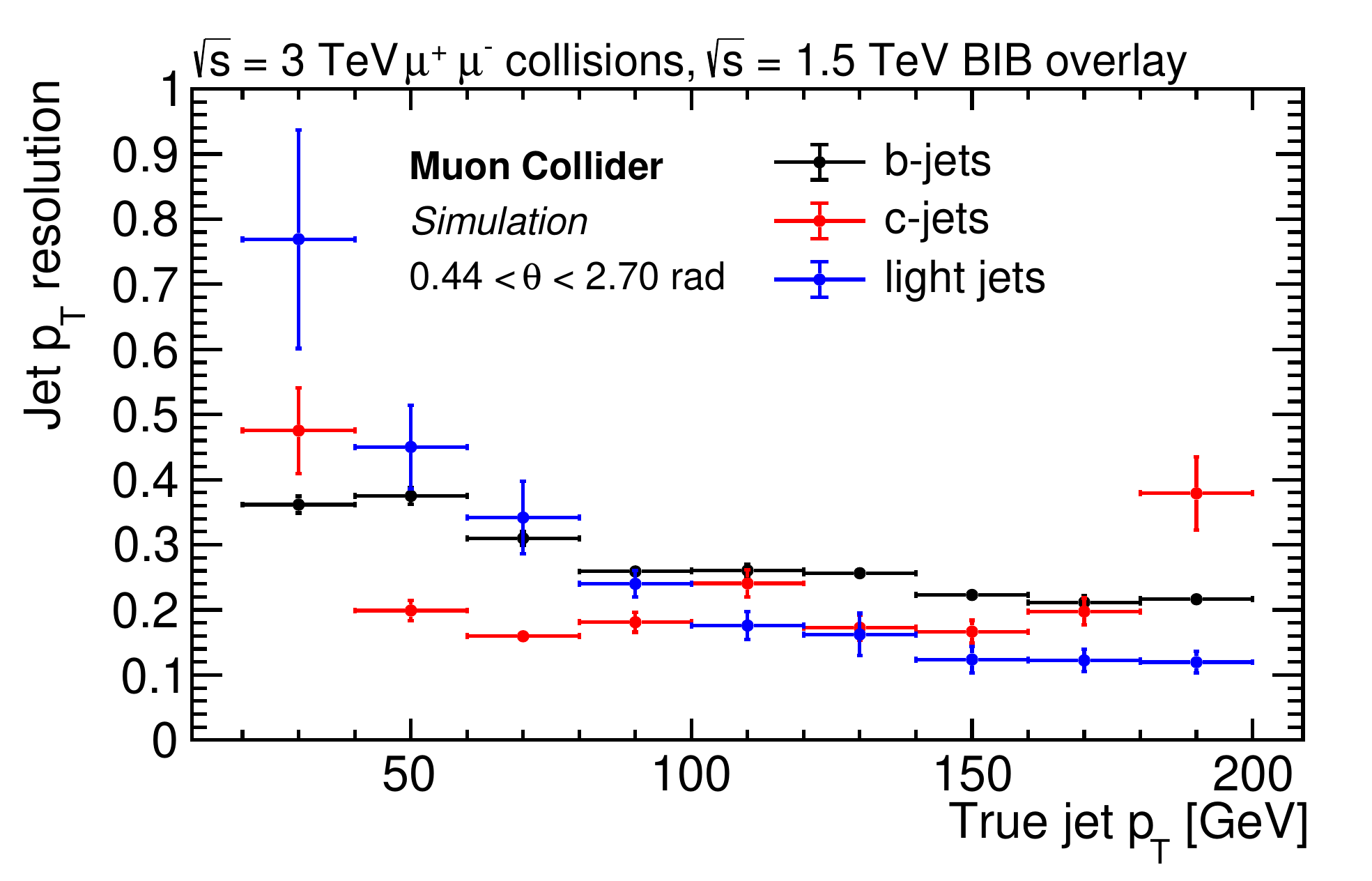}
\caption{Jet $p_{\rm T}$ resolution as a function jet $p_{\rm T}$ for $b$-jets, $c$-jets and light jets in the central region $0.44<\theta<2.70$. The differences between the jet flavours are mainly due to different jet $\theta$ distributions in the three samples.}
\label{fig:jet_flavour}
\end{figure}

The jet $p_{\rm T}$ resolution as a function of the true jet $p_{\rm T}$ is shown in Figure~\ref{fig:jet_flavour} for different jet flavours. The resolution goes from 35$\%$ for jet $p_{\rm T}$ around 20~GeV to 20$\%$ for high jet $p_{\rm T}$.

\begin{figure}[t]
\centering
\includegraphics[width=0.48\textwidth]{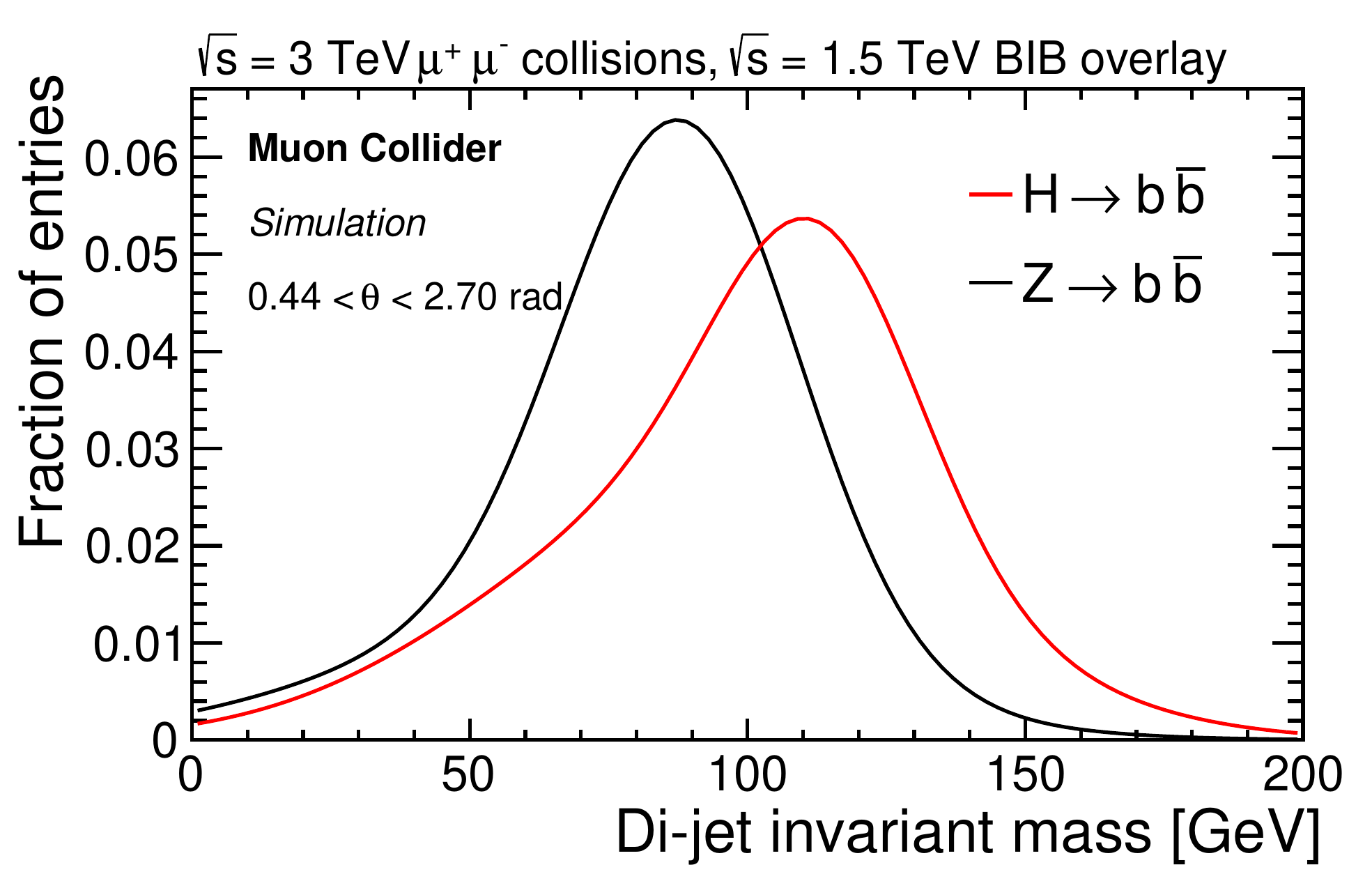}
\caption{Fitted dijet invariant mass distributions for $H \rightarrow b \bar{b}$ and $Z \rightarrow b \bar{b}$. The distributions are normalised to the same area.}
\label{fig:hbb}
\end{figure}

Simulated event samples of $H \rightarrow b \bar{b}$ and $Z \rightarrow b \bar{b}$ are used to evaluate the dijet invariant mass reconstruction.
The invariant mass separation between these two processes is of paramount importance for physics measurements at the muon collider. In this study both jets are required to have $p_{\rm T}>40$~GeV and $0.44<\theta<2.70$. The distributions for the two processes are fitted with double Gaussian functions, and the shapes are compared in Figure~\ref{fig:hbb}. A relative width, defined as the standard deviation divided by the average value of the mass distribution, of 27$\%$(29$\%$) for $H \rightarrow b \bar{b}$($Z \rightarrow b \bar{b}$) is found. 

\subsubsection*{Future prospects on jet reconstruction}
\label{sec:future_jets}
\begin{figure}[t]
    \centering
    \includegraphics[width=0.48\textwidth]{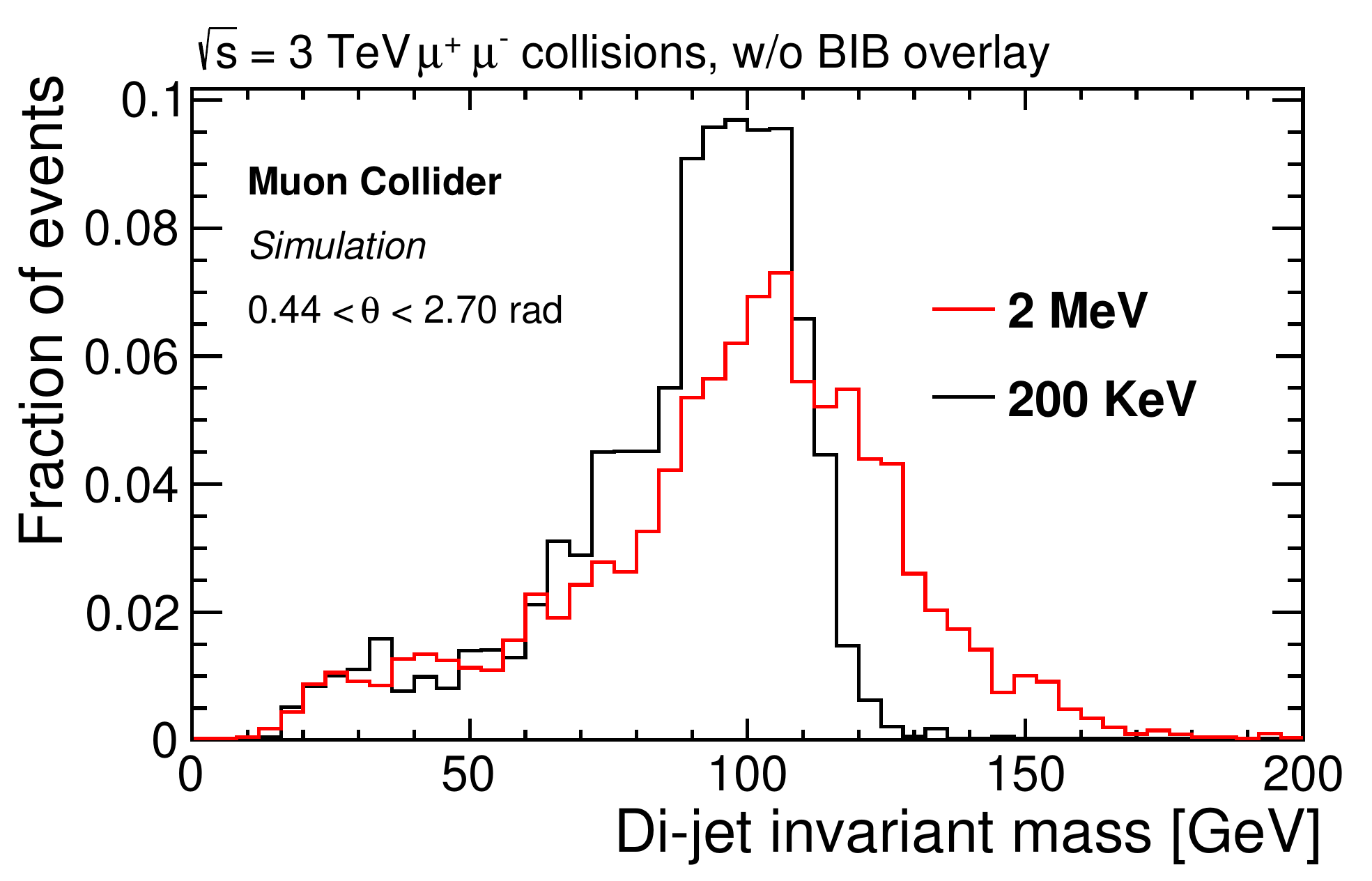}
    \caption{$H \rightarrow b \bar{b}$ dijet invariant mass, reconstructed without the presence of the BIB and with 2~MeV and 200~keV calorimeter hit energy thresholds. 
    }
    \label{fig:h_2_vs_1}
\end{figure}

\begin{figure}[t]
    \centering
    \includegraphics[width=0.5\textwidth]{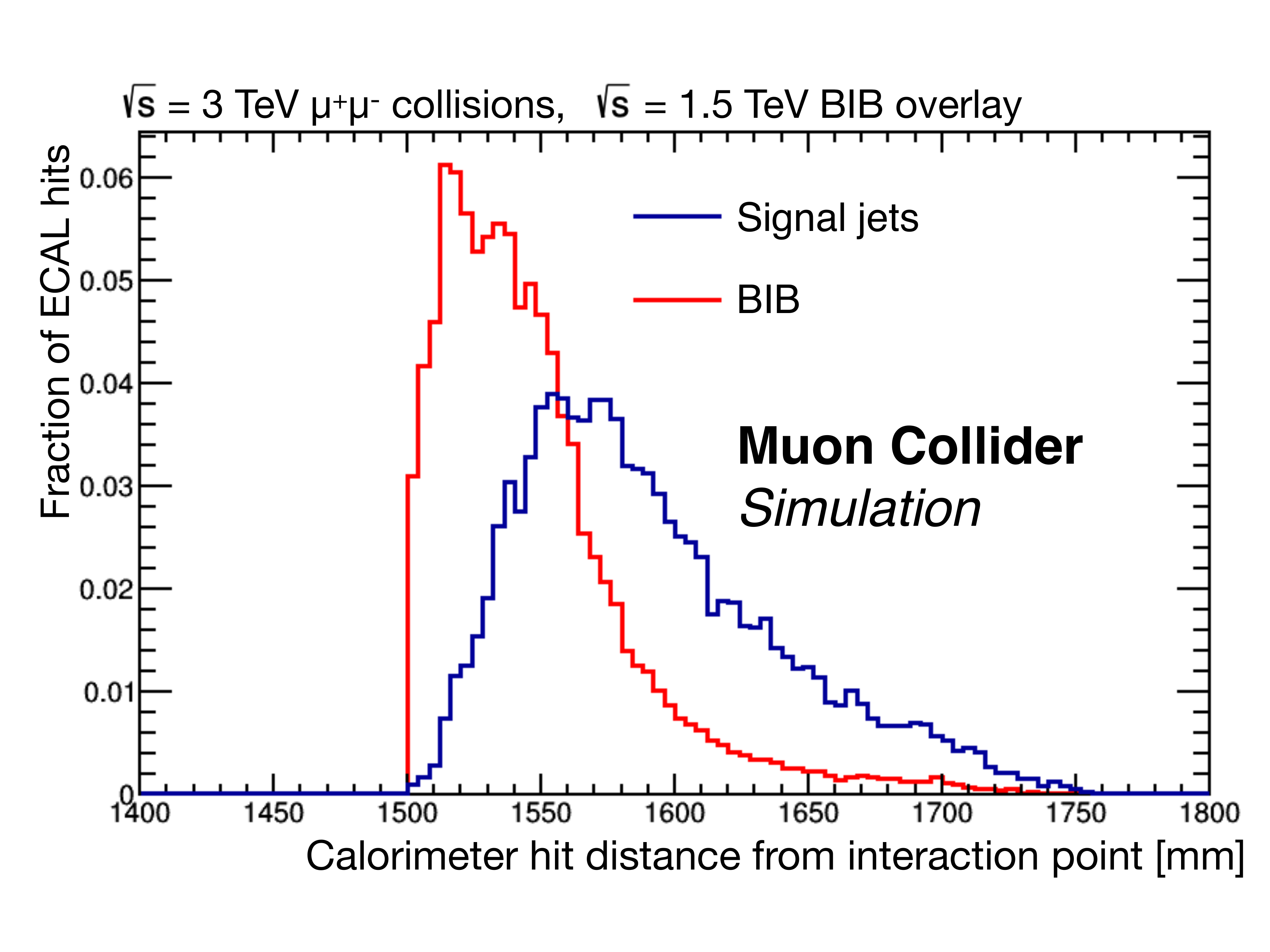}
    \caption{Distribution of the ECAL barrel hits distance from the interaction point (weighted for the hit energy), for both $b$-jets and BIB. Both distributions are normalised to the same area.}
    \label{fig:calo_hit}
\end{figure}

Several ongoing studies are aimed at improving the jet reconstruction performance targeting several aspects, such as:
\begin{itemize}
    \item \textbf{track filter}: the track filter has a different impact in the central and the forward region, in particular the efficiency in the forward region is lower. An optimised selection will be defined to mitigate the efficiency loss in the forward region;
    \item \textbf{cell energy threshold}: the hit energy threshold has been set to the relatively high value of 2~MeV, as a compromise between computing time and jet reconstruction performance. This is a major limitation in the jet performance as can be seen in Figure~\ref{fig:h_2_vs_1}, where the $H \rightarrow b \bar{b}$ dijet invariant mass, reconstructed without the BIB overlay, is compared between 2~MeV and 200~keV thresholds. Reducing this threshold is not an easy task, given the large number of calorimeter hits selected from the BIB that contaminate the jet reconstruction. 
    To tackle this problem an optimised algorithm should be developed: as an example thresholds that depend on the sensor depth could by applied, since the longitudinal energy distribution released by the BIB is different from the signal jets as shown in Figure~\ref{fig:calo_hit}. A generalisation of this idea could be the application of a multivariate-algorithm trained to select signal hits and reject BIB hits;
    \item \textbf{fake jet removal}: the fake jet removal applied in this study has an impact in reducing the jet efficiency in the forward region. Moreover this issue is highly dependent from the calorimeter thresholds. A fake removal tool based on machine learning and with jet sub-structure observables as input could be developed to solve this task.
\end{itemize}
    
\FloatBarrier
\subsubsection*{Jet flavour identification}

The $b$-jet identification algorithm described in this section relies on the reconstruction of the secondary vertices that are compatible with the decay of the heavy-flavour hadron in the geometrical proximity of a reconstructed jet.
 
For the secondary vertices reconstruction, tracks are reconstructed with the Conformal Tracking with Double Layers algorithm described in Section~\ref{sec:trk-roi}, within regions of interest defined by cones with $\Delta R=0.7$ around the jet axes. 

\subsubsection*{Secondary vertex tagging}
\label{sv_tag_algo}

The vertexing algorithm employed for the primary and secondary vertex reconstruction is described in detail in Ref.~\cite{Suehara:2015ura}. In order to reduce the amount of combinatorial tracks due to BIB, cuts are applied to the reconstructed tracks used as input to the algorithm. 
The algorithm proceeds as follows:
\begin{enumerate}
\item \textbf{primary vertex finding}: tracks with $|d_{0}| \leq \qty{0.1}{\mm}$ and $|z_{0}| \leq \qty{0.1}{\mm}$ are selected and used as inputs to the primary vertex fitter. Furthermore, each track is required to have at least four hits in the vertex detector in order to reduce the number of BIB tracks;
\item \textbf{tracks selection for secondary vertex finder}: tracks not used to reconstruct the primary vertex are used as input to the secondary vertex fitter. Figure~\ref{fig:b_c_vs_bib} shows the distributions of the total number of hits in the vertex detector for BIB tracks and for tracks matched at Monte Carlo level with particles generated by $b$ or $c$ hadrons decays. A minimum number of 4 hits in the vertex detector is required in order to keep most of the the tracks from $b$ and $c$ hadrons decay, while rejecting most of BIB tracks. 

\begin{figure}[t]
\centering
\includegraphics[width=0.48\textwidth]{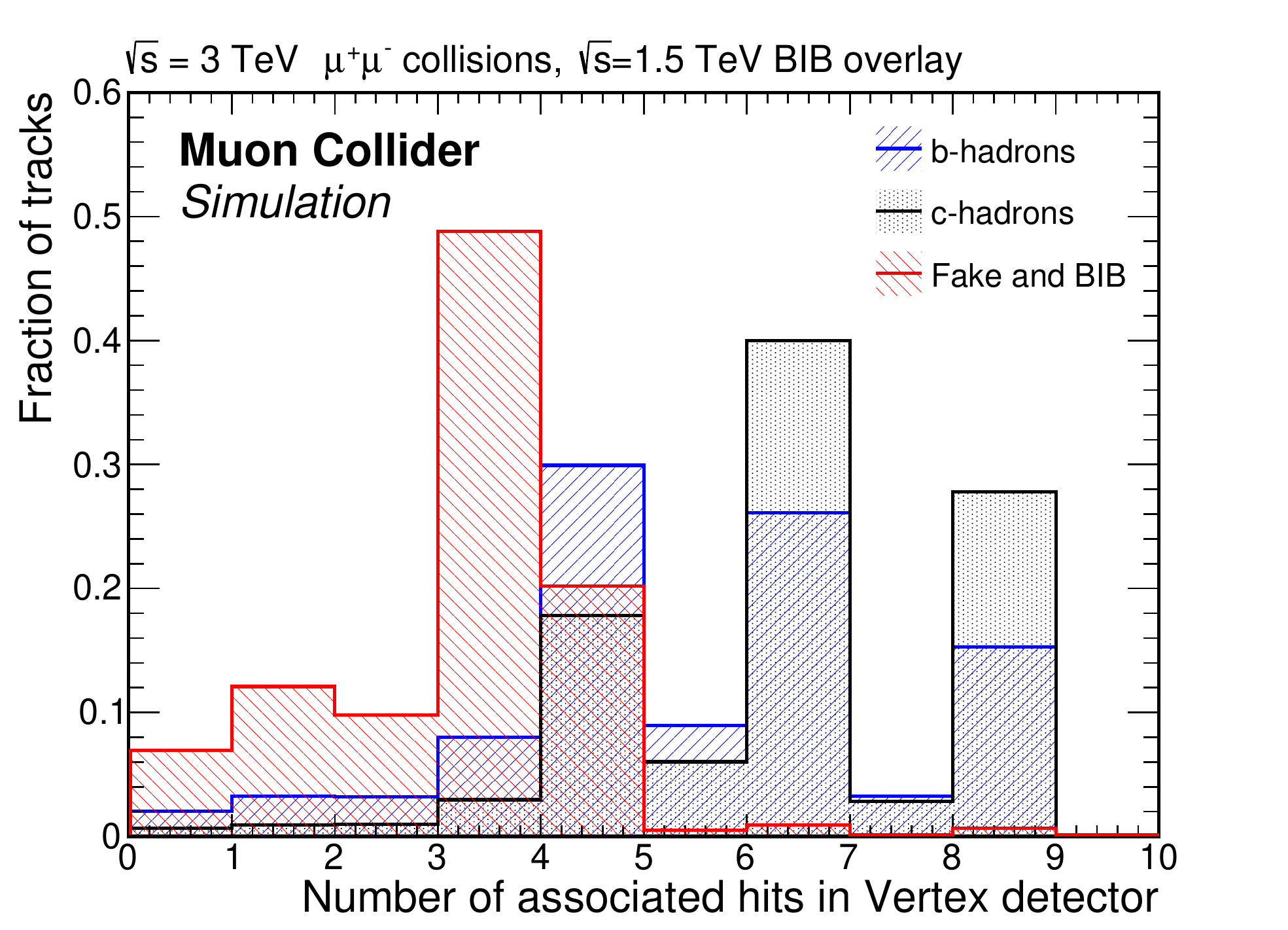}
\caption{Number of hits in the vertex detector of BIB combinatorial tracks (red) and of tracks matched with Monte Carlo truth particles coming from $b$ or $c$ hadrons (blue and black respectively) decay. The distributions are normalised to unit area.}
\label{fig:b_c_vs_bib}
\end{figure}

Further selections on the track $p_{\rm T}$, the maximum track $z_0$ and $d_0$, and the $d_0$ and $z_0$ errors are applied in order to further reduce the amount of BIB tracks. As an example, Figure~\ref{fig:b_c_vs_bib_pt} shows the $p_{\rm T}$ distributions for tracks from $b$- or $c$-hadrons decays and BIB. By requiring $p_{\rm T} > 0.8$~GeV about 80\% of the BIB tracks are rejected, while retaining an efficiency of approximately 85--90\% for the tracks from $b$- or $c$-hadrons.
The $z_0$ distributions for tracks coming from $b$- and $c$-hadrons and BIB are shown in Figure~\ref{fig:b_c_vs_bib_z0}. A requirement of $|z_0| \leq \qty{5}{mm}$ is applied to reject BIB tracks at large $z_0$;
\begin{figure}[t]
\centering
\includegraphics[width=0.48\textwidth]{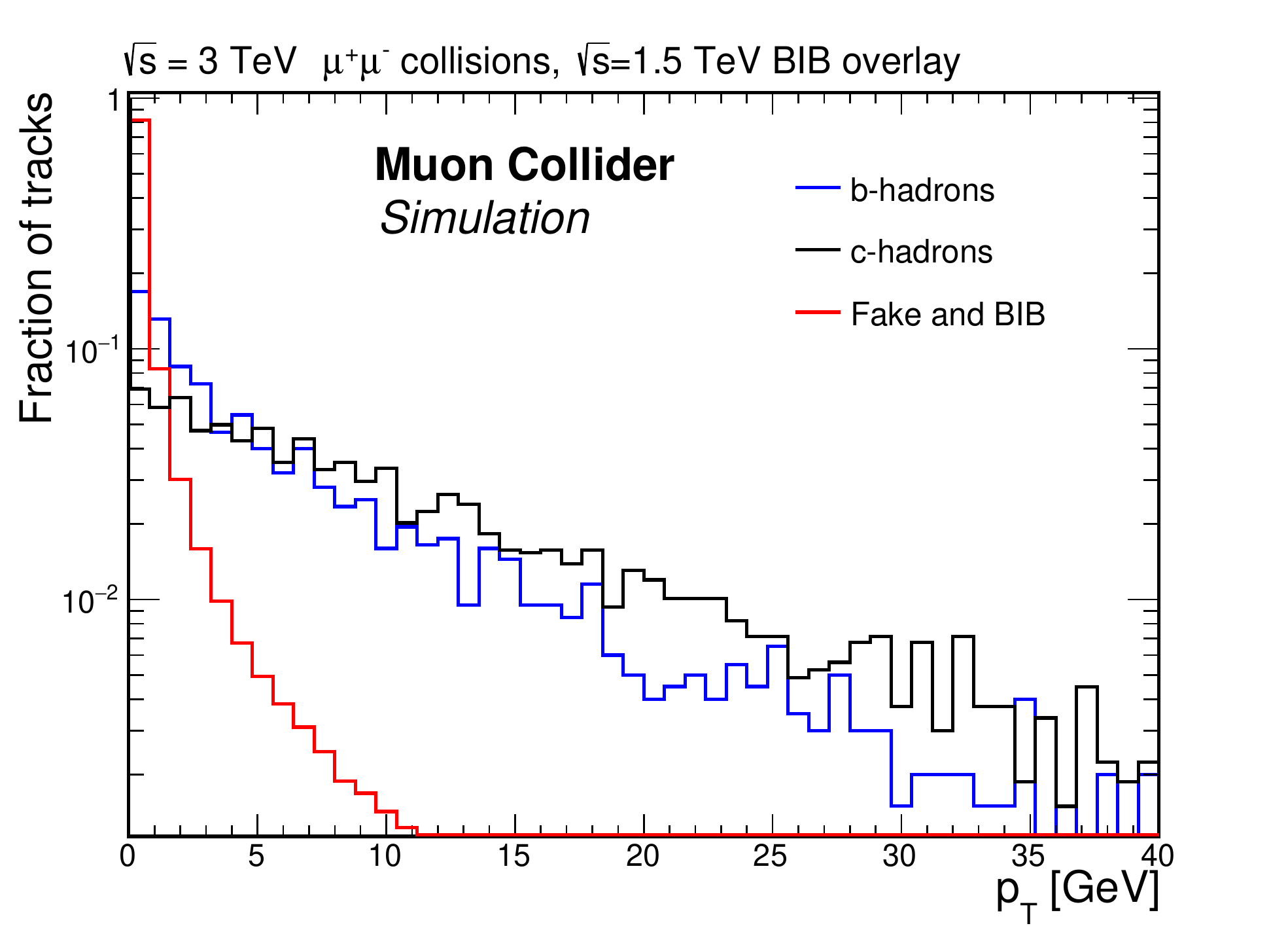}
 \caption{Transverse momentum distribution for tracks coming from $b$ (blue) or $c$ (black) hadrons decay and combinatorial BIB (red) tracks. The distributions are normalised to the number of tracks.}
\label{fig:b_c_vs_bib_pt}
\end{figure}

\begin{figure}[t]
\centering
\includegraphics[width=0.48\textwidth]{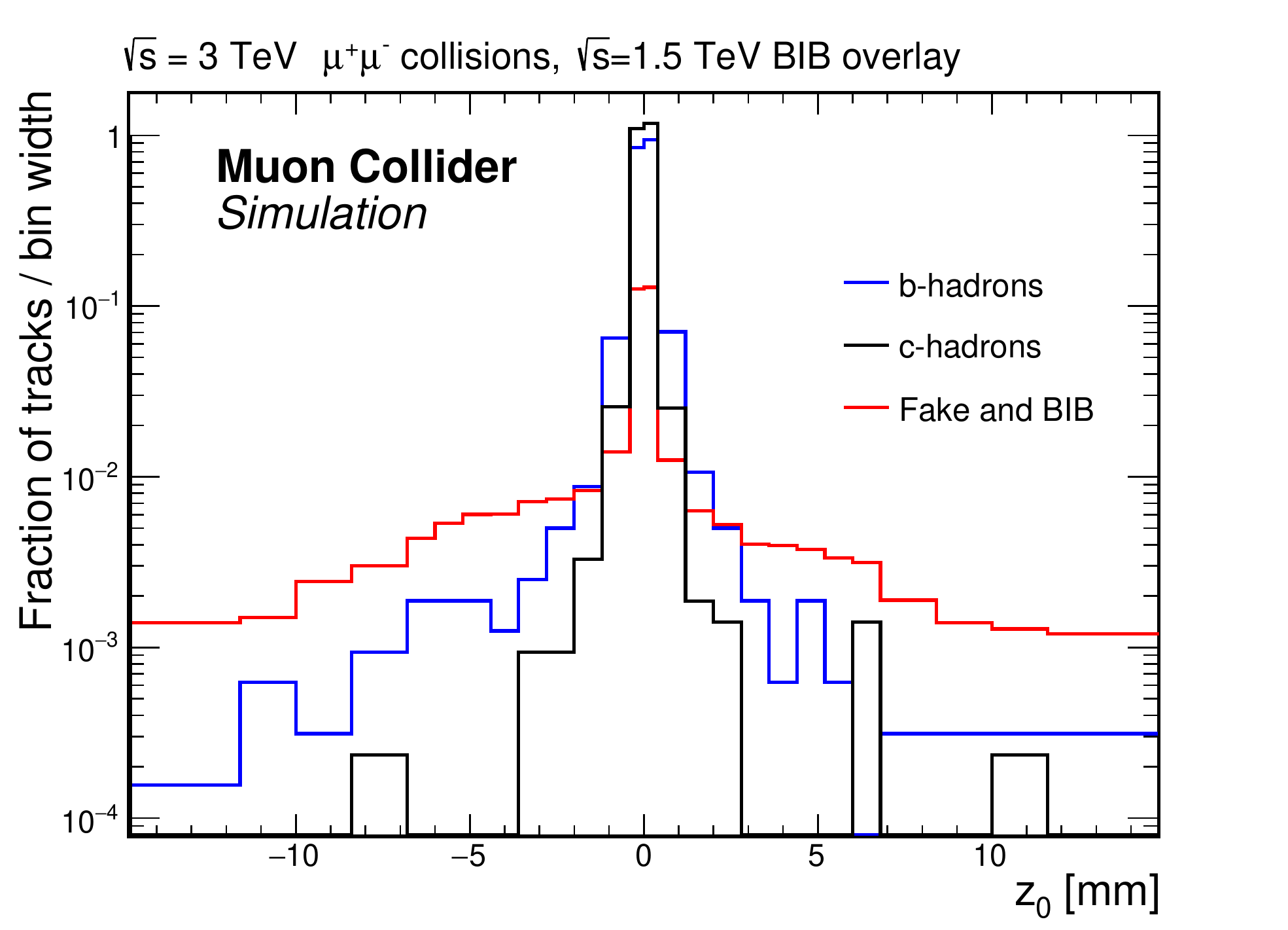}
 \caption{Longitudinal impact parameter distribution for tracks coming from $b$ (blue) or $c$ (black) hadrons decay and combinatorial BIB (red) tracks. The distributions are normalised to the number of tracks.}
\label{fig:b_c_vs_bib_z0}
\end{figure}

\item \textbf{secondary vertex finding}: the tracks passing the requirements are used to build two-tracks vertex candidates, that must satisfy the following requirements: the invariant mass must be below 10 GeV and must be smaller than the energy of each track; the position with respect to the primary vertex must lie in the same side of the sum of the tracks momenta; and the $\chi^2$ of the tracks with respect the secondary vertex position must be below 5. The track pairs are also required not to be compatible with coming from the decay of neutral long lived particles.
Additional tracks are iteratively added to the two-tracks vertices if they satisfy the above requirements. Tracks associated to more than one secondary vertex are assigned to the vertex with the lowest $\chi^2$ and removed from other vertices.
\end{enumerate}

\subsubsection*{Secondary vertex tagging performance}
\label{sv_tagging_eff}

The performance of the identification of jets arising from the hadronisation of $b$ quarks,  $b$-tagging, is evaluated using the secondary vertex tagging as only discriminator.
In order to proceed, a truth-level flavour is associated to the reconstructed jets with the following steps: first, the truth-level jets are matched with the quarks from Monte Carlo to determine its flavour, requiring a distance $\Delta R<0.5$ between the truth-level jet axis and quark momentum. If more than one truth-level jet is found to match with the same quark, the one with the lowest $\Delta R$ is chosen. Then, the flavour of the reconstructed jets is determined by matching them with the truth-level jets, by requiring a $\Delta R<0.5$. If the reconstructed jet does not match any truth-level jet, it is labelled as fake.

\begin{figure}[t]
\centering
\includegraphics[width=0.48\textwidth]{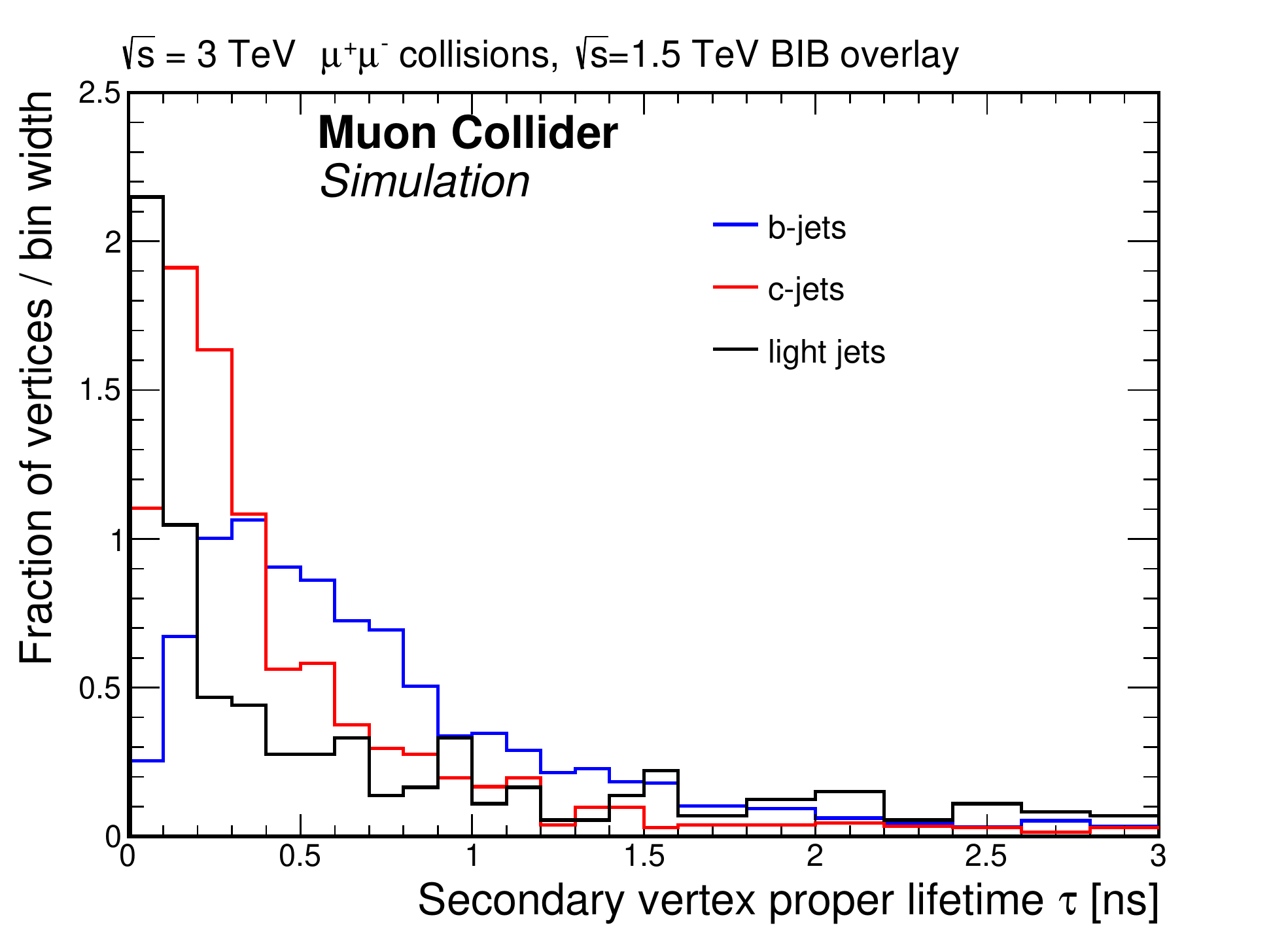}
\caption{Distribution of the secondary vertex proper lifetime for $b$, $c$ and light-tagged jets. Distributions are normalised to the unit area.}
\label{fig:tau}
\end{figure}

The characteristics of secondary vertices inside reconstructed jets have been studied in order to reduce the mis-identification of $c$, light and fake jets.
Figure~\ref{fig:tau} shows the distribution of the secondary vertices proper lifetime ($\tau$) for $b$-jets, $c$-jets and light+fake jets. A cut on $\tau>0.2$~ns rejects $\sim 30 \%$ of both $c$- and light+fake jets, while retaining 90\% of $b$-jets. 
A reconstructed jet is tagged as $b$-jet, if at least one secondary vertex with $\tau>$ 0.2~ns is found inside its cone ($\Delta R <$ 0.5).

The $b$-tagging efficiency is defined as the ratio between the number of tagged truth-matched $b$-jets and the total number of $b$-jets from the MC truth.
Analogously, the mistag of $c$- and light+fake jets is calculated as the ratio between the number of tagged truth-matched $c$ (light+fake)-jets and the total number of $c$ (light+fake)-jets from the MC truth.

\begin{figure}[b]
\centering
\includegraphics[width=0.5\textwidth]{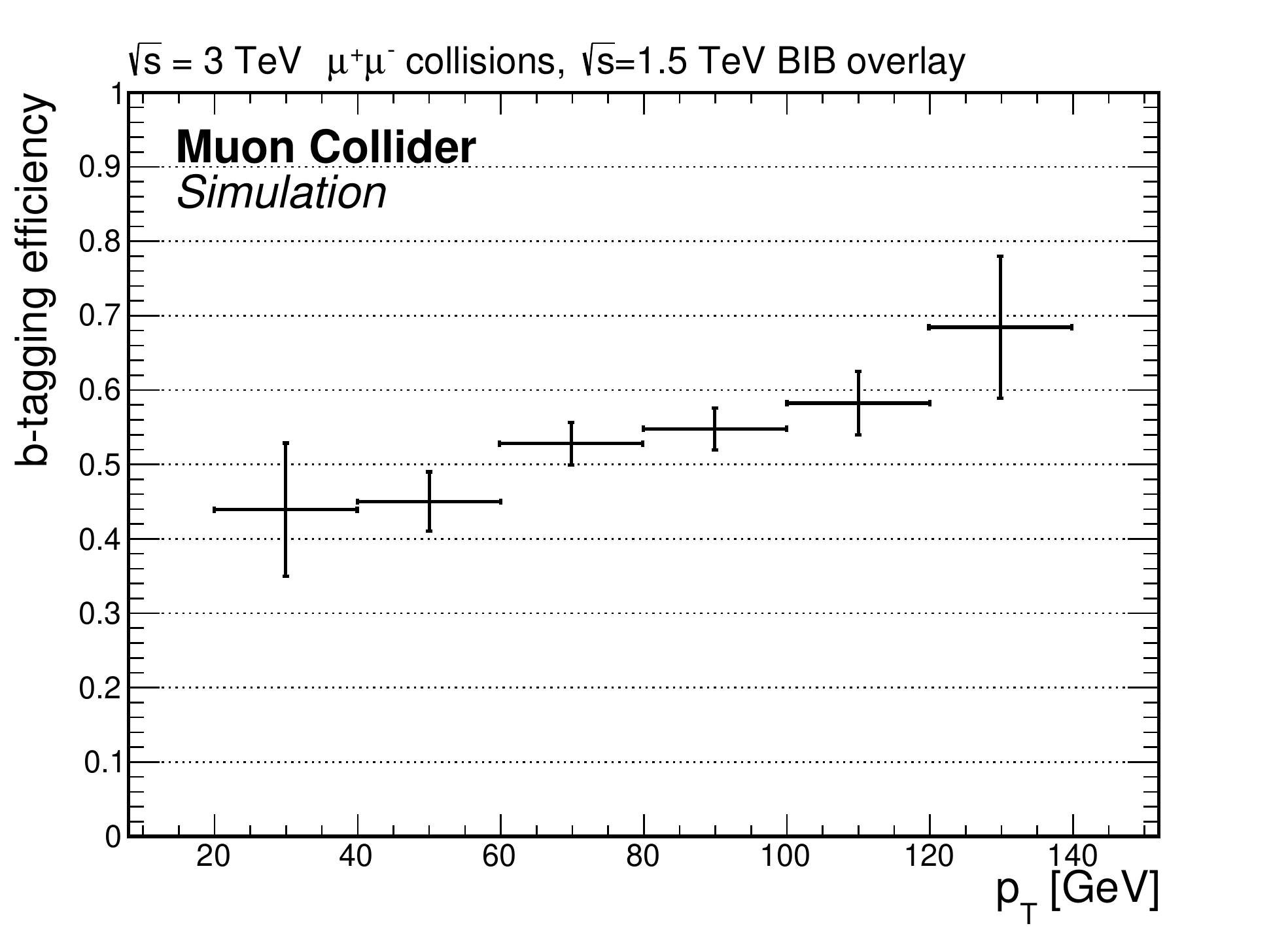}
\caption{Efficiency of the $b$-tagging algorithm as a function of the jet $p_{\rm T}$. The efficiency was evaluated in $b \bar{b}$ dijet events in $\mu^+\mu^-$ collisions at $\sqrt{s}=3$~TeV.}
\label{fig:b_eff_pt}
\end{figure}

\begin{figure}[t]
\centering
\includegraphics[width=0.5\textwidth]{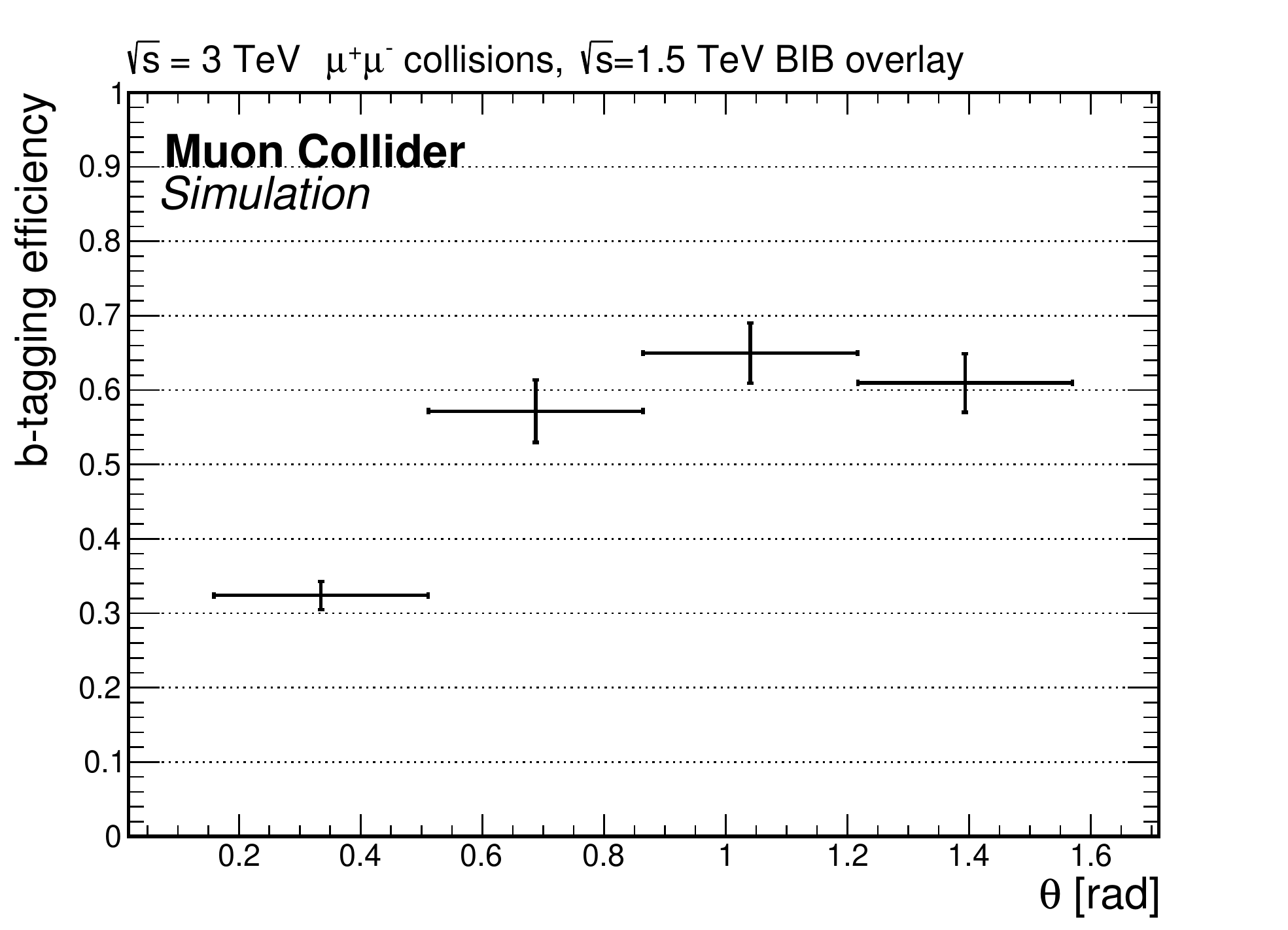}
\caption{Efficiency of the $b$-tagging algorithm as a function of the jet $\theta$. The efficiency was evaluated in $b \bar{b}$ dijet events in $\mu^+\mu^-$ collisions at $\sqrt{s}=3$~TeV.}
\label{fig:b_eff_theta}
\end{figure}

The effect of the Double Layer filter on the secondary vertex finding efficiency has been evaluated reconstructing $b\bar{b}$, $c\bar{c}$ and $q\bar{q}$ dijet samples without BIB, with and without the Double Layer filter. A correction for the impact of the Double Layer filter has been computed as a function of the jet $p_{\rm T}$ and $\theta$ as the ratio of the number of tagged jets without any double layer filtering and the number of tagged jets passing the double layer filter. The final tagging efficiencies are then corrected for this ratio, assuming that its value does not change in the presence of the BIB.
The $b$-tagging efficiency is shown as a function of the jet $p_{\rm T}$ in Figure~\ref{fig:b_eff_pt} and and as a function of the jet $\theta$ in Figure~\ref{fig:b_eff_theta}.

The efficiency is around 50\% at low $p_{\rm T}$ and increases up to 70\% at high $p_{\rm T}$.

\begin{figure}[t]
\centering
\includegraphics[width=0.5\textwidth]{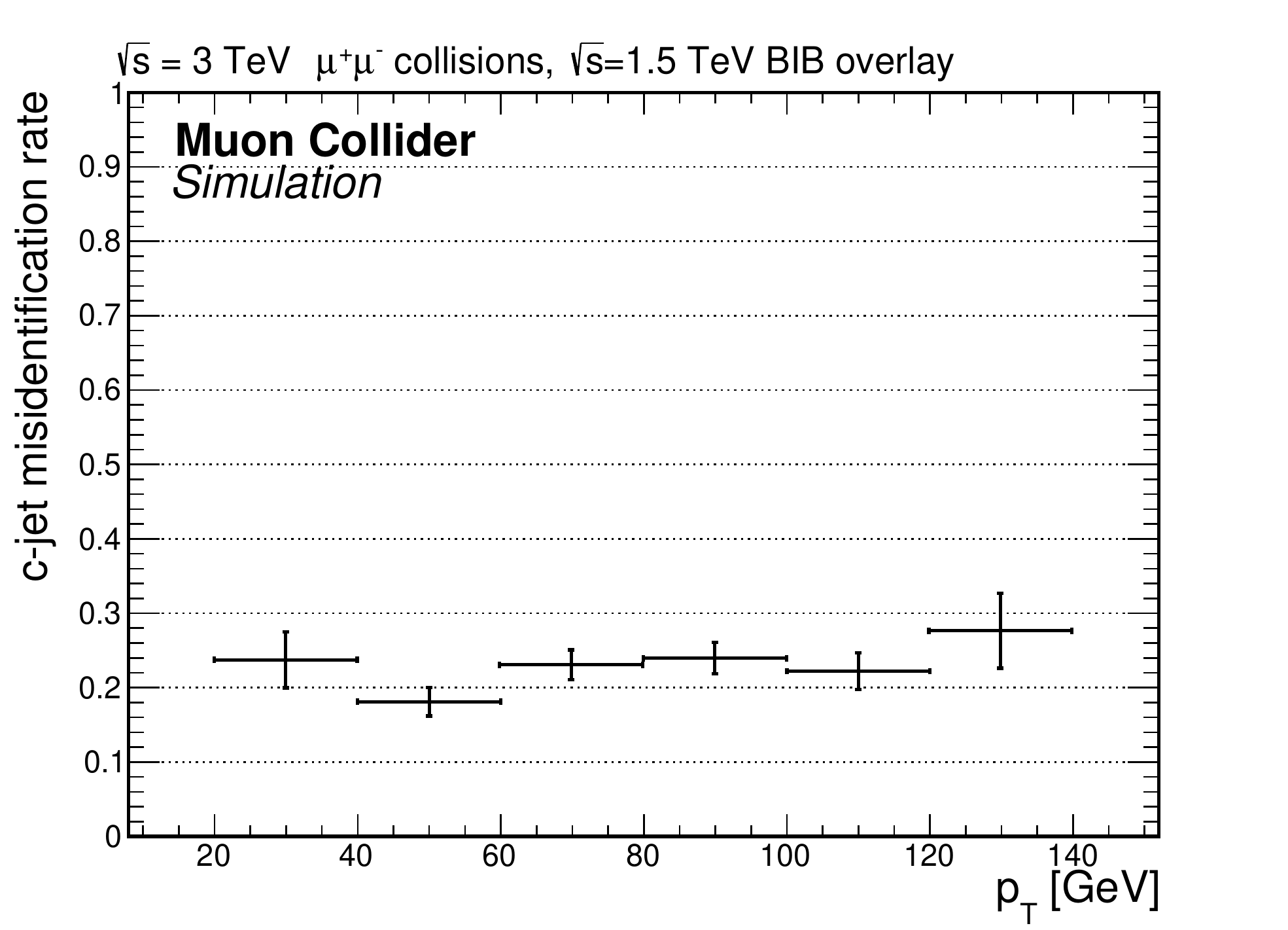}
\caption{Misidentification rate for $c$-jets as a function of the jet $p_{\rm T}$. The rate was evaluated in $c \bar{c}$ dijet events in $\mu^+\mu^-$ collisions at $\sqrt{s}=3$~TeV.}
\label{fig:c_eff_pt}
\end{figure}

\begin{figure}[t]
\centering
\includegraphics[width=0.5\textwidth]{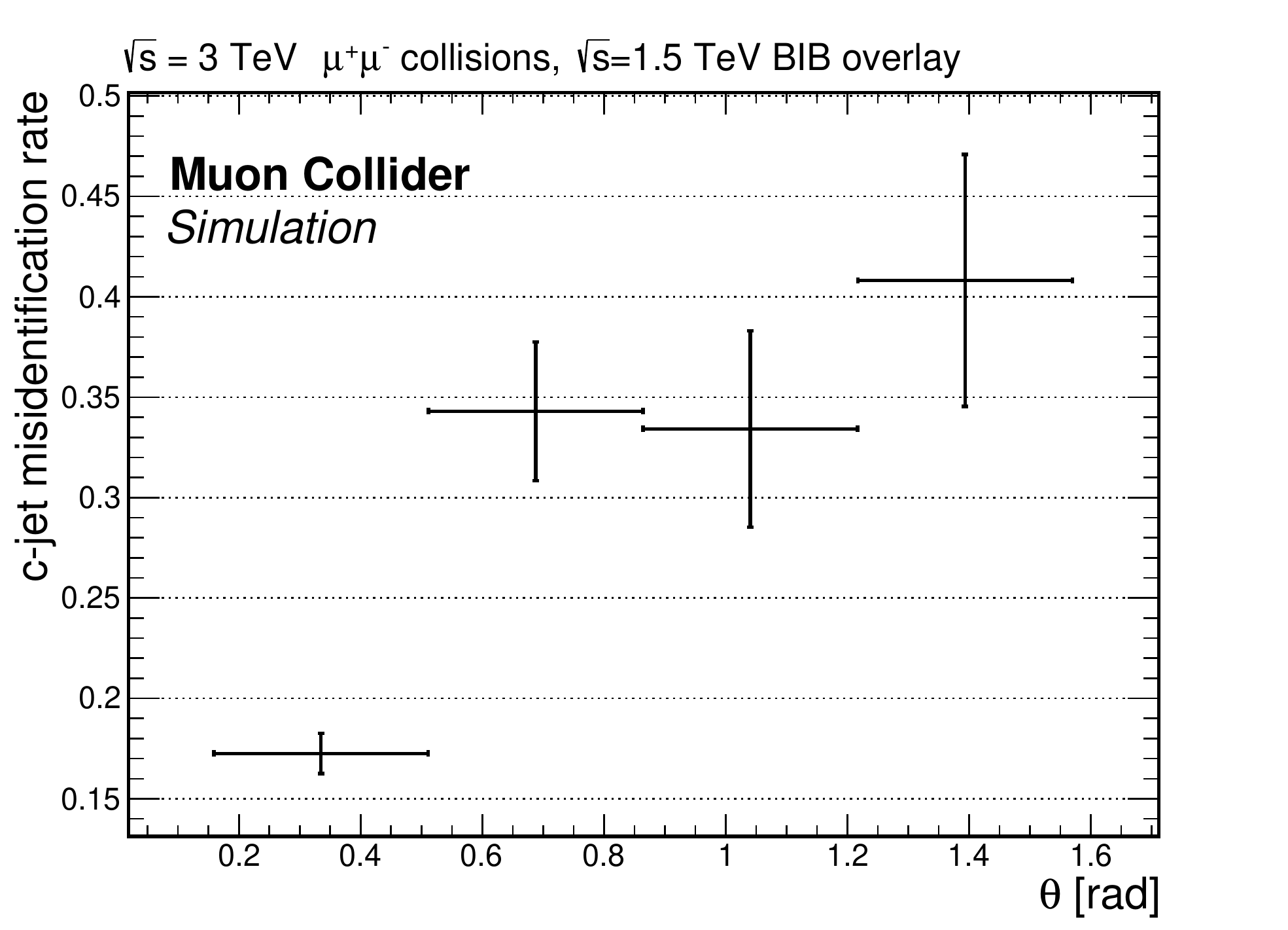}
\caption{Misidentification rate for $c$-jets as a function of the jet $\theta$. The rate was evaluated in $c \bar{c}$ dijet events in $\mu^+\mu^-$ collisions at $\sqrt{s}=3$~TeV.}
\label{fig:c_eff_theta}
\end{figure}

The mis-tagging rate for $c$-jets is shown in Figures~\ref{fig:c_eff_pt} and~\ref{fig:c_eff_theta} and is found to be around 20\%. As for $b$-jet efficiency, the $c$ mistag increases in the central region of the detector.

\begin{figure}[t]
\centering
\includegraphics[width=0.5\textwidth]{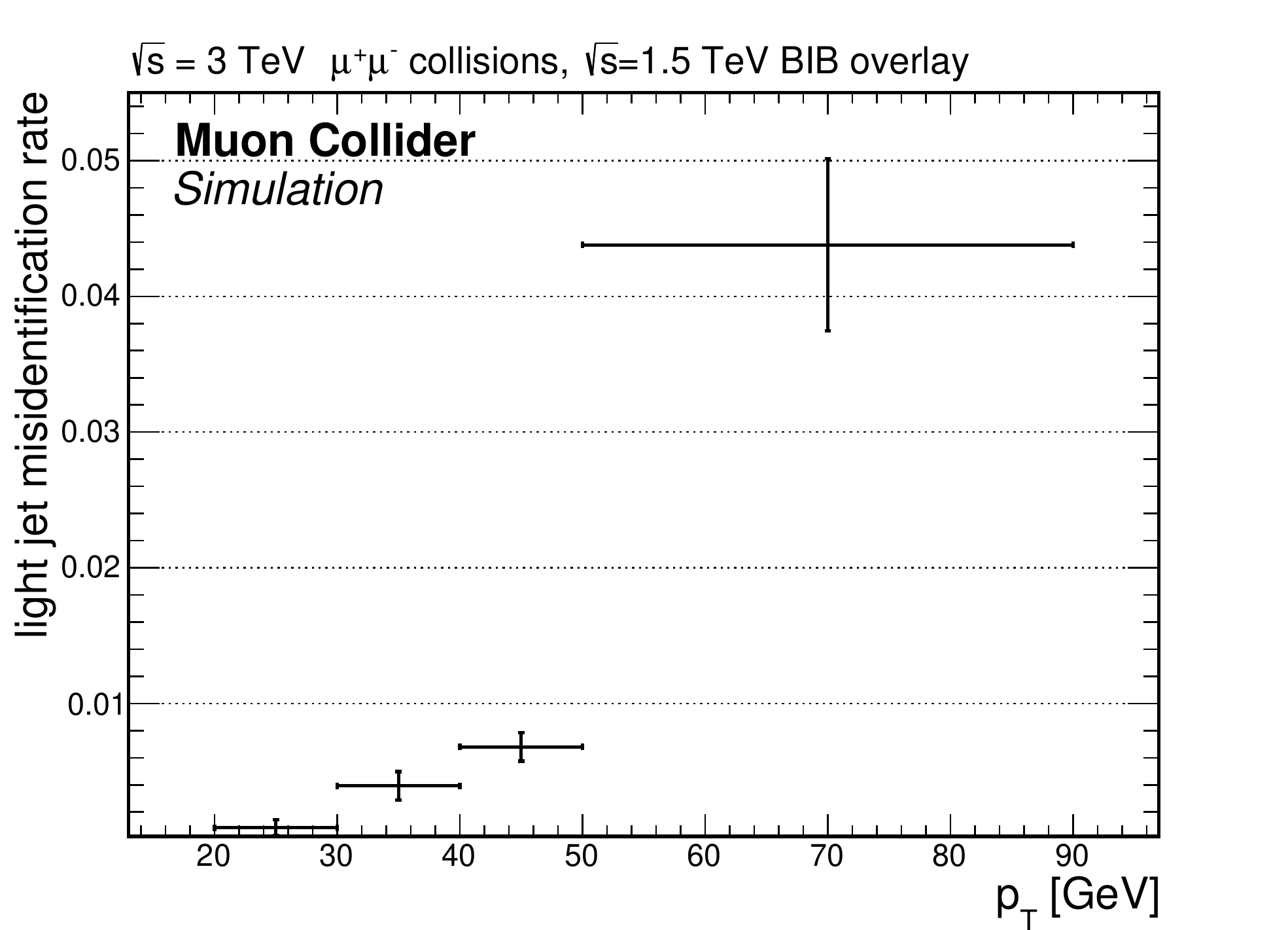}
\caption{Misidentification rate for light jets as a function of the jet $p_{\rm T}$. The rate was evaluated in $q \bar{q}$ dijet events in $\mu^+\mu^-$ collisions at $\sqrt{s}=3$~TeV.}
\label{fig:light_eff_pt}
\end{figure}

\begin{figure}[t]
\centering
\includegraphics[width=0.5\textwidth]{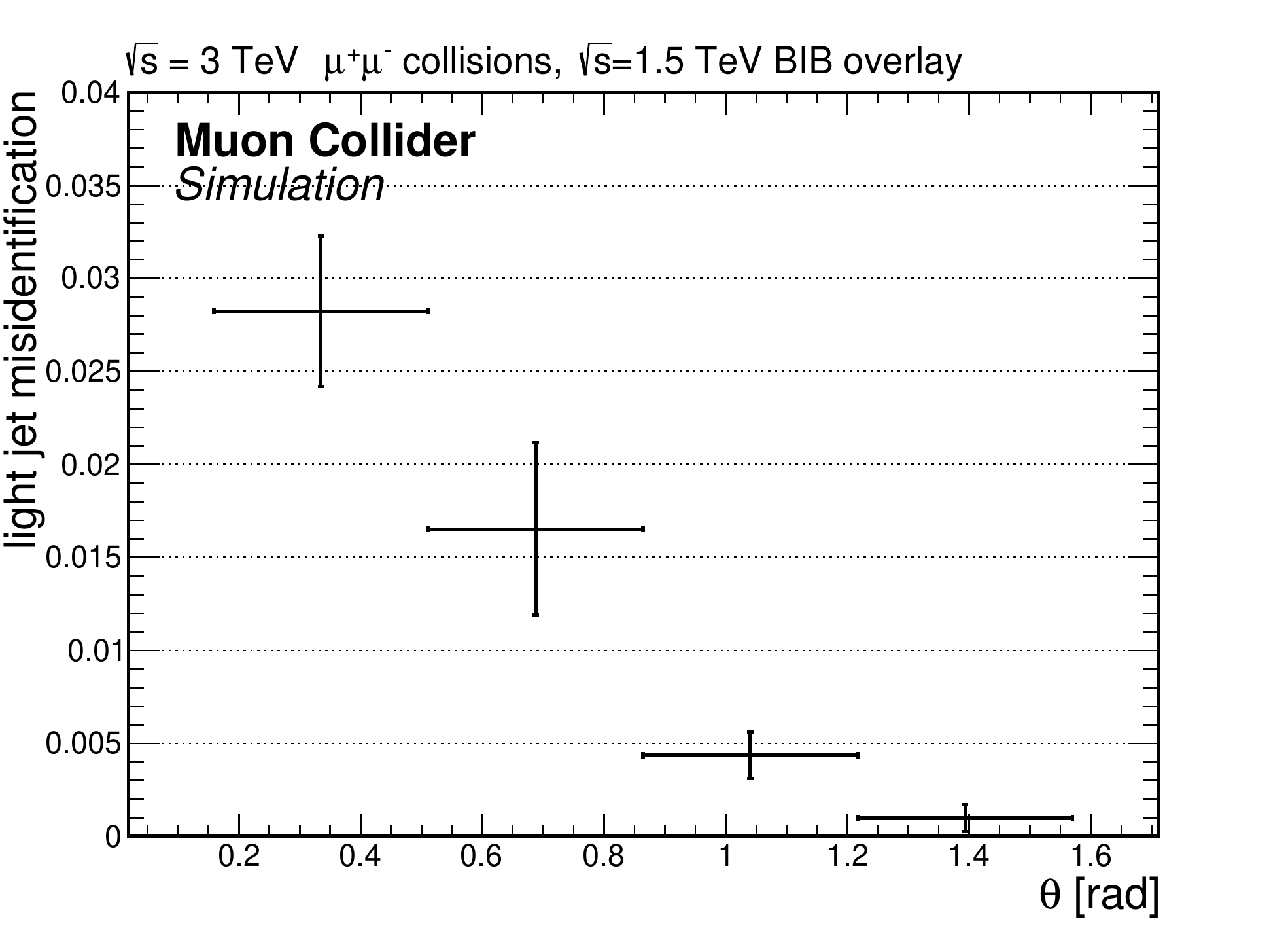}
\caption{Misidentification rate for light jets as a function of the jet $\theta$. The rate was evaluated in $q \bar{q}$ dijet events in $\mu^+\mu^-$ collisions at $\sqrt{s}=3$~TeV.}
\label{fig:light_eff_theta}
\end{figure}

Figures~\ref{fig:light_eff_pt} and~\ref{fig:light_eff_theta} show the mis-tagging rates for the light and fake jets as a function of the jet $p_{\rm T}$ and $\theta$. This rate is found to be lower than $1\%$ for a jet $p_{\rm T}$ below 50 GeV and increase to $5\%$ at higher jet $p_{\rm T}$.

An initial $b$-jet identification algorithm based exclusively on the identification of secondary vertices has been put in place, demonstrating the effectiveness of one of the basic components used in flavour tagging techniques in the complex muon collider environment. 
Further work is ongoing to take advantage of improved tracking methods, such as the CKF algorithm described in Section~\ref{sec:trk-ckf} to improve the inputs to secondary vertex finding, or on the exploitation of additional handles, such as the presence of charged leptons within the jet, or the impact parameter of the track associated to the jet.

\subsubsection*{Photons and electrons}

The photon reconstruction and identification performance of the muon collider detector is assessed in a sample of single photon events. The photons were generated in the nominal collision vertex at the centre of the detector, uniformly distributed in energy between 1 and 1500~GeV, in polar angle between $10^\circ$ and $170^\circ$, and in the full azimuthal angle range. The sample was then processed with the detector simulation and reconstruction software.

Prior to track reconstruction, the tracker hits were processed with the Double Layer filter. Moreover, to get rid of most of the fake tracks due to the spurious hits from the background, a track quality selection is applied before the track refitting step, which requires at least three hits in the vertex detector and at least two hits in the inner tracker.
To reject part of the background hits in the calorimeters, an energy threshold of 2~MeV is applied to both the ECAL and HCAL hits. Photons are reconstructed and identified with the Pandora Particle Flow algorithm~\cite{Thomson:2009rp}.

The energy threshold of the calorimeter hits and the presence of the beam-induced background affect the energy scale of the reconstructed photons.
A correction factor is applied to the reconstructed photon energy to make the detector response uniform as a function of the photon energy and the photon polar angle.
The correction was calculated from the ratio of the reconstructed photon energy with the photon energy at generator level in an independent set of events.

\begin{figure}[t]
\centering
\includegraphics[width=0.5\textwidth]{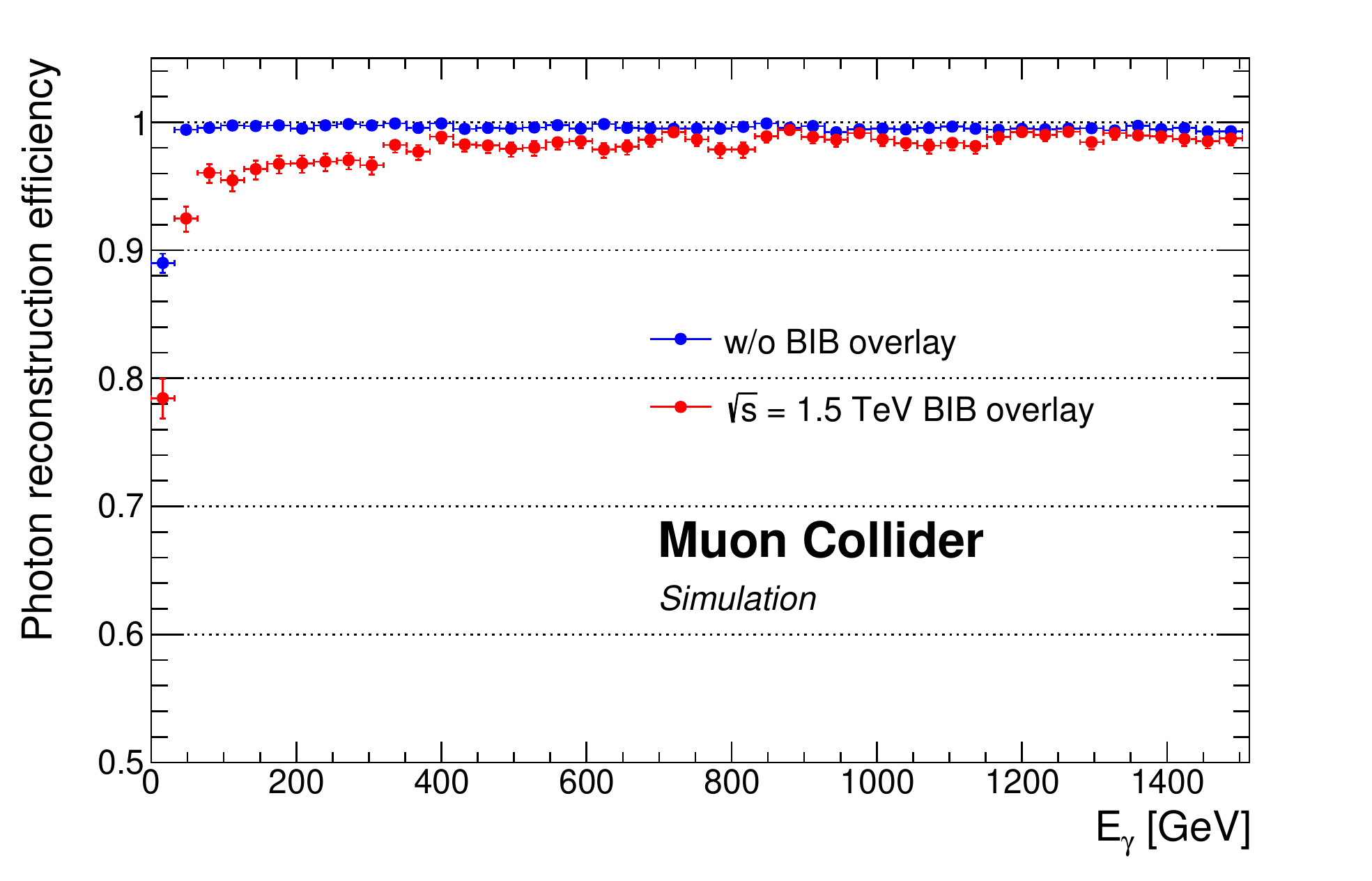}
\caption{Photon reconstruction efficiency as a function of the photon energy.
\label{fig:ph_efficiency}}
\end{figure}

\begin{figure}[t]
\centering
\includegraphics[width=0.5\textwidth]{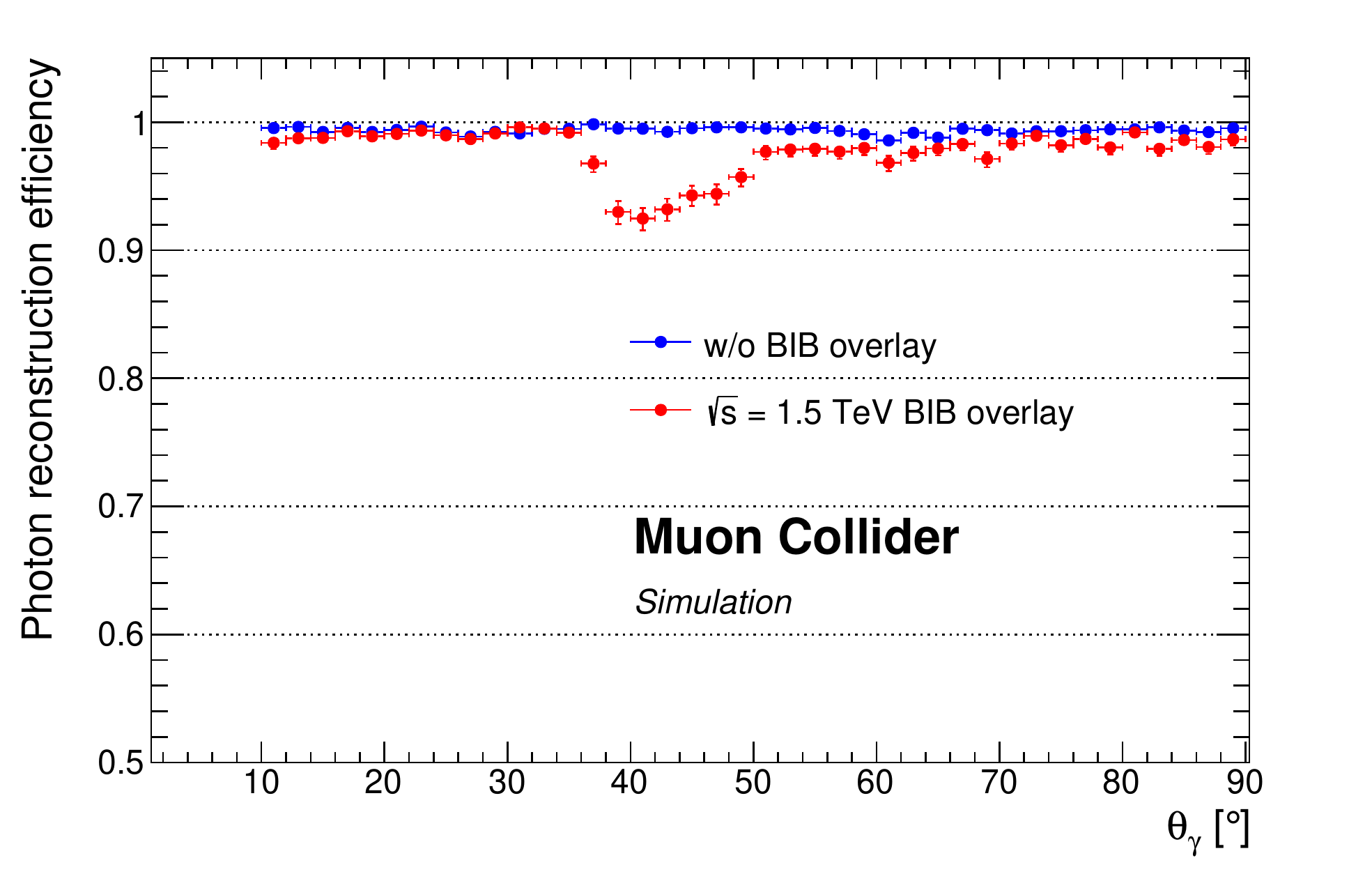}
\caption{Photon reconstruction efficiency as a function of the photon  polar angle $\theta$.
\label{fig:ph_efficiency_theta}}
\end{figure}

Figure~\ref{fig:ph_efficiency} shows a comparison of the photon reconstruction efficiency as a function of the generated photon energy, with and without the BIB overlay. The dependency on the  polar angle $\theta$ is shown instead in Figure~\ref{fig:ph_efficiency_theta}.

The efficiencies are defined as the fraction of generated photons in the range $10^\circ$ and $170^\circ$ that are matched to a reconstructed photon within $\Delta R<0.05$.
A decrease of about 10\% in the reconstruction efficiency in the presence of BIB is observed in the angular region corresponding to the transition between the barrel and endcap calorimeters, and is reflected in the efficiency below \qty{400}{\GeV}. 

\begin{figure}[t]
\centering
\includegraphics[width=0.5\textwidth]{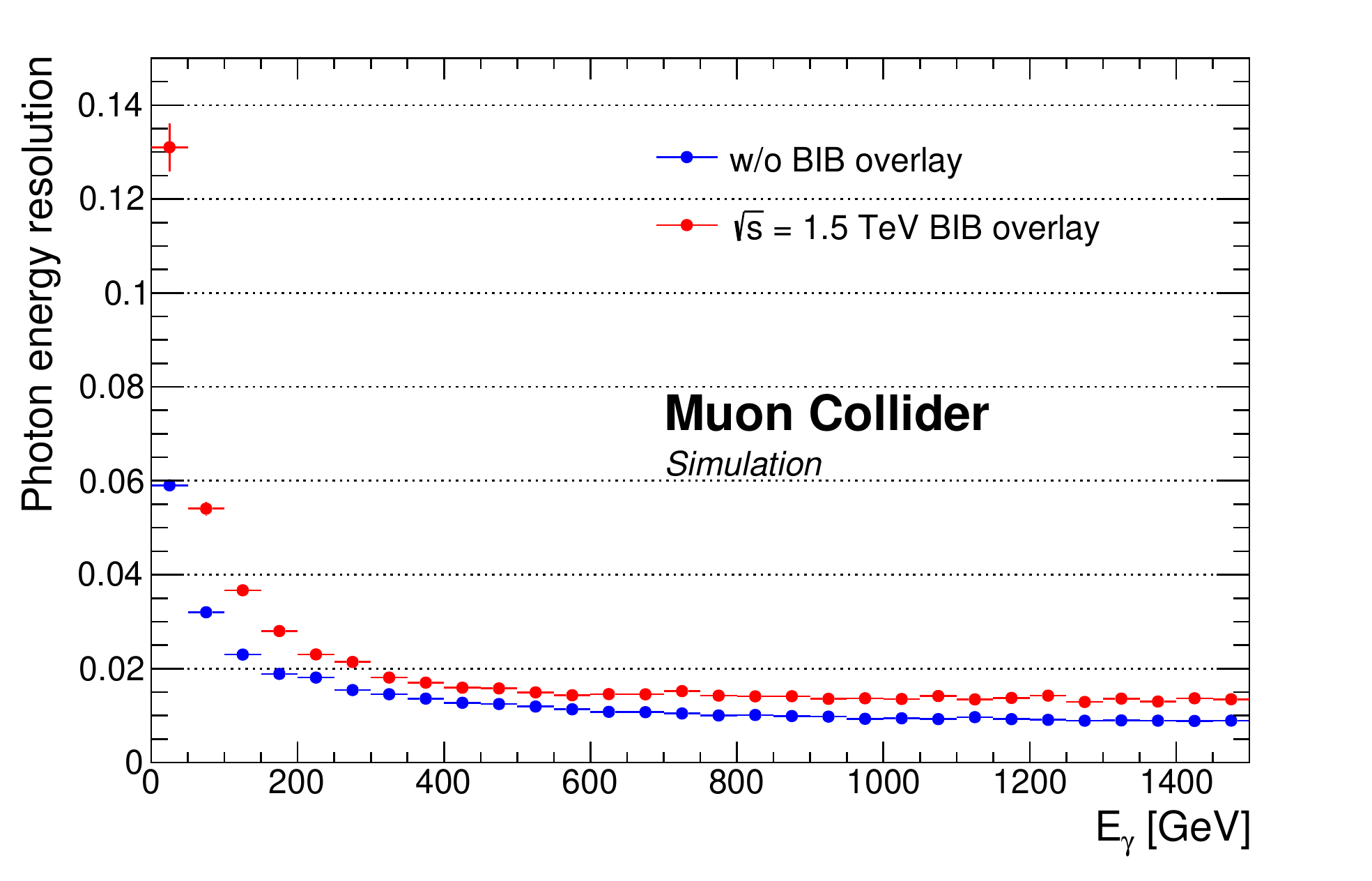}
\caption{Energy resolution of the reconstructed photon as a function of the photon energy.
\label{fig:ph_eneres}}
\end{figure}

\begin{figure}[t]
\centering
\includegraphics[width=0.5\textwidth]{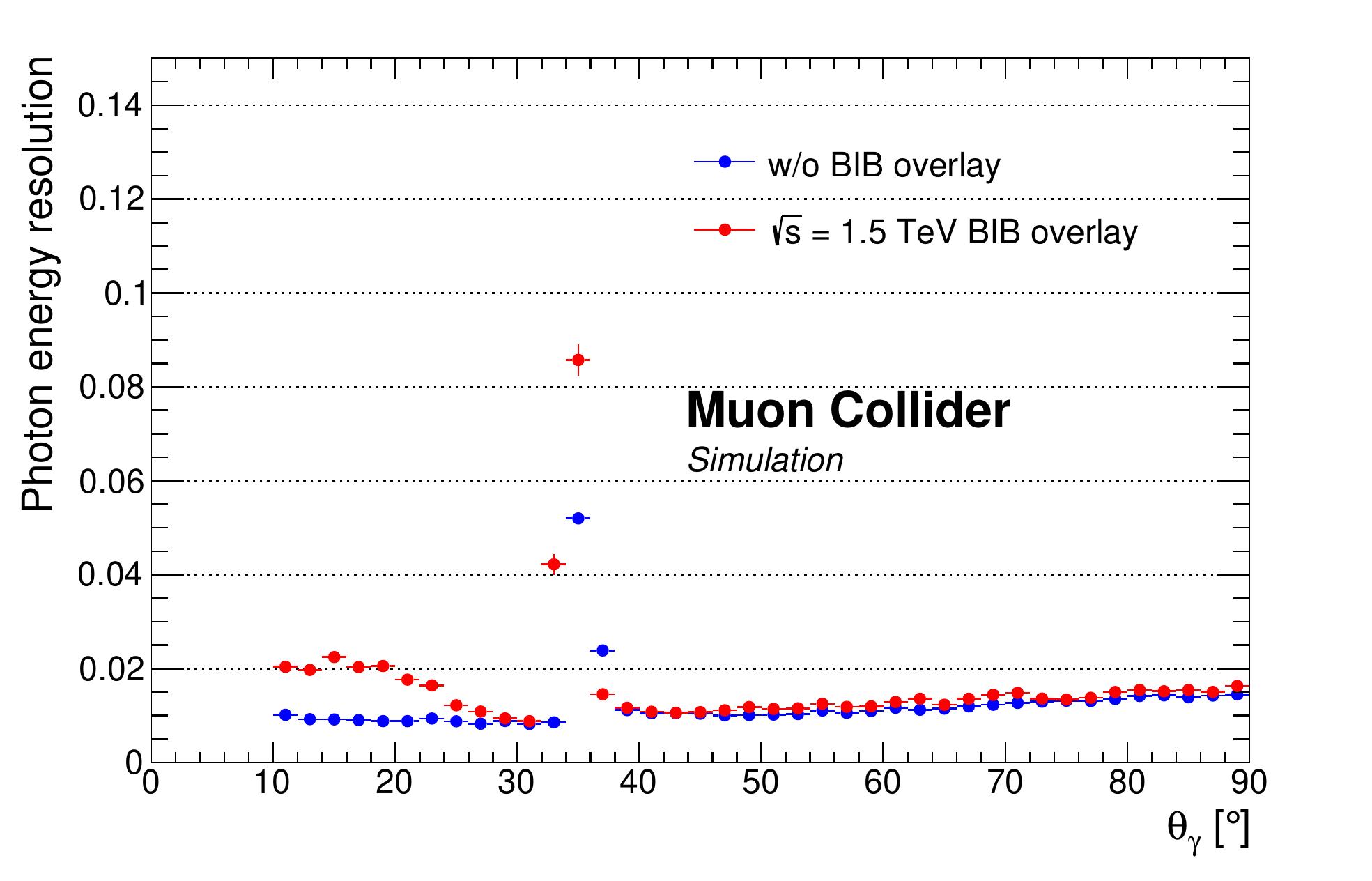}
\caption{Energy resolution of the reconstructed photon as a function of the photon polar angle $\theta$.
\label{fig:ph_eneres_theta}}
\end{figure}

The effect of BIB on the photon energy resolution has also been evaluated, and is shown in Figures~\ref{fig:ph_eneres} and~\ref{fig:ph_eneres_theta} as a function of the energy and polar angle of the photon. The BIB was found to affect more significantly the forward region, where the energy resolution is degraded by a factor of two, and the transition region between the barrel and the endcap calorimeters.

The development of a dedicated algorithm to recover the effects on both the reconstruction efficiency and energy resolution is ongoing. However, the results demonstrate an excellent level of expected performance across the full investigated energy spectrum.

\begin{figure}[t]
\centering
\includegraphics[width=0.5\textwidth]{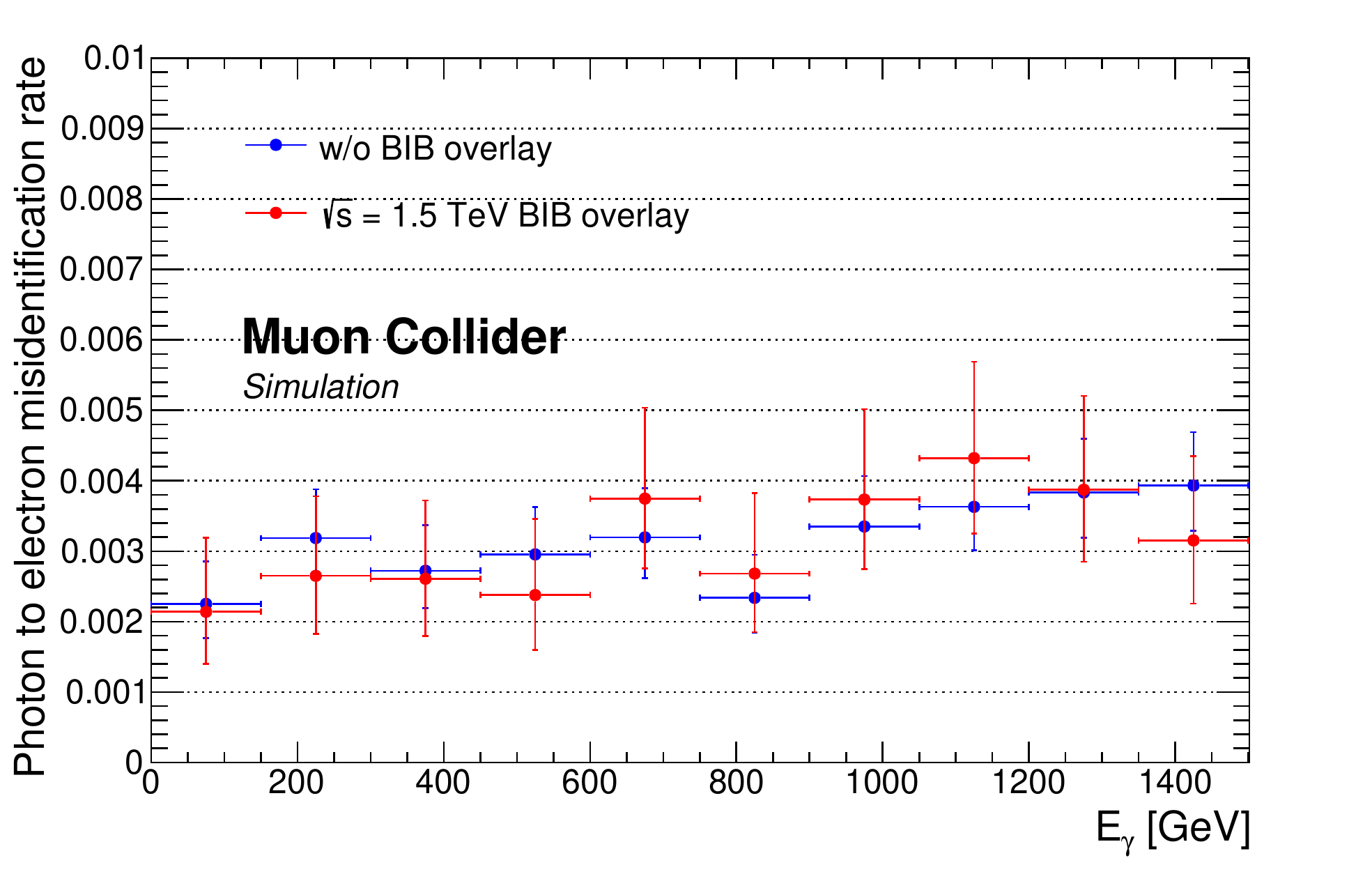}
\caption{Fraction of photons misidentified as electrons as a function of the photon energy. 
\label{fig:ph_elefake}}
\end{figure}

Figure~\ref{fig:ph_elefake} reports the fraction of photons that are reconstructed and identified as electrons. The resulting inefficiency from misidentifications was found to be at the level of 0.3\% and relatively unaffected by the presence of BIB.

The performance of electron reconstruction and identification was studied in single electron events, with the electrons produced at the nominal collision point. The generated electrons are uniformly distributed in energy between 1 and 1500~GeV, in polar angle between $10^\circ$ and $170^\circ$, and in azimuthal angle over the whole range. The sample was then processed with the detector simulation and reconstruction software.

Electrons are identified by means of an angular matching of the electromagnetic clusters with tracks reconstructed with the CKF algorithm, as described in Section~\ref{sec:trk-ckf} in a $R=0.1$ cone.
A Double Layer filter was used to reject BIB hits upstream of the track reconstruction and tracks are required to have $\chi^2/\textrm{ndof} < 10$.
In the presence of the beam-induced background, the energy thresholds of the calorimeter hits play a dominant role for an efficient and precise cluster reconstruction. In this study, a threshold of 5~MeV was set. 

\begin{figure}[t]
\centering
\includegraphics[width=0.48\textwidth]{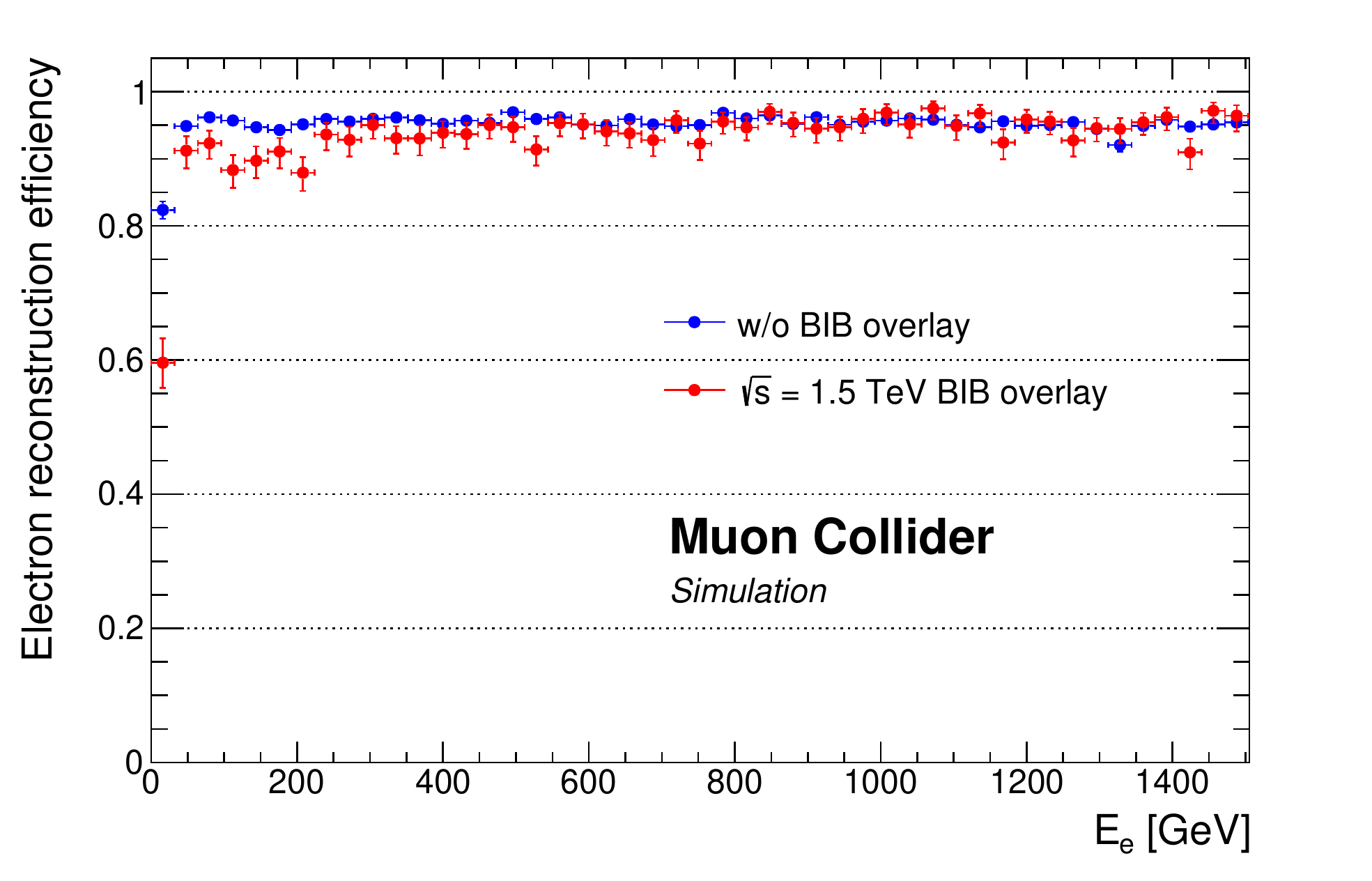}
\caption{Comparison of the electron reconstruction efficiency as a function of the electron energy in the cases of no beam-induced background and with the BIB added to the event.
\label{fig:ele_efficiency}}
\end{figure}

\begin{figure}[t]
\centering
\includegraphics[width=0.48\textwidth]{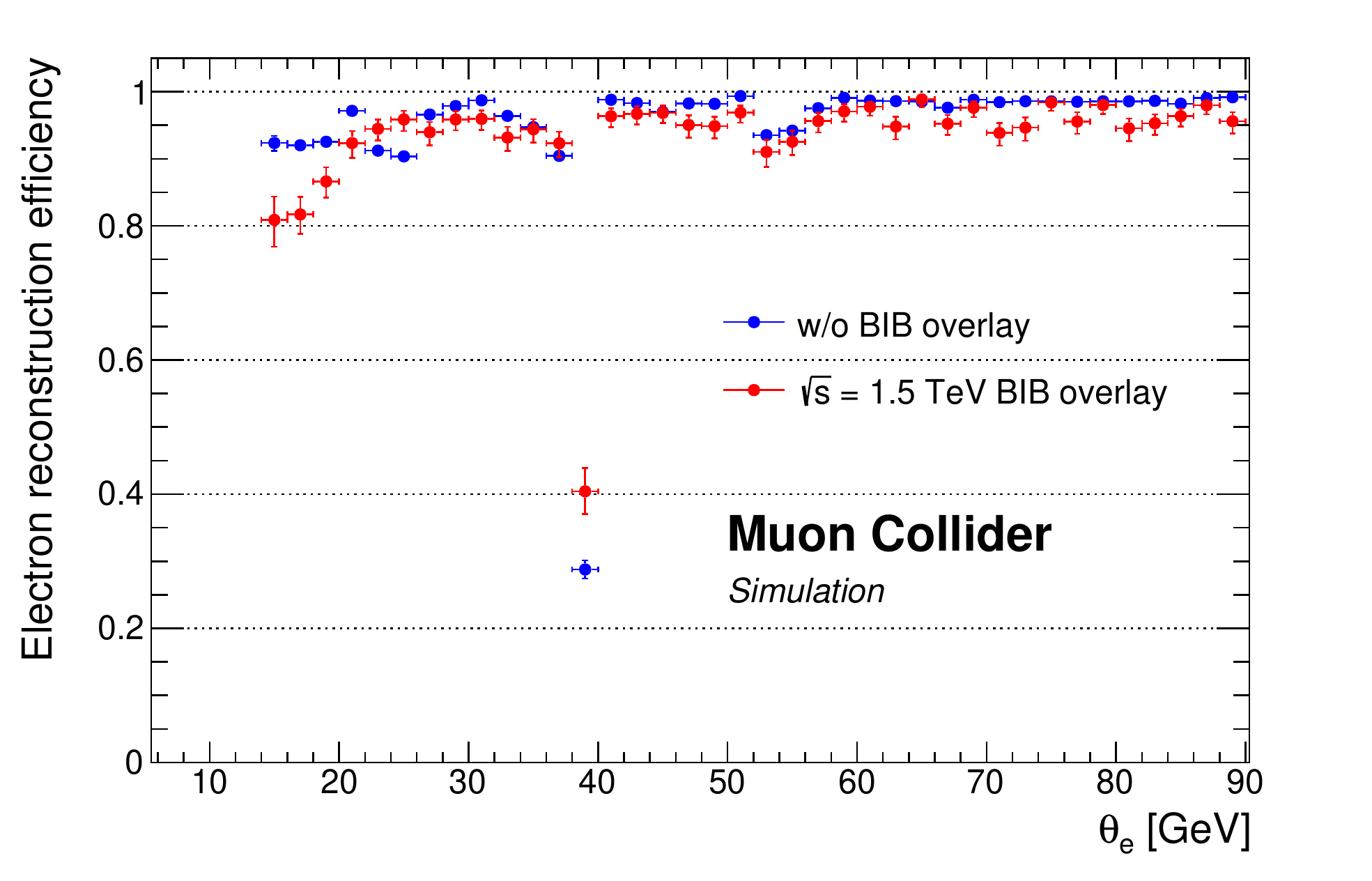}
\caption{Comparison of the electron reconstruction efficiency as a function of the electron polar angle in the cases of no beam-induced background and with the BIB added to the event.
\label{fig:ele_efficiency_theta}}
\end{figure}

The electron reconstruction and identification efficiencies as a function of the electron generated energy and polar angle are shown in Figures~\ref{fig:ele_efficiency} and~\ref{fig:ele_efficiency_theta}.

The efficiency drops at $\theta < 20^\circ$ and $\sim40^\circ$ are caused by tracking inefficiencies related to the application of the Double-Layer filter in the forward region and in the transition region between the VXD barrel and endcap. 

An excellent performance is observed across the investigated energy range. New developments and more sophisticated algorithms will be required to further improve the cluster reconstruction and cluster-to-track association, particularly in the lower energy regime.

\FloatBarrier
\subsubsection*{Muons}

The performance of muon reconstruction and identification was studied in single muon events. The muons were produced at the nominal collision point. The generated muons are uniformly distributed in energy in the range \qty{100}{\MeV}--\qty{50}{\GeV}, in polar angle between $8^\circ$ and $172^\circ$, and in azimuthal angle over the whole range.

The muons are reconstructed and identified with the Pandora Particle Flow algorithm~\cite{Thomson:2009rp}, by matching track in the inner detector reconstructed from the Combinatorial Kalman Filter approach with clusters of hits in the muon system. A cluster is defined as a combination of hits (one hit per layer) inside a cone extending to the neighbouring layers. A detailed description of the muon reconstruction algorithm is reported in~\cite{Linssen:2012hp}.

The cluster finding efficiency, defined as the ratio between generated particles associated with a cluster and total generated particles, was found to be higher than 99\% for $p_{\rm T}>\qty{10}{\GeV}$ and higher than 98\% for $\qty{8}{\degree} < \theta < \qty{172}{\degree}$.

\begin{figure}[t]
\center
\includegraphics[width=0.47\textwidth]{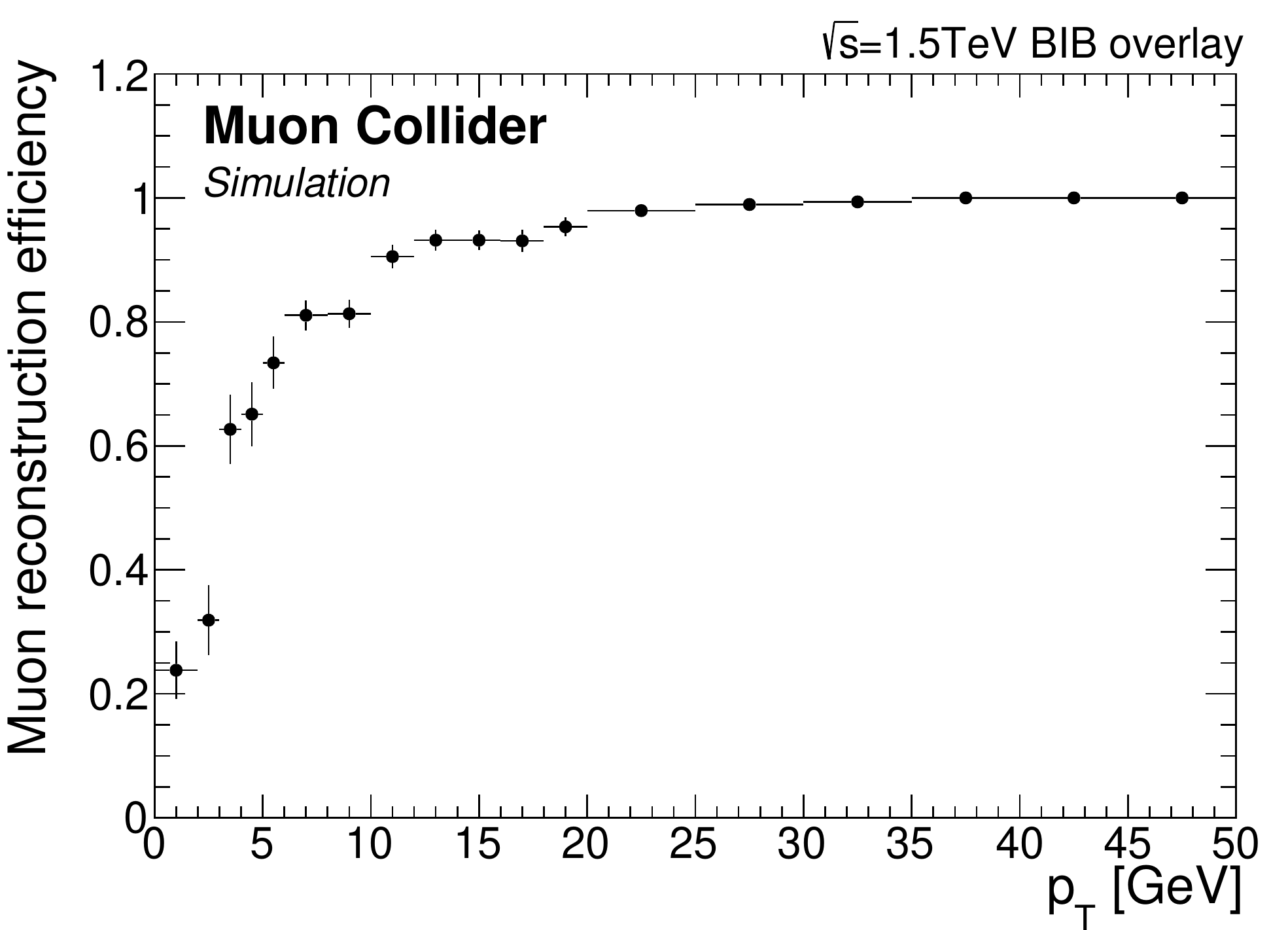}
\caption{Muon reconstruction efficiency as a function of transverse momentum in a sample of single muon events.}
\label{fig:muon_efficiency}
\end{figure}

\begin{figure}[t]
\center
\includegraphics[width=0.47\textwidth]{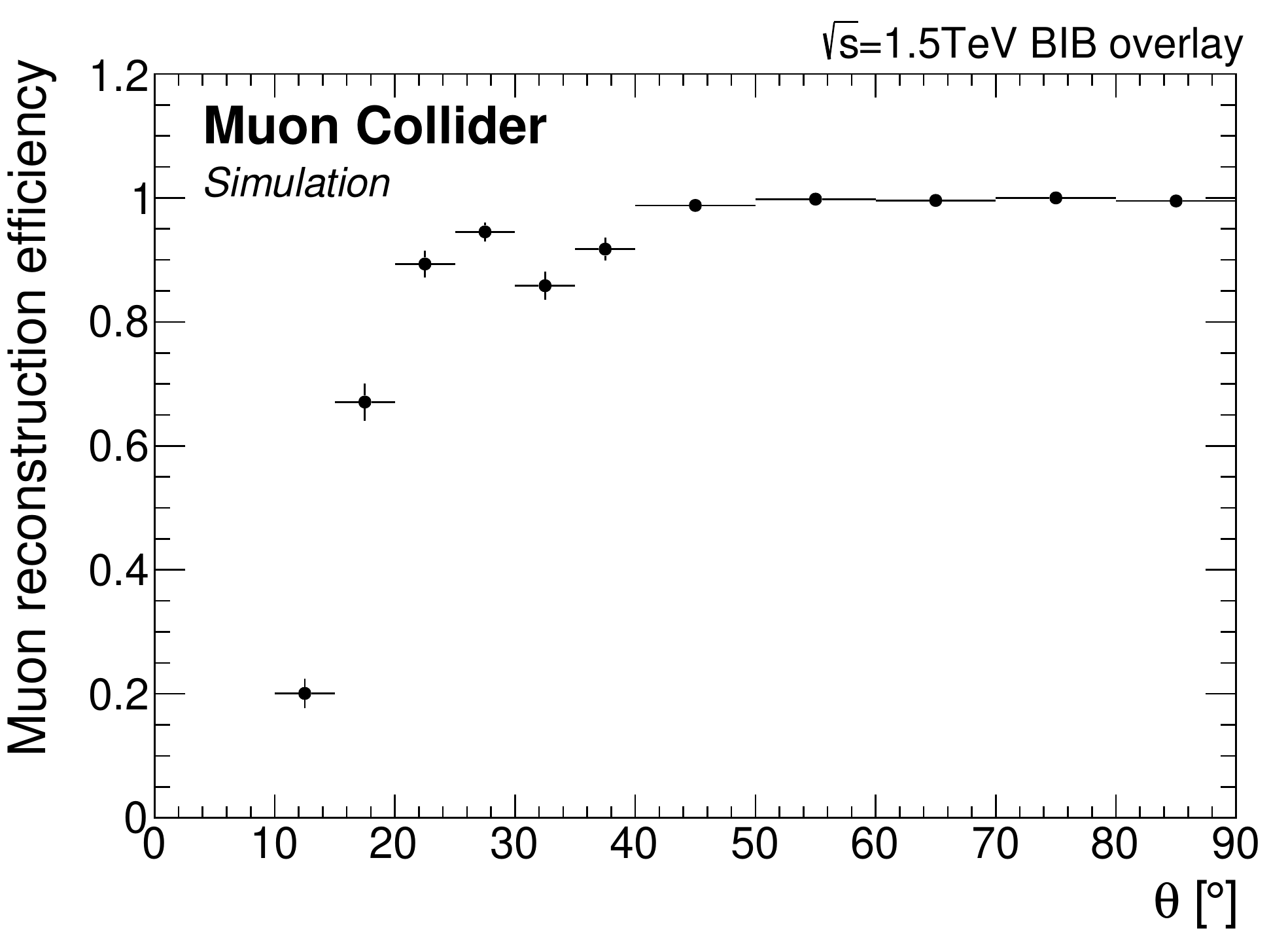}
\caption{Muon reconstruction efficiency as a function of the polar angle $\theta$ in a sample of single muon events.}
\label{fig:muon_thetaefficiency}
\end{figure}

Figures~\ref{fig:muon_efficiency} and~\ref{fig:muon_thetaefficiency} show the muon reconstruction efficiency respectively as a function of the muon $p_{\rm T}$ polar angle $\theta$.

\begin{figure}[t]
\center
\includegraphics[width=0.5\textwidth]{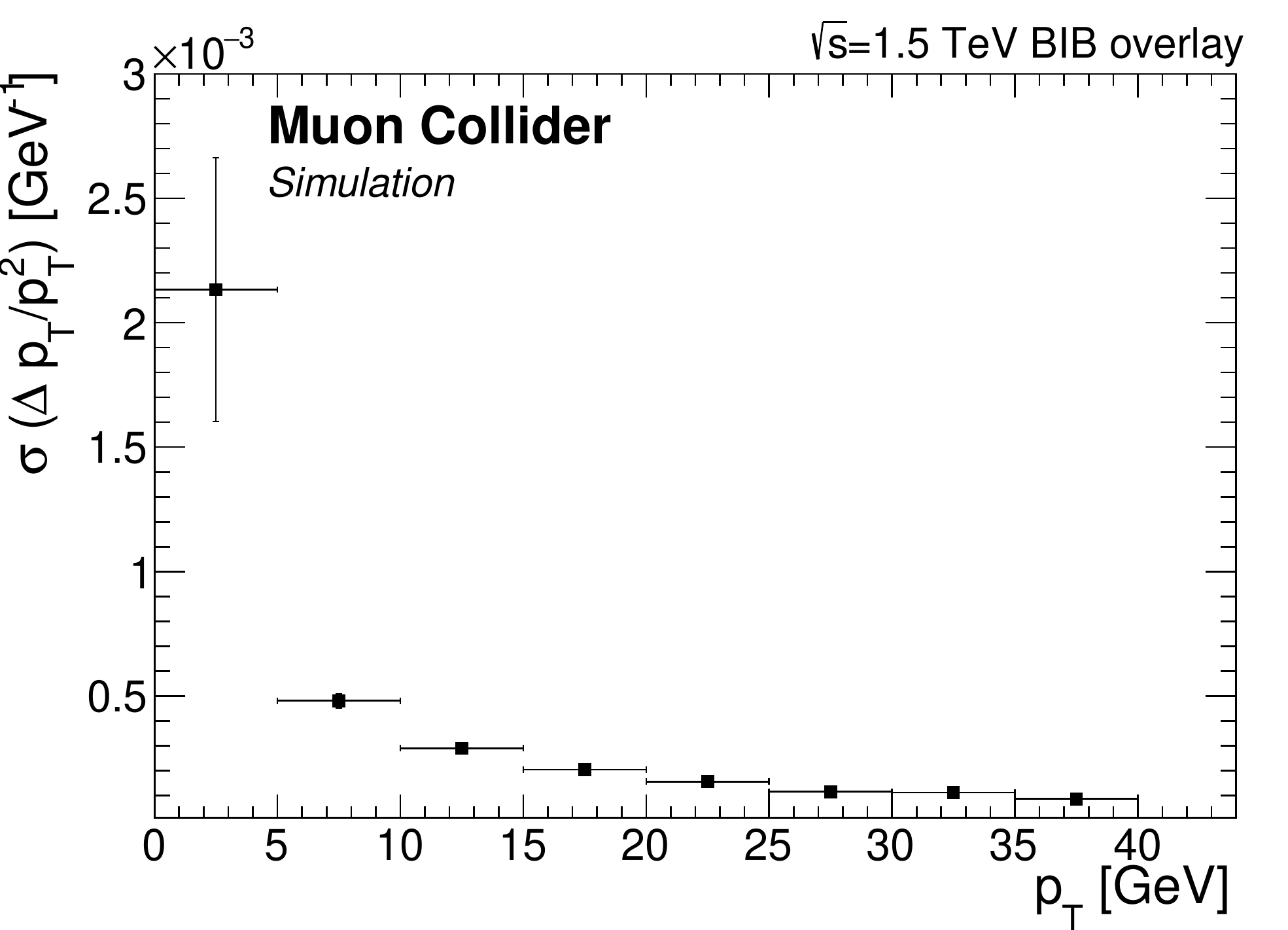}
 \caption{Muon track transverse momentum resolution as a function of the muon transverse momentum in a sample of single muon events.}
\label{fig:muon_ptresolution}
\end{figure}

The muon $p_{\rm T}$ resolution is shown in Figure~\ref{fig:muon_ptresolution} where $\Delta p_{\rm T}$ is the difference between the generated muon $p_{\rm T}$ and the $p_{\rm T}$ of the corresponding reconstructed particle. It results to be less than $10^{-4}$\,GeV$^{-1}$ for $p_{\rm T}>30$\,GeV and around a factor of 7 better in the barrel region compare to the endcap. 

The BIB was found not to strongly affect the muon reconstruction performance: the efficiency is lower only in the endcaps where all the BIB hits are concentrated, and the $p_{\rm T}$ resolution is just a few percent worse. Additional work is planned to improve the reconstruction efficiency for low transverse momenta. 

\FloatBarrier
\subsection{Forward detectors}
\label{sec:fwdandlumi}

Dedicated detectors in the forward region are required to fully harvest the physics potential of a muon collider. 

The ability to tag, or reconstruct muons down to $\theta \sim 0.1$ would provide a unique handle to distinguish events with a collinear emission of virtual $Z$ boson from the muon. This is a unique feature that can only be exploited at muon colliders, which would enable VBS and VBF measurements, as well as searches for new physics phenomena. Consequently, the design of a forward muon tagging system is considered to be of high-priority.
The low-interacting nature of muons, which can travel through hundreds of meters of material without being stopped, would allow them to traverse the shielding nozzles making this detector concept possible.
A detailed study of the expected muon trajectories as they propagate through the shielding material, final focusing magnets and other collider elements, as well as of the expected BIB rates in this region, still needs to be performed.
However, the outgoing muons are expected to have very large momenta and to be confined in a region relatively close to the beam axis.
The detector design and technology change depending on whether a simple muon tagging is required, a measurement of the muon momenta is also needed. Such measurement could be obtained with the measurement of the outgoing muon position, and scattering angle at the IP, in a high-precision forward tracking station. Alternative designs exploiting a dipole-based muon spectrometer, such as those considered for the FCC-hh detectors~\cite{FCC:2018vvp}, will be investigated, but would require a significantly larger detector volume.

A precise measurement of the collider luminosity could also be obtained using forward muons. The determination of the luminosity is of crucial or high importance for any absolute cross section measurement since it directly translates to its error. LHC experiments have dedicated detectors, so-called luminometers that are used in combination with the van der Meer scan method~\cite{vanderMeer:296752,ALICE:2022xir,ATLAS:2022hro,CMS:2021xjt,LHCb:2014vhh} to precisely measure the instantaneous luminosity. The $e^+e^-$ experiments like Belle-II~\cite{Belle-II:2019usr} and BESIII~\cite{BESIII:2015qfd} measure the integrated luminosity by counting the number of events of a process whose cross section is theoretically known with high precision. The most used one is the Bhabha scattering ($e^+e^- \rightarrow e^+e^-$) where, for example, the theoretical uncertainty on $\sigma$ at  $\sqrt{s}=1-10$ GeV is 0.1\% at large angle~\cite{CarloniCalame:2011aa}.
Due to the reduced acceptance in the forward region because of the nozzles shielding structure at muon collider, only the large angle muon Bhabha ($\mu^+$ $\mu^-$ $\rightarrow$ $\mu^+$ $\mu^-$) has been considered so far as a method for the luminosity~\cite{Giraldin:2021gxz}. 
By assuming an instantaneous luminosity $\mathcal{L} = 1.25 \times 10^{34} $ cm$^{-2}$s$^{-1}$ and considering a year of data taking ($10^7$ s) the statistical uncertainty obtained at $\sqrt{s}=1.5$~TeV is 0.2\%. The expected total uncertainty on the luminosity measurement using this method strongly depends on the accuracy on the theoretical cross section of $\mu$-Bhabha scattering at large angles, and at several TeV centre of mass energy. The addition of a forward muon detector could significantly improve these constraints.

The second major opportunity in the forward region arises from the decay of the high-energy muons, which produces a collimated beam of high-energy neutrinos very close to the muon beam axis that could enable precise measurements of neutrino cross-sections and dark sector studies. 
Such programme represents the ideal extension of the investigations carried out by dedicated LHC detectors such as FASER~\cite{FASER:2019dxq} and SND@LHC~\cite{SNDLHC:2022ihg}. The forward neutrinos at the LHC come from the decay of hadrons and they are complex to simulate, while the neutrino flux at the muon collider will be precisely known. 
The expected muon decay rates reported in Table~\ref{tab:BIB_com_energy} can be used to estimate the rate of neutrino interactions in a dedicated detector placed one kilometer away from the IP. At such distance, the neutrino beam would have a transverse size of the order of ten centimeters. Depending on the detector size, technology and material, up to several $10^{19}$ neutrino interactions per year can be observed.
The design and optimisation of such a detector, as well as a full characterisation of its physics potential, still needs to be performed.

\FloatBarrier
\subsection{Conclusions}
\label{sec:outlook}

Muon colliders can combine excellent discovery potential with high precision capabilities. For the purpose of event detection and reconstruction, the challenge that separates $\mu^{+}\mu^{-}$ with the $e^{+}e^{-}$ counterparts is the beam induced background. Because muons are unstable, they decay in flight, producing electrons that further interact with the accelerator and detector components. This creates very large multiplicities of mostly soft secondary particles, some of which end up in the detector. 
The hits produced by the secondary particles in the detectors lead to significant challenges for the particle detection and reconstruction. In this section, preliminary design and specifications for a muon collider detector were described. An assessment of the expected radiation levels was presented, showing that the radiation levels are similar to those at the HL-LHC experiments.
Possible methods to mitigate effects of the BIB were outlined.  The BIB imposes stringent requirements on the granularity, resolution, and timing properties of the muon collider detectors, and presented a number of emerging detector technologies that have a potential to address the challenge. Several R\&D efforts to further develop these technologies are needed to get them to the maturity level necessary for the detector construction, and a few of these were highlighted throughout the section.

The current performance of a detector design was also presented, including appropriate shielding near the interaction point. The expected performance for reconstructing charged particles, jets, electrons, muons and photons has been presented, in addition to preliminary results on tagging heavy-flavour jets and measuring the delivered integrated luminosity. The results demonstrate that it is possible to successfully cope with the expected BIB and reconstruct with high accuracy the main physics observables needed for carrying out the expected physics programme.

\clearpage

\section{Physics studies}
\label{sec:phys_studies}

Tentative energies and luminosities of future muon colliders, as well as a possible staged scenario that foresees a first collider with 3~TeV energy in the centre of mass followed by a \mbox{$10^+$\,TeV}~MuC, have been described in Section~\ref{IntroSect}. Based on these benchmark parameters, Section~\ref{ch2_phys_opp} offers a broad overview of the physics exploration opportunities of the muon collider project, as they emerge from the investigations performed so far. 

A selection of these studies is described in the present section in more details, focusing on indirect and direct opportunities for new physics searches in the Higgs and electroweak sector (Section~\ref{sec:Higgs}) and for dark matter searches (Section~\ref{secdm}). Studies outlining muon-specific exploration opportunities, namely advantages of colliding muons rather than electrons or protons, will be reviewed in Section~\ref{muonspec}. 

We do not aim at a complete assessment of the muon collider physics potential, which is still under development, nor at an exhaustive illustration of the opportunities for progress in these three directions of investigation. We describe concrete studies as a possible starting point for future work towards increasingly complete and realistic sensitivity projections. On top of strengthening and consolidating the physics case, and advancing the theoretical and experimental methodologies that will be required for the future exploitation of the facility, an extensive investigation of muon collider physics is a desirable input at this stage of the project also in view of a possible reassessment of the target design parameters and of the staging plan.

\subsection{Electroweak and Higgs physics \label{sec:Higgs}}

Muon colliders can probe the physics of electroweak interactions, including the nature of the Higgs sector responsible for the breaking of the electroweak symmetry, by employing three distinct and complementary strategies. 

First, as emphasised in Section~\ref{VBF}, one can exploit the large luminosity for effective vector bosons to produce a large number of Higgs bosons and study its on-shell couplings with precision. The Higgs trilinear coupling can be measured as well thanks to the relatively large rate for Higgs pair production. The quadrilinear coupling could be also accessible.

Second, one can probe short-distance electroweak and Higgs interactions directly by performing cross section measurements at the large available $\mu^+\mu^-$ energy. These measurements can shed light on heavy new physics potentially responsible for modifications of the on-shell Higgs couplings, or/and extend the muon collider sensitivity to much higher new physics scales than those probed by on-shell measurements or by the direct production of new particles. We anticipated in Section~\ref{HEM} that such high-energy measurements offer unique opportunities to probe the composite nature of the Higgs particle. 

Finally, the muon collider can probe directly new physics extensions of the Higgs sector that foresee relatively light new particles. 

The present section is devoted to an assessment of the muon colliders' perspectives for progress in these three directions.

\subsubsection*{Higgs couplings}
At high-energy muon colliders, as the electroweak gauge boson content of the muon beam becomes sizeable, vector boson fusion (VBF) becomes the most important channel for production of SM particles, as illustrated in Figure~\ref{fig:EV}. In particular, the VBF Higgs production rate neatly overcomes the Higgsstrahlung ($ZH$) process that is instead dominant at low-energy $e^+ e^-$ Higgs factories. A total of half a million Higgs bosons will be produced at the 3~TeV MuC, and around ten million at 10~TeV, with the integrated luminosities of $1~{\rm{ab}}^{-1}$ and $10~{\rm{ab}}^{-1}$, respectively. These figures qualify muon colliders as ``Higgs factories'' and call for detailed Higgs couplings sensitivity projections.

An initial estimate of the Higgs measurements precision attainable via $W$ boson fusion (WBF, $\mu^+ \mu^- \to H \overline{\nu} \nu$) and $Z$ boson fusion (ZBF, $\mu^+ \mu^- \to H \mu^+ \mu^-$) was presented in \cite{Han:2020pif,AlAli:2021let}. These results are obtained with a fast detector simulation and realistic detector acceptance~\footnote{Namely, $|\eta|<2.5$, or $\theta\in [10^\circ,170^\circ]$, due to radiation-absorbing tungsten nozzles expectedly needed to suppress the beam induced backgrounds, see Section~\ref{sec:environment}.}, but they do not include backgrounds for most channels. In what follows we employ instead the more recent estimates of Ref.~\cite{Forslund:2022xjq}, that do incorporate physics backgrounds as well as detector effects through the muon collider DELPHES card~\cite{deFavereau:2013fsa,delphes_card_mucol}. 

\begin{table*}[t]
  \centering
    \caption{68$\%$ probability sensitivity to the Higgs couplings, assuming no BSM Higgs decay channels.
    \label{t:Kappa_sensitivity}}
    {\vspace{4pt}
      \begin{tabular}{c|c|c|cc|c} 
        \hline
        &
        HL-LHC &
        HL-LHC &
        HL-LHC &
        HL-LHC &
        HL-LHC  \\
         &
         &
        + 125 GeV MuC  &
        
        + 3 TeV MuC &

        + 10 TeV MuC &
        
        + 10 TeV MuC
        \\
        Coupling &
        &
        5 / 20 fb$^{-1}$&
        1/2 ab$^{-1}$ &
        10 ab$^{-1}$&
         + FCC-ee\\ 
        
        \hline
        $\kappa_W~[\%]$ &
          1.7   & 
          
          1.3 / 0.9  & 

          0.4 / 0.3   & 
          
          0.1  & 
          
          0.1    \\ [0.1cm] 
        $\kappa_Z~[\%]$ &
          1.5 & 
          
          1.3 / 1.0 & 

          0.9 / 0.7 & 
          
          0.4 & 
            
          0.1   \\ [0.1cm] 
        $\kappa_g~[\%]$ &
          2.3 & 
          
          1.7 / 1.4 & 

          1.2 / 1.0  & 
          
          0.7  & 
          
          0.6   \\ [0.1cm] 
        $\kappa_\gamma~[\%]$ &
          1.9  & 
          
          1.6 / 1.5 & 

          1.3 / 1.2  & 
          
          0.8  & 
          
          0.8   \\ [0.1cm] 
        $\kappa_{Z\gamma}~[\%]$ &
          10  & 
          
          10 / 10 & 

          9.3 / 8.6  & 
          
          7.2 & 
          
          7.1 \\ [0.1cm] 
        $\kappa_c~[\%]$ &
          -& 
          
          12 / 5.9 & 

          6.2 / 4.4  & 
          
          2.3  & 
          
          1.1  \\ [0.1cm] 

        $\kappa_b~[\%]$ &
          3.6  & 
          
          1.6 / 1.0 & 

          0.8 / 0.7  & 
          
          0.4  & 
          
          0.4  \\ [0.1cm] 
        $\kappa_\mu~[\%]$ &
          4.6  & 
          
          0.6 / 0.3 & 

          4.2 / 4.0  & 
          
          3.4  & 
          
          3.2  \\ [0.1cm] 
        $\kappa_\tau~[\%]$ &
          1.9  & 
          
          1.4 / 1.2 & 

          1.2 / 1.0  & 
          
          0.6  & 
          
          0.4  \\ [0.1cm] 
        $\kappa_t^\dagger~[\%]$ &
          3.3  & 
          
          3.2 / 3.1 & 

          3.1 / 3.1  & 
          
          3.1  & 
          
          3.1 \\ [0.1cm] 
          %
          \hline
          $\Gamma_H^\ddag~[\%]$ &
          5.3  & 
          
          2.7 / 1.7 & 

          1.3 / 1.0  & 
          
          0.5 & 
          
          0.4 \\ [0.1cm]
          \hline
          \multicolumn{6}{l}{ {\scriptsize $^\dagger$ No input used for $\mu$ collider.}}\\
          \multicolumn{6}{l}{ {\scriptsize $^\ddag$ Prediction assuming only SM Higgs decay channels. Not a free parameter in the fits.}}\\
        \hline
      \end{tabular}
    }
\end{table*}

While realistic enough for a first quantitative assessment of the Higgs couplings precision, these results will have to be consolidated in the future by full detector simulation studies including beam induced backgrounds from the muon decays. A comparison between current full simulation results and DELPHES, performed in the $H\rightarrow b\bar{b}$ channel~\cite{Forslund:2022xjq}, displays a reduced precision for the $b$-jets energy determination. However the resolution degradation does not affect the $H\rightarrow b\bar{b}$ cross section measurement precision appreciably, leading to a good agreement between the fast and full simulation estimates in this channel. This is encouraging, taking also into account that detector and reconstruction design studies are at a very preliminary stage and present results definitely underestimate the attainable physics performances as described in Section~\ref{sec:detectorandreconstruction}. 

Further work is also needed for a robust assessment of the possibility, which we do assume in our estimates~\cite{Forslund:2022xjq}, to discriminate between the WBF and the ZBF Higgs production channels by tagging very forward muons well beyond $|\eta|\approx 2.5$. This would require a dedicated forward muon detector, as described in Section~\ref{sec:fwdandlumi}, that is still to be designed.

The projected sensitivities~\cite{Forslund:2022xjq} for the main Higgs decays ($b\bar{b}$, $WW^*$, etc) in single Higgs production are estimated at the few percent level at 3~TeV, whereas at 10~TeV with 10 ab$^{-1}$, sensitivities at the permille level are possible. Roughly these figures could be considered comparable or slightly superior to the HL-LHC measurements sensitivities in the 3~TeV case, and to those of future $e^+e^-$ Higgs factories~\cite{deBlas:2019rxi} for the 10~TeV MuC. Moreover the different production mechanisms make MuC results complementary to the other projects, as we will see in the Higgs couplings sensitivity projections presented below. 

It should also be noticed that some aspects of Higgs physics are challenging at muon colliders, and have not yet been investigated. For example the precision on the top Yukawa coupling ($y_t$) determination from the $t\bar{t}H$ measurement at 3 and 10~TeV is estimated to be $35\%$ and $53\%$~\cite{Forslund:2022xjq}, significantly below the LHC.  However muon colliders offer additional handles for $y_t$ determination, such as the measurement of $W^+W^-\rightarrow t\bar{t}$. Preliminary results in this channel are promising \cite{AlAli:2021let} but further study is needed.

\begin{figure*}[ht]
\centering
\includegraphics[width=\linewidth]{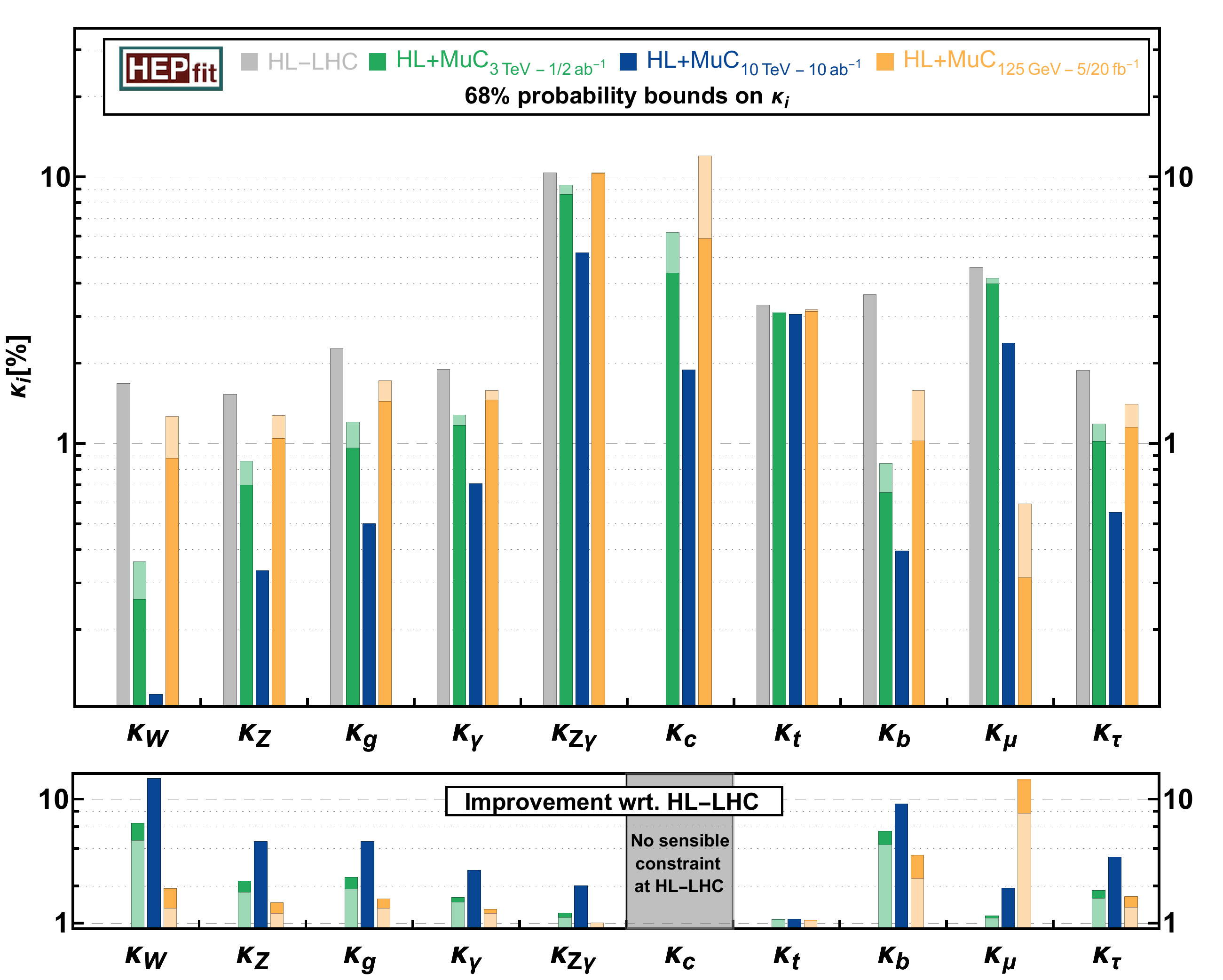}
\caption{\label{fig:kappa}
Sensitivity to modified Higgs couplings in the $\kappa$ framework. We show the marginalized 68\% probability reach for each coupling modifier. For the 125 GeV MuC, light (dark) shades correspond to a luminosity of 5 (20) fb$^{-1}$. The same color code is used for the 3~TeV MuC with 1 or 2 ab$^{-1}$. 
}
\end{figure*}

For a quantitative assessment of the muon collider potential to measure the properties of the Higgs boson, we perform here a series of fits to single-Higgs couplings in the so-called $\kappa$ framework~\cite{LHCHiggsCrossSectionWorkingGroup:2012nn,LHCHiggsCrossSectionWorkingGroup:2013rie}, where the interaction vertices predicted by the SM are modified by scaling factors $\kappa_i$. \footnote{For the effective 1-loop vertices to gluons, photons and $Z\gamma$ we use independent effective scaling parameters $\kappa_{g,\gamma,Z\gamma}$, to describe the possibility of extra particles running in the loops.}
In the $\kappa$ framework, the cross sections of the different production processes $i \to {H }$ at the different colliders, times the decay branching ratios
\begin{equation}
( \sigma \cdot \text{BR} ) ( i \to {H }\to f ) =  \frac{ \sigma_{i} \cdot \Gamma_{f}}{\Gamma_H},
\label{eq:kappa_zw}
\end{equation}
are parameterised as follows
\begin{eqnarray}
&& ( \sigma \cdot \text{BR} ) ( i \to {H }\to f )  =  \frac{ \sigma_{i}^{\text{SM}} \kappa_i^2 \cdot \Gamma_{f}^{\text{SM}}\kappa_f^2 }{\Gamma_H^{\text{SM}} \kappa_H^2} \\
&&  = \frac{\kappa_i^2\cdot\kappa_f^2}{\kappa_H^2} \left[ (\sigma \cdot \text{BR} ) ( i \to {H }\to f )\right]_{\text{SM}}\,, \nonumber
\end{eqnarray}
in terms of the SM predictions for cross sections and branching ratios. We are interested in studying the sensitivity to the couplings of the Higgs boson, not its putative decay to exotic final states. Therefore the Higgs width modifier $\kappa_{H}$ is determined by the other $\kappa$'s
\begin{equation}\label{kh}
\kappa_{H}^2 = \frac{\sum_{f} \kappa_f^2 \Gamma_{f}^\text{SM}}{
\Gamma_{H}^\text{SM}}=\sum_{f}\kappa_f^2 \text{BR}^{\text{SM}}(H\to f) \,,
\end{equation}
where the sum extends over the SM Higgs decay channels. We will comment later on the possibility of leaving the Higgs width as a free parameter in the fit. We further restrict our attention to the 10 coupling modifiers, listed in Table~\ref{t:Kappa_sensitivity}, that are most precisely determined in the different collider projects under examination.

The results of the fits, performed with the {\tt HEPfit} code~\cite{DeBlas:2019ehy}, are reported in Table~\ref{t:Kappa_sensitivity} and displayed in Figure~\ref{fig:kappa}. On top of the 3 and 10~TeV sensitivity projections previously described, the following input is considered for the other collider projects. For the HL-LHC, we employ the signal strengths measurement projections from Ref.~\cite{Cepeda:2019klc}, assuming the S2 scenario for the reduction of systematic uncertainties. We also consider single-Higgs measurements at the FCC-ee collider~\cite{FCC:2018byv,FCC:2018evy}, as a reference for the precision that is generically attainable at future low-energy $e^+e^-$ Higgs factories. These results assume that all intrinsic SM theory (and experimental) uncertainties are under control by the time any of these future colliders are built~\cite{Blondel:2019qlh,Freitas:2019bre}. These assumptions, and the inputs for the HL-LHC, are compatible with the ones employed in Ref.~\cite{deBlas:2019rxi} for a global comparative assessment of future collider sensitivities in preparation for the 2020 European Strategy Update process. Finally, the results employ sensitivity projections from Ref.~\cite{deBlas:2022aow} for the Higgs-pole muon collider operating at 125~GeV, to be discussed later in this section.

The second and the third columns of Table~\ref{t:Kappa_sensitivity} report the marginalised sensitivity projections for the 3 and 10~TeV MuC in combination with HL-LHC. A luminosity of 1 or of 2~ab$^{-1}$ is considered at 3~TeV, compatibly with the preliminary staging plan described in Section~\ref{IntroSect}. The 1~ab$^{-1}$ results represent the outcome of the first stage in the optimistic scenario where the 3~TeV MuC runs for a short time (of few years), leaving space soon to the 10~TeV collider that will rapidly supersede the 3~TeV measurements precision, as the table shows. The 2~ab$^{-1}$ sensitivities are attainable with a longer run (or possibly with the installation of two detectors) of the 3~TeV collider, in the event that the 10~TeV upgrade is delayed. In both scenarios, the 3~TeV MuC will advance the determination of several Higgs couplings relative to the HL-LHC. Furthermore it could play a crucial role in clarifying possible tensions with the SM of the HL-LHC  results by measuring the Higgs couplings with comparable or better statistical precision in a leptonic environment that is subject to different (and expectedly reduced) experimental and theoretical systematic uncertainties.

The 10~TeV stage of the muon collider will measure single-Higgs couplings at the sub-percent or permille level, enabling a jump ahead in the knowledge of Higgs physics that is comparable to the one of a dedicated $e^+e^-$ low-energy Higgs factory. Furthermore the muon collider measurements are complementary to those of $e^+e^-$ Higgs factories because different production modes ($ZH$ at $e^+e^-$ and WBF the MuC) dominate at the different colliders. This makes, for instance, the 10~TeV MuC more sensitive to $\kappa_W$, and the $e^+e^-$ Higgs factories more effective in the measurement of $\kappa_Z$. This complementarity is illustrated the fourth column of Table~\ref{t:Kappa_sensitivity}, where the 10~TeV MuC measurements are combined with those of an $e^+e^-$ Higgs factory. By comparing with the MuC-only results on the third column, we see that the muon collider dominates the combined sensitivity for several couplings. There would thus be space for improvements on single-Higgs physics at the muon collider even if constructed and operated after the completion of an $e^+e^-$ Higgs factory project.

\begin{table*}[t]
  \centering
    \caption{68$\%$ probability intervals for the Higgs trilinear coupling.
      \label{t:H3_sensitivity}
    }
  \vspace{4pt}
    {
      \begin{tabular}{c c c c c c} 
        \hline
         &
        HL-LHC
         &
        3 TeV MuC &

        10 TeV MuC &

        14 TeV MuC &

        30 TeV MuC 
        \\
         &
         &
        L$\approx$ 1 ab$^{-1}$ / 2 ab$^{-1}$ &
        L= 10 ab$^{-1}$ &
        L$\approx$ 20 ab$^{-1}$ &
        L= 90 ab$^{-1}$\\
        \hline
        $\delta \kappa_\lambda$ &
        
          [-0.5,0.5] & 
          [-0.27,0.35]~$\cup$~[0.85,0.94] / [-0.15,0.16] & 
          
          [-0.035, 0.037] &
          
          [-0.024, 0.025] & 
          
          [-0.011, 0.012] \\ [0.1cm] 
          \hline
          comb. w HL-LHC &
          -- &
          [-0.2,0.22] / [-0.13,0.14] & 
          [-0.035,0.036] & 
          [-0.024,0.025] & 
          [-0.011,0.012] \\ [0.05cm] 
        \hline
      \end{tabular}
    }
\end{table*}

\paragraph{The Higgs width and the 125~GeV muon collider} {\ } \\ 
\noindent
So far we studied new physics effects in the SM interaction vertices of the Higgs, excluding possible new vertices that mediate ``exotic'' decays of the Higgs particle either to light BSM states or to SM final states different from the decay channels (such as  $b\overline{b}$, $WW$, etc) foreseen by the SM. Specific exotic channels could be searched for individually. Or, the total Higgs branching ratio to exotic channels can be probed indirectly via its contribution to the Higgs total width, parametrised as~\cite{deBlas:2019rxi}
\begin{equation}
\Gamma_H = \Gamma_{H}^\text{SM} \frac{\kappa_H^2}{1-\text{BR}_{\rm exo}}\,,
\end{equation}
with $\kappa_H$ as in eq.~(\ref{kh}).

All the Higgs cross sections times branching ratios ($\sigma\cdot\text{BR}$) are sensitive to the total Higgs width. Therefore, precise measurements of these observables are powerful probes of new physics scenarios that foresee exotic Higgs decays. The precision on the determination of $\text{BR}_{\rm exo}$ can be estimated as the one of the most accurately measured coupling in Table~\ref{t:Kappa_sensitivity}, namely $\sim 0.1\,\%$ at the 10~TeV muon collider. However, it is technically possible to compensate the dependence of the $\sigma\cdot\text{BR}$ observables on  $\text{BR}_{\rm exo}$ by a correlated modification of the SM Higgs couplings. This defines a flat direction in the parameter space formed by the coupling modifiers $\kappa_i$ and the exotic $\text{BR}_{\rm exo}$, that requires additional measurements to be resolved. No (sufficently accurate) additional measurement seems possible at high-energy muon colliders and the flat direction can not be lifted. At low-energy $e^+e^-$ Higgs factories instead, an absolute measurement of the $ZH$ production cross section is possible by the missing mass method, resulting in a global sensitivity to $\text{BR}_{\rm exo}$ at the $1\,\%$ level even in an extended Higgs fit where $\text{BR}_{\rm exo}$ is added as a free parameter on top of the coupling modifiers. The possibility of performing this additional measurement is an additional element of complementarity between high-energy muon colliders and low-energy $e^+e^-$ Higgs factories.

Alternatively, or additionally, the flat direction could be resolved by a direct experimental determination of the Higgs boson width, performed at a muon collider operating close to the Higgs pole $\sqrt{s}=125$~GeV through the measurement of the Higgs resonance line shape~\cite{Han:2012rb,Greco:2016izi}. State-of-the art projections~\cite{deBlas:2022aow} (see also~\cite{Conway:2013lca}), that duly include initial state radiation (ISR) effects and physics backgrounds, foresee a sensitivity to $\Gamma_H$ ($\text{BR}_{\rm exo}$) at the $3\,\%$ ($2\,\%$) and the $2\,\%$ ($1.5\,\%$) level for, respectively, 5 and 20~ab$^{-1}$ integrated luminosity. These results ignore beam-induced backgrounds (that could be worse than at high energy), assume the feasibility of a collider with a beam energy spread as small as $R=0.003\%$, and an instantaneous luminosity well above the baseline scaling~in eq.~(\ref{lums}). These aspects should be investigated for a realistic assessment of the 125~GeV MuC physics potential and feasibility.

Apart from measuring $\Gamma_H$, the 125~GeV MuC can also study other aspects of Higgs physics, but with limited perspectives for progress in comparison with the HL-LHC.\footnote{The 125~GeV MuC would also enable an impressive determination of the Higgs mass with one part per million accuracy.} In fact, with a resonant $\mu^+ \mu^- \to H$ cross section of 70~pb, reduced to about 22~pb by the beam energy spread and ISR, a luminosity at the level of several fb$^{-1}$ would yield order $10^{5}$ Higgses, limiting a priori the statistical reach in terms of precision Higgs physics. This is illustrated by the 10-parameters couplings fit on the last column of Table~\ref{t:Kappa_sensitivity}. Some progress is possible, eminently in the coupling of the Higgs to muons, but globally the progress relative to the HL-LHC is mild, and inferior to the one attainable by the 3~TeV MuC.

\paragraph{The trilinear Higgs coupling}
{\ } \\ 
\noindent
Unlike low-energy $e^+e^-$ Higgs factories, high-energy muon colliders enable the direct measurement of the Higgs boson self-interactions, starting from the triple Higgs coupling $\lambda_3$. The relevant process is the WBF production of Higgs boson pairs, $\mu^+ \mu^- \to HH \bar{\nu}\nu$, that attains a total yield of $3\cdot 10^4$ events at the 10~TeV MuC with 10~ab$^{-1}$ as shown in Figure~\ref{fig:EV}. 

The single Higgs couplings are very precisely determined as previously discussed. Therefore the measurement of the differential double Higgs production cross section can be directly translated into the exclusive determination of the trilinear Higgs coupling, expressed in terms of $\kappa_\lambda\equiv \lambda_3/\lambda_3^{\rm SM}$. We employ the likelihood from Ref.~\cite{Buttazzo:2020uzc}, based on the differential distribution in the Higgs pair invariant mass. This analysis includes physics backgrounds and realistic detector effects, and the results nicely agree with previous studies (eminently, with Ref.~\cite{Han:2020pif}). The resulting projections for the $68\,\%$ confidence level regions for the measurement of $\delta\kappa_\lambda=\kappa_\lambda-1$ are reported in Table~\ref{t:H3_sensitivity} and illustrated in Figure~\ref{tab:higgscouplingfit}.

\begin{table*}[t]
\caption{\label{tab:eft-silh} 
68\% probability reach on the Wilson coefficients in the Lagrangian~(\ref{eq:LSilh}) from the global fit.
In parenthesis we give the corresponding results from a fit assuming only one operator is generated by the UV physics.\\
}
\centering
{

\begin{tabular}{c c}

\begin{tabular}{ c | c | c | c  }
\hline
        &       &\multicolumn{2}{c}{HL-LHC + MuC}  \\
      &HL-LHC   & 3 TeV   & 10 TeV    \\
      &   & (1 / 2 ab$^{-1}$)   & (10 ab$^{-1}$)    \\
\hline
$\frac{c_{\phi}}{\Lambda^2}[\mbox{TeV}^{-2}]$   & $0.51$   & $0.11 / 0.086$   & $0.037$  \\
   & $(0.28)^\dagger$   & $(0.1/0.074)$   & $(0.025)$  \\
$\frac{c_{T}}{\Lambda^2}[\mbox{TeV}^{-2}]$   & $0.0057$   & $0.0022 / 0.0021$   & $0.0019$  \\
   & $(0.0019)$   & $(0.0019 / 0.0019)$   & $(0.0018)$  \\
$\frac{c_{W}}{\Lambda^2}[\mbox{TeV}^{-2}]$   & $0.33$   & $0.0096 / 0.0068$   & $0.0011$  \\
   & $(0.021)$   & $(0.0033 / 0.0024)$   & $(0.00031)$  \\
$\frac{c_{B}}{\Lambda^2}[\mbox{TeV}^{-2}]$   & $0.33$   & $0.022 / 0.016$   & $0.0022$  \\
   & $(0.026)$   & $(0.0075 / 0.0054)$   & $(0.00065)$  \\
$\frac{c_{\phi W}}{\Lambda^2}[\mbox{TeV}^{-2}]$   & $0.31$   & $0.034 / 0.033$   & $0.024$  \\
   & $(0.033)$   & $(0.031 / 0.03)$   & $(0.017)$  \\
$\frac{c_{\phi B}}{\Lambda^2}[\mbox{TeV}^{-2}]$   & $0.32$   & $0.17 / 0.16$   & $0.1$  \\
   & $(0.19)$   & $(0.17 / 0.16)$   & $(0.097)$  \\
$\frac{c_{\gamma}}{\Lambda^2}[\mbox{TeV}^{-2}]$   & $0.0053$   & $0.0045 / 0.0042$   & $0.0026$  \\
   & $(0.0041)$   & $(0.0038 / 0.0036)$   & $(0.0022)$  \\
$\frac{c_{g}}{\Lambda^2}[\mbox{TeV}^{-2}]$   & $0.0012$   & $0.0011 / 0.00096$   & $0.00057$  \\
   & $(0.00052)$   & $(0.00037 / 0.0003)$   & $(0.00015)$  \\
\hline
\end{tabular}

&

\begin{tabular}{ c | c | c | c  }
\hline
        &       &\multicolumn{2}{c}{HL-LHC + MuC}  \\
      &HL-LHC   & 3 TeV   & 10 TeV    \\
      &   & (1 / 2 ab$^{-1}$)   & (10 ab$^{-1}$)    \\
\hline
$\frac{c_{y_{e}}}{\Lambda^2}[\mbox{TeV}^{-2}]$   & $0.25$   & $0.19 / 0.16$   & $0.086$  \\
   & $(0.2)$   & $(0.17 / 0.15)$   & $(0.081)$  \\
$\frac{c_{y_{u}}}{\Lambda^2}[\mbox{TeV}^{-2}]$   & $0.57$   & $0.47 / 0.42$   & $0.25$  \\
   & $(0.24)$   & $(0.16 / 0.13)$   & $(0.064)$  \\
$\frac{c_{y_{d}}}{\Lambda^2}[\mbox{TeV}^{-2}]$   & $0.46$   & $0.13 / 0.11$   & $0.059$  \\
   & $(0.25)$   & $(0.11 / 0.089)$   & $(0.052)$  \\
$\frac{c_{2B}}{\Lambda^2}[\mbox{TeV}^{-2}]$   & $0.087$   & $0.0036 / 0.0025$   & $0.00031$  \\
   & $(0.075)$   & $(0.0029 / 0.002)$   & $(0.00026)$  \\
$\frac{c_{2W}}{\Lambda^2}[\mbox{TeV}^{-2}]$   & $0.0087$   & $0.00097 / 0.00069$   & $0.000084$  \\
   & $(0.0075)$   & $(0.00078 / 0.00055)$   & $(0.00007)$  \\
$\frac{c_{3W}}{\Lambda^2}[\mbox{TeV}^{-2}]$   & $1.7$   & $1.7 / 1.7$   & $1.6$  \\
   & $(1.7)$   & $(1.7 / 1.7)$   & $(1.6)$  \\
$\frac{c_{6}}{\Lambda^2}[\mbox{TeV}^{-2}]$   & $8.4$   & $4.6 / 3.1$   & $0.64$  \\
   & $(7.8)$   & $(4.4 / 3.0)$   & $(0.6)$  \\
\hline 
\multicolumn{4}{c}{}\\[0.35cm]
\end{tabular}
\end{tabular}}\\
{ 
\footnotesize 
 \begin{flushleft}
 \justify{
$^\dagger$ As explained in~\cite{deBlas:2019rxi}, due to the treatment of systematic/theory uncertainties in the HL-LHC inputs, this number must be taken with caution, as it would correspond to an effect below the dominant theory uncertainties. A more conservative estimate accounting for 100$\%$ correlated theory errors would give $c_\phi/\Lambda^2\sim 0.42$ TeV$^{-2}$.}
\end{flushleft}
}

\end{table*}

Table~\ref{t:H3_sensitivity} shows that the 10~TeV MuC can measure the Higgs trilinear coupling with a precision of $4\,\%$, significantly better than the CLIC high-energy $e^+e^-$ future collider project, whose precision is limited to $12\,\%$~\cite{deBlas:2018mhx}. The comparison with the 100~TeV proton-proton collider (FCC-hh) is more uncertain because the FCC-hh sensitivity projections range from $3.5$ to $8\%$ depending on assumptions on the detector performances~\cite{Mangano:2020sao}. Muon colliders of even higher energy (14, or 30~TeV) could further improve the precision provided their integrated luminosity increases like the square of the centre of mass energy.

Our results for the 3~TeV stage are more structured. With an integrated luminosity of 1~ab$^{-1}$, the confidence region consists of two disjoint intervals, and it is significantly broader than the estimate (of around $18\,\%$ precision) one would naively obtain by cutting the likelihood at the Gaussian $1~\!\sigma$ level.  This is because the likelihood is highly non-Gaussian, due to a secondary local minimum at large $\delta\kappa_\lambda$. Recent projections~\cite{Cepeda:2019klc} suggest that the HL-LHC could offer a sufficiently accurate determination (at the $50\,\%$ level) of $\delta\kappa_\lambda$ to exclude the secondary minimum. Therefore the combination with HL-LHC projections produces a connected confidence region and a relative precise determination of $\delta\kappa_\lambda$ already with 1~ab$^{-1}$ luminosity. With 2~ab$^{-1}$ instead, the 3~TeV stage will not require combination with the HL-LHC for an accurate determination of $\delta\kappa_\lambda$ at the level of $16\,\%$.

Beyond double Higgs production, a multi-TeV muon collider could exploit triple-Higgs production to gain sensitivity to the quartic Higgs coupling, $\lambda_4$~\cite{Chiesa:2020awd}. The cubic and quartic Higgs interactions are related in the SM and in BSM scenarios where new physics is heavy and the couplings correction emerge from dimension-six effective operators. If this is the case, the measurement of $\lambda_4$ is irrelevant as it can not compete with the $\lambda_3$ determination. If this is not the case, for instance because new physics is light, $\lambda_4$ modifications are independent from those of $\lambda_3$, and possibly stronger. The quartic coupling is directly tested at leading order via, e.g. $\mu^+ \mu^- \to HHH\bar{\nu}\nu$, which has a cross section of 0.31 (4.18) ab at $\sqrt{s}=3$ (10) TeV~\cite{Chiesa:2020awd}. For realistic luminosities, this makes a 3 TeV option unable to probe the quartic coupling. At the 10~TeV MuC, $\lambda_4$ could be tested to a precision of tens of percent with integrated luminosities of several tens of ab, slightly above the current luminosity target.

\begin{figure*}[ht]
\centering
\includegraphics[width=0.95\linewidth]{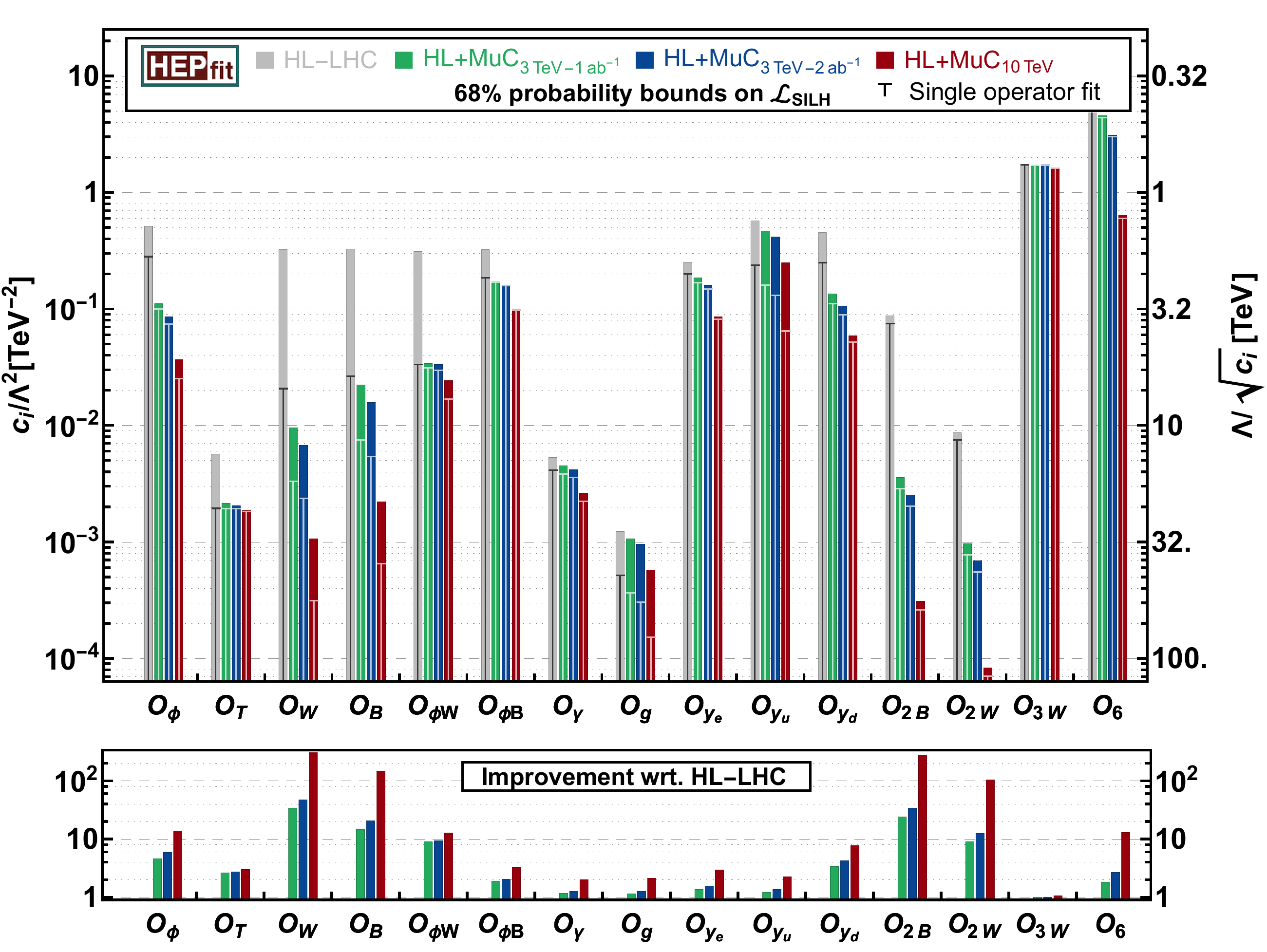}
\caption{\label{fig:SILH}
Global fit to the EFT operators in the Lagrangian (\ref{eq:LSilh}). We show the marginalised 68\% probability reach for each Wilson coefficient $c_i/\Lambda^2$ in eq.~(\ref{eq:LSilh}) from the global fit (solid bars). The reach of the vertical ``T'' lines indicate the results assuming only the corresponding operator is generated by the new physics.}
\end{figure*}

\subsubsection*{EFT probes of heavy new physics}

Measuring the properties of the Higgs boson is part of the broader endeavour to test the SM increasingly accurately and under unprecedented experimental conditions. Valid tests of the SM are those that can conceivably fail, revealing the presence of new physics effects. Theoretical BSM considerations thus provide a valuable guidance for the experimental exploration of the SM theory. This guidance becomes particularly strong and sharp under the hypothesis that all the new physics particles are heavy, such that their observable effects are encapsulated in Effective Field Theory (EFT) interaction operators of energy dimension larger than four. In this section we discuss the muon colliders sensitivity to putative EFT interactions beyond the SM, enabling a systematic and comprehensive exploration of high-scale new physics models.

Since hypothesising heavy new physic might seem a pessimistic attitude, for a collider project with great opportunities for direct discoveries, it is worth outlining the value of EFT studies like those of the present section if new physics is instead light. Even if not accurately described by the EFT, relatively light new physics could contribute to the same processes and observables, and be discovered by the same measurements that we are considering here to probe the EFT. Moreover even if the new light particles will be first discovered in other processes, probing their indirect effects will be an essential step for the characterisation of their properties and interactions with the SM particles. In other words, a program of characterisation for the newly discovered BSM physics would be similar to the program of SM characterisation based on the SM EFT.

For a first assessment of the muon colliders' potential to probe the SM EFT, we perform a global fit to the following dimension-6 EFT Lagrangian
\begin{eqnarray}
&{\cal L}_{\mathrm{SILH}}=
\frac{c_\phi }{\Lambda^2} \frac 12 \partial_\mu (\phi^\dagger \phi)\partial^\mu (\phi^\dagger \phi)+ 
\frac{c_T}{\Lambda^2} \frac 12(\phi^\dagger \lrD_\mu \phi)(\phi^\dagger \lrD^\mu \phi)\nonumber \\
&
- 
\frac{c_6}{\Lambda^2} \lambda (\phi^\dagger \phi)^3+\sum_f\left(\frac{c_{y_f} }{\Lambda^2} y^f_{ij} \phi^\dagger \phi  \bar{\psi}_{Li} \phi \psi_{Rj} + \hc \right)\nonumber \\
&
+\frac{c_W}{\Lambda^2} \frac{ig}{2}\phi^\dagger \lrDa_\mu \phi D_\nu W^{a~\!\mu\nu} 
+\frac{c_B}{\Lambda^2} \frac{ig^\prime}{2}\phi^\dagger \lrD_\mu \phi \partial_\nu B^{\mu\nu}
\nonumber \\
&+\frac{c_{\phi W}}{\Lambda^2} i g D_\mu\phi^\dagger \sigma_a D_\nu \phi W^{a~\!\mu\nu}+\frac{c_{\phi B} }{\Lambda^2} i g^\prime D_\mu\phi^\dagger \sigma_a D_\nu \phi B^{\mu\nu}
\nonumber \\
&+\frac{c_{\gamma} }{\Lambda^2} g^{\prime~\!2}  \phi^\dagger \phi B^{\mu\nu}B_{\mu\nu}+\frac{c_{g}}{\Lambda^2} g^{2}_s  \phi^\dagger \phi G^{A~\!\mu\nu}G^A_{\mu\nu}
\nonumber \\
&-\frac{c_{2W}}{\Lambda^2} \frac{g^2}{2} D^\mu W_{\mu\nu}^aD_\rho W^{a~\!\rho\nu}-\frac{c_{2B}}{\Lambda^2} \frac{g^{\prime~\!2}}{2}\partial^\mu B_{\mu\nu}\partial_\rho B^{\rho\nu}\nonumber \\
&
+\frac{c_{3W}}{ \Lambda^2}g^3\varepsilon_{abc}W_{\mu}^{a~\nu}W_\nu^{b~\rho}W_\rho^{c~\mu}. 
\label{eq:LSilh}
\end{eqnarray}
While this is only a subset of the operators of the more general dimension-6 SM EFT, the operators above are of special relevance for several BSM scenarios. Explicit examples are Composite Higgs scenarios and $U(1)$ extensions of the SM, to be discussed later in this section. The selection of operator in eq.~(\ref{eq:LSilh}) follows the one done as part of the 2020 European Strategy Group studies \cite{EuropeanStrategyforParticlePhysicsPreparatoryGroup:2019qin,deBlas:2019rxi}, enabling a direct comparison with the sensitivity projections of other future collider projects.

\begin{figure*}[t]
\centering
\begin{tabular}{c c}
\hspace{-0.2cm}\includegraphics[width=0.49\linewidth]{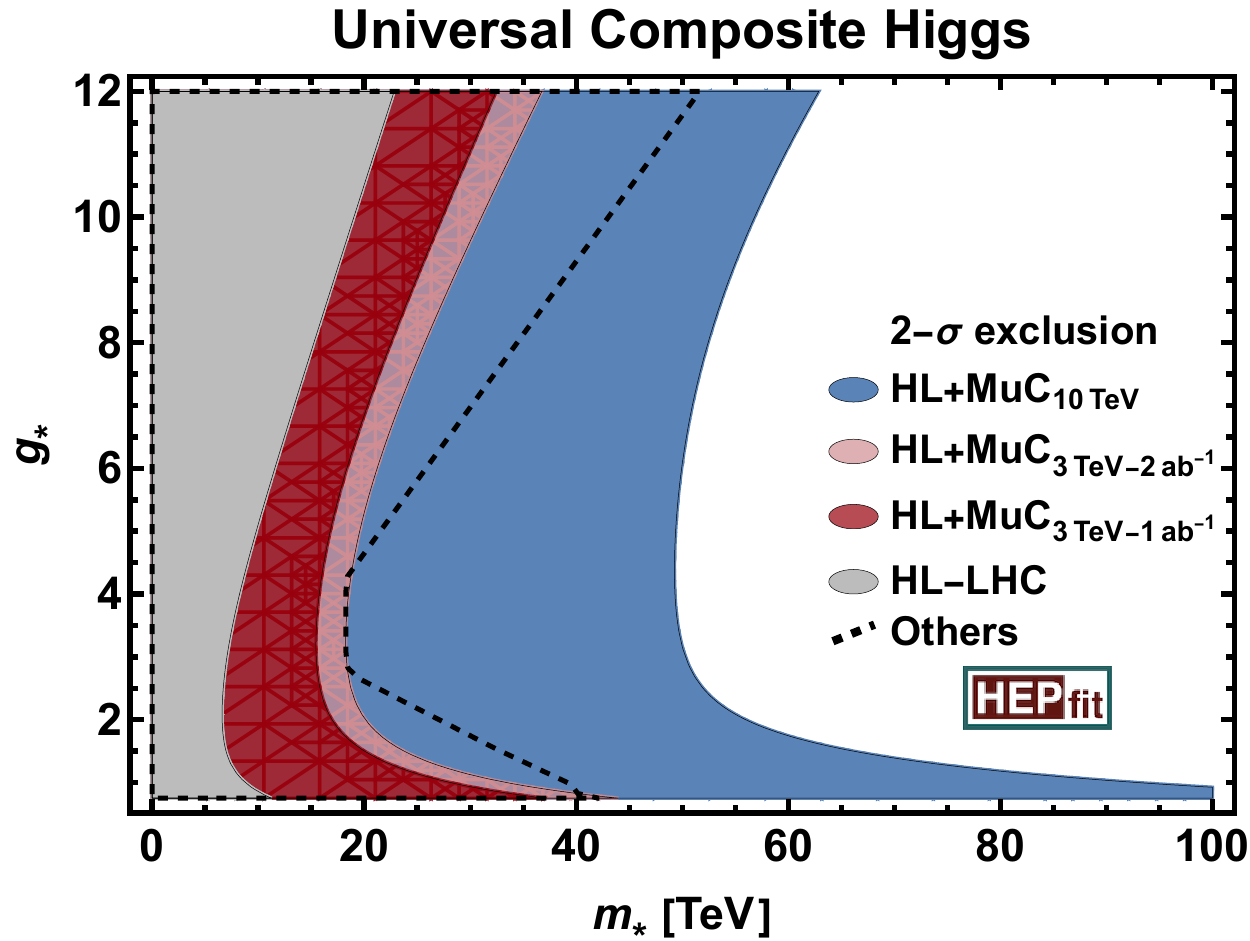}
&
\includegraphics[width=0.49\linewidth]{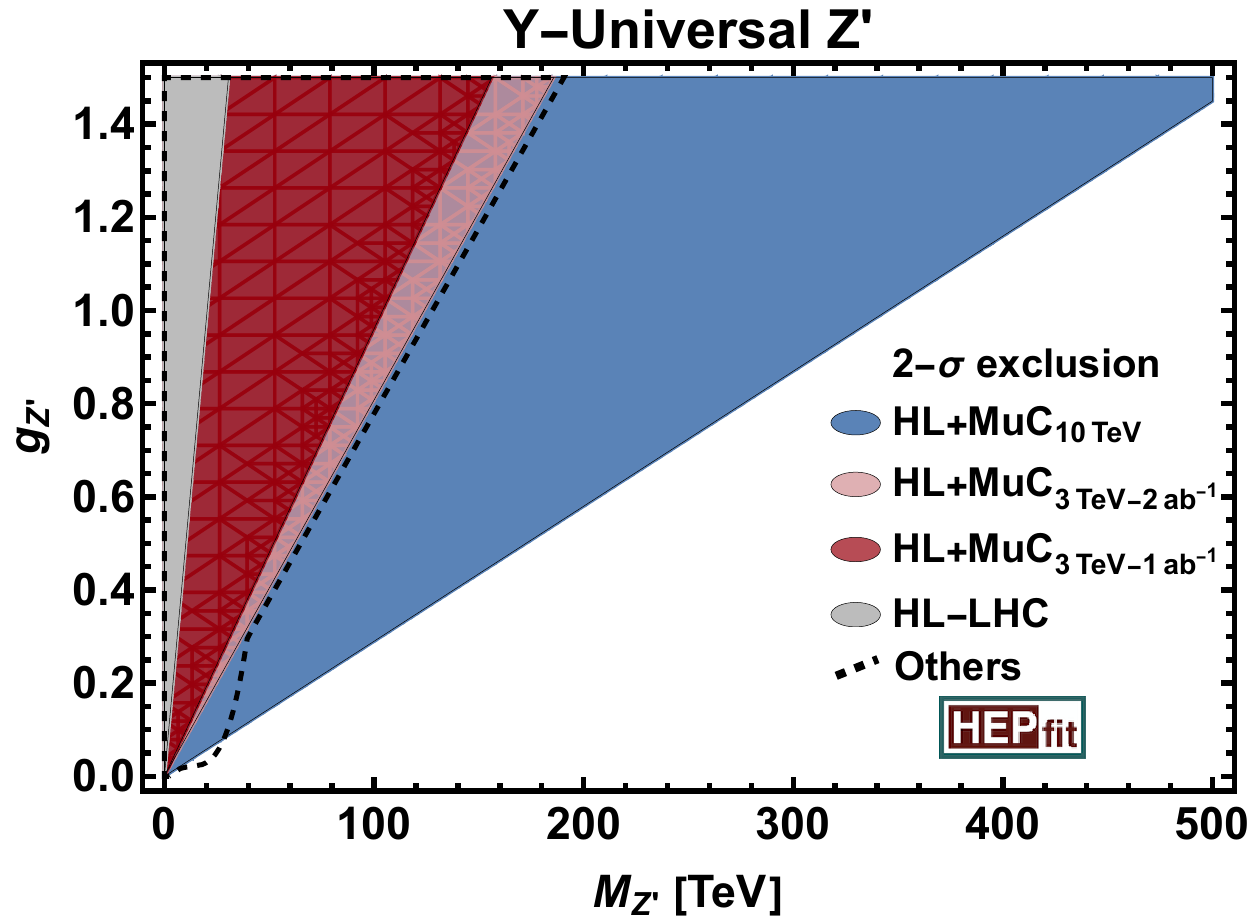}\\
\end{tabular}
\caption{\label{fig:CH_Zp}
(Left) The global reach for universal composite Higgs models at the HL-LHC and a high-energy muon collider. The figure compares the 2-$\sigma$ exclusion regions in the $(g_\star, m_\star)$ plane from the fit presented in Figure~\ref{fig:SILH}, using the power-counting in eq.~(\ref{eq:SILHpc}).
(Right) The same for a model featuring with an heavy replica of the $U(1)_Y$ gauge boson in the $(g_{Z^\prime}, m_{Z^\prime})$.
}
\end{figure*}

In the EFT fits to eq.~(\ref{eq:LSilh}) we include the following set of experimental inputs and projections:
\begin{itemize}
{\item The complete set of electroweak precision measurements from LEP/SLD~\cite{ALEPH:2005ab}, including the projected measurements of the $W$ mass at the HL-LHC~\cite{Azzi:2019yne}.  We also include the anomalous trilinear gauge coupling constraints from LEP2.}
{\item The HL-LHC projections for single Higgs measurements and for double Higgs production from~\cite{Cepeda:2019klc}. We assume the S2 scenario for the projected experimental and theoretical systematic uncertainties.}
{\item Also from the HL-LHC, the projections from two-to-two fermion processes, expressed in terms of the W and Y oblique parameters, from Ref.\cite{Farina:2016rws}, and the high energy diboson study from~\cite{Franceschini:2017xkh}.}
{\item The expected precision for single-Higgs observables at the 3 and 10 TeV muon colliders from the results of~\cite{Forslund:2022xjq}, previously described.}
{\item As in the HL-LHC case, we also include the projections from high-energy measurements in two-to-two fermion processes, expressed in terms of W and Y from~\cite{Chen:2022msz}, and in diboson processes such as $\mu^+\mu^-\to ZH, W^+W^-$, $\mu \nu \to WH,WZ$ from Ref.~\cite{Chen:2022msz,Buttazzo:2020uzc}.}
{\item The di-Higgs invariant mass distribution in $\mu^+ \mu^- \to \bar{\nu}\nu HH$ from Ref.~\cite{Buttazzo:2020uzc} (see also \cite{Han:2020pif}), as a probe of the $c_\phi$ and $c_6$ operators.
}
\end{itemize}
In all cases we assume the projected experimental measurements to be centred around the SM prediction. The assumptions in terms of theory uncertainties follow the same setup as in \cite{deBlas:2019rxi}.

Our analysis ignores the recent CDF determination of the $W$ mass~\cite{CDF:2022hxs}, which is in strong tension with the SM interpretation of the electroweak data and requires BSM physics. If confirmed, the CDF anomaly will become a major target for studies of new physics in the electroweak sector, and in particular for the SM EFT investigations described in the present section. The specific opportunities offered by the muon collider for exploring a possible BSM origin of the CDF anomaly have not been studied yet.

The results of these EFT fits are summarised in Table~\ref{tab:eft-silh} and Figure~\ref{fig:SILH}. Relative to the HL-LHC, a reach improvement of one order of magnitude is found at the 10~TeV~MuC for several operators, among which ${\cal O}_\phi$ and ${\cal O}_6$, as shown in the lower panel of the figure. The improvement on the ${\cal O}_\phi$ operator stems from the accurate measurements of the single-Higgs couplings (dominantly, $HZZ$ and $HWW$) and on the measurement of the invariant mass in VBF double-Higgs production, where ${\cal O}_\phi$ induces an energy-growing deformation. The improvement on the ${\cal O}_6$ operator simply follows from the accurate determination of $\delta\kappa_\lambda$. As previously discussed, these measurements exploit vector bosons initiated processes, and their accuracy reflects the effectiveness, emphasised in Section~\ref{VBF}, of the muon collider as a vector boson collider. 

A two orders of magnitude improvement is instead found for operators such as ${\cal O}_{W,B}$ and ${\cal O}_{2B, 2W}$. These operators induce growing with energy effects in diboson and difermion processes respectively, therefore they are very effectively probed by high-energy measurements as explained in Section~\ref{HEM}. 

As in the $\kappa$ analysis, we must also note that the muon collider potential on some operators such as ${\cal O}_{y_u}$ could be underestimated, due to the absence of detailed projections for the processes that could probe them most effectively. In the case of ${\cal O}_{y_u}$, a combination of different Higgs and top-quark measurements may improve the sensitivity~\cite{Franceschini:2021aqd}. 

Finally, we remark that all the sensitivity projections included in the fit assume unpolarised  muon beams. Polarised beams could bring extra information, allowing to test more directions in the SM EFT parameter space or resolving flat directions (see e.g.~\cite{Barklow:2017suo,DeBlas:2019qco}). For instance it was shown in~\cite{Chen:2022msz,Buttazzo:2020uzc} that the availability of polarised muon beams would further improve the reach on the ${\cal O}_{W,B}$ operators from diboson high-energy measurements, even for limited beam polarisation fraction. The  feasibility of polarised beams is not addressed by the current design studies, but it is not excluded.

The 3~TeV MuC is less performant than the 10~TeV one, but still enables a jump ahead of one order of magnitude or more, relative to HL-LHC, for several operators such as ${\cal O}_{W,B}$, ${\cal O}_{2B, 2W}$ and ${\cal O}_{\phi W,\phi B}$. The impact of these advances on the exploration of concrete BSM scenarios is detailed below and compared with the perspective for the progress attainable at all the other future collider projects that are currently under consideration. 

\begin{figure*}[ht]
\centering
\begin{tabular}{c c}
\hspace{-0.2cm}\includegraphics[width=0.49\linewidth]{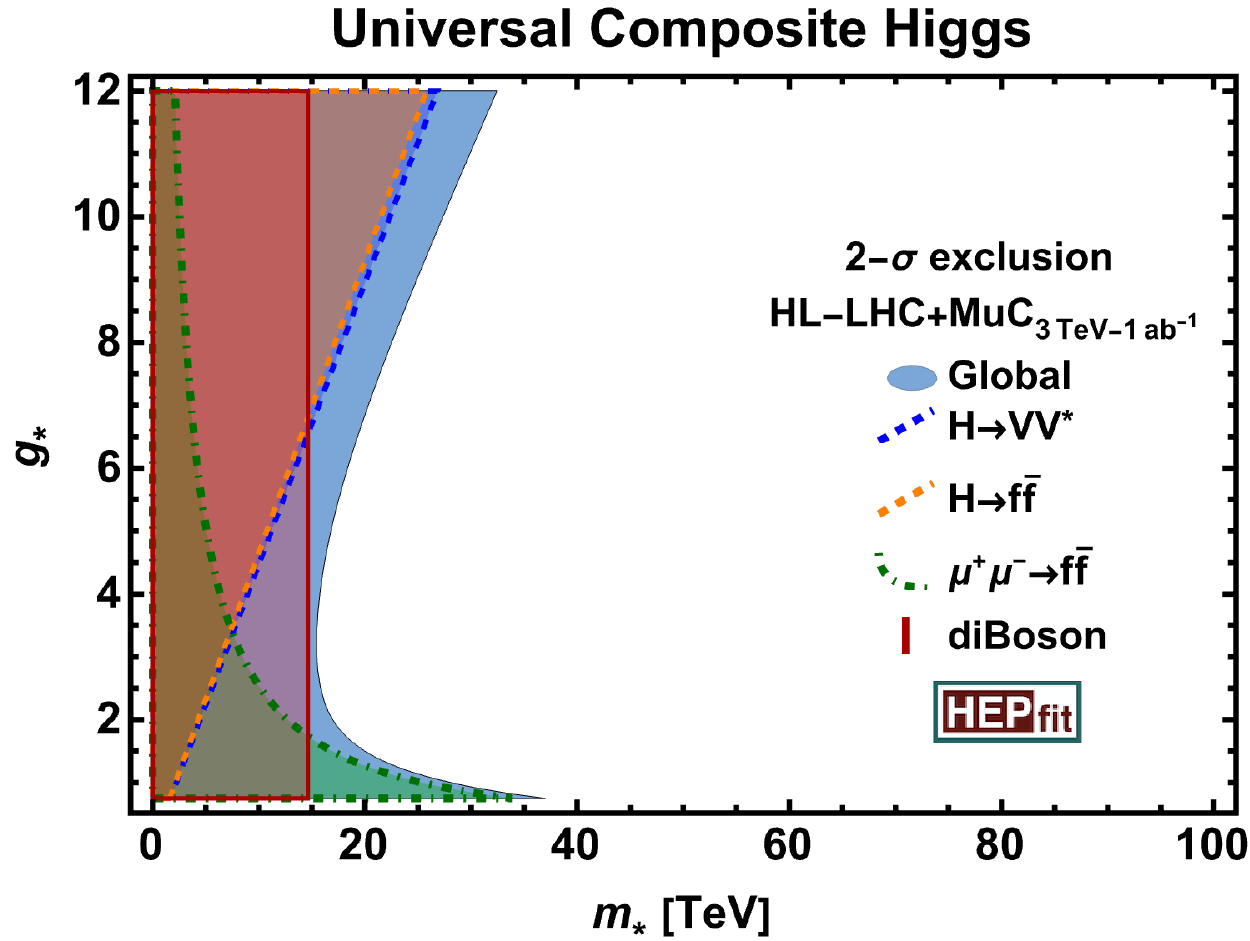}
&
\includegraphics[width=0.49\linewidth]{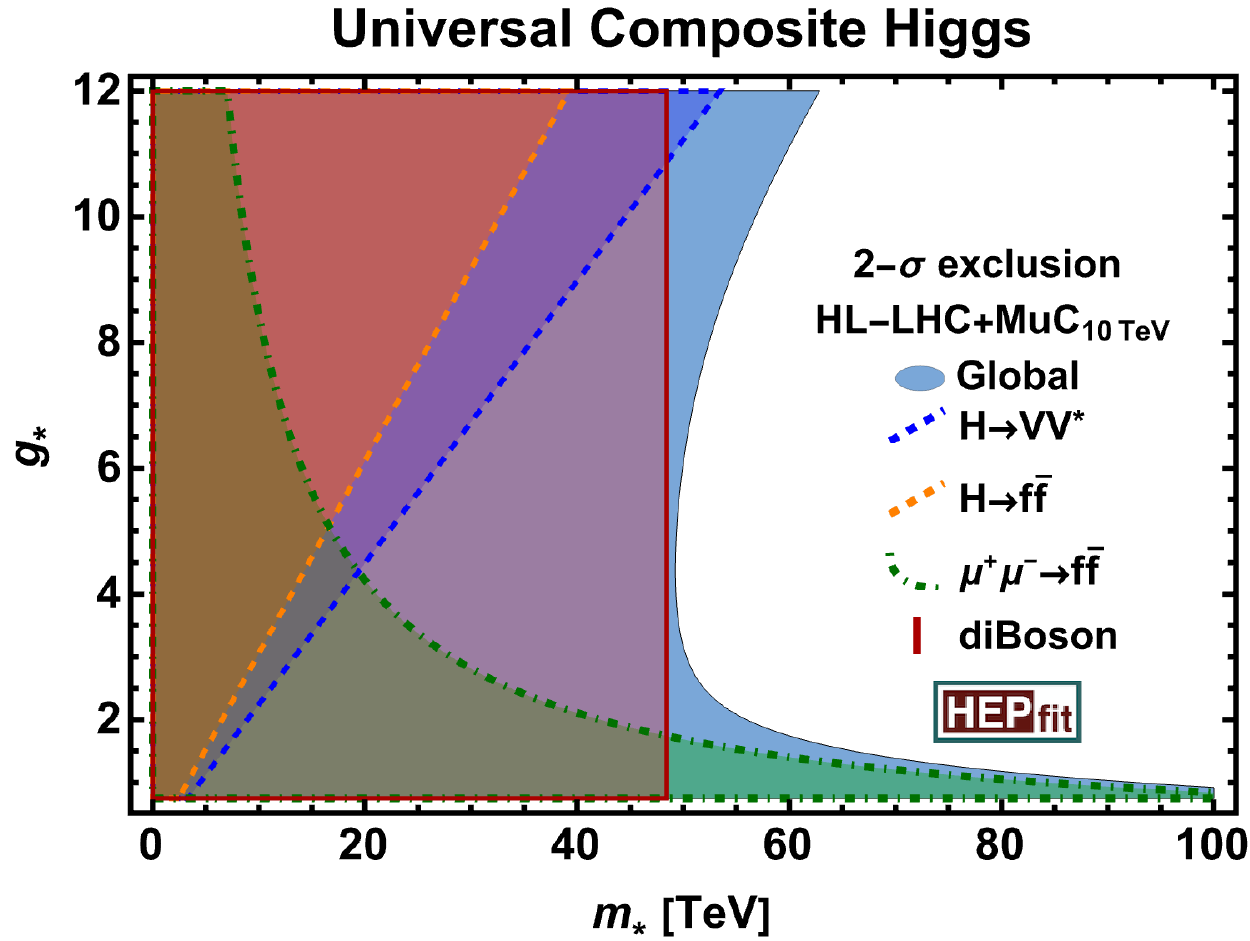}\\
\end{tabular}
\caption{\label{fig:SILH_gstar}
The breakdown of the global reach on composite Higgs, reported in Figure~\ref{fig:CH_Zp}, in the contribution of the individual processes. The 3 and the 10~TeV muon colliders are considered in the left and right panels, respectively.
}
\end{figure*}

\paragraph{BSM benchmark interpretations}{\ } \\ 
\noindent
The fit results can be used to perform sensitivity projections on specific new physics scenarios, enabling a concrete illustration of the muon colliders potential for indirect BSM searches. We consider a composite Higgs scenario and a simple $Z^\prime$ model, to be discussed in turn. 

Composite Higgs is described under the assumption that the new dynamics is parameterised in terms of a single coupling, $g_\star$, and mass, $m_\star$. Furthermore, as in \cite{deBlas:2019rxi}, we assume for definiteness that the new physics contributions to the operator Wilson coefficients in~(\ref{eq:LSilh}) follow the power-counting formula with unit numerical coefficients, namely we take
\begin{eqnarray}
&&\frac{c_{\phi,6,y_f}}{\Lambda^2}= \frac{g_\star^2}{ m_\star^2},\;\;\;\;\;
\frac{c_{W,B}}{\Lambda^2}= \frac{1}{m_\star^2},
\;\;\;\;\;
\frac{c_{2W,2B}}{\Lambda^2}= \frac{1}{g_\star^2}\frac{1}{m_\star^2},
\nonumber\\
&&
\frac{c_{T}}{\Lambda^2}= \frac{y_t^4}{16\pi^2}\frac{1}{m_\star^2},\;\;\;\;\;
\frac{c_{\gamma,g}}{\Lambda^2}= \frac{y_t^2}{16\pi^2}\frac{1}{m_\star^2},
\nonumber\\
&&
\frac{c_{\phi W,\phi B}}{\Lambda^2}= \frac{g_\star^2}{16\pi^2}\frac{1}{m_\star^2},
\frac{c_{3W}}{\Lambda^2}= \frac{1}{16\pi^2}\frac{1}{m_\star^2}.
%
\label{eq:SILHpc}
\end{eqnarray}
The operators above define the so-called ``Universal'' manifestation of the Composite Higgs scenarios. Additional effects that depend on the degree of compositeness of the top quark, not considered here, have been studied in Ref.~\cite{Chen:2022msz}.

By projecting the EFT likelihood onto the $(g_\star,m_\star)$ plane we obtain the exclusion regions in the right panel in Figure~\ref{fig:CH_Zp} for the different muon collider options, combined and in comparison with the HL-LHC reach. The results agree with those of Figure~\ref{fig:HEM}.\footnote{The HL-LHC exclusion is in fact stronger than in Figure~\ref{fig:HEM} at large $g_\star$. This is because the conservative $c_\phi$ limit (see the footnote in Table~\ref{tab:eft-silh}) is  employed in Figure~\ref{fig:HEM}.}

We also show the EFT fit results interpretation in a simple BSM model featuring a single $Z^\prime$ massive vector boson. As in \cite{EuropeanStrategyforParticlePhysicsPreparatoryGroup:2019qin}, we consider a $Z^\prime$ coupled to the hypercharge current. In this case the dimension-6 effective Lagrangian only receives tree-level contributions to the operator with coefficient $c_{2B}/\Lambda^2=g_{Z^\prime}^2/(g^{\prime~\!4} M_{Z^\prime}^2)$. The corresponding indirect constraints in the $(g_{Z^\prime},M_{Z^\prime})$ plane are shown in the right panel of Figure~\ref{fig:CH_Zp}. 

While the bounds on the $Z^\prime$ model is obviously dominated by the high-energy measurements of difermion process, and the resulting constraints on the Y parameter (i.e., the ${\cal{O}}_{2B}$ operator coefficient), the situation is more complex for composite Higgs. The contributions from the different processes in setting the limits are shown separately in Figure~\ref{fig:SILH_gstar}, highlighting (see also~\cite{Chen:2022msz,Buttazzo:2020uzc}) the complementarity of different probes. The diboson constraints set the overall mass reach, independently of $g_\star$. The reach gets extended for low values of $g_\star$ by the difermion measurements. For high $g_\star$, $c_\phi$ bounds from Higgs coupling determinations and from the di-Higgs mass distribution measurement dominate the sensitivity.

The muon colliders sensitivity to composite Higgs, and to a $Z^\prime$ model that is representative of new physics affecting the electroweak interactions, was emphasised already in Section~\ref{HEM}. In particular we pointed out that the 10~TeV muon collider is more effective than any other future collider project that is currently under design or consideration. This can be seen also in Figure~\ref{fig:CH_Zp}, by comparing with the dashed line labelled as ``Others''. This line is formed by the envelop of the contours probed in the same plane by the FCC programme (including FCC-ee and FCC-hh), by all stages of CLIC and ILC, and all other collider projects studied in~\cite{EuropeanStrategyforParticlePhysicsPreparatoryGroup:2019qin} for the 2020 updated of the European Strategy for Particle Physics. The mass reach would further improve in proportion to the muon collider energy, provided the luminosity scales quadratically with the energy as in eq.~(\ref{lums}). 

Figure~\ref{fig:CH_Zp} also shows that the 3~TeV MuC sensitivity is similar to the one of the most effective alternative project (namely the FCC, including the FCC-hh), and vastly superior to the one of the HL-LHC. The figures conservatively assumes 1~ab$^{-1}$ integrated luminosity. EFT studies like those described in this section thus provide strong physics motivations for the first stage of the muon collider with a centre of mass energy of around 3~TeV.

\begin{figure*}[ht]
\begin{centering}
\includegraphics[width=0.67\linewidth]{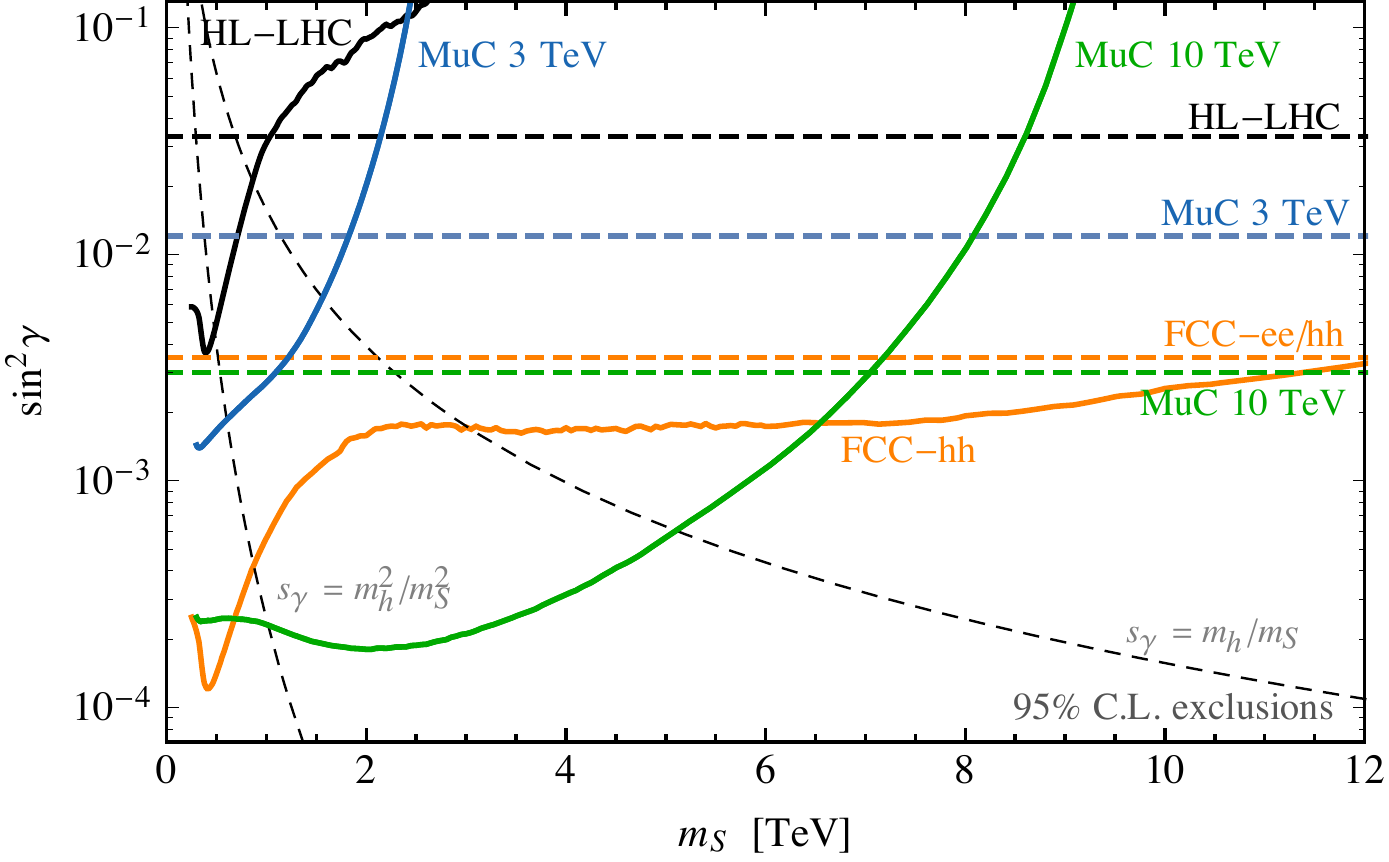}
\caption{95~\% 
C.L. reach (adapted from~\cite{Buttazzo:2018qqp}) on heavy singlet mixed with the 
SM Higgs doublet at the muon collider. The direct and indirect reach at other future projects and at the HL-LHC, documented in~\cite{EuropeanStrategyforParticlePhysicsPreparatoryGroup:2019qin}, is also shown for comparison.\label{fig:Singlet-reach}}
\par\end{centering}
\end{figure*}

\subsubsection*{Extended Higgs sectors}
The third exploration strategy by which muon colliders can advance knowledge of the electroweak and Higgs sector is to search directly for the resonant production of new particles. Few illustrative results are reported below, focusing in particular on BSM models that foresee an extension of the Higgs sector by additional scalar fields. The new particles in these scenarios do not carry QCD colour, therefore the mass reach of the LHC is intrinsically limited. The muon collider is thus generically expected to improve radically over the HL-LHC sensitivity projections already at the 3~TeV stage.

\paragraph{SM plus singlet}{\ } \\ 
\noindent
The simplest extension of the SM Higgs sector features one additional scalar field in the singlet of the SM gauge group. Its simplest interaction with the SM, if not forbidden by additional symmetries, is the ``Higgs portal'' (see also Section~\ref{dirr}) trilinear coupling $S\,H^\dagger H$ with the SM Higgs doublet $H$. By this coupling, the singlet mixes with the physical Higgs boson $h$. Following~\cite{Buttazzo:2018qqp}, we denote as $\sin\gamma$ the mixing parameter and as $m_S$ the physical mass of the singlet particle. We trade the fundamental Lagrangian parameter for $m_S$ and $\sin\gamma$, and we study the model in the $(m_S,\sin^2\gamma)$ plane as in Figure~\ref{fig:Singlet-reach}.

The mixing with the singlet scales down all the couplings of the Higgs particle by $\cos\gamma \simeq 1-1/2\,\sin^2\gamma$. It has the same effect as the ${\cal{O}}_\phi$ operator with $c_\phi / \Lambda^2 =\sin^2\gamma/v^2$. Therefore, the indirect sensitivity to $\sin^2\gamma$ can be read from Table~\ref{tab:eft-silh} and it is reported in Figure~\ref{fig:Singlet-reach} for the 3~TeV muon collider with 2~ab$^{-1}$ and for the 10~TeV MuC. The indirect sensitivity follows the general pattern previously described. The 3~TeV MuC improves Higgs coupling determination slightly and hence it slightly improves the HL-LHC sensitivity to $\sin^2\gamma$, by a factor 3 in this specific case. The 10~TeV MuC performances are comparable to the ones of an $e^+e^-$ Higgs factory, leading to the same reach on $\sin^2\gamma$.

Unlike low-energy Higgs factories, the muon collider also offers opportunities to produce the new scalar particle directly. The production occurs dominantly via the VBF $VV \rightarrow S$ process, that benefits from the large luminosity for effective vector bosons at the muon collider, as already discussed in Section~\ref{dirr}. The single production vertex $S\,V_L V_L$ involves longitudinally polarised vectors and it emerges directly from the $S\,H^\dagger H$ trilinear interaction via the Equivalence Theorem. The same interaction mediates the dominant decays of the singlet, to two massive vectors or to a pair of Higgs boson. The latter channel with Higgs decays to bottom quarks, namely $S \rightarrow hh \rightarrow 4b$, is the most sensitive one at high energy lepton colliders and it is employed~\cite{Buttazzo:2018qqp} in the direct reach estimates presented in Figure~\ref{fig:Singlet-reach}. 

\begin{figure*}
\centering{}\includegraphics[width=0.45\linewidth]{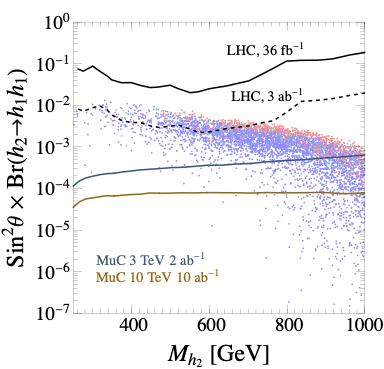}
\includegraphics[width=0.43\linewidth]{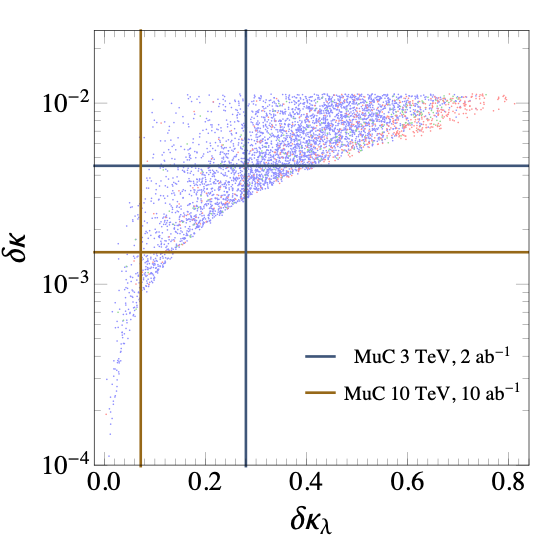}
\caption{\label{fig:EWpt-reach}Direct (left panel) and indirect (right panel) reach on the SM plus real scalar singlet scenario at the muon collider.  Dots indicate points with successful first-order EWPT, while red, green and blue dots represent signal-to-noise ratio for gravitational eave detection in the ranges $[50, +\infty)$, $[10, 50)$ and $[0, 10)$, respectively. Results adapted from~\cite{Liu:2021jyc}.
 }\end{figure*}

By combining the two blue lines (horizontal dashed and continuous), and comparing with the black lines, we can appreciate the potential progress of the 3~TeV MuC with respect to the HL-LHC.\footnote{The indirect HL-LHC sensitivity is most likely overestimated, as the aggressive bound on $c_\phi$ (see the footnote in Table~\ref{tab:eft-silh}) is employed in the figure.} We also notice a considerable region that can be only probed indirectly at the HL-LHC. In that region the presence of the singlet would produce tensions in the HL-LHC measurements of the Higgs couplings and the physics underlying such tension will be discovered directly at the 3~TeV MuC. 

The reach is greatly extended by the 10~TeV MuC. It covers almost all the region that could be probed by the combination (black dotted lines) of the FCC-ee and FCC-hh direct and indirect searches. Furthermore its reach extends to much higher mass in the region of small mixing angle. This improvement is of particular significance in light of the fact that a smaller mixing should be expected to emerge at higher mass, in concrete realisations of the singlet scenarios. This is illustrated in the figure by the grey dashed lines that correspond to two possible power-counting estimates for the scaling of the microscopic Lagrangian parameters. See~\cite{Buttazzo:2018qqp} for details.

The SM plus singlet model does not only provide a simple benchmark for future colliders comparison. It can mimic the signatures, and be reinterpreted, in strongly motivated BSM scenarios like Supersymmetry and Twin Higgs models~\cite{AlAli:2021let,Buttazzo:2018qqp}. ``Non-minimal'' (but plausible) Composite Higgs models could also feature a singlet or other scalar extensions of the Higgs sector~\cite{Gripaios:2009pe,Mrazek:2011iu}. SM plus a real singlet extension can also provide a strong first order ElectroWeak Phase Transition (EWPT), which is an essential ingredient for the electroweak baryogenesis mechanism potentially responsible for the matter-antimatter asymmetry in the Universe. Following~\cite{Liu:2021jyc} (see also~\cite{Ruhdorfer:2019utl}), we illustrate below the muon collider potential to probe this scenario.

In the left panel of Figure~\ref{fig:EWpt-reach}, the coloured solid curves show the muon collider 95\%~C.L. direct exclusion reach in the plane formed by the singlet mass and the product $\sin^2\gamma\times{\rm{BR}}(S\to hh)$.\footnote{The latter quantity, and not only $\sin^2\gamma$, is what controls the events yield in the di-Higgs final state. The branching ratio is set to $1/4$ in Figure~\ref{fig:Singlet-reach}, which is a good approximation when the singlet is heavy but not so in the mass range of Figure~\ref{fig:EWpt-reach}.} The points marked on the figure are obtained from a scan over the microscopic parameters of the specific model considered in Ref.~\cite{Liu:2021jyc}, and they correspond to configurations where the EWPT is of the first order and strong enough for electroweak baryogenesis. The 3~TeV MuC covers several of the relevant points, while the 10~TeV MuC enables an almost complete coverage. The points marked in red or in green (unlike those in blue) could perhaps also produce observable gravitation waves at LISA. Strong first order EWPT requires a modification of the Higgs potential. Therefore sizeable departures of the trilinear Higgs coupling with respect to the SM are expected in this scenario. This is shown on the right panel of Figure~\ref{fig:EWpt-reach}, in the plane formed by a universal modifier $\delta\kappa$ that affects all the single-Higgs couplings, and the trilinear coupling modifier $\delta\kappa_\lambda$. We see that the muon collider, already at the 3~TeV stage, has considerable chances to be sensitive to the predicted single- or triple-Higgs coupling modifications. It is in fact likely to observe correlated modifications in both couplings.

\begin{figure*}[ht]
\centering{}\includegraphics[width=0.45\linewidth]{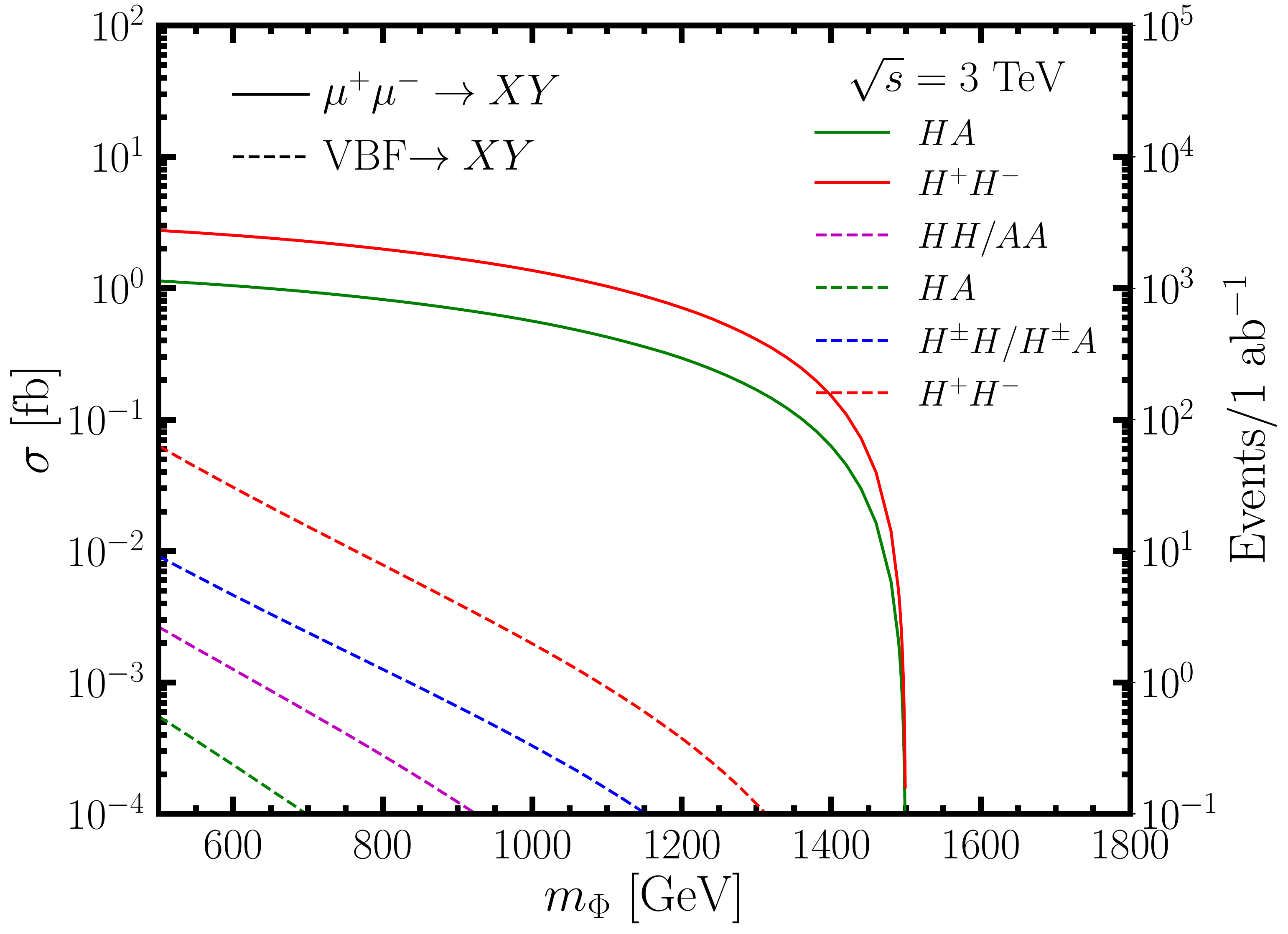}\hfill
\includegraphics[width=0.45\linewidth]{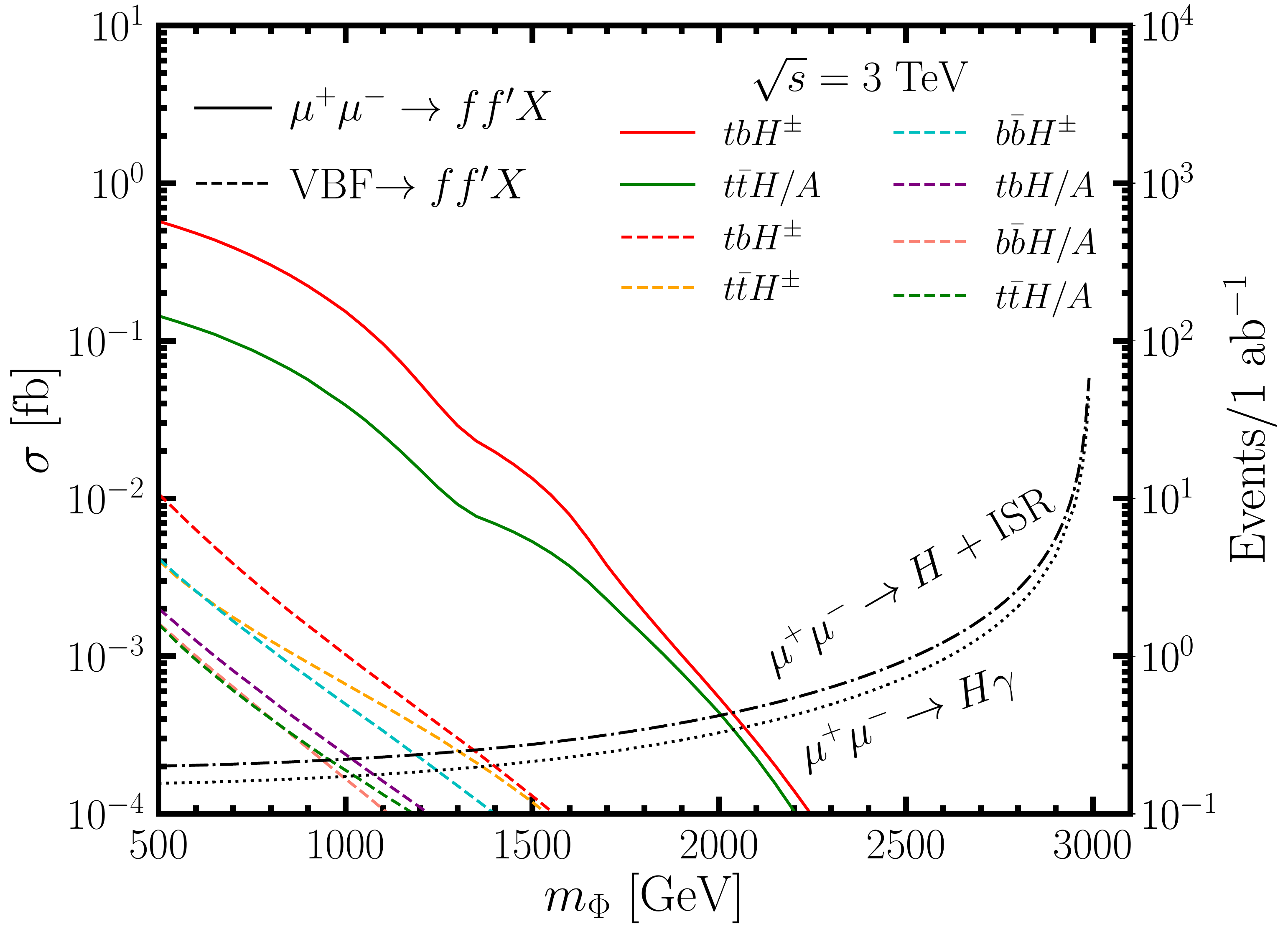}
\caption{\label{fig:Charged-reach}
Cross sections versus the  non-SM Higgs mass for pair production (left panel), single production with a pair of fermions and radiative return production (right panel) at the 3~TeV muon collider. The value of $\tan\beta=1$, and the alignment limit $\cos(\alpha-\beta)=0$, is considered in the figure.}
\end{figure*}

\paragraph{Two Higgs Doublet Model}{\ } \\ 
\noindent
Models with two Higgs doublets (2HDM) are another important target for muon colliders.~\footnote{Other extensions of the Higgs sector, in particular those featuring a doubly-charged scalar, were studied in~\cite{Rodejohann:2010jh,Li:2023ksw}} While much work is still to be done for the detailed assessment of the muon collider potential, a rather complete characterisation of the relevant phenomenology was provided in Ref.~\cite{Han:2021udl}, whose findings are briefly summarised below. Like in the case of the singlet model, very significant progress on the 2HDM parameters space is possible already at the first 3~TeV stage of the muon collider. In what follows we stick to this energy for definiteness. At the higher energies muon colliders, which are also considered in~\cite{Han:2021udl}, the performances improve.

The scalar sector of the 2HDM consists of 5 physical particles: the SM-like Higgs $h$ with  $m_h=125$~GeV and the non-SM ones $H,A,H^\pm$. The tree-level couplings of the Higgs bosons are determined by the mixing angle between the neutral CP-even Higgs bosons, $\alpha$, and by a second parameter $\tan\beta=v_2/v_1$, with $v_{1,2}$ being the vacuum expectation value for two Higgs doublets. The dominant couplings of the Higgses with the SM gauge bosons typically involve two non-SM Higgses, for example, $ZHA$ or $W^\pm H^\mp H$.  The Yukawa couplings of the non-SM like Higgses with the SM fermions depends on how the two Higgs  doublets  are  coupled  to  the  leptons  and  quarks via Yukawa couplings. Four different patterns of Yukawa couplings are typically considered in the literature, giving rise to four different types of 2HDMs that we denote as Type-I, Type-II, Type-L and Type-F, following the notation of Ref.~\cite{Branco:2011iw}. 

The couplings of the SM Higgs bosons are generically modified in the 2HDM, potentially producing observable effects at the HL-LHC. In this event, the 3~TeV MuC could access directly the new heavy bosons, likely too heavy to be observed at the LHC, and establish the origin of the putative discrepancy. If instead the HL-LHC Higgs couplings measurements will not deviate from the SM predictions, the masses of the extra scalars could still be within the reach of the 3~TeV MuC, and even well below if the model parameters are in the so-called ``alignment limit'' where $\cos(\alpha-\beta)\approx 0$ and the Higgs coupling modifications are suppressed. In both cases it will be interesting to search directly for the new scalars. In what follows we thus discuss the 3~TeV MuC direct discovery potential, focusing on nearly ``aligned'' parameters configurations.

The heavy Higgs bosons can be produced in pairs via the direct $\mu^+\mu^-$ annihilation mediated by the exchange of a virtual photon or $Z$ boson, or via Vector Boson Fusion (VBF)
\begin{eqnarray}
\mu^+\mu^-\to H^+ H^-
,\qquad\mu^+\mu^-\to HA,
&&\nonumber \\
 V_1V_2\to H^+H^-, HA, H^\pm H/H^\pm A, HH/AA. &&
\label{eq:anni}
\end{eqnarray}
In the latter case, the process proceeds by the collinear emission of $V_{1,2}=W,Z$, or $\gamma$, out of the incoming muons. The cross section for the different processes at the 3~TeV MuC as a function of the new scalars mass is shown on the left panel of Figure~\ref{fig:Charged-reach}. The annihilation processes dominates for masses above around 500~GeV. The VBF cross section steeply increases, and it would dominate at lower masses that however could be probed also at the HL-LHC and thus are not worth discussing here. It is worth emphasising that the situation is different at the muon collider with 10~TeV center of mass energy or more. In that setup, VBF channels become more important and dominate at low mass~\cite{Han:2021udl}.

The figure shows that a very large number of events is expected with 1~ab$^{-1}$ integrated luminosity. Furthermore, considering the dominant decay channel of non-SM Higgs into third generation fermions, the SM backgrounds can be easily suppressed. Reach up to pair production threshold $m_\phi<1.5$~TeV is thus generically possible. A detailed comparison of the 3~TeV muon collider reach with the HL-LHC sensitivity is not yet available. For Type-II 2HDM~\cite{Craig:2016ygr}, the 3~TeV muon collider reach is superior in the intermediate region of $\tan\beta \in [2, 10]$. For larger $\tan\beta$, a reach above 1.5~TeV mass has already been attained at the LHC~\cite{ATLAS:2022yvl}.

In the parameter region with enhanced Higgs Yukawa couplings, single production of non-SM Higgs with a pair of fermions could play an important role and potentially extend the sensitivity above the pair production threshold. The production cross section for fermion associated production are shown in the right panel of Figure~\ref{fig:Charged-reach} for both the annihilation and VBF processes, with $\tan\beta=1$ and $\cos(\alpha-\beta)=0$. The dominant channel is $tbH^\pm$, followed by $t\bar{t}H/A$. Note that there is a strong dependence on $\tan\beta$ of the production cross section, depending on the types of 2HDM~\cite{Han:2021udl}. Finally, the radiative return process $\mu^+\mu^- \to \gamma H$ offers another production channel for the non-SM Higgs, which is relevant in regions with enhanced $H\mu^+\mu^-$ coupling. The cross section increases as the heavy Higgs mass approaches the collider c.m.~energy, closer to the $s$-channel resonant production. The production cross section is shown as the black curves in the right panel of Figure~\ref{fig:Charged-reach}. 

\subsection{Dark matter}\label{secdm}
A global assessment of the MuC perspectives to search for Dark Matter (DM) is not yet available. The studies so far, reviewed below, investigated the MuC potential to probe scenarios where DM is a particle charged under the EW interactions and its observed abundance in the Universe emerges from the thermal freeze-out mechanism. This is a compelling possibility, and a one where the MuC will play a major role. However, the opportunities to probe other interesting scenarios for DM, either through muon collisions or with parasitical experiments, or as a byproduct of the muon collider demonstration program, should be also investigated.

If DM is a WIMP (i.e., a Weakly Interacting Massive Particle), it could be directly detected in ultra-low noise underground detectors by its interaction with the detector material. It can be also searched  in DM-rich astrophysical environments, where the DM pairs can annihilate and give rise to observable signals at cosmic ray observatories. These experimental investigation strategies have been and are being actively pursued, and are promising but suffer few potential roadblocks. Cosmic rays observation can be hampered by   large uncertainties on   astrophysical quantities and astrophysical processes that can mimic DM signals. Furthermore, these experiments exploit DM particles that are physically present in the Universe, whose local density is poorly known. This entails strong uncertainties on the expected signal. For instance, a recent analysis in Ref.~\cite{Rinchiuso:2020skh} quantifies the effect of density profile uncertainty for WIMP searches and finds order of magnitude effects on the (still very promising) cross section sensitivities of future generation experiments such as CTA~\cite{CTA:2020qlo}. Lab-based direct detection is less affected by profile uncertainties. But it suffers from being a very low momentum transfer process---even when DM is quite heavy---which makes background rejection very challenging. Future experiments like the Xenon's upgrades~\cite{Aalbers:2022dzr} will have sensitivity to a large variety of possible WIMP candidates, but also blind spots. See e.g.~\cite{Bottaro:2021snn,Bottaro:2022one} for an appraisal.

If DM is a WIMP, it can also be produced in the laboratory through EW interactions, provided a particle collider of sufficient energy and luminosity is available. The possibility of producing putative DM candidates and studying their signatures with precise particle detectors would, in the first place, firmly establish or conclusively exclude their existence. Furthermore, it will offer unique opportunities for the characterisation of the newly discovered DM particle. The vibrant forthcoming programme of non-collider DM searches, and the possible signals of WIMP DM that may emerge, thus provide additional motivations for collider studies. 

WIMP searches are also useful as sensitivity benchmarks to gauge the effectiveness of particle colliders to test DM, and to compare different collider projects. In fact, the WIMP scenario assumptions single out a relatively small set of compelling minimal benchmark models with no free parameters~(see~\cite{Bottaro:2021snn} and references therein). This is because the WIMP relic abundance is set by the (known) strength of the weak interactions coupling and the (unknown) mass of the WIMP. Therefore, for minimal models where the WIMP consists of a single ${\textrm{SU}}(2)_{L}$ $n$-plet it is possible to sharply predict the mass of the dark matter particle that produces the observed relic. Some examples are reported in Table~\ref{tab:thermal-masses}. As a general rule, the larger the $n$-plet the larger the mass of the WIMP. Smaller masses can be attained if the $n$-plet mixes with a state of lower multiplicity, e.g. a singlet. Therefore, testing the mass reach when only one ${\textrm{SU}}(2)_{L}$ $n$-plet is present as in the minimal models effectively demonstrates sensitivity to non-minimal models as well, as it demands to reach the highest mass for a given class of candidates.

\begin{table}[t]
\vspace{10pt}
\begin{centering}
\begin{tabular}{c|c||c|c|c|c}

$n$ & Y & Dirac & Majorana & Complex & Real\tabularnewline
\hline 
\hline 
2 & 1/2 & 1.08 & - & 0.58 & -\tabularnewline
\hline 
3 & 0 & -& 2.86
& - & 2.53
\tabularnewline
3 & $\epsilon$ & 2.0 \& 2.4 & -
& 1.6 \& 2.4 &-
\tabularnewline
3 & 1 & 2.84(14) & -
& 2.1 & -
\tabularnewline
\hline 
4 & 1/2 & 4.79(9) & - & 4.98(5) & -\tabularnewline
\hline 
5 & 0 & - & 13.6(5) & - & 15.4(7)\tabularnewline
5 & $\epsilon$ & 9.1(5) & - & 11.3(7) & -\tabularnewline
5 & 1 & 9.9(7) &- & 11.5(7) &-\tabularnewline
\hline 
\end{tabular}
\par\end{centering}
\caption{\label{tab:thermal-masses}Thermal mass, in TeV, for pure \mbox{SU$(2)_L$} $n$-plet dark matter WIMP, from Ref.~\cite{Bottaro:2021snn,Bottaro:2022one}. Some of the candidates are endowed with a tiny hypercharge $\epsilon$. Effects of bound states and Sommerfeld enhancement of the annihilation cross section are included in the calculation of the thermal mass.
}
\end{table}

A crucial phenomenological parameter for the detection of WIMPs at colliders is the mass splitting between the lightest, neutral component of the $n$-plet, which constitutes the actual DM particle, and the other electrically charged and neutral components of the multiplet. When this mass splitting is greater than some threshold, typically around 10~GeV, it is relatively easy to observe the production of the heavier states in the $n$-plet. These particles decay to the actual DM particle, which is invisible, accompanied however by easily detectable SM particles. If instead the mass splitting is below the threshold, one needs to rely on ``Mono-X'', on ``Indirect'' or on ``Disappearing Tracks'' strategies for detection. We describe these strategies in turn in the rest of this section.

\subsubsection*{Mono-X}

\begin{figure*}
\centering{}
\includegraphics[width=0.47\linewidth]{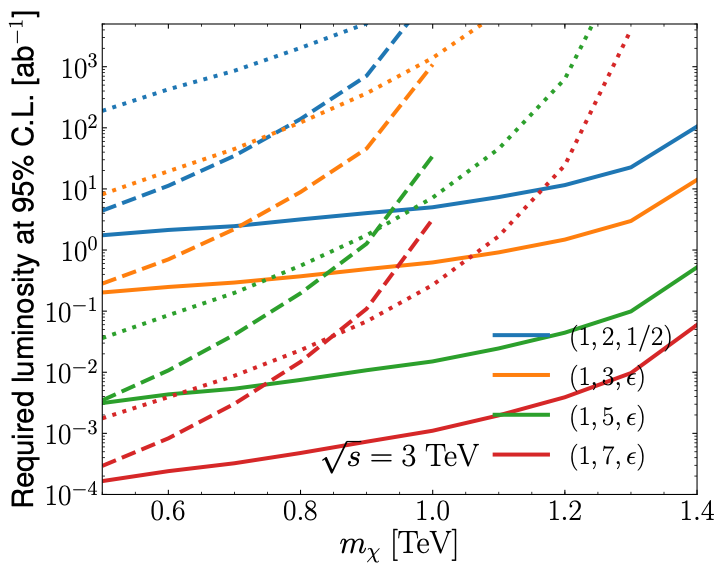}
\includegraphics[width=0.47\linewidth]{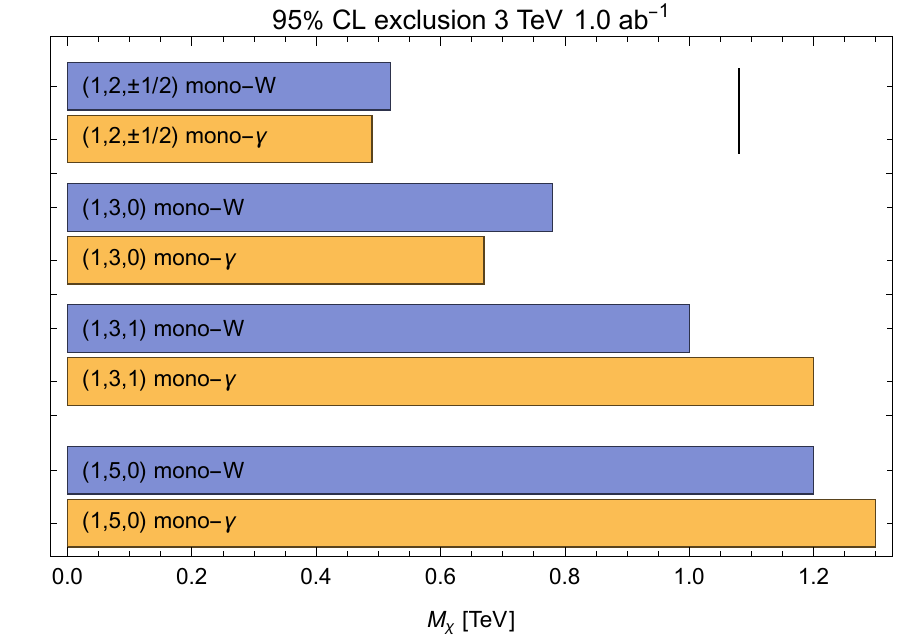}\\
\includegraphics[width=0.47\linewidth]{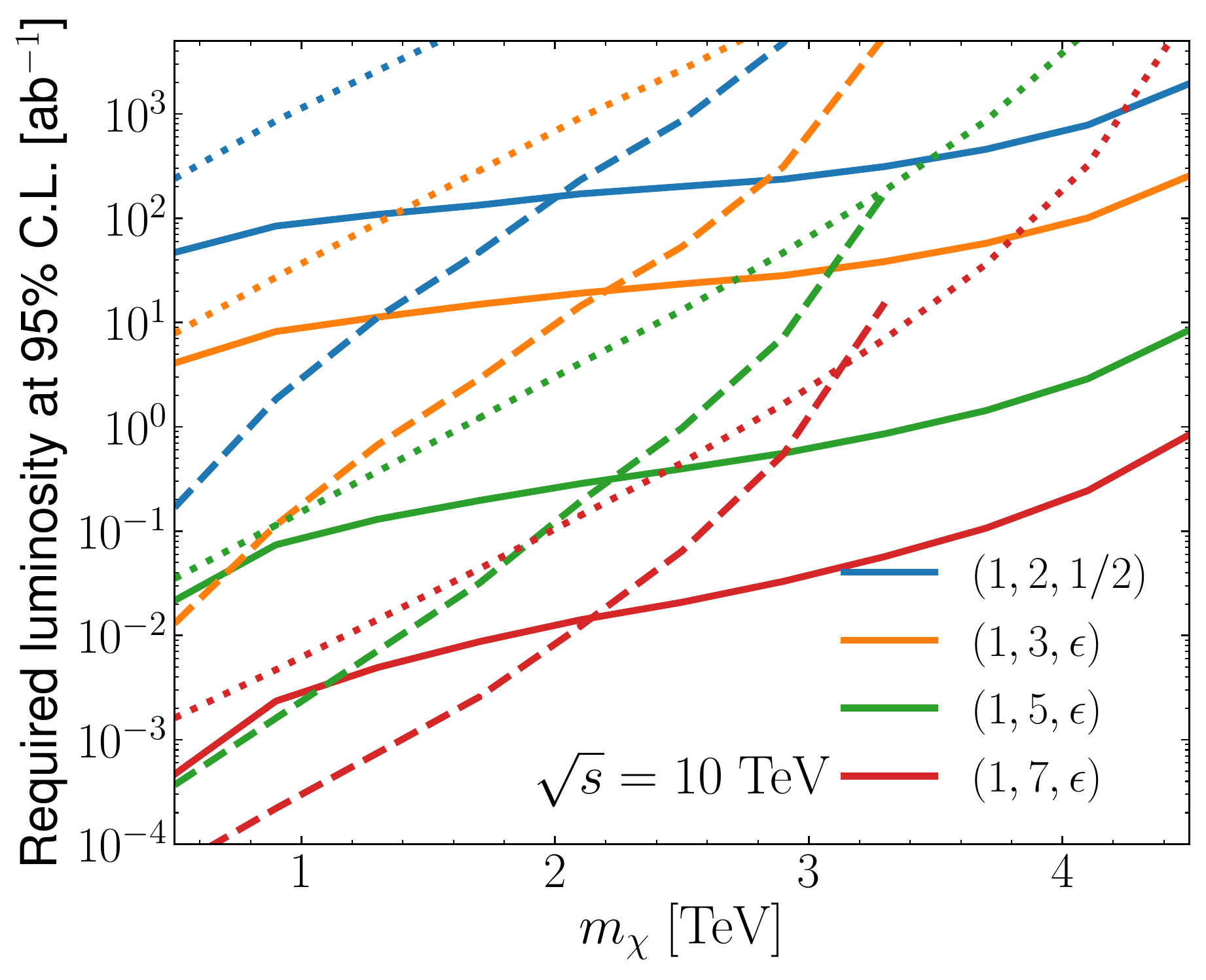}
\includegraphics[width=0.47\linewidth]{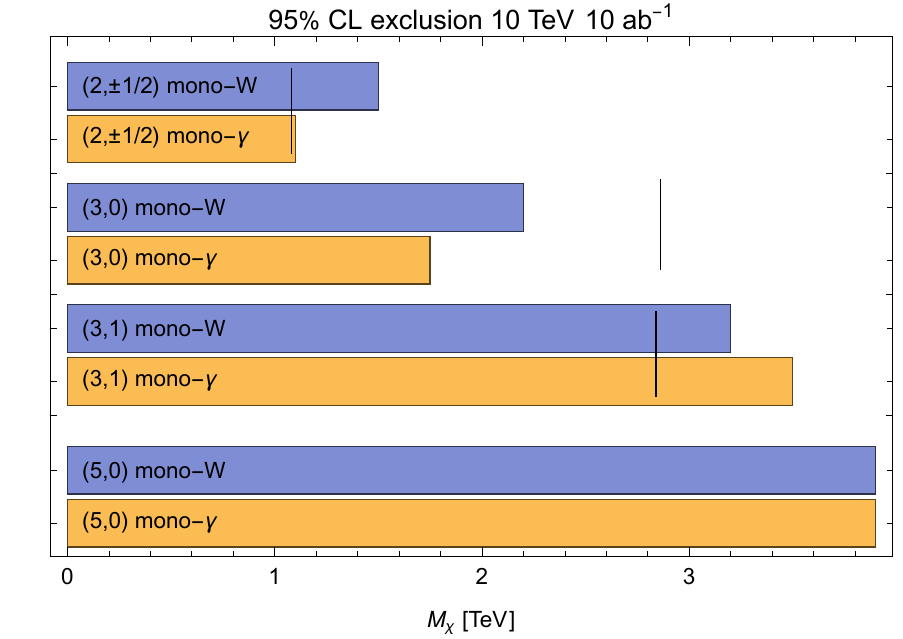}\\
\caption{\label{fig:EWstates}Direct reach on electroweak states in mono-$X$ signals. Left: Luminosity needed to exclude  a Dirac fermion DM candidate for zero systematics~\cite{Han:2020uak} for $X=\gamma$ (solid), $X=\mu$ (dotted), $X=\mu\mu$ (dashed). Right: Mass reach on a fermionic DM candidate (assumed Majorana when $Y=0$, Dirac otherwise) at fixed 1~${\rm{ab}^{-1}}$ luminosity for the 3 TeV and 10~${\rm{ab}^{-1}}$ for 10 TeV muon collider in the channels $X=\gamma$ and $X=W$ for 0.1\% systematics~\cite{Bottaro:2021snn,Bottaro:2022one}. Black vertical lines denote the thermal mass for each
DM candidate.}
\end{figure*}

The Mono-X strategy is to observe DM production ``by contrast'', i.e. to observe a bunch of particles apparently recoiling against nothing.
At a muon collider the relevant reaction is
\begin{equation}
    \mu^{+}\mu^{-} \to \chi \chi +X \,,\label{eq:monoX}
\end{equation}
where $X$ denotes any SM particle or set of particles allowed by the interactions and $\chi$ is either the DM particle or a generic state belonging to the DM $n$-plet. In both cases, $\chi$ is not seen if the mass splitting is low and no dedicated strategy is adopted to detect disappearing track as discussed later in this section. 

\begin{figure*}
    \centering
    \includegraphics[width=0.45\textwidth]{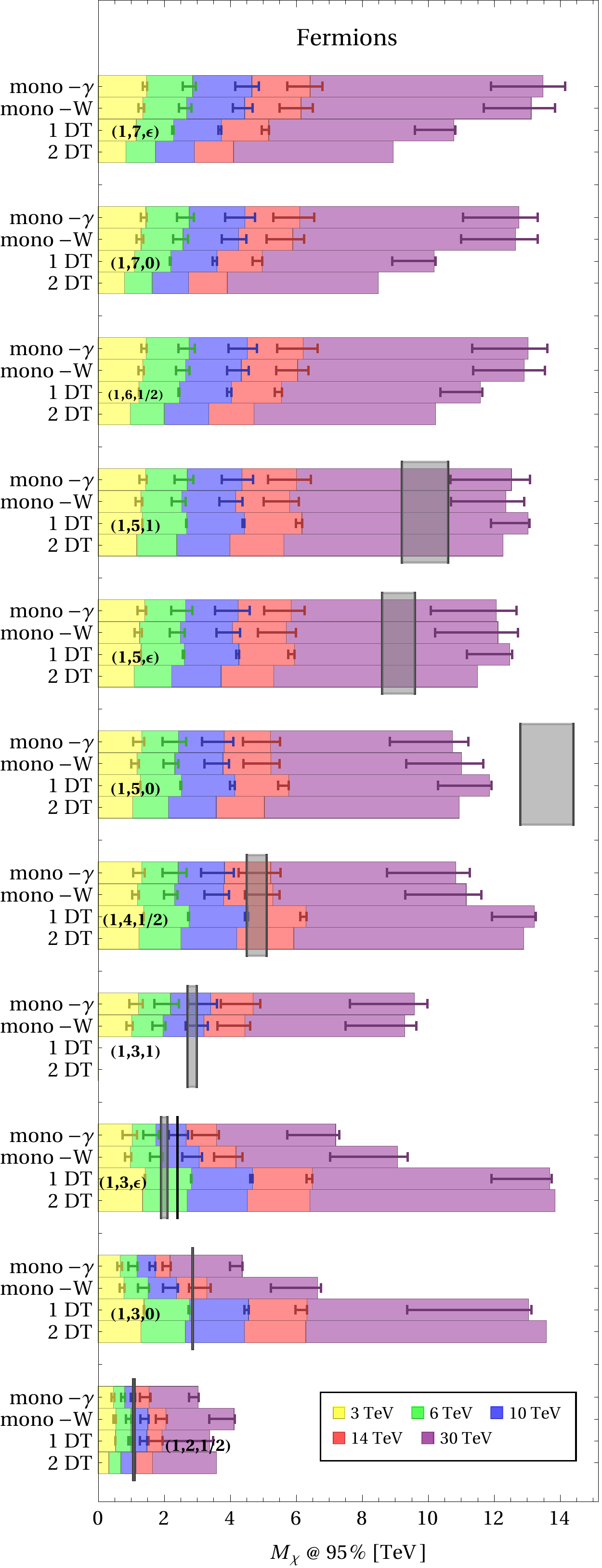}\hfill
        \includegraphics[width=0.45\textwidth]{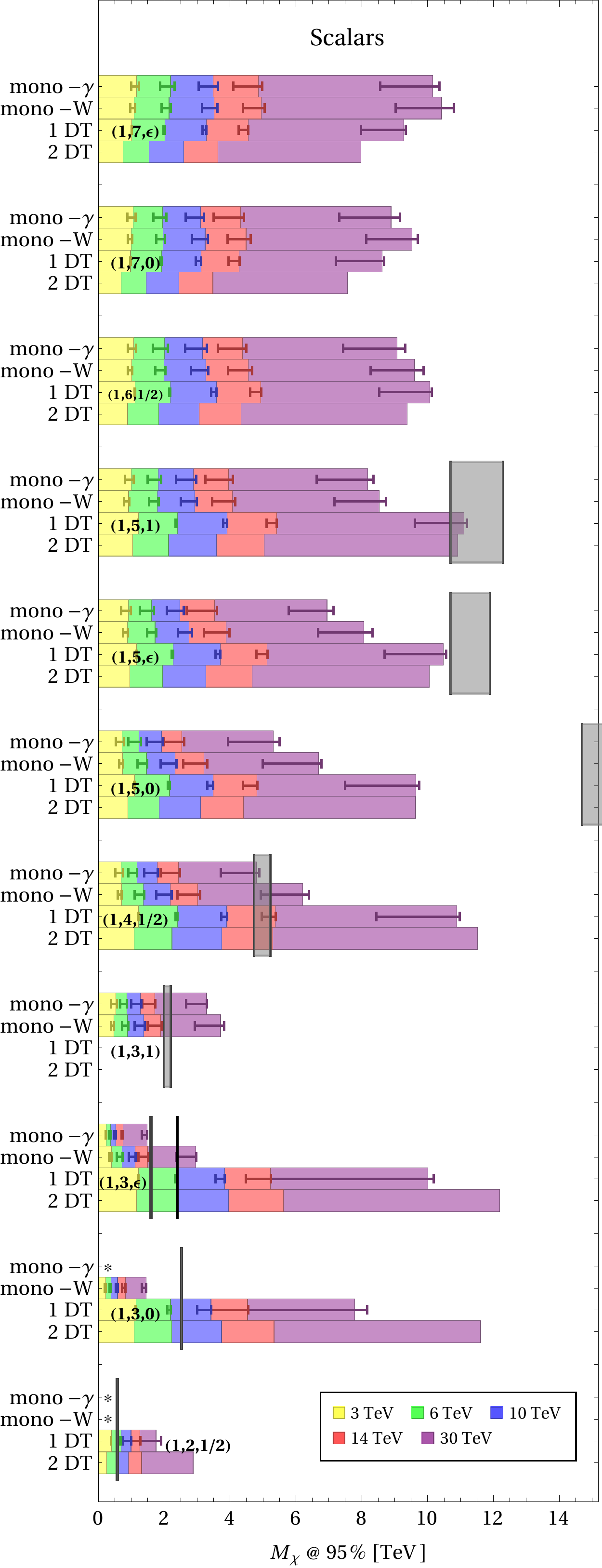}
    \caption{Mass reach in the mono-$\gamma$, mono-$W$ and DT channels with luminosity scaling with energy as in~\eqref{lums} at muon colliders of different energy $\sqrt{s}$, from Ref.~\cite{Bottaro:2022one}. Vertical bars display the thermal mass of the candidates with its uncertainty. In the mono-$W$ and mono-$\gamma$ searches we show an error bar, which covers the range of possible exclusion as the systematic uncertainties are varies from 0 to 1\%. For single displaced tracks the   error bar covers a possible systematic uncertainty from 0\% to 10\%. The coloured bars are for an intermediate choice of systematics at 0.1\% (1\% for 1DT). 
    Missing bars denoted by an asterisk * correspond to cases where no exclusion can be set in the mass range $M_\chi>0.1\sqrt{s}$. For such cases would be worth considering also the VBF production modes.}
    \label{fig:summary_all_colliders}
\end{figure*}

Searches for generic electroweak states have been studied for several types of observable particles $X$ accompanying the production of dark matter. The signal for $X=\gamma,W,Z,\mu^{\pm}$ and $\mu\mu$ have been studied in  \cite{Han:2020uak,Bottaro:2021snn}. Figure~\ref{fig:EWstates} summarises the reach illustrating on the left panels the luminosity needed to attain the 95\% CL exclusion of electroweak matter of a given mass in the production modes $X=\gamma,\mu,\mu\mu$. Among these, the mono-$\gamma$ search is the one placing the best bound for states heavier than about 500~GeV. The right panels summarise the reach with 1~${\rm{ab}^{-1}}$ and 10~${\rm{ab}^{-1}}$ for the 3 and the 10~TeV MuC (upper and lower panel). The mono-$W$ channel reach is reported on the right panel plots, together with mono-$\gamma$. This channel is effective for the same mass range in which mono-$\gamma$ leads the exclusion and in some cases exceeds mono-$\gamma$ results. All in all, the combination of mono-$\gamma$ and mono-$W$ dominates the mono-X strategy sensitivity and provide best mass reach for some DM candidates.

The 10~TeV MuC results show that at this energy it is possible to exclude fully a Dirac doublet DM at the thermal mass $1.1$~TeV in both the mono-$\gamma$ and the mono-$W$ channels. This specific WIMP candidate, known as ``higgsino'', is a widely-studied target for future colliders because of several reasons, including the fact that its detection is challenging for direct detection experiments based on scattering on Xenon nuclei~\cite{Bottaro:2022one}. Another popular candidate is the Majorana triplet with $2.9$~TeV mass, known as ``wino''. The 10~TeV MuC cannot access this candidate through Mono-X. Detection is instead possible through disappearing track searches as discussed later in this section.

A grand summary encompassing higher energies,  heavier thermal relic DM candidates and including the Disappearing Tracks (DT) reach is displayed in Figure~\ref{fig:summary_all_colliders}. We see that at energies above 10~TeV the muon collider starts probing  WIMP DM $n$-plets with $n>2$. 

It is worth noticing that the detection of the Mono-X processes in eq.~(\ref{eq:monoX}) is affected by large SM backgrounds that originate, for instance, from the production of invisible neutrinos in place of $\chi$. The sensitivity is thus limited by the accuracy of the SM background predictions that need to be subtracted from the observed data. The error bars in Figure~\ref{fig:summary_all_colliders} report the mass reach for a relative accuracy ranging from $0\%$ (perfect background prediction) to $1\%$. We see that $1\%$ systematic uncertainties on the background prediction significantly degrades the sensitivity. Uncertainties at the $0.1\%$ level would be desirable and are assumed in the coloured bars. The possibility to attain such accurate predictions deserves further investigation.

\subsubsection*{
Indirect reach through high-energy measurements}

Pure WIMP DM $n$-plets have a mass that scales roughly as $M_{\chi}\sim n^{5/2}$, as Table~\ref{tab:thermal-masses} shows, and can reach values above the kinematical threshold for resonant production of any collider we can imagine to build in the near and medium-term future. Indeed, masses around a fraction of 100~TeV  can be achieved, e.g. the thermal mass for a Majorana 7-plet is around 50~TeV. For large $n$ it becomes questionable if the weak interactions are still so weak, indeed scattering rates involving large $n$-plets end up being as large as 10\% of the  maximum allowed by unitarity for $n=9$ and a Landau pole for weak interactions emerges at energies less than two orders of magnitude from the mass of the $n=9$ dark matter candidate~\cite{Bottaro:2021snn,Bottaro:2022one}. Therefore, for $n\geq 9$ we do not possess a valid EFT description of the minimal scenario for DM in which the WIMP multiplet is the only new particle beyond the Standard Model. The SM plus a WIMP for $n\geq 9$ can be considered, but it should be interpreted as very rough sketch of some other non-minimal theory of dark matter that features more dynamics than what it is entailed by just adding one particle to the SM.

Keeping this caveat in mind, the overall picture is clear: WIMPs can be good dark matter candidates over a large range of masses from the TeV to the PeV. A muon collider programme where the energy is progressively raised in stages can probe the resonant production of heavier and heavier candidates, but a complete coverage of the heaviest WIMPs will have to exploit off-shell effects in precision measurements of SM processes, with a mass reach that is potentially above the direct production threshold. This indirect search strategy can be effectively pursued at the muon collider through the measurement of high energy cross sections~\cite{DiLuzio:2018jwd}. These measurements benefit from a boosted sensitivity to new physics above the collider reach as explained in Section~\ref{HEM}, and they fall in the same category of those employed in Section~\ref{sec:Higgs} for EFT searches.

In what follows we consider as  concrete examples the dark matter candidates with $Y=0,1/2,1$  listed in Table~\ref{tab:thermal-masses}. We refer to~\cite{Franceschini:2022sxc} for a study encompassing a larger set of candidates. The search strategy that we adopt leverages the observable effects that DM candidates can leave due to  their propagation as virtual states, which modify the rate and the distributions of SM processes such as 
\begin{eqnarray}
\mu^{+}\mu^{-} &\to& f \bar{f}\,, \label{proc:ffbar}\\
\mu^{+}\mu^{-} &\to& Z h \,, \label{proc:Zh} \\
\mu^{+}\mu^{-} &\to& W^{+}W^{-}\,, \label{proc:WW}
\end{eqnarray}
as well as $2\to3$ processes like
\begin{eqnarray}
    \mu^{+}\mu^{-}&\to& WWh\,,
    \label{proc:WWh}\\
\mu^{+}\mu^{-} &\to& f \bar{f^\prime} W\,. \label{proc:ffprimebar}
\end{eqnarray}
Measuring the total rate of eqs.(\ref{proc:ffbar}-\ref{proc:ffprimebar}) and using differential information on the angular distribution of the channels in which the charge of the final states, e.g. $f=e,\mu$,  can be tagged reliably, it is possible to probe the existence of new matter $n$-plets.

\begin{figure*}
\begin{centering}
\includegraphics[width=0.49\linewidth]{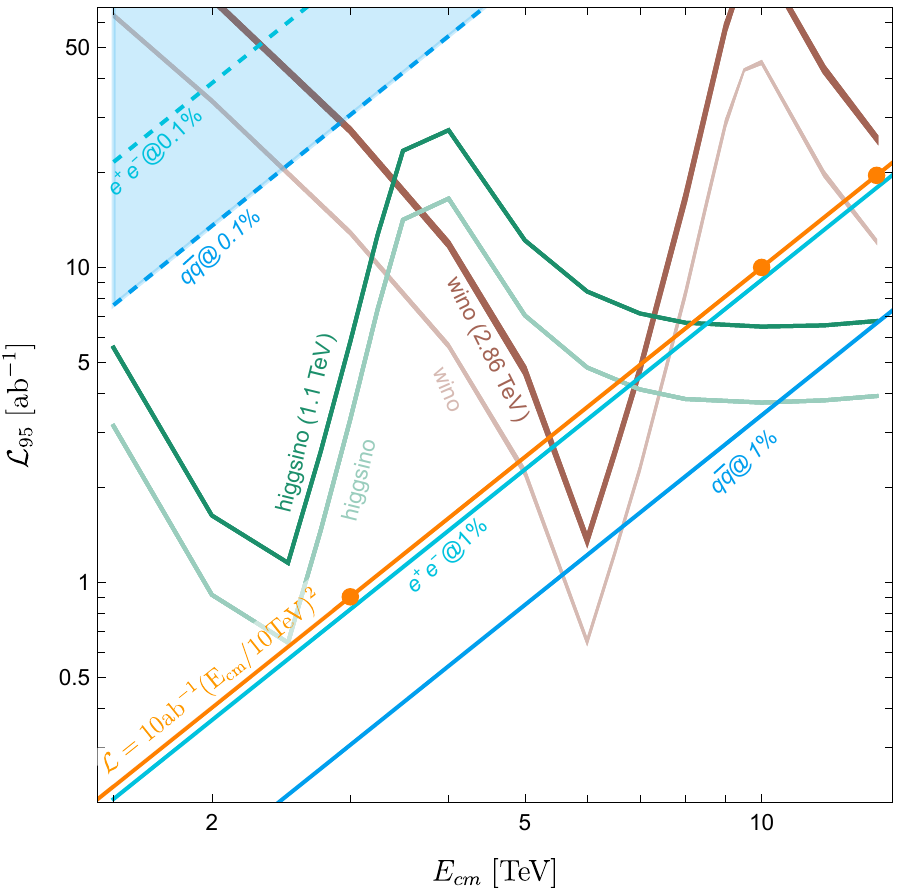}
\includegraphics[width=0.49\linewidth]{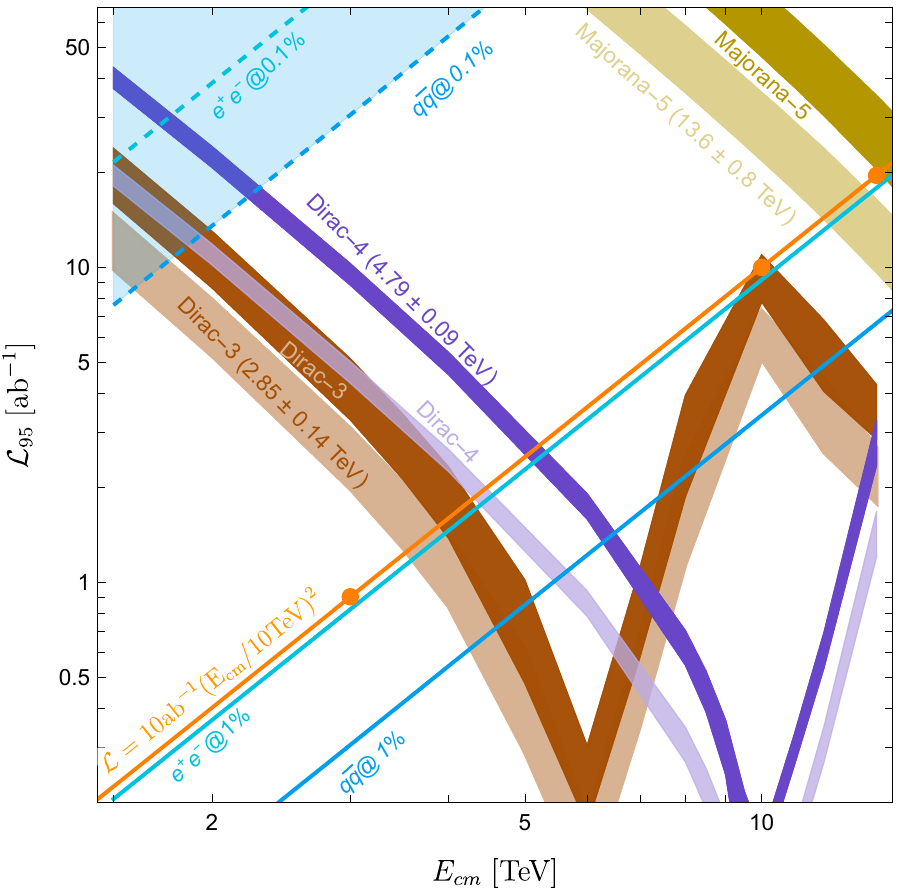}
\par\end{centering}
\caption{\label{fig:Indirect-reach-3plets} Minimal luminosity to exclude a thermal pure higgsino or wino dark matter (left panel) a 2.84~TeV Dirac triplet, 4.79~TeV Dirac 4-plet, a 13.6~TeV Majorana 5-plet (right panel) as function of the collider center of mass energy~\cite{Franceschini:2022sxc}. Lighter color lines are for polarized beams. The thickness of the wino, Dirac 3-plet, Dirac 4-plet, and Majoarana 5-plet bands covers the uncertainty on the thermal mass calculations. Diagonal lines mark the precision at which the total rate of the labeled channels are going to be measured. The shaded area indicates that at least one channel is going to be measured with 0.1\% uncertainty and systematic uncertainties need to be evaluated.}
\end{figure*}

It should be noted that the $2\to3$ processes cross sections, while formally of higher order in the EW loop expansion, are not suppressed relative to the $2\to 2$ cross sections, at the high energy MuC. This is a manifestation of the EW radiation enhancement that we described in Section~\ref{EWRadiation}. The enhancement emerges in the phase-space region where a $W$ boson is emitted with low energy and collinear to one of the initial muons or to one of the two other final state particles, which are instead energetic and central. The EW radiation enhancement offers novel opportunities to search for new physics. In the case at hand, it enables the high-rate production of new hard 2-body final states (namely $Wh$ and $f\bar{f^\prime}$ for eq.~(\ref{proc:WWh}) and (\ref{proc:ffprimebar}), respectively) to be exploited for WIMP searches. However, EW radiation effects also challenge theoretical predictions as they require not yet available systematic resummation techniques, as discussed in Section~\ref{EWRadiation}.  The estimates that follow do not include resummation. More work will thus be needed to turn them into fully quantitative sensitivity projections.

In Figure~\ref{fig:Indirect-reach-3plets} (left panel) we report the minimal luminosity needed to exclude a thermal pure wino DM (i.e., a Majorana triplet with $2.9$~TeV mass) and the higgsino (a Dirac doublet of $1.1$~TeV) as a function of the collider centre of mass energy. The luminosity curves feature a minimum around the direct production threshold $E_{{\rm{cm}}}= 2 M_{\chi}$, which provides the optimal energy for detection. For smaller $E_{{\rm{cm}}}$, more luminosity is needed as the effect of virtual $n$-plets decreases as $E_{{\rm{cm}}}^{2}/M_{\chi}^2$ below the production threshold. A larger luminosity is also needed moving above the threshold because the loop function (see~\cite{DiLuzio:2018jwd}) that describes the virtual DM exchange in the di-fermion final state happens to cross zero for some value of $E_{{\rm{cm}}}$ above $2\, M_{\chi}$. After crossing this second threshold, the required luminosity smoothly decreases with $E_{\rm{cm}}$ as the figure shows. A similar behaviour is observed for the other candidates considered in the right panel of Figure~\ref{fig:Indirect-reach-3plets}.

These studies are helped by the presence of left-handed fermions initial states, which source larger weak-boson mediated scattering. Therefore it is interesting to study the effect of beam polarization. In Figure~\ref{fig:Indirect-reach-3plets} the lighter colored lines give the necessary luminosity for an exclusion at a machine capable of 30\% left-handed polarization on the $\mu^{-}$ beam and -30\% for the $\mu^{+}$ beam. Even this modest polarization of the beams can reduce significantly the luminosity required for the exclusion.

When precision studies are involved it is important to keep in sight a possible bottleneck from systematic uncertainties. The origin of systematic uncertainties is difficult to assess at this stage, as there is not yet a fully developed experiment design. We identify the $0.1\%$ level as a possible reasonable level at which systematic uncertainties will need to be discussed. With this reference in mind we draw the shaded area of Figure~\ref{fig:Indirect-reach-3plets}, which indicates that the search for new electroweak matter is based on a high enough luminosity to have a statistical uncertainties of $0.1\%$ in the $q\bar{q}$ channel. As other channels are expected to be cleaner, and less statistically abundant, we reckon that in the shaded region of the plane a more careful evaluation of possible systematics needs to be performed before claiming a sensitivity. Along the baseline energy-luminosity line of eq.~(\ref{lums}), the required statistical precision is at the $1\%$ level, as the figure shows. The needs for accurate experimental measurements and theoretical predictions is thus reduced accordingly.

The left panel of Figure~\ref{fig:Indirect-reach-3plets} shows that the 10~TeV MuC with the baseline luminosity will probe the higgsino through indirect effects, on top of accessing it in Mono-X as we saw in the previous section. The wino is instead out of the indirect reach, but visible in mono-X. Interestingly, the 3~TeV MuC could access the higgsino with a luminosity slightly above $2~{\rm{ab}}^{-1}$, or by slightly lowering its centre of mass energy. The 3~TeV MuC seems instead unable to detect the higgsino with the Mono-X search strategy, according to the findings described in the previous section. 

The right panel of Figure~\ref{fig:Indirect-reach-3plets} displays the MuC sensitivity to higher-$n$ WIMP candidate multiplets. These are typically heavier than the higgsino and wino, but still they are easier to access indirectly because of the enhancement of the loop effects of large-$n$ electroweak multiplet. The 10~TeV MuC is thus sensitive to, for instance, a Dirac 4-plet with $4.8$~TeV thermal mass, very close to the kinematical threshold for resonant pair production. All fermionic WIMP candidates with $n=2$, 3 and 4 are also accessible with the baseline luminosity. Scalars in the same multiplets give dimmer signals and are hard to see unless higher luminosities are considered. A more promising way to observe the light scalars is the Mono-X search. Looking beyond 10~TeV, we see from  Figure~\ref{fig:Indirect-reach-3plets} that a collider in the 14~TeV ballpark can probe off-shell $n=5$ Majorana fermion dark matter with 14~TeV mass, way above the direct production threshold of 7~TeV. From Figure~\ref{fig:summary_all_colliders} we also see that a 14~TeV machine would be sensitive to  on-shell scalar $n=3$ $Y=1$ and to the scalar $n=4$ dark matter candidates. Looking at further high energies, results from Ref.~\cite{Franceschini:2022sxc} show that more candidates can be tested off-shell, e.g. the $n=7$ Majorana candidate, weighing about 50~TeV,  can be probed at a 30~TeV collider. Further dark matter candidate can be tested on-shell at larger $E_{{\rm{cm}}}$ as shown in Figure~\ref{fig:summary_all_colliders}. 

All in all the muon collider has a great potential to probe the idea of WIMP DM with several candidates that can be tested, some of which with multiple search strategies, strengthening the perspectives to establish a potential discovery. It should be noted however that both the strategies discussed so far would provide rather ``indirect'' signals. The Mono-X search is often considered as ``direct'' because the new particles are resonantly produced. However, neither the produced particles nor their decay products are detected. The signal would emerge as a small departure from the large SM background prediction in certain kinematical distributions, just as it would happen for an ``Indirect'' discovery based on loop effects. This is an additional motivation to investigate a truly direct detection strategy based on disappearing tracks, which is the subject of the following section. Another motivation is that disappearing tracks  searches display a better mass reach than Mono-X for several multiplets, as Figure~\ref{fig:summary_all_colliders} shows.

\subsubsection*{Unconventional signatures}

Disappearing tracks is one of the strategies to search for Long-Lived Particles (LLPs), which is among the priorities of the particle physics community~\cite{Curtin:2018mvb, Alimena:2019zri}. LLPs appear in a variety of new physics scenarios and yield a large range of signatures at colliders. Depending on the LLP quantum numbers and lifetime, these can span from LLP decay products appearing in the detector volume, even outside of the beam crossings, to metastable particles with anomalous ionisation disappearing after a short distance.

This wide range of ``unconventional'' experimental signatures is  intertwined with the development of detector technologies and their study can guide the design of the final detector layout. For example, the development of timing-sensitive detectors is crucial both to suppress the abundant beam-induced backgrounds and to detect the presence of heavy, slow-moving, particles that are travelling through the detector. A lively R\&D programme is ongoing to develop the reconstruction algorithms that will profit from these new technologies.

For heavy particles, whose production cross sections are dominated by the annihilation $s$-channel, there are two main features that make searches for unconventional signatures particularly competitive at a muon collider when compared to other future proposed machines like the FCC-hh. The produced particles tend to be more centrally distributed, impinging on the regions of the detector where reconstruction is comparatively easier, and furthermore they have a more sharply peaked Lorentz boost distribution, which can lead to effectively larger average observed lifetimes for the produced BSM states.

Searches for LLPs that decay within the volume of the tracking detectors (e.g. decay lengths between 1~mm and 500~mm) are particularly interesting as they directly probe the lifetime range motivated by compelling dark matter models. A rather detailed analysis including realistic simulation of the BIB from muon decays was performed in~\cite{Capdevilla:2021fmj} targeting higgsino and wino WIMPs, and it is summarised below. See~\cite{Han:2020uak} for simplified sensitivity estimates covering other candidates.

\paragraph{Search for disappearing tracks}{\ } \\ 
\noindent
The pure higgsino consists of a Dirac doublet with hypercharge $1/2$, with a thermal relic mass of 1.1~TeV. Due to loop radiative corrections, the charged state {\ensuremath{\tilde{\chi}^{\pm}}\xspace}\ splits from the neutral one {\ensuremath{\tilde{\chi}^{0}_{1}}\xspace}\ by 344~MeV, giving rise to a mean proper decay length of 6.6~mm for the charged state~\cite{Hisano:2006nn}. The {\ensuremath{\tilde{\chi}^{\pm}}\xspace}\ can thus travel a macroscopic distance before decaying into an invisible {\ensuremath{\tilde{\chi}^{0}_{1}}\xspace}\ and other low-energy Standard Model particles.

Searches at the LHC are actively targeting this scenario~\cite{ATLAS:2022rme,ATLAS:2017oal,ATLAS:2013ikw,CMS:2020atg,CMS:2014gxa}, but are not expected to cover the relic favoured mass~\cite{Saito:2019rtg, EuropeanStrategyforParticlePhysicsPreparatoryGroup:2019qin}. A muon collider operating at multi-TeV centre-of-mass energies could provide a perfect tool to look for these particles.

The production of {\ensuremath{\tilde{\chi}^{\pm}}\xspace}\ pairs  at a MuC proceeds via an s-channel photon or Z-boson, with other processes, such as vector boson fusion, being subdominant. The prospects to observe the disappearing track signal of {\ensuremath{\tilde{\chi}^{\pm}}\xspace}\ were investigated in detail in Ref.~\cite{Capdevilla:2021fmj} exploiting a detector simulation based on {\textsc{Geant}\xspace} 4~\cite{Agostinelli:2002hh} for the modeling of the response of the tracking detectors, which are crucial in the estimation of the backgrounds. The simulated events were overlaid with beam-induced background events simulated with MARS15~\cite{Mokhov:2017klc}.

\begin{figure}[t]
\centering
\includegraphics[width=0.5\textwidth]{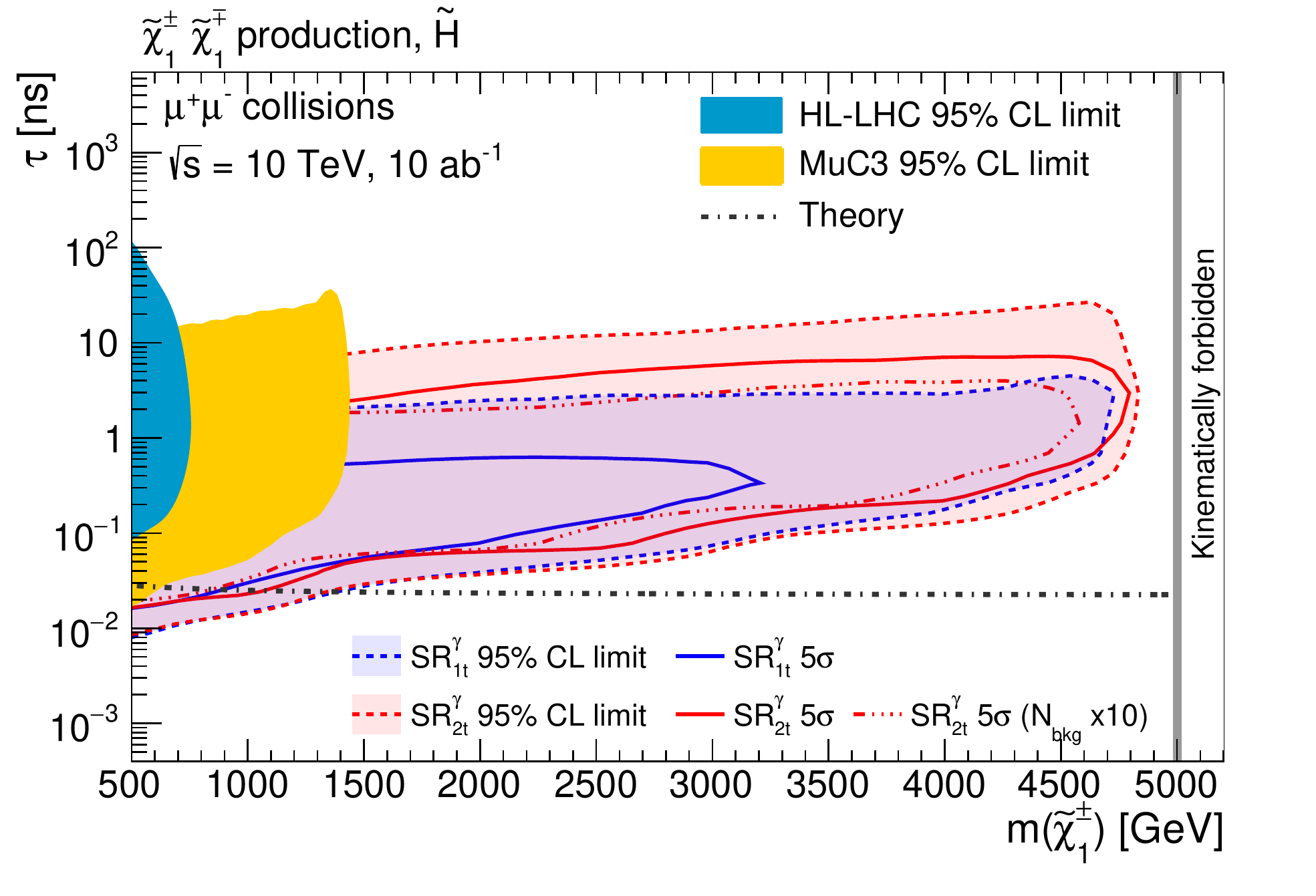}
\caption{Expected sensitivity~\cite{Capdevilla:2021fmj} to the higgsino, in the plane formed by the {\ensuremath{\tilde{\chi}^{\pm}}\xspace} mass and lifetime. The lifetime that corresponds to the thermal mass of 1.1~TeV, and to a mass-splitting of 344~MeV as in the pure-higgsino scenario, is reported as an horizontal dash-dotted line.}
\label{fig:higgsinoStubTracksReachMass}
\end{figure}

The analysis strategy relies on requiring one ({SR\ensuremath{^{\gamma}_{1t}}\xspace}) or two ({SR\ensuremath{^{\gamma}_{2t}}\xspace}) disappearing tracks in each event in addition to a 25~GeV ISR photon. Additional requirements are imposed on the transverse momentum and angular direction of the reconstructed tracklet and on the distance between the two tracklets along the beam axis in the case of events with two candidates. The expected backgrounds are extracted from the full detector simulation and the results are presented assuming a 30\% (100\%) systematic uncertainty on the total background yields for the single (double) tracklet selections.
The corresponding discovery prospects and 95\% CL exclusion reach are shown in Figure~\ref{fig:higgsinoStubTracksReachMass} for each of the two selection strategies discussed above, considering pure-higgsino production cross sections and 10~TeV $\mu^{+}\mu^{-}$ collisions. The expected limits at 95\% CL at the 3~TeV MuC  are also overlaid for comparison.

Both event selections are expected to cover a wide range of higgsino masses and lifetimes, well in excess of current and expected collider limits. In the most favourable scenarios, the analysis of 10~ab$^{-1}$ of 10~TeV muon collisions is expected to allow the discovery {\ensuremath{\tilde{\chi}^{\pm}}\xspace} masses up to a value close to the kinematic limit of 5~TeV. The interval of lifetimes covered by the experimental search directly depends on the layout of the tracking detector, i.e. the radial position of the tracking layers, and the choices made in the reconstruction and identification of the tracklets, i.e. the minimum number of measured space-points. Considering the current detector design~\cite{CLICdp:2017vju,CLICdp:2018vnx,ILDConceptGroup:2020sfq,ILC:2007vrf}, the 10~TeV MuC is expected to allow to discover the higgsino thermal target, though only by a narrow margin. 

\begin{figure}[t]
\centering
\includegraphics[width=0.5\textwidth]{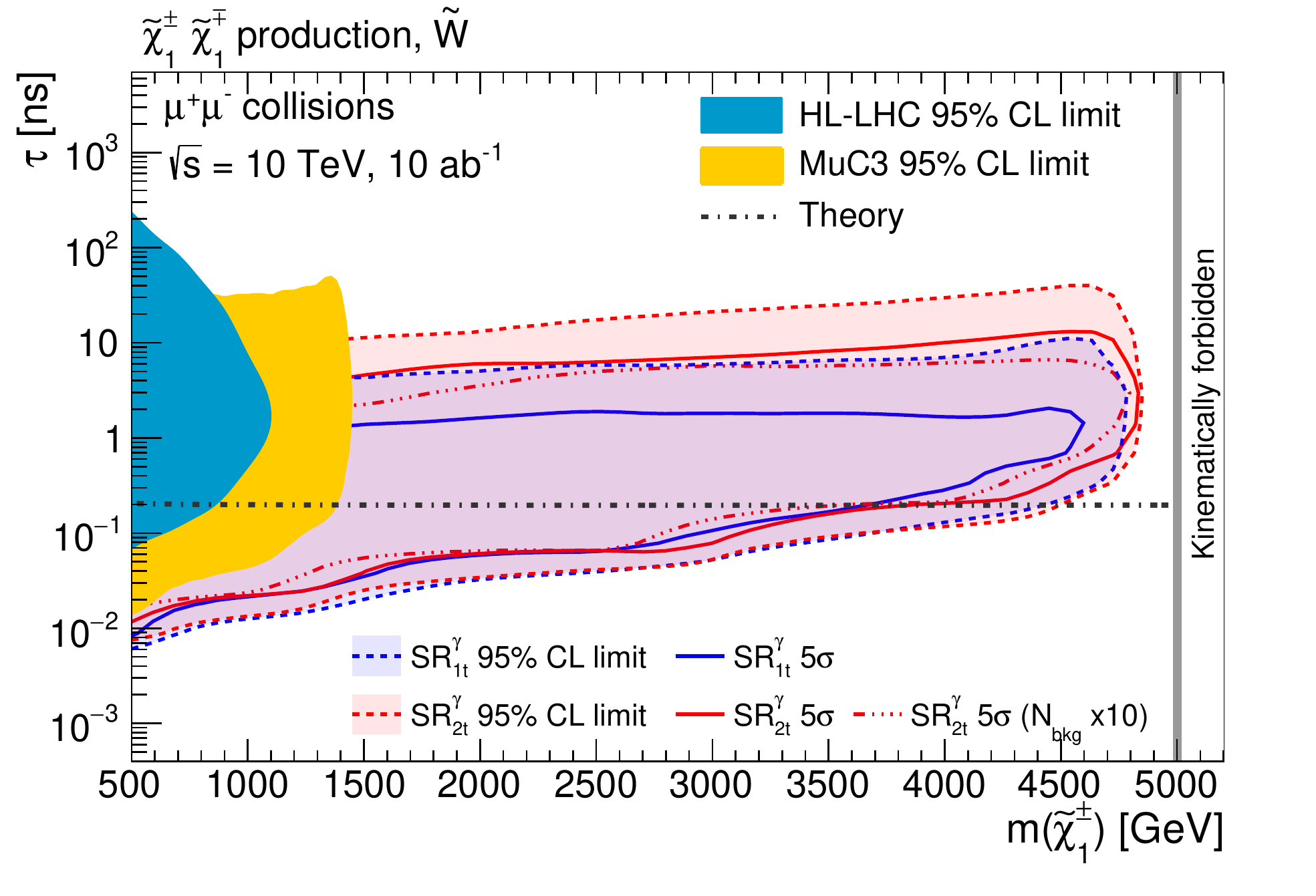}
\caption{Expected sensitivity~\cite{Capdevilla:2021fmj} to the wino, in the plane formed by the {\ensuremath{\tilde{\chi}^{\pm}}\xspace} mass and lifetime. The lifetime that corresponds to the thermal mass of 2.86~TeV, and to a mass-splitting of 166~MeV as in the pure-wino scenario, is reported as an horizontal dash-dotted line.}
\label{fig:winoStubTracksReachMass}
\end{figure}

An alternative tracking detector design, hard to realise in the presence of the BIB, with tracking layers significantly closer to the beam line would be needed to significantly boost the detection of such a signal. Other unconventional signatures, such as soft displaced tracks~\cite{Fukuda:2019kbp} detected in combination with an energetic ISR photon or kinked tracks should be investigated and have the potential to strongly enhance the sensitivity.  

Figure~\ref{fig:winoStubTracksReachMass} shows the expected sensitivity when considering a pure-wino scenario. The much longer predicted lifetime of the charged state significantly increases the likelihood of detecting at least one disappearing track, dramatically extending the reach.

In summary, the pure higgsino with thermal mass can be probed at the 5-$\sigma$ level by a 10~TeV MuC. Pure wino dark matter scenarios are well within the reach of a 10~TeV MuC and could be also probed at lower centre of mass energies.

\subsection{Muon-specific opportunities}
\label{muonspec}

We conclude our survey by reviewing a number a studies targeting new physics that preferentially couples to muons, entailing an inherent advantage of muon colliders over other facilities. As discussed in Section~\ref{muspec}, this scenario is, in the first place, a logical possibility that muon collisions will enable to probe for the first time systematically and effectively. Furthermore, it is motivated by the stronger coupling of second-generation particles to the Higgs, which typically results into a stronger coupling to new physics related with the breaking of the EW symmetry. New physics explaining the structure of leptonic flavour might also be probed more effectively with muons than with electrons.

We start by reviewing the work done in connection with the anomalous magnetic moment of the muon and its tension with the Standard Model (SM) prediction. Possible new physics explaining the anomaly definitely couples to muons and it preferentially couples to muons more strongly than to other particles like electrons because of existing constraints. The MuC thus is found to be a prime option for future  investigations of the possible new physics origin of the anomaly, if it will survive further scrutiny. Otherwise, the ones we present
will still define possible scenarios for new physics that the MuC can explore. Similar considerations hold for $B$-meson anomalies, which have been also studied extensively in connection with muon colliders~\cite{Huang:2021biu,Asadi:2021gah,Azatov:2022itm,Bause:2021prv,Qian:2021ihf,Allanach:2022blr,Bandyopadhyay:2021pld}. Since they do not incorporate the very recent LHCb results~\cite{LHCb:2022qnv,LHCb:2022zom}, these studies will not be reviewed here.

Next, we discuss the MuC opportunities to probe lepton flavour violation. High-energy muon collisions are found to be competitive and complementary with planned low-energy experiments, already at 3~TeV, entailing also opportunities to confirm and further investigate possible indirect evidences of new physics that might emerge from these very precise low-energy measurements. 

We will finally outline the opportunities to explore the Higgs sector by exploiting the relatively large muon Yukawa coupling. In particular, we consider the possibility of detecting and studying heavy Higgs bosons through the radiative return process. 

\subsubsection*{The muon anomalous magnetic moment}

The anomalous magnetic moment of the muon has provided, over the last ten years, an enduring hint for new physics. The experimental value of $a_\mu \!=\! (g_\mu \!-\! 2)/2$ from the E821 experiment at the Brookhaven National Lab~\cite{Muong-2:2006rrc} was recently confirmed by the E989 experiment at Fermilab~\cite{Muong-2:2021vma,Muong-2:2021ojo}, yielding the experimental average $a_\mu^{\scriptscriptstyle \rm EXP} \!=\! 116592061(41) \!\times\! 10^{-11}$. The comparison of this value with the SM prediction $a_\mu^{\scriptscriptstyle \rm SM} \!=\! 116591810(43) \times 10^{-11}$~\cite{Aoyama:2012wk,Gnendiger:2013pva,Keshavarzi:2018mgv,Davier:2019can,Kurz:2014wya,Melnikov:2003xd,Bijnens:2019ghy,Pauk:2014rta,Danilkin:2016hnh,Roig:2019reh,Colangelo:2014qya} 
shows a $4.2\,\sigma$ discrepancy
\begin{equation}
\Delta a_\mu = a_\mu^{\scriptscriptstyle \rm EXP}-a_\mu^{\scriptscriptstyle \rm SM} = 251 \, (59) \times 10^{-11}\,.
\label{eq:gmu}
\end{equation}
In the following, we refer to this as the $g$-2 anomaly. Recent lattice determinations of the hadronic vacuum polarization give a SM prediction that is more in agreement with the experimental result, but is in tension with the previous calculations based on dispersive methods~\cite{Borsanyi:2020mff}. Current and forthcoming plans to confirm the BSM origin of this anomaly include reducing the experimental uncertainty by a factor of four at E989, comparisons between phenomenological and Lattice determinations of the hadronic vacuum polarization contribution to $g$-2~\cite{Gerardin:2019vio,Chao:2021tvp,FermilabLattice:2017wgj,Budapest-Marseille-Wuppertal:2017okr,RBC:2018dos,Giusti:2019xct,Shintani:2019wai,FermilabLattice:2019ugu,Gerardin:2019rua,Giusti:2019hkz,Borsanyi:2020mff}, and new experiments aiming to probe the same physics~\cite{Sato:2017sdn,Abbiendi:2016xup}. 
If all of these efforts will confirm the presence of new physics, then the most urgent task at hand will be to probe this anomaly at higher energies, ultimately in order to discover and study the new BSM particles that give rise to the additional $\Delta a_\mu$ contributions. The $\mc$ is a uniquely well-suited machine for this endeavour, not least since it collides the actual particles displaying the anomaly, and hence the only particles guaranteed to couple to the new physics.

There are several ways in which a $\mc$ can provide a powerful high-energy test of the muon $g$-2 and discover the new physics responsible for the anomaly:
\begin{itemize}
\item If the physics responsible for $\Delta a_\mu$ is heavy enough, an Effective Field Theory (EFT) description holds up to the high $\mc$ energies. This was studied in \cite{Buttazzo:2020ibd}. In this case, scattering cross sections induced by the new physics effective operators grow at high energies (analogously to what discussed in Sections \ref{HEM}, \ref{sec:Higgs} and \ref{secdm}), so that a measurement with modest  precision at a sufficiently high energy will be sufficient to disentangle new physics effects from the SM background. These considerations are  independent from the specific underlying model.
\item In most motivated models of new physics, new particles responsible for $\Delta a_\mu$ are light enough to be directly produced in $\mu^+\mu^-$ collisions at attainable $\mc$ energies. Understanding such opportunities for direct production and discovery at a $\mc$ was studied in~\cite{Capdevilla:2020qel,Capdevilla:2021rwo}. It was found that a complete classification of perturbative BSM models that can give rise to the observed value of $\Delta a_\mu$, and of their experimental signatures, is possible. 
\item Additional effects in muon couplings to SM gauge and Higgs bosons, correlated with the muon $g$-2, can also be present at a level that can be probed by precision measurements at a $\mc$. Some of these effects can be predicted in a model-independent way, others arise in specific, motivated models.
\end{itemize}

These three strategies together make it possible to formulate a {\it no-lose theorem} for a high-energy $\mc$~\cite{Capdevilla:2020qel,Capdevilla:2021rwo}, if the experimental anomaly in the muon $g$-2 is really due to new physics.
The physics case of a high-energy determination of $\Delta a_\mu$, which is unique of a $\mc$, thus represents a striking example of the complementarity and interplay of the high-energy 
and high-intensity frontiers of particle physics, and it highlights the far reaching potential of a $\mc$ to probe new physics.

\paragraph{High-energy probes of the operators generating the \texorpdfstring{$g$}{g}-2}{\ }
\noindent

We start by reviewing the analysis of~\cite{Buttazzo:2020ibd}, which determined that precision measurements at high-energy $\mc$s can detect deviations in scattering rates that are generated by the same effective operators giving rise to the $g$-2 anomaly. Hence, they provide a powerful independent verification and detailed examination of the anomaly even if the responsible BSM degrees of freedom are too heavy to be produced on-shell at the collider. 
This would be a direct determination of the new physics contribution, not affected by the hadronic uncertainties that enter the SM prediction of $a_\mu$.

\begin{figure*}
\centering%
\includegraphics[width=0.55\textwidth]{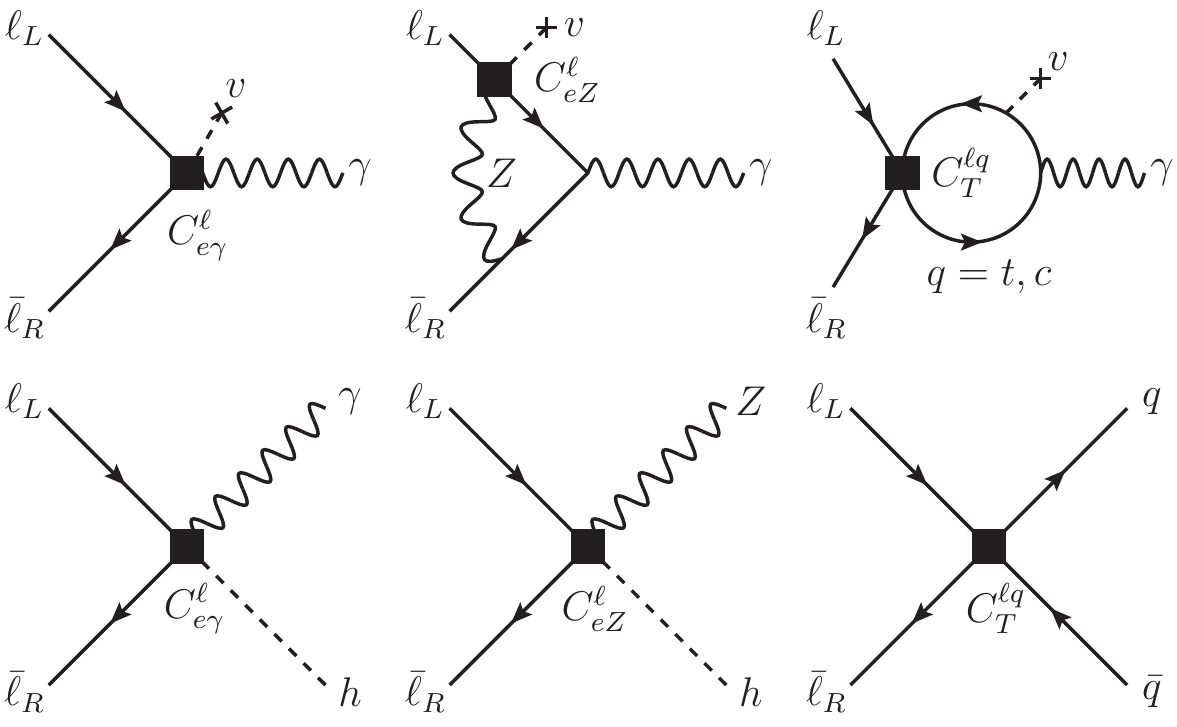}
\caption{{\it Upper row:} Feynman diagrams contributing to the leptonic $g$-2 up to one-loop order in the Standard Model EFT.
{\it Lower row:} Feynman diagrams of the corresponding high-energy scattering processes.
Dimension-6 effective interaction vertices are denoted by a square.}
\label{fig:feyn}
\end{figure*}

New interactions emerging at a scale $\Lambda$ larger than the EW scale can be described at energies $E \ll \Lambda$ 
by an effective Lagrangian containing non-renormalizable $SU(3)_c \otimes SU(2)_L \otimes U(1)_Y$ invariant operators.
The relevant effective Lagrangian contributing to $g$-2
reads~\cite{Buchmuller:1985jz}
\begin{eqnarray}
\mathcal{L} = 
&&\frac{C^\ell_{eB}}{\Lambda^2}
\left( \bar\ell_L \sigma^{\mu\nu}\ell_{R}\right) \! H B_{\mu\nu} + 
\frac{C^\ell_{eW}}{\Lambda^2}
\left( \bar\ell_L \sigma^{\mu\nu} \ell_{R} \right) \! \tau^I \! H W_{\mu\nu}^I  \nonumber \\ &&+ \frac{C^\ell_{T}}{\Lambda^2}( \overline{\ell}_L\sigma_{\mu\nu}\ell_{R}) (\overline{Q}_L\sigma^{\mu\nu} u_{R}) 
+ h.c..
\label{eq:L_SMEFT}
\end{eqnarray} 
It includes not only the interactions that generate the dipole operator at tree level, but also four-fermion operators that generate the dipole at one loop. The Feynman diagrams relevant for $g$-2 are displayed in Figure~\ref{fig:feyn}, top row. After EW symmetry breaking, $H$ is replaced by its vacuum expectation value $v$, and one obtains the prediction 
\begin{eqnarray}
\Delta a_\ell  \simeq &&\frac{4m_\ell v}{e\Lambda^2} \, 
\bigg(
C^\ell_{e\gamma} - \frac{3\alpha}{2\pi} \frac{c^2_W \!-\! s^2_W}{s_W c_W} \,C^\ell_{eZ} \log\frac{\Lambda}{m_Z}
\bigg) \nonumber \\
&&- \sum_{q=c,t} \frac{4m_\ell m_q}{\pi^2} \frac{C_T^{\ell q}}{\Lambda^2}\,
\log\frac{\Lambda}{m_q},
\label{eq:Delta_a_ell}
\end{eqnarray}
where $s_W$ and $c_W$ are the sine and cosine of the weak mixing angle, $C_{e\gamma}=c_W C_{eB} - s_W C_{eW}$ and
$C_{eZ} = -s_W C_{eB} - c_W C_{eW}$.
Additional radiative contributions from the three operators $H^\dag H W_{\mu\nu}^IW^{I\mu\nu}$, $H^\dag H B_{\mu\nu}B^{\mu\nu}$ and $H^\dag \tau^I H W_{\mu\nu}^I B^{\mu\nu}$ can be neglected because they are suppressed by the small lepton Yukawa couplings.
For simplicity, $C_{eB}$, $C_{eW}$ and $C_{T}$ are assumed to be real. The one-loop renormalization effects to $C^\ell_{e\gamma}$ 
\begin{align}
\!\!\! C^\ell_{e\gamma}(m_\ell) \simeq C^\ell_{e\gamma}\!(\Lambda)\left(1 \!-\! \frac{3y^2_t}{16\pi^2} \log\frac{\Lambda}{m_t}
\!-\! \frac{4 \alpha}{\pi} \log\frac{m_t}{m_\ell}\right)
\label{eq:running_Cegamma}
\end{align}
can be straightforwardly included. Numerically~\cite{Buttazzo:2020ibd}
\begin{eqnarray}
\frac{\Delta a_\mu}{3 \!\times\! 10^{-9}} \approx  &&\left( \frac{250 \, {\rm TeV}}{\Lambda} \right)^{2} \times\nonumber\\
&&\left(C^\mu_{e\gamma} - 0.2 C^{\mu t}_T - 0.001 C^{\mu c}_T - 0.05 C^{\mu}_{eZ}\right). 
\nonumber
\end{eqnarray}

\noindent A few comments are in order: 
\begin{itemize}
\item The $\Delta a_\mu$ discrepancy can be solved for a new physics scale up to $\Lambda\approx 250~$TeV.
This requires a strongly coupled new physics sector where $C^\mu_{e\gamma}$ and/or $C^{\mu t}_T \sim g^2_{\rm\scriptscriptstyle NP}/16\pi^2 \sim 1$ 
and a chiral enhancement $v/m_\mu$ compared with the weak SM contribution.
Directly producing new particles at such high scales is far beyond the capabilities of any foreseen collider.
Nevertheless, this new physics can be tested at a $\mc$ through high-energy processes such as $\mu^+\!\mu^- \!\to h\gamma$ or $\mu^+\!\mu^- \!\to q\bar{q}$ (with $q=c,t$), that are affected by the very same operators that generate $\Delta a_\mu$.
\item If the underlying new physics sector is weakly coupled, $g_{\rm\scriptscriptstyle NP}\lesssim 1$, then
$C^\mu_{e\gamma}$ and $C^{\mu t}_T \lesssim 1/16\pi^2$, implying $\Lambda\lesssim 20~$TeV to solve the $g$-2 anomaly.
In this case, a $\mc$ could still be able to directly produce new physics particles~\cite{Capdevilla:2020qel}. 
Even so, the study of the processes $\mu^+\mu^-\to h\gamma$
and $\mu^+\mu^-\to q\bar{q}$ could be crucial to reconstruct the effective dipole vertex $\mu^+\mu^-\gamma$, as has been explicitly shown in~\cite{Paradisi:2022vqp}.
\item If the new physics sector is weakly coupled, and further $\Delta a_\mu$ scales with lepton masses as the SM weak contribution,
then $\Delta a_\mu \sim m^2_\mu/16\pi^2\Lambda^2$.
Here, the experimental value of $\Delta a_\mu$ can be accommodated only provided that $\Lambda \lesssim 1~$TeV.
For such a low new physics scale the EFT description breaks down at the typical multi-TeV $\mc$ energies, and new resonances cannot escape from direct production.
\end{itemize}

\paragraph*{Dipole operator.}\;The main contribution to $\Delta a_\mu$ comes from the dipole operator $O_{e\gamma}=\left(\bar\ell_L \sigma_{\mu\nu} e_R\right) H F^{\mu\nu}$. The same operator also induces a contribution to the process $\mu^+\mu^- \to h \gamma$ 
that grows with energy, and thus can become dominant over the SM cross section at a very high energy collider. 
Neglecting all masses, the
total $\mu^+\mu^- \!\to  h\gamma$ cross section is
\begin{align}
\sigma_{h\gamma} \!=\! \frac{s}{48\pi}\frac{|C^{\mu}_{e\gamma}|^2}{\Lambda^4}\!\!
\approx\!  0.7 {\rm ab} \left(\frac{\sqrt{s}}{30\, {\rm TeV}}\right)^{\!\!2} \!\! \left(\frac{\Delta a_\mu}{3 \times 10^{-9}}, \right)^{\!2}
\label{xsec_HA}
\end{align}
where in the last equation no contribution to $\Delta a_\mu$ other than the one from $C^{\mu}_{e\gamma}$ was assumed, and running effects for $C^{\mu}_{e\gamma}$, see eq.~(\ref{eq:running_Cegamma}), from a scale $\Lambda \approx 100$~TeV have been included. Notice that there is an identical contribution also to the process $\mu^+\mu^- \!\to\! Z\gamma$ since $H$ contains the longitudinal polarisations of the $Z$. Given the scaling with energy of the baseline luminosity~(\ref{lums}), one gets about 60 total $h\gamma$ events at $\sqrt{s}=30~$TeV. As it is discussed below, this is a signal that the $\mc$ is sensitive to.

\begin{figure*}
\centering%
\includegraphics[width=0.6\textwidth]{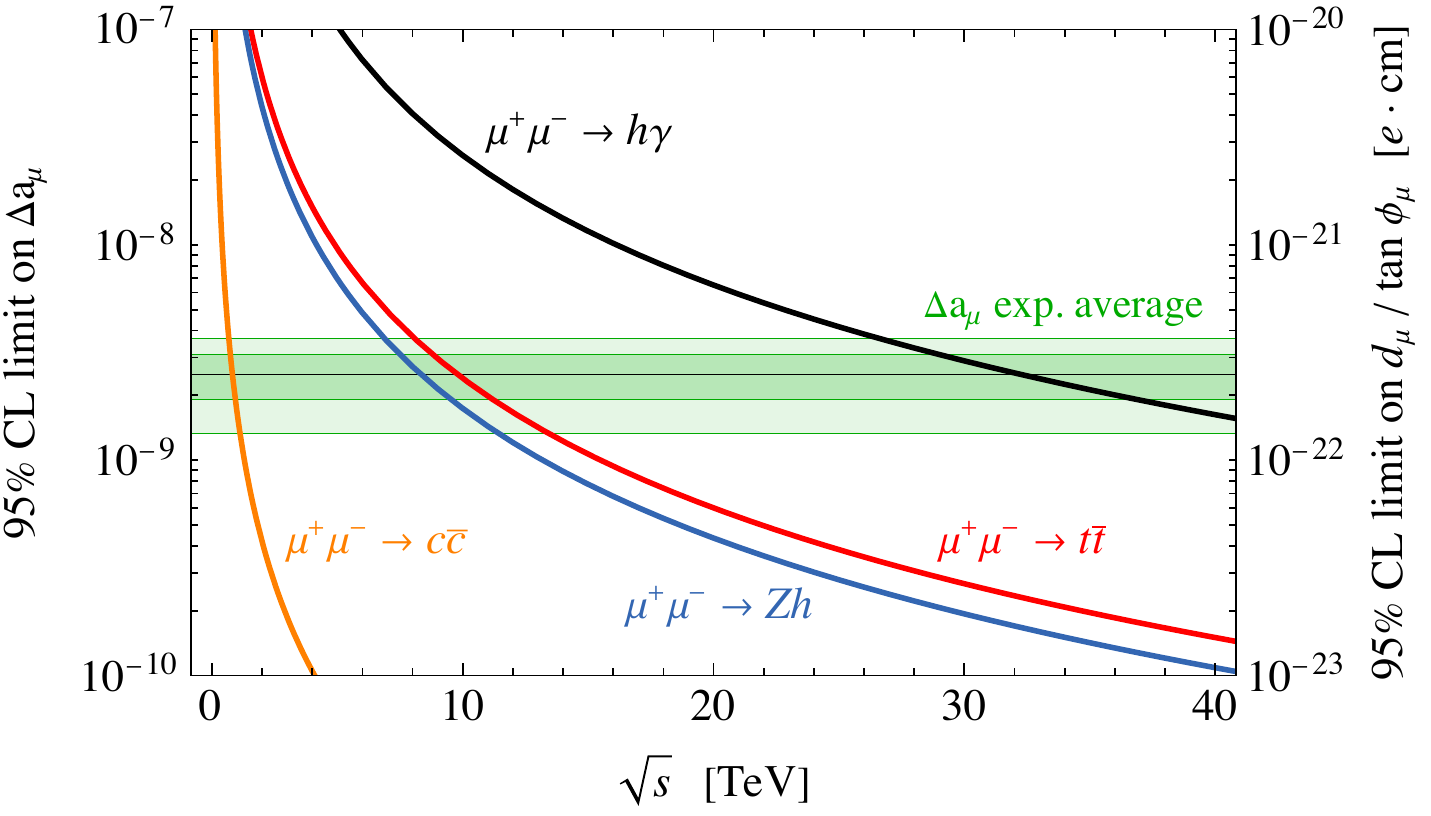}
\caption{Reach on the muon anomalous magnetic moment $\Delta a_\mu$ and muon EDM $d_\mu$, 
as a function of the $\mc$ collider center-of-mass energy $\sqrt{s}$, from the labeled processes. Figure taken from~\cite{Buttazzo:2020ibd}.}
\label{fig:hgamma}
\end{figure*}

The SM irreducible $\mu^+\mu^- \to h\gamma$ background is small, $\sigma_{h\gamma}^{\rm SM} \approx 2 \times 10^{-2}\,{\rm ab}\,\big({30\,{\rm TeV}}/{\sqrt{s}}\big)^{2}$,
with the dominant contribution arising at one-loop~\cite{Abbasabadi:1995rc} due to the muon Yukawa coupling suppression of the tree-level diagrams.
The main source of background comes from $Z\gamma$ events, where the $Z$ boson is incorrectly reconstructed as a Higgs. 
This cross section is large, due to the contribution from transverse polarisations.
There are two ways to isolate the $h\gamma$ signal from the background: by means of the different angular distributions of the two processes ---the SM $Z\gamma$ peaks in 
the forward region, while the signal is central---and by accurately distinguishing $h$ and $Z$ bosons from their decay products, e.g. by precisely reconstructing their invariant mass.
To estimate the reach on $\Delta a_\mu$ a cut-and-count experiment was considered in the $b\bar b$ final state, which has the highest signal yield. The significance of the signal is maximised in the central region $|\!\cos\theta| \lesssim 0.6$. At 30 TeV one gets
\begin{align}
\sigma_{h\gamma}^{\rm cut} &\approx 0.53\, {\rm ab} \,\bigg(\frac{\Delta a_\mu}{3 \times 10^{-9}} \bigg)^{\!2}, & ~~\sigma_{Z\gamma}^{\rm cut} &\approx 82\,{\rm ab}\,.
\end{align}
Requiring at least one jet to be tagged as a $b$, and assuming a $b$-tagging efficiency $\epsilon_b = 80\%$, one finds that a value $\Delta a_\mu = 3\!\times\! 10^{-9}$ can be tested at 95\% C.L.\ at a 30 TeV collider if the probability of reconstructing a $Z$ boson as a Higgs is less than 10\%. The resulting number of signal events is $N_S = 22$, and $N_S/N_B = 0.25$.
Figure~\ref{fig:hgamma} shows as a black line the 95\% C.L.\ reach from $\mu^+\mu^-\to h\gamma$ 
on the anomalous magnetic moment as a function of the collider energy. Note that since the number of signal events scales as the fourth power of the center-of-mass energy, only a collider with $\sqrt{s} \gtrsim 30$~TeV will have the sensitivity to test the $g$-2 anomaly in this channel.

\paragraph*{Semi-leptonic interactions.}\;
If the anomalous magnetic moment arises at one loop from one of the other operators in \eqref{eq:Delta_a_ell}, their Wilson coefficients must be larger to reproduce the observed signal, and the new physics will be easier to test at a $\mc$.
We now derive the constraints on the semi-leptonic operators. The operator $O_T^{\mu t}$ that enters 
$\Delta a_\mu$ at one loop can be probed by $\mu^+\mu^-\to t\bar t$ (Figure~\ref{fig:feyn}). Its contribution 
to the cross section is
\begin{align}
\label{eq:mumu_tt}
\!\sigma_{t\bar{t}} =\! \frac{s}{6\pi} \frac{|C^{\mu t}_{T}|^2}{\Lambda^4} N_c 
\approx 
58 {\rm ab} \left(\frac{\sqrt{s}}{10 \, {\rm TeV}}\right)^{\!\!2} \!\!\! \left(\frac{\Delta a_\mu}{3 \times 10^{-9}} \right)^{\!2}
\end{align}
where the last equality assumes $\Lambda \approx 100~$TeV and uses $|\Delta a_\mu| \approx 3 \times 10^{-9} \left(100 \,{\rm TeV}/\Lambda\right)^2 |C^{\mu t}_T|$.
We estimate the reach on $\Delta a_\mu$ assuming an overall 50\% efficiency for reconstructing the top quarks, 
and requiring a statistically significant deviation from the SM $\mu\mu\to t\bar{t}$ background, with cross section
$\sigma_{t\bar{t}}^{\rm SM} \approx 1.7 \,{\rm fb}\,\big({10\,{\rm TeV}}/{\sqrt{s}}\big)^2$.

A similar analysis can be performed for semi-leptonic operator involving charm quarks. If the contribution from the charm loop dominates, we can probe $|\Delta a_\mu| \approx 3 \times 10^{-9}(10\,{\rm TeV}/\Lambda)^2 |C^{\mu c}_T|$ through the process $\mu\mu \to \bar{c} c$.
In this case, unitarity constraints on the new physics coupling $C_T^{\mu c}$ require a much lower new physics scale $\Lambda \lesssim 10$~TeV, 
so that our effective theory analysis will only hold for lower centre of mass energies.
Combining eq.~(\ref{eq:Delta_a_ell}) and (\ref{eq:mumu_tt}), with $c \leftrightarrow t$, we find
\begin{align}
\sigma_{c\bar{c}}
\, \approx \, 100 \,{\rm fb} \left(\frac{\sqrt{s}}{3 \, {\rm TeV}}\right)^{\!2} \!\! \left(\frac{\Delta a_\mu}{3 \times 10^{-9}} \right)^{\!2}.
\label{eq:mumu_cc_numerics}
\end{align}
The SM cross section for $\mu^+\mu^- \!\to c\bar{c}$ 
at $\sqrt{s}= 3~$TeV is $\sim 19$~fb. In Figure~\ref{fig:hgamma} we show the 95\% C.L.\ constraints on the top and charm contributions to $\Delta a_\mu$ as red and orange lines, respectively, as functions of the collider energy. Notice that the charm contribution can be probed already at $\sqrt{s} = 1$~TeV, while the top contribution can be probed at $\sqrt{s} = 10$~TeV. 

\paragraph*{Electric dipole moments.}\;
So far, CP conservation has been assumed. If however the coefficients $C_{e\gamma}$, $C_{eZ}$ or $C_T$ are complex, an electric dipole moment (EDM) $d_\mu$ is unavoidably generated for the muon. Since the cross sections in eq.~\eqref{xsec_HA} and \eqref{eq:mumu_tt} are proportional to the absolute values of the same coefficients, a $\mc$ offers a unique opportunity to test also $d_\mu$. The current experimental limit $d_\mu < 1.9 \times 10^{-19}\,e\,$cm was set by the BNL E821 experiment~\cite{Muong-2:2008ebm} 
and the new E989 experiment at Fermilab aims to decrease this by two orders of magnitude~\cite{Chislett:2016jau}. 
Similar sensitivities could be reached also by the J-PARC $g$-2 experiment~\cite{Gorringe:2015cma}.

From the model-independent relation~\cite{Giudice:2012ms}
\begin{align}
\frac{d_\mu}{\tan\phi_\mu} =  \frac{\Delta a_\mu}{2 m_\mu} \,e \,\simeq\, 3 \times 10^{-22} \left(\frac{\Delta a_\mu}{3\!\times\! 10^{-9}}\right) e\, {\rm cm}\,,
\label{eq:d_mu}
\end{align}
where $\phi_\mu$ is the argument of the dipole amplitude, the bounds on $\Delta a_\mu$ in Figure~\ref{fig:hgamma} can be translated into a nearly model-independent constraint on $d_\mu$ by assuming ${\tan\phi_\mu}\approx1$. We find that a 10~TeV $\mc$ can reach a sensitivity comparable to the ones expected at Fermilab~\cite{Chislett:2016jau} and J-PARC~\cite{Gorringe:2015cma}, while at a 30 TeV collider one gets the bound $d_\mu \lesssim 3\times 10^{-22}\, e$~cm.

\begin{figure*}[t]
\center
\hspace{-0.75cm}
\includegraphics[width=0.5\textwidth]{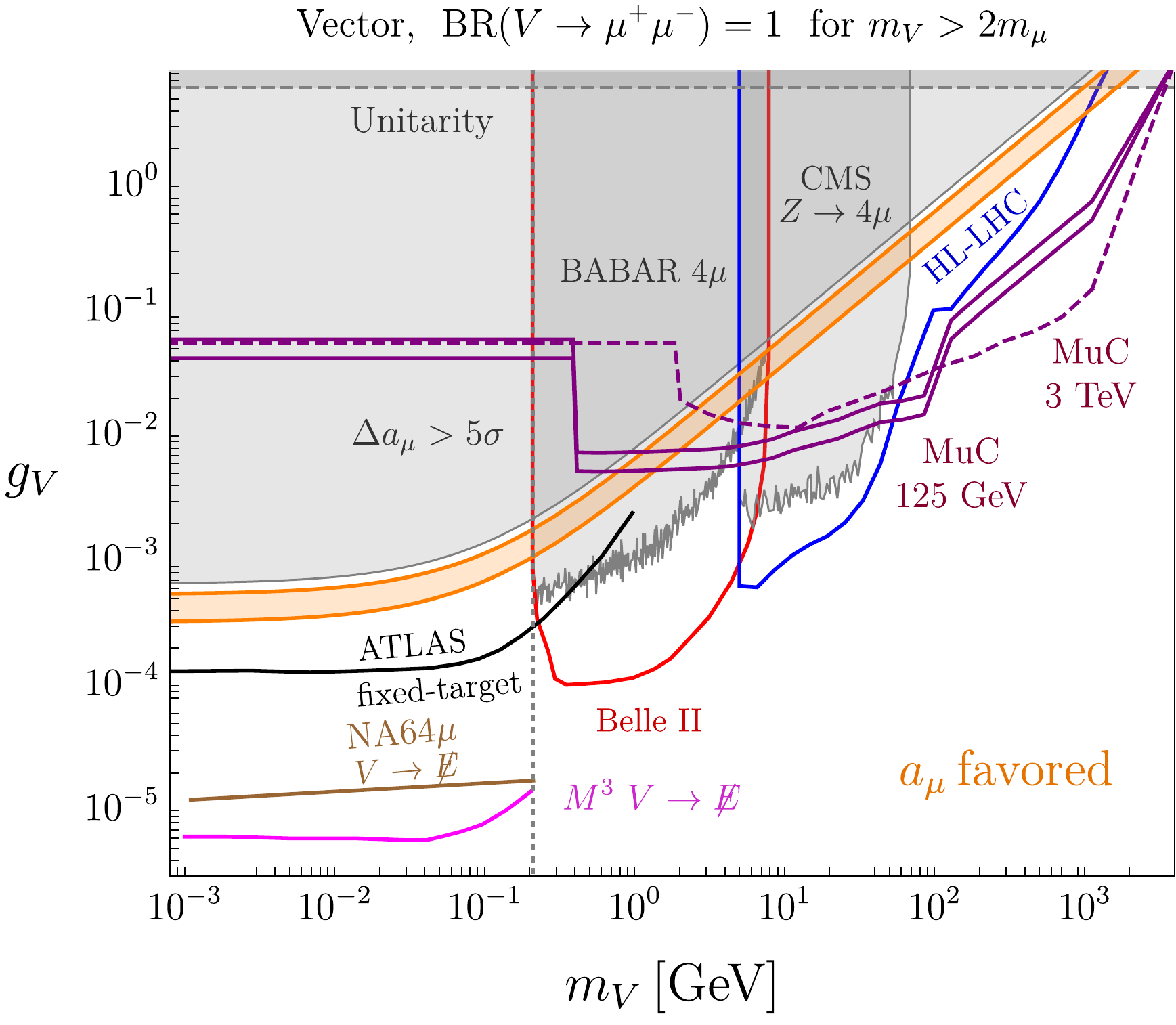} ~
\includegraphics[width=0.5\textwidth]{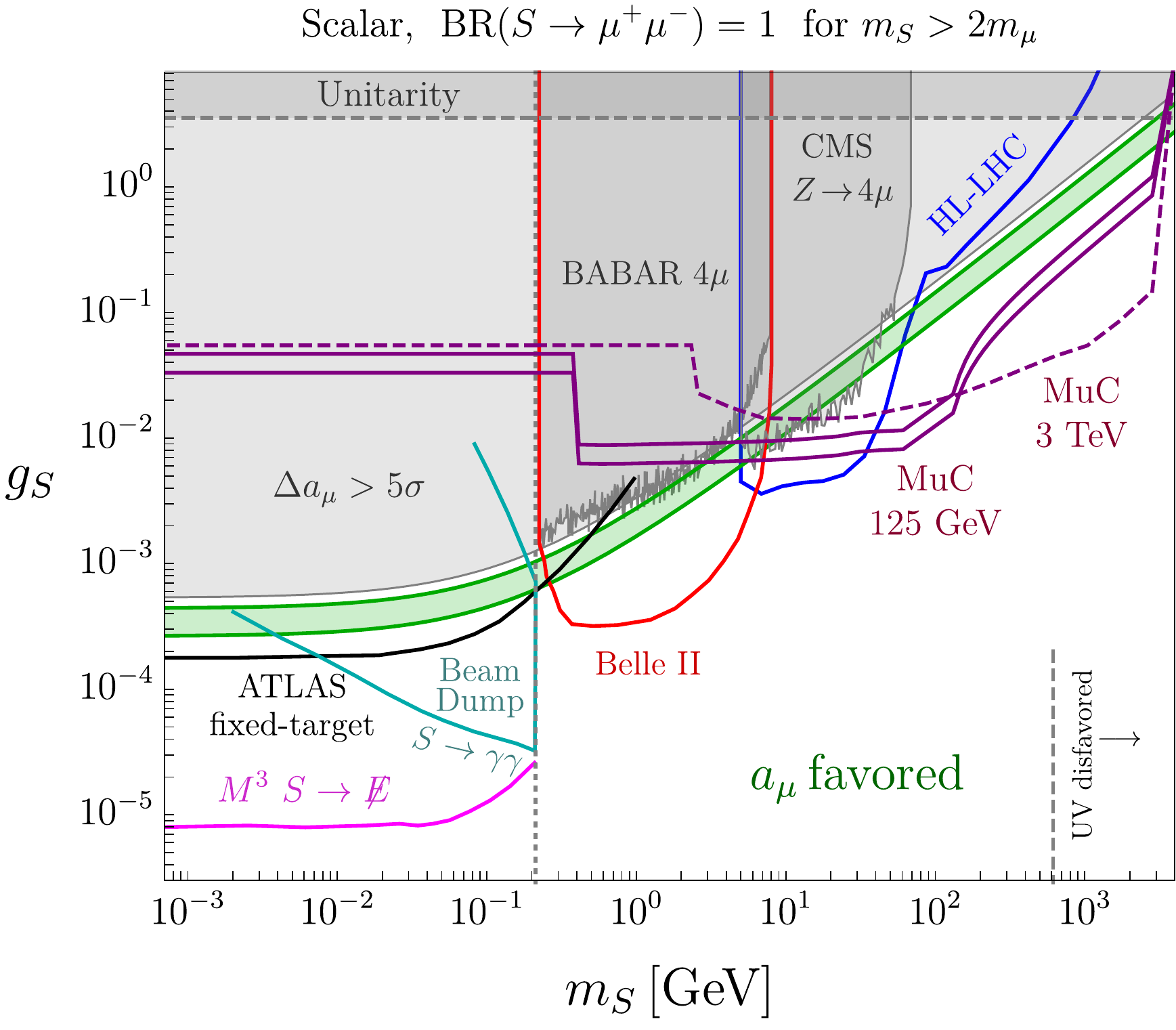}  \\ 
\hspace{-0.75cm}
\includegraphics[width=0.5\textwidth]{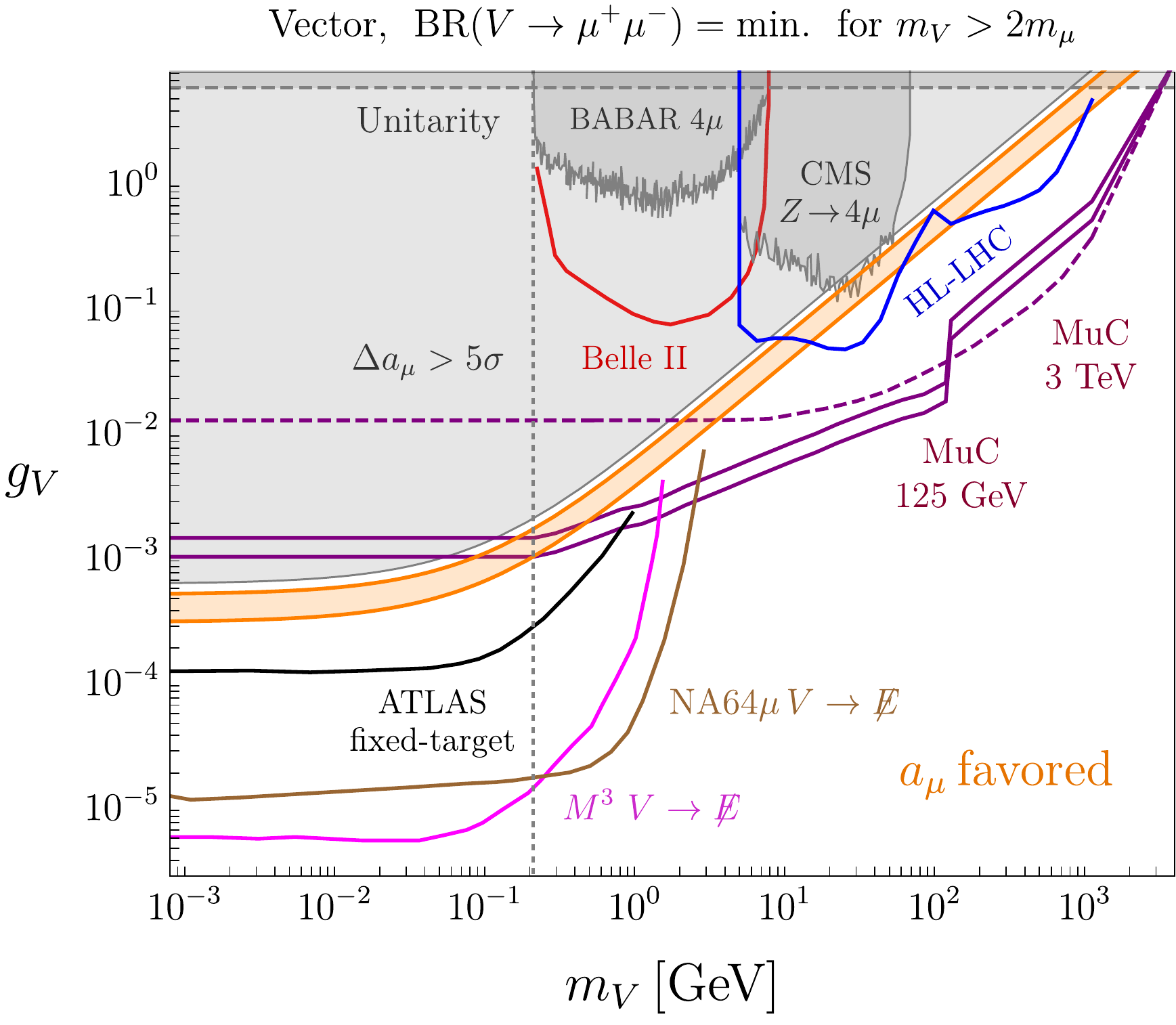} ~
\includegraphics[width=0.5\textwidth]{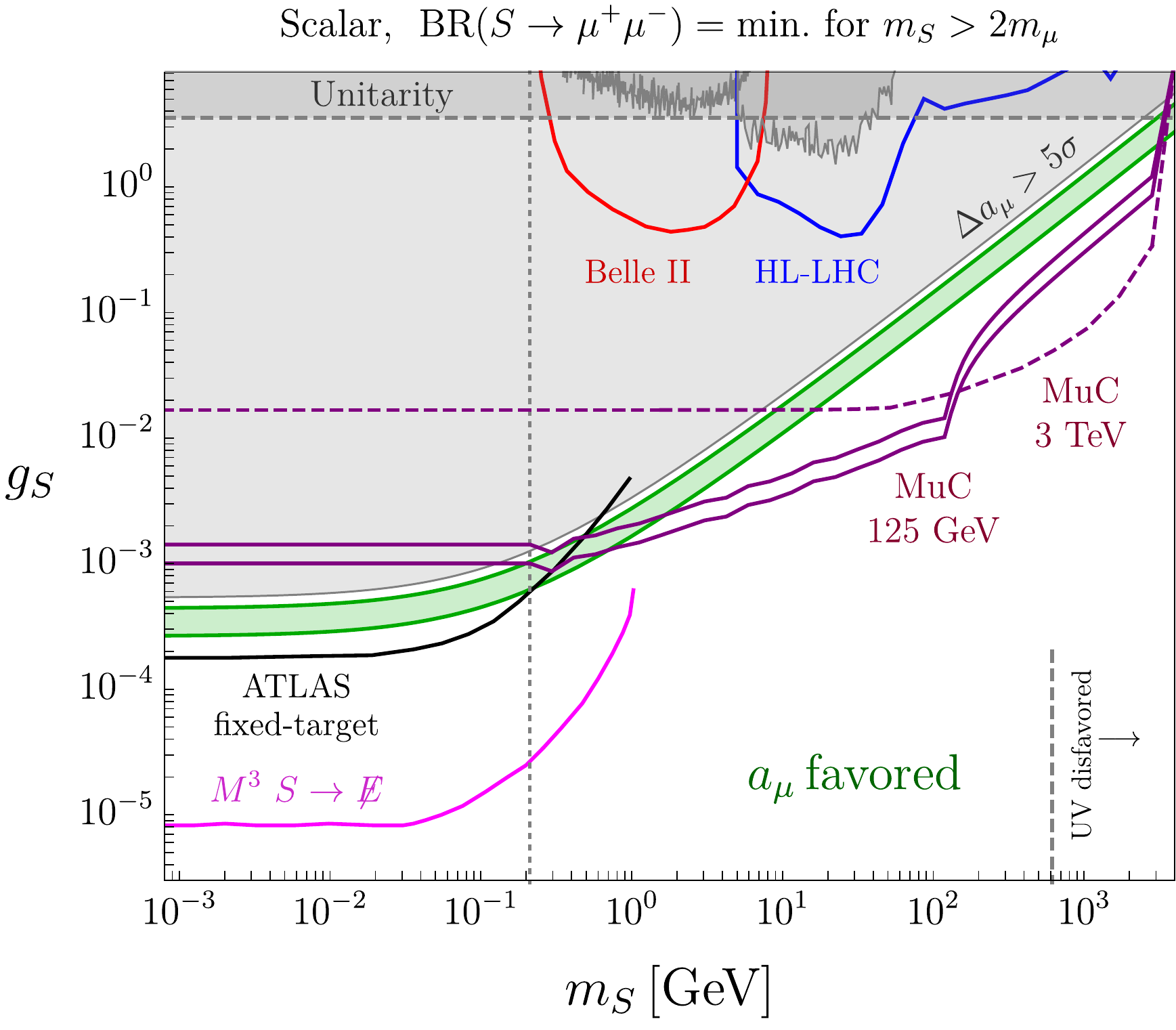} 
\caption{
Singlet models for $g$-2 and their probes at different masses, assuming 100\% branching ratio to di-muons (top) and the minimum branching ratio to di-muon allowed by perturbative unitarity. \cite{Capdevilla:2021kcf}.
 \label{babar_singlet}}
\end{figure*}

\paragraph{Direct searches for BSM particles generating the \texorpdfstring{$g$}{g}-2}{\ } \\ 
\noindent

We now review the model-exhaustive analyses conducted in~\cite{Capdevilla:2020qel,Capdevilla:2021rwo} and \cite{Capdevilla:2021kcf}, examining all possible perturbative BSM solutions to the $g$-2 anomaly to understand the associated direct production signatures of new states at future $\mc$s, and we summarise the related no-lose theorem. This model-exhaustive analysis first finds the highest possible mass scale of new physics subject only to perturbative unitarity, and optionally the requirements of minimum flavour violation and/or naturalness. It is assumed that one-loop effects involving BSM states are responsible for the anomaly. Scenarios where new contributions only appear at higher loop order require a lower BSM mass scale to generate the required new contribution.
All possible one-loop BSM contributions to $\Delta a_\mu$ can be organised into two classes: {\emph{Singlet Scenarios}}, in which the BSM $g$-2 contribution only involves a muon and a new SM singlet boson that couples to the muon, and {\emph{electroweak (EW) Scenarios}}, in which new states with EW quantum numbers contribute to $g$-2. 

\paragraph*{Singlet mediators.}
\;\;\;Throughout this section, ``Singlet Models'' refers to the family of models where $\Delta a_\mu$ is generated by a muon-philic singlet, either scalar or vector, through the couplings
\begin{equation}
 g_S S (\mu_L \mu^c~ + \mu^{c  \dagger} \mu_L^\dagger),~~~~
  g_V  V_\nu (\mu^\dagger_L \bar \sigma^\nu \mu_L + \mu^{c  \dagger}\bar \sigma^\nu \mu^{c})~,
\label{lag-singlets}
\end{equation}
where $\mu_L$ and $\mu^c$ are the muon Weyl spinors. Realisations of these scenarios appear in multiple contexts. For example, vector singlets can be classified either into dark photon or $L_\mu-L_\tau$ like scenarios. The former are solutions to $g$-2 where couplings between the vector and first generation fermions are generated via loop-induced kinetic mixing. These scenarios are all excluded~\cite{Bauer:2018onh,Ballett:2019xoj} or soon to be~\cite{Mohlabeng:2019vrz}. The second, $L_\mu-L_\tau$ like scenarios, are vectors that do not couple to first generation fermions. These are highly constrained and a combination near-future experiments might probe the remaining parameter space relevant for the $g$-2 anomaly~\cite{Altmannshofer:2014pba,Krnjaic:2019rsv}. The muon colliders perspectives to probe these scenarios will be discussed later in this section.
Singlet scalar models can be UV-completed by extra scalars and/or fermions that, after being integrated out, generate the dimension-5 operator $(S/\Lambda)\, H^\dagger L\mu^c$. Once the Higgs gets a vev one reproduces the interaction in eq.~\eqref{lag-singlets}. These models are disfavoured for large singlet masses~\cite{Capdevilla:2021kcf}.

Figure~\ref{babar_singlet} shows the limits and projections on muon-philic vector (left) and scalar (right) singlets. In the upper panels, $100\%$ branching ratio to muon is assumed when kinematically allowed. The green/orange bands represent the parameter space for which the singlet scalars/vectors resolve $g$-2 within $2\sigma$. Existing experimental limits are shaded in gray, while projections are indicated with coloured lines. The  $M^3$ \cite{Kahn:2018cqs}, NA64$\mu$ \cite{Gninenko:2018ter}, and ATLAS fixed-target \cite{Galon:2019owl} experiments probe invisibly-decaying singlets; projections here assume a 100\% invisible branching fraction. The LHC limits and HL-LHC projections were obtained from $3\mu$/$4\mu$ muon searches. The purple muon collider projections are obtained from a combination of searches in the singlet plus photon final state, and from deviations in angular observables of Bhabha scattering~\cite{Capdevilla:2021rwo}. 

For scalar singlets whose width is determined entirely by the muon coupling (top right), Figure~\ref{babar_singlet} also shows the projections for a search for $S \to \gamma\gamma$ at a muon beam dump experiment~\cite{Chen:2017awl}~\footnote{See~\cite{Cesarotti:2022ttv} for the opportunities of a beam dump experiment that exploits the high energy MuC beams.} under the minimal assumption that the scalar-photon 
coupling arises solely from integrating out the muon. 

The bottom row plots of the figure include the same experiments, but assume that for $m_{S,V}  > 2m_\mu$, the singlets have the minimal branching ratio to di-muon that is consistent with the di-muon $g_{S,V}$ coupling strength and with the upper bound on the total singlet with-over-mass ratio from unitarity. The curves that are unaffected by this change of the muonic branching fraction correspond to searches that are insensitive to the singlet's decay modes. Notice however that the projections for $M^3$, NA64$\mu$, and ATLAS fixed-target experiments assume a $\simeq 100\%$ invisible branching fraction for $m_{S/V} > 2m_\mu$, which is model-dependent.

The upshot is that a 3 TeV $\mc$ can directly and indirectly probe the entire space of possible singlet explanations for the $g$-2 anomaly for masses above a few or 10 GeV. Lower masses are accessible by other experiments, with the possible exception of the 1--10~GeV window of mass, in the case of scalars, when the branching ratio to muons is small (bottom right panel of Figure~\ref{babar_singlet}). This region could however be covered by a lower energy muon collider, for instance by a 125~GeV MuC with 5 or with 20~${\textrm{fb}}^{-1}$ that is considered in the figure. The 125~GeV MuC is advantageous in this case, because of the following. The sensitivity is dominated by the mono-photon search and, in the signal, the energy of the photon is peaked at $E_{\rm{cm}}/2$ if the singlet is light. The background instead emerges from the production of a massive $Z$-boson decaying invisibly, therefore  the peak moves below $E_{\rm{cm}}/2$ by an amount that is controlled by the $Z$ mass. This enables an effective background reduction at the 125~GeV MuC. At 3 TeV instead the one due to the $Z$ mass is a relatively small correction to the energy. The peak displacement cannot thus be exploited for background rejection due to the finite photon energy resolution and to the smearing of the initial muons energy due to photon radiation in the initial state.

\paragraph*{Electroweak mediators.}
\;Scenarios with electroweak mediators can generate the necessary $g$-2 contribution even for new physics much above the TeV scale. In particular, the analysis of~\cite{Capdevilla:2020qel,Capdevilla:2021rwo}  carefully studied simplified models featuring new scalars and fermions that yield the largest possible BSM mass scale able to account for the anomaly.
By systematically scanning over the entire parameter space of all these models, subject to the constraint that they resolve the $g$-2 anomaly while maintaining perturbative unitarity (as well as other optional constraints), it is possible to derive an upper bound on the mass of the lightest charged BSM particle that has to exist in order to generate the observed $\Delta a_\mu$. 
The possibility of a high multiplicity of BSM states was also considered by allowing $N_{\rm BSM}$ copies of each BSM model to be present simultaneously. The results show that, in order to contribute ${2.8 \times 10^{-9}}$ to ${\Delta a_\mu}$ explaining the anomaly, EW scenarios must always have at least one new charged state lighter than
\begin{equation}
\label{e.MchargedmaxXresultpreview}
M^\mathrm{max, X}_\mathrm{BSM, charged} 
\approx
\left\{
\begin{array}{l}
(100~{\textrm{TeV}})  \ N_{\rm BSM}^{1/2} \\   \mathrm{for\;}  X = \mbox{(unitarity*)}
\\[5pt]
(20~{\textrm{TeV}})  \ N_{\rm BSM}^{1/2}  \\ \mathrm{for\;}  X = \mbox{(unitarity+MFV)} 
\\[5pt]
(20~{\textrm{TeV}})  \ N_{\rm BSM}^{1/6} \\
\mathrm{for\;}  X = \mbox{(unit.+naturalness*)}
\\[5pt]
(9~{\textrm{TeV}})  \ N_{\rm BSM}^{1/6} \\
\mathrm{for\;}  X = \mbox{(unit.+nat.+MFV)}
\end{array}
\right.
\end{equation}
This upper bound is evaluated under four different assumptions for the BSM model solving the $g$-2 anomaly: perturbative unitarity only; unitarity  and MFV (Minimal Flavour Violation); unitarity and naturalness (i.e., specifically, avoiding fine-tuning in the Higgs and in the muon mass); and unitarity combined with naturalness and MFV. 

The unitarity-only bound represents the very upper limit of what is possible within quantum field theory at the edge of perturbativity, but realising such high masses requires severe alignment, tuning, or another unknown mechanism to avoid stringent constraints from charged lepton flavour-violating decays~\cite{Calibbi:2017uvl,Aubert:2009ag}.
Therefore, every scenario without MFV has been marked with a star (*) above, to indicate additional tuning or unknown flavour mechanisms that have to also be present.

These values of new physics particle masses provide a rough estimate of the maximal needed collider energy. It has been shown in \cite{Paradisi:2022vqp} that, in concrete models with new scalars and fermions, a MuC with center-of-mass energy in the 10 TeV range can discover the new physics responsible for the muon $g$-2 by means of both direct searches for the new states, and high-energy scattering of SM particles such as $\mu^+\mu^-\to h\gamma$. The combination of direct and indirect searches in different final states can be a powerful handle to disentangle among the underlying models accommodating the anomaly.

The results summarised in the previous two paragraphs show that a MuC with energies from the test-bed-scale $\mathcal{O}$(100 GeV) to $\mathcal{O}$(10 TeV) and beyond has excellent prospects to discover the new particles necessary to explain the $g$-2 anomaly. 

\begin{figure*}[t]
\begin{center}
\includegraphics[width=0.46\linewidth]{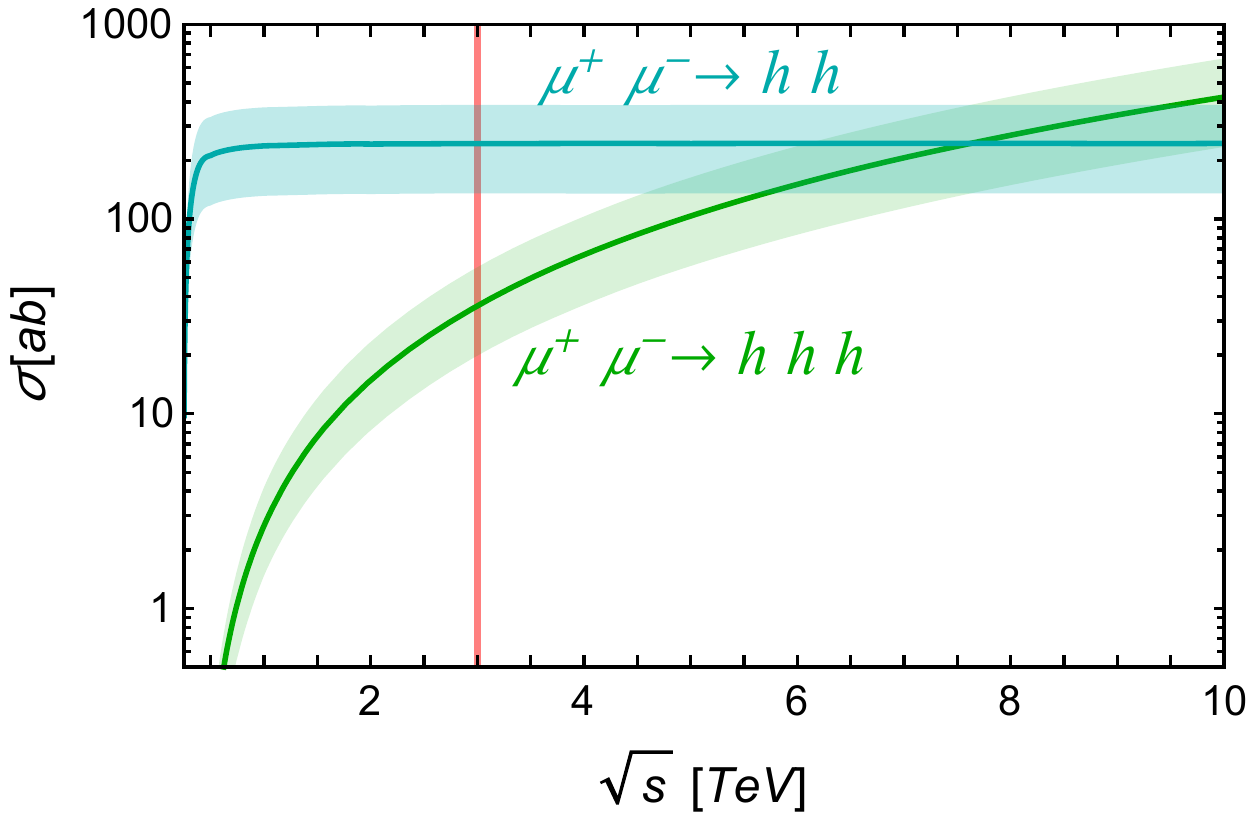}~\includegraphics[width=0.5\linewidth]{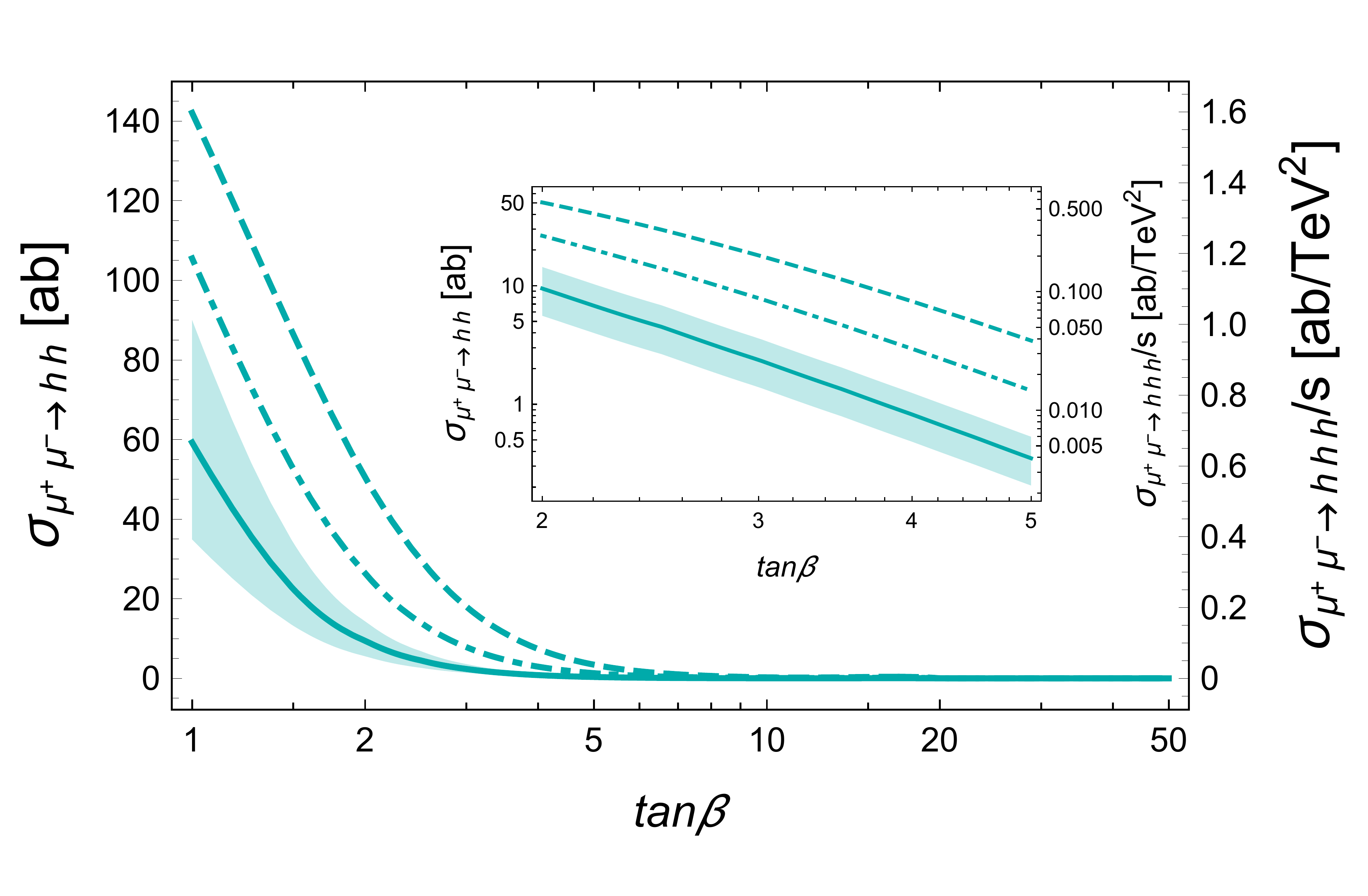}
\end{center}
\vspace{-0.5cm}
\caption{{\it Left:} Cross sections for $hh$ (cyan) and $hhh$ (green) production as a function of $\sqrt{s}$ in models with VLF. {\it Right:} Cross sections for $hh$ (left axis) and $hhh$ (right axis) production as a function of $\tan\beta$ in models with VLF and 2HDM for $M_{L,E}\simeq m_{H,A,H^{\pm}}$. The dot-dashed and dashed lines correspond to the predictions corresponding to the central value of $\Delta a_{\mu}$ and $m_{H,A,H^{\pm}}=3\times M_{L,E}$ and $m_{H,A,H^{\pm}}=5\times M_{L,E}$, respectively. Both panels assume $\Delta a_{\mu}$ is within $1\sigma$ of the measured value (shaded ranges)~\cite{Dermisek:2021mhi}.}
\label{fig:xsection_plot}
\end{figure*} 
%

\paragraph{Multi-Higgs Signatures from Vector-like Fermions}{\ } \\ 
\noindent
Simple explanations for $g$-2 involve extensions of the SM with new Vector-Like Fermions (VLF) where the corrections to the muon magnetic moment are mediated by the SM Higgs and gauge bosons~\cite{Kannike:2011ng,Dermisek:2013gta}. These models generate effective interactions between the muon and multiple Higgs bosons leading to predictions for di- and tri-Higgs production at a $\mc$ that are directly correlated with the corrections to $\Delta a_{\mu}$. This section reviews the findings of~\cite{Dermisek:2021mhi,Dermisek:2022aec} on this subject. The authors consider extensions of the SM with VLF doublets, $L_{L,R}$, and singlets $E_{L,R}$ with masses $M_{L,E}$, respectively. It will be assumed that new $L_{L}$ and $E_{R}$ have the same quantum numbers as the SM leptons, but other possibilities will also be commented upon later.

The Yukawa interactions of interest are
\begin{eqnarray}
\mathcal{L}\supset && - y_{\mu}\bar{l}_L\mu_{R}H - \lambda_{E}\bar{l}_{L}E_{R}H  - \lambda_{L}\bar{L}_{L}\mu_{R}H  
\nonumber\\
&&- \lambda\bar{L}_{L}E_{R}H  - \bar{\lambda}H^{\dagger}\bar{E}_{L}L_{R}  + h.c.,
\label{eq:lagrangian}	
\end{eqnarray}
where $l_{L}=( \nu_{\mu},  \mu_{L} )^T$,  $ L_{L,R}= ( L_{L,R}^{0}, L_{L,R}^{-})^T$, and $H=(0,\;v + h/\sqrt{2})^{T}$ with $v=174$ GeV.
In the limit $v\ll M_{L,E}$, after integrating out the heavy leptons at tree level, eq.~(\ref{eq:lagrangian}) becomes
\begin{equation}
\mathcal{L}\supset - y_{\mu}\bar{l}_L\mu_{R}H - \frac{m_{\mu}^{LE}}{v^{3}}\bar{l}_L\mu_{R}H(H^\dagger H) + h.c.,
\label{eq:eff_lagrangian}
\end{equation}
where
\begin{equation}
m_\mu^{LE} \equiv \frac{\lambda_{L} \bar{\lambda} \lambda_{E}}{M_{L}M_{E}} v^3\,,
\label{eq:m^LE}
\end{equation}
is the contribution to the muon mass from mixing with new leptons.
Mixing of the muon with heavy leptons also leads to modifications of the muon couplings to $W$, $Z$, and $h$, and generates new couplings of the muon to new leptons. 

Assuming that $v\ll M_{L,E}$, the total one-loop correction to $g$-2 induced by these effects is well approximated by~\cite{Kannike:2011ng,Dermisek:2013gta}
\begin{equation}
\Delta a_{\mu}
= - \frac{1}{16\pi^{2}}  \frac{m_\mu m_\mu^{LE}}{v^2}.
\label{eq:dela}
\end{equation}
The explanation of the measured value of $\Delta a_{\mu}$ within $1\sigma$ requires that
\begin{equation}
m_\mu^{LE}/m_\mu = -1.07 \pm 0.25.
\label{eq:mmu_LE_range}
\end{equation}
For couplings of $\mathcal{O}(1)$, eq.~(\ref{eq:mmu_LE_range}) can be achieved for new lepton masses even as heavy as 7~TeV while simultaneously satisfying current relevant constraints~\cite{Dermisek:2021ajd}. For couplings close to the limit of perturbativity, $\sqrt{4\pi}$, this range extends to close to 50 TeV. This far exceeds the reach of the LHC and even projected expectations of possible future proton-proton colliders, such as the FCC-hh. However, there are related signals that could be fully probed at, for example, a 3 TeV $\mc$ through the effective interactions generated between the muon and multiple Higgs bosons. These interactions are all generated by eq.~(\ref{eq:eff_lagrangian})~\cite{Dermisek:2021mhi} and they lead to the following predictions
\begin{eqnarray}
\sigma_{\mu^+\mu^- \to hh} &=&  \frac{\left|\lambda^{hh}_{\mu \mu}\right|^2}{64 \pi}   = \frac{9}{64 \pi} \left(\frac{m_\mu^{LE}}{v^2}\right)^2 \label{eq:sigma_hh},\label{eq:EFT_xsections_1}\\
\sigma_{\mu^+\mu^- \to hhh}  &=& \frac{\left|\lambda^{hhh}_{\mu \mu}\right|^2}{6144 \pi^3} s = \frac{3}{4096 \pi^3} \left(\frac{m_\mu^{LE}}{v^3}\right)^2 s \label{eq:sigma_hhh}.
\label{eq:EFT_xsections_2}
\end{eqnarray}

\begin{table}[t]
\begin{center}
\begin{tabular}{ |c||c||c|c| } 
\hline
$SU(2)\times U(1)_{Y}$& c & $\sigma_{hh}(3\;\text{TeV})$ & $\sigma_{hhh}(3\;\text{TeV})$ \\
\hline
$\mathbf{2}_{-1/2}\oplus\mathbf{1}_{-1}$ & 1 & $244^{+141}_{-109}$~[ab]& $35.8^{+20.8}_{-15.9}$~[ab] \\
\hline
$\mathbf{2}_{-1/2}\oplus\mathbf{3}_{-1}$  & 5 &$10^{+6}_{-4}$~[ab]&$1.43^{+0.8}_{-0.6}$~[ab]\\
\hline
$\mathbf{2}_{-3/2}\oplus\mathbf{1}_{-1}$ & 3 & $27^{+16}_{-12}$~[ab] & $4.0^{+2.3}_{-1.8}~[ab]$\\
\hline
$\mathbf{2}_{-3/2}\oplus\mathbf{3}_{-1}$ & 3 &$27^{+16}_{-12}$~[ab] & $4.0^{+2.3}_{-1.8}$~[ab]\\
\hline
$\mathbf{2}_{-1/2}\oplus\;\mathbf{3}_{0}$ &1 & $244^{+141}_{-109}$~[ab]& $35.7^{+20.7}_{-15.9}$~[ab]\\
\hline
\end{tabular}
\caption{Quantum numbers of $L_{L,R}\oplus E_{L,R}$ under $SU(2)\times U(1)_{Y}$, corresponding $c$-factor for $\Delta a_{\mu}$, and predictions for di- and tri-Higgs cross sections running at $\sqrt{s}=3$ TeV, assuming $\Delta a_{\mu}\pm 1\sigma$.}
\label{table:models}
\end{center}
\end{table}

Thus, considering eq~(\ref{eq:dela}), one can see that the effective interactions of the muon with the Higgs are completely fixed by the muon mass and the predicted value of $\Delta a_{\mu}$.
The left panel of Figure~\ref{fig:xsection_plot} shows the total $\mu^{+}\mu^{-}\to hh$ and $\mu^{+}\mu^{-}\to hhh$ cross sections at a $\mc$ as a function of $\sqrt{s}$ calculated from the effective lagrangian and assuming that $\Delta a_{\mu}$ is achieved within $1\sigma$ (shaded ranges). Cross sections for a 3 TeV $\mc$ are highlighted with the red line. One can see that, for example, a $\mc$ running at $\sqrt{s}= 3$ TeV with 1 ab$^{-1}$ of integrated luminosity would see about 240 di-Higgs events and about 35 tri-Higgs events. It should be noted that already at $\sqrt{s}=1$ TeV this is roughly 4 (3) orders of magnitude larger than $\mu^{+}\mu^{-}\to hh$ and $\mu^{+}\mu^{-}\to hhh$ in the SM. Di- and tri-Higgs final states produced from vector boson fusion in the SM are characterised by a low total invariant mass. They are therefore easily distinguishable from those from direct $\mu^{+}\mu^{-}$ annihilation, which carry the entire energy of the collider. Backgrounds involving the $Z$-boson such as $\mu^{+}\mu^{-}\to Zh$ or $\mu^{+}\mu^{-}\to ZZ$, which may be comparable at the level of cross sections, should be instead suppressed by an invariant-mass cut on the $Z$-boson decay products.

Models with more exotic quantum numbers can also generate a similar correction to $\Delta a_{\mu}$ and, hence, similar predictions for di- and tri-Higgs cross sections. In total there are 5 different combinations of new lepton fields that can lead to mass-enhanced corrections to $\Delta a_{\mu}$ mediated by the SM Higgs. In each case, the correction as given in eq.~(\ref{eq:dela}) is simply multiplied by a corresponding $c$-factor. The resulting cross sections are then rescaled by a factor of $1/c^{2}$ compared to those in Figure~\ref{fig:xsection_plot}. Table~\ref{table:models} lists the $c$-factor multiplying eq.~(\ref{eq:dela}), and the corresponding predictions for di- and tri-Higgs cross sections for a $\mc$ running at $\sqrt{s}=3$ TeV, assuming $\Delta a_{\mu}\pm 1\sigma$. A $\mc$ can fully probe these scenarios even with moderate energies $\sqrt{s}\sim 1-3$ TeV.


\begin{figure}[h]
\includegraphics[width=0.9\linewidth]{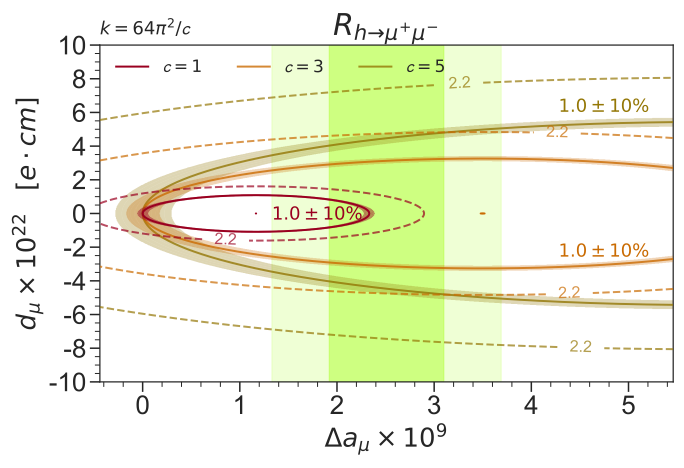} 
\caption{Contours of constant $R_{h\to \mu^+\mu^-} = 1$ (solid), $1\pm 0.1$ (shaded), and 2.2 (dashed) in the $\Delta a_{\mu}$ -- $d_{\mu}$ plane in models with $c=1$, 3, and 5.}
\label{fig:SM_ellipse}
\end{figure}

Allowing for complex couplings, a parameter-free correlation emerges between  $\Delta a_\mu$, $d_\mu$, and $R_{h\to \mu^+\mu^-} \equiv BR(h\to \mu^+\mu^-) / BR(h\to \mu^+\mu^-)_{SM}$, with $R_{h\to \mu^+\mu^-} $ carving an ellipse in the plane of dipole moments~\cite{Dermisek:2022aec}:
\begin{flalign}
R_{h\to \mu^+\mu^-}=\left(\frac{\Delta a_{\mu}}{2\omega} - 1\right)^{2} + \left(\frac{m_{\mu}d_{\mu}}{e\omega}\right)^{2},
\label{eq:ellipse}
\end{flalign}
where $\omega = m_{\mu}^{2}/kv^{2}$. The $k$-factor relates the dimension 6 mass and dipole operators and for models with VLF it is given by
\begin{equation}
k=\frac{64\pi^{2}}{c},
\label{eq:tree_X_SM}
\end{equation}
where $c$-factors are listed in Table~\ref{table:models}. In Figure~\ref{fig:SM_ellipse} we show contours of constant $R_{h\to \mu^+\mu^-} = 1$, $1\pm 0.1$, and 2.2 in the plane of the muon dipole moments for models with $c=1$, 3, and 5. Different explanations of $\Delta a_\mu$ and the SM-like $R_{h\to \mu^+\mu^-}$ (or any other fixed value) require specific values of $d_{\mu}$.  Non-zero $d_{\mu}$ can only  increase the quoted rates for $\mu^+ \mu^- \to hh$ and  $\mu^+ \mu^- \to hhh$ and similar ellipses can be shown for the corresponding production cross sections. Models can be efficiently distinguished by these correlations. 

\begin{figure*}[t]
\centering
\includegraphics[width=0.75\textwidth]{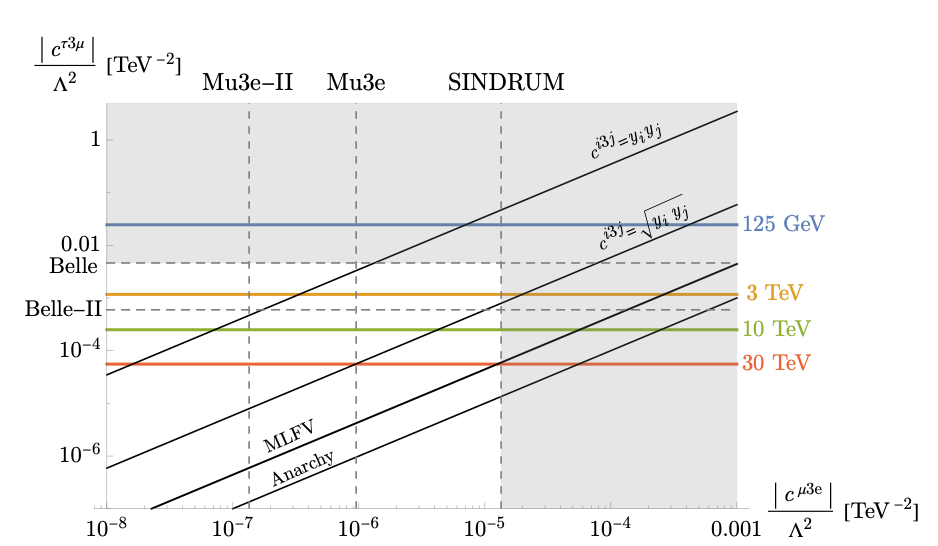}
\caption{Summary of $\mc$ and low-energy constraints on flavor-violating 3-body lepton decays. The colored horizontal lines show the sensitivity to the $\tau 3\mu$ operator at various energies, all assuming $1\text{ ab}^{-1}$ of data. The dashed horizontal (vertical) lines show the current or expected sensitivity from $\tau \to 3\mu$ ($\mu \to 3e$) decays for comparison. The diagonal black lines show the expected relationship between different Wilson coefficients with various ansatz for the scaling of the flavor-violating operators (e.g., ``Anarchy'' assumes that all Wilson coefficients are $\mathcal{O}(1)$).
}\label{fig:lfv-operator-summary}
\end{figure*}

\paragraph*{Vector-like fermions and Two-Higgs-Doublet models.}
\;
It is straightforward to extend the discussion from the previous section to a 2HDM~\cite{Dermisek:2020cod,Dermisek:2021ajd}. For instance, in a type-II 2HDM where charged leptons couple exclusively to one Higgs doublet, $H_{d}$, (which can be achieved by assuming a $Z_{2}$ symmetry) the lagrangian in eq.~(\ref{eq:lagrangian}) from the previous section, is simply modified with the replacement $H\to H_{d}$. In this case both Higgs doublets develop a vev $\left\langle H_{d}^{0} \right\rangle=v_{d}$ and $\left\langle H_{u}^{0} \right\rangle=v_{u}$, where $\sqrt{v_{d}^{2}+v_{u}^{2}}=v=174$ GeV and $\tan\beta = v_{u}/v_{d}$. The effective interactions generated by integrating out heavy leptons is then
\begin{equation}
\mathcal{L}\supset y_{\mu}\bar{\mu}_{L}\mu_{R}H_{d} - \frac{m_{\mu}^{LE}}{v_{d}^{3}}\bar{\mu}_{L}\mu_{R}H_{d}(H_{d}^{\dagger} H_{d}).
\label{eq:eff_lagrangian_2HDM}
\end{equation}
Similar modifications to $Z$, $W$, and the SM-like Higgs couplings to the muon are also generated after EWSB.
Including the additional corrections to $\Delta a_{\mu}$ from heavy charged and neutral Higgs bosons leads to~\cite{Dermisek:2020cod,Dermisek:2021ajd}
\begin{equation}
\Delta a_{\mu}=-\frac{1+\tan^{2}\beta}{16\pi^{2}}\frac{m_{\mu}m_{\mu}^{LE}}{v^{2}},\;\;m_\mu^{LE} \equiv \frac{\lambda_{L} \bar{\lambda} \lambda_{E}}{M_{L}M_{E}} v_{d}^3,
\label{eq:gm2}
\end{equation}
where $M_{L,E}\simeq m_{H,A,H^{\pm}}$ is assumed for simplicity. The first term in eq.~(\ref{eq:gm2}) results from the same loops as in the SM, i.e. involving the $Z$, $W$, and SM-like Higgs, whereas the second term, enhanced in comparison by $\tan^{2}\beta$, results from the additional contributions from the heavy Higgses. The corresponding requirement to satisfy $\Delta a_{\mu}$ within $1\sigma$ then becomes
\begin{equation}
m_{\mu}^{LE}/m_{\mu}=(-1.07\pm0.25)/(1+\tan^{2}\beta).
\label{eq:gm2_range}
\end{equation}
Just as in the previous section, effective interactions between the muon and multiple Higgs bosons are generated via the single dimension-six operator in eq.~(\ref{eq:eff_lagrangian}). Thus, predictions for di- and tri-Higgs cross sections follow in the same way simply by replacing $m_{\mu}^{LE}$ with the corresponding definition in eq.~(\ref{eq:gm2}). Considering eq.~(\ref{eq:gm2_range}), it follows that $\sigma_{\mu^{+}\mu^{-}\to hh}$ and $\sigma_{\mu^{+}\mu^{-}\to hhh}$ cross sections in a type-II 2HDM decrease as $1/\tan^{4}\beta$. 

Figure~\ref{fig:xsection_plot} shows the $\tan\beta$ dependence of $\sigma_{\mu^{+}\mu^{-}\to hh}$ and $\sigma_{\mu^{+}\mu^{-}\to hhh}/s$ as obtained from the effective lagrangian when $\Delta a_{\mu}$ is achieved within $1\sigma$ (shaded range) and $M_{L,E}\simeq m_{H,A,H^{\pm}}$. The dot-dashed and dashed lines correspond to the predictions corresponding to the central value of $\Delta a_{\mu}$ and $m_{H,A,H^{\pm}}=3\times M_{L,E}$ and $m_{H,A,H^{\pm}}=5\times M_{L,E}$, respectively. Its expected that future measurements of $h\to \mu^{+}\mu^{-}$ will probe $\tan\beta$ up to $\sim 5$ and the inset zooms into this region~\cite{Dermisek:2021mhi}.

For a $\mc$ running at centre-of-mass energy of 3 TeV with, for example, 1 ab$^{-1}$ of luminosity, 3 di-Higgs events are expected in these scenarios for $\tan\beta \simeq 3$. For tri-Higgs the same sensitivity does not extend much above $\tan\beta \simeq 1$. When $m_{H,A,H^{\pm}}=5\times M_{L,E}$, the corresponding sensitivities to $\tan\beta$ increase to about $\tan\beta\simeq 5$ and 2.5 for di-Higgs and tri-Higgs signals, respectively.

These conclusions also extend to models with additional scalars where the SM Higgs is only one component of the scalar sector responsible for EWSB. Mixing within the Higgs sector (e.g. $\tan\beta$ in a 2HDM) introduces a free parameter to the predictions and correlations between the muon magnetic moment and effective Higgs couplings. Thus, the corresponding predictions for di- and tri-Higgs signals at a $\mc$ are not as sharp in these scenarios as compared to the SM. Though in a 2HDM the observables parametrically interpolate between the SM and models with scalars that do not participate in EWSB.


Allowing for complex couplings, the correlation between $\Delta a_\mu$, $d_\mu$, and $R_{h\to \mu^+\mu^-}$ discussed in the previous paragraph emerges in all models with chiral enhancement. In 2HDM the $k$ factor is determined by $\tan\beta$, and in models with scalars not participating in EWSB, for example those discussed in~\cite{Capdevilla:2021rwo}, the $k$ factor is directly linked to the coupling responsible for chiral enhancement~\cite{Dermisek:2022aec}. Just like for the 2HDM, the corresponding rates for $\mu^+ \mu^- \to hh$ and  $\mu^+ \mu^- \to hhh$ can be calculated in any model with chiral enhancement. Furthermore, in all these models, the correlation between $\Delta a_\mu$, $d_\mu$, and $R_{h\to \mu^+\mu^-}$ allows to set the upper bound on masses of new particles able to explain  $\Delta a_\mu$.


\subsubsection*{Lepton Flavour Violation}

The SM exhibits a distinctive pattern of fermion masses and mixing angles, for which we currently have no deep explanation.
Delicate symmetries also lead to a strong suppression of flavor-changing processes in the quark and lepton sectors, which may be reintroduced by new particles or interactions.
The non-observation of such processes thus leads to some of the most stringent constraints on BSM physics, while a positive signal could give us insight into the observed structure of the SM.
A number of precision experiments searching for lepton flavor violating (LFV) processes such as $\mu \to 3e$, $\tau \to 3\mu$ or $\mu$-to-$e$ conversion within atomic nuclei will explore these processes with orders of magnitude more precision in the coming decades~\cite{Baldini:2018uhj}.
As we will see, a high-energy $\mc$ has the unique capability to explore the same physics --- either via measuring effective interactions or by directly producing new states with flavor-violating interactions --- at the TeV scale.

\begin{figure*}[t!]
\centering
\includegraphics[width=7cm]{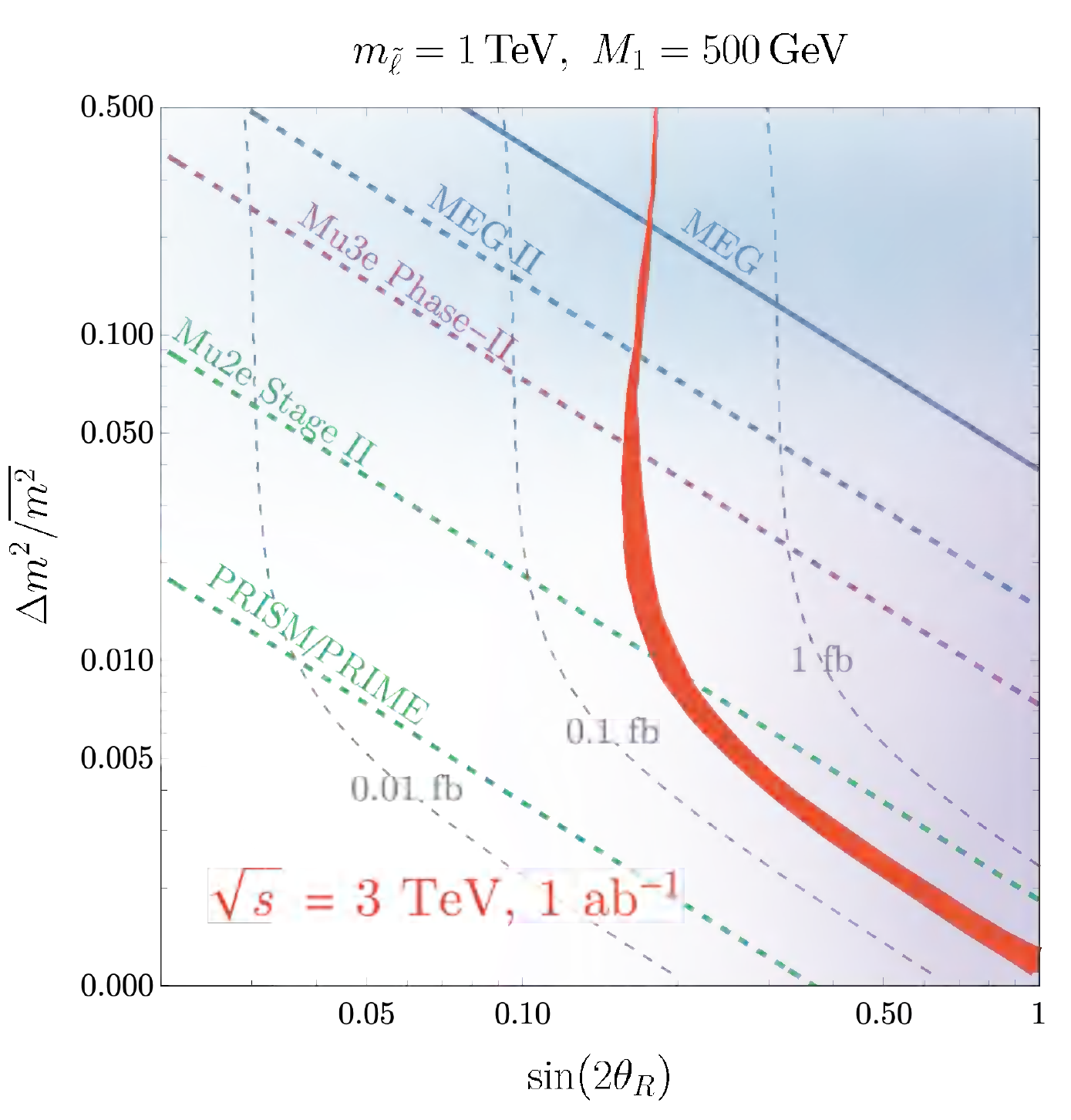}
~
\includegraphics[width=7cm]{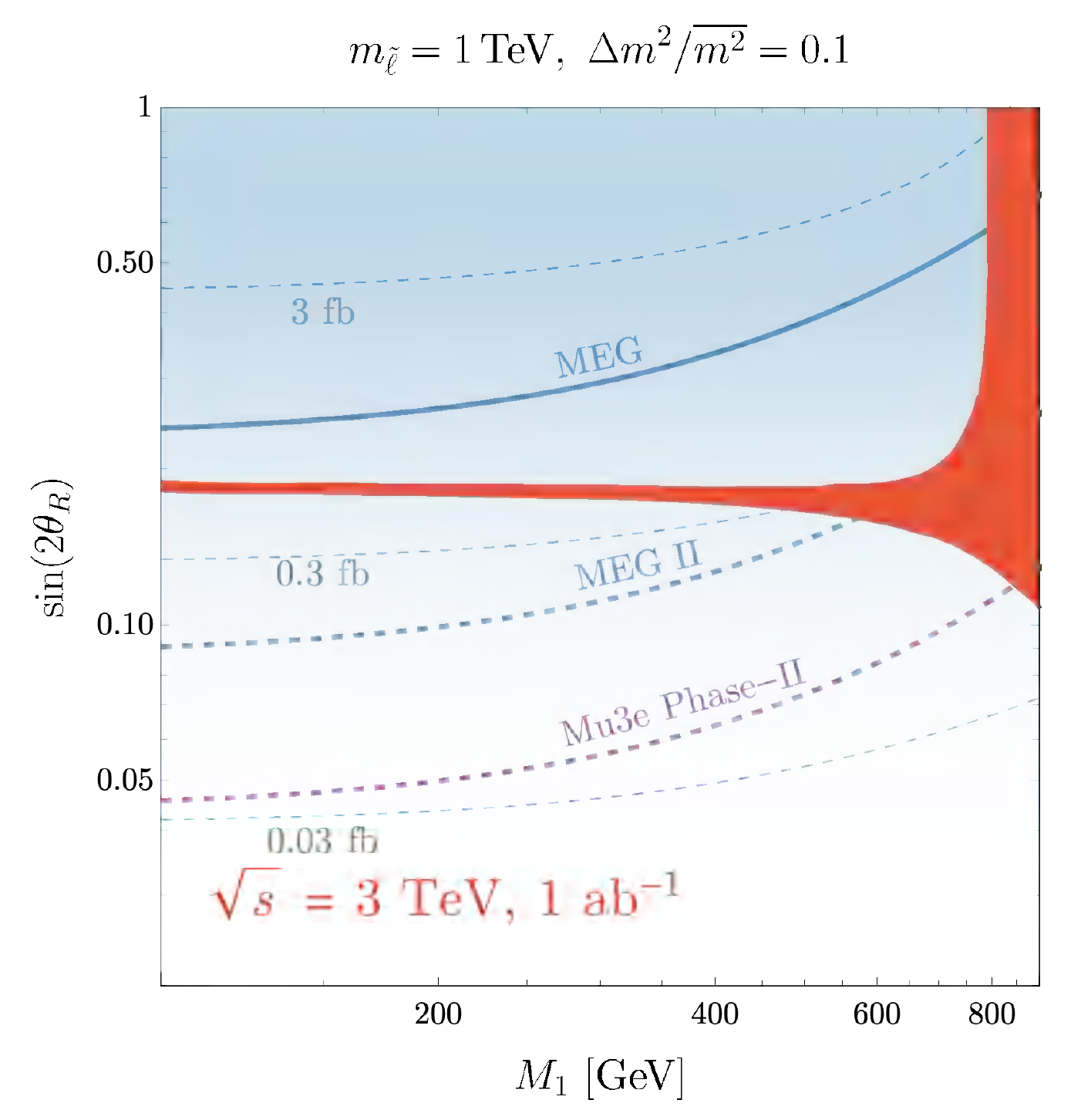}
\caption{Constraints on lepton flavor violation in the MSSM in the $\Delta m^2 / \bar{m}^2$ vs. $\sin 2\theta_R$ plane (left) and the $\sin 2\theta_R$ vs. $M_1$ plane (right) from measurements of the slepton pair production process with flavor-violating final states (red band) at a 3 TeV $\mc$, assuming $1\,\textrm{ab}^{-1}$ of luminosity.
The width of the band represents the uncertainty on the reach from the measurement of the slepton and neutralino masses in flavor-conserving channels.
The purple and blue shaded lightly shaded regions indicate parameters preferred in Gauge-Mediated Supersymmetry Breaking scenarios and flavor-dependent mediator scenarios, respectively.
Both plots assume a mean slepton mass of $1\,\textrm{TeV}$. In the left plot we fix the neutralino mass $M_1 = 500\,\textrm{GeV}$, while in the right figure $\Delta m^2 / \bar{m}^2$ is fixed to 0.1.
The current (solid) and expected (dashed, dotted) limits from low-energy lepton flavor violation experiments are indicated by the blue, purple and green lines.}
\label{fig:mssm-3tev-summary}
\end{figure*}

\paragraph{Effective LFV Contact Interactions}{\ } \\ 
\noindent
In this section, we study $\mc$ bounds on $\mu\mu \ell_i \ell_j$-type contact interactions, and demonstrate the complementarity with precision experiments looking for lepton-flavor violating decays, as first studied in~\cite{AlAli:2021let}. We will focus on $\tau 3\mu$ and $\mu 3e$ operators, since constraints on them can be compared directly with the sensitivity from $\tau \to 3\mu$ and $\mu \to 3 e$ decays. We parametrise the four-fermion operators relevant for the $\tau \to 3\mu$ decay via
\begin{eqnarray}
\mathcal{L} \supset &&
V_{LL}^{\tau 3\mu}\big( \bar{\mu} \gamma^{\mu} P_L \mu \big) \big(\bar{\tau} \gamma_{\mu} P_L \mu\big) \nonumber \\
&&+ V_{LR}^{\tau 3\mu} \big( \bar{\mu} \gamma^{\mu} P_L \mu\big) \big(\bar{\tau} \gamma_{\mu} P_R \mu\big) \nonumber \\
&&+ \big( L \leftrightarrow R \big) + \textrm{h.c.}\,,
\end{eqnarray}
with an equivalent set for the $\mu \to 3e$ decay. 
In what follows, we will assume all the $V_{ij}^{\tau 3\mu}$ coefficients are equal to $c^{\tau 3\mu} / \Lambda^2$, 
where $c^{\tau 3\mu}$ is a dimensionless coefficient and $\Lambda$ is to be interpreted as the scale of new physics, and similarly for $\mu 3 e$ coefficients.

At a $\mc$, the $\tau 3 \mu$ coefficients are probed via the $\mu^+\mu^- \to \mu \tau$ scattering process. 
Our analysis closely follows an analogous study at an $e^+ e^-$ collider in Ref.~\cite{Murakami:2014tna}. 
As discussed in~\cite{AlAli:2021let}, the SM backgrounds from $\tau^+\tau^-$ and $W^+W^-$ production can be substantially mitigated by a simple set of cuts, whereas the signal can be largely retained up to $\sim 10\%$ effects due to initial state radiation.
The resulting bounds, assuming fixed integrated luminosities of $1\,\textrm{ab}^{-1}$ at $0.125$, $3$, $10$ and $30\,\textrm{TeV}$ are shown in Figure~\ref{fig:lfv-operator-summary}, alongside current and future sensitivities of $\tau \to 3\mu$ and $\mu\to 3 e$ experiments.
A $3\,\textrm{TeV}$ machine would set a direct bound at the same level as the future Belle~II sensitivity. The sensitivity of higher energy MuCs is underestimated in Figure~\ref{fig:lfv-operator-summary} because the expected luminosity is higher, and vice versa for the 125~GeV MuC.

Given an ansatz regarding the flavour structure, the constraints on the $\tau 3\mu$ operators can be compared to the constraints on the analogous $\mu 3e$ operator in the $\mu \to 3 e$ decay. 
The diagonal lines in Figure~\ref{fig:lfv-operator-summary} show the expected relationship between the two Wilson coefficients for several different ansatz, including flavor anarchy (where all coefficients $\sim 1$), Minimal Leptonic Flavor Violation~\cite{Cirigliano:2005ck}, or scalings with different powers of the involved Yukawa couplings.
While muon decays set the strongest limits assuming anarchical coefficients, a $\mc$ could set competitive constraints for other ansatz: in the most extreme case, where the Wilson coefficients scale like the product of the Yukawas, a $3\,\textrm{TeV}$ machine would have sensitivity comparable to the final Mu3e sensitivity.

In addition to the $\tau 3\mu$ operators considered here, similar sensitivity should be attainable for the process $\mu^+ \mu^- \to \mu^{\pm} e^{\mp}$, as well as for the processes such as $\mu^+ \mu^- \to \tau^{\pm} e^{\mp}$ that violate lepton flavor by two units.
Overall, we see that a $\mc$ would be capable of directly probing flavor-violating interactions that are quite complementary to future precision constraints.

\paragraph{Direct Probes: LFV in the MSSM}{\ } \\ 
\noindent
An exciting possibility is that the flavor-changing processes that might be observed in low-energy experiments arise from loops of new particles near the TeV scale.
As a motivated example, consider the MSSM. The scalar superpartners of the SM leptons can have soft supersymmetry-breaking contributions to their mass- matrix that are off-diagonal in the SM lepton eigenbasis.
As a result, the slepton interactions with the leptons will be flavor-violating and lead to processes such as muon-to-electron conversion and rare muon decays at one loop. 
In well-motivated constructions, the mixing between the scalar partners of the electron and the muon states can be quite large, as the low-energy processes are protected by a ``Super-GIM'' mechanism~\cite{Arkani-Hamed:1996bxi}, allowing the new states to be near the TeV scale while consistent with current bounds.

A 3 TeV $\mc$ would dramatically extend the reach for electroweak-charged superpartners beyond a TeV, raising the possibility of directly producing the new states responsible for lepton flavor-violation. 
Moreover, the unique environment of a $\mc$ makes it possible to not only produce these new states, but measure their LFV interactions. 
This would provide detailed insight into both the mechanism of supersymmetry breaking and the origin of the flavor structure of the SM.
A detailed investigation of these prospects is carried out in~\cite{Homiller:2022iax};
here we briefly review results for 3 TeV. 

To understand the complementarity of low-energy LFV probes and the $\mc$ reach, we consider the scenario where only the right-handed selectron and smuon, along with one light neutralino (which we will assume to be a pure bino with mass $M_1$) are in the spectrum.
If the slepton masses $m_{\tilde{\ell}} > M_1$, the sleptons decay directly to a lepton and bino, and the LFV interactions can be measured directly via the pair-production process: $\mu^+ \mu^- \to \tilde{e}^+_{1,2} \tilde{e}^-_{1,2} \to \mu^{\pm} e^{\mp} \chi_1^0 \chi_1^0$, where the binos appear as missing momentum.
In this simplified scenario, both the low-energy LFV processes and the pair-production process at a $\mc$ depend only on the slepton masses and mixing angle, as well as $M_1$. 

\begin{figure}[t]
	\begin{center}
\includegraphics[width=0.45\textwidth]{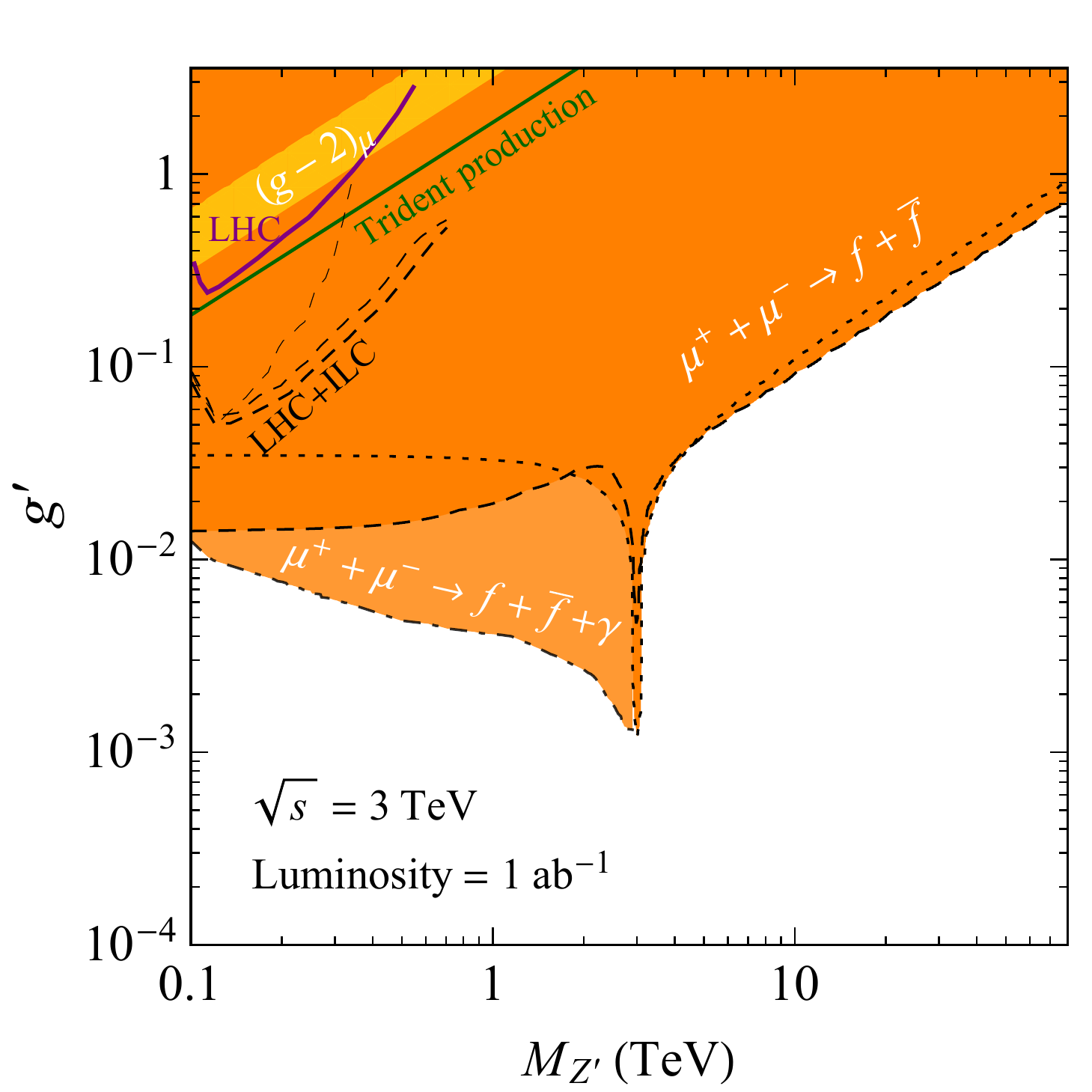}
	\end{center}
	\vspace{-0.3cm}
	\caption{The 2$\sigma$ sensitivity~\cite{Huang:2021nkl} of the 3~TeV MuC with $1~{\rm ab^{-1}}$ luminosity, given as orange regions. Other limits and projections are also shown for comparison. The region explaining the $(g-2)^{}_{\mu}$ anomaly is outlined by the yellow band.}
	\label{fig:contour}
\end{figure}

\begin{table*}[t]
\centering
\begin{tabular}{|c|c|c|c|}
  \hline
  Coupling & $\kappa\equiv g/g_{\rm SM}$ & Type-II \& lepton-specific& Type-I \& flipped\\ \hline
  $g_{H\mu^+\mu^-}$ & $\kappa_\mu$ &$\sin\alpha/\cos\beta$ & $\cos\alpha/\sin\beta$ \\ \hline
  $g_{A\mu^+\mu^-}$ & $\kappa_\mu$ &$\tan\beta$ & $-\cot\beta$ \\ \hline
  $g_{HZZ}$ & $\kappa_Z$ & $\cos(\beta-\alpha)$ & $\cos(\beta-\alpha)$ \\ \hline
  $g_{HAZ}$ & $1-\kappa^2_Z$ & $\sin(\beta-\alpha)$ & $\sin(\beta-\alpha)$ \\
  \hline
\end{tabular}
\caption[]{Coupling parameter values in different 2HDM setups.}
\label{tab:parameters}
\end{table*}

In Figure~\ref{fig:mssm-3tev-summary}, we show the $5\sigma$ reach for a $3\,\textrm{TeV}$ $\mc$, assuming an average slepton mass of $1\,\textrm{TeV}$.
The left panel shows the reach as a function of the mixing angle and mass-splitting, $\Delta m^2 = m_{\tilde{e},2}^2 - m_{\tilde{e},1}^2$, with $M_1 = 500\,\textrm{GeV}$. The right panel shows the constraints for fixed $\Delta m^2 / \bar{m}^2 = 0.1$ in the $M_1$ vs. $\sin 2\theta_R$ plane. Large mixing angles are motivated in models involving gauge-mediated supersymmetry breaking (GMSB), indicated by the purple region, while larger mass splittings are motivated in scenarios where the messengers carry flavor-dependent charges, such as $L_{\mu} - L_{\tau}$, indicated by the blue regions (see~\cite{Homiller:2022iax} for more details).
The complementary constraints from low-energy experiments searching for $\mu\to e\gamma$, $\mu\to 3e$ decays or $\mu$-to-$e$ transitions are shown in blue, purple and green, respectively.
We see that the $\mc$ reach extends to small mass splittings in the GMSB scenario, and covers a substantial part of the most well-motivated parameter space.

\paragraph{Gauge \texorpdfstring{$L_\mu-L_\tau$}{lmmt} Interactions}{\ } \\ 
\noindent
As a last example of new physics source of LFV, we consider the gauging of the $L_\mu-L_\tau$ charge. It is not straightforward to test this model at laboratories due to the preferred couplings to the second and third family leptons, unless we have a facility to directly collide muons. Here we summarise the findings of \cite{Huang:2021nkl} regarding searches of a gauged $L^{}_{\mu}{-}L^{}_{\tau}$ interaction at a MuC.
The discussion focuses on the 3~TeV MuC with $1~{\rm ab}^{-1}$ luminosity.

The relevant interactions of the new boson $Z'$ read
\begin{eqnarray} \label{eq:L}
	\mathcal{L} \supset  g^{\prime} \left( \overline{\ell^{}_{\rm L}} Q^{\prime} \gamma^{\mu} \ell^{}_{\rm L} + \overline{E^{}_{\rm R}}Q^{\prime} \gamma^{\mu} E^{}_{\rm R} \right) Z^{\prime}_{\mu}\;,
\end{eqnarray}
where $g'$ stands for the coupling constant of gauged $L^{}_{\mu}{-}L^{}_{\tau}$ symmetry, $\ell \equiv (\nu, E)^{\rm T}$ is the lepton doublet with $\nu$ and $E$ being the neutrino and the charged lepton, respectively, and $Q^{\prime} = {\rm Diag}(0,1,-1)$ represents the charge matrix in the basis of $(e,\mu,\tau)$.
The $Z'$ will inevitably mix with the SM gauge bosons,  i.e., $\gamma$ and $Z$. It is found that the mixing with $\gamma$ is strongly suppressed by the $Z'$ mass, while the mixing with $Z$ can be relevant if their masses are of the similar order.
For simplicity, we assume a negligible mixing in the following, which actually represents a conservative estimate of the sensitivity.

In such a setup, the relevant processes for the analysis include the final-state signatures of dimuon plus photon, ditau photon as well as monophoton. Even though the process with  initial photon radiation is of higher order compared to the trivial two-body scatterings, its impact is comparable and in some circumstances even larger than the two-body ones, due to the radiative return of resonant $Z'$ production~\cite{Chakrabarty:2014pja,Karliner:2015tga}.

The two-body scattering is very clean, as the final back-to-back dimuon or ditau carries all the energy delivered by the initial colliding muons. The only background of our concern should be the intrinsic SM processes, such as $\mu^+  \mu^- \to \gamma/Z \to l^+  l^-$ as well as $t$-channel exchanges.

Here one also benefits from the interference between the $Z'$ and SM-mediated diagrams. For instance, consider $\mu^+ \mu^- \to \tau^+ \tau^-$. The interference contribution to the cross section 
\begin{equation}
\sigma \sim e^2 g'^2/(4\pi s) \quad {\rm for} \quad s \gg M^{2}_{Z'} \,,
\end{equation}
and 
\begin{equation}
\sigma \sim -e^2 g'^2/(4\pi M^2_{Z'}) \quad {\rm for} \quad s \ll M^{2}_{Z'}\,,
\end{equation}
dominates over the $Z'$-only cross section, which is proportional to  $\propto g'^4$, when $g'$ is small.
Comparing with the SM cross section
\begin{equation}
\sigma
\sim e^4 /(8\pi s) \sim 10^4~{\rm ab}~(3~{\rm TeV}/\sqrt{s})^2\,,
\end{equation}
one can readily estimate the excellent sensitivity to the gauge coupling  
\begin{eqnarray} 
	g' < && 3.4 \times 10^{-2} \times \nonumber \\ && \left(\frac{\sqrt{s}}{3~{\rm TeV}} \right)^{\frac{1}{2}}  \left(\frac{1~{\rm ab^{-1}}}{\mathfrak{L}} \right)^{\frac{1}{4}}  {\rm max}\left(1, \frac{M^{}_{Z'}}{\sqrt{s}} \right)\,
\end{eqnarray}
for a collider with centre of mass energy $\sqrt{s}$ and integrated luminosity $\mathfrak{L}$.

\begin{figure*}[t]
\centering
\includegraphics[scale=0.45]{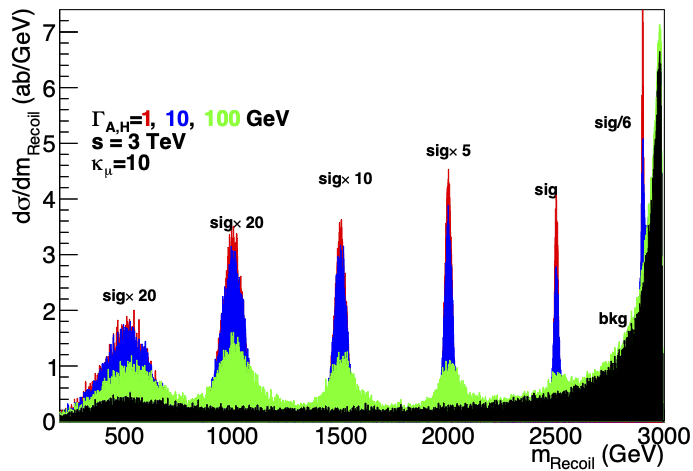}
\caption{Recoil mass distribution~\cite{Chakrabarty:2014pja} for heavy Higgs mass of 0.5, 1, 1.5, 2, 2.5, 2.9 TeV with a total width 1 (red), 10 (blue), and 100 (green) GeV at a 3 TeV $\mc$. Background (black shaded region) includes all events with a photon of $p_T>10~$GeV.}
\label{fig:sigbkg}
\end{figure*}

For more detailed sensitivity projections, one has to make a few assumptions about the particle identification and detection prospects.
For the two-body scatterings, we assume an efficiency for dimuon identification of $100\%$ and that for ditau of $70\%$.
The  search of resonance for the radiative return process severely relies on the energy resolution of photon or equivalently dilepton.
The energy resolution for photons, which is detailed in Ref.~\cite{Chatrchyan:2008aa}, has been taken from the current CMS detector with ${\rm PbWO}^{}_{4}$ crystals, while for dimuon we take $\Delta m^{}_{\mu^+\mu^-} \simeq 5\times 10^{-5}~{\rm GeV^{-1}}\cdot s$~\cite{Freitas:2004hq}.
Moreover, a systematic uncertainty of $0.1\%$ level has been assumed.

The projected sensitivity is presented in Figure~\ref{fig:contour}. The limits using $\mu^+  \mu^- \to \ell^+ \ell^-$ (dashed and dotted curves for $\ell=\mu$ and $\tau$, respectively) are given as the darker orange region, while the radiative return process yields the lighter orange region.
Other limits and projections are also shown for comparison, such as $e^{+}   e^{-} \to \mu^{+}  \mu^{-}   Z',~Z' \to \mu^{+}   \mu^{-} $ from the BaBar experiment \cite{TheBABAR:2016rlg}, the LHC searches~\cite{Ma:2001md,Heeck:2011wj}, and 
the trident production in neutrino scattering experiments~\cite{Altmannshofer:2014pba}.

The Gauge \texorpdfstring{$L_\mu-L_\tau$}{lmmt} model can explain the muon $g$-2 anomaly in a particular region of the parameter space, outlined by the yellow band in the figure. In the parameter space of our concern, with $M^{}_{Z'} > 100~{\rm GeV}$, the anomaly-favoured region will be completely covered by the 3~TeV MuC.

\subsubsection*{Heavy Higgses through the Radiative Return Process}

The relatively sizeable muon Yukawa coupling to the SM Higgs boson suggests that the muon coupling to new physics associated with the breaking of the EW symmetry could be directly observable at the MuC, or even be responsible for the discovery of new physics. This is illustrated below in an extended Higgs sector scenario featuring a second Higgs doublet.

As discussed in Section~\ref{sec:Higgs}, the muon collider has much better perspectives than the HL-LHC to observe new heavy Higgs bosons already at the 3~TeV stage. It would also enable a very detailed characterisation of the newly discovered extended scalar sector by a number of precision measurements, including line shape studies of the new resonances produced in the $s$-channel~\cite{Eichten:2013ckl,Alexahin:2013ojp}, once their mass will be known with sufficient precision. If the mass is unknown, one can exploit the ``radiative return'' (RR) process. Namely
\begin{equation}
\mu^{+} \mu^{-} \rightarrow  \gamma H, \gamma A ,
\label{eq:rr}
\end{equation}
where we indicate with $H$ ($A$) the neutral CP-even (CP-odd) new heavy scalar states. The process proceeds through the emission of a photon from the initial state muons, which enables them to collide at the heavy Higgs boson mass even if the MuC centre of mass energy was slightly above. In what follows we illustrate the main points of this approach in the context of a 2-Higgs-doublet model (2HDM), summarising the findings of~\cite{Chakrabarty:2014pja}.

\begin{figure*}[t]
\centering
\includegraphics[scale=0.35]{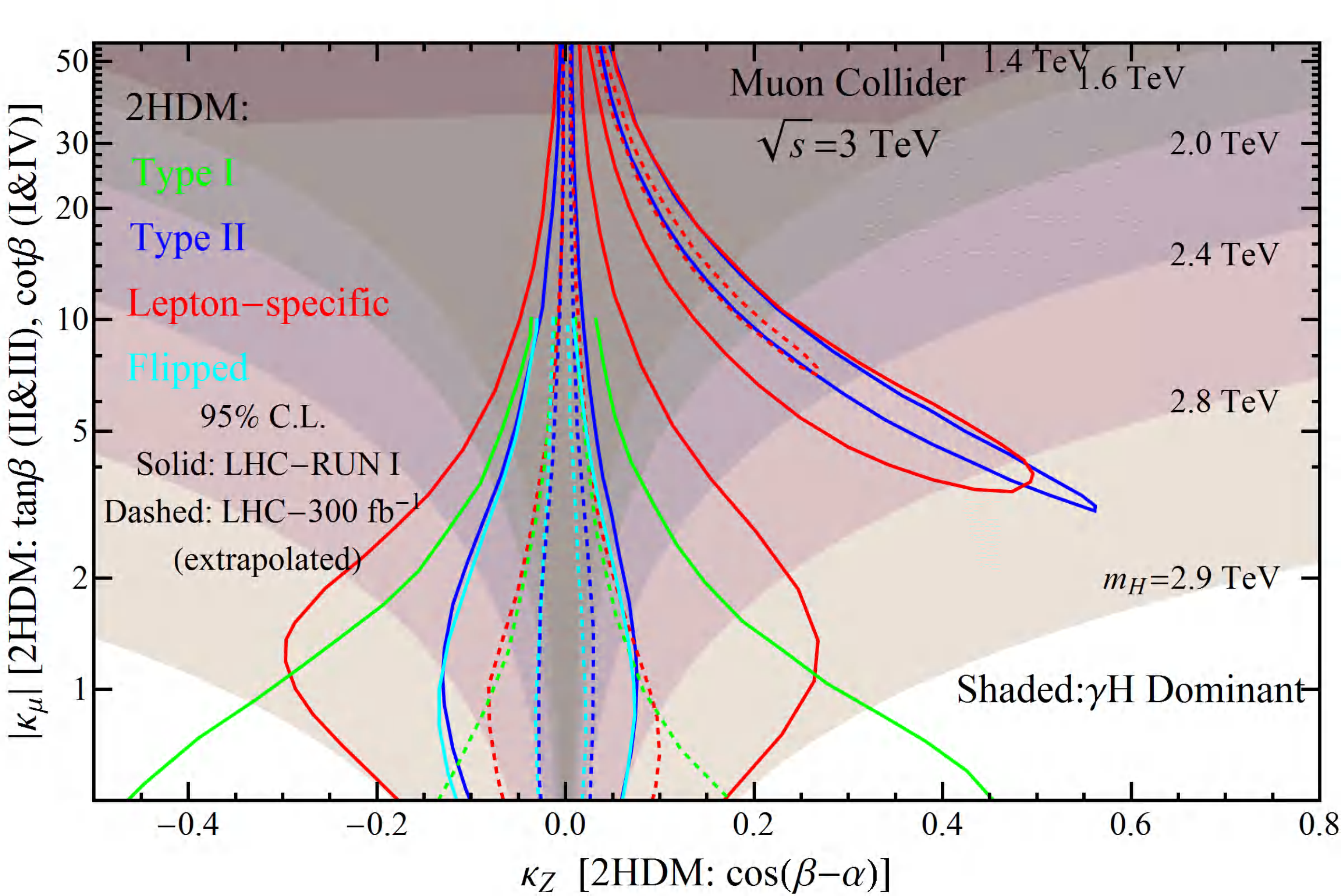}
\caption[5pt]{Comparison of sensitivities between different production mechanisms in the parameter plane $\kappa_\mu$\nobreakdash-$\kappa_Z$ for different masses of the heavy Higgs boson at the $3~{\textrm{TeV}}$ $\mc$. The shaded regions show a higher direct signal rate from the RR process than the $ZH$ associated production and $HA$ pair production channels. 
One can also see the allowed parameter regions (extracted from Ref.~\cite{Barger:2013ofa}). From~\cite{Chakrabarty:2014pja}.}
\label{fig:kappas}
\end{figure*}

In this model, the relevant heavy Higgs boson couplings can be parametrised as
\begin{eqnarray}
\label{eq:int}
\mathcal{L}_{int} = && -\kappa_\mu \frac {m_\mu} {v} H \bar \mu \mu + i \kappa_\mu \frac {m_\mu} {v} A \bar \mu \gamma_5 \mu 
\nonumber \\
    && 
    + \kappa_Z \frac {m_Z^2} {v} H Z^\mu Z_\mu \\ \nonumber
    && 
    +\frac {g \sqrt{ (1-\kappa_Z^2)}} {2\cos\theta_W} (H\partial^\mu A-A\partial^\mu H) Z_\mu\,,
\end{eqnarray}
where the two parameters $\kappa_\mu$ and $\kappa_Z$ characterise the coupling strength relative to the SM Higgs boson couplings. The coupling $\kappa_\mu$ controls the heavy Higgs resonant production and the radiative return cross sections while $\kappa_Z$ controls the cross sections for $ZH$ associated production and heavy Higgs pair $HA$ production.
We choose for simplicity equal coupling to muons of the CP-even $H$ and the CP-odd $A$. For the $HAZ$ coupling, the generic 2HDM relation is used where $\kappa_Z$ is proportional to $\cos(\beta-\alpha)$ and the $HAZ$ coupling is proportional to 
$\sin(\beta-\alpha)$. In the decoupling limit of the 2HDM at large $m_A$, $\kappa_{Z}\equiv \cos(\beta-\alpha)\sim m_{Z}^{2}/m_{A}^{2}$ is highly suppressed and $\kappa_\mu \approx \tan\beta\ (-\cot\beta)$ in Type-II and lepton-specific (Type-I and flipped) 2HDM. The value of the parameters in different 2HDM setups is shown in Table~\ref{tab:parameters}. 

 The signature of the RR process is quite striking as it results in  a monochromatic photon. For narrow scalars, the ``recoil mass'' is a sharp resonant peak at $m_{H/A}$, standing out of the continuous SM background. The reconstruction of the heavy Higgs boson from its decay product could provide an extra handle and a clean way to determine the heavy scalar branching ratios in a later stage after discovery. 

The characteristic of the RR signal is a photon energy
\begin{equation}
E_\gamma= \frac {\hat s - m_{H/A}^2} {2\sqrt{\hat s}},
\end{equation}
from which a recoil mass peaked at the heavy Higgs mass $m_{H/A}$ can be reconstructed. The energy of this photon is broadened by both detector effects, e.g.,  photon energy resolution, beam energy spread and physics effects, e.g., additional (soft) ISR/FSR, and the heavy Higgs width. 
The beam energy spread and additional soft ISR/FSR are expected to be at the GeV level~\cite{Delahaye:2013jla}. The recoil mass reconstruction uncertainty is dominated by the photon energy resolution, at least if the Higgs boson mass is significantly below the collider centre of mass energy. The heavy Higgs width, if sizeable, could effectively smear the energy of the photon. In the 2HDM it ranges between order 1 and 100~GeV for TeV-sized heavy Higgs masses. 

The inclusive cross section for the mono-photon background is substantial compared to the radiative return signal. The background can be estimated  from the M\"oller scattering with initial or final state photon emission, $\mu^+\mu^- \to \mu^+\mu^- \gamma$, and from the $W$ $t$-channel exchange with initial photon radiation, $\mu^+\mu^- \to \nu\nu \gamma$. 
The signal to background ratio is typically of order $10^{-3}$ for a 3 TeV $\mc$. Consequently, to discover through RR, we rely on exclusive processes and specific final states.

For concrete illustration, a Type-II 2HDM has been adopted with the $b\bar b$ final state with $80\%$ decaying branching ratio. An $80\%$ $b$-tagging efficiency is assumed and at least one $b$-jet tagged required in the analysis. Madgraph5~\cite{Alwall:2011uj} has been used for parton level signal and background simulations and then Pythia~\cite{Sjostrand:2006za} for initial and final state photon radiation. Detector smearing and beam energy spread have been also implemented.

Figure~\ref{fig:sigbkg} shows the recoil mass distribution at a 3 TeV $\mc$. Both the signal and the background cross sections at fixed beam energy increase as the recoil mass increase from the photon emission. One can clearly distinguish the pronounced mass peaks.

In order to assess the discovery perspectives of the RR process, in Figure~\ref{fig:kappas} we compare its reach with the one from the $ZH$ associated production and $HA$ pair production, in the $\kappa_\mu$\nobreakdash-$\kappa_Z$ plane. The RR production mode covers a large region of the plane, which expands when the heavy Higgs mass gets close to the MuC energy of 3~TeV. The region where RR dominates the discovery obviously shrinks when the heavy Higgs mass crosses the threshold for pair production. This is shown by the darker contour in the figure, for $1.4$~TeV mass. The figure also displays the LHC constraints from the measurements of the coupling of the SM Higgs. A more detailed discussions can be found in Ref.~\cite{Chakrabarty:2014pja}.

\clearpage

\section{Outlook}\label{ConclSect}

In this Review we summarised the motivations, status and plans of the ongoing multi-disciplinary effort towards a muon collider.

These studies are relevant and timely for several reasons. The outcome of the first part of the LHC experimental programme suggests that an ambitious jump ahead in energy will be needed for a fruitful exploration of fundamental interaction physics. Also the extension to higher energies of established $ee$ and $pp$ collider concepts faces severe feasibility challenges in terms of size, cost and power consumption. We saw in Sections~\ref{ch2_phys_opp} and~\ref{sec:phys_studies} that a muon collider with 10~TeV energy or more in the centre of mass offers tremendous and varied exploration opportunities. The muon collider at 10~TeV centre of mass energy offers equivalent or superior physics reach to a $pp$ collider with the highest envisageable centre of mass energy, 100~TeV. The energy limit for a muon collider is not yet known, but it is expected to be above 10~TeV.

The second reason for the renewed interest in muon colliders stems from recent advances in technology and muon collider design. These include the outcome of the MAP studies, which demonstrated the feasibility of many critical components of the facility, as well as several proof-of-principle experiments and component tests like MICE and the MUCOOL RF programme. Previously-considered limits for cooling such as operation of RF cavities in very high magnetic fields and the feasibility of a 30~T solenoid for the final cooling are now demonstrated. Several other advances are detailed in Section~\ref{sec_fac}. Muon colliders are now included in the European Roadmap for Accelerator R\&{D}. No showstopper has been identified and an R\&{D} and demonstration plan has been defined to address the remaining challenges in the next few years. The first assessment of the experimental conditions of a muon collider also contributed to enhance the global confidence in the project. It demonstrated the possibility of running a comprehensive experimental programme coping with the BIB from muon decay and identified wide margins for progress as discussed in Section~\ref{sec:detectorandreconstruction}.

The technically limited timeline for the muon collider R\&{D} programme has been described in Section~\ref{sec_fac} and it is displayed in Figure~\ref{fig_muon:RDtimeline}. The next Update of the European Strategy for Particle Physics (ESPPU) in 2026/2027 is an important milestone. By then, the design of the facility and of the cooling demonstrator will be consolidated. If the necessary investments will be supported by a favourable ESPPU recommendation, the preparation of a Conceptual Design Report and the construction and operation of the demonstrator and hardware prototypes will then initiate. The muon collider is a long-term project, which offers immediate opportunities for R\&D.



The muon collider technical feasibility is only one of the key aspects of the project that must be studied intensively in the next few years. The muon collider will be the first facility to collide leptons at such high energies, the first facility to collide second-generation leptons, and the first facility to collide particles that are unstable. These innovations entail novel opportunities and challenges for the exploitation of the muon collider facility, once built. Advances are thus necessary also in experimental and theoretical  physics and phenomenology in order to fully assess and consolidate the physics potential of the muon collider project. 

We saw in Section~\ref{sec:detectorandreconstruction} that the design of experiments at the muon collider will require an extensive investigation and development of detector technologies and of innovative reconstruction algorithms tailored to the suppression of the BIB from the decaying muons. 

We outlined in Sections~\ref{ch2_phys_opp} and~\ref{sec:phys_studies} that the current assessment of the potential of the muon collider for the exploration of fundamental interactions is incomplete. Preliminary sensitivity estimates have not been performed for many promising new physics search channels. It is likely that many novel physics channels have not been identified. For example, no comprehensive assessment is available of the muon collider perspectives to probe structured new physics scenarios such as for instance Supersymmetry, Composite Higgs, or extended Higgs sectors. Work is also needed towards a global analysis of the model-independent perspectives to probe new physics indirectly through the SM EFT.

Theoretical predictions at the muon collider will require major methodological advances, arguably comparable to those that were needed and achieved for the exploitation of the LHC. The principal challenges and opportunities stem from the copious emission of EW radiation that requires resummation for sufficiently accurate predictions, calls for the development of novel EW showering Monte Carlo codes and also challenges fixed-order calculations. While there are good perspectives for progress, building upon the vast experience with QED and QCD radiation, we remarked in Section~\ref{EWRadiation} a number of unique theoretical aspects of the challenges surrounding the treatment of the EW radiation. Novel theoretical ideas are thus needed, on top of the adaptation and implementation of existing methodologies. 

Increasingly complete and accurate theoretical predictions will have to be progressively integrated with detector design and simulation advances, towards a fully realistic assessment of the muon collider physics potential. Among the most urgent questions at the interface between theory and experiment one could mention for instance the detectability of very boosted SM particles emerging either from heavy resonance decays or from high energy scattering processes, the observability of unconventional new physics signatures and the feasibility of accurate per mil level cross-section measurements needed for Higgs coupling studies. Investigating these and other questions, related for example to muon-specific exploration opportunities, will provide theory targets to guide the design of the muon collider experiments and, conversely, drive the development and assess the adequacy of the theoretical predictions.

On top of ensuring a robust assessment of the muon collider project perspectives, the simultaneous advancement on accelerator, experimental and theoretical physics is needed also to exploit and develop inter-disciplinary synergies. The optimisation of the Machine Detector Interface described in Section~\ref{sec:environment} is only one of the tasks where an inter-disciplinary approach is mandatory. Others include the design of a possible forward detector for muons and of a very forward detector to exploit the collimated beam of energetic neutrinos from the decay of the colliding muons as described in Section~\ref{sec:fwdandlumi}. Inter-disciplinary work will be also needed in order to define the energy staging plan ensuring a good balance between physics case, technical risk and cost. The current baseline plan appears adequate so far, but it will need to be reconsidered as the design and the physics studies advance. Designing the cooling demonstrator such as to maximise the synergies with other projects, exploiting for example the opportunities for neutrino physics, offers additional inter-disciplinary opportunities.

Muon collider physics is still in its infancy. The studies presented in this Review represent the first exploration of the topic, but they enable us to identify the relevant questions and directions for rapid progress in the next few years. The muon collider programme offers appealing perspectives for ambitious innovative research that will advance particle collider physics as a whole.

\clearpage

\begin{acknowledgements}
This work was triggered by the Snowmass 2021 Community Planning Exercise~\cite{snowmass}. It is based on---and in some cases significantly extends---the Snowmass white papers~\cite{Aime:2022flm,MuonCollider:2022xlm,MuonCollider:2022ded,MuonCollider:2022glg,MuonCollider:2022nsa,Bottura:2022qqf} that have been prepared under the coordination of the IMCC.

This work was supported by the EU HORIZON Research and Innovation Actions under the grant agreement number 101094300.
Funded by the European Union (EU). Views and opinions expressed are however those
of the author(s) only and do not necessarily reflect those of the EU or European Research Executive Agency (REA). Neither the EU nor the REA can be held responsible for them.
The work has been supported by the Fermi Research Alliance, LLC under Contract No. DE-AC02-07CH11359 with the U.S. Department of Energy, Office of Science, Office of High Energy Physics.
This work is supported by the Atracci\'on de Talento Grant n. 2022-T1/TIC-24176 of the Comunidad Aut\'onoma de Madrid, Spain.
G.~Stark is supported by the Department of Energy Office of Science grant DE-SC0010107.
The work of R.~Dermisek was supported in part by the U.S. Department of Energy under Award No. {DE}-SC0010120.
This work is supported by the Deutsche Forschungsgemeinschaft under Germany’s Excellence Strategy EXC 2121 ``Quantum Universe'' - 390833306, as well as by the grant 491245950.
Contribution by T.~Holmes and her group are supported by U.S. Department of Energy, Office of Science, Office of Basic Energy Sciences Energy Frontier Research Centers program under Award Number DE-SC0023122.
This manuscript has been authored by employees of Brookhaven Science Associates, LLC under Contract No. DE-SC0012704 with the U.S. Department of Energy.
M.~Gallinaro and G.~Da~Molin acknowledge the support from the Funda\c{c}\~ao para a Ci\^encia e a Tecnologia (FCT), Portugal.
J.~Zurita is supported by the {\it Generalitat Valenciana} (Spain) through the 
{\it plan GenT} (CIDEGENT/2019/068) program, by the Spanish Government 
(Agencia Estatal de
Investigaci{\'o}n) and ERDF funds from European Commission 
(MCIN/AEI/10.13039/501100011033, Grant No. PID2020-\-114473GB-I00).
J.~Reuter acknowledges the support by the Deutsche
Forschungsgemeinschaft (DFG, German Research Association) under
Germany's Excellence Strategy-EXC 2121 "Quantum Universe"-3908333. 
The work of L.~Reina has been supported by the U.S. Department of Energy under grant DE-SC0010102.
This work was supported by the EU Horizon 2020 Research and Innovation Programmes: AIDAinnova under Grant Agreement No 101004761, I.FAST under Grant Agreement No 101004730, and the Marie Sklodowska-Curie grant agreement number 101006726.
The work of N.~Kumar is supported by Department of Science and Technology, Government of India under the SRG grant, Grant Agreement Number SRG/2022/000363.
We acknowledge financial support for this research from the United Kingdom Science and Technology Facilities Council via the John Adams Institute, University of Oxford.
R.~Ruiz acknowledges the support of Narodowe Centrum Nauki under Grant No. 2019/34/E /ST2/00186. R.Ruiz also acknowledges the support of the Polska Akademia Nauk (grant agreement PAN.BFD.S.BDN. 613. 022. 2021 - PASIFIC 1, POPSICLE).
S.~Trifinopoulos is supported by the Swiss National Science Foundation - project n. P500PT 203156 and by the Center of Theoretical Physics at MIT (MIT-
CTP/5538).
W.~Su is supported by the Junior Foundation of Sun Yat-sen University and Shenzhen Science and Technology Program (Grant No. 202206193000001, 20220816094256002).
W.~Kilian was supported by the Deutsche Forschungsgemeinschaft
(DFG, German Research Foundation) under grant  396021762 - TRR 257.
The work of Y.~Kahn was supported in part by U.S. Department of Energy grant DE-SC0015655.
The work of G.~Chachamis was supported by the Funda\c{c}{\~ a}o para a Ci{\^ e}ncia e a Tecnologia (Portugal) under project CERN/FIS-PAR\slash 0024\slash 2019 and contract `Investigador FCT - Individual Call\slash 03216\slash 2017'.
The work of J.~de Blas has been supported by the FEDER/Junta de Andaluc\'ia project grant P18-FRJ-3735.
This work is Supported in part by the NSF under Grant No. PHY-2210361 and by the Maryland Center for Fundamental Physics (MCFP).
The research activities of K.~R.~Dienes are supported in part by the U.S. Department of Energy
under Grant DE-FG02-13ER41976 DE-SC0009913, and also by the U.S. National Science Foundation through its employee IR/D program.
The research activities of B.~Thomas are supported in part by the U.S. National Science Foundation under Grant PHY-2014104.
\end{acknowledgements}

\bibliographystyle{report}       
\bibliography{references}

\end{document}